%% file: Direct-Tracking-TSP-2026-V12.tex
\documentclass[journal]{IEEEtran}
\usepackage{cite}

\ifCLASSINFOpdf
   \usepackage[pdftex]{graphicx}
\else
   \usepackage[dvips]{graphicx}
  \makeatletter
  \define@key{Gin}{interpolate}[true]{}
  \makeatother
\fi
\usepackage{amsmath}
\usepackage{algorithmic}

\usepackage{array}

\ifCLASSOPTIONcompsoc
 \usepackage[caption=false,font=normalsize,labelfont=sf,textfont=sf]{subfig}
\else
 \usepackage[caption=false,font=footnotesize]{subfig}
\fi

\usepackage{stfloats}
\usepackage{url}

\usepackage[utf8]{inputenc}
\usepackage[T1]{fontenc}
\usepackage{pgf}
\usepackage{xcolor}
\usepackage{physics}
\usepackage{notation}
\usepackage{multirow}
\usepackage{amssymb}
\usepackage{acronym}
\usepackage{psfrag}
\usepackage{soul}
\usepackage{import}
\usepackage[inline]{enumitem}
\usepackage[level=3]{LatexInclusion/messages}
\usepackage{auto-pst-pdf}
\usepackage{tikz}

\acrodef{bp}[BP]{belief propagation}
\acrodef{slam}[SLAM]{simultaneous localization and mapping}
\acrodef{2d}[2-D]{two-dimensional}
\acrodef{po}[PO]{potential object}
\acrodef{mmse}[MMSE]{minimum mean square error}
\acrodef{pdf}[PDF]{probability density function}
\acrodef{gospa}[GOSPA]{generalized optimal sub-pattern assignment}
\acrodef{roi}[ROI]{region of interest}
\acrodef{sbl}[SBL]{sparse Bayesian learning}
\acrodef{mp}[MP]{matching pursuit}
\acrodef{vla}[VLA]{vertical linear array}
\acrodef{tbd}[TkBD]{track-before-detect}
\acrodef{dtt}[DTT]{detect-then-track}
\acrodef{sot}[SOT]{\textcolor{black}{single-object tracking}}
\acrodef{mot}[MOT]{multiobject tracking}
\acrodef{stft}[STFT]{short-time Fourier transform}
\acrodef{dft}[DFT]{discrete Fourier transform}
\acrodef{snr}[SNR]{signal-to-noise ratio}
\acrodef{aoa}[AOA]{angle-of-arrival}
\acrodef{lfmcw}[LFMCW]{\textcolor{black}{linear frequency-modulated continuous-wave}}
\acrodef{rfs}[RFS]{random finite set}
\acrodef{hpmht}[H-PMHT]{histogram probabilistic multihypothesis tracking}

\newcommand{\rd}{\textcolor{black}}

\newcommand{\ist}{\hspace*{.3mm}}
\newcommand{\rmv}{\hspace*{-.3mm}}

\newcommand{\nn}{\nonumber}
\newcommand{\T}{\mathrm{T}}
\newcommand{\CH}{\mathrm{H}}

\definecolor{myblue}{RGB}{79, 129, 189}
\definecolor{myorange}{RGB}{247, 150, 70}
\definecolor{olivegreen}{RGB}{107,142,35}

\DeclareMathAlphabet{\mathpzc}{OT1}{pzc}{m}{it}

\begin{document}

\setlength{\abovedisplayskip}{4.5pt plus 1pt minus 1pt}
\setlength{\belowdisplayskip}{4.5pt plus 1pt minus 1pt}
\setlength{\abovedisplayshortskip}{0pt plus 1pt}
\setlength{\belowdisplayshortskip}{4.5pt plus 1pt minus 1pt}
\setlength{\textfloatsep}{12pt plus 2pt minus 2pt}
\setlength{\dbltextfloatsep}{12pt plus 2pt minus 2pt}
\setlength{\floatsep}{8pt plus 2pt minus 2pt}
\setlength{\dblfloatsep}{8pt plus 2pt minus 2pt}
\setlength{\intextsep}{8pt plus 2pt minus 2pt}
\setlength{\abovecaptionskip}{4pt}
\setlength{\jot}{2pt}
\interdisplaylinepenalty=2500

\title{\rd{Sparse Signal Reconstruction\\ for Track-Before-Detect}}

\author{Mingchao~Liang,~\IEEEmembership{Student Member,~IEEE,}
        and~Florian~Meyer,~\IEEEmembership{Member,~IEEE,} \vspace{-4mm}
        
                \thanks{This work was supported by the National Science Foundation (NSF) under Awards No. 2146261 and 2506477 as well as the Qualcomm Innovation Fellowship No. 492866. }
\thanks{Mingchao~Liang is with the Department of Electrical and Computer Engineering, University of California San Diego, La Jolla, CA 92093, USA (e-mail: \texttt{m3liang@ucsd.edu}).}
\thanks{Florian~Meyer is with the Scripps Institution of Oceanography and the Department of Electrical and Computer Engineering, University of California San Diego, La Jolla, CA 92093, USA (e-mail: \texttt{flmeyer@ucsd.edu}).}
          }



\maketitle

\begin{abstract}
Accurately tracking an unknown and time-varying number of objects in complex environments is \textcolor{black}{a challenging problem in radar and sonar sensing}. \rd{The \ac{tbd} paradigm aims to avoid a detection stage by operating directly on sensor data to prevent information loss. \textcolor{black}{This paper introduces a novel \ac{tbd} method based on a superpositional signal model adopted from the sparse signal reconstruction literature.} The resulting \ac{tbd} measurement model accounts for sensor data correlations, amplitude \textcolor{black}{fluctuations}, and unknown measurement noise variance.} \textcolor{black}{By retaining relative phase information in complex-valued sensor data, the proposed measurement model supports phase-coherent processing directly within the \ac{tbd} method.} It also enables the accurate representation of the physics of the data-generating process in the statistical model used for \ac{tbd}\textcolor{black}{, including the complex acoustic wavefields used in matched-field processing}. \textcolor{black}{To accommodate measurements from multiple sensors or frequencies, we introduce a general multi-snapshot, multi-dictionary formulation of this model.} The proposed statistical model is suitable for a wide range of problems in active and passive radar, active and passive sonar, as well as integrated sensing and communication systems. Based on a factor graph representation of the proposed model, a scalable \ac{bp} method is developed to perform efficient Bayesian inference. Experimental results \textcolor{black}{obtained} with both synthetic \textcolor{black}{radar data} and real \textcolor{black}{acoustic} data \textcolor{black}{for matched-field processing} demonstrate that the proposed method outperforms state-of-the-art conventional \ac{mot} methods.
\end{abstract}

\begin{IEEEkeywords}
Multiobject tracking, track-before-detect, Bayesian estimation, belief propagation, factor graphs
\vspace{-0mm}
\end{IEEEkeywords}

%
\IEEEpeerreviewmaketitle

\acresetall

\section{Introduction} \label{sec:intro}

\textcolor{black}{Tracking an unknown and time-varying number of objects is a fundamental task in radar and sonar sensing.} Multiobject tracking (\acs{mot})\acused{mot} \cite{BarWilTia:B11,Mah:B07,Bla:J04,Wil:J15,SauCoaRab:17,VoSinDou:J05,VoMa:J06,VoVoCan:J07,GarWilGraSve:J18,SchBenRosKriGra:18,MeyKroWilLauHlaBraWin:J18,MeyWil:J21,OrtFit:J02,StrGraWal:J02,BoeDri:J04,PapKim:J15,DavGae:B18,LiaMey:J23,ItoGod:J20,LiaKroMey:J23,VoVoPhaSut:10,RisVoVoFar:J13,Mah:C18,RisRosKimWanWil:J20,KroWilMey:21,KimRisGuaRos:21,DavGar:J24,VoVoNguShim:J24} \textcolor{black}{is an active area of research devoted to this task.} In \ac{mot}, the time-varying states of multiple objects are estimated based on measurements from sensing technologies such as radar or sonar. However, accurately and reliably tracking objects in dynamic environments and in the presence of complex object interactions (e.g., occlusions and close proximity) remains challenging. In particular, it is typically unknown when and where objects appear and disappear. This necessitates robust mechanisms for object initialization and termination, also known as track management, which adds further complexity to the \ac{mot} task.

\subsection{State-of-the-Art MOT} \label{sec:intro_mot}

Conventional methods address the \ac{mot} task following a \ac{dtt} framework that operates in two stages. In the first stage, a nonlinear detector preprocesses the raw sensor data in order to reduce data flow and computational complexity. \textcolor{black}{In the second stage,} ``point measurements'' generated by the object detector \textcolor{black}{are used} as input for tracking \cite{BarWilTia:B11,Mah:B07,Bla:J04,Wil:J15,MeyBraWilHla:J17,SauCoaRab:17,MeyKroWilLauHlaBraWin:J18,GarWilGraSve:J18,SchBenRosKriGra:18,MeyWil:J21}. The extracted ``point measurements'' are prone to \textcolor{black}{measurement-origin uncertainty}. State-of-the-art \ac{dtt}-based methods solve the \ac{mot} problem in a Bayesian estimation framework, utilizing statistical models to describe object birth, object motion, and measurement generation. Based on the statistical models, data association can be performed to address measurement-origin uncertainty. The Bayesian inference engine sequentially estimates object states and dynamically introduces new object states in the state space. Prominent examples include joint probabilistic data association filters \cite{BarWilTia:B11}, multi-hypothesis trackers \cite{Bla:J04}, \textcolor{black}{\ac{rfs}}-based filters \cite{Wil:J15,SauCoaRab:17,GarWilGraSve:J18,SchBenRosKriGra:18}, and \ac{bp}-based \ac{mot} methods \cite{MeyBraWilHla:J17,MeyKroWilLauHlaBraWin:J18,MeyWil:J21}. \rd{Popular RFS-based methods include the probability hypothesis density (PHD) filter 
\cite{Mah:J03,Mah:B07,VoSinDou:J05, VoMa:J06}, the cardinalized PHD (CPHD) filter \cite{Mah:B07,VoVoCan:J07}, 
and \textcolor{black}{multi-Bernoulli} filters \cite{VoVoCan:J09,VoVoPhaSut:10, RisVoVoFar:J13, Wil:J15, SauCoaRab:17, SchBenRosKriGra:18, GarWilGraSve:J18}. RFS-based methods  avoid a specific ordering of objects and measurements and explicitly model the fact that the number of objects is unknown.}

 Furthermore, neural networks have recently been integrated to refine the statistical models employed in \ac{mot} \cite{LiaMey:J23,WeiLiaMey:24,WeiLiaMey:J26}. However, a significant limitation of \ac{dtt}-based \ac{mot} is its potential for suboptimal performance, stemming from the irreversible loss of valuable information during the initial detection stage.

To address the limitations of \ac{dtt}, the \ac{tbd} approach operates directly on received signals that have been preprocessed by a linear mapping. The linear mapping typically transforms the received signals into data cells, e.g., into angle and distance cells, which are then used as measurements for \ac{tbd}. The potential advantage of the \textcolor{black}{\ac{tbd}} strategy is that it can significantly reduce or completely avoid information loss related to irreversible object detection. \rd{\textcolor{black}{In addition, the use of a Kalman filter can limit the performance of \ac{dtt} at low \acp{snr} \cite{Dav:J13}.}} Sequential Bayesian estimation techniques for \ac{tbd} are of particular interest \textcolor{black}{for online processing}. Early \ac{tbd} research focused on the simpler \ac{sot} problem, where the task is to determine the existence and state of at most one object \cite{RisVoVoFar:J13,SalBir:01,RabRicLep:12,DavRutChe:J12,KimUneMul:J20,UneHorMulMas:J23}. Building on this, subsequent work has extended the \ac{tbd} paradigm to the more general and challenging \ac{mot} problem, which is the focus of this paper.

For \ac{mot} scenarios, various \ac{tbd} approaches have been developed. The resulting methods represent object states using either random vectors \cite{OrtFit:J02,StrGraWal:J02,BoeDri:J04,DavGae:B18,ItoGod:J20,LiLeiVenTuf:J22,LiaKroMey:J23} or \textcolor{black}{\acp{rfs}} \cite{VoVoPhaSut:10,RisVoVoFar:J13,PapKim:J15,Mah:C18,RisRosKimWanWil:J20,KroWilMey:21,KimRisGuaRos:21,DavGar:J24}. \textcolor{black}{Despite the use of similar Bayesian frameworks,} a fundamental distinction from \ac{dtt} lies in the statistical measurement model: whereas \ac{dtt} relates ``point measurements'' from detectors to object states, \ac{tbd} establishes a statistical relationship between the sensor data and the object states. However, accurately modeling this relationship for sensor data is challenging, \textcolor{black}{and therefore simplifying modeling assumptions are often required}. For instance, \ac{hpmht} \cite{StrGraWal:J02,DavGae:B18} treats the sensor data as if they are point measurements and requires heuristics for track management. Others, like \cite{VoVoPhaSut:10,KroWilMey:21}, assume that a measurement cannot be affected by more than one object. 

A more physically motivated direction characterizes the raw sensor data using a superpositional signal model \cite{ PapKim:J15,Mah:C18,RisRosKimWanWil:J20,ItoGod:J20,KimRisGuaRos:21,LiaKroMey:J23,DavGar:J24}, which represents signals as a summation of contributions from different objects and noise. Crucially, the measurement likelihood depends not only on object states but also on their amplitudes, which describe the strength of signal contributions. These amplitudes vary between objects and fluctuate over time, significantly complicating the computation of the measurement likelihood. To reduce complexity, existing methods often rely on assumptions that simplify the signal model. Some assume amplitudes are known \cite{ItoGod:J20,DavGar:J24}, while others \textcolor{black}{model amplitude fluctuations and exploit conditional independence across measurement cells in their likelihood formulations} \cite{RisRosKimWanWil:J20,KimRisGuaRos:21,LiaKroMey:J23}. \textcolor{black}{Fluctuating complex amplitudes shared across measurement components can produce correlations among those components even when conditioned on the kinematic, existence, signal power, and noise power states.} \rd{\textcolor{black}{Another class of methods} stacks all object states and propagates the high-dimensional joint object vector in time using particle-filtering techniques \cite{PapKim:J15}. This approach avoids the need to reconstruct the contributions of individual objects to the superpositional measurement, thereby facilitating the evaluation of the measurement likelihood. However, due to the curse of dimensionality in particle filtering  \cite{BenBicLi:B08}, this \ac{tbd} method is limited to scenarios with few objects \cite{PapKim:J15}.}

 \textcolor{black}{Assuming known amplitudes or conditional independence across measurement cells simplifies inference, but neglecting correlations induced by fluctuating amplitudes can introduce model mismatch.} \rd{\textcolor{black}{A} more general measurement model accounting for various amplitude fluctuations of complex-valued sensor data was introduced in \cite{LepRabGla:J16}\textcolor{black}{.}}\footnote{\rd{Note that single-Bernoulli filters for TkBD \cite{RisVoVoFar:J13} and the RFS-based TkBD method in \cite{PapKim:J15} could incorporate this more general measurement model in a straightforward manner. \textcolor{black}{However, single-Bernoulli filters are restricted to at most one object, whereas the joint-state particle filtering approach faces the curse of dimensionality as the number of objects increases.}}} \textcolor{black}{To address these limitations, our preliminary conference publication \cite{LiaMey:C24} introduced a \ac{bp} approach for \ac{mot} based on a correlated complex-valued superpositional measurement model.}
 
Recently, promising \textcolor{black}{advances in} sequential estimation methods that rely on superpositional signal modeling have emerged. In particular, \cite{LiaLeiMey:J25} proposes direct multipath-based \ac{slam}.  \ac{slam} is a problem that is closely related to \ac{mot}. \textcolor{black}{In} \cite{LiaLeiMey:C23,LiaLeiMey:J25}, a \ac{bp} method has been developed based on a superpositional model similar to the one proposed in \cite{LepRabGla:J16}. \textcolor{black}{Following our preliminary conference publication \cite{LiaMey:C24}, a related variational message passing method for \ac{mot} using raw radar signals was presented by Westerkam et al.~\cite{WesMoeLeiPed:C25}.} These methods demonstrate superior tracking performance compared with their \ac{dtt}-based counterparts, underscoring the substantial benefits achievable by leveraging more general signal models.

\subsection{Contributions and Paper Organization} \label{subsec:intro_contr}

\rd{In this paper, we propose a novel Bayesian method for \ac{mot} that directly operates on sensor data, performing sequential Bayesian estimation \textcolor{black}{using a likelihood of the sensor data rather than detector-generated point measurements}. Our proposed method adopts \textcolor{black}{a superpositional signal model} from the sparse signal reconstruction literature \cite{Tip:J01,WipRao:J07,GerMecChrXenNan:J16,ParMeyGer:J21,SanLeiRao:J22} for \ac{tbd}. In contrast to existing \ac{tbd} methods, our approach \textcolor{black}{accounts for} correlations among sensor measurements, time-varying signal amplitudes, and unknown measurement noise \textcolor{black}{variances}.} \rd{\textcolor{black}{By retaining the complex covariance structure of sensor measurements, our model exploits relative phase information, which is essential in a range of sensing applications, including matched-field processing \cite{BagKupMik:J93}.} Computationally efficient Bayesian inference is performed based on a factor graph representation of the statistical model \cite{KscFreLoe:01,Del:T12, MeyKroWilLauHlaBraWin:J18}. In particular, we use BP \textcolor{black}{with Gaussian moment matching} to avoid \textcolor{black}{joint integration over all object states} and \textcolor{black}{a particle-based implementation} to efficiently update individual object states in low dimensions.} \textcolor{black}{Iterative sparse signal reconstruction in our method can be viewed as a counterpart to iterative data association in \ac{bp}-based \ac{dtt} \cite{MeyKroWilLauHlaBraWin:J18}, resolving uncertainty about the contributions of individual objects directly from superpositional sensor data.}

We further extend the signal model in \cite{LepRabGla:J16} to accommodate multiple \textcolor{black}{conditionally independent} \textcolor{black}{measurement vectors}, or \textcolor{black}{``snapshots,''} as well as multiple concurrent signal models, or \textcolor{black}{``dictionaries.''} This multi-dictionary formulation offers flexibility for multi-sensor fusion \cite{CarKilPotVan:07,ModWesVenLei:25}, multi-frequency beamforming \cite{BooAbaSchHod:J00, NanGemGerHodMec:J19}, and multi-waveform localization \cite{TemGriRit:J22}. \textcolor{black}{Each dictionary describes the sensor response as a function of an object's kinematic state.} The amplitudes are modeled as zero-mean \textcolor{black}{complex} Gaussian variables, with variances governed by the object's signal power state. \textcolor{black}{The signal power state consists of a binary existence variable shared across dictionaries and a vector of dictionary-specific continuous power parameters.} The resulting Bernoulli-Gaussian model is common in the sparse signal reconstruction literature \cite{KorMen:J82,BadHanThoFle:J17,HanFleRao:J18}, where it is similarly used to describe the individual components of superpositional signals.

\textcolor{black}{At each time step, the variance of the zero-mean additive noise for each dictionary is modeled as a random noise power state and inferred jointly with the objects' kinematic and signal power states.} 
In the Bayesian setting, the goal of our \ac{tbd} method is to compute the marginal posterior \acp{pdf} of object states. \textcolor{black}{On this factor graph \cite{KscFreLoe:01,Loe:04}, we develop a scalable \ac{bp} method that \textcolor{black}{approximates the desired marginal posterior densities}.} \ac{bp} operates by computing local ``messages'', which are \textcolor{black}{real-valued} functions of random variables, along the edges of the factor graph. \textcolor{black}{We approximate intractable measurement-update messages by moment-matched Gaussian expressions in the measurement vector and evaluate the required state integrals using weighted particles \cite{AruMasGorCla:02,IhlMca:09}.} 

The key contributions of this paper are as follows:
\begin{itemize}
    \item \textcolor{black}{We introduce a multi-snapshot, multi-dictionary \ac{tbd} model accounting for measurement correlations, fluctuating amplitudes, and unknown noise variances.}
    \item \textcolor{black}{We derive a factor graph and develop a scalable particle-based \ac{bp} implementation for estimating the existence and states of individual objects.}
    \item \textcolor{black}{We demonstrate superior tracking performance over state-of-the-art reference methods in radar simulations and matched-field processing with real acoustic data.}
\end{itemize}

This paper advances over the preliminary account of our method provided in the conference publication \cite{LiaMey:C24} by (i)~presenting a detailed discussion of the statistical model, (ii)~introducing a derivation of the corresponding factor graph and resulting \ac{bp} method\textcolor{black}{,} and (iii)~conducting extensive experiments in synthetic radar tracking scenarios as well as a challenging real-world acoustic tracking application.

The organization of this paper is as follows. Section \ref{sec:sig} presents the superpositional signal model and the statistical model. Section \ref{sec:bp} develops the proposed \ac{bp} method for \ac{tbd}. Section \ref{sec:exp} discusses the results of the \textcolor{black}{numerical study}. Finally, Section \ref{sec:conclusion} concludes the paper.


\section{Signal Model and Problem Formulation} \label{sec:sig}

In this section, we introduce the statistical model of the considered \ac{tbd} problem\vspace{-2mm}. 

\subsection{Measurement Model} \label{subsec:meas_model}

At each time step $k$, we assume that there are $L_k$ objects in the environment. \textcolor{black}{We consider $I$ dictionaries with a common measurement dimension $M$ and $J$ snapshots per dictionary.} The individual measurement vector $\V{z}_{k, j}^{(i)} = [z_{k, j, 1}^{(i)} \cdots z_{k, j, M}^{(i)}]^\T \in \mathbb{C}^M, i \in \{1, \dots, I\}, j \in \{1, \dots, J\}$ is then modeled as  



\begin{equation}
    \V{z}_{k, j}^{(i)} = \sum_{l = 1}^{L_k} \varrho_{k, l, j}^{(i)} \V{a}^{(i)}(\V{x}_{k, l}) + \V{\epsilon}_{k, j}^{(i)} \label{eq:signal_model}
\end{equation}
where \textcolor{black}{$\V{a}^{(i)}(\V{x}_{k, l}) \in \mathbb{C}^M$} is the known basis function for the $i$-th dictionary that encodes a forward model. The forward model maps the object's kinematic state $\V{x}_{k, l}$, e.g., position and velocity, \textcolor{black}{to its sensor response}. Multiple dictionaries may arise from distinct sensors \cite{CarKilPotVan:07,ModWesVenLei:25} or scenarios where objects transmit \textcolor{black}{signals at multiple frequencies} \cite{BooAbaSchHod:J00,NanGemGerHodMec:J19} or multiple waveforms \cite{TemGriRit:J22}. The complex amplitude $\varrho_{k, l, j}^{(i)} \in \mathbb{C}$ quantifies the contribution of object $l$ to snapshot $j$ in dictionary $i$, while $\V{\epsilon}_{k, j}^{(i)} \in \mathbb{C}^M$ is the noise vector. The number of objects $L_k$, complex amplitudes $\varrho_{k, l, j}^{(i)}$, and \textcolor{black}{kinematic states} $\V{x}_{k, l}$ are unknown.

\rd{For example, consider the localization of a source that transmits $I$ narrowband signals using an array of $M$ synchronized sensors. \textcolor{black}{Here, $\V{a}^{(i)}(\V{x}_{k, l})$ is the frequency-dependent steering vector of the array corresponding to the source position encoded in $\V{x}_{k, l}$. For each frequency $i$, the measurements $\V{z}_{k, j}^{(i)}$, $j \in \{1, \dots, J\}$, are $J$ noisy array snapshots.}}

\rd{Further concrete examples for superpositional measurement models that follow \eqref{eq:signal_model} as well as the corresponding dictionaries for \ac{lfmcw} radar \cite{LevMoz:B04,ZhuXuGuoWanZheQu:24} and a passive sonar scenario \cite{JenKupPorSchTol:B11,BagKupMik:J93} are presented in Sections \ref{subsec:exp_radar} and \ref{subsec:exp_s96_high}, respectively.}

The goal of \ac{mot} is to estimate the time-varying number of objects as well as their kinematic states. To facilitate this estimation, at each time step $k$, we employ a model that has $N_k$ \acp{po} as proposed in \cite{MeyKroWilLauHlaBraWin:J18}. Each \ac{po} $n \in \{1, \dots, N_k\}$ is characterized by \textcolor{black}{a kinematic state $\V{x}_{k, n}$} and a binary existence random variable $r_{k, n} \in \{0, 1\}$, where $r_{k, n} = 1$ indicates that the \ac{po} corresponds to a physical object. \textcolor{black}{The number of physical objects is $L_k = \sum_{n = 1}^{N_k} r_{k, n}$.} The signal model in \eqref{eq:signal_model} can thus be reformulated to account for all \acp{po}, i.e.,

\begin{equation}
    \V{z}_{k, j}^{(i)} = \sum_{n = 1}^{N_k} \ist r_{k, n} \rho_{k, n, j}^{(i)} \ist \V{a}_{k, n}^{(i)} + \V{\epsilon}_{k, j}^{(i)} \label{eq:signal_model_po}
\end{equation}
where $\V{a}_{k, n}^{(i)} \triangleq \V{a}^{(i)}(\V{x}_{k, n})$ \textcolor{black}{is the basis vector of \ac{po}} $n \in \{1, \dots, N_k\}$ at dictionary $i \in \{1, \dots, I\}$. \textcolor{black}{Conditioned on the signal power $\gamma_{k, n}^{(i)} > 0$, the complex amplitude $\rho_{k, n, j}^{(i)}$ is assumed zero-mean circularly symmetric complex Gaussian with a common variance across snapshots, i.e., $f(\rho_{k, n, j}^{(i)} | \gamma_{k, n}^{(i)}) = \mathcal{CN}(\rho_{k, n, j}^{(i)}; 0, \gamma_{k, n}^{(i)}), \forall j \in \{1, \dots, J\}$}, \textcolor{black}{so the magnitude of the complex amplitude follows a Rayleigh distribution} \cite{Sko:B00, LepRabGla:J16}. \textcolor{black}{Conditioned on all signal and noise powers, the amplitudes are assumed mutually independent across \acp{po}, dictionaries, and snapshots, and are independent of all $\V{x}_{k, n}$, $r_{k, n}$, and $\V{\epsilon}_{k, j}^{(i)}$.} Note that $r_{k, n} \gamma_{k, n}^{(i)}$ is the ``effective'' signal power for each \ac{po} $n$ on the $i$-th dictionary, where $r_{k, n} \in \{0, 1\}$ controls the \ac{po}'s existence and $\gamma_{k, n}^{(i)}$ scales the strength of its contribution. Hence, we define $\V{\phi}_{k, n} = [\V{\gamma}_{k, n}^\T \hspace{1mm} r_{k, n}]^\T$ as the signal power state of \ac{po} $n$ with $\V{\gamma}_{k, n} = [\gamma_{k, n}^{(1)} \cdots \gamma_{k, n}^{(I)}]^\T$. \textcolor{black}{Conditioned on the noise power $\eta_{k}^{(i)}$, the noise vector $\V{\epsilon}_{k, j}^{(i)}$ is modeled as zero-mean circularly symmetric complex Gaussian with a common power across snapshots, i.e., $f(\V{\epsilon}_{k, j}^{(i)} | \eta_{k}^{(i)}) = \mathcal{CN}(\V{\epsilon}_{k, j}^{(i)}; \V{0}, \eta_{k}^{(i)} \M{I}_M), \forall j \in \{1, \dots, J\}$.} \textcolor{black}{Conditioned on all signal and noise powers, the measurement noise vectors are mutually independent across dictionaries and snapshots, and are independent of all $\V{x}_{k, n}$, $r_{k, n}$, and $\rho_{k, n, j}^{(i)}$.} \textcolor{black}{The signal and noise powers, $\gamma_{k, n}^{(i)} \ge 0$ and $\eta_{k}^{(i)} > 0$, are unknown random variables. If the signal power is zero, the corresponding amplitude is zero.} 

\rd{The superpositional Bernoulli-Gaussian model in  \eqref{eq:signal_model_po} is adopted from the sparse signal reconstruction literature \cite{KorMen:J82,BadHanThoFle:J17,HanFleRao:J18}. \textcolor{black}{Compared with \ac{tbd} models that assume conditionally independent measurement components, this superpositional model accounts for correlations among sensor measurements, fluctuating signal amplitudes, and unknown measurement noise powers.} \textcolor{black}{A key difference from such models} is that even conditioned on the random state variables \textcolor{black}{$\V{x}_k$, $\V{\phi}_k$, and $\eta_{k}^{(i)}\rmv$}, the elements of the measurement vector $ \V{z}_{k, j}^{(i)}$ can be correlated. \textcolor{black}{In particular, conditioned on these states, the complex amplitudes $\rho_{k, n, j}^{(i)}$ and thus the $N_k$ contributions $r_{k, n} \rho_{k, n, j}^{(i)} \ist \V{a}_{k, n}^{(i)}$ in \eqref{eq:signal_model_po} can remain random. These contributions determine the correlation structure of $ \V{z}_{k, j}^{(i)}$.} While the \textcolor{black}{basis vectors} $ \V{a}_{k, n}^{(i)}$ can describe different sensor and propagation models, modeling $ \rho_{k, n, j}^{(i)}$ as random captures amplitude fluctuations.}
The likelihood function of the individual measurement vector $\V{z}_{k, j}^{(i)}$ is zero-mean complex Gaussian, i.e.,
\begin{align}
    f(\V{z}_{k, j}^{(i)} | \V{x}_k, \V{\phi}_{k}, \eta_{k}^{(i)}) = \mathcal{CN}(\V{z}_{k, j}^{(i)}; \V{0}, \M{C}_{k}^{(i)}) \nn
\end{align}
with \rd{\textcolor{black}{a complex, generally non-diagonal covariance matrix}} given by 
\begin{equation}
\label{eq:cov}
 \M{C}_{k}^{(i)} = \sum_{n = 1}^{N_k} r_{k, n} \gamma_{k, n}^{(i)}  \V{a}_{k, n}^{(i)} \V{a}_{k, n}^{(i) \CH} + \eta_{k}^{(i)} \M{I}_M.
 \end{equation}
  We have $\V{x}_k = [\V{x}_{k , 1}^\T \cdots \V{x}_{k, N_k}^\T ]^\T$ and $\V{\phi}_{k} = [\V{\phi}_{k, 1}^\T \cdots \V{\phi}_{k, N_k}^{\T}]^\T$. From the independence assumptions, the likelihood of the joint measurement vector can also be obtained as 
\begin{align}
    f(\V{z}_{k} | \V{x}_k, \V{\phi}_{k}, \V{\eta}_{k}) = \prod_{i = 1}^I \prod_{j = 1}^J f(\V{z}_{k, j}^{(i)} | \V{x}_k, \V{\phi}_{k}, \eta_{k}^{(i)}) \label{eq:likelihood_indep}
\end{align}
where we have $\V{z}_{k} = [\V{z}_{k}^{(1) \T}  \cdots \V{z}_{k}^{(I) \T} ]^\T$, $\V{z}_{k}^{(i)} = [\V{z}_{k, 1}^{(i) \T} \cdots \V{z}_{k, J}^{(i) \T} ]^\T$, and $\V{\eta}_{k} = [\eta_{k}^{(1)} \cdots \eta_{k}^{(I)}]^\T$.

\subsection{Remarks on the Measurement Model} \label{subsec:disc_meas_model}

 When there is a single snapshot and single dictionary, i.e., $J \rmv\rmv=\rmv\rmv 1$ and $I \rmv\rmv=\rmv\rmv 1$, our signal model in \eqref{eq:signal_model_po} is similar to the one in \cite{LiaKroMey:J23}. However, a key difference is that the model in \cite{LiaKroMey:J23} relies on the assumption that the elements within the measurement vectors\vspace{0mm}, i.e., $z_{k, j, m}^{(i)}, m \in \{1, \dots, M\}$\textcolor{black}{,} are independent conditioned on random states $\V{x}_k, \V{\phi}_{k},$ and $\eta_{k}^{(i)}$. \rd{This common assumption in \ac{tbd} \cite{BoeDri:J04,RisRosKimWanWil:J20,KimRisGuaRos:21}  \textcolor{black}{can introduce model mismatch when amplitudes $\varrho_{k, l, j}^{(i)}$ remain random conditioned on the states to be estimated and the basis functions $\V{a}^{(i)}(\V{x}_{k, l})$ distribute each object's contribution across multiple measurement components.} As discussed above, this combination \textcolor{black}{can result} in correlated entries in the measurement vector $\V{z}_{k}$ \textcolor{black}{and ignoring these correlations can degrade tracking performance} \cite{LepRabGla:J16}.} 

\rd{\textcolor{black}{Common strategies for simplifying the measurement likelihood based on the conditional independence assumption} \cite{BoeDri:J04,RisRosKimWanWil:J20,KimRisGuaRos:21} \textcolor{black}{include (a) modeling amplitudes $\varrho_{k, l, j}^{(i)}$ as nonrandom conditioned on the states to be estimated and (b) applying preprocessing such as matched filtering, beamforming, or similar steps} \cite[Sec.~4.2]{RisRosKimWanWil:J20}. \textcolor{black}{Strategy (a) can describe state-dependent amplitude variations but omits residual fluctuations conditioned on those states, as can occur with multipath propagation or angle-dependent target cross-sections. Strategy (b) often leaves residual correlations among measurements.}}

\rd{Existing  \ac{tbd} methods typically also only exploit phase information in the preprocessing step and are based on models that describe real-valued measurements \cite{BoeDri:J04,RisRosKimWanWil:J20,KimRisGuaRos:21}. For example, in the source localization problem discussed in Section \ref{subsec:meas_model}, a common \ac{tbd} approach \cite{NorNar:J19} following \textcolor{black}{strategy (b) is to} (i) perform conventional beamforming \cite{Van:B01} to transform the complex-valued array observations $\V{z}_{k, j}^{(i)} $ to \textcolor{black}{the discrete angle domain}, (ii) \textcolor{black}{take the magnitudes of the beamformer outputs}, and (iii) use the resulting real-valued ``beamformed data'' as measurements for \ac{tbd} that are assumed independent across angle bins conditioned on object states to be estimated. While the required independence assumption is much more appropriate after the beamforming step than before, it still leads to a certain degree of model mismatch due to unavoidable side lobes and spectral leakage effects \cite{Van:B01}.}

 In contrast, our model does account for correlations across measurements and leverages phase information, thereby reducing reliance on preprocessing steps. \rd{In the discussed source localization problem, for example, this makes it possible to perform beamforming implicitly within the proposed \ac{tbd} method presented in Sec.~\ref{sec:bp}. The fundamental principle of our method is to infer the combination of \textcolor{black}{kinematic states} $\V{x}_k$, \textcolor{black}{signal power states} $\V{\phi}_{k}$, and \textcolor{black}{noise power states} $\eta_{k}^{(i)}$, for which the resulting complex covariance matrices $\M{C}_{k}^{(i)}$ in \eqref{eq:cov} \textcolor{black}{best reconstruct the coherence patterns within the measurement vectors $\V{z}_{k, j}^{(i)}$, jointly across all $I$ dictionaries}.}

The multi-dictionary measurement model has been successfully applied \rd{to traditional signal reconstruction tasks} such as image reconstruction \cite{JiDunCar:J09}, multimodal classification \cite{FedRaoNgu:17}, underwater source localization \cite{GemNanGer:J19}, and \ac{aoa} estimation \cite{NanGemGerHodMec:J19}. However, the aforementioned existing methods focus on static scenarios, neglecting state transitions inherent to dynamic tracking problems. Furthermore, existing methods typically model the complex amplitude $\varrho_{k, l, j}^{(i)}$ using Gaussian scale mixture priors, \textcolor{black}{whereas we adopt} a Bernoulli-Gaussian model for signal amplitude. The Bernoulli-Gaussian model \cite{KorMen:J82,BadHanThoFle:J17,HanFleRao:J18} can describe objects that dynamically appear and disappear over time, together with time-varying signal strengths.

\subsection{State-Transition Models} \label{subsec:state_tran_model}

In addition to the measurement model, we also establish state-transition models to describe the dynamics of the kinematic states and the signal power states of \acp{po} as well as noise power states. A first-order Markov model is considered, i.e., we assume that conditioned on $\V{x}_{k - 1}, \V{\phi}_{k - 1}, \V{\eta}_{k - 1}$, the states $\underline{\V{x}}{}_{k}, \underline{\V{\phi}}{}_{k}, \V{\eta}_{k}$ are independent of all previous states $\V{x}_{k'}, \V{\phi}_{k'}, \V{\eta}_{k'}, k' < k - 1$ with $\underline{\V{x}}{}_k = [\V{x}_{k, 1}^\T \cdots \V{x}_{k, N_{k - 1}}^\T]^\T$ and $\underline{\V{\phi}}{}_k = [\V{\phi}_{k, 1}^\T \cdots \V{\phi}_{k, N_{k - 1}}^\T]^\T$. The joint state-transition process is then governed by the conditional \ac{pdf} $f(\underline{\V{x}}{}_{k}, \underline{\V{\phi}}{}_{k}, \V{\eta}_{k} | \V{x}_{k - 1}, \V{\phi}_{k - 1}, \V{\eta}_{k - 1})$. Furthermore, we assume that the joint Markov process consists of multiple independent Markov processes. Specifically, $\V{x}_{k, n}, \V{\phi}_{k, n}, \eta_{k}^{(i)}, n \in \{1, \dots, N_{k - 1}\}, i \in \{1, \dots, I\}$ are assumed to evolve independently, i.e., the joint state-transition \ac{pdf} can be written as the product of individual state-transition \acp{pdf} as follows
\begin{align}
    f(\underline{\V{x}}{}_{k}, \underline{\V{\phi}}{}_{k}, \V{\eta}_{k} | \V{x}_{k - 1}, \V{\phi}_{k - 1}, \V{\eta}_{k - 1}) \nn \\
    &\hspace{-45.5mm}= \textcolor{black}{\Big( \prod_{n = 1}^{N_{k - 1}} f(\V{x}_{k, n} | \V{x}_{k - 1, n}) f(\V{\phi}_{k, n} | \V{\phi}_{k - 1, n}) \Big) \prod_{i = 1}^I f(\eta_k^{(i)} | \eta_{k - 1}^{(i)})}. \nn 
\end{align}
Here, the kinematic state evolution of each \ac{po} $n \in \{1, \dots, N_{k - 1}\}$ is described by  the conditional \ac{pdf} $f(\V{x}_{k, n}$ $| \V{x}_{k - 1, n})$. A common choice for this conditional \ac{pdf} relies on a constant velocity model \cite[Ch. 4]{ShaKirLi:B02}. The conditional \ac{pdf}  $f(\eta_k^{(i)} | \eta_{k - 1}^{(i)})$ describes the state-transition of the noise power states. A suitable choice for this conditional \ac{pdf} is the Gamma PDF \cite{GraOrg:J14}. \textcolor{black}{The state transitions} of \ac{po} existence and signal power are also assumed independent, i.e., $f(\V{\phi}_{k, n} | \V{\phi}_{k - 1, n}) = p(r_{k, n} | r_{k - 1, n})  f(\V{\gamma}_{k, n} | \V{\gamma}_{k - 1, n})$.   For the existence of \ac{po} $n \in \{1, \dots, N_{k - 1}\}$, if it does not exist in the previous time step $k - 1$, i.e., $r_{k - 1, n} = 0$, then it also does not exist at time $k$, i.e., $p(r_{k, n} = 1 | r_{k - 1, n} = 0) = 0$. If it exists at time $k - 1$, i.e., $r_{k - 1, n} = 1$, then it continues to exist at time $k$ with survival probability $p_{\mathrm{s}}$, i.e., $p(r_{k, n} = 1 | r_{k - 1, n} = 1) = p_{\mathrm{s}}$.

\subsection{Object Birth Models} \label{subsec:birth_model}

The joint prior \ac{pdf} of new \acp{po} is modeled as the product of prior \acp{pdf} of individual new \acp{po}, i.e., $f(\overline{\V{x}}{}_{k}, \overline{\V{\phi}}{}_{k}) = \prod_{n = N_{k - 1} + 1}^{N_k} f(\V{x}_{k, n}, \V{\phi}_{k, n})$ with $\overline{\V{x}}{}_k = [\V{x}_{k, N_{k - 1} + 1}^\T \cdots \V{x}_{k, N_{k}}^\T]^\T$ and $\overline{\V{\phi}}{}_k = [\V{\phi}_{k, N_{k - 1} + 1}^\T \cdots \V{\phi}_{k, N_{k}}^\T]^\T$. This corresponds to a multi-Bernoulli birth model \cite{VoVoNguShim:J24} with $N_k - N_{k - 1}$ components.

This model can approximately represent newly appearing objects that follow a Poisson point process with mean $\mu_{\text{B}}$ and joint birth-state \ac{pdf} $f_{\text{B}}(\V{x}, \V{\gamma}) = f_{\text{B}}(\V{x}) f_{\text{B}}(\V{\gamma})$, where we assume the kinematic states and signal powers of new \acp{po} are independent a priori. To systematically introduce new \acp{po} across the \ac{roi} $\Set{X}$ with $\V{x} \in \Set{X}$, we partition the \ac{roi} into $Q$ non-overlapping regions denoted by $\Set{X}_q, q \in \{1, \dots, Q\}$ with $\uplus_{q = 1}^Q \Set{X}_q = \Set{X}$. Each subregion $\Set{X}_q$ spawns one new \ac{po}, resulting in $N_k = N_{k - 1} + Q$ total \acp{po} at time $k$. The reference Poisson process restricted to each subregion $\Set{X}_q$ has mean $\mu_{\text{B}, n} = \mu_{\text{B}} \int_{\Set{X}_q} f_{\text{B}}(\V{x}) \hspace{1mm} \mathrm{d} \V{x}$, $n = N_{k - 1} + q$. Its kinematic birth \ac{pdf} is given by
\begin{equation}
    f_{\text{B}, n}(\V{x}) \propto \begin{cases}
        f_{\text{B}}(\V{x}), & \V{x} \in \Set{X}_q \\
        0, & \V{x} \not\in \Set{X}_q
    \end{cases} \nn
\end{equation}
where $\propto$ indicates equality up to a multiplicative constant. 

For $\mu_{\text{B}, n} \ll 1$, we approximate the regional Poisson birth count by conditioning on at most one new object. The original Poisson probabilities of zero and one new object in $\Set{X}_q$ are $\mathrm{e}^{-\mu_{\text{B}, n}}$ and $\mu_{\text{B}, n} \mathrm{e}^{-\mu_{\text{B}, n}}$, respectively, with $n = N_{k - 1} + q$. Renormalizing these two probabilities gives the existence probability of the corresponding new \ac{po} as
\begin{equation}
    p_{\text{B}, n} = \frac{\mu_{\text{B}, n} \mathrm{e}^{-\mu_{\text{B}, n}}}{\mu_{\text{B}, n} \mathrm{e}^{-\mu_{\text{B}, n}} + \mathrm{e}^{-\mu_{\text{B}, n}}} = \frac{\mu_{\text{B}, n}}{\mu_{\text{B}, n} + 1}. \nn
\end{equation}

As a result, the \ac{pdf} of new \ac{po} $n \in \{N_{k - 1} + 1, \dots, N_k\}$ is $f(\V{x}_{k, n}, \V{\phi}_{k, n}) = f(\V{x}_{k, n}) f(\V{\phi}_{k, n})$ with
\begin{align}
    f(\V{x}_{k, n}) &= f_{\text{B}, n}(\V{x}_{k, n}) \nn \\
    f(\V{\phi}_{k, n}) &= p_{\text{B}}(r_{k, n}) f_{\text{B}}(\V{\gamma}_{k, n}) \nn
\end{align}
where $p_{\text{B}}(r_{k, n} = 1) = p_{\text{B}, n}$. The kinematic and signal power states $\V{x}_{k, n}, \V{\phi}_{k, n}, n \in \{N_{k - 1} + 1, \dots, N_k\}$ of a new \ac{po} are assumed independent a priori of all current and previous states of other \acp{po} and of all current and previous noise power states.

At time $k = 0$\vspace{-.5mm}, the prior PDFs $f(\V{x}_{0, n})$, $f(\V{\phi}_{0, n})$, $n \in \{1, \dots, N_0\}$, and $f(\eta_{0}^{(i)}), i \in \{1, \dots, I\}$ are assumed known. The random variables $\V{x}_{0, n}, \V{\phi}_{0, n}, n \in \{1, \dots, N_0\}$ and $\eta_{0}^{(i)}, i \in \{1, \dots, I\}$ are all independent of each other\vspace{-2mm}.

\subsection{Joint Posterior PDF and Factor Graph} \label{subsec:factor_graph}

In the Bayesian setting, estimating the number of objects reduces to the computation of existence probabilities $\textcolor{black}{p(r_{k , n} = 1 | \V{z}_{1 : k})}$ that involve all the measurements $\V{z}_{1 : k} = [\V{z}_{1}^\T \cdots \V{z}_{k}^\T]^\T$ up to time $k$. Object declarations are made by thresholding existence probabilities based on a threshold $T_{\mathrm{dec}}$, i.e., a \ac{po} $n$ is declared to exist if $\textcolor{black}{p(r_{k , n} = 1 | \V{z}_{1 : k})} > T_{\mathrm{dec}}$. The number of \acp{po} increases over time. To manage computational complexity, we remove a \ac{po} from the state space if its existence probability is lower than a threshold $T_{\mathrm{pru}}$. \textcolor{black}{To estimate the kinematic state of \ac{po} $n$, we use the \ac{mmse} estimator given by}
\begin{equation}
    \hat{\V{x}}_{k, n} = \int \V{x}_{k, n} f(\V{x}_{k , n} | \V{z}_{1 : k}) \mathrm{d} \V{x}_{k, n}. \nn
\end{equation}
\textcolor{black}{The estimator uses the marginal posterior \ac{pdf} $f(\V{x}_{k , n} | \V{z}_{1 : k})$, which averages over both values of the \ac{po} existence variable $r_{k, n}$.} 

\begin{figure*}[!t]
    \centering
    \psfrag{da1}[c][c][0.75]{\raisebox{-2mm}{\hspace{.8mm}$\V{x}_{1}$}}
    \psfrag{daI}[c][c][0.75]{\raisebox{-2.5mm}{\hspace{.3mm}$\V{x}_{\underline{N}}$}}
    \psfrag{db1}[c][c][0.75]{\raisebox{-2mm}{\hspace{.2mm}$\V{x}_{\scriptscriptstyle \underline{N} + 1}$}}
    \psfrag{dbJ}[c][c][0.75]{\raisebox{1mm}{$\V{x}_{N}$}}
    \psfrag{q1}[c][c][0.75]{\raisebox{-1mm}{$f_{1}^{\V{x}}$}}
    \psfrag{qI}[c][c][0.75]{\raisebox{-3mm}{$f_{\underline{N}}^{\V{x}}$}}
    \psfrag{v1}[c][c][0.75]{\raisebox{-3.3mm}{\hspace{.15mm}$f_{\scriptscriptstyle \underline{N} + 1}^{\V{x}}$}}
    \psfrag{vJ}[c][c][0.75]{\raisebox{-1mm}{\hspace{.3mm}$f_{N}^{\V{x}}$}}
    \psfrag{g1}[c][c][0.75]{\raisebox{-1mm}{\hspace{.3mm}$f_{1}^{\V{z}}$}}
    \psfrag{gJ}[c][c][0.75]{\raisebox{-1.7mm}{\hspace{.5mm}$f_{J}^{\V{z}}$}}
    \psfrag{g1I}[c][c][0.75]{\raisebox{-1mm}{\hspace{.3mm}$f_{1}^{\V{z}}$}}
    \psfrag{gJI}[c][c][0.75]{\raisebox{-1.7mm}{\hspace{.5mm}$f_{J}^{\V{z}}$}}
    \psfrag{ma1}[r][r][0.75]{\color{black}{$\alpha_{1}$}}
    \psfrag{maJ}[l][l][0.75]{\color{black}{$\alpha_{N}$}}
    \psfrag{mb11}[l][l][0.75]{\color{black}{\hspace{-0.7mm}$\beta_{1, 1}^{(1)}$}}
    \psfrag{mk11}[r][r][0.75]{\color{black}{\raisebox{1.5mm}{$\kappa_{1, 1}^{(1)}$}}}
    \psfrag{fn}[c][c][0.75]{\raisebox{-3mm}{$f^{\eta}$}}
    \psfrag{fnI}[c][c][0.75]{\raisebox{1mm}{$f^{\eta}$}}
    \psfrag{dn}[c][c][0.75]{$\eta$}
    \psfrag{dnI}[c][c][0.75]{$\eta$}
    \psfrag{me1}[r][r][0.75]{\color{black}{$\xi$}}
    \psfrag{meI}[r][r][0.75]{\color{black}{$\xi$}}
    \psfrag{mxi1J}[l][l][0.75]{\raisebox{-3mm}{\color{black}{\hspace{0mm}$\iota_J$}}}
    \psfrag{mnu11}[r][r][0.75]{\raisebox{-3mm}{\color{black}{\hspace{0mm}$\nu_1$}}}
    \psfrag{mxiIJ}[l][l][0.75]{\raisebox{-3mm}{\color{black}{\hspace{0mm}$\iota_J$}}}
    \psfrag{mnuI1}[r][r][0.75]{\raisebox{-3mm}{\color{black}{\hspace{0mm}$\nu_1$}}}
    \psfrag{dc1}[c][c][0.75]{\raisebox{-2mm}{\hspace{.8mm}$\V{\phi}_{1}$}}
    \psfrag{dcI}[c][c][0.75]{\raisebox{-2.5mm}{\hspace{.3mm}$\V{\phi}_{\underline{N}}$}}
    \psfrag{dd1}[c][c][0.75]{\raisebox{-2mm}{\hspace{.2mm}$\V{\phi}_{\scriptscriptstyle \underline{N} + 1}$}}
    \psfrag{ddJ}[c][c][0.75]{\raisebox{1mm}{$\V{\phi}_{N}$}}
    \psfrag{qp1}[c][c][0.75]{\raisebox{-1mm}{$f_{1}^{\V{\phi}}$}}
    \psfrag{qpI}[c][c][0.75]{\raisebox{-3mm}{$f_{\underline{N}}^{\V{\phi}}$}}
    \psfrag{vp1}[c][c][0.75]{\raisebox{-3.3mm}{\hspace{.15mm}$f_{\scriptscriptstyle \underline{N} + 1}^{\V{\phi}}$}}
    \psfrag{vpJ}[c][c][0.75]{\raisebox{-1mm}{\hspace{.3mm}$f_{N}^{\V{\phi}}$}}
    \psfrag{mp1}[r][r][0.75]{\color{black}{$\psi_{1}$}}
    \psfrag{mpJ}[l][l][0.75]{\color{black}{$\psi_{N}$}}
    \psfrag{ms11}[l][l][0.75]{\color{black}{$\lambda_{N, J}^{(I)}$}}
    \psfrag{mz11}[c][c][0.75]{\color{black}{\raisebox{1.0mm}{\hspace{6mm}$\zeta_{N, J}^{(I)}$}}}
    \psfrag{D1}[r][r][0.95]{\raisebox{1mm}{\hspace{-3mm} {Dictionary} $i = 1$}}
    \psfrag{DI}[r][r][0.95]{\raisebox{1mm}{\hspace{-3mm} {Dictionary} $i = I$}}
    \includegraphics[scale=0.85]{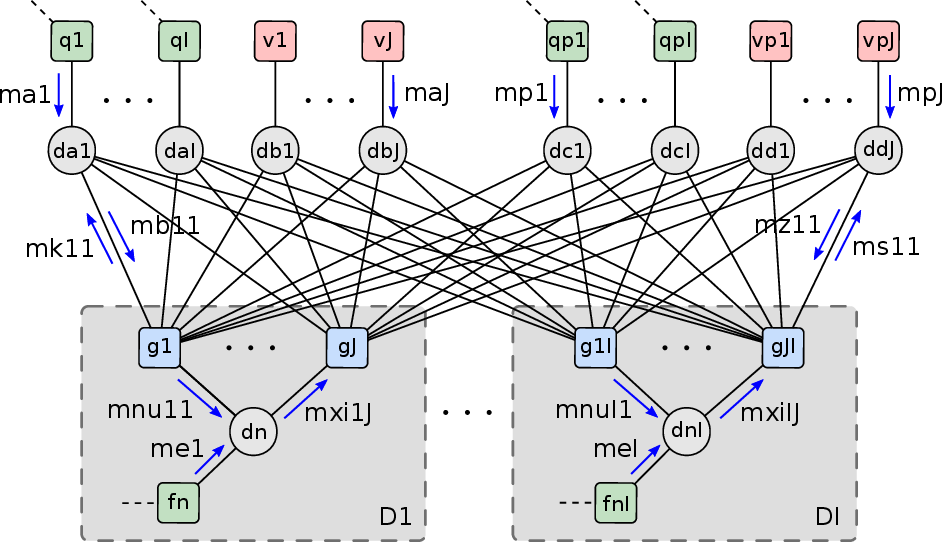}
    \caption{Factor graph representing the joint posterior \ac{pdf} in \eqref{eq:factorization} and the BP messages for a single time step $k$. \textcolor{black}{The time index $k$ and message-passing iteration index $t$ are omitted from the graph labels. For dictionary-local labels without an explicit dictionary index, the corresponding block identifies the dictionary.} BP messages are shown as blue arrows. The following shorthand notations are used: $\underline{N} = N_{k - 1}$, $N = N_k$, $\V{x}_n = \V{x}_{k, n}$, $\V{\phi}_n = \V{\phi}_{k, n}$, $\eta = \eta_k^{(i)}$, \textcolor{black}{$f^{\eta} = f(\eta_k^{(i)} | \eta_{k - 1}^{(i)})$}, $f_{j}^{\V{z}} = f(\V{z}_{k, j}^{(i)} | \V{x}_k, \V{\phi}_{k}, \eta_{k}^{(i)})$, $f_{n}^{\V{x}} = f(\V{x}_{k, n} | \V{x}_{k - 1, n})$, $f_{n}^{\V{\phi}} = f(\V{\phi}_{k, n} | \V{\phi}_{k - 1, n})$ for $n \in \{1, \dots, N_{k - 1}\}$,  $f_{n}^{\V{x}} = f(\V{x}_{k, n})$, $f_{n}^{\V{\phi}} = f(\V{\phi}_{k, n})$ for $n \in \{N_{k - 1} + 1, \dots, N_k\}$, $\alpha_n = \alpha(\V{x}_{k, n})$, $\psi_n = \psi(\V{\phi}_{k, n})$, $\xi = \xi(\eta_k^{(i)})$, $\beta_{n, j}^{(i)} = \beta_{j}^{(i, t)}(\V{x}_{k, n})$, $\zeta_{n, j}^{(i)} = \zeta_{j}^{(i, t)}(\V{\phi}_{k, n})$, $\iota_j = \iota_{j}^{(t)}(\eta_k^{(i)})$, $\kappa_{n, j}^{(i)} = \kappa_{j}^{(i, t)}(\V{x}_{k, n}; \V{z}_{k, j}^{(i)})$, $\lambda_{n, j}^{(i)} = \lambda_{j}^{(i, t)}(\V{\phi}_{k, n}; \V{z}_{k, j}^{(i)})$, and $\nu_j = \nu_{j}^{(t)}(\eta_k^{(i)}; \V{z}_{k, j}^{(i)})$. }
    \label{fig:factor_graph}
    \vspace{-4mm}
\end{figure*}

\textcolor{black}{Given the conditional independence assumptions in Sections}~\ref{subsec:meas_model} \textcolor{black}{through} \ref{subsec:birth_model}, the joint posterior \ac{pdf} can be factorized\vspace{0mm} as
\begin{align}
    &f(\V{x}_{0 : k}, \V{\phi}_{0 : k}, \V{\eta}_{0 : k} | \V{z}_{1 : k})  \nn \\
    &\propto \Big( \prod_{n = 1}^{N_0} f(\V{x}_{0, n}) f(\V{\phi}_{0, n}) \Big) \Big( \prod_{i = 1}^I f(\eta_0^{(i)}) \Big) \nn \\
    &\textcolor{black}{\times \prod_{k' = 1}^{k} \Bigg( \prod_{i = 1}^I \Big( f(\eta_{k'}^{(i)} | \eta_{k' - 1}^{(i)})} \nn \\
    &\textcolor{black}{\qquad\qquad \times \prod_{j = 1}^J f(\V{z}_{k', j}^{(i)} | \V{x}_{k'}, \V{\phi}_{k'}, \eta_{k'}^{(i)}) \Big)} \nn \\
    &\textcolor{black}{\quad \times \Big( \prod_{n = 1}^{N_{k' - 1}} f(\V{x}_{k', n} | \V{x}_{k' - 1, n}) f(\V{\phi}_{k', n} | \V{\phi}_{k' - 1, n}) \Big)} \nn \\
    &\textcolor{black}{\quad \times \Big( \prod_{n = N_{k' - 1} + 1}^{N_{k'}} f(\V{x}_{k', n}) f(\V{\phi}_{k', n}) \Big) \Bigg)} \label{eq:factorization}
\end{align}
where we have $\V{x}_{0 : k} = [\V{x}_{0}^\T \cdots \V{x}_{k}^\T]^\T$, $\V{\phi}_{0 : k} = [\V{\phi}_{0}^\T \cdots \V{\phi}_{k}^\T]^\T$, and $\V{\eta}_{0 : k} = [\V{\eta}_{0}^\T \cdots \V{\eta}_{k}^\T]^\T$. With the factorization \eqref{eq:factorization}, the joint posterior \ac{pdf} can be represented by a factor graph \cite{KscFreLoe:01,Loe:04,KolFri:B09}. This factor graph comprises factor nodes encapsulating statistical models introduced in Sections~\ref{subsec:meas_model} to \ref{subsec:birth_model}, and variable nodes representing random variables, including the kinematic state, signal power state of \acp{po}, and the noise power states. Edges in the graph connect factor nodes and variable nodes, describing their statistical dependencies. The factor graph that represents \eqref{eq:factorization} \textcolor{black}{is shown} in Fig.~\ref{fig:factor_graph}. Based on this factor graph, we develop the proposed Bayesian \ac{bp} method \cite{KscFreLoe:01,Loe:04,KolFri:B09}, which enables the \textcolor{black}{efficient approximation of marginal posterior \acp{pdf}} for object declaration and state estimation\vspace{-2mm}.

\section{Proposed Method} \label{sec:bp}

In this section, we introduce the proposed \ac{bp} method \cite{KscFreLoe:01,Loe:04,KolFri:B09} for \ac{tbd}. Based on the statistical model introduced in Section~\ref{subsec:factor_graph}, we aim to compute the existence probabilities $\textcolor{black}{p(r_{k , n} = 1 | \V{z}_{1 : k})}$ and the marginal posterior \acp{pdf} $f(\V{x}_{k, n} | \V{z}_{1 : k})$ for each \ac{po} $n \in \{1, \dots, N_k\}$. However, directly marginalizing the joint posterior $f(\V{x}_{0 : k}, \V{\phi}_{0 : k}, \V{\eta}_{0 : k} | \V{z}_{1 : k})$ is computationally infeasible due to high-dimensional integrations over kinematic states, signal power states, and noise power states. Instead, we make use of \ac{bp}. \ac{bp} exploits the conditional independence structure represented by the factor graph for efficient and scalable computation of existence probabilities and marginal posterior \acp{pdf}. \ac{bp} operates by computing so-called ``messages'' that are \textcolor{black}{real-valued} functions of the random variables over the edges of the factor graph, propagating probabilistic information between variable and factor nodes. \textcolor{black}{When the factor graph is a tree and all required messages are computed using exact sum-product rules, the existence probabilities and marginal posterior \acp{pdf} provided by \ac{bp}, called ``beliefs'', agree with their true counterparts.} If the factor graph has loops, as in our \ac{tbd} problem, beliefs serve as approximations of the true marginal posterior \acp{pdf} and theoretical performance guarantees are typically unavailable. \textcolor{black}{\textcolor{black}{Gaussian moment matching and finite particle sets can introduce additional approximation errors.} We evaluate the resulting detection and tracking performance in Section~\ref{sec:exp}.} Loopy \ac{bp} has been applied in a wide range of applications \cite{RicUrb:B08,WymLieWin:09,MeyKroWilLauHlaBraWin:J18,LeiMeyHlaWitTufWin:J19}. 

The computation order of \ac{bp} messages is flexible on loopy factor graphs. Depending on the chosen order, the resulting beliefs may vary. In this work, we adopt the following message passing order: (i) messages are only sent forward in time, (ii) within each time step, messages of the signal power states, noise power states, as well as kinematic states are computed sequentially and iteratively, (iii) \textcolor{black}{within each update stage, message computations for different dictionaries can be performed in parallel}. \ac{bp} messages involved in our considered \ac{mot} problem are shown in Fig.~\ref{fig:factor_graph} and can be computed using standard sum-product rules \cite{KscFreLoe:01,Loe:04,KolFri:B09}. We detail the computation of the messages in the rest of the section\vspace{-2mm}.

\subsection{Prediction and Birth Messages} \label{subsec:bp_pred}

\textcolor{black}{In the message notation below, the time index $k$ and, where applicable, the \ac{po} index $n$ are omitted from the message symbols and identified by their state arguments. }

At each time step $k$, the prediction messages are computed based on the beliefs from the previous time step $k - 1$ and the state-transition model. For legacy \ac{po} $n \in \{1, \dots, N_{k - 1}\}$, the prediction messages of the kinematic states sent from ``$f(\V{x}_{k, n} | \V{x}_{k - 1, n})$'' to ``$\V{x}_{k, n}$'' are given by
\begin{equation}
    \alpha(\V{x}_{k, n}) = \int f(\V{x}_{k, n} | \V{x}_{k - 1, n}) \tilde{f}(\V{x}_{k - 1, n}) \hspace{1mm} \mathrm{d} \V{x}_{k - 1, n} \nn
\end{equation}
and the prediction messages of the signal power states sent from ``$f(\V{\phi}_{k, n} | \V{\phi}_{k - 1, n})$'' to ``$\V{\phi}_{k, n}$'' can be expressed as
\begin{align}
    \psi(\V{\phi}_{k, n}) &= \sum_{r_{k - 1, n} \in \{0, 1\}} \int f(\V{\gamma}_{k, n}, r_{k, n} | \V{\gamma}_{k - 1, n}, r_{k - 1, n}) \nn \\[-2mm]
    & \hspace{22mm} \times \tilde{f}(\V{\gamma}_{k - 1, n}, r_{k - 1, n}) \hspace{1mm} \mathrm{d} \V{\gamma}_{k - 1, n} \label{eq:bp_pred_power}
\end{align}
where $\tilde{f}(\V{x}_{k - 1, n})$ and $\tilde{f}(\V{\phi}_{k - 1, n}) = \tilde{f}(\V{\gamma}_{k - 1, n}, r_{k - 1, n})$ are the beliefs of the kinematic and signal power state at time $k - 1$, respectively. Specifically, inserting the state-transition model defined in Section~\ref{subsec:state_tran_model} into \eqref{eq:bp_pred_power}, we obtain 
\begin{equation}
    \psi(\V{\gamma}_{k, n}, 1) = \int p_{\mathrm{s}} \ist\ist f(\V{\gamma}_{k, n} | \V{\gamma}_{k - 1, n}) \tilde{f}(\V{\gamma}_{k - 1, n}, 1) \hspace{1mm} \mathrm{d} \V{\gamma}_{k - 1, n}. \nn
\end{equation}
\textcolor{black}{For $r_{k, n} = 0$, we replace the conditional predicted signal power density by a normalized dummy \ac{pdf} while preserving the nonexistence probability, i.e.,} $\psi(\V{\gamma}_{k, n}, 0) = \psi_{k, n} f_{\mathrm{D}}(\V{\gamma}_{k, n})$. Here, the constant $ \psi_{k, n} $ is given\vspace{-2mm} by
\begin{equation}
    \psi_{k, n} = \int (1 - p_{\mathrm{s}}) \tilde{f}(\V{\gamma}_{k - 1, n}, 1) + \tilde{f}(\V{\gamma}_{k - 1, n}, 0) \hspace{1mm} \mathrm{d} \V{\gamma}_{k - 1, n} \nn
\end{equation}
\textcolor{black}{Here, $f_{\mathrm{D}}(\V{\gamma}_{k, n})$ is an arbitrary normalized \ac{pdf}. Its choice does not affect existence probabilities or kinematic state estimates, since a nonexistent legacy \ac{po} cannot reappear and the measurement likelihood does not depend on its signal power when $r_{k, n} = 0$} \cite{MeyKroWilLauHlaBraWin:J18}.

\acp{po} with indices $n \in \{N_{k - 1} + 1, \dots, N_k\}$ are newly introduced at time $k$. The birth messages of their kinematic and signal power states are the ones sent from ``$f(\V{x}_{k, n})$'' to ``$\V{x}_{k, n}$'', and from ``$f(\V{\phi}_{k, n})$'' to ``$\V{\phi}_{k, n}$'', respectively. Since $f(\V{x}_{k, n})$ and $f(\V{\phi}_{k, n})$ are singleton factors, the corresponding messages are the factors \textcolor{black}{themselves}, i.e., for $n \in \{N_{k - 1} + 1, \dots, N_k\}$ we \vspace{-4mm}have 
\begin{equation}
    \alpha(\V{x}_{k, n}) = f(\V{x}_{k, n}) \hspace{8mm} \psi(\V{\phi}_{k, n}) = f(\V{\phi}_{k, n}) \nn
    \vspace{0mm} 
\end{equation}
where $f(\V{x}_{k, n})$ and $f(\V{\phi}_{k, n})$ have been introduced in Section~\ref{subsec:birth_model}.

Similarly, for each dictionary $i \in \{1, \dots, I\}$, the \textcolor{black}{prediction message for the noise power state} from ``$f(\eta_{k}^{(i)} | \eta_{k - 1}^{(i)})$'' to ``$\eta_{k}^{(i)}$'' \textcolor{black}{is} computed\vspace{-2mm} as
\begin{equation}
    \xi(\eta_k^{(i)}) = \int f(\eta_k^{(i)}| \eta_{k - 1}^{(i)}) \tilde{f}(\eta_{k - 1}^{(i)}) \hspace{1mm} \mathrm{d} \eta_{k - 1}^{(i)} \nn
    \vspace{-2mm}
\end{equation}
where $\tilde{f}(\eta_{k - 1}^{(i)})$ is the belief of the \textcolor{black}{noise power state} at time $k - 1$\vspace{-2mm}.

\subsection{Iterative Measurement Update Messages} \label{subsec:bp_meas}

After the prediction messages are computed, iterative message passing is performed to incorporate the information from measurements $\V{z}_k$. At iteration $t \in \{1, \dots, T\}$, the messages that pass from the variable nodes ``$\V{x}_{k, n}, \V{\phi}_{k, n}, \eta_k^{(i)}$'' to the factor nodes ``$f(\V{z}_{k, j}^{(i)} | \V{x}_k, \V{\phi}_{k}, \eta_{k}^{(i)})$'' $n \in \{1, \dots, N_k\}, j \in \{1, \dots, J\}, i \in \{1, \dots, I\}$ are given as follows.

\noindent \textit{Measurement Update Signal Power State -- Incoming}
\begin{align}
    \zeta_{j}^{(i, t)}(\V{\phi}_{k, n}) = \psi(\V{\phi}_{k, n}) \underset{(i', j') \ne (i, j)}{\prod_{i' = 1}^I \prod_{j' = 1}^J} \lambda_{j'}^{(i', t)}(\V{\phi}_{k, n}; \V{z}_{k, j'}^{(i')}).  \label{eq:bp_power_meas_e}\\[-3mm]
    \nn
\end{align}

\noindent \textit{Measurement Update Noise Power State -- Incoming}
\begin{equation}
    \iota_{j}^{(t)}(\eta_k^{(i)}) =   \xi(\eta_k^{(i)})  \prod_{\substack{j' = 1 \\ j' \ne j}}^J \nu_{j'}^{(t)}(\eta_k^{(i)}; \V{z}_{k, j'}^{(i)}).  \label{eq:bp_noise_meas_e}
\end{equation}

\noindent \textit{Measurement Update Kinematic State -- Incoming}
\begin{align}
    \beta_{j}^{(i, t)}(\V{x}_{k, n}) = \alpha(\V{x}_{k, n}) \underset{(i', j') \ne (i, j)}{\prod_{i' = 1}^I \prod_{j' = 1}^J} \kappa_{j'}^{(i', t)}(\V{x}_{k, n}; \V{z}_{k, j'}^{(i')}).  \label{eq:bp_kinematic_meas_e}
\end{align}

The messages sent from factor node ``$f(\V{z}_{k, j}^{(i)} | \V{x}_k, \V{\phi}_{k}, \eta_{k}^{(i)})$'' to the variable nodes are $\lambda_{j}^{(i, t)}(\V{\phi}_{k, n}; \V{z}_{k, j}^{(i)})$, $\nu_{j}^{(t)}(\eta_k^{(i)}; \V{z}_{k, j}^{(i)})$, and $\kappa_{j}^{(i, t)}(\V{x}_{k, n}; \V{z}_{k, j}^{(i)})$. These messages are presented below.

\noindent \textit{Measurement Update Signal Power State -- Outgoing}
\begin{align}
    \hspace{-10mm}& \lambda_{j}^{(i, t)}(\V{\phi}_{k, n}; \V{z}_{k, j}^{(i)}) = \nn \\[0.5mm]
    &\hspace{3mm} \sum_{r_{k, 1} \in \{0, 1\}} \cdots \sum_{r_{k, n - 1} \in \{0, 1\}} \sum_{r_{k, n + 1} \in \{0, 1\}} \cdots \sum_{r_{k, N_k} \in \{0, 1\}}    \nn \\[0.75mm]
    &\hspace{0mm} \int \cdots \int f( \V{z}_{k, j}^{(i)} | \V{x}_k, \V{\phi}_k, \eta_k^{(i)} ) \ist\ist \iota_j^{(t - 1)}(\eta_k^{(i)}) \hspace{.5mm} \mathrm{d} \eta_{k}^{(i)} \nn \\[0.5mm]
    &\hspace{0mm} \times\hspace{0mm} \prod_{\substack{n' = 1 \\ n' \ne n}}^{N_k} \zeta_{j}^{(i, t - 1)}(\V{\phi}_{k, n'}) \hspace{0mm} \mathrm{d} \V{\gamma}_{k, 1} \cdots \mathrm{d} \V{\gamma}_{k, n - 1} \hspace{1mm} \mathrm{d} \V{\gamma}_{k, n + 1} \cdots \mathrm{d} \V{\gamma}_{k, N_k}  \nn \\[0.5mm]
    &\hspace{0mm} \times \prod_{n' = 1}^{N_k} \beta_{j}^{(i, t - 1)}(\V{x}_{k, n'})  \hspace{1mm} \mathrm{d} \V{x}_{k, 1} \cdots \mathrm{d} \V{x}_{k, N_k}.  \label{eq:bp_power_meas_update} 
\end{align}

\noindent \textit{Measurement Update Noise Power State -- Outgoing}
\begin{align}
    & \hspace{-15mm}\nu_{j}^{(t)}(\eta_k^{(i)}; \V{z}_{k, j}^{(i)}) =  \sum_{r_{k, 1} \in \{0, 1\}} \cdots \sum_{r_{k, N_k} \in \{0, 1\}} \nn \\
    &\hspace{3mm} \int \cdots \int f( \V{z}_{k, j}^{(i)} | \V{x}_k, \V{\phi}_k, \eta_k^{(i)} ) \nn \\
    &\hspace{2mm} \times \prod_{n' = 1}^{N_k} \beta_{j}^{(i, t - 1)}(\V{x}_{k, n'})  \hspace{1mm} \mathrm{d} \V{x}_{k, 1} \cdots \mathrm{d} \V{x}_{k, N_k}  \nn \\
    &\hspace{2mm} \times\hspace{0mm} \prod_{n' = 1}^{N_k} \zeta_{j}^{(i, t)}(\V{\phi}_{k, n'}) \hspace{1mm} \mathrm{d} \V{\gamma}_{k, 1} \cdots \mathrm{d} \V{\gamma}_{k, N_k}.  \label{eq:bp_noise_meas_update} 
\end{align}

\noindent \textit{Measurement Update Kinematic State -- Outgoing}
\begin{align}
    \hspace{-10mm}& \kappa_{j}^{(i, t)}(\V{x}_{k, n}; \V{z}_{k, j}^{(i)}) = \sum_{r_{k, 1} \in \{0, 1\}} \cdots \sum_{r_{k, N_k} \in \{0, 1\}} \nn \\[0.5mm]
    &\hspace{3mm} \int \dots \int f( \V{z}_{k, j}^{(i)} | \V{x}_k, \V{\phi}_k, \eta_k^{(i)} ) \iota_j^{(t)}(\eta_k^{(i)}) \hspace{1mm} \mathrm{d} \eta_{k}^{(i)} \nn \\[0.5mm]
    &\hspace{2mm} \times \prod_{\substack{n' = 1 \\ n' \ne n}}^{N_k} \beta_{j}^{(i, t - 1)}(\V{x}_{k, n'}) \hspace{1mm} \mathrm{d} \V{x}_{k, 1} \cdots \mathrm{d} \V{x}_{k, n - 1} \hspace{1mm} \mathrm{d} \V{x}_{k, n + 1} \cdots \mathrm{d} \V{x}_{k, N_k} \nn \\[0.5mm]
    &\hspace{2mm} \times \prod_{n' = 1}^{N_k} \zeta_{j}^{(i, t)}(\V{\phi}_{k, n'}) \hspace{1mm} \mathrm{d} \V{\gamma}_{k, 1} \cdots \mathrm{d} \V{\gamma}_{k, N_k}. \label{eq:bp_kinematic_meas_update} \\[-3mm]
    \nn
\end{align}

Note that the messages in \eqref{eq:bp_power_meas_update}-\eqref{eq:bp_kinematic_meas_update} depend on fixed sensor measurements $\V{z}_{k, j}^{(i)}$, which is indicated by the semicolon notation ``$(\cdot ; \V{z}_{k, j}^{(i)})$''. The messages in \eqref{eq:bp_power_meas_update}-\eqref{eq:bp_kinematic_meas_update} contain information related to measurements $\V{z}_{k, j}^{(i)}$ while also taking \textcolor{black}{the} most recent statistical information for the random signal power states, noise power states, and kinematic states into account. This most recent statistical information is given by \eqref{eq:bp_power_meas_e}-\eqref{eq:bp_kinematic_meas_e}. For example, the computation of $\nu_{j}^{(t)}(\eta_k^{(i)}; \V{z}_{k, j}^{(i)})$ includes the likelihood function $f( \V{z}_{k, j}^{(i)} | \V{x}_k, \V{\phi}_k, \eta_k^{(i)} )$ of measurement $ \V{z}_{k, j}^{(i)}$, the product $\prod_{n' = 1}^{N_k} \beta_{j}^{(i, t - 1)}(\V{x}_{k, n'})$ involving all kinematic states, and the product $\prod_{n' = 1}^{N_k} \zeta_{j}^{(i, t)}(\V{\phi}_{k, n'})$ \textcolor{black}{involving} all signal power states. 

The iteration is initialized by $\zeta_{j}^{(i, 0)}(\V{\phi}_{k, n}) = \psi(\V{\phi}_{k, n})$, $\iota_{j}^{(0)}(\eta_k^{(i)}) = \xi(\eta_k^{(i)})$, and $\beta_{j}^{(i, 0)}(\V{x}_{k, n}) = \alpha(\V{x}_{k, n})$. As defined in \eqref{eq:bp_power_meas_update}-\eqref{eq:bp_kinematic_meas_update}, within each iteration, \textcolor{black}{message families are updated in the following order:} $\lambda_{j}^{(i, t)}(\V{\phi}_{k, n}; \V{z}_{k, j}^{(i)})$, $\zeta_{j}^{(i, t)}(\V{\phi}_{k, n})$, $\nu_{j}^{(t)}(\eta_k^{(i)}; \V{z}_{k, j}^{(i)})$, $\iota_{j}^{(t)}(\eta_k^{(i)})$, $\kappa_{j}^{(i, t)}(\V{x}_{k, n}; \V{z}_{k, j}^{(i)})$, and $\beta_{j}^{(i, t)}(\V{x}_{k, n})$. \textcolor{black}{Each update stage is completed for all applicable \acp{po}, dictionaries, and snapshots before the next stage begins.}

\subsection{Message Approximation and Belief Calculation} \label{subsec:bp_approx}

Messages in \eqref{eq:bp_power_meas_update}-\eqref{eq:bp_kinematic_meas_update} require \textcolor{black}{marginalization over the kinematic, signal power, and noise power states connected to the measurement factor, except for the state retained as each message's argument}. This marginalization cannot be performed in closed form. \textcolor{black}{A naive numerical integration approach would lead to a computational complexity that grows exponentially with the number of \acp{po} $N_k$.} Since $f( \V{z}_{k, j}^{(i)} | \V{x}_k, \V{\phi}_k, \eta_k^{(i)} )$ is zero-mean complex Gaussian with respect to $\V{z}_{k, j}^{(i)}$, \textcolor{black}{each of the three messages} in \eqref{eq:bp_power_meas_update}-\eqref{eq:bp_kinematic_meas_update} \textcolor{black}{is} a Gaussian mixture with an uncountably infinite number of components with respect to $\V{z}_{k, j}^{(i)}$. All components have a zero mean. To ensure the scalability of the proposed method with respect to $N_k$, we employ moment matching to approximate these Gaussian mixtures by a single Gaussian component. The approximated messages read
\begin{align}
    \tilde{\lambda}_{j}^{(i, t)}(\V{\phi}_{k, n}; \V{z}_{k, j}^{(i)}) &= \mathcal{CN} \big(\V{z}_{k, j}^{(i)}; \V{0}, \M{C}_{\lambda, k, n, j}^{(i, t)} \big) \nn \\[1mm]
    \tilde{\nu}_{j}^{(t)}(\eta_k^{(i)}; \V{z}_{k, j}^{(i)}) &= \mathcal{CN} \big(\V{z}_{k, j}^{(i)}; \V{0}, \textcolor{black}{\M{C}_{\nu, k, j}^{(i, t)}} \big) \nn \\[1mm]
    \tilde{\kappa}_{j}^{(i, t)}(\V{x}_{k, n}; \V{z}_{k, j}^{(i)}) &= \mathcal{CN} \big(\V{z}_{k, j}^{(i)}; \V{0}, \M{C}_{\kappa, k, n, j}^{(i, t)} \big)  \nn\\[-4.5mm]
    \nn
\end{align}
where the covariances $\M{C}_{\lambda, k, \textcolor{black}{n}, j}^{(i, t)}$, $\M{C}_{\nu, k, j}^{(i, t)}$, and $\M{C}_{\kappa, k, \textcolor{black}{n}, j}^{(i, t)}$ preserve the second-order statistics of their respective original Gaussian mixture \acp{pdf}. Based on derivations previously presented in \cite{LiaKroMey:J23,DavGar:J24,LiaLeiMey:J25}, covariance matrices can be computed as
\vspace{-4mm} 


\begin{align}
    \M{C}_{\lambda, k, n, j}^{(i, t)} &= r_{k, n} \gamma_{k, n}^{(i)} \M{\Sigma}_{k, n, j}^{(i, \textcolor{black}{t - 1})} + \sum_{\substack{n' = 1 \\ n' \ne n}}^{N_k} \hat{\gamma}_{k, n', j}^{(i, \textcolor{black}{t - 1})} \M{\Sigma}_{k, n', j}^{(i, \textcolor{black}{t - 1})} \nn\\[-5mm]
    &\quad{} + \hat{\eta}_{k, j}^{(i, \textcolor{black}{t - 1})} \M{I}_M \label{eq:cov_lambda} \\
    \textcolor{black}{\M{C}_{\nu, k, j}^{(i, t)}} &= \sum_{n' = 1}^{N_k} \hat{\gamma}_{k, n', j}^{(i, t)} \M{\Sigma}_{k, n', j}^{(i, \textcolor{black}{t - 1})} + \eta_{k}^{(i)} \M{I}_M \label{eq:cov_nu} \\[0.5mm]
    \M{C}_{\kappa, k, n, j}^{(i, t)} &= \hat{\gamma}_{k, n, j}^{(i, t)} \V{a}_{k, n}^{\textcolor{black}{(i)}} \V{a}_{k, n}^{\textcolor{black}{(i)} \CH} + \sum_{\substack{n' = 1 \\ n' \ne n}}^{N_k} \hat{\gamma}_{k, n', j}^{(i, t)} \M{\Sigma}_{k, n', j}^{(i, \textcolor{black}{t - 1})} + \hat{\eta}_{k, j}^{(i, t)} \M{I}_M \label{eq:cov_kappa} \nn\\[-5mm]
\end{align}
where
\begin{align}
    \hat{\gamma}_{k, n, j}^{(i, t)} &= \sum_{r_{k, n} \in \{0, 1\}} \int \textcolor{black}{r_{k, n}} \gamma_{k, n}^{(i)} \zeta_j^{(i, t)}(\V{\phi}_{k, n})  \hspace{1mm} \mathrm{d} \V{\gamma}_{k, n} \label{eq:power_approx} \\[0.75mm]
    \M{\Sigma}_{k, n, j}^{(i, t)} &= \int \V{a}_{k, n}^{\textcolor{black}{(i)}} \V{a}_{k, n}^{\textcolor{black}{(i)} \CH} \beta_{j}^{(i, t)}(\V{x}_{k, n}) \hspace{1mm} \mathrm{d} \V{x}_{k, n} \label{eq:kinematic_approx} \\[1mm]
    \hat{\eta}_{k, j}^{(i, t)} &= b_\eta \cdot \int \eta_{k}^{(i)} \iota_{j}^{(t)}(\eta_k^{(i)}) \hspace{1mm} \mathrm{d} \eta_{k}^{(i)}. \label{eq:noise_approx}\\[-3mm]
    \nn
\end{align}
\textcolor{black}{In \eqref{eq:power_approx}, $\hat{\gamma}_{k, n, j}^{(i, t)}$ is the mean effective signal power; the $r_{k, n} = 0$ term vanishes.} Here, $b_\eta$ is the \textcolor{black}{noise inflation factor}. When $b_\eta = 1$, the matrices $\M{C}_{\lambda, k, \textcolor{black}{n}, j}^{(i, t)}$, $\M{C}_{\nu, k, j}^{(i, t)}$, and $\M{C}_{\kappa, k, \textcolor{black}{n}, j}^{(i, t)}$ align with the covariances of the original messages. \textcolor{black}{Setting $b_\eta > 1$ inflates the isotropic noise contribution in the signal power and kinematic state updates \eqref{eq:cov_lambda} and \eqref{eq:cov_kappa}. The term $\eta_k^{(i)} \M{I}_M$ in \eqref{eq:cov_nu} remains unscaled. This scalar inflation is a heuristic that can improve robustness against model mismatch, including colored measurement noise, while retaining an isotropic noise covariance model.} \textcolor{black}{These approximated outgoing messages replace the corresponding exact messages in \eqref{eq:bp_power_meas_e}-\eqref{eq:bp_kinematic_meas_e} and are also used in the belief calculations below.}

After $T$ iterations, the beliefs for each random variable can be obtained by multiplying all the incoming messages at the corresponding variable node \cite{KscFreLoe:01}, i.e.,
\begin{align}
    \tilde{f}(\V{x}_{k, n}) &\propto \alpha(\V{x}_{k, n}) \prod_{i = 1}^I \prod_{j = 1}^J \tilde{\kappa}_{j}^{(i, T)}(\V{x}_{k, n}; \V{z}_{k, j}^{(i)}) \nn \\[.5mm]
    \tilde{f}(\V{\phi}_{k, n}) &\propto \psi(\V{\phi}_{k, n}) \prod_{i = 1}^I \prod_{j = 1}^J \tilde{\lambda}_{j}^{(i, T)}(\V{\phi}_{k, n}; \V{z}_{k, j}^{(i)}) \nn \\[.5mm]
    \tilde{f}(\eta_k^{(i)}) &\propto \xi(\eta_k^{(i)}) \prod_{j = 1}^J \tilde{\nu}_{j}^{(T)}(\eta_k^{(i)}; \V{z}_{k, j}^{(i)}). \nn
\end{align}
\textcolor{black}{We normalize these beliefs to unit mass over their respective state spaces. For $\tilde{f}(\V{\phi}_{k, n})$, normalization includes summation over $r_{k, n} \in \{0, 1\}$ and integration over $\V{\gamma}_{k, n}$. The existence probability is then approximated by $p(r_{k, n} = 1 | \V{z}_{1 : k}) \approx \int \tilde{f}(\V{\gamma}_{k, n}, 1) \hspace{.5mm} \mathrm{d} \V{\gamma}_{k, n}$.}

\textcolor{black}{The beliefs approximate the corresponding marginal posterior \acp{pdf}; for example, $\tilde{f}(\V{x}_{k, n}) \approx f(\V{x}_{k , n} | \V{z}_{1 : k})$. They can then be used for state estimation as detailed in Section~\ref{subsec:factor_graph} and for inference at subsequent time steps.}

\subsection{Scalability and Relationship to Bernoulli Filtering } \label{subsec:bern}

The integration in \eqref{eq:power_approx}-\textcolor{black}{\eqref{eq:noise_approx}} is done individually for each random variable, i.e., a joint integration as in \eqref{eq:bp_power_meas_update}-\eqref{eq:bp_kinematic_meas_update} can be avoided. \textcolor{black}{A linear scaling with the number of \acp{po} can be obtained by reusing a total sum of covariance contributions. The sums excluding one \ac{po} in \eqref{eq:cov_lambda}-\eqref{eq:cov_kappa} can be obtained by subtracting its contribution from the corresponding total sum. Likewise, the products excluding one measurement factor in \eqref{eq:bp_power_meas_e}-\eqref{eq:bp_kinematic_meas_e} can be computed by reusing aggregate products at shared particle locations.} \textcolor{black}{Evaluating the Gaussian likelihood expressions in the approximated messages} $\tilde{\lambda}_{j}^{(i, t)}(\V{\phi}_{k, n}; \V{z}_{k, j}^{(i)})$, $\tilde{\nu}_{j}^{(t)}(\eta_k^{(i)}; \V{z}_{k, j}^{(i)})$, and $\tilde{\kappa}_{j}^{(i, t)}(\V{x}_{k, n}; \V{z}_{k, j}^{(i)})$ results in a computational complexity that scales as $\mathcal{O}(M^3)$.

Since there are $2IJN_k + IJ$ such messages, \textcolor{black}{a message-update complexity of $\mathcal{O}(M^3 I J N_k)$ per iteration can be achieved with fixed particle counts, the reuse described above, and basis-vector evaluation costs that do not dominate the matrix calculations. The cost of $T$ iterations is then $\mathcal{O}(T M^3 I J N_k)$, which is linear in $N_k$, $I$, and $J$ for fixed $T$.} \rd{\textcolor{black}{As an optional additional approximation, the kinematic second-moment matrices in \eqref{eq:kinematic_approx} can be replaced by outer products of the corresponding mean basis vectors. The resulting low-rank covariance representations permit use of the Woodbury matrix identity and the matrix determinant lemma, following steps similar to} \cite[Sec.~3]{LiaLeiMey:J24SM}\textcolor{black}{. With the assumptions above and reuse of inner products between these vectors, a per-iteration message-update complexity of $\mathcal{O}((M^2 + M N_k^2 + N_k^4) I J)$ can be achieved. This alternative neglects basis-vector variability and can be advantageous when $N_k$ is small relative to $M$.}}

\textcolor{black}{Following the pruning rule in Section~\ref{subsec:factor_graph}, \acp{po} whose approximate existence probability is below $T_{\mathrm{pru}}$ can be removed after belief calculation and before prediction at time $k + 1$. The practical message-update cost then depends on the number of retained legacy \acp{po} plus newly introduced \acp{po}.}

 \textcolor{black}{The expressions in \eqref{eq:power_approx}-\eqref{eq:noise_approx} are computed by means of Monte Carlo integration using weighted particles. We approximate the continuous integrals by weighted sums and retain the summation over the binary existence state in \eqref{eq:power_approx}. Given weighted particles representing the prediction or birth messages, normalized incoming-message weights can be obtained by multiplying their weights by the corresponding products of approximated outgoing messages in \eqref{eq:bp_power_meas_e}-\eqref{eq:bp_kinematic_meas_e} and renormalizing.}\rd{\textcolor{black}{In this way, we avoid a fixed grid discretization\vspace{-1mm} of the continuous state variables $\V{x}_{k, n}$, $\V{\gamma}_{k, n}$, and $\eta_k^{(i)}$.}}
 
 \rd{\textcolor{black}{When we retain a single initial \ac{po} and introduce no birth \acp{po}, i.e., $N_0 = 1$ and $Q = 0$, we have $N_k = 1$ for all $k$. With $T = 1$, our method can be interpreted as an approximate particle-based single-Bernoulli filter without object births that uses the proposed multi-dictionary and multi-snapshot measurement model with random signal and noise powers. Compared with a direct particle-based implementation of a single-Bernoulli filter} \cite{RisVoVoFar:J13}, our method \textcolor{black}{\textcolor{black}{uses Gaussian moment matching and}} maintains separate particle clouds for object kinematics, signal powers, and noise powers to alleviate the curse of dimensionality inherent to particle-based processing \cite{DauHua:C03}. }

\section{Experimental Results} \label{sec:exp}

In this section, we evaluate the performance of our proposed method based on both synthetic and real data. \textcolor{black}{Tracking performance is quantified using the \ac{gospa} metric.} In the \textcolor{black}{considered scenarios}, the entries in the measurement vectors $\V{z}_{k, j}^{(i)} = [z_{k, j, 1}^{(i)} \cdots z_{k, j, M}^{(i)}]^\T \in \mathbb{C}^M, i \in \{1, \dots, I\}, j \in \{1, \dots, J\}$ are strongly correlated across measurement index $m$\textcolor{black}{. When used with likelihoods that assume independent measurement components, conventional multiobject \ac{tbd} methods such as \cite{SalBir:01,OrtFit:J02,StrGraWal:J02,BoeDri:J04,DavGae:B18,ItoGod:J20,LiaKroMey:J23,VoVoPhaSut:10,RisVoVoFar:J13,RisRosKimWanWil:J20,KroWilMey:21,KimRisGuaRos:21,DavGar:J24} do not represent these correlations.} 

\subsection{Synthetic Radar Tracking Scenario} \label{subsec:exp_radar}

We consider a \ac{2d} multi-sensor radar system with $I$ sensors equipped with a uniform linear array. Sensors are located at positions $\V{p}^{(i)}\rmv, i \in \{1, \dots, I\}$. A linear chirp signal is transmitted, and reflections from objects are received. \rd{This type of system is referred to as \ac{lfmcw} radar \cite{LevMoz:B04,ZhuXuGuoWanZheQu:24} as used in applications such as tracking individuals in shopping mall environments.} At each time step $k$, the radar system demodulates, downconverts, and discretizes the analog received signal to obtain the baseband signal $\V{z}_{k, j}^{(i)}$ in the time domain \cite[Sec.~9.2.3]{LevMoz:B04}.  \rd{The measurement vectors $\V{z}_{k, j}^{(i)}$ adhere to the signal model in \eqref{eq:signal_model} through the \textcolor{black}{state-dependent} basis function, i.e.,
\begin{equation}
    \V{a}^{(i)}(\V{x}) = \mathrm{vec} \big(\V{a}_{\mathrm{d}}^{(i)}\rmv\rmv(\V{x}) \otimes \V{a}_{\theta}^{(i)}\rmv(\V{x}) \otimes \V{a}^{(i)}_{\mathrm{s}}(\V{x}) \big) \label{eq:dict_radar}
\end{equation}
 with $\otimes$ being the Kronecker product and $\mathrm{vec}(\cdot)$ the vectorization operation. Here, $\V{x} = [\V{p}^\T \hspace{1mm} \V{v}^\T]^\T$ with $\V{p}$ being the position and $\V{v}$ being the velocity of the object. In addition, fast-time (delay), angular, and slow-time (Doppler) steering vectors are defined as
\begin{align}
    \textcolor{black}{\V{a}_{\mathrm{d}}^{(i)}(\V{x})} &= \big[1 \hspace{2mm} \mathrm{e}^{j 4 \pi \mu T_{\mathrm{s}} d^{(i)} / c} \cdots \mathrm{e}^{j 4 \pi (\textcolor{black}{M_{\mathrm{d}}} - 1) \mu T_{\mathrm{s}} d^{(i)} / c}\big]^\T \nn \\[0.5mm]
    \V{a}_{\theta}^{(i)}(\V{x}) &= \big[1 \hspace{2mm} \mathrm{e}^{j \pi \sin(\theta^{(i)})} \cdots \mathrm{e}^{j \pi (M_{\theta} - 1) \sin(\theta^{(i)})}\big]^\T \nn
    \\[0.5mm]
  \V{a}^{(i)}_{\mathrm{s}}(\V{x})	&= \big[1\hspace{2mm} e^{j4\pi T_c v^{(i)}/\lambda}  \cdots e^{j4\pi (M_{\mathrm{s}}-1)  T_c v^{(i)}/\lambda}\big]^\T. \label{eq:radarDictionary}
	\\[-4.5mm]
	\nn
\end{align}}
\rd{Here, $d^{(i)} = \Vert \V{p} - \V{p}^{(i)} \Vert$ denotes the distance between object at position $\V{p}$ and sensor $i$, $\theta^{(i)} = \angle (\V{p} - \V{p}^{(i)}) - \varphi^{(i)}$ \textcolor{black}{is the \ac{aoa} relative to the array broadside orientation $\varphi^{(i)}$. The bearing $\angle(\V{p} - \V{p}^{(i)})$ and $\varphi^{(i)}$ are measured counterclockwise from the first coordinate axis. The radial velocity is $v^{(i)} = (\V{p} - \V{p}^{(i)})^\T \V{v}/d^{(i)}$, positive for increasing range.} The \textcolor{black}{broadside} orientation of both arrays is $\varphi^{(i)} = \frac{\pi}{4}$. \textcolor{black}{The array elements are spaced by $\lambda/2$.} The parameters $c$, $\mu$, $\lambda$, $T_{\mathrm{s}}$, and $T_{\mathrm{c}}$ represent the speed of light, chirp rate, wavelength, sampling interval, and \textcolor{black}{chirp repetition interval}, respectively. The total measurement size is $M = M_\mathrm{d} M_\theta  M_{\mathrm{s}}$, combining $M_\mathrm{d}$ fast time samples, $M_\theta$ array elements, and $M_\mathrm{s}$ slow time samples. Note that $ \V{a}_{\mathrm{d}}^{(i)}(\V{x})$ is a complex narrowband signal $\mathrm{e}^{j \omega^{(i)} \tau} $ with distance-dependent frequency $\omega^{(i)} = 2 \pi ( 2 \ist \mu \ist d^{(i)} / c)$ sampled at times $\tau \in \big\{0, T_{\mathrm{s}}, \dots, (M_{\mathrm{d}} - 1) T_{\mathrm{s}} \big\}$.} \rd{\textcolor{black}{The interval between consecutive tracking time steps is one second. Within each snapshot, we assume a common complex amplitude across the $M_{\mathrm{s}}$ chirps, with the Doppler-induced phase evolution described by \eqref{eq:radarDictionary}. } Objects are considered to be in the sensor's far field.}

For data generation, \textcolor{black}{we set the power of the complex amplitude $\varrho^{(i)}$ using the simplified radar-range}\vspace{-.5mm} equation
\begin{equation}
    |\varrho^{(i)}|^2 = \frac{c^2 }{f_\mathrm{c}^2 (4 \pi)^3 (d^{(i)})^4} \label{eq:amp_radar}
\end{equation}
where $f_\mathrm{c} = 77$~GHz is the carrier frequency. \textcolor{black}{In \eqref{eq:amp_radar}, the combined prefactor accounting for transmit power, element gains, radar cross section, and losses is set to one.} \textcolor{black}{The phase of $\varrho^{(i)}$ is uniformly distributed over $[0, 2\pi)$. This generator has a fixed amplitude magnitude at a given range and differs from the complex Gaussian amplitude model used for inference.}  \rd{Note that we model the propagation path for each array element individually. Thus, the array gain does not appear in \eqref{eq:amp_radar}. This is because beamforming, which yields the array gain, is performed in the measurement update step of the proposed \ac{tbd} method.}

In this experiment, there are $I = 2$ radar sensors at positions $\V{p}^{(1)} = [0 \hspace{1mm} 0]^\T$ and $\V{p}^{(2)} = [50 \hspace{1mm} 50]^\T$. We set the chirp rate to $\mu = 8\text{MHz/$\mu$s}$, the sampling interval to $T_{\mathrm{s}} = 0.2\mu$s, the \textcolor{black}{chirp repetition interval} to $T_{\mathrm{c}} = 3$ms,  the number of samples in time domain to $M_\mathrm{d} = 20$, the number of array elements to $M_\theta = 10$, \rd{and the number of slow time samples to $M_\mathrm{s} = 4$}. The kinematic state $\V{x}_{k, n} = [\V{p}_{k, n}^\T \hspace{1mm} \V{v}_{k, n}^\T]^\T$ of \ac{po} $n \in \{1, \dots, N_k\}$ contains its \ac{2d} position $\V{p}_{k, n}$ and \ac{2d} velocity $\V{v}_{k, n}$. A constant velocity model $\V{x}_{k, n} = \M{F}\V{x}_{k - 1, n} + \M{W}\textcolor{black}{\V{q}_{k, n}}$ \cite[Ch. 4]{ShaKirLi:B02} is employed to describe the dynamics of the kinematic states, where $\textcolor{black}{\V{q}_{k, n}} \in \mathbb{R}^2$ is the zero-mean Gaussian driving noise with covariance $10^{-4}\M{I}_2$. \textcolor{black}{The driving noises are independent across objects and time.} \textcolor{black}{The transition matrices are
\begin{equation*}
    \M{F} = \begin{bmatrix} \M{I}_2 & \Delta t\,\M{I}_2 \\ \M{0} & \M{I}_2 \end{bmatrix}
    \qquad
    \M{W} = \begin{bmatrix} \frac{\Delta t^2}{2}\M{I}_2 \\ \Delta t\,\M{I}_2 \end{bmatrix}
\end{equation*}
where $\Delta t$ is the interval between consecutive tracking time steps, equal to one second in this experiment.}

The state-transition of the signal power for each dictionary is assumed to be independent, i.e., $f(\V{\gamma}_{k, n} | \V{\gamma}_{k - 1, n} ) = \prod_{i = 1}^I f(\gamma_{k, n}^{(i)} | \gamma_{k - 1, n}^{(i)} )$, and the individual state-transition \acp{pdf} are given by $f(\gamma_{k, n}^{(i)} | \gamma_{k - 1, n}^{(i)}) = \mathcal{G}(\gamma_{k, n}^{(i)}; \gamma_{k - 1, n}^{(i)} / c_{\gamma}, c_{\gamma})$ with $c_{\gamma} = 10^3$. Here, $\mathcal{G}(\cdot; a, b)$ is the Gamma \ac{pdf} with the shape parameter $a$ and the scale parameter $b$. The surveillance region spans $[0, 50] \times [0, 50]$, \textcolor{black}{with the \ac{roi}} defined as $\Set{X} = [0, 50] \times [0, 50] \times \mathbb{R}^2$. The \textcolor{black}{kinematic birth \ac{pdf}} is given by $f_{\text{B}}(\V{x}) = f_{\text{B}}(\V{p}) f_{\text{B}}(\V{v})$ with $f_{\text{B}}(\V{p})$ uniformly distributed over the surveillance region and $f_{\text{B}}(\V{v})$ being zero-mean Gaussian with covariance $0.25\M{I}_2$. The expected number of new objects is set to $\mu_{\text{B}} = 10^{-6}$. As discussed in Section~\ref{subsec:birth_model}, we discretize the surveillance region using a step size of 1~m in both dimensions, leading to $Q = 2500$ non-overlapping regions $\Set{X}_q, q \in \{1, \dots, Q\}$ for the initialization of new \acp{po}. The signal power of new \acp{po} is uniformly distributed from 0 to 1 and independent across different sensors. \textcolor{black}{This interval specifies the birth-power prior. The Gamma state transitions have unbounded support.} The dynamics of the noise power states are modeled by a Gamma \ac{pdf} $f(\eta_{k}^{(i)} | \eta_{k - 1}^{(i)}) = \mathcal{G}(\eta_{k}^{(i)}; \eta_{k - 1}^{(i)} / c_{\eta}, c_{\eta})$ with $c_{\eta} = 100$. \textcolor{black}{The \textcolor{black}{noise inflation factor} in \eqref{eq:noise_approx} is set to $b_\eta = 1$ in this synthetic experiment.} At the initial time step $k = 0$, \textcolor{black}{the noise-power priors for $\eta_{0}^{(i)}, i \in \{1, 2\}$ are uniform on} $[0, 10^{-3}]$ and \textcolor{black}{the tracker is initialized with no \acp{po}}, i.e., $N_0 = 0$. We set the survival probability to $p_{\text{s}} = 0.9$, the declaration threshold to $T_{\text{dec}} = 0.5$, and the pruning threshold to $T_{\text{pru}} = 10^{-2}$.

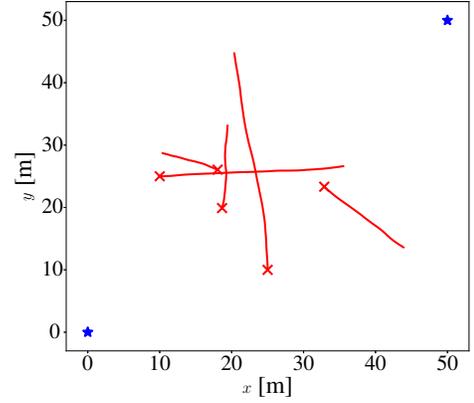
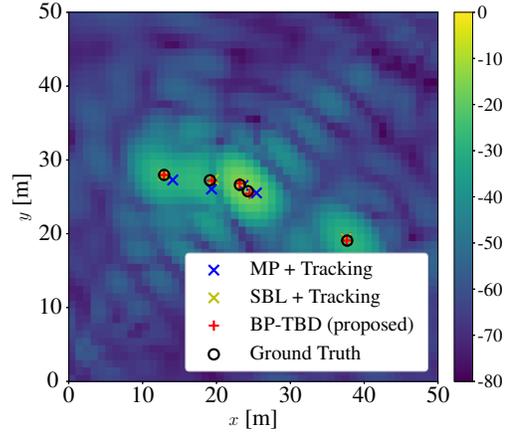
\begin{figure}[!tbp]
    \centering \hspace{-5mm}
    \subfloat[ \hspace{-7mm} ]{\resizebox{0.7\linewidth}{!}{\input{Figs/gt.pgf}}\label{fig:trajectory_syn}} \\
    \subfloat[]{\resizebox{0.75\linewidth}{!}{\import{Figs/rawsignal/}{results_041_eps.pgf}}\label{fig:spectrum_syn}}
    \caption{\rd{Visualization of the synthetic radar tracking scenario in an example simulation run: (a) Trajectories of the objects with their initial positions marked as red crosses and sensor positions marked as blue stars. (b) Tracking results of different methods at time step $k = 41$. \textcolor{black}{A nominal} SNR of $-6$~dB was considered. The Bartlett spectrum obtained by matched filtering is shown as a background.}}
    \label{fig:syn}
    \vspace{-4mm}
\end{figure}

We simulate with 80 time steps, i.e., $k \in \{1, \dots, 80\}$ and five objects. Two of the objects appear at time $k = 10$ and disappear at time $k = 70$. The two objects are initialized at positions $[10 \hspace{1mm} 25]^\T$ and $[25 \hspace{1mm} 10]^\T$ with velocities $[0.5 \hspace{1mm} 0]^\T$ and $[0 \hspace{1mm} 0.5]^\T$, respectively, creating intersecting trajectories at the surveillance region center. The other three objects appear at $k =15$, $20$, and $25$, and disappear at $k = 55$, $60$, and $65$, respectively. Their initial positions are uniformly distributed over $[15, 35] \times [15, 35]$ and velocities are drawn from a zero-mean Gaussian PDF with covariance $0.04\M{I}_2$. Objects also disappear if they leave the surveillance region. We conduct 100 simulation runs. Fig.~\ref{fig:syn}(a) shows representative trajectories. The measurements $\V{z}_{k, j}^{(i)}$ are generated based on \eqref{eq:signal_model}, \eqref{eq:dict_radar}, and \eqref{eq:amp_radar} for $J = 3$ snapshots at each time step. \textcolor{black}{For data generation, we consider two noise cases with $\eta_{k}^{(i)} = 2 \times 10^{-14}$ and $\eta_{k}^{(i)} = 5 \times 10^{-14}$, respectively, for all sensors and time steps. For an object at the surveillance region center $[25 \hspace{1mm} 25]^\T$, the corresponding single-object \acp{snr} before array gain, $10\log_{10}(|\varrho^{(i)}|^2/\eta_k^{(i)})$, are approximately $-6$ and $-10$~dB, respectively.} \rd{As shown in Fig.~\ref{fig:snr_syn}, the \ac{snr} is dependent on the positions of objects.}

\begin{figure}[!tbp]
    \centering
    \subfloat[ \hspace{-5mm} ]{\resizebox{0.45\linewidth}{!}{\import{Figs/snr/}{snr_sensor1_eps.pgf}}} \hspace{1mm}
    \subfloat[ \hspace{-7mm} ]{\resizebox{0.45\linewidth}{!}{\import{Figs/snr/}{snr_vs_time_sensor1.pgf}}} \\
    \subfloat[ \hspace{-5mm} ]{\resizebox{0.45\linewidth}{!}{\import{Figs/snr/}{snr_sensor2_eps.pgf}}} \hspace{1mm}
    \subfloat[ \hspace{-7mm} ]{\resizebox{0.45\linewidth}{!}{\import{Figs/snr/}{snr_vs_time_sensor2.pgf}}} 
    \caption{\rd{Visualization of the SNR levels for the synthetic radar tracking scenario with a nominal SNR of $-6$~dB: (a) SNR map of sensor 1, (b) SNR levels at the true object positions over time for sensor 1, (c) SNR map of sensor 2, (d) SNR levels at the true object positions over time for sensor 2. Crosses in (a) and (c) mark the starting point of the objects.}}
    \label{fig:snr_syn}
    \vspace{-5mm}
\end{figure}

\begin{figure}[!tbp]
    \centering
    \subfloat[]{\resizebox{0.9\linewidth}{!}{\input{Figs/gospa_-6dB/gospa.pgf}}} \\
    \subfloat[]{\resizebox{0.9\linewidth}{!}{\input{Figs/gospa_-10dB/gospa.pgf}}}
    \caption{GOSPA error of different tracking methods for synthetic radar tracking averaged over 100 simulation runs at (a) $-6$~dB and (b) $-10$~dB. \rd{Note that SNR levels refer to the center of the \ac{roi}.}}
    \label{fig:gospa_syn}
    \vspace{-2mm}
\end{figure}
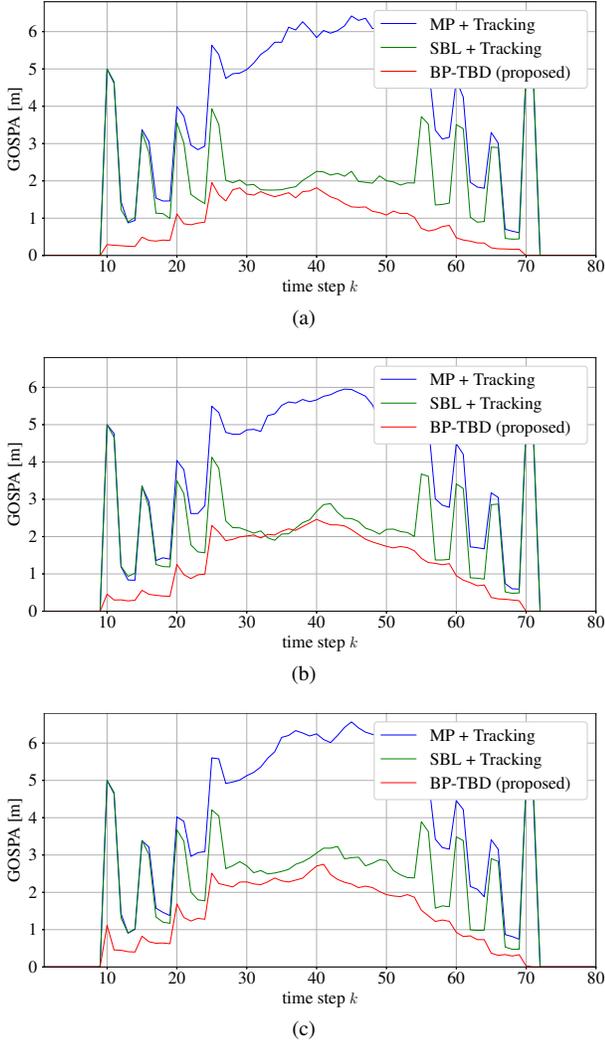

\begin{figure}[!tbp]
    \centering
    \subfloat[ \hspace{-5mm} ]{\resizebox{0.45\linewidth}{!}{\input{Figs/gospa_-10dB/gospa_total.pgf}}} \hspace{1mm}
    \subfloat[ \hspace{-7mm} ]{\resizebox{0.45\linewidth}{!}{\input{Figs/gospa_-10dB/gospa_loc.pgf}}} \\
    \subfloat[ \hspace{-5mm} ]{\resizebox{0.45\linewidth}{!}{\input{Figs/gospa_-10dB/gospa_false.pgf}}} \hspace{1mm}
    \subfloat[ \hspace{-7mm} ]{\resizebox{0.45\linewidth}{!}{\input{Figs/gospa_-10dB/gospa_miss.pgf}}} 
    \caption{\rd{\textcolor{black}{GOSPA and its decomposition for synthetic radar tracking, averaged over 100 simulation runs (nominal SNR $=-10$~dB):}
    (a) GOSPA, (b) \textcolor{black}{Localization error}, (c) \textcolor{black}{False-object error}, (d) \textcolor{black}{Missed-object error}.}}
    \label{fig:gospa_syn_contributions}
    \vspace{-5mm}
\end{figure}

We employ methods based on the \ac{dtt} strategy, comprising frontend detectors paired with a conventional tracking method as a reference. For object detection, we utilize two distinct detectors: \ac{mp} \cite{MalZha:J93} and \ac{sbl} \cite{GemNanGer:J19,NanGemGerHodMec:J19,SanLeiRao:J22}. \rd{These detectors provide \textcolor{black}{point measurements} to the \ac{bp}-based \ac{mot} method \cite{MeyKroWilLauHlaBraWin:J18} that has been implemented using particles. In the scenario with \textcolor{black}{nominal \ac{snr} $-6$~dB}, the detection probability is 0.763 for matching pursuit as well as 0.925 for SBL, and the false alarm rate is  0.502 for matching pursuit and 0.046 for SBL.  In the scenario with \textcolor{black}{nominal \ac{snr} $-10$~dB}, the detection probability is 0.759 for matching pursuit as well as 0.917 for SBL, and the false alarm rate is  0.515 for matching pursuit and 0.079 for SBL. Statistics were obtained by assigning detections to ground truth using the Hungarian assignment algorithm with a distance threshold of 1.} 

We refer to the two variants as ``MP + BP-DTT'' and ``SBL + BP-DTT'', respectively. \textcolor{black}{The pairing of \ac{sbl} with detect-then-track \ac{mot} in \cite{MeyKroWilLauHlaBraWin:J18} was first introduced in \cite{LiLeiVenTuf:J22}.} Our proposed ``BP-\ac{tbd}'' method is implemented using particle-based \ac{bp} \cite{IhlMca:09}, \textcolor{black}{with 30,000 kinematic particles per \ac{po}, 1000 signal-power particles per \ac{po} and dictionary, and 300 noise-power particles per dictionary. For all three scenarios, prediction particles are propagated using the stated transition models, and birth particles are drawn within the selected grid cells. Particle locations remain fixed during the \ac{bp} measurement iterations; only their weights are updated. Systematic resampling is performed after belief calculation at each tracking time step. We perform at most 10 \ac{bp} iterations, with an early-stopping threshold of $10^{-6}$ on the mean squared change in the incoming-message expectations of $r_{k,n}\gamma_{k,n}^{(i)}$, averaged over \acp{po}, dictionaries, and snapshots.} \rd{We also compare to the particle-based \textcolor{black}{labeled multi-Bernoulli (LMB)} \ac{tbd} \ac{mot} method \cite{PapKim:J15}, implemented here with 30,000 particles and referred to as ``LMB-TkBD''.} Conventional BP-based tracking \cite{MeyKroWilLauHlaBraWin:J18} is also implemented using particle-based \ac{bp} \cite{IhlMca:09} with 30,000 particles. \textcolor{black}{To further reduce complexity, we initialize new \acp{po} only on grid cells selected using \ac{mp} detections. The number of retained \acp{po} can differ from that of MP + BP-DTT because inference and pruning are performed separately.} The tracking performance is evaluated using the \ac{gospa} \cite{RahGarSve:17} metric \textcolor{black}{using Euclidean distance between the \ac{2d} positions as the base distance, with cutoff parameter $c = 1$~m, order $p = 1$, and $\alpha = 2$. The resulting \ac{gospa} values are expressed in meters.} \textcolor{black}{At each time step, we average \ac{gospa} over the 100 simulation runs; the resulting curve is shown in Fig.~\ref{fig:gospa_syn}.} \textcolor{black}{In this scenario, BP-\ac{tbd} achieves the lowest time-averaged \ac{gospa} for both noise levels.}  \rd{Fig.~\ref{fig:gospa_syn_contributions} shows different error contributions together with the \textcolor{black}{total GOSPA}. \textcolor{black}{The proposed method reduces each error contribution on average over the evaluated tracking interval. These results are consistent with the benefit of retaining sensor-data information in the tracking updates, whereas \ac{dtt} operates on detector-generated point measurements. Our implementation uses \ac{mp} detections to select birth cells.}} In the radar scenario with nominal \ac{snr} $-10$~dB, increasing $c$ from $1$~m to $5$~m or $10$~m does lead to very similar time-averaged \ac{gospa} curves (results not shown)\vspace{-.5mm}.

The qualitative tracking result for \textcolor{black}{a nominal \ac{snr} of $-6$~dB} at a single time step is shown in Fig.~\ref{fig:syn}(b). Here, the background in Fig.~\ref{fig:syn}(b) is the ``Bartlett'' spectrum, i.e., \textcolor{black}{a product of sensor-specific matched-filter powers, averaged over trial velocities and defined as} \rd{$\frac{1}{|\Set{V}|}\sum_{\V{v} \in \Set{V}} \prod_{i = 1}^I \sum_{j = 1}^J \Vert \big( \V{a}^{(i)}(\V{p}, \V{v}) \big)^\CH \V{z}_{k, j}^{(i)} \Vert_2^2$} \rd{\textcolor{black}{evaluated at $50 \times 50 = 2500$ discrete positions $\V{p}$. Here, $\Set{V} \subset \mathbb{R}^2$ is the finite set of trial velocities used for visualization, and $|\Set{V}|$ denotes its cardinality. Snapshot powers are summed for each sensor before the product across sensors is formed.}}

\begin{figure}[!tbp]
    \centering
    \subfloat[]{\resizebox{0.49\linewidth}{!}{\input{Figs/noise_var_sensor1.pgf}}}
    \subfloat[]{\resizebox{0.49\linewidth}{!}{\input{Figs/noise_var_sensor2.pgf}}}
    \caption{\rd{\textcolor{black}{Estimated noise variances $\hat{\eta}_{k}^{(i)}$ and true noise variances $\eta_{k}^{(i)}$} averaged over \textcolor{black}{100 simulation runs} for (a) the sensor at $[0 \hspace{1mm} 0]^\T$ and (b) the sensor at $[50 \hspace{1mm} 50]^\T$. \textcolor{black}{A nominal} SNR of $-6$~dB was considered.}}
    \label{fig:noise_syn}
    \vspace{-2mm}
\end{figure}
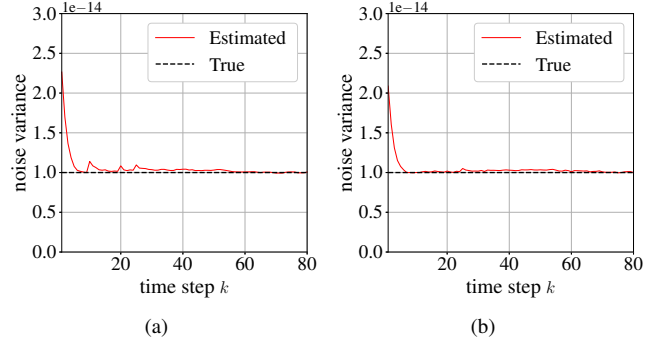

\textcolor{black}{In the example shown in Fig.~\ref{fig:syn}(b), BP-\ac{tbd} estimates both closely spaced objects, whereas MP + BP-DTT misses one of them.} \textcolor{black}{Fig.~\ref{fig:noise_syn} shows the noise-variance estimates $\hat{\eta}_k^{(i)} = \int \eta_k^{(i)} \tilde{f}(\eta_k^{(i)}) \mathrm{d}\eta_k^{(i)}$, averaged over 100 simulation runs. After the initial transient, the mean estimates approach the true noise variances.}



\subsection{High-SNR Passive Acoustic Scenario} \label{subsec:exp_s96_high}

In this section, we evaluate the proposed method using real data in a shallow-water underwater scenario following the matched-field processing paradigm \cite{JenKupPorSchTol:B11,BagKupMik:J93}. \textcolor{black}{We consider a 20-minute segment from Event S5 of the SWellEx-96 passive acoustic dataset} \cite{BooAbaSchHod:J00}\textcolor{black}{, sampled at 1500~Hz.} A mobile source transmits seven tonals at frequencies $f_i \in \{49, 79, 112, 148, 201, 283, 388\}$~Hz simultaneously. \textcolor{black}{Signals are recorded by a \ac{vla} subarray with $M = 21$ elements at a known position, with nominal depths from $94.125$~m to $212.25$~m.} The seven tones define $I = 7$ distinct dictionaries, each corresponding to one tonal transmission. This set of tones is referred to as the ``high tonal set''. The measurements $\V{z}_{k, j}^{(i)}$ are extracted from \textcolor{black}{an 8192-point \ac{stft} using a Hamming window and 75\% overlap} applied to the received acoustic signals. In particular, frequency bins for $i \in \{1, \dots, I\}$ are extracted by taking the Doppler shift into account. Doppler compensation is performed by selecting the maximum-power bin among the $\pm 5$ \ac{dft} bins adjacent to the nominal \textcolor{black}{DFT} bin. \textcolor{black}{We group three consecutive \ac{stft} frames into $J = 3$ snapshots per time step, resulting in a total of 292 time steps. The frames have distinct indices across tracking steps, but their overlapping windows share received-data samples and can introduce statistical dependence across snapshots and steps. We neglect this dependence in the factorized measurement likelihood.} \textcolor{black}{The interval between consecutive tracking time steps is $4.096$ seconds.} We make use of a range-independent geoacoustic model \cite{JenKupPorSchTol:B11} of the considered shallow water waveguide \cite{BooAbaSchHod:J00}. Here, $\V{a}^{(i)}(x_{\mathrm{r}}, x_{\mathrm{d}}) \in \mathbb{C}^{M}$ corresponds to the complex wavefield of a source at frequency $f_i$, located at a range of $x_{\mathrm{r}}$, and a depth of $x_{\mathrm{d}}$, received by the $M$-element \ac{vla}. This wavefield of the underwater environment was computed based on the range-independent geoacoustic model and by using the KRAKEN normal mode program \cite{Por:91}. In particular, we use KRAKEN to precompute the propagating modes of the range-independent acoustic waveguide and then, for each relevant $(x_{\mathrm{r}}, x_{\mathrm{d}})$ value, synthesize the wavefield from modes during runtime. The mirage effect in shallow water \cite{DspMurHodBooSch:J99} as well as an array tilt of $1^\circ$ has been taken into account. \textcolor{black}{For the tracking likelihood, we normalize each basis vector $\V{a}^{(i)}(x_{\mathrm{r}}, x_{\mathrm{d}})$ to unit Euclidean norm.} \textcolor{black}{In this matched-field processing approach, the received array data are coherently compared with modeled wavefields for candidate source positions, exploiting the relative phases and amplitudes across array elements \cite{JenKupPorSchTol:B11,GemNanGerHod:J17,GemNanGer:J19}.} A similar concept is referred to as field focusing in the radar literature. \textcolor{black}{In radar field focusing \cite{Van:B01}, an analytic dictionary can account for array geometry and spherical-wave propagation. In contrast, matched-field processing uses a numerical propagation model to represent multipath, refraction, and boundary interactions in the underwater environment \cite{JenKupPorSchTol:B11}.}



\begin{figure}[!tbp]
    \centering
    \resizebox{0.9\linewidth}{!}{\import{Figs/rawsignal/}{results_k246_eps.pgf}} 
    \caption{Bartlett spectrum of \textcolor{black}{the high tonal set in the artificially augmented two-source scenario} in dB at time $k = 246$. The red circles are the true positions of the two sources.}
    \label{fig:spectrum_high}
    \vspace{-2mm}
\end{figure}
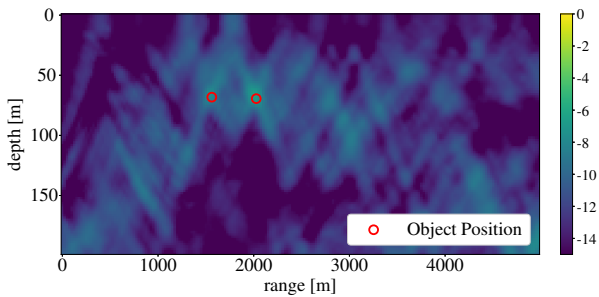

To illustrate that our proposed method is capable of tracking multiple objects, \textcolor{black}{we construct an artificially augmented two-source scenario from the field measurements.} \textcolor{black}{For each tone, we combine the original moving-source measurements with a repeated copy of the final \ac{stft} snapshot in the frequency domain, \textcolor{black}{applying independent random phases, uniform on $[0,2\pi)$, to each component at every \ac{stft} frame and tone. Each phase is common across array elements.} We scale the moving component using the data-matrix norms so that its received power, averaged over all \ac{stft} snapshots and summed over array elements, is $3$~dB below that of the static component \cite{GemNanGerHod:J17}. This creates a scenario with one static and one moving source transmitting at the same frequencies.}  \textcolor{black}{At time step $k$,} the goal is to estimate the number of sources (two in this case) as well as their ranges from the \ac{vla} and their depths using measurements $\V{z}_{1 : k}$. \textcolor{black}{The product-fused Bartlett spectrum, defined as} $\frac{1}{IJ} \prod_{i = 1}^I \sum_{j = 1}^J \Vert \big( \V{a}^{(i)}(x_{\mathrm{r}}, x_{\mathrm{d}}) \big)^\CH \V{z}_{k, j}^{(i)} \Vert_2^2 $, at time $k = 246$, is shown in Fig.~\ref{fig:spectrum_high}. \rd{The Bartlett spectrum of the measurement is shown for a grid of 200 discrete values of $x_{\mathrm{r}}$ and 100 discrete values of $x_{\mathrm{d}}$.}





\begin{figure}[!tbp]
    \centering
    \hspace{-5mm} \resizebox{0.9\linewidth}{!}{\input{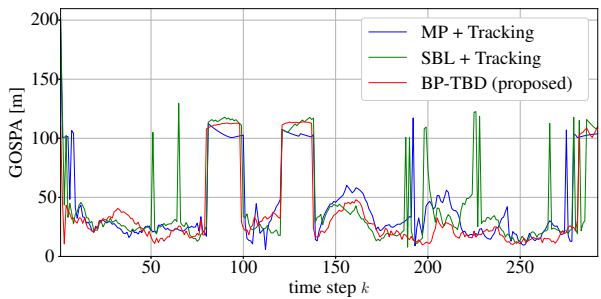}} 
    \caption{GOSPA error in the \textcolor{black}{artificially augmented two-source high-SNR passive acoustic scenario}. Different tracking methods based on the high tonal set are considered.}
    \label{fig:gospa_high}
    \vspace{-2mm}
\end{figure}




\begin{figure}[!tbp]
    \centering
    \resizebox{0.9\linewidth}{!}{\input{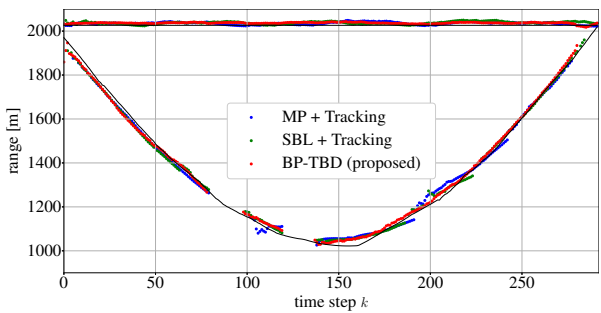}} \\
    \caption{Range estimates of different methods in the \textcolor{black}{artificially augmented two-source high-SNR passive acoustic scenario}. The true range of the two sources is indicated by black lines.}
    \label{fig:qualitative_high}
    \vspace{-2mm}
\end{figure}

The kinematic state $\V{x}_{k, n} \in \mathbb{R}^4$ of \ac{po} $n \in \{1, \dots, N_k\}$ consists of its range $x_{\mathrm{r}, k, n}$, depth $x_{\mathrm{d}, k, n}$, range rate $v_{\mathrm{r}, k, n}$, and depth rate $v_{\mathrm{d}, k, n}$. A constant rate model $\V{x}_{k, n} = \M{F}\V{x}_{k - 1, n} + \M{W}\textcolor{black}{\V{q}_{k, n}}$ \cite[Ch. 4]{ShaKirLi:B02} is employed to describe the dynamics of the kinematic states, where $\textcolor{black}{\V{q}_{k, n}} \in \mathbb{R}^2$ is the zero-mean Gaussian driving noise with covariance $\text{diag}(10^{-2}, 10^{-5})$. \textcolor{black}{The driving noises are independent across objects and time.} \textcolor{black}{The matrices $\M{F}$ and $\M{W}$ have the forms specified in Section~\ref{subsec:exp_radar}, with $\Delta t$ given by the interval between consecutive tracking time steps.} The state-transition of the signal power for each dictionary is assumed to be independent, i.e., $f(\V{\gamma}_{k, n} | \V{\gamma}_{k - 1, n} ) = \prod_{i = 1}^I f(\gamma_{k, n}^{(i)} | \gamma_{k - 1, n}^{(i)} )$, and the individual state-transition \acp{pdf} are given by a Gamma \ac{pdf} $f(\gamma_{k, n}^{(i)} | \gamma_{k - 1, n}^{(i)}) = \textcolor{black}{\mathcal{G}(\gamma_{k, n}^{(i)}; c_{\gamma}, \gamma_{k - 1, n}^{(i)} / c_{\gamma})}$ with $c_{\gamma} = 10^4$. The \ac{roi} is defined as $\Set{X} = [0, 5000] \times [0, 200] \times \mathbb{R}^2$, corresponding to a maximum range of 5000~m and depth of 200~m. The expected number of new objects is set to $\mu_{\text{B}} = 10^{-4}$, with a spatial \ac{pdf} for range and depth that is uniform on $[0, 5000] \times [0, 200]$. The range rate and depth rate are modeled as independent zero-mean Gaussian \acp{pdf}, with standard deviations of $4 \text{m/s}$  and $1 \text{m/s}$, respectively. The disjoint sets $\Set{X}_q, q \in \{1, \dots, Q\}$ are constructed by a range and depth discretization with intervals 25~m and 2~m, resulting in a 200-by-100 grid over the \ac{roi}. \textcolor{black}{We choose a uniform birth-power prior on $[0,1]$, independently across tones.} \textcolor{black}{The evolution of the measurement noise variance is modeled} by a Gamma \ac{pdf} $f(\eta_{k}^{(i)} | \eta_{k - 1}^{(i)}) = \textcolor{black}{\mathcal{G}(\eta_{k}^{(i)}; c_{\eta}, \eta_{k - 1}^{(i)} / c_{\eta})}$ with $c_{\eta} = 10^2$. The \textcolor{black}{noise inflation factor} in \eqref{eq:noise_approx} is set to $b_\eta = 10$ \textcolor{black}{to improve robustness to model mismatch}. \textcolor{black}{The initial priors for the noise variances $\eta_{0}^{(i)}, i \in \{1, \dots, I\}$ are uniform on} $[0, 2 \times 10^{-4}]$, and we set $N_0 = 0$. Particle-based \ac{bp} \cite{IhlMca:09} is used for the implementation of the proposed method, \textcolor{black}{using 10,000 kinematic particles per \ac{po}, 1000 signal-power particles per \ac{po} and dictionary, and 1000 noise-power particles per dictionary.} In addition, we set the survival probability to $p_{\text{s}} = 0.95$, the declaration threshold to $T_{\text{dec}} = 0.5$, and the pruning threshold to $T_{\text{pru}} = 10^{-2}$.



Consistent with our synthetic experiment, we adopt \ac{dtt} as our baseline, with \ac{mp} \cite{MalZha:J93} and \ac{sbl} \cite{GemNanGerHod:J17,NanGemGerHodMec:J19} as object detectors combined with \ac{bp}-based \ac{mot} \cite{MeyKroWilLauHlaBraWin:J18}, termed MP + BP-DTT and SBL + BP-DTT, respectively. Also here, \textcolor{black}{the measurements are strongly correlated. When used with likelihoods that assume independent measurement components, conventional \ac{tbd} methods such as \cite{SalBir:01,OrtFit:J02,StrGraWal:J02,BoeDri:J04,RisVoVoFar:J13,RabRicLep:12,DavRutChe:J12,KimUneMul:J20,UneHorMulMas:J23,DavGae:B18,ItoGod:J20,LiaKroMey:J23,VoVoPhaSut:10,RisRosKimWanWil:J20,KroWilMey:21,KimRisGuaRos:21,DavGar:J24} do not represent these correlations.} \textcolor{black}{For the acoustic \ac{dtt} baselines, the assumed detection probability is 0.9 and the mean number of false point detections per tracking time step is 1.}

The object detectors provide range and depth measurements.  \ac{bp}-based \ac{mot} \cite{MeyKroWilLauHlaBraWin:J18} is implemented using 30,000 particles. For the proposed BP-\ac{tbd} method, we only initialize new \acp{po} for grid cells with an \ac{mp} detection to reduce computational complexity. The tracking performance is evaluated using the \ac{gospa} \cite{RahGarSve:17} metric \textcolor{black}{using Euclidean distance between range-depth vectors as the base distance, with cutoff parameter $c = 200$~m, order $p = 1$, and $\alpha = 2$. The resulting \ac{gospa} values are expressed in meters.} \ac{gospa} vs time is shown in Fig.~\ref{fig:gospa_high}. \rd{\textcolor{black}{The time-averaged \ac{gospa} is} $44.6$ for MP + BP-DTT, $44.6$ for SBL + BP-DTT, and $38.0$ for BP-\ac{tbd}. \textcolor{black}{Our results demonstrate that the proposed method achieves the lowest time-averaged \ac{gospa} in this scenario.}} A visualization of tracking results is shown in Fig.~\ref{fig:qualitative_high}. \textcolor{black}{The proposed method achieves lower \ac{gospa} than the \ac{dtt} baselines around $k = 200$. Sidelobes and model mismatch can degrade detector performance and may contribute to the observed differences. BP-\ac{tbd} retains the sensor-data likelihood during tracking updates, although \ac{mp} detections are used to initialize new \acp{po}.} Note that the \ac{gospa} error peaks around time steps 90 and 140 are related to the time steps where sources stop transmitting the aforementioned tones.

\subsection{Low-SNR Passive Acoustic Scenario} \label{subsec:exp_s96_low}

We further extend our experiment to a more challenging low-\ac{snr} scenario. A 10-minute segment is extracted from the same dataset as in Section~\ref{subsec:exp_s96_high}. The considered tonal signals are at frequencies $f_i \in \{52, 82, 115, 151, 204, 286, 391\}$~Hz. \textcolor{black}{The transmitted source level of the ``low tonal set'' is approximately $26$~dB below that of the ``high tonal set'' used in Section~\ref{subsec:exp_s96_high}.} To increase SNR, we adopt the range-coherent model in \cite{AkiKup:J21} that relies on a \textcolor{black}{synthetic (virtual) array formulation}. \textcolor{black}{Let $\V{y}_{m}^{(i)} \in \mathbb{C}^{M}$ denote the processed tonal vector for a selected frequency bin in \ac{stft} frame $m$. At tracking step $k$, synthetic snapshot $j$ stacks $N_{\mathrm{syn}}$ such vectors as $\V{z}_{\mathrm{syn},k,j}^{(i)} = [\V{y}_{m_{k,j,1}}^{(i)\T} \cdots \V{y}_{m_{k,j,N_{\mathrm{syn}}}}^{(i)\T}]^\T$, with $m_{k,j,\ell}=2(k-1)+j+J(\ell-1)$, $j\in\{1,\dots,J\}$, and $\ell\in\{1,\dots,N_{\mathrm{syn}}\}$. We use $J=4$ and $N_{\mathrm{syn}}=5$. The \ac{stft} frame hop is $8192/1500$~s, so adjacent subarrays are separated by $\tau_{\mathrm{syn}}=4 \times (8192/1500)$~s, approximately $21.845$~s. For a source at range $x_{\mathrm{r}}$ in the initial subarray, with constant range rate $v_{\mathrm{r}}$ and constant depth $x_{\mathrm{d}}$, the synthetic array response is}
\begin{equation}
    \V{a}_{\mathrm{syn}}^{(i)}(x_{\mathrm{r}}, x_{\mathrm{d}}, v_{\mathrm{r}}) = \begin{bmatrix}
        \V{a}^{(i)}(x_{\mathrm{r}}, x_{\mathrm{d}}) \\
        \V{a}^{(i)}(x_{\mathrm{r}} + v_{\mathrm{r}} \tau_{\mathrm{syn}}, x_{\mathrm{d}}) \\
        \vdots \\
        \V{a}^{(i)} \big(x_{\mathrm{r}} + (N_{\mathrm{syn}} - 1) v_{\mathrm{r}} \tau_{\mathrm{syn}}, x_{\mathrm{d}} \big)
    \end{bmatrix} \nn
\end{equation}
which generates a planar synthetic array with the source at the endfire. Note that  $\V{a}^{(i)}(x_{\mathrm{r}}, x_{\mathrm{d}}) \in \mathbb{C}^{M}$ is the wavefield model discussed in Sec.~\ref{subsec:exp_s96_high}. \textcolor{black}{Each $M$-element basis vector is normalized to unit Euclidean norm before stacking. The resulting stacked basis vector therefore has squared Euclidean norm $N_{\mathrm{syn}}$.} \textcolor{black}{Before stacking, we reference the processed tonal coefficients to a common phase origin by multiplying the coefficient at frame time $\tau$ by $\mathrm{e}^{-j2\pi f_i(\tau-\tau_0)}$, where $\tau_0$ is the first frame time. Propagation phase is retained in the modeled wavefields. Applying \eqref{eq:signal_model_po} to each synthetic snapshot assumes a common complex source amplitude across its $N_{\mathrm{syn}}$ subarrays after this phase correction. This enables coherent integration using the \textcolor{black}{synthetic planar array}.} \textcolor{black}{\textcolor{black}{Depth} is held constant within the aperture, although the depth-rate state governs its evolution between tracking steps. The stacked vectors have dimension $M_{\mathrm{syn}}=MN_{\mathrm{syn}}=105$; this replaces $M$ in \eqref{eq:signal_model_po} and its covariance matrices, including the noise term $\eta_k^{(i)}\M{I}_{M_{\mathrm{syn}}}$. Processing each aperture requires all its constituent frames to be available.} The segment that we consider for the experiment contains 45 time steps with a time interval of \textcolor{black}{approximately $10.9$ seconds}.

The statistical model and parameter setup are similar to those used for the high tonal set discussed in Section~\ref{subsec:exp_s96_high}. We set the power and noise state-transition parameters to $c_\gamma = 1000$ and $c_\eta = 10$. The \textcolor{black}{noise inflation factor} is set to $b_\eta = 5$. The expected number of new objects is set to $\mu_{\text{B}} = 10^{-4}$. \textcolor{black}{The initial priors for the noise variances $\eta_{0}^{(i)}, i \in \{1, \dots, I\}$ are uniform over} $[0, 6 \times 10^{-3}]$ and $N_0 = 0$. \textcolor{black}{We use 3000 kinematic particles per \ac{po}, 1000 signal-power particles per \ac{po} and dictionary, and 300 noise-power particles per dictionary.} Other parameters remain unchanged compared to Sec.~\ref{subsec:exp_s96_high}.

To extract measurements from the received acoustic signals, we first compute \textcolor{black}{the 16384-point \ac{stft} spectrum using a Hamming window and 50\% overlap at the same 1500~Hz sampling rate.} \textcolor{black}{Each update uses 20 consecutive \ac{stft} frames arranged into $J = 4$ synthetic snapshots with $N_{\mathrm{syn}} = 5$ subarrays each. The aperture advances by two frames per tracking step, reusing 18 frames between consecutive updates.} Since the \ac{snr} in the low tonal set is too low, we are not able to obtain $\V{z}_{k, j}^{(i)}$ by choosing \ac{dft} bins based on power as in the high \ac{snr} case (cf. Section~\ref{subsec:exp_s96_high}). Instead, we propose selecting bins for each \ac{po} separately based on their estimated range rate. \textcolor{black}{At each update, the range-rate estimate used for bin selection is the weighted mean of the predicted kinematic particles for a legacy \ac{po}. For a newly introduced \ac{po}, it is the mean of the birth particles, whose range-rate values are sampled around the corresponding \ac{mp} detection. Denoting this estimate by $\bar{v}_{\mathrm{r},k,n}$, the predicted received frequency is approximately $(1-\bar{v}_{\mathrm{r},k,n}/c)f_i$, where $c$ is the sound speed and positive range rate corresponds to increasing range. We select the nearest \ac{dft} bin.} \textcolor{black}{In this approach, an outgoing signal-power or kinematic-state message uses the bin selected for its recipient \ac{po}. The recipient's interference covariance in \eqref{eq:cov_lambda} and \eqref{eq:cov_kappa} retains contributions only from other \acp{po} selecting the same bin on tone $i$. This is a recipient-dependent mask on the covariance sums; the second-moment matrices in \eqref{eq:kinematic_approx} remain unchanged. The noise-power state $\eta_k^{(i)}$ remains shared across the selected bins of tone $i$.}

The tracking performance is again evaluated using the \ac{gospa} \cite{RahGarSve:17} metric \textcolor{black}{with cutoff parameter $c = 500$~m, order $p = 1$, and $\alpha = 2$, using the same base distance as in Section~\ref{subsec:exp_s96_high}.} \textcolor{black}{The MP + BP-DTT and SBL + BP-DTT baselines use the same detector families and \ac{bp}-based tracking backend as in the high tonal scenario, with \textcolor{black}{synthetic-array processing} for the low tonal data. Their tracking-model parameters remain unchanged.}  \rd{\textcolor{black}{For single-object tracking, we also include an adapted particle-based single-Bernoulli reference \cite{RisVoVoFar:J13}. It uses the proposed multi-dictionary and multi-snapshot measurement model, \textcolor{black}{Gaussian moment matching}, and separate particle clouds for kinematic states, signal power states, and noise power states. We refer to this method as SOT-TkBD.}}

\begin{figure}[!tbp]
    \centering
    \resizebox{0.9\linewidth}{!}{\import{Figs/rawsignal/}{results_k001_eps.pgf}} 
    \caption{Bartlett spectrum in dB at time $k = 1$ in the low-SNR passive acoustic scenario with \textcolor{black}{synthetic array processing}. The red circle is the true position of the source.}
    \label{fig:spectrum_low}
    \vspace{0mm}
\end{figure}
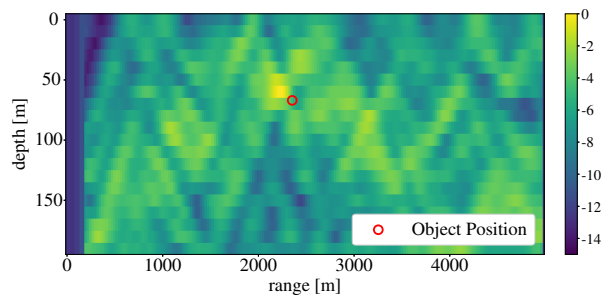

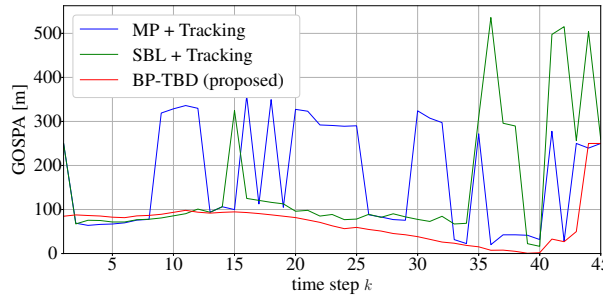
\begin{figure}[!tbp]
    \centering
    \resizebox{0.9\linewidth}{!}{\input{Figs/gospa_sot.pgf}}
    \caption{GOSPA error of different tracking methods in the low-SNR passive acoustic scenario.}
    \label{fig:gospa_low}
    \vspace{-3mm}
\end{figure}

 To reduce computational complexity, the proposed BP-\ac{tbd} method only initializes new \acp{po} for grid cells with a \ac{mp} detection. The \ac{roi} is defined as $\Set{X} = [0, 5000] \times [0, 200] \times [-3, -1] \times \mathbb{R}$. \textcolor{black}{The kinematic birth \ac{pdf} $f_{\mathrm{B}}(\V{x})$ factors into independent uniform densities for range, depth, and range rate over their bounded intervals in $\Set{X}$, and a zero-mean Gaussian density for depth rate with standard deviation $1$~m/s.} The disjoint sets $\Set{X}_q, q \in \{1, \dots, Q\}$ are constructed by discretizing range, depth, and range rate with intervals 25~m, 10~m, and 0.1~m/s, resulting in a $200 \times 20 \times 20$ voxel grid covering the \ac{roi}. The Bartlett spectrum and results are shown in Figs.~\ref{fig:spectrum_low} and \ref{fig:gospa_low}, respectively. \textcolor{black}{BP-\ac{tbd} achieves the lowest time-averaged \ac{gospa} in this segment.} \textcolor{black}{The multiobject \ac{dtt} baselines can produce extra-object declarations. The \ac{gospa} curves jointly penalize localization errors and false or missed objects.} \rd{\textcolor{black}{SOT-TkBD incurs larger \ac{gospa} errors than BP-\ac{tbd} at several later tracking time steps.}}
 \textcolor{black}{All methods exhibit large \ac{gospa} errors near the end of the segment. The source's varying range rate is inconsistent with the constant-rate assumption of the range-coherent model and may contribute to these errors.}\vspace{0mm}


\section{Conclusion} \label{sec:conclusion}

This paper introduced a \textcolor{black}{\ac{bp}}-based method for \textcolor{black}{\ac{tbd}}. \textcolor{black}{BP-\ac{tbd} processes correlated sensor data directly in its tracking updates.} The proposed method introduces a novel measurement model \textcolor{black}{adopted from the sparse signal reconstruction literature} for \ac{tbd}. The new \ac{tbd} measurement model supports multiple concurrent signal models, or \textcolor{black}{``dictionaries,''} as well as \textcolor{black}{multiple conditionally independent ``snapshots''}.  This multi-dictionary formulation offers high flexibility in characterizing \textcolor{black}{sensor-data generation}, e.g., the proposed model can be used in multi-sensor fusion, multi-frequency beamforming, and multi-waveform localization problems. The new measurement model is combined with statistical models for object dynamics, object birth, and \textcolor{black}{time-varying noise powers}. The joint statistical model is represented by a factor graph, enabling efficient \ac{bp}-inference for object declaration and state estimation. \textcolor{black}{Within our \ac{bp} method, we approximate intractable measurement-update messages by moment-matched Gaussian expressions in the measurement vector and evaluate the required state integrals using weighted particles.} Our numerical results in a synthetic radar tracking scenario demonstrate that the proposed BP-\ac{tbd} approach can achieve better tracking performance compared to \textcolor{black}{\ac{dtt}}-based methods that rely on object detectors. Performance is also evaluated in a passive acoustic source localization scenario. Results show that the proposed \ac{tbd} approach is robust to model mismatch (e.g., colored noise) and outperforms reference methods\vspace{0mm}. 

The Gaussian likelihood used in this work follows from the assumed conditionally Gaussian amplitude and noise models. Future work will consider a model for heavy-tailed clutter consisting of a complex Gaussian speckle vector scaled by the square root of an independent Gamma-distributed texture variable \cite{RisRosKimWanWil:J20}. This texture variable can be incorporated into the factor graph as a latent variable that scales the clutter covariance, with the existing noise covariance accounting separately for additive Gaussian noise. The likelihood remains Gaussian conditioned on the texture and the existing states. This provides a basis for developing corresponding BP message updates.

The \acs{mp}-based birth selection used in our implementation can delay or prevent initialization of weak objects in unselected regions. Future work will therefore investigate scalable birth mechanisms that exploit sensor-data evidence without requiring point detections.





\ifCLASSOPTIONcaptionsoff
  \newpage
\fi


\renewcommand{\baselinestretch}{1.02}
\begingroup
\color{black}
\renewcommand{\IEEEbibitemsep}{-0.4pt}
\bibliographystyle{ieeetr}
\bibliography{IEEEabrv,StringDefinitions,SALBooks,SALPapers,Temp}
\endgroup

\end{document}

%% file: Figs/gt.pgf
\begingroup%
\makeatletter%
\begin{pgfpicture}%
\pgfpathrectangle{\pgfpointorigin}{\pgfqpoint{9.600000in}{8.400000in}}%
\pgfusepath{use as bounding box, clip}%
\begin{pgfscope}%
\pgfsetbuttcap%
\pgfsetmiterjoin%
\definecolor{currentfill}{rgb}{1.000000,1.000000,1.000000}%
\pgfsetfillcolor{currentfill}%
\pgfsetlinewidth{0.000000pt}%
\definecolor{currentstroke}{rgb}{1.000000,1.000000,1.000000}%
\pgfsetstrokecolor{currentstroke}%
\pgfsetdash{}{0pt}%
\pgfpathmoveto{\pgfqpoint{0.000000in}{0.000000in}}%
\pgfpathlineto{\pgfqpoint{9.600000in}{0.000000in}}%
\pgfpathlineto{\pgfqpoint{9.600000in}{8.400000in}}%
\pgfpathlineto{\pgfqpoint{0.000000in}{8.400000in}}%
\pgfpathlineto{\pgfqpoint{0.000000in}{0.000000in}}%
\pgfpathclose%
\pgfusepath{fill}%
\end{pgfscope}%
\begin{pgfscope}%
\pgfsetbuttcap%
\pgfsetmiterjoin%
\definecolor{currentfill}{rgb}{1.000000,1.000000,1.000000}%
\pgfsetfillcolor{currentfill}%
\pgfsetlinewidth{0.000000pt}%
\definecolor{currentstroke}{rgb}{0.000000,0.000000,0.000000}%
\pgfsetstrokecolor{currentstroke}%
\pgfsetstrokeopacity{0.000000}%
\pgfsetdash{}{0pt}%
\pgfpathmoveto{\pgfqpoint{1.268682in}{1.165881in}}%
\pgfpathlineto{\pgfqpoint{9.561111in}{1.165881in}}%
\pgfpathlineto{\pgfqpoint{9.561111in}{8.361111in}}%
\pgfpathlineto{\pgfqpoint{1.268682in}{8.361111in}}%
\pgfpathlineto{\pgfqpoint{1.268682in}{1.165881in}}%
\pgfpathclose%
\pgfusepath{fill}%
\end{pgfscope}%
\begin{pgfscope}%
\pgfpathrectangle{\pgfqpoint{1.268682in}{1.165881in}}{\pgfqpoint{8.292429in}{7.195230in}}%
\pgfusepath{clip}%
\pgfsetbuttcap%
\pgfsetroundjoin%
\definecolor{currentfill}{rgb}{1.000000,0.000000,0.000000}%
\pgfsetfillcolor{currentfill}%
\pgfsetlinewidth{3.011250pt}%
\definecolor{currentstroke}{rgb}{1.000000,0.000000,0.000000}%
\pgfsetstrokecolor{currentstroke}%
\pgfsetdash{}{0pt}%
\pgfsys@defobject{currentmarker}{\pgfqpoint{-0.098209in}{-0.098209in}}{\pgfqpoint{0.098209in}{0.098209in}}{%
\pgfpathmoveto{\pgfqpoint{-0.098209in}{-0.098209in}}%
\pgfpathlineto{\pgfqpoint{0.098209in}{0.098209in}}%
\pgfpathmoveto{\pgfqpoint{-0.098209in}{0.098209in}}%
\pgfpathlineto{\pgfqpoint{0.098209in}{-0.098209in}}%
\pgfusepath{stroke,fill}%
}%
\begin{pgfscope}%
\pgfsys@transformshift{5.414897in}{2.836202in}%
\pgfsys@useobject{currentmarker}{}%
\end{pgfscope}%
\end{pgfscope}%
\begin{pgfscope}%
\pgfpathrectangle{\pgfqpoint{1.268682in}{1.165881in}}{\pgfqpoint{8.292429in}{7.195230in}}%
\pgfusepath{clip}%
\pgfsetbuttcap%
\pgfsetroundjoin%
\definecolor{currentfill}{rgb}{1.000000,0.000000,0.000000}%
\pgfsetfillcolor{currentfill}%
\pgfsetlinewidth{3.011250pt}%
\definecolor{currentstroke}{rgb}{1.000000,0.000000,0.000000}%
\pgfsetstrokecolor{currentstroke}%
\pgfsetdash{}{0pt}%
\pgfsys@defobject{currentmarker}{\pgfqpoint{-0.098209in}{-0.098209in}}{\pgfqpoint{0.098209in}{0.098209in}}{%
\pgfpathmoveto{\pgfqpoint{-0.098209in}{-0.098209in}}%
\pgfpathlineto{\pgfqpoint{0.098209in}{0.098209in}}%
\pgfpathmoveto{\pgfqpoint{-0.098209in}{0.098209in}}%
\pgfpathlineto{\pgfqpoint{0.098209in}{-0.098209in}}%
\pgfusepath{stroke,fill}%
}%
\begin{pgfscope}%
\pgfsys@transformshift{3.193710in}{4.763496in}%
\pgfsys@useobject{currentmarker}{}%
\end{pgfscope}%
\end{pgfscope}%
\begin{pgfscope}%
\pgfpathrectangle{\pgfqpoint{1.268682in}{1.165881in}}{\pgfqpoint{8.292429in}{7.195230in}}%
\pgfusepath{clip}%
\pgfsetbuttcap%
\pgfsetroundjoin%
\definecolor{currentfill}{rgb}{1.000000,0.000000,0.000000}%
\pgfsetfillcolor{currentfill}%
\pgfsetlinewidth{3.011250pt}%
\definecolor{currentstroke}{rgb}{1.000000,0.000000,0.000000}%
\pgfsetstrokecolor{currentstroke}%
\pgfsetdash{}{0pt}%
\pgfsys@defobject{currentmarker}{\pgfqpoint{-0.098209in}{-0.098209in}}{\pgfqpoint{0.098209in}{0.098209in}}{%
\pgfpathmoveto{\pgfqpoint{-0.098209in}{-0.098209in}}%
\pgfpathlineto{\pgfqpoint{0.098209in}{0.098209in}}%
\pgfpathmoveto{\pgfqpoint{-0.098209in}{0.098209in}}%
\pgfpathlineto{\pgfqpoint{0.098209in}{-0.098209in}}%
\pgfusepath{stroke,fill}%
}%
\begin{pgfscope}%
\pgfsys@transformshift{4.477571in}{4.108399in}%
\pgfsys@useobject{currentmarker}{}%
\end{pgfscope}%
\end{pgfscope}%
\begin{pgfscope}%
\pgfpathrectangle{\pgfqpoint{1.268682in}{1.165881in}}{\pgfqpoint{8.292429in}{7.195230in}}%
\pgfusepath{clip}%
\pgfsetbuttcap%
\pgfsetroundjoin%
\definecolor{currentfill}{rgb}{1.000000,0.000000,0.000000}%
\pgfsetfillcolor{currentfill}%
\pgfsetlinewidth{3.011250pt}%
\definecolor{currentstroke}{rgb}{1.000000,0.000000,0.000000}%
\pgfsetstrokecolor{currentstroke}%
\pgfsetdash{}{0pt}%
\pgfsys@defobject{currentmarker}{\pgfqpoint{-0.098209in}{-0.098209in}}{\pgfqpoint{0.098209in}{0.098209in}}{%
\pgfpathmoveto{\pgfqpoint{-0.098209in}{-0.098209in}}%
\pgfpathlineto{\pgfqpoint{0.098209in}{0.098209in}}%
\pgfpathmoveto{\pgfqpoint{-0.098209in}{0.098209in}}%
\pgfpathlineto{\pgfqpoint{0.098209in}{-0.098209in}}%
\pgfusepath{stroke,fill}%
}%
\begin{pgfscope}%
\pgfsys@transformshift{4.385390in}{4.899399in}%
\pgfsys@useobject{currentmarker}{}%
\end{pgfscope}%
\end{pgfscope}%
\begin{pgfscope}%
\pgfpathrectangle{\pgfqpoint{1.268682in}{1.165881in}}{\pgfqpoint{8.292429in}{7.195230in}}%
\pgfusepath{clip}%
\pgfsetbuttcap%
\pgfsetroundjoin%
\definecolor{currentfill}{rgb}{1.000000,0.000000,0.000000}%
\pgfsetfillcolor{currentfill}%
\pgfsetlinewidth{3.011250pt}%
\definecolor{currentstroke}{rgb}{1.000000,0.000000,0.000000}%
\pgfsetstrokecolor{currentstroke}%
\pgfsetdash{}{0pt}%
\pgfsys@defobject{currentmarker}{\pgfqpoint{-0.098209in}{-0.098209in}}{\pgfqpoint{0.098209in}{0.098209in}}{%
\pgfpathmoveto{\pgfqpoint{-0.098209in}{-0.098209in}}%
\pgfpathlineto{\pgfqpoint{0.098209in}{0.098209in}}%
\pgfpathmoveto{\pgfqpoint{-0.098209in}{0.098209in}}%
\pgfpathlineto{\pgfqpoint{0.098209in}{-0.098209in}}%
\pgfusepath{stroke,fill}%
}%
\begin{pgfscope}%
\pgfsys@transformshift{6.580013in}{4.548406in}%
\pgfsys@useobject{currentmarker}{}%
\end{pgfscope}%
\end{pgfscope}%
\begin{pgfscope}%
\pgfpathrectangle{\pgfqpoint{1.268682in}{1.165881in}}{\pgfqpoint{8.292429in}{7.195230in}}%
\pgfusepath{clip}%
\pgfsetbuttcap%
\pgfsetroundjoin%
\definecolor{currentfill}{rgb}{0.000000,0.000000,1.000000}%
\pgfsetfillcolor{currentfill}%
\pgfsetlinewidth{3.011250pt}%
\definecolor{currentstroke}{rgb}{0.000000,0.000000,1.000000}%
\pgfsetstrokecolor{currentstroke}%
\pgfsetdash{}{0pt}%
\pgfsys@defobject{currentmarker}{\pgfqpoint{-0.093403in}{-0.079453in}}{\pgfqpoint{0.093403in}{0.098209in}}{%
\pgfpathmoveto{\pgfqpoint{0.000000in}{0.098209in}}%
\pgfpathlineto{\pgfqpoint{-0.022049in}{0.030348in}}%
\pgfpathlineto{\pgfqpoint{-0.093403in}{0.030348in}}%
\pgfpathlineto{\pgfqpoint{-0.035677in}{-0.011592in}}%
\pgfpathlineto{\pgfqpoint{-0.057726in}{-0.079453in}}%
\pgfpathlineto{\pgfqpoint{-0.000000in}{-0.037513in}}%
\pgfpathlineto{\pgfqpoint{0.057726in}{-0.079453in}}%
\pgfpathlineto{\pgfqpoint{0.035677in}{-0.011592in}}%
\pgfpathlineto{\pgfqpoint{0.093403in}{0.030348in}}%
\pgfpathlineto{\pgfqpoint{0.022049in}{0.030348in}}%
\pgfpathlineto{\pgfqpoint{0.000000in}{0.098209in}}%
\pgfpathclose%
\pgfusepath{stroke,fill}%
}%
\begin{pgfscope}%
\pgfsys@transformshift{1.712919in}{1.551340in}%
\pgfsys@useobject{currentmarker}{}%
\end{pgfscope}%
\begin{pgfscope}%
\pgfsys@transformshift{9.116874in}{7.975652in}%
\pgfsys@useobject{currentmarker}{}%
\end{pgfscope}%
\end{pgfscope}%
\begin{pgfscope}%
\pgfsetbuttcap%
\pgfsetroundjoin%
\definecolor{currentfill}{rgb}{0.000000,0.000000,0.000000}%
\pgfsetfillcolor{currentfill}%
\pgfsetlinewidth{0.803000pt}%
\definecolor{currentstroke}{rgb}{0.000000,0.000000,0.000000}%
\pgfsetstrokecolor{currentstroke}%
\pgfsetdash{}{0pt}%
\pgfsys@defobject{currentmarker}{\pgfqpoint{0.000000in}{-0.048611in}}{\pgfqpoint{0.000000in}{0.000000in}}{%
\pgfpathmoveto{\pgfqpoint{0.000000in}{0.000000in}}%
\pgfpathlineto{\pgfqpoint{0.000000in}{-0.048611in}}%
\pgfusepath{stroke,fill}%
}%
\begin{pgfscope}%
\pgfsys@transformshift{1.712919in}{1.165881in}%
\pgfsys@useobject{currentmarker}{}%
\end{pgfscope}%
\end{pgfscope}%
\begin{pgfscope}%
\definecolor{textcolor}{rgb}{0.000000,0.000000,0.000000}%
\pgfsetstrokecolor{textcolor}%
\pgfsetfillcolor{textcolor}%
\pgftext[x=1.712919in,y=1.068659in,,top]{\color{textcolor}\rmfamily\fontsize{32.000000}{38.400000}\selectfont 0}%
\end{pgfscope}%
\begin{pgfscope}%
\pgfsetbuttcap%
\pgfsetroundjoin%
\definecolor{currentfill}{rgb}{0.000000,0.000000,0.000000}%
\pgfsetfillcolor{currentfill}%
\pgfsetlinewidth{0.803000pt}%
\definecolor{currentstroke}{rgb}{0.000000,0.000000,0.000000}%
\pgfsetstrokecolor{currentstroke}%
\pgfsetdash{}{0pt}%
\pgfsys@defobject{currentmarker}{\pgfqpoint{0.000000in}{-0.048611in}}{\pgfqpoint{0.000000in}{0.000000in}}{%
\pgfpathmoveto{\pgfqpoint{0.000000in}{0.000000in}}%
\pgfpathlineto{\pgfqpoint{0.000000in}{-0.048611in}}%
\pgfusepath{stroke,fill}%
}%
\begin{pgfscope}%
\pgfsys@transformshift{3.193710in}{1.165881in}%
\pgfsys@useobject{currentmarker}{}%
\end{pgfscope}%
\end{pgfscope}%
\begin{pgfscope}%
\definecolor{textcolor}{rgb}{0.000000,0.000000,0.000000}%
\pgfsetstrokecolor{textcolor}%
\pgfsetfillcolor{textcolor}%
\pgftext[x=3.193710in,y=1.068659in,,top]{\color{textcolor}\rmfamily\fontsize{32.000000}{38.400000}\selectfont 10}%
\end{pgfscope}%
\begin{pgfscope}%
\pgfsetbuttcap%
\pgfsetroundjoin%
\definecolor{currentfill}{rgb}{0.000000,0.000000,0.000000}%
\pgfsetfillcolor{currentfill}%
\pgfsetlinewidth{0.803000pt}%
\definecolor{currentstroke}{rgb}{0.000000,0.000000,0.000000}%
\pgfsetstrokecolor{currentstroke}%
\pgfsetdash{}{0pt}%
\pgfsys@defobject{currentmarker}{\pgfqpoint{0.000000in}{-0.048611in}}{\pgfqpoint{0.000000in}{0.000000in}}{%
\pgfpathmoveto{\pgfqpoint{0.000000in}{0.000000in}}%
\pgfpathlineto{\pgfqpoint{0.000000in}{-0.048611in}}%
\pgfusepath{stroke,fill}%
}%
\begin{pgfscope}%
\pgfsys@transformshift{4.674501in}{1.165881in}%
\pgfsys@useobject{currentmarker}{}%
\end{pgfscope}%
\end{pgfscope}%
\begin{pgfscope}%
\definecolor{textcolor}{rgb}{0.000000,0.000000,0.000000}%
\pgfsetstrokecolor{textcolor}%
\pgfsetfillcolor{textcolor}%
\pgftext[x=4.674501in,y=1.068659in,,top]{\color{textcolor}\rmfamily\fontsize{32.000000}{38.400000}\selectfont 20}%
\end{pgfscope}%
\begin{pgfscope}%
\pgfsetbuttcap%
\pgfsetroundjoin%
\definecolor{currentfill}{rgb}{0.000000,0.000000,0.000000}%
\pgfsetfillcolor{currentfill}%
\pgfsetlinewidth{0.803000pt}%
\definecolor{currentstroke}{rgb}{0.000000,0.000000,0.000000}%
\pgfsetstrokecolor{currentstroke}%
\pgfsetdash{}{0pt}%
\pgfsys@defobject{currentmarker}{\pgfqpoint{0.000000in}{-0.048611in}}{\pgfqpoint{0.000000in}{0.000000in}}{%
\pgfpathmoveto{\pgfqpoint{0.000000in}{0.000000in}}%
\pgfpathlineto{\pgfqpoint{0.000000in}{-0.048611in}}%
\pgfusepath{stroke,fill}%
}%
\begin{pgfscope}%
\pgfsys@transformshift{6.155292in}{1.165881in}%
\pgfsys@useobject{currentmarker}{}%
\end{pgfscope}%
\end{pgfscope}%
\begin{pgfscope}%
\definecolor{textcolor}{rgb}{0.000000,0.000000,0.000000}%
\pgfsetstrokecolor{textcolor}%
\pgfsetfillcolor{textcolor}%
\pgftext[x=6.155292in,y=1.068659in,,top]{\color{textcolor}\rmfamily\fontsize{32.000000}{38.400000}\selectfont 30}%
\end{pgfscope}%
\begin{pgfscope}%
\pgfsetbuttcap%
\pgfsetroundjoin%
\definecolor{currentfill}{rgb}{0.000000,0.000000,0.000000}%
\pgfsetfillcolor{currentfill}%
\pgfsetlinewidth{0.803000pt}%
\definecolor{currentstroke}{rgb}{0.000000,0.000000,0.000000}%
\pgfsetstrokecolor{currentstroke}%
\pgfsetdash{}{0pt}%
\pgfsys@defobject{currentmarker}{\pgfqpoint{0.000000in}{-0.048611in}}{\pgfqpoint{0.000000in}{0.000000in}}{%
\pgfpathmoveto{\pgfqpoint{0.000000in}{0.000000in}}%
\pgfpathlineto{\pgfqpoint{0.000000in}{-0.048611in}}%
\pgfusepath{stroke,fill}%
}%
\begin{pgfscope}%
\pgfsys@transformshift{7.636083in}{1.165881in}%
\pgfsys@useobject{currentmarker}{}%
\end{pgfscope}%
\end{pgfscope}%
\begin{pgfscope}%
\definecolor{textcolor}{rgb}{0.000000,0.000000,0.000000}%
\pgfsetstrokecolor{textcolor}%
\pgfsetfillcolor{textcolor}%
\pgftext[x=7.636083in,y=1.068659in,,top]{\color{textcolor}\rmfamily\fontsize{32.000000}{38.400000}\selectfont 40}%
\end{pgfscope}%
\begin{pgfscope}%
\pgfsetbuttcap%
\pgfsetroundjoin%
\definecolor{currentfill}{rgb}{0.000000,0.000000,0.000000}%
\pgfsetfillcolor{currentfill}%
\pgfsetlinewidth{0.803000pt}%
\definecolor{currentstroke}{rgb}{0.000000,0.000000,0.000000}%
\pgfsetstrokecolor{currentstroke}%
\pgfsetdash{}{0pt}%
\pgfsys@defobject{currentmarker}{\pgfqpoint{0.000000in}{-0.048611in}}{\pgfqpoint{0.000000in}{0.000000in}}{%
\pgfpathmoveto{\pgfqpoint{0.000000in}{0.000000in}}%
\pgfpathlineto{\pgfqpoint{0.000000in}{-0.048611in}}%
\pgfusepath{stroke,fill}%
}%
\begin{pgfscope}%
\pgfsys@transformshift{9.116874in}{1.165881in}%
\pgfsys@useobject{currentmarker}{}%
\end{pgfscope}%
\end{pgfscope}%
\begin{pgfscope}%
\definecolor{textcolor}{rgb}{0.000000,0.000000,0.000000}%
\pgfsetstrokecolor{textcolor}%
\pgfsetfillcolor{textcolor}%
\pgftext[x=9.116874in,y=1.068659in,,top]{\color{textcolor}\rmfamily\fontsize{32.000000}{38.400000}\selectfont 50}%
\end{pgfscope}%
\begin{pgfscope}%
\definecolor{textcolor}{rgb}{0.000000,0.000000,0.000000}%
\pgfsetstrokecolor{textcolor}%
\pgfsetfillcolor{textcolor}%
\pgftext[x=5.414897in,y=0.622587in,,top]{\color{textcolor}\rmfamily\fontsize{36.000000}{43.200000}\selectfont \(\displaystyle x\) [m]}%
\end{pgfscope}%
\begin{pgfscope}%
\pgfsetbuttcap%
\pgfsetroundjoin%
\definecolor{currentfill}{rgb}{0.000000,0.000000,0.000000}%
\pgfsetfillcolor{currentfill}%
\pgfsetlinewidth{0.803000pt}%
\definecolor{currentstroke}{rgb}{0.000000,0.000000,0.000000}%
\pgfsetstrokecolor{currentstroke}%
\pgfsetdash{}{0pt}%
\pgfsys@defobject{currentmarker}{\pgfqpoint{-0.048611in}{0.000000in}}{\pgfqpoint{-0.000000in}{0.000000in}}{%
\pgfpathmoveto{\pgfqpoint{-0.000000in}{0.000000in}}%
\pgfpathlineto{\pgfqpoint{-0.048611in}{0.000000in}}%
\pgfusepath{stroke,fill}%
}%
\begin{pgfscope}%
\pgfsys@transformshift{1.268682in}{1.551340in}%
\pgfsys@useobject{currentmarker}{}%
\end{pgfscope}%
\end{pgfscope}%
\begin{pgfscope}%
\definecolor{textcolor}{rgb}{0.000000,0.000000,0.000000}%
\pgfsetstrokecolor{textcolor}%
\pgfsetfillcolor{textcolor}%
\pgftext[x=0.922704in, y=1.431355in, left, base]{\color{textcolor}\rmfamily\fontsize{32.000000}{38.400000}\selectfont 0}%
\end{pgfscope}%
\begin{pgfscope}%
\pgfsetbuttcap%
\pgfsetroundjoin%
\definecolor{currentfill}{rgb}{0.000000,0.000000,0.000000}%
\pgfsetfillcolor{currentfill}%
\pgfsetlinewidth{0.803000pt}%
\definecolor{currentstroke}{rgb}{0.000000,0.000000,0.000000}%
\pgfsetstrokecolor{currentstroke}%
\pgfsetdash{}{0pt}%
\pgfsys@defobject{currentmarker}{\pgfqpoint{-0.048611in}{0.000000in}}{\pgfqpoint{-0.000000in}{0.000000in}}{%
\pgfpathmoveto{\pgfqpoint{-0.000000in}{0.000000in}}%
\pgfpathlineto{\pgfqpoint{-0.048611in}{0.000000in}}%
\pgfusepath{stroke,fill}%
}%
\begin{pgfscope}%
\pgfsys@transformshift{1.268682in}{2.836202in}%
\pgfsys@useobject{currentmarker}{}%
\end{pgfscope}%
\end{pgfscope}%
\begin{pgfscope}%
\definecolor{textcolor}{rgb}{0.000000,0.000000,0.000000}%
\pgfsetstrokecolor{textcolor}%
\pgfsetfillcolor{textcolor}%
\pgftext[x=0.764226in, y=2.716218in, left, base]{\color{textcolor}\rmfamily\fontsize{32.000000}{38.400000}\selectfont 10}%
\end{pgfscope}%
\begin{pgfscope}%
\pgfsetbuttcap%
\pgfsetroundjoin%
\definecolor{currentfill}{rgb}{0.000000,0.000000,0.000000}%
\pgfsetfillcolor{currentfill}%
\pgfsetlinewidth{0.803000pt}%
\definecolor{currentstroke}{rgb}{0.000000,0.000000,0.000000}%
\pgfsetstrokecolor{currentstroke}%
\pgfsetdash{}{0pt}%
\pgfsys@defobject{currentmarker}{\pgfqpoint{-0.048611in}{0.000000in}}{\pgfqpoint{-0.000000in}{0.000000in}}{%
\pgfpathmoveto{\pgfqpoint{-0.000000in}{0.000000in}}%
\pgfpathlineto{\pgfqpoint{-0.048611in}{0.000000in}}%
\pgfusepath{stroke,fill}%
}%
\begin{pgfscope}%
\pgfsys@transformshift{1.268682in}{4.121065in}%
\pgfsys@useobject{currentmarker}{}%
\end{pgfscope}%
\end{pgfscope}%
\begin{pgfscope}%
\definecolor{textcolor}{rgb}{0.000000,0.000000,0.000000}%
\pgfsetstrokecolor{textcolor}%
\pgfsetfillcolor{textcolor}%
\pgftext[x=0.764226in, y=4.001080in, left, base]{\color{textcolor}\rmfamily\fontsize{32.000000}{38.400000}\selectfont 20}%
\end{pgfscope}%
\begin{pgfscope}%
\pgfsetbuttcap%
\pgfsetroundjoin%
\definecolor{currentfill}{rgb}{0.000000,0.000000,0.000000}%
\pgfsetfillcolor{currentfill}%
\pgfsetlinewidth{0.803000pt}%
\definecolor{currentstroke}{rgb}{0.000000,0.000000,0.000000}%
\pgfsetstrokecolor{currentstroke}%
\pgfsetdash{}{0pt}%
\pgfsys@defobject{currentmarker}{\pgfqpoint{-0.048611in}{0.000000in}}{\pgfqpoint{-0.000000in}{0.000000in}}{%
\pgfpathmoveto{\pgfqpoint{-0.000000in}{0.000000in}}%
\pgfpathlineto{\pgfqpoint{-0.048611in}{0.000000in}}%
\pgfusepath{stroke,fill}%
}%
\begin{pgfscope}%
\pgfsys@transformshift{1.268682in}{5.405927in}%
\pgfsys@useobject{currentmarker}{}%
\end{pgfscope}%
\end{pgfscope}%
\begin{pgfscope}%
\definecolor{textcolor}{rgb}{0.000000,0.000000,0.000000}%
\pgfsetstrokecolor{textcolor}%
\pgfsetfillcolor{textcolor}%
\pgftext[x=0.764226in, y=5.285943in, left, base]{\color{textcolor}\rmfamily\fontsize{32.000000}{38.400000}\selectfont 30}%
\end{pgfscope}%
\begin{pgfscope}%
\pgfsetbuttcap%
\pgfsetroundjoin%
\definecolor{currentfill}{rgb}{0.000000,0.000000,0.000000}%
\pgfsetfillcolor{currentfill}%
\pgfsetlinewidth{0.803000pt}%
\definecolor{currentstroke}{rgb}{0.000000,0.000000,0.000000}%
\pgfsetstrokecolor{currentstroke}%
\pgfsetdash{}{0pt}%
\pgfsys@defobject{currentmarker}{\pgfqpoint{-0.048611in}{0.000000in}}{\pgfqpoint{-0.000000in}{0.000000in}}{%
\pgfpathmoveto{\pgfqpoint{-0.000000in}{0.000000in}}%
\pgfpathlineto{\pgfqpoint{-0.048611in}{0.000000in}}%
\pgfusepath{stroke,fill}%
}%
\begin{pgfscope}%
\pgfsys@transformshift{1.268682in}{6.690790in}%
\pgfsys@useobject{currentmarker}{}%
\end{pgfscope}%
\end{pgfscope}%
\begin{pgfscope}%
\definecolor{textcolor}{rgb}{0.000000,0.000000,0.000000}%
\pgfsetstrokecolor{textcolor}%
\pgfsetfillcolor{textcolor}%
\pgftext[x=0.764226in, y=6.570805in, left, base]{\color{textcolor}\rmfamily\fontsize{32.000000}{38.400000}\selectfont 40}%
\end{pgfscope}%
\begin{pgfscope}%
\pgfsetbuttcap%
\pgfsetroundjoin%
\definecolor{currentfill}{rgb}{0.000000,0.000000,0.000000}%
\pgfsetfillcolor{currentfill}%
\pgfsetlinewidth{0.803000pt}%
\definecolor{currentstroke}{rgb}{0.000000,0.000000,0.000000}%
\pgfsetstrokecolor{currentstroke}%
\pgfsetdash{}{0pt}%
\pgfsys@defobject{currentmarker}{\pgfqpoint{-0.048611in}{0.000000in}}{\pgfqpoint{-0.000000in}{0.000000in}}{%
\pgfpathmoveto{\pgfqpoint{-0.000000in}{0.000000in}}%
\pgfpathlineto{\pgfqpoint{-0.048611in}{0.000000in}}%
\pgfusepath{stroke,fill}%
}%
\begin{pgfscope}%
\pgfsys@transformshift{1.268682in}{7.975652in}%
\pgfsys@useobject{currentmarker}{}%
\end{pgfscope}%
\end{pgfscope}%
\begin{pgfscope}%
\definecolor{textcolor}{rgb}{0.000000,0.000000,0.000000}%
\pgfsetstrokecolor{textcolor}%
\pgfsetfillcolor{textcolor}%
\pgftext[x=0.764226in, y=7.855668in, left, base]{\color{textcolor}\rmfamily\fontsize{32.000000}{38.400000}\selectfont 50}%
\end{pgfscope}%
\begin{pgfscope}%
\definecolor{textcolor}{rgb}{0.000000,0.000000,0.000000}%
\pgfsetstrokecolor{textcolor}%
\pgfsetfillcolor{textcolor}%
\pgftext[x=0.625337in,y=4.763496in,,bottom,rotate=90.000000]{\color{textcolor}\rmfamily\fontsize{36.000000}{43.200000}\selectfont \(\displaystyle y\) [m]}%
\end{pgfscope}%
\begin{pgfscope}%
\pgfpathrectangle{\pgfqpoint{1.268682in}{1.165881in}}{\pgfqpoint{8.292429in}{7.195230in}}%
\pgfusepath{clip}%
\pgfsetrectcap%
\pgfsetroundjoin%
\pgfsetlinewidth{3.011250pt}%
\definecolor{currentstroke}{rgb}{1.000000,0.000000,0.000000}%
\pgfsetstrokecolor{currentstroke}%
\pgfsetdash{}{0pt}%
\pgfpathmoveto{\pgfqpoint{5.414897in}{2.836202in}}%
\pgfpathlineto{\pgfqpoint{5.414353in}{2.900053in}}%
\pgfpathlineto{\pgfqpoint{5.411927in}{2.963017in}}%
\pgfpathlineto{\pgfqpoint{5.408318in}{3.026084in}}%
\pgfpathlineto{\pgfqpoint{5.403906in}{3.089096in}}%
\pgfpathlineto{\pgfqpoint{5.399131in}{3.151905in}}%
\pgfpathlineto{\pgfqpoint{5.394972in}{3.215871in}}%
\pgfpathlineto{\pgfqpoint{5.390885in}{3.280019in}}%
\pgfpathlineto{\pgfqpoint{5.388023in}{3.343799in}}%
\pgfpathlineto{\pgfqpoint{5.386088in}{3.408844in}}%
\pgfpathlineto{\pgfqpoint{5.382670in}{3.474178in}}%
\pgfpathlineto{\pgfqpoint{5.378037in}{3.539463in}}%
\pgfpathlineto{\pgfqpoint{5.372229in}{3.606305in}}%
\pgfpathlineto{\pgfqpoint{5.367063in}{3.673851in}}%
\pgfpathlineto{\pgfqpoint{5.362941in}{3.741846in}}%
\pgfpathlineto{\pgfqpoint{5.357015in}{3.811160in}}%
\pgfpathlineto{\pgfqpoint{5.348563in}{3.881543in}}%
\pgfpathlineto{\pgfqpoint{5.338309in}{3.951702in}}%
\pgfpathlineto{\pgfqpoint{5.327897in}{4.022004in}}%
\pgfpathlineto{\pgfqpoint{5.316969in}{4.092773in}}%
\pgfpathlineto{\pgfqpoint{5.305229in}{4.163555in}}%
\pgfpathlineto{\pgfqpoint{5.292301in}{4.234239in}}%
\pgfpathlineto{\pgfqpoint{5.277332in}{4.305401in}}%
\pgfpathlineto{\pgfqpoint{5.262100in}{4.376277in}}%
\pgfpathlineto{\pgfqpoint{5.247916in}{4.447614in}}%
\pgfpathlineto{\pgfqpoint{5.233780in}{4.520366in}}%
\pgfpathlineto{\pgfqpoint{5.218472in}{4.593819in}}%
\pgfpathlineto{\pgfqpoint{5.202672in}{4.668870in}}%
\pgfpathlineto{\pgfqpoint{5.187568in}{4.744967in}}%
\pgfpathlineto{\pgfqpoint{5.172962in}{4.821215in}}%
\pgfpathlineto{\pgfqpoint{5.158601in}{4.897463in}}%
\pgfpathlineto{\pgfqpoint{5.144388in}{4.972711in}}%
\pgfpathlineto{\pgfqpoint{5.128528in}{5.047276in}}%
\pgfpathlineto{\pgfqpoint{5.110929in}{5.122610in}}%
\pgfpathlineto{\pgfqpoint{5.092273in}{5.199026in}}%
\pgfpathlineto{\pgfqpoint{5.072123in}{5.276495in}}%
\pgfpathlineto{\pgfqpoint{5.050997in}{5.354928in}}%
\pgfpathlineto{\pgfqpoint{5.030362in}{5.434724in}}%
\pgfpathlineto{\pgfqpoint{5.011596in}{5.515665in}}%
\pgfpathlineto{\pgfqpoint{4.993804in}{5.597379in}}%
\pgfpathlineto{\pgfqpoint{4.976473in}{5.679363in}}%
\pgfpathlineto{\pgfqpoint{4.960325in}{5.760709in}}%
\pgfpathlineto{\pgfqpoint{4.945462in}{5.842380in}}%
\pgfpathlineto{\pgfqpoint{4.931311in}{5.924845in}}%
\pgfpathlineto{\pgfqpoint{4.918809in}{6.006872in}}%
\pgfpathlineto{\pgfqpoint{4.907238in}{6.089845in}}%
\pgfpathlineto{\pgfqpoint{4.895289in}{6.173680in}}%
\pgfpathlineto{\pgfqpoint{4.883283in}{6.257654in}}%
\pgfpathlineto{\pgfqpoint{4.871641in}{6.342548in}}%
\pgfpathlineto{\pgfqpoint{4.861520in}{6.427498in}}%
\pgfpathlineto{\pgfqpoint{4.852661in}{6.512909in}}%
\pgfpathlineto{\pgfqpoint{4.842089in}{6.599170in}}%
\pgfpathlineto{\pgfqpoint{4.828837in}{6.685674in}}%
\pgfpathlineto{\pgfqpoint{4.814785in}{6.772909in}}%
\pgfpathlineto{\pgfqpoint{4.799933in}{6.860741in}}%
\pgfpathlineto{\pgfqpoint{4.784457in}{6.947768in}}%
\pgfpathlineto{\pgfqpoint{4.769274in}{7.034121in}}%
\pgfpathlineto{\pgfqpoint{4.754948in}{7.120936in}}%
\pgfpathlineto{\pgfqpoint{4.742131in}{7.208026in}}%
\pgfpathlineto{\pgfqpoint{4.729834in}{7.295411in}}%
\pgfusepath{stroke}%
\end{pgfscope}%
\begin{pgfscope}%
\pgfpathrectangle{\pgfqpoint{1.268682in}{1.165881in}}{\pgfqpoint{8.292429in}{7.195230in}}%
\pgfusepath{clip}%
\pgfsetrectcap%
\pgfsetroundjoin%
\pgfsetlinewidth{3.011250pt}%
\definecolor{currentstroke}{rgb}{1.000000,0.000000,0.000000}%
\pgfsetstrokecolor{currentstroke}%
\pgfsetdash{}{0pt}%
\pgfpathmoveto{\pgfqpoint{3.193710in}{4.763496in}}%
\pgfpathlineto{\pgfqpoint{3.267970in}{4.764161in}}%
\pgfpathlineto{\pgfqpoint{3.342242in}{4.766635in}}%
\pgfpathlineto{\pgfqpoint{3.416808in}{4.770821in}}%
\pgfpathlineto{\pgfqpoint{3.490939in}{4.776734in}}%
\pgfpathlineto{\pgfqpoint{3.563751in}{4.784221in}}%
\pgfpathlineto{\pgfqpoint{3.635546in}{4.790955in}}%
\pgfpathlineto{\pgfqpoint{3.706366in}{4.796244in}}%
\pgfpathlineto{\pgfqpoint{3.776562in}{4.800875in}}%
\pgfpathlineto{\pgfqpoint{3.845875in}{4.804502in}}%
\pgfpathlineto{\pgfqpoint{3.914752in}{4.807685in}}%
\pgfpathlineto{\pgfqpoint{3.982888in}{4.810635in}}%
\pgfpathlineto{\pgfqpoint{4.049087in}{4.813476in}}%
\pgfpathlineto{\pgfqpoint{4.115499in}{4.816279in}}%
\pgfpathlineto{\pgfqpoint{4.182915in}{4.818751in}}%
\pgfpathlineto{\pgfqpoint{4.250408in}{4.821112in}}%
\pgfpathlineto{\pgfqpoint{4.318377in}{4.823558in}}%
\pgfpathlineto{\pgfqpoint{4.385957in}{4.826638in}}%
\pgfpathlineto{\pgfqpoint{4.453719in}{4.830428in}}%
\pgfpathlineto{\pgfqpoint{4.520612in}{4.834002in}}%
\pgfpathlineto{\pgfqpoint{4.585493in}{4.836269in}}%
\pgfpathlineto{\pgfqpoint{4.650200in}{4.838498in}}%
\pgfpathlineto{\pgfqpoint{4.714832in}{4.842143in}}%
\pgfpathlineto{\pgfqpoint{4.779886in}{4.845501in}}%
\pgfpathlineto{\pgfqpoint{4.844404in}{4.847318in}}%
\pgfpathlineto{\pgfqpoint{4.908831in}{4.848848in}}%
\pgfpathlineto{\pgfqpoint{4.974901in}{4.850419in}}%
\pgfpathlineto{\pgfqpoint{5.042555in}{4.852229in}}%
\pgfpathlineto{\pgfqpoint{5.109119in}{4.854260in}}%
\pgfpathlineto{\pgfqpoint{5.173720in}{4.856766in}}%
\pgfpathlineto{\pgfqpoint{5.238342in}{4.859432in}}%
\pgfpathlineto{\pgfqpoint{5.303169in}{4.861661in}}%
\pgfpathlineto{\pgfqpoint{5.368751in}{4.864108in}}%
\pgfpathlineto{\pgfqpoint{5.434191in}{4.866935in}}%
\pgfpathlineto{\pgfqpoint{5.499076in}{4.871236in}}%
\pgfpathlineto{\pgfqpoint{5.562914in}{4.875765in}}%
\pgfpathlineto{\pgfqpoint{5.625197in}{4.878583in}}%
\pgfpathlineto{\pgfqpoint{5.687703in}{4.879415in}}%
\pgfpathlineto{\pgfqpoint{5.750260in}{4.879609in}}%
\pgfpathlineto{\pgfqpoint{5.811625in}{4.879746in}}%
\pgfpathlineto{\pgfqpoint{5.870676in}{4.879955in}}%
\pgfpathlineto{\pgfqpoint{5.927402in}{4.880785in}}%
\pgfpathlineto{\pgfqpoint{5.983795in}{4.881727in}}%
\pgfpathlineto{\pgfqpoint{6.040272in}{4.883615in}}%
\pgfpathlineto{\pgfqpoint{6.096903in}{4.886592in}}%
\pgfpathlineto{\pgfqpoint{6.154050in}{4.891012in}}%
\pgfpathlineto{\pgfqpoint{6.211014in}{4.895780in}}%
\pgfpathlineto{\pgfqpoint{6.268450in}{4.899387in}}%
\pgfpathlineto{\pgfqpoint{6.326985in}{4.902772in}}%
\pgfpathlineto{\pgfqpoint{6.385152in}{4.906571in}}%
\pgfpathlineto{\pgfqpoint{6.443269in}{4.910501in}}%
\pgfpathlineto{\pgfqpoint{6.502222in}{4.914750in}}%
\pgfpathlineto{\pgfqpoint{6.562503in}{4.919843in}}%
\pgfpathlineto{\pgfqpoint{6.623201in}{4.925871in}}%
\pgfpathlineto{\pgfqpoint{6.682584in}{4.932458in}}%
\pgfpathlineto{\pgfqpoint{6.740990in}{4.939464in}}%
\pgfpathlineto{\pgfqpoint{6.798958in}{4.946838in}}%
\pgfpathlineto{\pgfqpoint{6.855980in}{4.954459in}}%
\pgfpathlineto{\pgfqpoint{6.913334in}{4.962512in}}%
\pgfpathlineto{\pgfqpoint{6.971256in}{4.970875in}}%
\pgfusepath{stroke}%
\end{pgfscope}%
\begin{pgfscope}%
\pgfpathrectangle{\pgfqpoint{1.268682in}{1.165881in}}{\pgfqpoint{8.292429in}{7.195230in}}%
\pgfusepath{clip}%
\pgfsetrectcap%
\pgfsetroundjoin%
\pgfsetlinewidth{3.011250pt}%
\definecolor{currentstroke}{rgb}{1.000000,0.000000,0.000000}%
\pgfsetstrokecolor{currentstroke}%
\pgfsetdash{}{0pt}%
\pgfpathmoveto{\pgfqpoint{4.477571in}{4.108399in}}%
\pgfpathlineto{\pgfqpoint{4.488902in}{4.156715in}}%
\pgfpathlineto{\pgfqpoint{4.501331in}{4.204441in}}%
\pgfpathlineto{\pgfqpoint{4.511527in}{4.251931in}}%
\pgfpathlineto{\pgfqpoint{4.518663in}{4.299505in}}%
\pgfpathlineto{\pgfqpoint{4.524682in}{4.347658in}}%
\pgfpathlineto{\pgfqpoint{4.530351in}{4.395437in}}%
\pgfpathlineto{\pgfqpoint{4.535634in}{4.442773in}}%
\pgfpathlineto{\pgfqpoint{4.540003in}{4.489848in}}%
\pgfpathlineto{\pgfqpoint{4.543482in}{4.535160in}}%
\pgfpathlineto{\pgfqpoint{4.546967in}{4.579709in}}%
\pgfpathlineto{\pgfqpoint{4.550480in}{4.623825in}}%
\pgfpathlineto{\pgfqpoint{4.554631in}{4.666147in}}%
\pgfpathlineto{\pgfqpoint{4.558492in}{4.707506in}}%
\pgfpathlineto{\pgfqpoint{4.561207in}{4.748290in}}%
\pgfpathlineto{\pgfqpoint{4.563621in}{4.789180in}}%
\pgfpathlineto{\pgfqpoint{4.563859in}{4.830058in}}%
\pgfpathlineto{\pgfqpoint{4.560397in}{4.871610in}}%
\pgfpathlineto{\pgfqpoint{4.555130in}{4.914253in}}%
\pgfpathlineto{\pgfqpoint{4.549979in}{4.958442in}}%
\pgfpathlineto{\pgfqpoint{4.545622in}{5.004919in}}%
\pgfpathlineto{\pgfqpoint{4.542773in}{5.051861in}}%
\pgfpathlineto{\pgfqpoint{4.540569in}{5.097582in}}%
\pgfpathlineto{\pgfqpoint{4.538304in}{5.142428in}}%
\pgfpathlineto{\pgfqpoint{4.537776in}{5.185940in}}%
\pgfpathlineto{\pgfqpoint{4.538767in}{5.228952in}}%
\pgfpathlineto{\pgfqpoint{4.540431in}{5.272516in}}%
\pgfpathlineto{\pgfqpoint{4.543019in}{5.314633in}}%
\pgfpathlineto{\pgfqpoint{4.546229in}{5.356390in}}%
\pgfpathlineto{\pgfqpoint{4.549682in}{5.398954in}}%
\pgfpathlineto{\pgfqpoint{4.554974in}{5.440835in}}%
\pgfpathlineto{\pgfqpoint{4.562052in}{5.482007in}}%
\pgfpathlineto{\pgfqpoint{4.568142in}{5.523053in}}%
\pgfpathlineto{\pgfqpoint{4.572699in}{5.563796in}}%
\pgfpathlineto{\pgfqpoint{4.576161in}{5.604260in}}%
\pgfpathlineto{\pgfqpoint{4.579152in}{5.644571in}}%
\pgfpathlineto{\pgfqpoint{4.581869in}{5.684897in}}%
\pgfpathlineto{\pgfqpoint{4.584284in}{5.724625in}}%
\pgfpathlineto{\pgfqpoint{4.586872in}{5.764306in}}%
\pgfpathlineto{\pgfqpoint{4.589658in}{5.804544in}}%
\pgfusepath{stroke}%
\end{pgfscope}%
\begin{pgfscope}%
\pgfpathrectangle{\pgfqpoint{1.268682in}{1.165881in}}{\pgfqpoint{8.292429in}{7.195230in}}%
\pgfusepath{clip}%
\pgfsetrectcap%
\pgfsetroundjoin%
\pgfsetlinewidth{3.011250pt}%
\definecolor{currentstroke}{rgb}{1.000000,0.000000,0.000000}%
\pgfsetstrokecolor{currentstroke}%
\pgfsetdash{}{0pt}%
\pgfpathmoveto{\pgfqpoint{4.385390in}{4.899399in}}%
\pgfpathlineto{\pgfqpoint{4.355232in}{4.910338in}}%
\pgfpathlineto{\pgfqpoint{4.326944in}{4.922449in}}%
\pgfpathlineto{\pgfqpoint{4.300095in}{4.934190in}}%
\pgfpathlineto{\pgfqpoint{4.273545in}{4.944403in}}%
\pgfpathlineto{\pgfqpoint{4.246430in}{4.954660in}}%
\pgfpathlineto{\pgfqpoint{4.219953in}{4.965054in}}%
\pgfpathlineto{\pgfqpoint{4.194726in}{4.976162in}}%
\pgfpathlineto{\pgfqpoint{4.168908in}{4.988528in}}%
\pgfpathlineto{\pgfqpoint{4.141643in}{5.001058in}}%
\pgfpathlineto{\pgfqpoint{4.113293in}{5.012432in}}%
\pgfpathlineto{\pgfqpoint{4.083789in}{5.022645in}}%
\pgfpathlineto{\pgfqpoint{4.053123in}{5.031446in}}%
\pgfpathlineto{\pgfqpoint{4.023149in}{5.039509in}}%
\pgfpathlineto{\pgfqpoint{3.994176in}{5.047870in}}%
\pgfpathlineto{\pgfqpoint{3.963976in}{5.056113in}}%
\pgfpathlineto{\pgfqpoint{3.932497in}{5.064010in}}%
\pgfpathlineto{\pgfqpoint{3.900489in}{5.071507in}}%
\pgfpathlineto{\pgfqpoint{3.868852in}{5.078648in}}%
\pgfpathlineto{\pgfqpoint{3.838239in}{5.085579in}}%
\pgfpathlineto{\pgfqpoint{3.807660in}{5.092553in}}%
\pgfpathlineto{\pgfqpoint{3.776018in}{5.100209in}}%
\pgfpathlineto{\pgfqpoint{3.743654in}{5.108952in}}%
\pgfpathlineto{\pgfqpoint{3.712283in}{5.118038in}}%
\pgfpathlineto{\pgfqpoint{3.682308in}{5.126752in}}%
\pgfpathlineto{\pgfqpoint{3.653537in}{5.135661in}}%
\pgfpathlineto{\pgfqpoint{3.624597in}{5.144738in}}%
\pgfpathlineto{\pgfqpoint{3.595287in}{5.152733in}}%
\pgfpathlineto{\pgfqpoint{3.566585in}{5.160750in}}%
\pgfpathlineto{\pgfqpoint{3.539511in}{5.167487in}}%
\pgfpathlineto{\pgfqpoint{3.513904in}{5.172271in}}%
\pgfpathlineto{\pgfqpoint{3.487824in}{5.177657in}}%
\pgfpathlineto{\pgfqpoint{3.460254in}{5.182927in}}%
\pgfpathlineto{\pgfqpoint{3.432271in}{5.189091in}}%
\pgfpathlineto{\pgfqpoint{3.405646in}{5.197500in}}%
\pgfpathlineto{\pgfqpoint{3.379541in}{5.206387in}}%
\pgfpathlineto{\pgfqpoint{3.350983in}{5.214501in}}%
\pgfpathlineto{\pgfqpoint{3.319757in}{5.222237in}}%
\pgfpathlineto{\pgfqpoint{3.286915in}{5.230606in}}%
\pgfpathlineto{\pgfqpoint{3.252755in}{5.239350in}}%
\pgfusepath{stroke}%
\end{pgfscope}%
\begin{pgfscope}%
\pgfpathrectangle{\pgfqpoint{1.268682in}{1.165881in}}{\pgfqpoint{8.292429in}{7.195230in}}%
\pgfusepath{clip}%
\pgfsetrectcap%
\pgfsetroundjoin%
\pgfsetlinewidth{3.011250pt}%
\definecolor{currentstroke}{rgb}{1.000000,0.000000,0.000000}%
\pgfsetstrokecolor{currentstroke}%
\pgfsetdash{}{0pt}%
\pgfpathmoveto{\pgfqpoint{6.580013in}{4.548406in}}%
\pgfpathlineto{\pgfqpoint{6.620759in}{4.512313in}}%
\pgfpathlineto{\pgfqpoint{6.661376in}{4.476407in}}%
\pgfpathlineto{\pgfqpoint{6.702705in}{4.440581in}}%
\pgfpathlineto{\pgfqpoint{6.744995in}{4.404813in}}%
\pgfpathlineto{\pgfqpoint{6.787123in}{4.370322in}}%
\pgfpathlineto{\pgfqpoint{6.829048in}{4.337201in}}%
\pgfpathlineto{\pgfqpoint{6.871965in}{4.305120in}}%
\pgfpathlineto{\pgfqpoint{6.916290in}{4.273278in}}%
\pgfpathlineto{\pgfqpoint{6.961126in}{4.241316in}}%
\pgfpathlineto{\pgfqpoint{7.006675in}{4.209214in}}%
\pgfpathlineto{\pgfqpoint{7.053141in}{4.176570in}}%
\pgfpathlineto{\pgfqpoint{7.100299in}{4.142902in}}%
\pgfpathlineto{\pgfqpoint{7.148533in}{4.108202in}}%
\pgfpathlineto{\pgfqpoint{7.197270in}{4.072864in}}%
\pgfpathlineto{\pgfqpoint{7.246776in}{4.036965in}}%
\pgfpathlineto{\pgfqpoint{7.295923in}{4.001477in}}%
\pgfpathlineto{\pgfqpoint{7.342692in}{3.965652in}}%
\pgfpathlineto{\pgfqpoint{7.387244in}{3.929108in}}%
\pgfpathlineto{\pgfqpoint{7.429932in}{3.895560in}}%
\pgfpathlineto{\pgfqpoint{7.471954in}{3.864543in}}%
\pgfpathlineto{\pgfqpoint{7.514013in}{3.833106in}}%
\pgfpathlineto{\pgfqpoint{7.556029in}{3.801567in}}%
\pgfpathlineto{\pgfqpoint{7.598803in}{3.770419in}}%
\pgfpathlineto{\pgfqpoint{7.642914in}{3.739321in}}%
\pgfpathlineto{\pgfqpoint{7.687371in}{3.707387in}}%
\pgfpathlineto{\pgfqpoint{7.731464in}{3.674069in}}%
\pgfpathlineto{\pgfqpoint{7.773681in}{3.639828in}}%
\pgfpathlineto{\pgfqpoint{7.812712in}{3.606028in}}%
\pgfpathlineto{\pgfqpoint{7.850094in}{3.571690in}}%
\pgfpathlineto{\pgfqpoint{7.887097in}{3.536837in}}%
\pgfpathlineto{\pgfqpoint{7.923353in}{3.503810in}}%
\pgfpathlineto{\pgfqpoint{7.958732in}{3.473257in}}%
\pgfpathlineto{\pgfqpoint{7.993758in}{3.445624in}}%
\pgfpathlineto{\pgfqpoint{8.028780in}{3.419893in}}%
\pgfpathlineto{\pgfqpoint{8.064533in}{3.394630in}}%
\pgfpathlineto{\pgfqpoint{8.100645in}{3.369859in}}%
\pgfpathlineto{\pgfqpoint{8.137325in}{3.346081in}}%
\pgfpathlineto{\pgfqpoint{8.174219in}{3.323355in}}%
\pgfpathlineto{\pgfqpoint{8.209771in}{3.300279in}}%
\pgfusepath{stroke}%
\end{pgfscope}%
\begin{pgfscope}%
\pgfsetrectcap%
\pgfsetmiterjoin%
\pgfsetlinewidth{0.803000pt}%
\definecolor{currentstroke}{rgb}{0.000000,0.000000,0.000000}%
\pgfsetstrokecolor{currentstroke}%
\pgfsetdash{}{0pt}%
\pgfpathmoveto{\pgfqpoint{1.268682in}{1.165881in}}%
\pgfpathlineto{\pgfqpoint{1.268682in}{8.361111in}}%
\pgfusepath{stroke}%
\end{pgfscope}%
\begin{pgfscope}%
\pgfsetrectcap%
\pgfsetmiterjoin%
\pgfsetlinewidth{0.803000pt}%
\definecolor{currentstroke}{rgb}{0.000000,0.000000,0.000000}%
\pgfsetstrokecolor{currentstroke}%
\pgfsetdash{}{0pt}%
\pgfpathmoveto{\pgfqpoint{9.561111in}{1.165881in}}%
\pgfpathlineto{\pgfqpoint{9.561111in}{8.361111in}}%
\pgfusepath{stroke}%
\end{pgfscope}%
\begin{pgfscope}%
\pgfsetrectcap%
\pgfsetmiterjoin%
\pgfsetlinewidth{0.803000pt}%
\definecolor{currentstroke}{rgb}{0.000000,0.000000,0.000000}%
\pgfsetstrokecolor{currentstroke}%
\pgfsetdash{}{0pt}%
\pgfpathmoveto{\pgfqpoint{1.268682in}{1.165881in}}%
\pgfpathlineto{\pgfqpoint{9.561111in}{1.165881in}}%
\pgfusepath{stroke}%
\end{pgfscope}%
\begin{pgfscope}%
\pgfsetrectcap%
\pgfsetmiterjoin%
\pgfsetlinewidth{0.803000pt}%
\definecolor{currentstroke}{rgb}{0.000000,0.000000,0.000000}%
\pgfsetstrokecolor{currentstroke}%
\pgfsetdash{}{0pt}%
\pgfpathmoveto{\pgfqpoint{1.268682in}{8.361111in}}%
\pgfpathlineto{\pgfqpoint{9.561111in}{8.361111in}}%
\pgfusepath{stroke}%
\end{pgfscope}%
\end{pgfpicture}%
\makeatother%
\endgroup%

%% file: Figs/rawsignal/results_041_eps.pgf
\begingroup%
\makeatletter%
\begin{pgfpicture}%
\pgfpathrectangle{\pgfpointorigin}{\pgfqpoint{9.600000in}{8.400000in}}%
\pgfusepath{use as bounding box, clip}%
\begin{pgfscope}%
\pgfsetbuttcap%
\pgfsetmiterjoin%
\definecolor{currentfill}{rgb}{1.000000,1.000000,1.000000}%
\pgfsetfillcolor{currentfill}%
\pgfsetlinewidth{0.000000pt}%
\definecolor{currentstroke}{rgb}{1.000000,1.000000,1.000000}%
\pgfsetstrokecolor{currentstroke}%
\pgfsetdash{}{0pt}%
\pgfpathmoveto{\pgfqpoint{0.000000in}{0.000000in}}%
\pgfpathlineto{\pgfqpoint{9.600000in}{0.000000in}}%
\pgfpathlineto{\pgfqpoint{9.600000in}{8.400000in}}%
\pgfpathlineto{\pgfqpoint{0.000000in}{8.400000in}}%
\pgfpathlineto{\pgfqpoint{0.000000in}{0.000000in}}%
\pgfpathclose%
\pgfusepath{fill}%
\end{pgfscope}%
\begin{pgfscope}%
\pgfsetbuttcap%
\pgfsetmiterjoin%
\definecolor{currentfill}{rgb}{1.000000,1.000000,1.000000}%
\pgfsetfillcolor{currentfill}%
\pgfsetlinewidth{0.000000pt}%
\definecolor{currentstroke}{rgb}{0.000000,0.000000,0.000000}%
\pgfsetstrokecolor{currentstroke}%
\pgfsetstrokeopacity{0.000000}%
\pgfsetdash{}{0pt}%
\pgfpathmoveto{\pgfqpoint{1.201204in}{1.067125in}}%
\pgfpathlineto{\pgfqpoint{8.365915in}{1.067125in}}%
\pgfpathlineto{\pgfqpoint{8.365915in}{8.220391in}}%
\pgfpathlineto{\pgfqpoint{1.201204in}{8.220391in}}%
\pgfpathlineto{\pgfqpoint{1.201204in}{1.067125in}}%
\pgfpathclose%
\pgfusepath{fill}%
\end{pgfscope}%
\begin{pgfscope}%
\pgfpathrectangle{\pgfqpoint{1.201204in}{1.067125in}}{\pgfqpoint{7.164711in}{7.153266in}}%
\pgfusepath{clip}%
\pgfsys@transformshift{1.201204in}{1.067125in}%
\pgftext[left,bottom]{\includegraphics[interpolate=true,width=7.170000in,height=7.160000in]{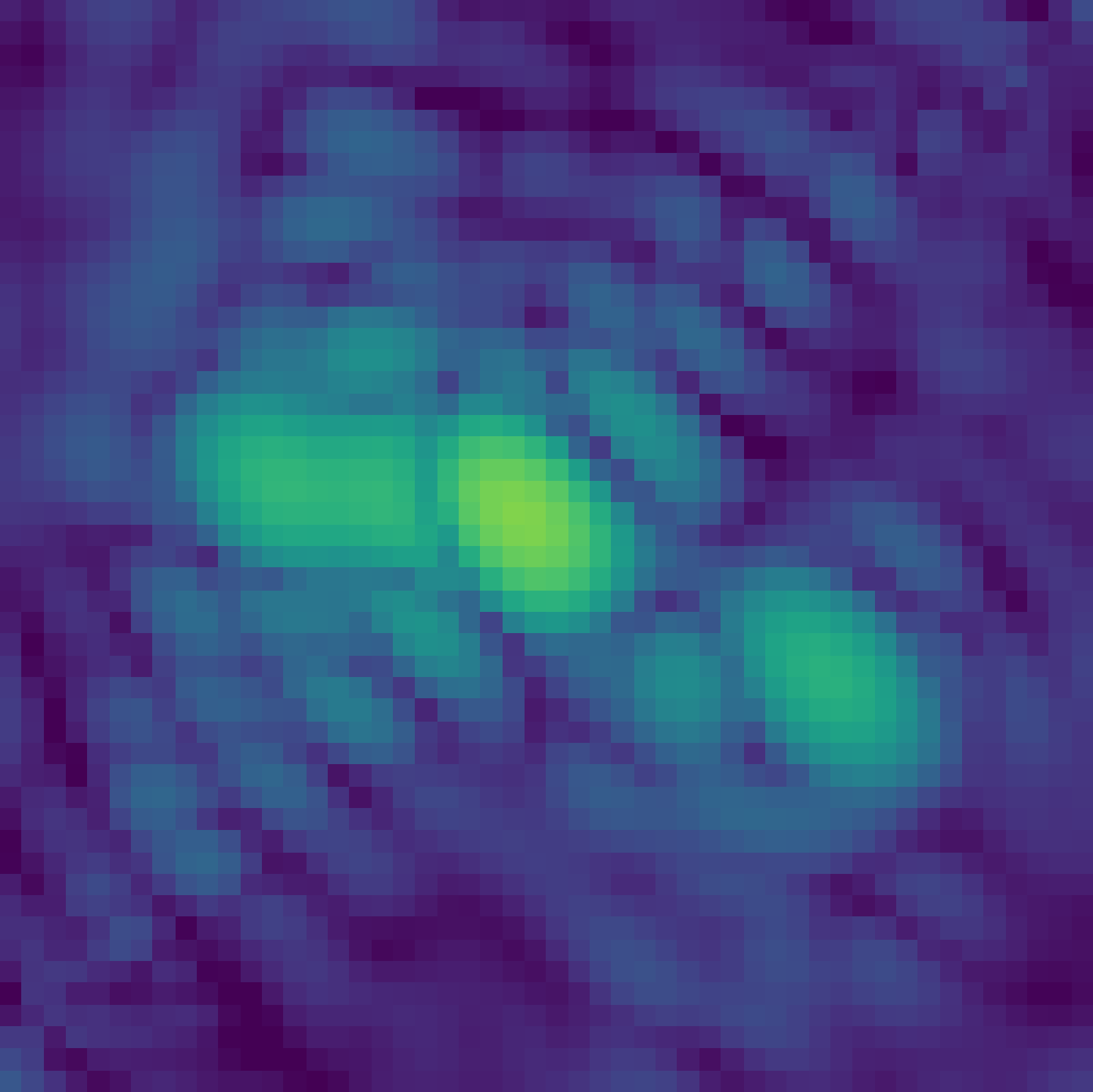}}%
\end{pgfscope}%
\begin{pgfscope}%
\pgfpathrectangle{\pgfqpoint{1.201204in}{1.067125in}}{\pgfqpoint{7.164711in}{7.153266in}}%
\pgfusepath{clip}%
\pgfsetbuttcap%
\pgfsetroundjoin%
\definecolor{currentfill}{rgb}{0.750000,0.750000,0.000000}%
\pgfsetfillcolor{currentfill}%
\pgfsetlinewidth{3.011250pt}%
\definecolor{currentstroke}{rgb}{0.750000,0.750000,0.000000}%
\pgfsetstrokecolor{currentstroke}%
\pgfsetdash{}{0pt}%
\pgfsys@defobject{currentmarker}{\pgfqpoint{-0.098209in}{-0.098209in}}{\pgfqpoint{0.098209in}{0.098209in}}{%
\pgfpathmoveto{\pgfqpoint{-0.098209in}{-0.098209in}}%
\pgfpathlineto{\pgfqpoint{0.098209in}{0.098209in}}%
\pgfpathmoveto{\pgfqpoint{-0.098209in}{0.098209in}}%
\pgfpathlineto{\pgfqpoint{0.098209in}{-0.098209in}}%
\pgfusepath{stroke,fill}%
}%
\begin{pgfscope}%
\pgfsys@transformshift{4.142288in}{4.939224in}%
\pgfsys@useobject{currentmarker}{}%
\end{pgfscope}%
\begin{pgfscope}%
\pgfsys@transformshift{3.341067in}{5.181948in}%
\pgfsys@useobject{currentmarker}{}%
\end{pgfscope}%
\begin{pgfscope}%
\pgfsys@transformshift{4.592832in}{4.817046in}%
\pgfsys@useobject{currentmarker}{}%
\end{pgfscope}%
\begin{pgfscope}%
\pgfsys@transformshift{6.677447in}{3.902020in}%
\pgfsys@useobject{currentmarker}{}%
\end{pgfscope}%
\begin{pgfscope}%
\pgfsys@transformshift{4.619303in}{4.911322in}%
\pgfsys@useobject{currentmarker}{}%
\end{pgfscope}%
\end{pgfscope}%
\begin{pgfscope}%
\pgfpathrectangle{\pgfqpoint{1.201204in}{1.067125in}}{\pgfqpoint{7.164711in}{7.153266in}}%
\pgfusepath{clip}%
\pgfsetbuttcap%
\pgfsetroundjoin%
\definecolor{currentfill}{rgb}{0.000000,0.000000,1.000000}%
\pgfsetfillcolor{currentfill}%
\pgfsetlinewidth{3.011250pt}%
\definecolor{currentstroke}{rgb}{0.000000,0.000000,1.000000}%
\pgfsetstrokecolor{currentstroke}%
\pgfsetdash{}{0pt}%
\pgfsys@defobject{currentmarker}{\pgfqpoint{-0.098209in}{-0.098209in}}{\pgfqpoint{0.098209in}{0.098209in}}{%
\pgfpathmoveto{\pgfqpoint{-0.098209in}{-0.098209in}}%
\pgfpathlineto{\pgfqpoint{0.098209in}{0.098209in}}%
\pgfpathmoveto{\pgfqpoint{-0.098209in}{0.098209in}}%
\pgfpathlineto{\pgfqpoint{0.098209in}{-0.098209in}}%
\pgfusepath{stroke,fill}%
}%
\begin{pgfscope}%
\pgfsys@transformshift{4.662684in}{4.600562in}%
\pgfsys@useobject{currentmarker}{}%
\end{pgfscope}%
\begin{pgfscope}%
\pgfsys@transformshift{6.598102in}{3.837232in}%
\pgfsys@useobject{currentmarker}{}%
\end{pgfscope}%
\begin{pgfscope}%
\pgfsys@transformshift{3.241936in}{4.994174in}%
\pgfsys@useobject{currentmarker}{}%
\end{pgfscope}%
\begin{pgfscope}%
\pgfsys@transformshift{3.849971in}{4.973170in}%
\pgfsys@useobject{currentmarker}{}%
\end{pgfscope}%
\end{pgfscope}%
\begin{pgfscope}%
\pgfpathrectangle{\pgfqpoint{1.201204in}{1.067125in}}{\pgfqpoint{7.164711in}{7.153266in}}%
\pgfusepath{clip}%
\pgfsetbuttcap%
\pgfsetroundjoin%
\definecolor{currentfill}{rgb}{0.000000,0.500000,0.000000}%
\pgfsetfillcolor{currentfill}%
\pgfsetlinewidth{3.011250pt}%
\definecolor{currentstroke}{rgb}{0.000000,0.500000,0.000000}%
\pgfsetstrokecolor{currentstroke}%
\pgfsetdash{}{0pt}%
\pgfsys@defobject{currentmarker}{\pgfqpoint{-0.098209in}{-0.098209in}}{\pgfqpoint{0.098209in}{0.098209in}}{%
\pgfpathmoveto{\pgfqpoint{-0.098209in}{-0.098209in}}%
\pgfpathlineto{\pgfqpoint{0.098209in}{0.098209in}}%
\pgfpathmoveto{\pgfqpoint{-0.098209in}{0.098209in}}%
\pgfpathlineto{\pgfqpoint{0.098209in}{-0.098209in}}%
\pgfusepath{stroke,fill}%
}%
\begin{pgfscope}%
\pgfsys@transformshift{4.552044in}{4.876891in}%
\pgfsys@useobject{currentmarker}{}%
\end{pgfscope}%
\begin{pgfscope}%
\pgfsys@transformshift{4.664151in}{4.718166in}%
\pgfsys@useobject{currentmarker}{}%
\end{pgfscope}%
\begin{pgfscope}%
\pgfsys@transformshift{3.043954in}{5.066609in}%
\pgfsys@useobject{currentmarker}{}%
\end{pgfscope}%
\begin{pgfscope}%
\pgfsys@transformshift{4.016152in}{4.929217in}%
\pgfsys@useobject{currentmarker}{}%
\end{pgfscope}%
\begin{pgfscope}%
\pgfsys@transformshift{6.613924in}{3.828561in}%
\pgfsys@useobject{currentmarker}{}%
\end{pgfscope}%
\end{pgfscope}%
\begin{pgfscope}%
\pgfpathrectangle{\pgfqpoint{1.201204in}{1.067125in}}{\pgfqpoint{7.164711in}{7.153266in}}%
\pgfusepath{clip}%
\pgfsetbuttcap%
\pgfsetroundjoin%
\definecolor{currentfill}{rgb}{1.000000,0.000000,0.000000}%
\pgfsetfillcolor{currentfill}%
\pgfsetlinewidth{3.011250pt}%
\definecolor{currentstroke}{rgb}{1.000000,0.000000,0.000000}%
\pgfsetstrokecolor{currentstroke}%
\pgfsetdash{}{0pt}%
\pgfsys@defobject{currentmarker}{\pgfqpoint{-0.098209in}{-0.098209in}}{\pgfqpoint{0.098209in}{0.098209in}}{%
\pgfpathmoveto{\pgfqpoint{-0.098209in}{0.000000in}}%
\pgfpathlineto{\pgfqpoint{0.098209in}{0.000000in}}%
\pgfpathmoveto{\pgfqpoint{0.000000in}{-0.098209in}}%
\pgfpathlineto{\pgfqpoint{0.000000in}{0.098209in}}%
\pgfusepath{stroke,fill}%
}%
\begin{pgfscope}%
\pgfsys@transformshift{4.675828in}{4.749273in}%
\pgfsys@useobject{currentmarker}{}%
\end{pgfscope}%
\begin{pgfscope}%
\pgfsys@transformshift{4.501507in}{4.908685in}%
\pgfsys@useobject{currentmarker}{}%
\end{pgfscope}%
\begin{pgfscope}%
\pgfsys@transformshift{3.057642in}{5.068276in}%
\pgfsys@useobject{currentmarker}{}%
\end{pgfscope}%
\begin{pgfscope}%
\pgfsys@transformshift{3.942133in}{4.962544in}%
\pgfsys@useobject{currentmarker}{}%
\end{pgfscope}%
\begin{pgfscope}%
\pgfsys@transformshift{6.608997in}{3.794366in}%
\pgfsys@useobject{currentmarker}{}%
\end{pgfscope}%
\end{pgfscope}%
\begin{pgfscope}%
\pgfpathrectangle{\pgfqpoint{1.201204in}{1.067125in}}{\pgfqpoint{7.164711in}{7.153266in}}%
\pgfusepath{clip}%
\pgfsetbuttcap%
\pgfsetroundjoin%
\pgfsetlinewidth{3.011250pt}%
\definecolor{currentstroke}{rgb}{0.000000,0.000000,0.000000}%
\pgfsetstrokecolor{currentstroke}%
\pgfsetdash{}{0pt}%
\pgfpathmoveto{\pgfqpoint{4.521792in}{4.778503in}}%
\pgfpathcurveto{\pgfqpoint{4.547837in}{4.778503in}}{\pgfqpoint{4.572819in}{4.788851in}}{\pgfqpoint{4.591236in}{4.807267in}}%
\pgfpathcurveto{\pgfqpoint{4.609653in}{4.825684in}}{\pgfqpoint{4.620001in}{4.850666in}}{\pgfqpoint{4.620001in}{4.876712in}}%
\pgfpathcurveto{\pgfqpoint{4.620001in}{4.902757in}}{\pgfqpoint{4.609653in}{4.927739in}}{\pgfqpoint{4.591236in}{4.946156in}}%
\pgfpathcurveto{\pgfqpoint{4.572819in}{4.964573in}}{\pgfqpoint{4.547837in}{4.974921in}}{\pgfqpoint{4.521792in}{4.974921in}}%
\pgfpathcurveto{\pgfqpoint{4.495746in}{4.974921in}}{\pgfqpoint{4.470764in}{4.964573in}}{\pgfqpoint{4.452347in}{4.946156in}}%
\pgfpathcurveto{\pgfqpoint{4.433930in}{4.927739in}}{\pgfqpoint{4.423582in}{4.902757in}}{\pgfqpoint{4.423582in}{4.876712in}}%
\pgfpathcurveto{\pgfqpoint{4.423582in}{4.850666in}}{\pgfqpoint{4.433930in}{4.825684in}}{\pgfqpoint{4.452347in}{4.807267in}}%
\pgfpathcurveto{\pgfqpoint{4.470764in}{4.788851in}}{\pgfqpoint{4.495746in}{4.778503in}}{\pgfqpoint{4.521792in}{4.778503in}}%
\pgfpathlineto{\pgfqpoint{4.521792in}{4.778503in}}%
\pgfpathclose%
\pgfusepath{stroke}%
\end{pgfscope}%
\begin{pgfscope}%
\pgfpathrectangle{\pgfqpoint{1.201204in}{1.067125in}}{\pgfqpoint{7.164711in}{7.153266in}}%
\pgfusepath{clip}%
\pgfsetbuttcap%
\pgfsetroundjoin%
\pgfsetlinewidth{3.011250pt}%
\definecolor{currentstroke}{rgb}{0.000000,0.000000,0.000000}%
\pgfsetstrokecolor{currentstroke}%
\pgfsetdash{}{0pt}%
\pgfpathmoveto{\pgfqpoint{4.675442in}{4.654852in}}%
\pgfpathcurveto{\pgfqpoint{4.701487in}{4.654852in}}{\pgfqpoint{4.726470in}{4.665200in}}{\pgfqpoint{4.744886in}{4.683617in}}%
\pgfpathcurveto{\pgfqpoint{4.763303in}{4.702033in}}{\pgfqpoint{4.773651in}{4.727016in}}{\pgfqpoint{4.773651in}{4.753061in}}%
\pgfpathcurveto{\pgfqpoint{4.773651in}{4.779106in}}{\pgfqpoint{4.763303in}{4.804089in}}{\pgfqpoint{4.744886in}{4.822505in}}%
\pgfpathcurveto{\pgfqpoint{4.726470in}{4.840922in}}{\pgfqpoint{4.701487in}{4.851270in}}{\pgfqpoint{4.675442in}{4.851270in}}%
\pgfpathcurveto{\pgfqpoint{4.649397in}{4.851270in}}{\pgfqpoint{4.624414in}{4.840922in}}{\pgfqpoint{4.605998in}{4.822505in}}%
\pgfpathcurveto{\pgfqpoint{4.587581in}{4.804089in}}{\pgfqpoint{4.577233in}{4.779106in}}{\pgfqpoint{4.577233in}{4.753061in}}%
\pgfpathcurveto{\pgfqpoint{4.577233in}{4.727016in}}{\pgfqpoint{4.587581in}{4.702033in}}{\pgfqpoint{4.605998in}{4.683617in}}%
\pgfpathcurveto{\pgfqpoint{4.624414in}{4.665200in}}{\pgfqpoint{4.649397in}{4.654852in}}{\pgfqpoint{4.675442in}{4.654852in}}%
\pgfpathlineto{\pgfqpoint{4.675442in}{4.654852in}}%
\pgfpathclose%
\pgfusepath{stroke}%
\end{pgfscope}%
\begin{pgfscope}%
\pgfpathrectangle{\pgfqpoint{1.201204in}{1.067125in}}{\pgfqpoint{7.164711in}{7.153266in}}%
\pgfusepath{clip}%
\pgfsetbuttcap%
\pgfsetroundjoin%
\pgfsetlinewidth{3.011250pt}%
\definecolor{currentstroke}{rgb}{0.000000,0.000000,0.000000}%
\pgfsetstrokecolor{currentstroke}%
\pgfsetdash{}{0pt}%
\pgfpathmoveto{\pgfqpoint{3.939617in}{4.866634in}}%
\pgfpathcurveto{\pgfqpoint{3.965662in}{4.866634in}}{\pgfqpoint{3.990645in}{4.876982in}}{\pgfqpoint{4.009061in}{4.895398in}}%
\pgfpathcurveto{\pgfqpoint{4.027478in}{4.913815in}}{\pgfqpoint{4.037826in}{4.938797in}}{\pgfqpoint{4.037826in}{4.964843in}}%
\pgfpathcurveto{\pgfqpoint{4.037826in}{4.990888in}}{\pgfqpoint{4.027478in}{5.015870in}}{\pgfqpoint{4.009061in}{5.034287in}}%
\pgfpathcurveto{\pgfqpoint{3.990645in}{5.052704in}}{\pgfqpoint{3.965662in}{5.063052in}}{\pgfqpoint{3.939617in}{5.063052in}}%
\pgfpathcurveto{\pgfqpoint{3.913572in}{5.063052in}}{\pgfqpoint{3.888589in}{5.052704in}}{\pgfqpoint{3.870173in}{5.034287in}}%
\pgfpathcurveto{\pgfqpoint{3.851756in}{5.015870in}}{\pgfqpoint{3.841408in}{4.990888in}}{\pgfqpoint{3.841408in}{4.964843in}}%
\pgfpathcurveto{\pgfqpoint{3.841408in}{4.938797in}}{\pgfqpoint{3.851756in}{4.913815in}}{\pgfqpoint{3.870173in}{4.895398in}}%
\pgfpathcurveto{\pgfqpoint{3.888589in}{4.876982in}}{\pgfqpoint{3.913572in}{4.866634in}}{\pgfqpoint{3.939617in}{4.866634in}}%
\pgfpathlineto{\pgfqpoint{3.939617in}{4.866634in}}%
\pgfpathclose%
\pgfusepath{stroke}%
\end{pgfscope}%
\begin{pgfscope}%
\pgfpathrectangle{\pgfqpoint{1.201204in}{1.067125in}}{\pgfqpoint{7.164711in}{7.153266in}}%
\pgfusepath{clip}%
\pgfsetbuttcap%
\pgfsetroundjoin%
\pgfsetlinewidth{3.011250pt}%
\definecolor{currentstroke}{rgb}{0.000000,0.000000,0.000000}%
\pgfsetstrokecolor{currentstroke}%
\pgfsetdash{}{0pt}%
\pgfpathmoveto{\pgfqpoint{3.051110in}{4.970050in}}%
\pgfpathcurveto{\pgfqpoint{3.077156in}{4.970050in}}{\pgfqpoint{3.102138in}{4.980398in}}{\pgfqpoint{3.120555in}{4.998815in}}%
\pgfpathcurveto{\pgfqpoint{3.138972in}{5.017231in}}{\pgfqpoint{3.149320in}{5.042214in}}{\pgfqpoint{3.149320in}{5.068259in}}%
\pgfpathcurveto{\pgfqpoint{3.149320in}{5.094304in}}{\pgfqpoint{3.138972in}{5.119287in}}{\pgfqpoint{3.120555in}{5.137703in}}%
\pgfpathcurveto{\pgfqpoint{3.102138in}{5.156120in}}{\pgfqpoint{3.077156in}{5.166468in}}{\pgfqpoint{3.051110in}{5.166468in}}%
\pgfpathcurveto{\pgfqpoint{3.025065in}{5.166468in}}{\pgfqpoint{3.000083in}{5.156120in}}{\pgfqpoint{2.981666in}{5.137703in}}%
\pgfpathcurveto{\pgfqpoint{2.963249in}{5.119287in}}{\pgfqpoint{2.952901in}{5.094304in}}{\pgfqpoint{2.952901in}{5.068259in}}%
\pgfpathcurveto{\pgfqpoint{2.952901in}{5.042214in}}{\pgfqpoint{2.963249in}{5.017231in}}{\pgfqpoint{2.981666in}{4.998815in}}%
\pgfpathcurveto{\pgfqpoint{3.000083in}{4.980398in}}{\pgfqpoint{3.025065in}{4.970050in}}{\pgfqpoint{3.051110in}{4.970050in}}%
\pgfpathlineto{\pgfqpoint{3.051110in}{4.970050in}}%
\pgfpathclose%
\pgfusepath{stroke}%
\end{pgfscope}%
\begin{pgfscope}%
\pgfpathrectangle{\pgfqpoint{1.201204in}{1.067125in}}{\pgfqpoint{7.164711in}{7.153266in}}%
\pgfusepath{clip}%
\pgfsetbuttcap%
\pgfsetroundjoin%
\pgfsetlinewidth{3.011250pt}%
\definecolor{currentstroke}{rgb}{0.000000,0.000000,0.000000}%
\pgfsetstrokecolor{currentstroke}%
\pgfsetdash{}{0pt}%
\pgfpathmoveto{\pgfqpoint{6.603805in}{3.697065in}}%
\pgfpathcurveto{\pgfqpoint{6.629850in}{3.697065in}}{\pgfqpoint{6.654832in}{3.707413in}}{\pgfqpoint{6.673249in}{3.725830in}}%
\pgfpathcurveto{\pgfqpoint{6.691666in}{3.744247in}}{\pgfqpoint{6.702014in}{3.769229in}}{\pgfqpoint{6.702014in}{3.795274in}}%
\pgfpathcurveto{\pgfqpoint{6.702014in}{3.821320in}}{\pgfqpoint{6.691666in}{3.846302in}}{\pgfqpoint{6.673249in}{3.864719in}}%
\pgfpathcurveto{\pgfqpoint{6.654832in}{3.883136in}}{\pgfqpoint{6.629850in}{3.893483in}}{\pgfqpoint{6.603805in}{3.893483in}}%
\pgfpathcurveto{\pgfqpoint{6.577759in}{3.893483in}}{\pgfqpoint{6.552777in}{3.883136in}}{\pgfqpoint{6.534360in}{3.864719in}}%
\pgfpathcurveto{\pgfqpoint{6.515944in}{3.846302in}}{\pgfqpoint{6.505596in}{3.821320in}}{\pgfqpoint{6.505596in}{3.795274in}}%
\pgfpathcurveto{\pgfqpoint{6.505596in}{3.769229in}}{\pgfqpoint{6.515944in}{3.744247in}}{\pgfqpoint{6.534360in}{3.725830in}}%
\pgfpathcurveto{\pgfqpoint{6.552777in}{3.707413in}}{\pgfqpoint{6.577759in}{3.697065in}}{\pgfqpoint{6.603805in}{3.697065in}}%
\pgfpathlineto{\pgfqpoint{6.603805in}{3.697065in}}%
\pgfpathclose%
\pgfusepath{stroke}%
\end{pgfscope}%
\begin{pgfscope}%
\pgfsetbuttcap%
\pgfsetroundjoin%
\definecolor{currentfill}{rgb}{0.000000,0.000000,0.000000}%
\pgfsetfillcolor{currentfill}%
\pgfsetlinewidth{0.803000pt}%
\definecolor{currentstroke}{rgb}{0.000000,0.000000,0.000000}%
\pgfsetstrokecolor{currentstroke}%
\pgfsetdash{}{0pt}%
\pgfsys@defobject{currentmarker}{\pgfqpoint{0.000000in}{-0.048611in}}{\pgfqpoint{0.000000in}{0.000000in}}{%
\pgfpathmoveto{\pgfqpoint{0.000000in}{0.000000in}}%
\pgfpathlineto{\pgfqpoint{0.000000in}{-0.048611in}}%
\pgfusepath{stroke,fill}%
}%
\begin{pgfscope}%
\pgfsys@transformshift{1.201204in}{1.067125in}%
\pgfsys@useobject{currentmarker}{}%
\end{pgfscope}%
\end{pgfscope}%
\begin{pgfscope}%
\definecolor{textcolor}{rgb}{0.000000,0.000000,0.000000}%
\pgfsetstrokecolor{textcolor}%
\pgfsetfillcolor{textcolor}%
\pgftext[x=1.201204in,y=0.969903in,,top]{\color{textcolor}\rmfamily\fontsize{30.000000}{36.000000}\selectfont 0}%
\end{pgfscope}%
\begin{pgfscope}%
\pgfsetbuttcap%
\pgfsetroundjoin%
\definecolor{currentfill}{rgb}{0.000000,0.000000,0.000000}%
\pgfsetfillcolor{currentfill}%
\pgfsetlinewidth{0.803000pt}%
\definecolor{currentstroke}{rgb}{0.000000,0.000000,0.000000}%
\pgfsetstrokecolor{currentstroke}%
\pgfsetdash{}{0pt}%
\pgfsys@defobject{currentmarker}{\pgfqpoint{0.000000in}{-0.048611in}}{\pgfqpoint{0.000000in}{0.000000in}}{%
\pgfpathmoveto{\pgfqpoint{0.000000in}{0.000000in}}%
\pgfpathlineto{\pgfqpoint{0.000000in}{-0.048611in}}%
\pgfusepath{stroke,fill}%
}%
\begin{pgfscope}%
\pgfsys@transformshift{2.634146in}{1.067125in}%
\pgfsys@useobject{currentmarker}{}%
\end{pgfscope}%
\end{pgfscope}%
\begin{pgfscope}%
\definecolor{textcolor}{rgb}{0.000000,0.000000,0.000000}%
\pgfsetstrokecolor{textcolor}%
\pgfsetfillcolor{textcolor}%
\pgftext[x=2.634146in,y=0.969903in,,top]{\color{textcolor}\rmfamily\fontsize{30.000000}{36.000000}\selectfont 10}%
\end{pgfscope}%
\begin{pgfscope}%
\pgfsetbuttcap%
\pgfsetroundjoin%
\definecolor{currentfill}{rgb}{0.000000,0.000000,0.000000}%
\pgfsetfillcolor{currentfill}%
\pgfsetlinewidth{0.803000pt}%
\definecolor{currentstroke}{rgb}{0.000000,0.000000,0.000000}%
\pgfsetstrokecolor{currentstroke}%
\pgfsetdash{}{0pt}%
\pgfsys@defobject{currentmarker}{\pgfqpoint{0.000000in}{-0.048611in}}{\pgfqpoint{0.000000in}{0.000000in}}{%
\pgfpathmoveto{\pgfqpoint{0.000000in}{0.000000in}}%
\pgfpathlineto{\pgfqpoint{0.000000in}{-0.048611in}}%
\pgfusepath{stroke,fill}%
}%
\begin{pgfscope}%
\pgfsys@transformshift{4.067089in}{1.067125in}%
\pgfsys@useobject{currentmarker}{}%
\end{pgfscope}%
\end{pgfscope}%
\begin{pgfscope}%
\definecolor{textcolor}{rgb}{0.000000,0.000000,0.000000}%
\pgfsetstrokecolor{textcolor}%
\pgfsetfillcolor{textcolor}%
\pgftext[x=4.067089in,y=0.969903in,,top]{\color{textcolor}\rmfamily\fontsize{30.000000}{36.000000}\selectfont 20}%
\end{pgfscope}%
\begin{pgfscope}%
\pgfsetbuttcap%
\pgfsetroundjoin%
\definecolor{currentfill}{rgb}{0.000000,0.000000,0.000000}%
\pgfsetfillcolor{currentfill}%
\pgfsetlinewidth{0.803000pt}%
\definecolor{currentstroke}{rgb}{0.000000,0.000000,0.000000}%
\pgfsetstrokecolor{currentstroke}%
\pgfsetdash{}{0pt}%
\pgfsys@defobject{currentmarker}{\pgfqpoint{0.000000in}{-0.048611in}}{\pgfqpoint{0.000000in}{0.000000in}}{%
\pgfpathmoveto{\pgfqpoint{0.000000in}{0.000000in}}%
\pgfpathlineto{\pgfqpoint{0.000000in}{-0.048611in}}%
\pgfusepath{stroke,fill}%
}%
\begin{pgfscope}%
\pgfsys@transformshift{5.500031in}{1.067125in}%
\pgfsys@useobject{currentmarker}{}%
\end{pgfscope}%
\end{pgfscope}%
\begin{pgfscope}%
\definecolor{textcolor}{rgb}{0.000000,0.000000,0.000000}%
\pgfsetstrokecolor{textcolor}%
\pgfsetfillcolor{textcolor}%
\pgftext[x=5.500031in,y=0.969903in,,top]{\color{textcolor}\rmfamily\fontsize{30.000000}{36.000000}\selectfont 30}%
\end{pgfscope}%
\begin{pgfscope}%
\pgfsetbuttcap%
\pgfsetroundjoin%
\definecolor{currentfill}{rgb}{0.000000,0.000000,0.000000}%
\pgfsetfillcolor{currentfill}%
\pgfsetlinewidth{0.803000pt}%
\definecolor{currentstroke}{rgb}{0.000000,0.000000,0.000000}%
\pgfsetstrokecolor{currentstroke}%
\pgfsetdash{}{0pt}%
\pgfsys@defobject{currentmarker}{\pgfqpoint{0.000000in}{-0.048611in}}{\pgfqpoint{0.000000in}{0.000000in}}{%
\pgfpathmoveto{\pgfqpoint{0.000000in}{0.000000in}}%
\pgfpathlineto{\pgfqpoint{0.000000in}{-0.048611in}}%
\pgfusepath{stroke,fill}%
}%
\begin{pgfscope}%
\pgfsys@transformshift{6.932973in}{1.067125in}%
\pgfsys@useobject{currentmarker}{}%
\end{pgfscope}%
\end{pgfscope}%
\begin{pgfscope}%
\definecolor{textcolor}{rgb}{0.000000,0.000000,0.000000}%
\pgfsetstrokecolor{textcolor}%
\pgfsetfillcolor{textcolor}%
\pgftext[x=6.932973in,y=0.969903in,,top]{\color{textcolor}\rmfamily\fontsize{30.000000}{36.000000}\selectfont 40}%
\end{pgfscope}%
\begin{pgfscope}%
\pgfsetbuttcap%
\pgfsetroundjoin%
\definecolor{currentfill}{rgb}{0.000000,0.000000,0.000000}%
\pgfsetfillcolor{currentfill}%
\pgfsetlinewidth{0.803000pt}%
\definecolor{currentstroke}{rgb}{0.000000,0.000000,0.000000}%
\pgfsetstrokecolor{currentstroke}%
\pgfsetdash{}{0pt}%
\pgfsys@defobject{currentmarker}{\pgfqpoint{0.000000in}{-0.048611in}}{\pgfqpoint{0.000000in}{0.000000in}}{%
\pgfpathmoveto{\pgfqpoint{0.000000in}{0.000000in}}%
\pgfpathlineto{\pgfqpoint{0.000000in}{-0.048611in}}%
\pgfusepath{stroke,fill}%
}%
\begin{pgfscope}%
\pgfsys@transformshift{8.365915in}{1.067125in}%
\pgfsys@useobject{currentmarker}{}%
\end{pgfscope}%
\end{pgfscope}%
\begin{pgfscope}%
\definecolor{textcolor}{rgb}{0.000000,0.000000,0.000000}%
\pgfsetstrokecolor{textcolor}%
\pgfsetfillcolor{textcolor}%
\pgftext[x=8.365915in,y=0.969903in,,top]{\color{textcolor}\rmfamily\fontsize{30.000000}{36.000000}\selectfont 50}%
\end{pgfscope}%
\begin{pgfscope}%
\definecolor{textcolor}{rgb}{0.000000,0.000000,0.000000}%
\pgfsetstrokecolor{textcolor}%
\pgfsetfillcolor{textcolor}%
\pgftext[x=4.783560in,y=0.523831in,,top]{\color{textcolor}\rmfamily\fontsize{34.000000}{40.800000}\selectfont \(\displaystyle x\) [m]}%
\end{pgfscope}%
\begin{pgfscope}%
\pgfsetbuttcap%
\pgfsetroundjoin%
\definecolor{currentfill}{rgb}{0.000000,0.000000,0.000000}%
\pgfsetfillcolor{currentfill}%
\pgfsetlinewidth{0.803000pt}%
\definecolor{currentstroke}{rgb}{0.000000,0.000000,0.000000}%
\pgfsetstrokecolor{currentstroke}%
\pgfsetdash{}{0pt}%
\pgfsys@defobject{currentmarker}{\pgfqpoint{-0.048611in}{0.000000in}}{\pgfqpoint{-0.000000in}{0.000000in}}{%
\pgfpathmoveto{\pgfqpoint{-0.000000in}{0.000000in}}%
\pgfpathlineto{\pgfqpoint{-0.048611in}{0.000000in}}%
\pgfusepath{stroke,fill}%
}%
\begin{pgfscope}%
\pgfsys@transformshift{1.201204in}{1.067125in}%
\pgfsys@useobject{currentmarker}{}%
\end{pgfscope}%
\end{pgfscope}%
\begin{pgfscope}%
\definecolor{textcolor}{rgb}{0.000000,0.000000,0.000000}%
\pgfsetstrokecolor{textcolor}%
\pgfsetfillcolor{textcolor}%
\pgftext[x=0.855226in, y=0.947140in, left, base]{\color{textcolor}\rmfamily\fontsize{30.000000}{36.000000}\selectfont 0}%
\end{pgfscope}%
\begin{pgfscope}%
\pgfsetbuttcap%
\pgfsetroundjoin%
\definecolor{currentfill}{rgb}{0.000000,0.000000,0.000000}%
\pgfsetfillcolor{currentfill}%
\pgfsetlinewidth{0.803000pt}%
\definecolor{currentstroke}{rgb}{0.000000,0.000000,0.000000}%
\pgfsetstrokecolor{currentstroke}%
\pgfsetdash{}{0pt}%
\pgfsys@defobject{currentmarker}{\pgfqpoint{-0.048611in}{0.000000in}}{\pgfqpoint{-0.000000in}{0.000000in}}{%
\pgfpathmoveto{\pgfqpoint{-0.000000in}{0.000000in}}%
\pgfpathlineto{\pgfqpoint{-0.048611in}{0.000000in}}%
\pgfusepath{stroke,fill}%
}%
\begin{pgfscope}%
\pgfsys@transformshift{1.201204in}{2.497778in}%
\pgfsys@useobject{currentmarker}{}%
\end{pgfscope}%
\end{pgfscope}%
\begin{pgfscope}%
\definecolor{textcolor}{rgb}{0.000000,0.000000,0.000000}%
\pgfsetstrokecolor{textcolor}%
\pgfsetfillcolor{textcolor}%
\pgftext[x=0.696748in, y=2.377793in, left, base]{\color{textcolor}\rmfamily\fontsize{30.000000}{36.000000}\selectfont 10}%
\end{pgfscope}%
\begin{pgfscope}%
\pgfsetbuttcap%
\pgfsetroundjoin%
\definecolor{currentfill}{rgb}{0.000000,0.000000,0.000000}%
\pgfsetfillcolor{currentfill}%
\pgfsetlinewidth{0.803000pt}%
\definecolor{currentstroke}{rgb}{0.000000,0.000000,0.000000}%
\pgfsetstrokecolor{currentstroke}%
\pgfsetdash{}{0pt}%
\pgfsys@defobject{currentmarker}{\pgfqpoint{-0.048611in}{0.000000in}}{\pgfqpoint{-0.000000in}{0.000000in}}{%
\pgfpathmoveto{\pgfqpoint{-0.000000in}{0.000000in}}%
\pgfpathlineto{\pgfqpoint{-0.048611in}{0.000000in}}%
\pgfusepath{stroke,fill}%
}%
\begin{pgfscope}%
\pgfsys@transformshift{1.201204in}{3.928431in}%
\pgfsys@useobject{currentmarker}{}%
\end{pgfscope}%
\end{pgfscope}%
\begin{pgfscope}%
\definecolor{textcolor}{rgb}{0.000000,0.000000,0.000000}%
\pgfsetstrokecolor{textcolor}%
\pgfsetfillcolor{textcolor}%
\pgftext[x=0.696748in, y=3.808447in, left, base]{\color{textcolor}\rmfamily\fontsize{30.000000}{36.000000}\selectfont 20}%
\end{pgfscope}%
\begin{pgfscope}%
\pgfsetbuttcap%
\pgfsetroundjoin%
\definecolor{currentfill}{rgb}{0.000000,0.000000,0.000000}%
\pgfsetfillcolor{currentfill}%
\pgfsetlinewidth{0.803000pt}%
\definecolor{currentstroke}{rgb}{0.000000,0.000000,0.000000}%
\pgfsetstrokecolor{currentstroke}%
\pgfsetdash{}{0pt}%
\pgfsys@defobject{currentmarker}{\pgfqpoint{-0.048611in}{0.000000in}}{\pgfqpoint{-0.000000in}{0.000000in}}{%
\pgfpathmoveto{\pgfqpoint{-0.000000in}{0.000000in}}%
\pgfpathlineto{\pgfqpoint{-0.048611in}{0.000000in}}%
\pgfusepath{stroke,fill}%
}%
\begin{pgfscope}%
\pgfsys@transformshift{1.201204in}{5.359084in}%
\pgfsys@useobject{currentmarker}{}%
\end{pgfscope}%
\end{pgfscope}%
\begin{pgfscope}%
\definecolor{textcolor}{rgb}{0.000000,0.000000,0.000000}%
\pgfsetstrokecolor{textcolor}%
\pgfsetfillcolor{textcolor}%
\pgftext[x=0.696748in, y=5.239100in, left, base]{\color{textcolor}\rmfamily\fontsize{30.000000}{36.000000}\selectfont 30}%
\end{pgfscope}%
\begin{pgfscope}%
\pgfsetbuttcap%
\pgfsetroundjoin%
\definecolor{currentfill}{rgb}{0.000000,0.000000,0.000000}%
\pgfsetfillcolor{currentfill}%
\pgfsetlinewidth{0.803000pt}%
\definecolor{currentstroke}{rgb}{0.000000,0.000000,0.000000}%
\pgfsetstrokecolor{currentstroke}%
\pgfsetdash{}{0pt}%
\pgfsys@defobject{currentmarker}{\pgfqpoint{-0.048611in}{0.000000in}}{\pgfqpoint{-0.000000in}{0.000000in}}{%
\pgfpathmoveto{\pgfqpoint{-0.000000in}{0.000000in}}%
\pgfpathlineto{\pgfqpoint{-0.048611in}{0.000000in}}%
\pgfusepath{stroke,fill}%
}%
\begin{pgfscope}%
\pgfsys@transformshift{1.201204in}{6.789738in}%
\pgfsys@useobject{currentmarker}{}%
\end{pgfscope}%
\end{pgfscope}%
\begin{pgfscope}%
\definecolor{textcolor}{rgb}{0.000000,0.000000,0.000000}%
\pgfsetstrokecolor{textcolor}%
\pgfsetfillcolor{textcolor}%
\pgftext[x=0.696748in, y=6.669753in, left, base]{\color{textcolor}\rmfamily\fontsize{30.000000}{36.000000}\selectfont 40}%
\end{pgfscope}%
\begin{pgfscope}%
\pgfsetbuttcap%
\pgfsetroundjoin%
\definecolor{currentfill}{rgb}{0.000000,0.000000,0.000000}%
\pgfsetfillcolor{currentfill}%
\pgfsetlinewidth{0.803000pt}%
\definecolor{currentstroke}{rgb}{0.000000,0.000000,0.000000}%
\pgfsetstrokecolor{currentstroke}%
\pgfsetdash{}{0pt}%
\pgfsys@defobject{currentmarker}{\pgfqpoint{-0.048611in}{0.000000in}}{\pgfqpoint{-0.000000in}{0.000000in}}{%
\pgfpathmoveto{\pgfqpoint{-0.000000in}{0.000000in}}%
\pgfpathlineto{\pgfqpoint{-0.048611in}{0.000000in}}%
\pgfusepath{stroke,fill}%
}%
\begin{pgfscope}%
\pgfsys@transformshift{1.201204in}{8.220391in}%
\pgfsys@useobject{currentmarker}{}%
\end{pgfscope}%
\end{pgfscope}%
\begin{pgfscope}%
\definecolor{textcolor}{rgb}{0.000000,0.000000,0.000000}%
\pgfsetstrokecolor{textcolor}%
\pgfsetfillcolor{textcolor}%
\pgftext[x=0.696748in, y=8.100406in, left, base]{\color{textcolor}\rmfamily\fontsize{30.000000}{36.000000}\selectfont 50}%
\end{pgfscope}%
\begin{pgfscope}%
\definecolor{textcolor}{rgb}{0.000000,0.000000,0.000000}%
\pgfsetstrokecolor{textcolor}%
\pgfsetfillcolor{textcolor}%
\pgftext[x=0.557859in,y=4.643758in,,bottom,rotate=90.000000]{\color{textcolor}\rmfamily\fontsize{34.000000}{40.800000}\selectfont \(\displaystyle y\) [m]}%
\end{pgfscope}%
\begin{pgfscope}%
\pgfsetrectcap%
\pgfsetmiterjoin%
\pgfsetlinewidth{0.803000pt}%
\definecolor{currentstroke}{rgb}{0.000000,0.000000,0.000000}%
\pgfsetstrokecolor{currentstroke}%
\pgfsetdash{}{0pt}%
\pgfpathmoveto{\pgfqpoint{1.201204in}{1.067125in}}%
\pgfpathlineto{\pgfqpoint{1.201204in}{8.220391in}}%
\pgfusepath{stroke}%
\end{pgfscope}%
\begin{pgfscope}%
\pgfsetrectcap%
\pgfsetmiterjoin%
\pgfsetlinewidth{0.803000pt}%
\definecolor{currentstroke}{rgb}{0.000000,0.000000,0.000000}%
\pgfsetstrokecolor{currentstroke}%
\pgfsetdash{}{0pt}%
\pgfpathmoveto{\pgfqpoint{8.365915in}{1.067125in}}%
\pgfpathlineto{\pgfqpoint{8.365915in}{8.220391in}}%
\pgfusepath{stroke}%
\end{pgfscope}%
\begin{pgfscope}%
\pgfsetrectcap%
\pgfsetmiterjoin%
\pgfsetlinewidth{0.803000pt}%
\definecolor{currentstroke}{rgb}{0.000000,0.000000,0.000000}%
\pgfsetstrokecolor{currentstroke}%
\pgfsetdash{}{0pt}%
\pgfpathmoveto{\pgfqpoint{1.201204in}{1.067125in}}%
\pgfpathlineto{\pgfqpoint{8.365915in}{1.067125in}}%
\pgfusepath{stroke}%
\end{pgfscope}%
\begin{pgfscope}%
\pgfsetrectcap%
\pgfsetmiterjoin%
\pgfsetlinewidth{0.803000pt}%
\definecolor{currentstroke}{rgb}{0.000000,0.000000,0.000000}%
\pgfsetstrokecolor{currentstroke}%
\pgfsetdash{}{0pt}%
\pgfpathmoveto{\pgfqpoint{1.201204in}{8.220391in}}%
\pgfpathlineto{\pgfqpoint{8.365915in}{8.220391in}}%
\pgfusepath{stroke}%
\end{pgfscope}%
\begin{pgfscope}%
\pgfsetbuttcap%
\pgfsetmiterjoin%
\definecolor{currentfill}{rgb}{1.000000,1.000000,1.000000}%
\pgfsetfillcolor{currentfill}%
\pgfsetfillopacity{0.800000}%
\pgfsetlinewidth{1.003750pt}%
\definecolor{currentstroke}{rgb}{0.800000,0.800000,0.800000}%
\pgfsetstrokecolor{currentstroke}%
\pgfsetstrokeopacity{0.800000}%
\pgfsetdash{}{0pt}%
\pgfpathmoveto{\pgfqpoint{1.453982in}{1.247681in}}%
\pgfpathlineto{\pgfqpoint{6.138717in}{1.247681in}}%
\pgfpathquadraticcurveto{\pgfqpoint{6.210939in}{1.247681in}}{\pgfqpoint{6.210939in}{1.319903in}}%
\pgfpathlineto{\pgfqpoint{6.210939in}{3.811490in}}%
\pgfpathquadraticcurveto{\pgfqpoint{6.210939in}{3.883712in}}{\pgfqpoint{6.138717in}{3.883712in}}%
\pgfpathlineto{\pgfqpoint{1.453982in}{3.883712in}}%
\pgfpathquadraticcurveto{\pgfqpoint{1.381760in}{3.883712in}}{\pgfqpoint{1.381760in}{3.811490in}}%
\pgfpathlineto{\pgfqpoint{1.381760in}{1.319903in}}%
\pgfpathquadraticcurveto{\pgfqpoint{1.381760in}{1.247681in}}{\pgfqpoint{1.453982in}{1.247681in}}%
\pgfpathlineto{\pgfqpoint{1.453982in}{1.247681in}}%
\pgfpathclose%
\pgfusepath{stroke,fill}%
\end{pgfscope}%
\begin{pgfscope}%
\pgfsetbuttcap%
\pgfsetroundjoin%
\definecolor{currentfill}{rgb}{0.750000,0.750000,0.000000}%
\pgfsetfillcolor{currentfill}%
\pgfsetlinewidth{3.011250pt}%
\definecolor{currentstroke}{rgb}{0.750000,0.750000,0.000000}%
\pgfsetstrokecolor{currentstroke}%
\pgfsetdash{}{0pt}%
\pgfsys@defobject{currentmarker}{\pgfqpoint{-0.098209in}{-0.098209in}}{\pgfqpoint{0.098209in}{0.098209in}}{%
\pgfpathmoveto{\pgfqpoint{-0.098209in}{-0.098209in}}%
\pgfpathlineto{\pgfqpoint{0.098209in}{0.098209in}}%
\pgfpathmoveto{\pgfqpoint{-0.098209in}{0.098209in}}%
\pgfpathlineto{\pgfqpoint{0.098209in}{-0.098209in}}%
\pgfusepath{stroke,fill}%
}%
\begin{pgfscope}%
\pgfsys@transformshift{1.887315in}{3.581282in}%
\pgfsys@useobject{currentmarker}{}%
\end{pgfscope}%
\end{pgfscope}%
\begin{pgfscope}%
\definecolor{textcolor}{rgb}{0.000000,0.000000,0.000000}%
\pgfsetstrokecolor{textcolor}%
\pgfsetfillcolor{textcolor}%
\pgftext[x=2.537315in,y=3.486490in,left,base]{\color{textcolor}\rmfamily\fontsize{26.000000}{31.200000}\selectfont LMB-TkBD}%
\end{pgfscope}%
\begin{pgfscope}%
\pgfsetbuttcap%
\pgfsetroundjoin%
\definecolor{currentfill}{rgb}{0.000000,0.000000,1.000000}%
\pgfsetfillcolor{currentfill}%
\pgfsetlinewidth{3.011250pt}%
\definecolor{currentstroke}{rgb}{0.000000,0.000000,1.000000}%
\pgfsetstrokecolor{currentstroke}%
\pgfsetdash{}{0pt}%
\pgfsys@defobject{currentmarker}{\pgfqpoint{-0.098209in}{-0.098209in}}{\pgfqpoint{0.098209in}{0.098209in}}{%
\pgfpathmoveto{\pgfqpoint{-0.098209in}{-0.098209in}}%
\pgfpathlineto{\pgfqpoint{0.098209in}{0.098209in}}%
\pgfpathmoveto{\pgfqpoint{-0.098209in}{0.098209in}}%
\pgfpathlineto{\pgfqpoint{0.098209in}{-0.098209in}}%
\pgfusepath{stroke,fill}%
}%
\begin{pgfscope}%
\pgfsys@transformshift{1.887315in}{3.080735in}%
\pgfsys@useobject{currentmarker}{}%
\end{pgfscope}%
\end{pgfscope}%
\begin{pgfscope}%
\definecolor{textcolor}{rgb}{0.000000,0.000000,0.000000}%
\pgfsetstrokecolor{textcolor}%
\pgfsetfillcolor{textcolor}%
\pgftext[x=2.537315in,y=2.985943in,left,base]{\color{textcolor}\rmfamily\fontsize{26.000000}{31.200000}\selectfont MP + BP-DTT}%
\end{pgfscope}%
\begin{pgfscope}%
\pgfsetbuttcap%
\pgfsetroundjoin%
\definecolor{currentfill}{rgb}{0.000000,0.500000,0.000000}%
\pgfsetfillcolor{currentfill}%
\pgfsetlinewidth{3.011250pt}%
\definecolor{currentstroke}{rgb}{0.000000,0.500000,0.000000}%
\pgfsetstrokecolor{currentstroke}%
\pgfsetdash{}{0pt}%
\pgfsys@defobject{currentmarker}{\pgfqpoint{-0.098209in}{-0.098209in}}{\pgfqpoint{0.098209in}{0.098209in}}{%
\pgfpathmoveto{\pgfqpoint{-0.098209in}{-0.098209in}}%
\pgfpathlineto{\pgfqpoint{0.098209in}{0.098209in}}%
\pgfpathmoveto{\pgfqpoint{-0.098209in}{0.098209in}}%
\pgfpathlineto{\pgfqpoint{0.098209in}{-0.098209in}}%
\pgfusepath{stroke,fill}%
}%
\begin{pgfscope}%
\pgfsys@transformshift{1.887315in}{2.580188in}%
\pgfsys@useobject{currentmarker}{}%
\end{pgfscope}%
\end{pgfscope}%
\begin{pgfscope}%
\definecolor{textcolor}{rgb}{0.000000,0.000000,0.000000}%
\pgfsetstrokecolor{textcolor}%
\pgfsetfillcolor{textcolor}%
\pgftext[x=2.537315in,y=2.485396in,left,base]{\color{textcolor}\rmfamily\fontsize{26.000000}{31.200000}\selectfont SBL + BP-DTT}%
\end{pgfscope}%
\begin{pgfscope}%
\pgfsetbuttcap%
\pgfsetroundjoin%
\definecolor{currentfill}{rgb}{1.000000,0.000000,0.000000}%
\pgfsetfillcolor{currentfill}%
\pgfsetlinewidth{3.011250pt}%
\definecolor{currentstroke}{rgb}{1.000000,0.000000,0.000000}%
\pgfsetstrokecolor{currentstroke}%
\pgfsetdash{}{0pt}%
\pgfsys@defobject{currentmarker}{\pgfqpoint{-0.098209in}{-0.098209in}}{\pgfqpoint{0.098209in}{0.098209in}}{%
\pgfpathmoveto{\pgfqpoint{-0.098209in}{0.000000in}}%
\pgfpathlineto{\pgfqpoint{0.098209in}{0.000000in}}%
\pgfpathmoveto{\pgfqpoint{0.000000in}{-0.098209in}}%
\pgfpathlineto{\pgfqpoint{0.000000in}{0.098209in}}%
\pgfusepath{stroke,fill}%
}%
\begin{pgfscope}%
\pgfsys@transformshift{1.887315in}{2.073563in}%
\pgfsys@useobject{currentmarker}{}%
\end{pgfscope}%
\end{pgfscope}%
\begin{pgfscope}%
\definecolor{textcolor}{rgb}{0.000000,0.000000,0.000000}%
\pgfsetstrokecolor{textcolor}%
\pgfsetfillcolor{textcolor}%
\pgftext[x=2.537315in,y=1.978772in,left,base]{\color{textcolor}\rmfamily\fontsize{26.000000}{31.200000}\selectfont BP-TkBD (proposed)~~~~~}%
\end{pgfscope}%
\begin{pgfscope}%
\pgfsetbuttcap%
\pgfsetroundjoin%
\pgfsetlinewidth{3.011250pt}%
\definecolor{currentstroke}{rgb}{0.000000,0.000000,0.000000}%
\pgfsetstrokecolor{currentstroke}%
\pgfsetdash{}{0pt}%
\pgfpathmoveto{\pgfqpoint{1.887315in}{1.455921in}}%
\pgfpathcurveto{\pgfqpoint{1.913361in}{1.455921in}}{\pgfqpoint{1.938343in}{1.466269in}}{\pgfqpoint{1.956760in}{1.484686in}}%
\pgfpathcurveto{\pgfqpoint{1.975177in}{1.503103in}}{\pgfqpoint{1.985525in}{1.528085in}}{\pgfqpoint{1.985525in}{1.554130in}}%
\pgfpathcurveto{\pgfqpoint{1.985525in}{1.580176in}}{\pgfqpoint{1.975177in}{1.605158in}}{\pgfqpoint{1.956760in}{1.623575in}}%
\pgfpathcurveto{\pgfqpoint{1.938343in}{1.641991in}}{\pgfqpoint{1.913361in}{1.652339in}}{\pgfqpoint{1.887315in}{1.652339in}}%
\pgfpathcurveto{\pgfqpoint{1.861270in}{1.652339in}}{\pgfqpoint{1.836288in}{1.641991in}}{\pgfqpoint{1.817871in}{1.623575in}}%
\pgfpathcurveto{\pgfqpoint{1.799454in}{1.605158in}}{\pgfqpoint{1.789106in}{1.580176in}}{\pgfqpoint{1.789106in}{1.554130in}}%
\pgfpathcurveto{\pgfqpoint{1.789106in}{1.528085in}}{\pgfqpoint{1.799454in}{1.503103in}}{\pgfqpoint{1.817871in}{1.484686in}}%
\pgfpathcurveto{\pgfqpoint{1.836288in}{1.466269in}}{\pgfqpoint{1.861270in}{1.455921in}}{\pgfqpoint{1.887315in}{1.455921in}}%
\pgfpathlineto{\pgfqpoint{1.887315in}{1.455921in}}%
\pgfpathclose%
\pgfusepath{stroke}%
\end{pgfscope}%
\begin{pgfscope}%
\definecolor{textcolor}{rgb}{0.000000,0.000000,0.000000}%
\pgfsetstrokecolor{textcolor}%
\pgfsetfillcolor{textcolor}%
\pgftext[x=2.537315in,y=1.459338in,left,base]{\color{textcolor}\rmfamily\fontsize{26.000000}{31.200000}\selectfont Ground Truth~~~~}%
\end{pgfscope}%
\begin{pgfscope}%
\pgfsetbuttcap%
\pgfsetmiterjoin%
\definecolor{currentfill}{rgb}{1.000000,1.000000,1.000000}%
\pgfsetfillcolor{currentfill}%
\pgfsetlinewidth{0.000000pt}%
\definecolor{currentstroke}{rgb}{0.000000,0.000000,0.000000}%
\pgfsetstrokecolor{currentstroke}%
\pgfsetstrokeopacity{0.000000}%
\pgfsetdash{}{0pt}%
\pgfpathmoveto{\pgfqpoint{8.679470in}{1.067125in}}%
\pgfpathlineto{\pgfqpoint{9.037133in}{1.067125in}}%
\pgfpathlineto{\pgfqpoint{9.037133in}{8.220391in}}%
\pgfpathlineto{\pgfqpoint{8.679470in}{8.220391in}}%
\pgfpathlineto{\pgfqpoint{8.679470in}{1.067125in}}%
\pgfpathclose%
\pgfusepath{fill}%
\end{pgfscope}%
\begin{pgfscope}%
\pgfpathrectangle{\pgfqpoint{8.679470in}{1.067125in}}{\pgfqpoint{0.357663in}{7.153266in}}%
\pgfusepath{clip}%
\pgfsetbuttcap%
\pgfsetmiterjoin%
\definecolor{currentfill}{rgb}{1.000000,1.000000,1.000000}%
\pgfsetfillcolor{currentfill}%
\pgfsetlinewidth{0.010037pt}%
\definecolor{currentstroke}{rgb}{1.000000,1.000000,1.000000}%
\pgfsetstrokecolor{currentstroke}%
\pgfsetdash{}{0pt}%
\pgfusepath{stroke,fill}%
\end{pgfscope}%
\begin{pgfscope}%
\pgfsys@transformshift{8.680000in}{1.070000in}%
\pgftext[left,bottom]{\includegraphics[interpolate=true,width=0.360000in,height=7.150000in]{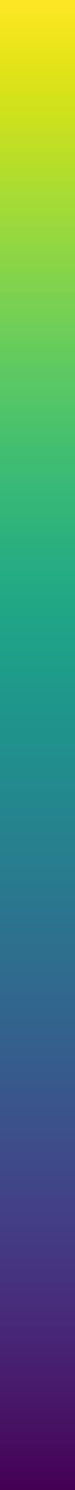}}%
\end{pgfscope}%
\begin{pgfscope}%
\pgfsetbuttcap%
\pgfsetroundjoin%
\definecolor{currentfill}{rgb}{0.000000,0.000000,0.000000}%
\pgfsetfillcolor{currentfill}%
\pgfsetlinewidth{0.803000pt}%
\definecolor{currentstroke}{rgb}{0.000000,0.000000,0.000000}%
\pgfsetstrokecolor{currentstroke}%
\pgfsetdash{}{0pt}%
\pgfsys@defobject{currentmarker}{\pgfqpoint{0.000000in}{0.000000in}}{\pgfqpoint{0.048611in}{0.000000in}}{%
\pgfpathmoveto{\pgfqpoint{0.000000in}{0.000000in}}%
\pgfpathlineto{\pgfqpoint{0.048611in}{0.000000in}}%
\pgfusepath{stroke,fill}%
}%
\begin{pgfscope}%
\pgfsys@transformshift{9.037133in}{1.067125in}%
\pgfsys@useobject{currentmarker}{}%
\end{pgfscope}%
\end{pgfscope}%
\begin{pgfscope}%
\definecolor{textcolor}{rgb}{0.000000,0.000000,0.000000}%
\pgfsetstrokecolor{textcolor}%
\pgfsetfillcolor{textcolor}%
\pgftext[x=9.134355in, y=0.947140in, left, base]{\color{textcolor}\rmfamily\fontsize{24.000000}{28.800000}\selectfont -80}%
\end{pgfscope}%
\begin{pgfscope}%
\pgfsetbuttcap%
\pgfsetroundjoin%
\definecolor{currentfill}{rgb}{0.000000,0.000000,0.000000}%
\pgfsetfillcolor{currentfill}%
\pgfsetlinewidth{0.803000pt}%
\definecolor{currentstroke}{rgb}{0.000000,0.000000,0.000000}%
\pgfsetstrokecolor{currentstroke}%
\pgfsetdash{}{0pt}%
\pgfsys@defobject{currentmarker}{\pgfqpoint{0.000000in}{0.000000in}}{\pgfqpoint{0.048611in}{0.000000in}}{%
\pgfpathmoveto{\pgfqpoint{0.000000in}{0.000000in}}%
\pgfpathlineto{\pgfqpoint{0.048611in}{0.000000in}}%
\pgfusepath{stroke,fill}%
}%
\begin{pgfscope}%
\pgfsys@transformshift{9.037133in}{1.961283in}%
\pgfsys@useobject{currentmarker}{}%
\end{pgfscope}%
\end{pgfscope}%
\begin{pgfscope}%
\definecolor{textcolor}{rgb}{0.000000,0.000000,0.000000}%
\pgfsetstrokecolor{textcolor}%
\pgfsetfillcolor{textcolor}%
\pgftext[x=9.134355in, y=1.841299in, left, base]{\color{textcolor}\rmfamily\fontsize{24.000000}{28.800000}\selectfont -70}%
\end{pgfscope}%
\begin{pgfscope}%
\pgfsetbuttcap%
\pgfsetroundjoin%
\definecolor{currentfill}{rgb}{0.000000,0.000000,0.000000}%
\pgfsetfillcolor{currentfill}%
\pgfsetlinewidth{0.803000pt}%
\definecolor{currentstroke}{rgb}{0.000000,0.000000,0.000000}%
\pgfsetstrokecolor{currentstroke}%
\pgfsetdash{}{0pt}%
\pgfsys@defobject{currentmarker}{\pgfqpoint{0.000000in}{0.000000in}}{\pgfqpoint{0.048611in}{0.000000in}}{%
\pgfpathmoveto{\pgfqpoint{0.000000in}{0.000000in}}%
\pgfpathlineto{\pgfqpoint{0.048611in}{0.000000in}}%
\pgfusepath{stroke,fill}%
}%
\begin{pgfscope}%
\pgfsys@transformshift{9.037133in}{2.855441in}%
\pgfsys@useobject{currentmarker}{}%
\end{pgfscope}%
\end{pgfscope}%
\begin{pgfscope}%
\definecolor{textcolor}{rgb}{0.000000,0.000000,0.000000}%
\pgfsetstrokecolor{textcolor}%
\pgfsetfillcolor{textcolor}%
\pgftext[x=9.134355in, y=2.735457in, left, base]{\color{textcolor}\rmfamily\fontsize{24.000000}{28.800000}\selectfont -60}%
\end{pgfscope}%
\begin{pgfscope}%
\pgfsetbuttcap%
\pgfsetroundjoin%
\definecolor{currentfill}{rgb}{0.000000,0.000000,0.000000}%
\pgfsetfillcolor{currentfill}%
\pgfsetlinewidth{0.803000pt}%
\definecolor{currentstroke}{rgb}{0.000000,0.000000,0.000000}%
\pgfsetstrokecolor{currentstroke}%
\pgfsetdash{}{0pt}%
\pgfsys@defobject{currentmarker}{\pgfqpoint{0.000000in}{0.000000in}}{\pgfqpoint{0.048611in}{0.000000in}}{%
\pgfpathmoveto{\pgfqpoint{0.000000in}{0.000000in}}%
\pgfpathlineto{\pgfqpoint{0.048611in}{0.000000in}}%
\pgfusepath{stroke,fill}%
}%
\begin{pgfscope}%
\pgfsys@transformshift{9.037133in}{3.749600in}%
\pgfsys@useobject{currentmarker}{}%
\end{pgfscope}%
\end{pgfscope}%
\begin{pgfscope}%
\definecolor{textcolor}{rgb}{0.000000,0.000000,0.000000}%
\pgfsetstrokecolor{textcolor}%
\pgfsetfillcolor{textcolor}%
\pgftext[x=9.134355in, y=3.629615in, left, base]{\color{textcolor}\rmfamily\fontsize{24.000000}{28.800000}\selectfont -50}%
\end{pgfscope}%
\begin{pgfscope}%
\pgfsetbuttcap%
\pgfsetroundjoin%
\definecolor{currentfill}{rgb}{0.000000,0.000000,0.000000}%
\pgfsetfillcolor{currentfill}%
\pgfsetlinewidth{0.803000pt}%
\definecolor{currentstroke}{rgb}{0.000000,0.000000,0.000000}%
\pgfsetstrokecolor{currentstroke}%
\pgfsetdash{}{0pt}%
\pgfsys@defobject{currentmarker}{\pgfqpoint{0.000000in}{0.000000in}}{\pgfqpoint{0.048611in}{0.000000in}}{%
\pgfpathmoveto{\pgfqpoint{0.000000in}{0.000000in}}%
\pgfpathlineto{\pgfqpoint{0.048611in}{0.000000in}}%
\pgfusepath{stroke,fill}%
}%
\begin{pgfscope}%
\pgfsys@transformshift{9.037133in}{4.643758in}%
\pgfsys@useobject{currentmarker}{}%
\end{pgfscope}%
\end{pgfscope}%
\begin{pgfscope}%
\definecolor{textcolor}{rgb}{0.000000,0.000000,0.000000}%
\pgfsetstrokecolor{textcolor}%
\pgfsetfillcolor{textcolor}%
\pgftext[x=9.134355in, y=4.523773in, left, base]{\color{textcolor}\rmfamily\fontsize{24.000000}{28.800000}\selectfont -40}%
\end{pgfscope}%
\begin{pgfscope}%
\pgfsetbuttcap%
\pgfsetroundjoin%
\definecolor{currentfill}{rgb}{0.000000,0.000000,0.000000}%
\pgfsetfillcolor{currentfill}%
\pgfsetlinewidth{0.803000pt}%
\definecolor{currentstroke}{rgb}{0.000000,0.000000,0.000000}%
\pgfsetstrokecolor{currentstroke}%
\pgfsetdash{}{0pt}%
\pgfsys@defobject{currentmarker}{\pgfqpoint{0.000000in}{0.000000in}}{\pgfqpoint{0.048611in}{0.000000in}}{%
\pgfpathmoveto{\pgfqpoint{0.000000in}{0.000000in}}%
\pgfpathlineto{\pgfqpoint{0.048611in}{0.000000in}}%
\pgfusepath{stroke,fill}%
}%
\begin{pgfscope}%
\pgfsys@transformshift{9.037133in}{5.537916in}%
\pgfsys@useobject{currentmarker}{}%
\end{pgfscope}%
\end{pgfscope}%
\begin{pgfscope}%
\definecolor{textcolor}{rgb}{0.000000,0.000000,0.000000}%
\pgfsetstrokecolor{textcolor}%
\pgfsetfillcolor{textcolor}%
\pgftext[x=9.134355in, y=5.417931in, left, base]{\color{textcolor}\rmfamily\fontsize{24.000000}{28.800000}\selectfont -30}%
\end{pgfscope}%
\begin{pgfscope}%
\pgfsetbuttcap%
\pgfsetroundjoin%
\definecolor{currentfill}{rgb}{0.000000,0.000000,0.000000}%
\pgfsetfillcolor{currentfill}%
\pgfsetlinewidth{0.803000pt}%
\definecolor{currentstroke}{rgb}{0.000000,0.000000,0.000000}%
\pgfsetstrokecolor{currentstroke}%
\pgfsetdash{}{0pt}%
\pgfsys@defobject{currentmarker}{\pgfqpoint{0.000000in}{0.000000in}}{\pgfqpoint{0.048611in}{0.000000in}}{%
\pgfpathmoveto{\pgfqpoint{0.000000in}{0.000000in}}%
\pgfpathlineto{\pgfqpoint{0.048611in}{0.000000in}}%
\pgfusepath{stroke,fill}%
}%
\begin{pgfscope}%
\pgfsys@transformshift{9.037133in}{6.432074in}%
\pgfsys@useobject{currentmarker}{}%
\end{pgfscope}%
\end{pgfscope}%
\begin{pgfscope}%
\definecolor{textcolor}{rgb}{0.000000,0.000000,0.000000}%
\pgfsetstrokecolor{textcolor}%
\pgfsetfillcolor{textcolor}%
\pgftext[x=9.134355in, y=6.312090in, left, base]{\color{textcolor}\rmfamily\fontsize{24.000000}{28.800000}\selectfont -20}%
\end{pgfscope}%
\begin{pgfscope}%
\pgfsetbuttcap%
\pgfsetroundjoin%
\definecolor{currentfill}{rgb}{0.000000,0.000000,0.000000}%
\pgfsetfillcolor{currentfill}%
\pgfsetlinewidth{0.803000pt}%
\definecolor{currentstroke}{rgb}{0.000000,0.000000,0.000000}%
\pgfsetstrokecolor{currentstroke}%
\pgfsetdash{}{0pt}%
\pgfsys@defobject{currentmarker}{\pgfqpoint{0.000000in}{0.000000in}}{\pgfqpoint{0.048611in}{0.000000in}}{%
\pgfpathmoveto{\pgfqpoint{0.000000in}{0.000000in}}%
\pgfpathlineto{\pgfqpoint{0.048611in}{0.000000in}}%
\pgfusepath{stroke,fill}%
}%
\begin{pgfscope}%
\pgfsys@transformshift{9.037133in}{7.326233in}%
\pgfsys@useobject{currentmarker}{}%
\end{pgfscope}%
\end{pgfscope}%
\begin{pgfscope}%
\definecolor{textcolor}{rgb}{0.000000,0.000000,0.000000}%
\pgfsetstrokecolor{textcolor}%
\pgfsetfillcolor{textcolor}%
\pgftext[x=9.134355in, y=7.206248in, left, base]{\color{textcolor}\rmfamily\fontsize{24.000000}{28.800000}\selectfont -10}%
\end{pgfscope}%
\begin{pgfscope}%
\pgfsetbuttcap%
\pgfsetroundjoin%
\definecolor{currentfill}{rgb}{0.000000,0.000000,0.000000}%
\pgfsetfillcolor{currentfill}%
\pgfsetlinewidth{0.803000pt}%
\definecolor{currentstroke}{rgb}{0.000000,0.000000,0.000000}%
\pgfsetstrokecolor{currentstroke}%
\pgfsetdash{}{0pt}%
\pgfsys@defobject{currentmarker}{\pgfqpoint{0.000000in}{0.000000in}}{\pgfqpoint{0.048611in}{0.000000in}}{%
\pgfpathmoveto{\pgfqpoint{0.000000in}{0.000000in}}%
\pgfpathlineto{\pgfqpoint{0.048611in}{0.000000in}}%
\pgfusepath{stroke,fill}%
}%
\begin{pgfscope}%
\pgfsys@transformshift{9.037133in}{8.220391in}%
\pgfsys@useobject{currentmarker}{}%
\end{pgfscope}%
\end{pgfscope}%
\begin{pgfscope}%
\definecolor{textcolor}{rgb}{0.000000,0.000000,0.000000}%
\pgfsetstrokecolor{textcolor}%
\pgfsetfillcolor{textcolor}%
\pgftext[x=9.134355in, y=8.100406in, left, base]{\color{textcolor}\rmfamily\fontsize{24.000000}{28.800000}\selectfont 0}%
\end{pgfscope}%
\begin{pgfscope}%
\pgfsetrectcap%
\pgfsetmiterjoin%
\pgfsetlinewidth{0.803000pt}%
\definecolor{currentstroke}{rgb}{0.000000,0.000000,0.000000}%
\pgfsetstrokecolor{currentstroke}%
\pgfsetdash{}{0pt}%
\pgfpathmoveto{\pgfqpoint{8.679470in}{1.067125in}}%
\pgfpathlineto{\pgfqpoint{8.858301in}{1.067125in}}%
\pgfpathlineto{\pgfqpoint{9.037133in}{1.067125in}}%
\pgfpathlineto{\pgfqpoint{9.037133in}{8.220391in}}%
\pgfpathlineto{\pgfqpoint{8.858301in}{8.220391in}}%
\pgfpathlineto{\pgfqpoint{8.679470in}{8.220391in}}%
\pgfpathlineto{\pgfqpoint{8.679470in}{1.067125in}}%
\pgfpathclose%
\pgfusepath{stroke}%
\end{pgfscope}%
\end{pgfpicture}%
\makeatother%
\endgroup%

%% file: Figs/snr/snr_sensor1_eps.pgf
\begingroup%
\makeatletter%
\begin{pgfpicture}%
\pgfpathrectangle{\pgfpointorigin}{\pgfqpoint{7.600000in}{6.400000in}}%
\pgfusepath{use as bounding box, clip}%
\begin{pgfscope}%
\pgfsetbuttcap%
\pgfsetmiterjoin%
\definecolor{currentfill}{rgb}{1.000000,1.000000,1.000000}%
\pgfsetfillcolor{currentfill}%
\pgfsetlinewidth{0.000000pt}%
\definecolor{currentstroke}{rgb}{1.000000,1.000000,1.000000}%
\pgfsetstrokecolor{currentstroke}%
\pgfsetdash{}{0pt}%
\pgfpathmoveto{\pgfqpoint{0.000000in}{0.000000in}}%
\pgfpathlineto{\pgfqpoint{7.600000in}{0.000000in}}%
\pgfpathlineto{\pgfqpoint{7.600000in}{6.400000in}}%
\pgfpathlineto{\pgfqpoint{0.000000in}{6.400000in}}%
\pgfpathlineto{\pgfqpoint{0.000000in}{0.000000in}}%
\pgfpathclose%
\pgfusepath{fill}%
\end{pgfscope}%
\begin{pgfscope}%
\pgfsetbuttcap%
\pgfsetmiterjoin%
\definecolor{currentfill}{rgb}{1.000000,1.000000,1.000000}%
\pgfsetfillcolor{currentfill}%
\pgfsetlinewidth{0.000000pt}%
\definecolor{currentstroke}{rgb}{0.000000,0.000000,0.000000}%
\pgfsetstrokecolor{currentstroke}%
\pgfsetstrokeopacity{0.000000}%
\pgfsetdash{}{0pt}%
\pgfpathmoveto{\pgfqpoint{1.214743in}{1.077744in}}%
\pgfpathlineto{\pgfqpoint{6.680908in}{1.077744in}}%
\pgfpathlineto{\pgfqpoint{6.680908in}{6.249898in}}%
\pgfpathlineto{\pgfqpoint{1.214743in}{6.249898in}}%
\pgfpathlineto{\pgfqpoint{1.214743in}{1.077744in}}%
\pgfpathclose%
\pgfusepath{fill}%
\end{pgfscope}%
\begin{pgfscope}%
\pgfpathrectangle{\pgfqpoint{1.214743in}{1.077744in}}{\pgfqpoint{5.466165in}{5.172154in}}%
\pgfusepath{clip}%
\pgfsys@transformshift{1.214743in}{1.077744in}%
\pgftext[left,bottom]{\includegraphics[interpolate=true,width=5.470000in,height=5.180000in]{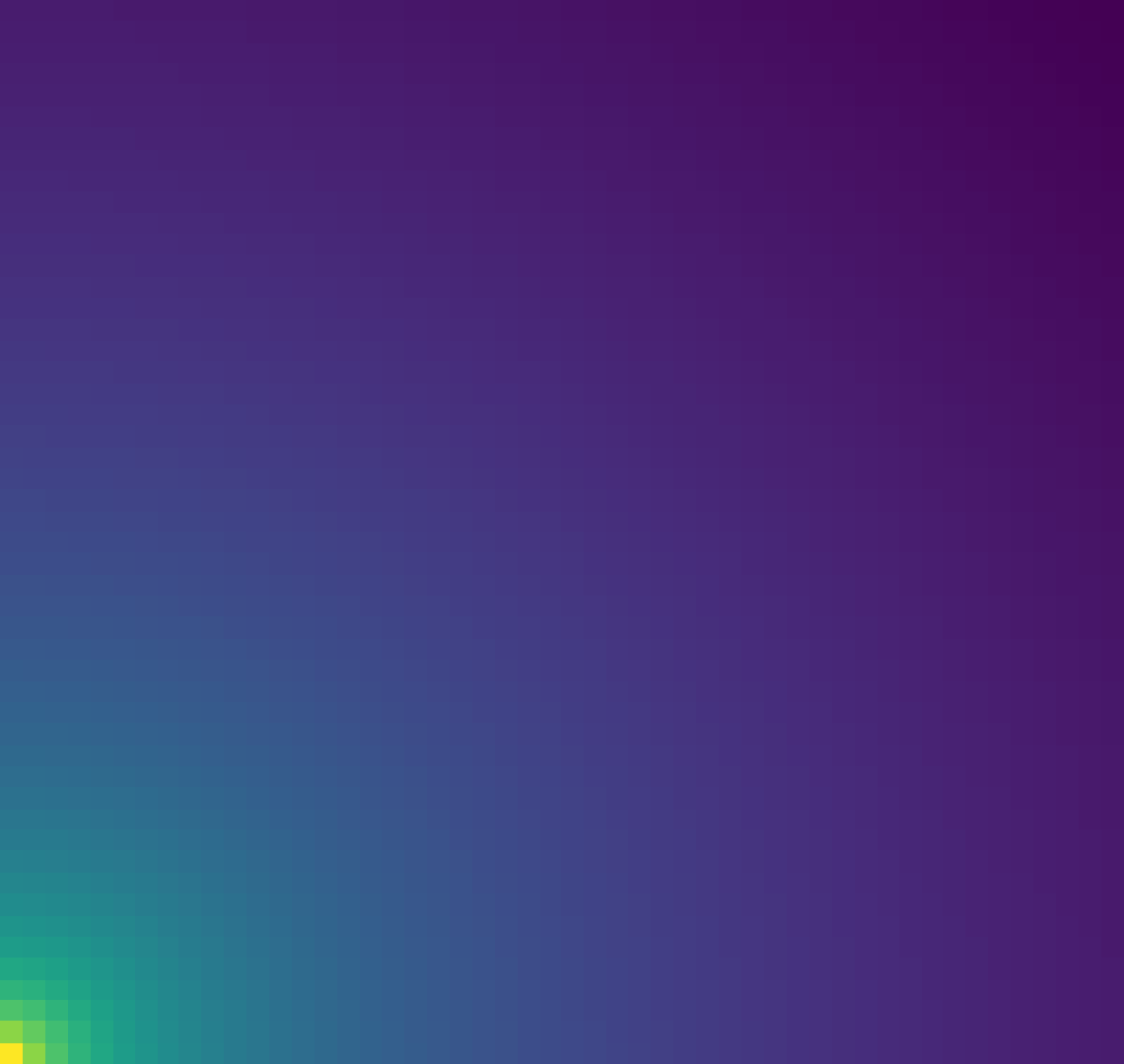}}%
\end{pgfscope}%
\begin{pgfscope}%
\pgfpathrectangle{\pgfqpoint{1.214743in}{1.077744in}}{\pgfqpoint{5.466165in}{5.172154in}}%
\pgfusepath{clip}%
\pgfsetbuttcap%
\pgfsetroundjoin%
\definecolor{currentfill}{rgb}{0.839216,0.152941,0.156863}%
\pgfsetfillcolor{currentfill}%
\pgfsetlinewidth{3.011250pt}%
\definecolor{currentstroke}{rgb}{0.839216,0.152941,0.156863}%
\pgfsetstrokecolor{currentstroke}%
\pgfsetdash{}{0pt}%
\pgfsys@defobject{currentmarker}{\pgfqpoint{-0.098209in}{-0.098209in}}{\pgfqpoint{0.098209in}{0.098209in}}{%
\pgfpathmoveto{\pgfqpoint{-0.098209in}{-0.098209in}}%
\pgfpathlineto{\pgfqpoint{0.098209in}{0.098209in}}%
\pgfpathmoveto{\pgfqpoint{-0.098209in}{0.098209in}}%
\pgfpathlineto{\pgfqpoint{0.098209in}{-0.098209in}}%
\pgfusepath{stroke,fill}%
}%
\begin{pgfscope}%
\pgfsys@transformshift{3.947826in}{2.112175in}%
\pgfsys@useobject{currentmarker}{}%
\end{pgfscope}%
\end{pgfscope}%
\begin{pgfscope}%
\pgfpathrectangle{\pgfqpoint{1.214743in}{1.077744in}}{\pgfqpoint{5.466165in}{5.172154in}}%
\pgfusepath{clip}%
\pgfsetbuttcap%
\pgfsetroundjoin%
\definecolor{currentfill}{rgb}{0.121569,0.466667,0.705882}%
\pgfsetfillcolor{currentfill}%
\pgfsetlinewidth{3.011250pt}%
\definecolor{currentstroke}{rgb}{0.121569,0.466667,0.705882}%
\pgfsetstrokecolor{currentstroke}%
\pgfsetdash{}{0pt}%
\pgfsys@defobject{currentmarker}{\pgfqpoint{-0.098209in}{-0.098209in}}{\pgfqpoint{0.098209in}{0.098209in}}{%
\pgfpathmoveto{\pgfqpoint{-0.098209in}{-0.098209in}}%
\pgfpathlineto{\pgfqpoint{0.098209in}{0.098209in}}%
\pgfpathmoveto{\pgfqpoint{-0.098209in}{0.098209in}}%
\pgfpathlineto{\pgfqpoint{0.098209in}{-0.098209in}}%
\pgfusepath{stroke,fill}%
}%
\begin{pgfscope}%
\pgfsys@transformshift{2.307976in}{3.663821in}%
\pgfsys@useobject{currentmarker}{}%
\end{pgfscope}%
\end{pgfscope}%
\begin{pgfscope}%
\pgfpathrectangle{\pgfqpoint{1.214743in}{1.077744in}}{\pgfqpoint{5.466165in}{5.172154in}}%
\pgfusepath{clip}%
\pgfsetbuttcap%
\pgfsetroundjoin%
\definecolor{currentfill}{rgb}{0.172549,0.627451,0.172549}%
\pgfsetfillcolor{currentfill}%
\pgfsetlinewidth{3.011250pt}%
\definecolor{currentstroke}{rgb}{0.172549,0.627451,0.172549}%
\pgfsetstrokecolor{currentstroke}%
\pgfsetdash{}{0pt}%
\pgfsys@defobject{currentmarker}{\pgfqpoint{-0.098209in}{-0.098209in}}{\pgfqpoint{0.098209in}{0.098209in}}{%
\pgfpathmoveto{\pgfqpoint{-0.098209in}{-0.098209in}}%
\pgfpathlineto{\pgfqpoint{0.098209in}{0.098209in}}%
\pgfpathmoveto{\pgfqpoint{-0.098209in}{0.098209in}}%
\pgfpathlineto{\pgfqpoint{0.098209in}{-0.098209in}}%
\pgfusepath{stroke,fill}%
}%
\begin{pgfscope}%
\pgfsys@transformshift{3.255820in}{3.136409in}%
\pgfsys@useobject{currentmarker}{}%
\end{pgfscope}%
\end{pgfscope}%
\begin{pgfscope}%
\pgfpathrectangle{\pgfqpoint{1.214743in}{1.077744in}}{\pgfqpoint{5.466165in}{5.172154in}}%
\pgfusepath{clip}%
\pgfsetbuttcap%
\pgfsetroundjoin%
\definecolor{currentfill}{rgb}{1.000000,0.498039,0.054902}%
\pgfsetfillcolor{currentfill}%
\pgfsetlinewidth{3.011250pt}%
\definecolor{currentstroke}{rgb}{1.000000,0.498039,0.054902}%
\pgfsetstrokecolor{currentstroke}%
\pgfsetdash{}{0pt}%
\pgfsys@defobject{currentmarker}{\pgfqpoint{-0.098209in}{-0.098209in}}{\pgfqpoint{0.098209in}{0.098209in}}{%
\pgfpathmoveto{\pgfqpoint{-0.098209in}{-0.098209in}}%
\pgfpathlineto{\pgfqpoint{0.098209in}{0.098209in}}%
\pgfpathmoveto{\pgfqpoint{-0.098209in}{0.098209in}}%
\pgfpathlineto{\pgfqpoint{0.098209in}{-0.098209in}}%
\pgfusepath{stroke,fill}%
}%
\begin{pgfscope}%
\pgfsys@transformshift{3.187765in}{3.773235in}%
\pgfsys@useobject{currentmarker}{}%
\end{pgfscope}%
\end{pgfscope}%
\begin{pgfscope}%
\pgfpathrectangle{\pgfqpoint{1.214743in}{1.077744in}}{\pgfqpoint{5.466165in}{5.172154in}}%
\pgfusepath{clip}%
\pgfsetbuttcap%
\pgfsetroundjoin%
\definecolor{currentfill}{rgb}{0.580392,0.403922,0.741176}%
\pgfsetfillcolor{currentfill}%
\pgfsetlinewidth{3.011250pt}%
\definecolor{currentstroke}{rgb}{0.580392,0.403922,0.741176}%
\pgfsetstrokecolor{currentstroke}%
\pgfsetdash{}{0pt}%
\pgfsys@defobject{currentmarker}{\pgfqpoint{-0.098209in}{-0.098209in}}{\pgfqpoint{0.098209in}{0.098209in}}{%
\pgfpathmoveto{\pgfqpoint{-0.098209in}{-0.098209in}}%
\pgfpathlineto{\pgfqpoint{0.098209in}{0.098209in}}%
\pgfpathmoveto{\pgfqpoint{-0.098209in}{0.098209in}}%
\pgfpathlineto{\pgfqpoint{0.098209in}{-0.098209in}}%
\pgfusepath{stroke,fill}%
}%
\begin{pgfscope}%
\pgfsys@transformshift{4.808004in}{3.490654in}%
\pgfsys@useobject{currentmarker}{}%
\end{pgfscope}%
\end{pgfscope}%
\begin{pgfscope}%
\pgfsetbuttcap%
\pgfsetroundjoin%
\definecolor{currentfill}{rgb}{0.000000,0.000000,0.000000}%
\pgfsetfillcolor{currentfill}%
\pgfsetlinewidth{0.803000pt}%
\definecolor{currentstroke}{rgb}{0.000000,0.000000,0.000000}%
\pgfsetstrokecolor{currentstroke}%
\pgfsetdash{}{0pt}%
\pgfsys@defobject{currentmarker}{\pgfqpoint{0.000000in}{-0.048611in}}{\pgfqpoint{0.000000in}{0.000000in}}{%
\pgfpathmoveto{\pgfqpoint{0.000000in}{0.000000in}}%
\pgfpathlineto{\pgfqpoint{0.000000in}{-0.048611in}}%
\pgfusepath{stroke,fill}%
}%
\begin{pgfscope}%
\pgfsys@transformshift{1.214743in}{1.077744in}%
\pgfsys@useobject{currentmarker}{}%
\end{pgfscope}%
\end{pgfscope}%
\begin{pgfscope}%
\definecolor{textcolor}{rgb}{0.000000,0.000000,0.000000}%
\pgfsetstrokecolor{textcolor}%
\pgfsetfillcolor{textcolor}%
\pgftext[x=1.214743in,y=0.980522in,,top]{\color{textcolor}\rmfamily\fontsize{32.000000}{38.400000}\selectfont 0}%
\end{pgfscope}%
\begin{pgfscope}%
\pgfsetbuttcap%
\pgfsetroundjoin%
\definecolor{currentfill}{rgb}{0.000000,0.000000,0.000000}%
\pgfsetfillcolor{currentfill}%
\pgfsetlinewidth{0.803000pt}%
\definecolor{currentstroke}{rgb}{0.000000,0.000000,0.000000}%
\pgfsetstrokecolor{currentstroke}%
\pgfsetdash{}{0pt}%
\pgfsys@defobject{currentmarker}{\pgfqpoint{0.000000in}{-0.048611in}}{\pgfqpoint{0.000000in}{0.000000in}}{%
\pgfpathmoveto{\pgfqpoint{0.000000in}{0.000000in}}%
\pgfpathlineto{\pgfqpoint{0.000000in}{-0.048611in}}%
\pgfusepath{stroke,fill}%
}%
\begin{pgfscope}%
\pgfsys@transformshift{2.307976in}{1.077744in}%
\pgfsys@useobject{currentmarker}{}%
\end{pgfscope}%
\end{pgfscope}%
\begin{pgfscope}%
\definecolor{textcolor}{rgb}{0.000000,0.000000,0.000000}%
\pgfsetstrokecolor{textcolor}%
\pgfsetfillcolor{textcolor}%
\pgftext[x=2.307976in,y=0.980522in,,top]{\color{textcolor}\rmfamily\fontsize{32.000000}{38.400000}\selectfont 10}%
\end{pgfscope}%
\begin{pgfscope}%
\pgfsetbuttcap%
\pgfsetroundjoin%
\definecolor{currentfill}{rgb}{0.000000,0.000000,0.000000}%
\pgfsetfillcolor{currentfill}%
\pgfsetlinewidth{0.803000pt}%
\definecolor{currentstroke}{rgb}{0.000000,0.000000,0.000000}%
\pgfsetstrokecolor{currentstroke}%
\pgfsetdash{}{0pt}%
\pgfsys@defobject{currentmarker}{\pgfqpoint{0.000000in}{-0.048611in}}{\pgfqpoint{0.000000in}{0.000000in}}{%
\pgfpathmoveto{\pgfqpoint{0.000000in}{0.000000in}}%
\pgfpathlineto{\pgfqpoint{0.000000in}{-0.048611in}}%
\pgfusepath{stroke,fill}%
}%
\begin{pgfscope}%
\pgfsys@transformshift{3.401209in}{1.077744in}%
\pgfsys@useobject{currentmarker}{}%
\end{pgfscope}%
\end{pgfscope}%
\begin{pgfscope}%
\definecolor{textcolor}{rgb}{0.000000,0.000000,0.000000}%
\pgfsetstrokecolor{textcolor}%
\pgfsetfillcolor{textcolor}%
\pgftext[x=3.401209in,y=0.980522in,,top]{\color{textcolor}\rmfamily\fontsize{32.000000}{38.400000}\selectfont 20}%
\end{pgfscope}%
\begin{pgfscope}%
\pgfsetbuttcap%
\pgfsetroundjoin%
\definecolor{currentfill}{rgb}{0.000000,0.000000,0.000000}%
\pgfsetfillcolor{currentfill}%
\pgfsetlinewidth{0.803000pt}%
\definecolor{currentstroke}{rgb}{0.000000,0.000000,0.000000}%
\pgfsetstrokecolor{currentstroke}%
\pgfsetdash{}{0pt}%
\pgfsys@defobject{currentmarker}{\pgfqpoint{0.000000in}{-0.048611in}}{\pgfqpoint{0.000000in}{0.000000in}}{%
\pgfpathmoveto{\pgfqpoint{0.000000in}{0.000000in}}%
\pgfpathlineto{\pgfqpoint{0.000000in}{-0.048611in}}%
\pgfusepath{stroke,fill}%
}%
\begin{pgfscope}%
\pgfsys@transformshift{4.494442in}{1.077744in}%
\pgfsys@useobject{currentmarker}{}%
\end{pgfscope}%
\end{pgfscope}%
\begin{pgfscope}%
\definecolor{textcolor}{rgb}{0.000000,0.000000,0.000000}%
\pgfsetstrokecolor{textcolor}%
\pgfsetfillcolor{textcolor}%
\pgftext[x=4.494442in,y=0.980522in,,top]{\color{textcolor}\rmfamily\fontsize{32.000000}{38.400000}\selectfont 30}%
\end{pgfscope}%
\begin{pgfscope}%
\pgfsetbuttcap%
\pgfsetroundjoin%
\definecolor{currentfill}{rgb}{0.000000,0.000000,0.000000}%
\pgfsetfillcolor{currentfill}%
\pgfsetlinewidth{0.803000pt}%
\definecolor{currentstroke}{rgb}{0.000000,0.000000,0.000000}%
\pgfsetstrokecolor{currentstroke}%
\pgfsetdash{}{0pt}%
\pgfsys@defobject{currentmarker}{\pgfqpoint{0.000000in}{-0.048611in}}{\pgfqpoint{0.000000in}{0.000000in}}{%
\pgfpathmoveto{\pgfqpoint{0.000000in}{0.000000in}}%
\pgfpathlineto{\pgfqpoint{0.000000in}{-0.048611in}}%
\pgfusepath{stroke,fill}%
}%
\begin{pgfscope}%
\pgfsys@transformshift{5.587675in}{1.077744in}%
\pgfsys@useobject{currentmarker}{}%
\end{pgfscope}%
\end{pgfscope}%
\begin{pgfscope}%
\definecolor{textcolor}{rgb}{0.000000,0.000000,0.000000}%
\pgfsetstrokecolor{textcolor}%
\pgfsetfillcolor{textcolor}%
\pgftext[x=5.587675in,y=0.980522in,,top]{\color{textcolor}\rmfamily\fontsize{32.000000}{38.400000}\selectfont 40}%
\end{pgfscope}%
\begin{pgfscope}%
\pgfsetbuttcap%
\pgfsetroundjoin%
\definecolor{currentfill}{rgb}{0.000000,0.000000,0.000000}%
\pgfsetfillcolor{currentfill}%
\pgfsetlinewidth{0.803000pt}%
\definecolor{currentstroke}{rgb}{0.000000,0.000000,0.000000}%
\pgfsetstrokecolor{currentstroke}%
\pgfsetdash{}{0pt}%
\pgfsys@defobject{currentmarker}{\pgfqpoint{0.000000in}{-0.048611in}}{\pgfqpoint{0.000000in}{0.000000in}}{%
\pgfpathmoveto{\pgfqpoint{0.000000in}{0.000000in}}%
\pgfpathlineto{\pgfqpoint{0.000000in}{-0.048611in}}%
\pgfusepath{stroke,fill}%
}%
\begin{pgfscope}%
\pgfsys@transformshift{6.680908in}{1.077744in}%
\pgfsys@useobject{currentmarker}{}%
\end{pgfscope}%
\end{pgfscope}%
\begin{pgfscope}%
\definecolor{textcolor}{rgb}{0.000000,0.000000,0.000000}%
\pgfsetstrokecolor{textcolor}%
\pgfsetfillcolor{textcolor}%
\pgftext[x=6.680908in,y=0.980522in,,top]{\color{textcolor}\rmfamily\fontsize{32.000000}{38.400000}\selectfont 50}%
\end{pgfscope}%
\begin{pgfscope}%
\definecolor{textcolor}{rgb}{0.000000,0.000000,0.000000}%
\pgfsetstrokecolor{textcolor}%
\pgfsetfillcolor{textcolor}%
\pgftext[x=3.947826in,y=0.534450in,,top]{\color{textcolor}\rmfamily\fontsize{36.000000}{43.200000}\selectfont \(\displaystyle x\) [m]}%
\end{pgfscope}%
\begin{pgfscope}%
\pgfsetbuttcap%
\pgfsetroundjoin%
\definecolor{currentfill}{rgb}{0.000000,0.000000,0.000000}%
\pgfsetfillcolor{currentfill}%
\pgfsetlinewidth{0.803000pt}%
\definecolor{currentstroke}{rgb}{0.000000,0.000000,0.000000}%
\pgfsetstrokecolor{currentstroke}%
\pgfsetdash{}{0pt}%
\pgfsys@defobject{currentmarker}{\pgfqpoint{-0.048611in}{0.000000in}}{\pgfqpoint{-0.000000in}{0.000000in}}{%
\pgfpathmoveto{\pgfqpoint{-0.000000in}{0.000000in}}%
\pgfpathlineto{\pgfqpoint{-0.048611in}{0.000000in}}%
\pgfusepath{stroke,fill}%
}%
\begin{pgfscope}%
\pgfsys@transformshift{1.214743in}{1.077744in}%
\pgfsys@useobject{currentmarker}{}%
\end{pgfscope}%
\end{pgfscope}%
\begin{pgfscope}%
\definecolor{textcolor}{rgb}{0.000000,0.000000,0.000000}%
\pgfsetstrokecolor{textcolor}%
\pgfsetfillcolor{textcolor}%
\pgftext[x=0.868765in, y=0.957759in, left, base]{\color{textcolor}\rmfamily\fontsize{32.000000}{38.400000}\selectfont 0}%
\end{pgfscope}%
\begin{pgfscope}%
\pgfsetbuttcap%
\pgfsetroundjoin%
\definecolor{currentfill}{rgb}{0.000000,0.000000,0.000000}%
\pgfsetfillcolor{currentfill}%
\pgfsetlinewidth{0.803000pt}%
\definecolor{currentstroke}{rgb}{0.000000,0.000000,0.000000}%
\pgfsetstrokecolor{currentstroke}%
\pgfsetdash{}{0pt}%
\pgfsys@defobject{currentmarker}{\pgfqpoint{-0.048611in}{0.000000in}}{\pgfqpoint{-0.000000in}{0.000000in}}{%
\pgfpathmoveto{\pgfqpoint{-0.000000in}{0.000000in}}%
\pgfpathlineto{\pgfqpoint{-0.048611in}{0.000000in}}%
\pgfusepath{stroke,fill}%
}%
\begin{pgfscope}%
\pgfsys@transformshift{1.214743in}{2.112175in}%
\pgfsys@useobject{currentmarker}{}%
\end{pgfscope}%
\end{pgfscope}%
\begin{pgfscope}%
\definecolor{textcolor}{rgb}{0.000000,0.000000,0.000000}%
\pgfsetstrokecolor{textcolor}%
\pgfsetfillcolor{textcolor}%
\pgftext[x=0.710287in, y=1.992190in, left, base]{\color{textcolor}\rmfamily\fontsize{32.000000}{38.400000}\selectfont 10}%
\end{pgfscope}%
\begin{pgfscope}%
\pgfsetbuttcap%
\pgfsetroundjoin%
\definecolor{currentfill}{rgb}{0.000000,0.000000,0.000000}%
\pgfsetfillcolor{currentfill}%
\pgfsetlinewidth{0.803000pt}%
\definecolor{currentstroke}{rgb}{0.000000,0.000000,0.000000}%
\pgfsetstrokecolor{currentstroke}%
\pgfsetdash{}{0pt}%
\pgfsys@defobject{currentmarker}{\pgfqpoint{-0.048611in}{0.000000in}}{\pgfqpoint{-0.000000in}{0.000000in}}{%
\pgfpathmoveto{\pgfqpoint{-0.000000in}{0.000000in}}%
\pgfpathlineto{\pgfqpoint{-0.048611in}{0.000000in}}%
\pgfusepath{stroke,fill}%
}%
\begin{pgfscope}%
\pgfsys@transformshift{1.214743in}{3.146606in}%
\pgfsys@useobject{currentmarker}{}%
\end{pgfscope}%
\end{pgfscope}%
\begin{pgfscope}%
\definecolor{textcolor}{rgb}{0.000000,0.000000,0.000000}%
\pgfsetstrokecolor{textcolor}%
\pgfsetfillcolor{textcolor}%
\pgftext[x=0.710287in, y=3.026621in, left, base]{\color{textcolor}\rmfamily\fontsize{32.000000}{38.400000}\selectfont 20}%
\end{pgfscope}%
\begin{pgfscope}%
\pgfsetbuttcap%
\pgfsetroundjoin%
\definecolor{currentfill}{rgb}{0.000000,0.000000,0.000000}%
\pgfsetfillcolor{currentfill}%
\pgfsetlinewidth{0.803000pt}%
\definecolor{currentstroke}{rgb}{0.000000,0.000000,0.000000}%
\pgfsetstrokecolor{currentstroke}%
\pgfsetdash{}{0pt}%
\pgfsys@defobject{currentmarker}{\pgfqpoint{-0.048611in}{0.000000in}}{\pgfqpoint{-0.000000in}{0.000000in}}{%
\pgfpathmoveto{\pgfqpoint{-0.000000in}{0.000000in}}%
\pgfpathlineto{\pgfqpoint{-0.048611in}{0.000000in}}%
\pgfusepath{stroke,fill}%
}%
\begin{pgfscope}%
\pgfsys@transformshift{1.214743in}{4.181037in}%
\pgfsys@useobject{currentmarker}{}%
\end{pgfscope}%
\end{pgfscope}%
\begin{pgfscope}%
\definecolor{textcolor}{rgb}{0.000000,0.000000,0.000000}%
\pgfsetstrokecolor{textcolor}%
\pgfsetfillcolor{textcolor}%
\pgftext[x=0.710287in, y=4.061052in, left, base]{\color{textcolor}\rmfamily\fontsize{32.000000}{38.400000}\selectfont 30}%
\end{pgfscope}%
\begin{pgfscope}%
\pgfsetbuttcap%
\pgfsetroundjoin%
\definecolor{currentfill}{rgb}{0.000000,0.000000,0.000000}%
\pgfsetfillcolor{currentfill}%
\pgfsetlinewidth{0.803000pt}%
\definecolor{currentstroke}{rgb}{0.000000,0.000000,0.000000}%
\pgfsetstrokecolor{currentstroke}%
\pgfsetdash{}{0pt}%
\pgfsys@defobject{currentmarker}{\pgfqpoint{-0.048611in}{0.000000in}}{\pgfqpoint{-0.000000in}{0.000000in}}{%
\pgfpathmoveto{\pgfqpoint{-0.000000in}{0.000000in}}%
\pgfpathlineto{\pgfqpoint{-0.048611in}{0.000000in}}%
\pgfusepath{stroke,fill}%
}%
\begin{pgfscope}%
\pgfsys@transformshift{1.214743in}{5.215467in}%
\pgfsys@useobject{currentmarker}{}%
\end{pgfscope}%
\end{pgfscope}%
\begin{pgfscope}%
\definecolor{textcolor}{rgb}{0.000000,0.000000,0.000000}%
\pgfsetstrokecolor{textcolor}%
\pgfsetfillcolor{textcolor}%
\pgftext[x=0.710287in, y=5.095483in, left, base]{\color{textcolor}\rmfamily\fontsize{32.000000}{38.400000}\selectfont 40}%
\end{pgfscope}%
\begin{pgfscope}%
\pgfsetbuttcap%
\pgfsetroundjoin%
\definecolor{currentfill}{rgb}{0.000000,0.000000,0.000000}%
\pgfsetfillcolor{currentfill}%
\pgfsetlinewidth{0.803000pt}%
\definecolor{currentstroke}{rgb}{0.000000,0.000000,0.000000}%
\pgfsetstrokecolor{currentstroke}%
\pgfsetdash{}{0pt}%
\pgfsys@defobject{currentmarker}{\pgfqpoint{-0.048611in}{0.000000in}}{\pgfqpoint{-0.000000in}{0.000000in}}{%
\pgfpathmoveto{\pgfqpoint{-0.000000in}{0.000000in}}%
\pgfpathlineto{\pgfqpoint{-0.048611in}{0.000000in}}%
\pgfusepath{stroke,fill}%
}%
\begin{pgfscope}%
\pgfsys@transformshift{1.214743in}{6.249898in}%
\pgfsys@useobject{currentmarker}{}%
\end{pgfscope}%
\end{pgfscope}%
\begin{pgfscope}%
\definecolor{textcolor}{rgb}{0.000000,0.000000,0.000000}%
\pgfsetstrokecolor{textcolor}%
\pgfsetfillcolor{textcolor}%
\pgftext[x=0.710287in, y=6.129914in, left, base]{\color{textcolor}\rmfamily\fontsize{32.000000}{38.400000}\selectfont 50}%
\end{pgfscope}%
\begin{pgfscope}%
\definecolor{textcolor}{rgb}{0.000000,0.000000,0.000000}%
\pgfsetstrokecolor{textcolor}%
\pgfsetfillcolor{textcolor}%
\pgftext[x=0.571398in,y=3.663821in,,bottom,rotate=90.000000]{\color{textcolor}\rmfamily\fontsize{36.000000}{43.200000}\selectfont \(\displaystyle y\) [m]}%
\end{pgfscope}%
\begin{pgfscope}%
\pgfpathrectangle{\pgfqpoint{1.214743in}{1.077744in}}{\pgfqpoint{5.466165in}{5.172154in}}%
\pgfusepath{clip}%
\pgfsetrectcap%
\pgfsetroundjoin%
\pgfsetlinewidth{3.011250pt}%
\definecolor{currentstroke}{rgb}{0.839216,0.152941,0.156863}%
\pgfsetstrokecolor{currentstroke}%
\pgfsetdash{}{0pt}%
\pgfpathmoveto{\pgfqpoint{3.947826in}{2.112175in}}%
\pgfpathlineto{\pgfqpoint{3.947425in}{2.163581in}}%
\pgfpathlineto{\pgfqpoint{3.945633in}{2.214272in}}%
\pgfpathlineto{\pgfqpoint{3.942969in}{2.265047in}}%
\pgfpathlineto{\pgfqpoint{3.939711in}{2.315778in}}%
\pgfpathlineto{\pgfqpoint{3.936186in}{2.366344in}}%
\pgfpathlineto{\pgfqpoint{3.933116in}{2.417843in}}%
\pgfpathlineto{\pgfqpoint{3.930098in}{2.469488in}}%
\pgfpathlineto{\pgfqpoint{3.927985in}{2.520836in}}%
\pgfpathlineto{\pgfqpoint{3.926557in}{2.573204in}}%
\pgfpathlineto{\pgfqpoint{3.924033in}{2.625803in}}%
\pgfpathlineto{\pgfqpoint{3.920613in}{2.678364in}}%
\pgfpathlineto{\pgfqpoint{3.916325in}{2.732177in}}%
\pgfpathlineto{\pgfqpoint{3.912511in}{2.786558in}}%
\pgfpathlineto{\pgfqpoint{3.909468in}{2.841300in}}%
\pgfpathlineto{\pgfqpoint{3.905093in}{2.897104in}}%
\pgfpathlineto{\pgfqpoint{3.898853in}{2.953769in}}%
\pgfpathlineto{\pgfqpoint{3.891283in}{3.010253in}}%
\pgfpathlineto{\pgfqpoint{3.883596in}{3.066853in}}%
\pgfpathlineto{\pgfqpoint{3.875528in}{3.123828in}}%
\pgfpathlineto{\pgfqpoint{3.866860in}{3.180814in}}%
\pgfpathlineto{\pgfqpoint{3.857316in}{3.237721in}}%
\pgfpathlineto{\pgfqpoint{3.846265in}{3.295013in}}%
\pgfpathlineto{\pgfqpoint{3.835019in}{3.352074in}}%
\pgfpathlineto{\pgfqpoint{3.824548in}{3.409508in}}%
\pgfpathlineto{\pgfqpoint{3.814112in}{3.468079in}}%
\pgfpathlineto{\pgfqpoint{3.802810in}{3.527216in}}%
\pgfpathlineto{\pgfqpoint{3.791145in}{3.587638in}}%
\pgfpathlineto{\pgfqpoint{3.779994in}{3.648903in}}%
\pgfpathlineto{\pgfqpoint{3.769211in}{3.710290in}}%
\pgfpathlineto{\pgfqpoint{3.758609in}{3.771677in}}%
\pgfpathlineto{\pgfqpoint{3.748115in}{3.832258in}}%
\pgfpathlineto{\pgfqpoint{3.736406in}{3.892289in}}%
\pgfpathlineto{\pgfqpoint{3.723413in}{3.952940in}}%
\pgfpathlineto{\pgfqpoint{3.709640in}{4.014463in}}%
\pgfpathlineto{\pgfqpoint{3.694764in}{4.076832in}}%
\pgfpathlineto{\pgfqpoint{3.679167in}{4.139978in}}%
\pgfpathlineto{\pgfqpoint{3.663933in}{4.204220in}}%
\pgfpathlineto{\pgfqpoint{3.650079in}{4.269386in}}%
\pgfpathlineto{\pgfqpoint{3.636943in}{4.335173in}}%
\pgfpathlineto{\pgfqpoint{3.624148in}{4.401177in}}%
\pgfpathlineto{\pgfqpoint{3.612226in}{4.466668in}}%
\pgfpathlineto{\pgfqpoint{3.601253in}{4.532420in}}%
\pgfpathlineto{\pgfqpoint{3.590806in}{4.598812in}}%
\pgfpathlineto{\pgfqpoint{3.581576in}{4.664851in}}%
\pgfpathlineto{\pgfqpoint{3.573033in}{4.731653in}}%
\pgfpathlineto{\pgfqpoint{3.564212in}{4.799147in}}%
\pgfpathlineto{\pgfqpoint{3.555348in}{4.866753in}}%
\pgfpathlineto{\pgfqpoint{3.546753in}{4.935101in}}%
\pgfpathlineto{\pgfqpoint{3.539281in}{5.003494in}}%
\pgfpathlineto{\pgfqpoint{3.532740in}{5.072257in}}%
\pgfpathlineto{\pgfqpoint{3.524935in}{5.141705in}}%
\pgfpathlineto{\pgfqpoint{3.515152in}{5.211349in}}%
\pgfpathlineto{\pgfqpoint{3.504777in}{5.281581in}}%
\pgfpathlineto{\pgfqpoint{3.493813in}{5.352294in}}%
\pgfpathlineto{\pgfqpoint{3.482387in}{5.422358in}}%
\pgfpathlineto{\pgfqpoint{3.471177in}{5.491881in}}%
\pgfpathlineto{\pgfqpoint{3.460601in}{5.561774in}}%
\pgfpathlineto{\pgfqpoint{3.451139in}{5.631890in}}%
\pgfpathlineto{\pgfqpoint{3.442060in}{5.702242in}}%
\pgfusepath{stroke}%
\end{pgfscope}%
\begin{pgfscope}%
\pgfpathrectangle{\pgfqpoint{1.214743in}{1.077744in}}{\pgfqpoint{5.466165in}{5.172154in}}%
\pgfusepath{clip}%
\pgfsetrectcap%
\pgfsetroundjoin%
\pgfsetlinewidth{3.011250pt}%
\definecolor{currentstroke}{rgb}{0.121569,0.466667,0.705882}%
\pgfsetstrokecolor{currentstroke}%
\pgfsetdash{}{0pt}%
\pgfpathmoveto{\pgfqpoint{2.307976in}{3.663821in}}%
\pgfpathlineto{\pgfqpoint{2.362800in}{3.664357in}}%
\pgfpathlineto{\pgfqpoint{2.417634in}{3.666348in}}%
\pgfpathlineto{\pgfqpoint{2.472684in}{3.669718in}}%
\pgfpathlineto{\pgfqpoint{2.527413in}{3.674479in}}%
\pgfpathlineto{\pgfqpoint{2.581168in}{3.680507in}}%
\pgfpathlineto{\pgfqpoint{2.634173in}{3.685928in}}%
\pgfpathlineto{\pgfqpoint{2.686457in}{3.690186in}}%
\pgfpathlineto{\pgfqpoint{2.738282in}{3.693915in}}%
\pgfpathlineto{\pgfqpoint{2.789454in}{3.696835in}}%
\pgfpathlineto{\pgfqpoint{2.840304in}{3.699397in}}%
\pgfpathlineto{\pgfqpoint{2.890607in}{3.701772in}}%
\pgfpathlineto{\pgfqpoint{2.939481in}{3.704059in}}%
\pgfpathlineto{\pgfqpoint{2.988511in}{3.706316in}}%
\pgfpathlineto{\pgfqpoint{3.038283in}{3.708307in}}%
\pgfpathlineto{\pgfqpoint{3.088111in}{3.710207in}}%
\pgfpathlineto{\pgfqpoint{3.138291in}{3.712176in}}%
\pgfpathlineto{\pgfqpoint{3.188184in}{3.714656in}}%
\pgfpathlineto{\pgfqpoint{3.238211in}{3.717708in}}%
\pgfpathlineto{\pgfqpoint{3.287596in}{3.720584in}}%
\pgfpathlineto{\pgfqpoint{3.335496in}{3.722410in}}%
\pgfpathlineto{\pgfqpoint{3.383268in}{3.724204in}}%
\pgfpathlineto{\pgfqpoint{3.430985in}{3.727139in}}%
\pgfpathlineto{\pgfqpoint{3.479012in}{3.729843in}}%
\pgfpathlineto{\pgfqpoint{3.526645in}{3.731306in}}%
\pgfpathlineto{\pgfqpoint{3.574209in}{3.732537in}}%
\pgfpathlineto{\pgfqpoint{3.622987in}{3.733802in}}%
\pgfpathlineto{\pgfqpoint{3.672935in}{3.735259in}}%
\pgfpathlineto{\pgfqpoint{3.722077in}{3.736894in}}%
\pgfpathlineto{\pgfqpoint{3.769770in}{3.738912in}}%
\pgfpathlineto{\pgfqpoint{3.817479in}{3.741058in}}%
\pgfpathlineto{\pgfqpoint{3.865339in}{3.742853in}}%
\pgfpathlineto{\pgfqpoint{3.913757in}{3.744823in}}%
\pgfpathlineto{\pgfqpoint{3.962070in}{3.747099in}}%
\pgfpathlineto{\pgfqpoint{4.009973in}{3.750561in}}%
\pgfpathlineto{\pgfqpoint{4.057103in}{3.754208in}}%
\pgfpathlineto{\pgfqpoint{4.103085in}{3.756477in}}%
\pgfpathlineto{\pgfqpoint{4.149232in}{3.757146in}}%
\pgfpathlineto{\pgfqpoint{4.195417in}{3.757303in}}%
\pgfpathlineto{\pgfqpoint{4.240721in}{3.757413in}}%
\pgfpathlineto{\pgfqpoint{4.284317in}{3.757581in}}%
\pgfpathlineto{\pgfqpoint{4.326196in}{3.758249in}}%
\pgfpathlineto{\pgfqpoint{4.367830in}{3.759008in}}%
\pgfpathlineto{\pgfqpoint{4.409525in}{3.760528in}}%
\pgfpathlineto{\pgfqpoint{4.451335in}{3.762924in}}%
\pgfpathlineto{\pgfqpoint{4.493525in}{3.766483in}}%
\pgfpathlineto{\pgfqpoint{4.535580in}{3.770322in}}%
\pgfpathlineto{\pgfqpoint{4.577984in}{3.773225in}}%
\pgfpathlineto{\pgfqpoint{4.621199in}{3.775951in}}%
\pgfpathlineto{\pgfqpoint{4.664142in}{3.779010in}}%
\pgfpathlineto{\pgfqpoint{4.707049in}{3.782174in}}%
\pgfpathlineto{\pgfqpoint{4.750572in}{3.785595in}}%
\pgfpathlineto{\pgfqpoint{4.795076in}{3.789695in}}%
\pgfpathlineto{\pgfqpoint{4.839888in}{3.794548in}}%
\pgfpathlineto{\pgfqpoint{4.883729in}{3.799851in}}%
\pgfpathlineto{\pgfqpoint{4.926849in}{3.805491in}}%
\pgfpathlineto{\pgfqpoint{4.969646in}{3.811428in}}%
\pgfpathlineto{\pgfqpoint{5.011744in}{3.817563in}}%
\pgfpathlineto{\pgfqpoint{5.054087in}{3.824047in}}%
\pgfpathlineto{\pgfqpoint{5.096849in}{3.830780in}}%
\pgfusepath{stroke}%
\end{pgfscope}%
\begin{pgfscope}%
\pgfpathrectangle{\pgfqpoint{1.214743in}{1.077744in}}{\pgfqpoint{5.466165in}{5.172154in}}%
\pgfusepath{clip}%
\pgfsetrectcap%
\pgfsetroundjoin%
\pgfsetlinewidth{3.011250pt}%
\definecolor{currentstroke}{rgb}{0.172549,0.627451,0.172549}%
\pgfsetstrokecolor{currentstroke}%
\pgfsetdash{}{0pt}%
\pgfpathmoveto{\pgfqpoint{3.255820in}{3.136409in}}%
\pgfpathlineto{\pgfqpoint{3.264186in}{3.175307in}}%
\pgfpathlineto{\pgfqpoint{3.273362in}{3.213731in}}%
\pgfpathlineto{\pgfqpoint{3.280889in}{3.251965in}}%
\pgfpathlineto{\pgfqpoint{3.286157in}{3.290266in}}%
\pgfpathlineto{\pgfqpoint{3.290601in}{3.329034in}}%
\pgfpathlineto{\pgfqpoint{3.294786in}{3.367501in}}%
\pgfpathlineto{\pgfqpoint{3.298687in}{3.405610in}}%
\pgfpathlineto{\pgfqpoint{3.301912in}{3.443510in}}%
\pgfpathlineto{\pgfqpoint{3.304481in}{3.479990in}}%
\pgfpathlineto{\pgfqpoint{3.307053in}{3.515856in}}%
\pgfpathlineto{\pgfqpoint{3.309647in}{3.551373in}}%
\pgfpathlineto{\pgfqpoint{3.312712in}{3.585446in}}%
\pgfpathlineto{\pgfqpoint{3.315562in}{3.618744in}}%
\pgfpathlineto{\pgfqpoint{3.317567in}{3.651579in}}%
\pgfpathlineto{\pgfqpoint{3.319349in}{3.684499in}}%
\pgfpathlineto{\pgfqpoint{3.319525in}{3.717409in}}%
\pgfpathlineto{\pgfqpoint{3.316969in}{3.750863in}}%
\pgfpathlineto{\pgfqpoint{3.313080in}{3.785194in}}%
\pgfpathlineto{\pgfqpoint{3.309277in}{3.820771in}}%
\pgfpathlineto{\pgfqpoint{3.306060in}{3.858189in}}%
\pgfpathlineto{\pgfqpoint{3.303957in}{3.895981in}}%
\pgfpathlineto{\pgfqpoint{3.302330in}{3.932791in}}%
\pgfpathlineto{\pgfqpoint{3.300658in}{3.968896in}}%
\pgfpathlineto{\pgfqpoint{3.300268in}{4.003927in}}%
\pgfpathlineto{\pgfqpoint{3.301000in}{4.038556in}}%
\pgfpathlineto{\pgfqpoint{3.302229in}{4.073628in}}%
\pgfpathlineto{\pgfqpoint{3.304139in}{4.107536in}}%
\pgfpathlineto{\pgfqpoint{3.306509in}{4.141154in}}%
\pgfpathlineto{\pgfqpoint{3.309058in}{4.175423in}}%
\pgfpathlineto{\pgfqpoint{3.312965in}{4.209141in}}%
\pgfpathlineto{\pgfqpoint{3.318191in}{4.242287in}}%
\pgfpathlineto{\pgfqpoint{3.322687in}{4.275333in}}%
\pgfpathlineto{\pgfqpoint{3.326051in}{4.308135in}}%
\pgfpathlineto{\pgfqpoint{3.328607in}{4.340713in}}%
\pgfpathlineto{\pgfqpoint{3.330815in}{4.373166in}}%
\pgfpathlineto{\pgfqpoint{3.332821in}{4.405632in}}%
\pgfpathlineto{\pgfqpoint{3.334604in}{4.437617in}}%
\pgfpathlineto{\pgfqpoint{3.336514in}{4.469564in}}%
\pgfpathlineto{\pgfqpoint{3.338572in}{4.501959in}}%
\pgfusepath{stroke}%
\end{pgfscope}%
\begin{pgfscope}%
\pgfpathrectangle{\pgfqpoint{1.214743in}{1.077744in}}{\pgfqpoint{5.466165in}{5.172154in}}%
\pgfusepath{clip}%
\pgfsetrectcap%
\pgfsetroundjoin%
\pgfsetlinewidth{3.011250pt}%
\definecolor{currentstroke}{rgb}{1.000000,0.498039,0.054902}%
\pgfsetstrokecolor{currentstroke}%
\pgfsetdash{}{0pt}%
\pgfpathmoveto{\pgfqpoint{3.187765in}{3.773235in}}%
\pgfpathlineto{\pgfqpoint{3.165500in}{3.782042in}}%
\pgfpathlineto{\pgfqpoint{3.144616in}{3.791793in}}%
\pgfpathlineto{\pgfqpoint{3.124794in}{3.801246in}}%
\pgfpathlineto{\pgfqpoint{3.105192in}{3.809468in}}%
\pgfpathlineto{\pgfqpoint{3.085175in}{3.817725in}}%
\pgfpathlineto{\pgfqpoint{3.065627in}{3.826093in}}%
\pgfpathlineto{\pgfqpoint{3.047002in}{3.835036in}}%
\pgfpathlineto{\pgfqpoint{3.027942in}{3.844992in}}%
\pgfpathlineto{\pgfqpoint{3.007813in}{3.855080in}}%
\pgfpathlineto{\pgfqpoint{2.986882in}{3.864237in}}%
\pgfpathlineto{\pgfqpoint{2.965100in}{3.872460in}}%
\pgfpathlineto{\pgfqpoint{2.942461in}{3.879546in}}%
\pgfpathlineto{\pgfqpoint{2.920331in}{3.886036in}}%
\pgfpathlineto{\pgfqpoint{2.898941in}{3.892768in}}%
\pgfpathlineto{\pgfqpoint{2.876645in}{3.899404in}}%
\pgfpathlineto{\pgfqpoint{2.853405in}{3.905762in}}%
\pgfpathlineto{\pgfqpoint{2.829774in}{3.911798in}}%
\pgfpathlineto{\pgfqpoint{2.806417in}{3.917547in}}%
\pgfpathlineto{\pgfqpoint{2.783816in}{3.923127in}}%
\pgfpathlineto{\pgfqpoint{2.761241in}{3.928742in}}%
\pgfpathlineto{\pgfqpoint{2.737880in}{3.934905in}}%
\pgfpathlineto{\pgfqpoint{2.713987in}{3.941944in}}%
\pgfpathlineto{\pgfqpoint{2.690826in}{3.949259in}}%
\pgfpathlineto{\pgfqpoint{2.668697in}{3.956275in}}%
\pgfpathlineto{\pgfqpoint{2.647455in}{3.963447in}}%
\pgfpathlineto{\pgfqpoint{2.626090in}{3.970756in}}%
\pgfpathlineto{\pgfqpoint{2.604451in}{3.977192in}}%
\pgfpathlineto{\pgfqpoint{2.583261in}{3.983646in}}%
\pgfpathlineto{\pgfqpoint{2.563272in}{3.989071in}}%
\pgfpathlineto{\pgfqpoint{2.544368in}{3.992922in}}%
\pgfpathlineto{\pgfqpoint{2.525113in}{3.997258in}}%
\pgfpathlineto{\pgfqpoint{2.504759in}{4.001501in}}%
\pgfpathlineto{\pgfqpoint{2.484100in}{4.006463in}}%
\pgfpathlineto{\pgfqpoint{2.464443in}{4.013234in}}%
\pgfpathlineto{\pgfqpoint{2.445171in}{4.020388in}}%
\pgfpathlineto{\pgfqpoint{2.424087in}{4.026921in}}%
\pgfpathlineto{\pgfqpoint{2.401034in}{4.033149in}}%
\pgfpathlineto{\pgfqpoint{2.376787in}{4.039887in}}%
\pgfpathlineto{\pgfqpoint{2.351567in}{4.046927in}}%
\pgfusepath{stroke}%
\end{pgfscope}%
\begin{pgfscope}%
\pgfpathrectangle{\pgfqpoint{1.214743in}{1.077744in}}{\pgfqpoint{5.466165in}{5.172154in}}%
\pgfusepath{clip}%
\pgfsetrectcap%
\pgfsetroundjoin%
\pgfsetlinewidth{3.011250pt}%
\definecolor{currentstroke}{rgb}{0.580392,0.403922,0.741176}%
\pgfsetstrokecolor{currentstroke}%
\pgfsetdash{}{0pt}%
\pgfpathmoveto{\pgfqpoint{4.808004in}{3.490654in}}%
\pgfpathlineto{\pgfqpoint{4.838085in}{3.461596in}}%
\pgfpathlineto{\pgfqpoint{4.868072in}{3.432689in}}%
\pgfpathlineto{\pgfqpoint{4.898584in}{3.403845in}}%
\pgfpathlineto{\pgfqpoint{4.929806in}{3.375048in}}%
\pgfpathlineto{\pgfqpoint{4.960908in}{3.347280in}}%
\pgfpathlineto{\pgfqpoint{4.991860in}{3.320615in}}%
\pgfpathlineto{\pgfqpoint{5.023545in}{3.294787in}}%
\pgfpathlineto{\pgfqpoint{5.056269in}{3.269151in}}%
\pgfpathlineto{\pgfqpoint{5.089370in}{3.243419in}}%
\pgfpathlineto{\pgfqpoint{5.122998in}{3.217574in}}%
\pgfpathlineto{\pgfqpoint{5.157303in}{3.191292in}}%
\pgfpathlineto{\pgfqpoint{5.192118in}{3.164187in}}%
\pgfpathlineto{\pgfqpoint{5.227728in}{3.136250in}}%
\pgfpathlineto{\pgfqpoint{5.263710in}{3.107800in}}%
\pgfpathlineto{\pgfqpoint{5.300259in}{3.078898in}}%
\pgfpathlineto{\pgfqpoint{5.336543in}{3.050327in}}%
\pgfpathlineto{\pgfqpoint{5.371072in}{3.021485in}}%
\pgfpathlineto{\pgfqpoint{5.403963in}{2.992063in}}%
\pgfpathlineto{\pgfqpoint{5.435479in}{2.965054in}}%
\pgfpathlineto{\pgfqpoint{5.466503in}{2.940083in}}%
\pgfpathlineto{\pgfqpoint{5.497554in}{2.914773in}}%
\pgfpathlineto{\pgfqpoint{5.528573in}{2.889381in}}%
\pgfpathlineto{\pgfqpoint{5.560152in}{2.864304in}}%
\pgfpathlineto{\pgfqpoint{5.592718in}{2.839267in}}%
\pgfpathlineto{\pgfqpoint{5.625540in}{2.813558in}}%
\pgfpathlineto{\pgfqpoint{5.658092in}{2.786734in}}%
\pgfpathlineto{\pgfqpoint{5.689260in}{2.759166in}}%
\pgfpathlineto{\pgfqpoint{5.718076in}{2.731954in}}%
\pgfpathlineto{\pgfqpoint{5.745674in}{2.704309in}}%
\pgfpathlineto{\pgfqpoint{5.772993in}{2.676250in}}%
\pgfpathlineto{\pgfqpoint{5.799760in}{2.649659in}}%
\pgfpathlineto{\pgfqpoint{5.825879in}{2.625062in}}%
\pgfpathlineto{\pgfqpoint{5.851738in}{2.602814in}}%
\pgfpathlineto{\pgfqpoint{5.877594in}{2.582099in}}%
\pgfpathlineto{\pgfqpoint{5.903989in}{2.561760in}}%
\pgfpathlineto{\pgfqpoint{5.930650in}{2.541817in}}%
\pgfpathlineto{\pgfqpoint{5.957730in}{2.522674in}}%
\pgfpathlineto{\pgfqpoint{5.984968in}{2.504377in}}%
\pgfpathlineto{\pgfqpoint{6.011216in}{2.485798in}}%
\pgfusepath{stroke}%
\end{pgfscope}%
\begin{pgfscope}%
\pgfsetrectcap%
\pgfsetmiterjoin%
\pgfsetlinewidth{0.803000pt}%
\definecolor{currentstroke}{rgb}{0.000000,0.000000,0.000000}%
\pgfsetstrokecolor{currentstroke}%
\pgfsetdash{}{0pt}%
\pgfpathmoveto{\pgfqpoint{1.214743in}{1.077744in}}%
\pgfpathlineto{\pgfqpoint{1.214743in}{6.249898in}}%
\pgfusepath{stroke}%
\end{pgfscope}%
\begin{pgfscope}%
\pgfsetrectcap%
\pgfsetmiterjoin%
\pgfsetlinewidth{0.803000pt}%
\definecolor{currentstroke}{rgb}{0.000000,0.000000,0.000000}%
\pgfsetstrokecolor{currentstroke}%
\pgfsetdash{}{0pt}%
\pgfpathmoveto{\pgfqpoint{6.680908in}{1.077744in}}%
\pgfpathlineto{\pgfqpoint{6.680908in}{6.249898in}}%
\pgfusepath{stroke}%
\end{pgfscope}%
\begin{pgfscope}%
\pgfsetrectcap%
\pgfsetmiterjoin%
\pgfsetlinewidth{0.803000pt}%
\definecolor{currentstroke}{rgb}{0.000000,0.000000,0.000000}%
\pgfsetstrokecolor{currentstroke}%
\pgfsetdash{}{0pt}%
\pgfpathmoveto{\pgfqpoint{1.214743in}{1.077744in}}%
\pgfpathlineto{\pgfqpoint{6.680908in}{1.077744in}}%
\pgfusepath{stroke}%
\end{pgfscope}%
\begin{pgfscope}%
\pgfsetrectcap%
\pgfsetmiterjoin%
\pgfsetlinewidth{0.803000pt}%
\definecolor{currentstroke}{rgb}{0.000000,0.000000,0.000000}%
\pgfsetstrokecolor{currentstroke}%
\pgfsetdash{}{0pt}%
\pgfpathmoveto{\pgfqpoint{1.214743in}{6.249898in}}%
\pgfpathlineto{\pgfqpoint{6.680908in}{6.249898in}}%
\pgfusepath{stroke}%
\end{pgfscope}%
\begin{pgfscope}%
\pgfsetbuttcap%
\pgfsetmiterjoin%
\definecolor{currentfill}{rgb}{1.000000,1.000000,1.000000}%
\pgfsetfillcolor{currentfill}%
\pgfsetlinewidth{0.000000pt}%
\definecolor{currentstroke}{rgb}{0.000000,0.000000,0.000000}%
\pgfsetstrokecolor{currentstroke}%
\pgfsetstrokeopacity{0.000000}%
\pgfsetdash{}{0pt}%
\pgfpathmoveto{\pgfqpoint{6.920127in}{1.077744in}}%
\pgfpathlineto{\pgfqpoint{7.178735in}{1.077744in}}%
\pgfpathlineto{\pgfqpoint{7.178735in}{6.249898in}}%
\pgfpathlineto{\pgfqpoint{6.920127in}{6.249898in}}%
\pgfpathlineto{\pgfqpoint{6.920127in}{1.077744in}}%
\pgfpathclose%
\pgfusepath{fill}%
\end{pgfscope}%
\begin{pgfscope}%
\pgfpathrectangle{\pgfqpoint{6.920127in}{1.077744in}}{\pgfqpoint{0.258608in}{5.172154in}}%
\pgfusepath{clip}%
\pgfsetbuttcap%
\pgfsetmiterjoin%
\definecolor{currentfill}{rgb}{1.000000,1.000000,1.000000}%
\pgfsetfillcolor{currentfill}%
\pgfsetlinewidth{0.010037pt}%
\definecolor{currentstroke}{rgb}{1.000000,1.000000,1.000000}%
\pgfsetstrokecolor{currentstroke}%
\pgfsetdash{}{0pt}%
\pgfusepath{stroke,fill}%
\end{pgfscope}%
\begin{pgfscope}%
\pgfsys@transformshift{6.920000in}{1.080000in}%
\pgftext[left,bottom]{\includegraphics[interpolate=true,width=0.260000in,height=5.170000in]{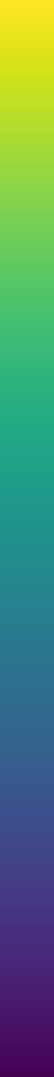}}%
\end{pgfscope}%
\begin{pgfscope}%
\pgfsetbuttcap%
\pgfsetroundjoin%
\definecolor{currentfill}{rgb}{0.000000,0.000000,0.000000}%
\pgfsetfillcolor{currentfill}%
\pgfsetlinewidth{0.803000pt}%
\definecolor{currentstroke}{rgb}{0.000000,0.000000,0.000000}%
\pgfsetstrokecolor{currentstroke}%
\pgfsetdash{}{0pt}%
\pgfsys@defobject{currentmarker}{\pgfqpoint{0.000000in}{0.000000in}}{\pgfqpoint{0.048611in}{0.000000in}}{%
\pgfpathmoveto{\pgfqpoint{0.000000in}{0.000000in}}%
\pgfpathlineto{\pgfqpoint{0.048611in}{0.000000in}}%
\pgfusepath{stroke,fill}%
}%
\begin{pgfscope}%
\pgfsys@transformshift{7.178735in}{2.243039in}%
\pgfsys@useobject{currentmarker}{}%
\end{pgfscope}%
\end{pgfscope}%
\begin{pgfscope}%
\definecolor{textcolor}{rgb}{0.000000,0.000000,0.000000}%
\pgfsetstrokecolor{textcolor}%
\pgfsetfillcolor{textcolor}%
\pgftext[x=7.275957in, y=2.123055in, left, base]{\color{textcolor}\rmfamily\fontsize{24.000000}{28.800000}\selectfont 0}%
\end{pgfscope}%
\begin{pgfscope}%
\pgfsetbuttcap%
\pgfsetroundjoin%
\definecolor{currentfill}{rgb}{0.000000,0.000000,0.000000}%
\pgfsetfillcolor{currentfill}%
\pgfsetlinewidth{0.803000pt}%
\definecolor{currentstroke}{rgb}{0.000000,0.000000,0.000000}%
\pgfsetstrokecolor{currentstroke}%
\pgfsetdash{}{0pt}%
\pgfsys@defobject{currentmarker}{\pgfqpoint{0.000000in}{0.000000in}}{\pgfqpoint{0.048611in}{0.000000in}}{%
\pgfpathmoveto{\pgfqpoint{0.000000in}{0.000000in}}%
\pgfpathlineto{\pgfqpoint{0.048611in}{0.000000in}}%
\pgfusepath{stroke,fill}%
}%
\begin{pgfscope}%
\pgfsys@transformshift{7.178735in}{3.538906in}%
\pgfsys@useobject{currentmarker}{}%
\end{pgfscope}%
\end{pgfscope}%
\begin{pgfscope}%
\definecolor{textcolor}{rgb}{0.000000,0.000000,0.000000}%
\pgfsetstrokecolor{textcolor}%
\pgfsetfillcolor{textcolor}%
\pgftext[x=7.275957in, y=3.418921in, left, base]{\color{textcolor}\rmfamily\fontsize{24.000000}{28.800000}\selectfont 20}%
\end{pgfscope}%
\begin{pgfscope}%
\pgfsetbuttcap%
\pgfsetroundjoin%
\definecolor{currentfill}{rgb}{0.000000,0.000000,0.000000}%
\pgfsetfillcolor{currentfill}%
\pgfsetlinewidth{0.803000pt}%
\definecolor{currentstroke}{rgb}{0.000000,0.000000,0.000000}%
\pgfsetstrokecolor{currentstroke}%
\pgfsetdash{}{0pt}%
\pgfsys@defobject{currentmarker}{\pgfqpoint{0.000000in}{0.000000in}}{\pgfqpoint{0.048611in}{0.000000in}}{%
\pgfpathmoveto{\pgfqpoint{0.000000in}{0.000000in}}%
\pgfpathlineto{\pgfqpoint{0.048611in}{0.000000in}}%
\pgfusepath{stroke,fill}%
}%
\begin{pgfscope}%
\pgfsys@transformshift{7.178735in}{4.834773in}%
\pgfsys@useobject{currentmarker}{}%
\end{pgfscope}%
\end{pgfscope}%
\begin{pgfscope}%
\definecolor{textcolor}{rgb}{0.000000,0.000000,0.000000}%
\pgfsetstrokecolor{textcolor}%
\pgfsetfillcolor{textcolor}%
\pgftext[x=7.275957in, y=4.714788in, left, base]{\color{textcolor}\rmfamily\fontsize{24.000000}{28.800000}\selectfont 40}%
\end{pgfscope}%
\begin{pgfscope}%
\pgfsetbuttcap%
\pgfsetroundjoin%
\definecolor{currentfill}{rgb}{0.000000,0.000000,0.000000}%
\pgfsetfillcolor{currentfill}%
\pgfsetlinewidth{0.803000pt}%
\definecolor{currentstroke}{rgb}{0.000000,0.000000,0.000000}%
\pgfsetstrokecolor{currentstroke}%
\pgfsetdash{}{0pt}%
\pgfsys@defobject{currentmarker}{\pgfqpoint{0.000000in}{0.000000in}}{\pgfqpoint{0.048611in}{0.000000in}}{%
\pgfpathmoveto{\pgfqpoint{0.000000in}{0.000000in}}%
\pgfpathlineto{\pgfqpoint{0.048611in}{0.000000in}}%
\pgfusepath{stroke,fill}%
}%
\begin{pgfscope}%
\pgfsys@transformshift{7.178735in}{6.130639in}%
\pgfsys@useobject{currentmarker}{}%
\end{pgfscope}%
\end{pgfscope}%
\begin{pgfscope}%
\definecolor{textcolor}{rgb}{0.000000,0.000000,0.000000}%
\pgfsetstrokecolor{textcolor}%
\pgfsetfillcolor{textcolor}%
\pgftext[x=7.275957in, y=6.010655in, left, base]{\color{textcolor}\rmfamily\fontsize{24.000000}{28.800000}\selectfont 60}%
\end{pgfscope}%
\begin{pgfscope}%
\pgfsetrectcap%
\pgfsetmiterjoin%
\pgfsetlinewidth{0.803000pt}%
\definecolor{currentstroke}{rgb}{0.000000,0.000000,0.000000}%
\pgfsetstrokecolor{currentstroke}%
\pgfsetdash{}{0pt}%
\pgfpathmoveto{\pgfqpoint{6.920127in}{1.077744in}}%
\pgfpathlineto{\pgfqpoint{7.049431in}{1.077744in}}%
\pgfpathlineto{\pgfqpoint{7.178735in}{1.077744in}}%
\pgfpathlineto{\pgfqpoint{7.178735in}{6.249898in}}%
\pgfpathlineto{\pgfqpoint{7.049431in}{6.249898in}}%
\pgfpathlineto{\pgfqpoint{6.920127in}{6.249898in}}%
\pgfpathlineto{\pgfqpoint{6.920127in}{1.077744in}}%
\pgfpathclose%
\pgfusepath{stroke}%
\end{pgfscope}%
\end{pgfpicture}%
\makeatother%
\endgroup%

%% file: Figs/snr/snr_vs_time_sensor1.pgf
\begingroup%
\makeatletter%
\begin{pgfpicture}%
\pgfpathrectangle{\pgfpointorigin}{\pgfqpoint{7.600000in}{6.400000in}}%
\pgfusepath{use as bounding box, clip}%
\begin{pgfscope}%
\pgfsetbuttcap%
\pgfsetmiterjoin%
\definecolor{currentfill}{rgb}{1.000000,1.000000,1.000000}%
\pgfsetfillcolor{currentfill}%
\pgfsetlinewidth{0.000000pt}%
\definecolor{currentstroke}{rgb}{1.000000,1.000000,1.000000}%
\pgfsetstrokecolor{currentstroke}%
\pgfsetdash{}{0pt}%
\pgfpathmoveto{\pgfqpoint{0.000000in}{0.000000in}}%
\pgfpathlineto{\pgfqpoint{7.600000in}{0.000000in}}%
\pgfpathlineto{\pgfqpoint{7.600000in}{6.400000in}}%
\pgfpathlineto{\pgfqpoint{0.000000in}{6.400000in}}%
\pgfpathlineto{\pgfqpoint{0.000000in}{0.000000in}}%
\pgfpathclose%
\pgfusepath{fill}%
\end{pgfscope}%
\begin{pgfscope}%
\pgfsetbuttcap%
\pgfsetmiterjoin%
\definecolor{currentfill}{rgb}{1.000000,1.000000,1.000000}%
\pgfsetfillcolor{currentfill}%
\pgfsetlinewidth{0.000000pt}%
\definecolor{currentstroke}{rgb}{0.000000,0.000000,0.000000}%
\pgfsetstrokecolor{currentstroke}%
\pgfsetstrokeopacity{0.000000}%
\pgfsetdash{}{0pt}%
\pgfpathmoveto{\pgfqpoint{1.362189in}{1.077744in}}%
\pgfpathlineto{\pgfqpoint{7.378608in}{1.077744in}}%
\pgfpathlineto{\pgfqpoint{7.378608in}{6.400000in}}%
\pgfpathlineto{\pgfqpoint{1.362189in}{6.400000in}}%
\pgfpathlineto{\pgfqpoint{1.362189in}{1.077744in}}%
\pgfpathclose%
\pgfusepath{fill}%
\end{pgfscope}%
\begin{pgfscope}%
\pgfpathrectangle{\pgfqpoint{1.362189in}{1.077744in}}{\pgfqpoint{6.016419in}{5.322256in}}%
\pgfusepath{clip}%
\pgfsetrectcap%
\pgfsetroundjoin%
\pgfsetlinewidth{0.803000pt}%
\definecolor{currentstroke}{rgb}{0.690196,0.690196,0.690196}%
\pgfsetstrokecolor{currentstroke}%
\pgfsetdash{}{0pt}%
\pgfpathmoveto{\pgfqpoint{2.809176in}{1.077744in}}%
\pgfpathlineto{\pgfqpoint{2.809176in}{6.400000in}}%
\pgfusepath{stroke}%
\end{pgfscope}%
\begin{pgfscope}%
\pgfsetbuttcap%
\pgfsetroundjoin%
\definecolor{currentfill}{rgb}{0.000000,0.000000,0.000000}%
\pgfsetfillcolor{currentfill}%
\pgfsetlinewidth{0.803000pt}%
\definecolor{currentstroke}{rgb}{0.000000,0.000000,0.000000}%
\pgfsetstrokecolor{currentstroke}%
\pgfsetdash{}{0pt}%
\pgfsys@defobject{currentmarker}{\pgfqpoint{0.000000in}{-0.048611in}}{\pgfqpoint{0.000000in}{0.000000in}}{%
\pgfpathmoveto{\pgfqpoint{0.000000in}{0.000000in}}%
\pgfpathlineto{\pgfqpoint{0.000000in}{-0.048611in}}%
\pgfusepath{stroke,fill}%
}%
\begin{pgfscope}%
\pgfsys@transformshift{2.809176in}{1.077744in}%
\pgfsys@useobject{currentmarker}{}%
\end{pgfscope}%
\end{pgfscope}%
\begin{pgfscope}%
\definecolor{textcolor}{rgb}{0.000000,0.000000,0.000000}%
\pgfsetstrokecolor{textcolor}%
\pgfsetfillcolor{textcolor}%
\pgftext[x=2.809176in,y=0.980522in,,top]{\color{textcolor}\rmfamily\fontsize{32.000000}{38.400000}\selectfont 20}%
\end{pgfscope}%
\begin{pgfscope}%
\pgfpathrectangle{\pgfqpoint{1.362189in}{1.077744in}}{\pgfqpoint{6.016419in}{5.322256in}}%
\pgfusepath{clip}%
\pgfsetrectcap%
\pgfsetroundjoin%
\pgfsetlinewidth{0.803000pt}%
\definecolor{currentstroke}{rgb}{0.690196,0.690196,0.690196}%
\pgfsetstrokecolor{currentstroke}%
\pgfsetdash{}{0pt}%
\pgfpathmoveto{\pgfqpoint{4.332320in}{1.077744in}}%
\pgfpathlineto{\pgfqpoint{4.332320in}{6.400000in}}%
\pgfusepath{stroke}%
\end{pgfscope}%
\begin{pgfscope}%
\pgfsetbuttcap%
\pgfsetroundjoin%
\definecolor{currentfill}{rgb}{0.000000,0.000000,0.000000}%
\pgfsetfillcolor{currentfill}%
\pgfsetlinewidth{0.803000pt}%
\definecolor{currentstroke}{rgb}{0.000000,0.000000,0.000000}%
\pgfsetstrokecolor{currentstroke}%
\pgfsetdash{}{0pt}%
\pgfsys@defobject{currentmarker}{\pgfqpoint{0.000000in}{-0.048611in}}{\pgfqpoint{0.000000in}{0.000000in}}{%
\pgfpathmoveto{\pgfqpoint{0.000000in}{0.000000in}}%
\pgfpathlineto{\pgfqpoint{0.000000in}{-0.048611in}}%
\pgfusepath{stroke,fill}%
}%
\begin{pgfscope}%
\pgfsys@transformshift{4.332320in}{1.077744in}%
\pgfsys@useobject{currentmarker}{}%
\end{pgfscope}%
\end{pgfscope}%
\begin{pgfscope}%
\definecolor{textcolor}{rgb}{0.000000,0.000000,0.000000}%
\pgfsetstrokecolor{textcolor}%
\pgfsetfillcolor{textcolor}%
\pgftext[x=4.332320in,y=0.980522in,,top]{\color{textcolor}\rmfamily\fontsize{32.000000}{38.400000}\selectfont 40}%
\end{pgfscope}%
\begin{pgfscope}%
\pgfpathrectangle{\pgfqpoint{1.362189in}{1.077744in}}{\pgfqpoint{6.016419in}{5.322256in}}%
\pgfusepath{clip}%
\pgfsetrectcap%
\pgfsetroundjoin%
\pgfsetlinewidth{0.803000pt}%
\definecolor{currentstroke}{rgb}{0.690196,0.690196,0.690196}%
\pgfsetstrokecolor{currentstroke}%
\pgfsetdash{}{0pt}%
\pgfpathmoveto{\pgfqpoint{5.855464in}{1.077744in}}%
\pgfpathlineto{\pgfqpoint{5.855464in}{6.400000in}}%
\pgfusepath{stroke}%
\end{pgfscope}%
\begin{pgfscope}%
\pgfsetbuttcap%
\pgfsetroundjoin%
\definecolor{currentfill}{rgb}{0.000000,0.000000,0.000000}%
\pgfsetfillcolor{currentfill}%
\pgfsetlinewidth{0.803000pt}%
\definecolor{currentstroke}{rgb}{0.000000,0.000000,0.000000}%
\pgfsetstrokecolor{currentstroke}%
\pgfsetdash{}{0pt}%
\pgfsys@defobject{currentmarker}{\pgfqpoint{0.000000in}{-0.048611in}}{\pgfqpoint{0.000000in}{0.000000in}}{%
\pgfpathmoveto{\pgfqpoint{0.000000in}{0.000000in}}%
\pgfpathlineto{\pgfqpoint{0.000000in}{-0.048611in}}%
\pgfusepath{stroke,fill}%
}%
\begin{pgfscope}%
\pgfsys@transformshift{5.855464in}{1.077744in}%
\pgfsys@useobject{currentmarker}{}%
\end{pgfscope}%
\end{pgfscope}%
\begin{pgfscope}%
\definecolor{textcolor}{rgb}{0.000000,0.000000,0.000000}%
\pgfsetstrokecolor{textcolor}%
\pgfsetfillcolor{textcolor}%
\pgftext[x=5.855464in,y=0.980522in,,top]{\color{textcolor}\rmfamily\fontsize{32.000000}{38.400000}\selectfont 60}%
\end{pgfscope}%
\begin{pgfscope}%
\pgfpathrectangle{\pgfqpoint{1.362189in}{1.077744in}}{\pgfqpoint{6.016419in}{5.322256in}}%
\pgfusepath{clip}%
\pgfsetrectcap%
\pgfsetroundjoin%
\pgfsetlinewidth{0.803000pt}%
\definecolor{currentstroke}{rgb}{0.690196,0.690196,0.690196}%
\pgfsetstrokecolor{currentstroke}%
\pgfsetdash{}{0pt}%
\pgfpathmoveto{\pgfqpoint{7.378608in}{1.077744in}}%
\pgfpathlineto{\pgfqpoint{7.378608in}{6.400000in}}%
\pgfusepath{stroke}%
\end{pgfscope}%
\begin{pgfscope}%
\pgfsetbuttcap%
\pgfsetroundjoin%
\definecolor{currentfill}{rgb}{0.000000,0.000000,0.000000}%
\pgfsetfillcolor{currentfill}%
\pgfsetlinewidth{0.803000pt}%
\definecolor{currentstroke}{rgb}{0.000000,0.000000,0.000000}%
\pgfsetstrokecolor{currentstroke}%
\pgfsetdash{}{0pt}%
\pgfsys@defobject{currentmarker}{\pgfqpoint{0.000000in}{-0.048611in}}{\pgfqpoint{0.000000in}{0.000000in}}{%
\pgfpathmoveto{\pgfqpoint{0.000000in}{0.000000in}}%
\pgfpathlineto{\pgfqpoint{0.000000in}{-0.048611in}}%
\pgfusepath{stroke,fill}%
}%
\begin{pgfscope}%
\pgfsys@transformshift{7.378608in}{1.077744in}%
\pgfsys@useobject{currentmarker}{}%
\end{pgfscope}%
\end{pgfscope}%
\begin{pgfscope}%
\definecolor{textcolor}{rgb}{0.000000,0.000000,0.000000}%
\pgfsetstrokecolor{textcolor}%
\pgfsetfillcolor{textcolor}%
\pgftext[x=7.378608in,y=0.980522in,,top]{\color{textcolor}\rmfamily\fontsize{32.000000}{38.400000}\selectfont 80}%
\end{pgfscope}%
\begin{pgfscope}%
\definecolor{textcolor}{rgb}{0.000000,0.000000,0.000000}%
\pgfsetstrokecolor{textcolor}%
\pgfsetfillcolor{textcolor}%
\pgftext[x=4.370398in,y=0.534450in,,top]{\color{textcolor}\rmfamily\fontsize{36.000000}{43.200000}\selectfont time step \(\displaystyle k\)}%
\end{pgfscope}%
\begin{pgfscope}%
\pgfpathrectangle{\pgfqpoint{1.362189in}{1.077744in}}{\pgfqpoint{6.016419in}{5.322256in}}%
\pgfusepath{clip}%
\pgfsetrectcap%
\pgfsetroundjoin%
\pgfsetlinewidth{0.803000pt}%
\definecolor{currentstroke}{rgb}{0.690196,0.690196,0.690196}%
\pgfsetstrokecolor{currentstroke}%
\pgfsetdash{}{0pt}%
\pgfpathmoveto{\pgfqpoint{1.362189in}{2.168897in}}%
\pgfpathlineto{\pgfqpoint{7.378608in}{2.168897in}}%
\pgfusepath{stroke}%
\end{pgfscope}%
\begin{pgfscope}%
\pgfsetbuttcap%
\pgfsetroundjoin%
\definecolor{currentfill}{rgb}{0.000000,0.000000,0.000000}%
\pgfsetfillcolor{currentfill}%
\pgfsetlinewidth{0.803000pt}%
\definecolor{currentstroke}{rgb}{0.000000,0.000000,0.000000}%
\pgfsetstrokecolor{currentstroke}%
\pgfsetdash{}{0pt}%
\pgfsys@defobject{currentmarker}{\pgfqpoint{-0.048611in}{0.000000in}}{\pgfqpoint{-0.000000in}{0.000000in}}{%
\pgfpathmoveto{\pgfqpoint{-0.000000in}{0.000000in}}%
\pgfpathlineto{\pgfqpoint{-0.048611in}{0.000000in}}%
\pgfusepath{stroke,fill}%
}%
\begin{pgfscope}%
\pgfsys@transformshift{1.362189in}{2.168897in}%
\pgfsys@useobject{currentmarker}{}%
\end{pgfscope}%
\end{pgfscope}%
\begin{pgfscope}%
\definecolor{textcolor}{rgb}{0.000000,0.000000,0.000000}%
\pgfsetstrokecolor{textcolor}%
\pgfsetfillcolor{textcolor}%
\pgftext[x=0.753414in, y=2.048913in, left, base]{\color{textcolor}\rmfamily\fontsize{32.000000}{38.400000}\selectfont -10}%
\end{pgfscope}%
\begin{pgfscope}%
\pgfpathrectangle{\pgfqpoint{1.362189in}{1.077744in}}{\pgfqpoint{6.016419in}{5.322256in}}%
\pgfusepath{clip}%
\pgfsetrectcap%
\pgfsetroundjoin%
\pgfsetlinewidth{0.803000pt}%
\definecolor{currentstroke}{rgb}{0.690196,0.690196,0.690196}%
\pgfsetstrokecolor{currentstroke}%
\pgfsetdash{}{0pt}%
\pgfpathmoveto{\pgfqpoint{1.362189in}{3.326760in}}%
\pgfpathlineto{\pgfqpoint{7.378608in}{3.326760in}}%
\pgfusepath{stroke}%
\end{pgfscope}%
\begin{pgfscope}%
\pgfsetbuttcap%
\pgfsetroundjoin%
\definecolor{currentfill}{rgb}{0.000000,0.000000,0.000000}%
\pgfsetfillcolor{currentfill}%
\pgfsetlinewidth{0.803000pt}%
\definecolor{currentstroke}{rgb}{0.000000,0.000000,0.000000}%
\pgfsetstrokecolor{currentstroke}%
\pgfsetdash{}{0pt}%
\pgfsys@defobject{currentmarker}{\pgfqpoint{-0.048611in}{0.000000in}}{\pgfqpoint{-0.000000in}{0.000000in}}{%
\pgfpathmoveto{\pgfqpoint{-0.000000in}{0.000000in}}%
\pgfpathlineto{\pgfqpoint{-0.048611in}{0.000000in}}%
\pgfusepath{stroke,fill}%
}%
\begin{pgfscope}%
\pgfsys@transformshift{1.362189in}{3.326760in}%
\pgfsys@useobject{currentmarker}{}%
\end{pgfscope}%
\end{pgfscope}%
\begin{pgfscope}%
\definecolor{textcolor}{rgb}{0.000000,0.000000,0.000000}%
\pgfsetstrokecolor{textcolor}%
\pgfsetfillcolor{textcolor}%
\pgftext[x=0.911892in, y=3.206775in, left, base]{\color{textcolor}\rmfamily\fontsize{32.000000}{38.400000}\selectfont -7}%
\end{pgfscope}%
\begin{pgfscope}%
\pgfpathrectangle{\pgfqpoint{1.362189in}{1.077744in}}{\pgfqpoint{6.016419in}{5.322256in}}%
\pgfusepath{clip}%
\pgfsetrectcap%
\pgfsetroundjoin%
\pgfsetlinewidth{0.803000pt}%
\definecolor{currentstroke}{rgb}{0.690196,0.690196,0.690196}%
\pgfsetstrokecolor{currentstroke}%
\pgfsetdash{}{0pt}%
\pgfpathmoveto{\pgfqpoint{1.362189in}{4.484623in}}%
\pgfpathlineto{\pgfqpoint{7.378608in}{4.484623in}}%
\pgfusepath{stroke}%
\end{pgfscope}%
\begin{pgfscope}%
\pgfsetbuttcap%
\pgfsetroundjoin%
\definecolor{currentfill}{rgb}{0.000000,0.000000,0.000000}%
\pgfsetfillcolor{currentfill}%
\pgfsetlinewidth{0.803000pt}%
\definecolor{currentstroke}{rgb}{0.000000,0.000000,0.000000}%
\pgfsetstrokecolor{currentstroke}%
\pgfsetdash{}{0pt}%
\pgfsys@defobject{currentmarker}{\pgfqpoint{-0.048611in}{0.000000in}}{\pgfqpoint{-0.000000in}{0.000000in}}{%
\pgfpathmoveto{\pgfqpoint{-0.000000in}{0.000000in}}%
\pgfpathlineto{\pgfqpoint{-0.048611in}{0.000000in}}%
\pgfusepath{stroke,fill}%
}%
\begin{pgfscope}%
\pgfsys@transformshift{1.362189in}{4.484623in}%
\pgfsys@useobject{currentmarker}{}%
\end{pgfscope}%
\end{pgfscope}%
\begin{pgfscope}%
\definecolor{textcolor}{rgb}{0.000000,0.000000,0.000000}%
\pgfsetstrokecolor{textcolor}%
\pgfsetfillcolor{textcolor}%
\pgftext[x=0.911892in, y=4.364638in, left, base]{\color{textcolor}\rmfamily\fontsize{32.000000}{38.400000}\selectfont -5}%
\end{pgfscope}%
\begin{pgfscope}%
\pgfpathrectangle{\pgfqpoint{1.362189in}{1.077744in}}{\pgfqpoint{6.016419in}{5.322256in}}%
\pgfusepath{clip}%
\pgfsetrectcap%
\pgfsetroundjoin%
\pgfsetlinewidth{0.803000pt}%
\definecolor{currentstroke}{rgb}{0.690196,0.690196,0.690196}%
\pgfsetstrokecolor{currentstroke}%
\pgfsetdash{}{0pt}%
\pgfpathmoveto{\pgfqpoint{1.362189in}{5.642486in}}%
\pgfpathlineto{\pgfqpoint{7.378608in}{5.642486in}}%
\pgfusepath{stroke}%
\end{pgfscope}%
\begin{pgfscope}%
\pgfsetbuttcap%
\pgfsetroundjoin%
\definecolor{currentfill}{rgb}{0.000000,0.000000,0.000000}%
\pgfsetfillcolor{currentfill}%
\pgfsetlinewidth{0.803000pt}%
\definecolor{currentstroke}{rgb}{0.000000,0.000000,0.000000}%
\pgfsetstrokecolor{currentstroke}%
\pgfsetdash{}{0pt}%
\pgfsys@defobject{currentmarker}{\pgfqpoint{-0.048611in}{0.000000in}}{\pgfqpoint{-0.000000in}{0.000000in}}{%
\pgfpathmoveto{\pgfqpoint{-0.000000in}{0.000000in}}%
\pgfpathlineto{\pgfqpoint{-0.048611in}{0.000000in}}%
\pgfusepath{stroke,fill}%
}%
\begin{pgfscope}%
\pgfsys@transformshift{1.362189in}{5.642486in}%
\pgfsys@useobject{currentmarker}{}%
\end{pgfscope}%
\end{pgfscope}%
\begin{pgfscope}%
\definecolor{textcolor}{rgb}{0.000000,0.000000,0.000000}%
\pgfsetstrokecolor{textcolor}%
\pgfsetfillcolor{textcolor}%
\pgftext[x=0.911892in, y=5.522501in, left, base]{\color{textcolor}\rmfamily\fontsize{32.000000}{38.400000}\selectfont -2}%
\end{pgfscope}%
\begin{pgfscope}%
\definecolor{textcolor}{rgb}{0.000000,0.000000,0.000000}%
\pgfsetstrokecolor{textcolor}%
\pgfsetfillcolor{textcolor}%
\pgftext[x=0.614525in,y=3.738872in,,bottom,rotate=90.000000]{\color{textcolor}\rmfamily\fontsize{36.000000}{43.200000}\selectfont SNR [dB]}%
\end{pgfscope}%
\begin{pgfscope}%
\pgfpathrectangle{\pgfqpoint{1.362189in}{1.077744in}}{\pgfqpoint{6.016419in}{5.322256in}}%
\pgfusepath{clip}%
\pgfsetrectcap%
\pgfsetroundjoin%
\pgfsetlinewidth{3.011250pt}%
\definecolor{currentstroke}{rgb}{0.839216,0.152941,0.156863}%
\pgfsetstrokecolor{currentstroke}%
\pgfsetdash{}{0pt}%
\pgfpathmoveto{\pgfqpoint{2.047604in}{6.158079in}}%
\pgfpathlineto{\pgfqpoint{2.123761in}{6.102958in}}%
\pgfpathlineto{\pgfqpoint{2.199918in}{6.050165in}}%
\pgfpathlineto{\pgfqpoint{2.276075in}{5.997520in}}%
\pgfpathlineto{\pgfqpoint{2.352232in}{5.944490in}}%
\pgfpathlineto{\pgfqpoint{2.428390in}{5.890481in}}%
\pgfpathlineto{\pgfqpoint{2.504547in}{5.832476in}}%
\pgfpathlineto{\pgfqpoint{2.580704in}{5.772475in}}%
\pgfpathlineto{\pgfqpoint{2.656861in}{5.709230in}}%
\pgfpathlineto{\pgfqpoint{2.733018in}{5.641646in}}%
\pgfpathlineto{\pgfqpoint{2.809176in}{5.574779in}}%
\pgfpathlineto{\pgfqpoint{2.885333in}{5.508531in}}%
\pgfpathlineto{\pgfqpoint{2.961490in}{5.440974in}}%
\pgfpathlineto{\pgfqpoint{3.037647in}{5.370242in}}%
\pgfpathlineto{\pgfqpoint{3.113804in}{5.296155in}}%
\pgfpathlineto{\pgfqpoint{3.189962in}{5.221987in}}%
\pgfpathlineto{\pgfqpoint{3.266119in}{5.148951in}}%
\pgfpathlineto{\pgfqpoint{3.342276in}{5.077448in}}%
\pgfpathlineto{\pgfqpoint{3.418433in}{5.004708in}}%
\pgfpathlineto{\pgfqpoint{3.494590in}{4.930876in}}%
\pgfpathlineto{\pgfqpoint{3.570748in}{4.856945in}}%
\pgfpathlineto{\pgfqpoint{3.646905in}{4.783561in}}%
\pgfpathlineto{\pgfqpoint{3.723062in}{4.711043in}}%
\pgfpathlineto{\pgfqpoint{3.799219in}{4.638094in}}%
\pgfpathlineto{\pgfqpoint{3.875377in}{4.562226in}}%
\pgfpathlineto{\pgfqpoint{3.951534in}{4.483519in}}%
\pgfpathlineto{\pgfqpoint{4.027691in}{4.404373in}}%
\pgfpathlineto{\pgfqpoint{4.103848in}{4.322809in}}%
\pgfpathlineto{\pgfqpoint{4.180005in}{4.238283in}}%
\pgfpathlineto{\pgfqpoint{4.256163in}{4.152328in}}%
\pgfpathlineto{\pgfqpoint{4.332320in}{4.065530in}}%
\pgfpathlineto{\pgfqpoint{4.408477in}{3.979421in}}%
\pgfpathlineto{\pgfqpoint{4.484634in}{3.895426in}}%
\pgfpathlineto{\pgfqpoint{4.560791in}{3.811631in}}%
\pgfpathlineto{\pgfqpoint{4.636949in}{3.726911in}}%
\pgfpathlineto{\pgfqpoint{4.713106in}{3.641695in}}%
\pgfpathlineto{\pgfqpoint{4.789263in}{3.555597in}}%
\pgfpathlineto{\pgfqpoint{4.865420in}{3.466824in}}%
\pgfpathlineto{\pgfqpoint{4.941577in}{3.374623in}}%
\pgfpathlineto{\pgfqpoint{5.017735in}{3.280428in}}%
\pgfpathlineto{\pgfqpoint{5.093892in}{3.185444in}}%
\pgfpathlineto{\pgfqpoint{5.170049in}{3.090408in}}%
\pgfpathlineto{\pgfqpoint{5.246206in}{2.994094in}}%
\pgfpathlineto{\pgfqpoint{5.322363in}{2.896435in}}%
\pgfpathlineto{\pgfqpoint{5.398521in}{2.798470in}}%
\pgfpathlineto{\pgfqpoint{5.474678in}{2.699027in}}%
\pgfpathlineto{\pgfqpoint{5.550835in}{2.599189in}}%
\pgfpathlineto{\pgfqpoint{5.626992in}{2.499697in}}%
\pgfpathlineto{\pgfqpoint{5.703149in}{2.399320in}}%
\pgfpathlineto{\pgfqpoint{5.779307in}{2.298506in}}%
\pgfpathlineto{\pgfqpoint{5.855464in}{2.197001in}}%
\pgfpathlineto{\pgfqpoint{5.931621in}{2.096130in}}%
\pgfpathlineto{\pgfqpoint{6.007778in}{1.997133in}}%
\pgfpathlineto{\pgfqpoint{6.083935in}{1.898284in}}%
\pgfpathlineto{\pgfqpoint{6.160093in}{1.799741in}}%
\pgfpathlineto{\pgfqpoint{6.236250in}{1.703090in}}%
\pgfpathlineto{\pgfqpoint{6.312407in}{1.607681in}}%
\pgfpathlineto{\pgfqpoint{6.388564in}{1.511905in}}%
\pgfpathlineto{\pgfqpoint{6.464722in}{1.415730in}}%
\pgfpathlineto{\pgfqpoint{6.540879in}{1.319665in}}%
\pgfusepath{stroke}%
\end{pgfscope}%
\begin{pgfscope}%
\pgfpathrectangle{\pgfqpoint{1.362189in}{1.077744in}}{\pgfqpoint{6.016419in}{5.322256in}}%
\pgfusepath{clip}%
\pgfsetrectcap%
\pgfsetroundjoin%
\pgfsetlinewidth{3.011250pt}%
\definecolor{currentstroke}{rgb}{0.121569,0.466667,0.705882}%
\pgfsetstrokecolor{currentstroke}%
\pgfsetdash{}{0pt}%
\pgfpathmoveto{\pgfqpoint{2.047604in}{6.158079in}}%
\pgfpathlineto{\pgfqpoint{2.123761in}{6.100016in}}%
\pgfpathlineto{\pgfqpoint{2.199918in}{6.036264in}}%
\pgfpathlineto{\pgfqpoint{2.276075in}{5.967063in}}%
\pgfpathlineto{\pgfqpoint{2.352232in}{5.893240in}}%
\pgfpathlineto{\pgfqpoint{2.428390in}{5.816166in}}%
\pgfpathlineto{\pgfqpoint{2.504547in}{5.740458in}}%
\pgfpathlineto{\pgfqpoint{2.580704in}{5.667473in}}%
\pgfpathlineto{\pgfqpoint{2.656861in}{5.595383in}}%
\pgfpathlineto{\pgfqpoint{2.733018in}{5.525109in}}%
\pgfpathlineto{\pgfqpoint{2.809176in}{5.455217in}}%
\pgfpathlineto{\pgfqpoint{2.885333in}{5.385655in}}%
\pgfpathlineto{\pgfqpoint{2.961490in}{5.317418in}}%
\pgfpathlineto{\pgfqpoint{3.037647in}{5.248404in}}%
\pgfpathlineto{\pgfqpoint{3.113804in}{5.178386in}}%
\pgfpathlineto{\pgfqpoint{3.189962in}{5.107904in}}%
\pgfpathlineto{\pgfqpoint{3.266119in}{5.036271in}}%
\pgfpathlineto{\pgfqpoint{3.342276in}{4.963503in}}%
\pgfpathlineto{\pgfqpoint{3.418433in}{4.889010in}}%
\pgfpathlineto{\pgfqpoint{3.494590in}{4.815429in}}%
\pgfpathlineto{\pgfqpoint{3.570748in}{4.745657in}}%
\pgfpathlineto{\pgfqpoint{3.646905in}{4.675838in}}%
\pgfpathlineto{\pgfqpoint{3.723062in}{4.603713in}}%
\pgfpathlineto{\pgfqpoint{3.799219in}{4.531430in}}%
\pgfpathlineto{\pgfqpoint{3.875377in}{4.461810in}}%
\pgfpathlineto{\pgfqpoint{3.951534in}{4.392520in}}%
\pgfpathlineto{\pgfqpoint{4.027691in}{4.321294in}}%
\pgfpathlineto{\pgfqpoint{4.103848in}{4.247942in}}%
\pgfpathlineto{\pgfqpoint{4.180005in}{4.175331in}}%
\pgfpathlineto{\pgfqpoint{4.256163in}{4.104092in}}%
\pgfpathlineto{\pgfqpoint{4.332320in}{4.032612in}}%
\pgfpathlineto{\pgfqpoint{4.408477in}{3.961501in}}%
\pgfpathlineto{\pgfqpoint{4.484634in}{3.889340in}}%
\pgfpathlineto{\pgfqpoint{4.560791in}{3.816909in}}%
\pgfpathlineto{\pgfqpoint{4.636949in}{3.743345in}}%
\pgfpathlineto{\pgfqpoint{4.713106in}{3.670773in}}%
\pgfpathlineto{\pgfqpoint{4.789263in}{3.602045in}}%
\pgfpathlineto{\pgfqpoint{4.865420in}{3.535528in}}%
\pgfpathlineto{\pgfqpoint{4.941577in}{3.469757in}}%
\pgfpathlineto{\pgfqpoint{5.017735in}{3.405357in}}%
\pgfpathlineto{\pgfqpoint{5.093892in}{3.343363in}}%
\pgfpathlineto{\pgfqpoint{5.170049in}{3.283195in}}%
\pgfpathlineto{\pgfqpoint{5.246206in}{3.223353in}}%
\pgfpathlineto{\pgfqpoint{5.322363in}{3.162531in}}%
\pgfpathlineto{\pgfqpoint{5.398521in}{3.100557in}}%
\pgfpathlineto{\pgfqpoint{5.474678in}{3.036746in}}%
\pgfpathlineto{\pgfqpoint{5.550835in}{2.973013in}}%
\pgfpathlineto{\pgfqpoint{5.626992in}{2.910214in}}%
\pgfpathlineto{\pgfqpoint{5.703149in}{2.846737in}}%
\pgfpathlineto{\pgfqpoint{5.779307in}{2.783463in}}%
\pgfpathlineto{\pgfqpoint{5.855464in}{2.720364in}}%
\pgfpathlineto{\pgfqpoint{5.931621in}{2.656367in}}%
\pgfpathlineto{\pgfqpoint{6.007778in}{2.590513in}}%
\pgfpathlineto{\pgfqpoint{6.083935in}{2.523687in}}%
\pgfpathlineto{\pgfqpoint{6.160093in}{2.458012in}}%
\pgfpathlineto{\pgfqpoint{6.236250in}{2.393282in}}%
\pgfpathlineto{\pgfqpoint{6.312407in}{2.329008in}}%
\pgfpathlineto{\pgfqpoint{6.388564in}{2.265807in}}%
\pgfpathlineto{\pgfqpoint{6.464722in}{2.202255in}}%
\pgfpathlineto{\pgfqpoint{6.540879in}{2.138243in}}%
\pgfusepath{stroke}%
\end{pgfscope}%
\begin{pgfscope}%
\pgfpathrectangle{\pgfqpoint{1.362189in}{1.077744in}}{\pgfqpoint{6.016419in}{5.322256in}}%
\pgfusepath{clip}%
\pgfsetrectcap%
\pgfsetroundjoin%
\pgfsetlinewidth{3.011250pt}%
\definecolor{currentstroke}{rgb}{0.172549,0.627451,0.172549}%
\pgfsetstrokecolor{currentstroke}%
\pgfsetdash{}{0pt}%
\pgfpathmoveto{\pgfqpoint{2.809176in}{6.050556in}}%
\pgfpathlineto{\pgfqpoint{2.885333in}{5.954617in}}%
\pgfpathlineto{\pgfqpoint{2.961490in}{5.858935in}}%
\pgfpathlineto{\pgfqpoint{3.037647in}{5.767274in}}%
\pgfpathlineto{\pgfqpoint{3.113804in}{5.679986in}}%
\pgfpathlineto{\pgfqpoint{3.189962in}{5.593641in}}%
\pgfpathlineto{\pgfqpoint{3.266119in}{5.508802in}}%
\pgfpathlineto{\pgfqpoint{3.342276in}{5.425601in}}%
\pgfpathlineto{\pgfqpoint{3.418433in}{5.344332in}}%
\pgfpathlineto{\pgfqpoint{3.494590in}{5.267328in}}%
\pgfpathlineto{\pgfqpoint{3.570748in}{5.191895in}}%
\pgfpathlineto{\pgfqpoint{3.646905in}{5.117474in}}%
\pgfpathlineto{\pgfqpoint{3.723062in}{5.045563in}}%
\pgfpathlineto{\pgfqpoint{3.799219in}{4.975844in}}%
\pgfpathlineto{\pgfqpoint{3.875377in}{4.908580in}}%
\pgfpathlineto{\pgfqpoint{3.951534in}{4.841764in}}%
\pgfpathlineto{\pgfqpoint{4.027691in}{4.777484in}}%
\pgfpathlineto{\pgfqpoint{4.103848in}{4.716108in}}%
\pgfpathlineto{\pgfqpoint{4.180005in}{4.654941in}}%
\pgfpathlineto{\pgfqpoint{4.256163in}{4.591341in}}%
\pgfpathlineto{\pgfqpoint{4.332320in}{4.523549in}}%
\pgfpathlineto{\pgfqpoint{4.408477in}{4.453792in}}%
\pgfpathlineto{\pgfqpoint{4.484634in}{4.385549in}}%
\pgfpathlineto{\pgfqpoint{4.560791in}{4.318944in}}%
\pgfpathlineto{\pgfqpoint{4.636949in}{4.253065in}}%
\pgfpathlineto{\pgfqpoint{4.713106in}{4.186895in}}%
\pgfpathlineto{\pgfqpoint{4.789263in}{4.119634in}}%
\pgfpathlineto{\pgfqpoint{4.865420in}{4.054107in}}%
\pgfpathlineto{\pgfqpoint{4.941577in}{3.988950in}}%
\pgfpathlineto{\pgfqpoint{5.017735in}{3.922757in}}%
\pgfpathlineto{\pgfqpoint{5.093892in}{3.856461in}}%
\pgfpathlineto{\pgfqpoint{5.170049in}{3.790190in}}%
\pgfpathlineto{\pgfqpoint{5.246206in}{3.725319in}}%
\pgfpathlineto{\pgfqpoint{5.322363in}{3.662491in}}%
\pgfpathlineto{\pgfqpoint{5.398521in}{3.601275in}}%
\pgfpathlineto{\pgfqpoint{5.474678in}{3.540977in}}%
\pgfpathlineto{\pgfqpoint{5.550835in}{3.481188in}}%
\pgfpathlineto{\pgfqpoint{5.626992in}{3.422791in}}%
\pgfpathlineto{\pgfqpoint{5.703149in}{3.364639in}}%
\pgfpathlineto{\pgfqpoint{5.779307in}{3.305869in}}%
\pgfusepath{stroke}%
\end{pgfscope}%
\begin{pgfscope}%
\pgfpathrectangle{\pgfqpoint{1.362189in}{1.077744in}}{\pgfqpoint{6.016419in}{5.322256in}}%
\pgfusepath{clip}%
\pgfsetrectcap%
\pgfsetroundjoin%
\pgfsetlinewidth{3.011250pt}%
\definecolor{currentstroke}{rgb}{1.000000,0.498039,0.054902}%
\pgfsetstrokecolor{currentstroke}%
\pgfsetdash{}{0pt}%
\pgfpathmoveto{\pgfqpoint{2.428390in}{4.845459in}}%
\pgfpathlineto{\pgfqpoint{2.504547in}{4.856949in}}%
\pgfpathlineto{\pgfqpoint{2.580704in}{4.864358in}}%
\pgfpathlineto{\pgfqpoint{2.656861in}{4.870660in}}%
\pgfpathlineto{\pgfqpoint{2.733018in}{4.878893in}}%
\pgfpathlineto{\pgfqpoint{2.809176in}{4.887285in}}%
\pgfpathlineto{\pgfqpoint{2.885333in}{4.894552in}}%
\pgfpathlineto{\pgfqpoint{2.961490in}{4.899167in}}%
\pgfpathlineto{\pgfqpoint{3.037647in}{4.901903in}}%
\pgfpathlineto{\pgfqpoint{3.113804in}{4.905342in}}%
\pgfpathlineto{\pgfqpoint{3.189962in}{4.911370in}}%
\pgfpathlineto{\pgfqpoint{3.266119in}{4.920053in}}%
\pgfpathlineto{\pgfqpoint{3.342276in}{4.931828in}}%
\pgfpathlineto{\pgfqpoint{3.418433in}{4.943930in}}%
\pgfpathlineto{\pgfqpoint{3.494590in}{4.954321in}}%
\pgfpathlineto{\pgfqpoint{3.570748in}{4.965636in}}%
\pgfpathlineto{\pgfqpoint{3.646905in}{4.978284in}}%
\pgfpathlineto{\pgfqpoint{3.723062in}{4.991712in}}%
\pgfpathlineto{\pgfqpoint{3.799219in}{5.005097in}}%
\pgfpathlineto{\pgfqpoint{3.875377in}{5.017667in}}%
\pgfpathlineto{\pgfqpoint{3.951534in}{5.029788in}}%
\pgfpathlineto{\pgfqpoint{4.027691in}{5.041171in}}%
\pgfpathlineto{\pgfqpoint{4.103848in}{5.050761in}}%
\pgfpathlineto{\pgfqpoint{4.180005in}{5.058541in}}%
\pgfpathlineto{\pgfqpoint{4.256163in}{5.065536in}}%
\pgfpathlineto{\pgfqpoint{4.332320in}{5.070897in}}%
\pgfpathlineto{\pgfqpoint{4.408477in}{5.075716in}}%
\pgfpathlineto{\pgfqpoint{4.484634in}{5.082451in}}%
\pgfpathlineto{\pgfqpoint{4.560791in}{5.088354in}}%
\pgfpathlineto{\pgfqpoint{4.636949in}{5.095144in}}%
\pgfpathlineto{\pgfqpoint{4.713106in}{5.104268in}}%
\pgfpathlineto{\pgfqpoint{4.789263in}{5.112341in}}%
\pgfpathlineto{\pgfqpoint{4.865420in}{5.121386in}}%
\pgfpathlineto{\pgfqpoint{4.941577in}{5.128725in}}%
\pgfpathlineto{\pgfqpoint{5.017735in}{5.130584in}}%
\pgfpathlineto{\pgfqpoint{5.093892in}{5.130881in}}%
\pgfpathlineto{\pgfqpoint{5.170049in}{5.133922in}}%
\pgfpathlineto{\pgfqpoint{5.246206in}{5.139018in}}%
\pgfpathlineto{\pgfqpoint{5.322363in}{5.143491in}}%
\pgfpathlineto{\pgfqpoint{5.398521in}{5.147584in}}%
\pgfusepath{stroke}%
\end{pgfscope}%
\begin{pgfscope}%
\pgfpathrectangle{\pgfqpoint{1.362189in}{1.077744in}}{\pgfqpoint{6.016419in}{5.322256in}}%
\pgfusepath{clip}%
\pgfsetrectcap%
\pgfsetroundjoin%
\pgfsetlinewidth{3.011250pt}%
\definecolor{currentstroke}{rgb}{0.580392,0.403922,0.741176}%
\pgfsetstrokecolor{currentstroke}%
\pgfsetdash{}{0pt}%
\pgfpathmoveto{\pgfqpoint{3.189962in}{2.912727in}}%
\pgfpathlineto{\pgfqpoint{3.266119in}{2.900023in}}%
\pgfpathlineto{\pgfqpoint{3.342276in}{2.886578in}}%
\pgfpathlineto{\pgfqpoint{3.418433in}{2.871563in}}%
\pgfpathlineto{\pgfqpoint{3.494590in}{2.854719in}}%
\pgfpathlineto{\pgfqpoint{3.570748in}{2.836294in}}%
\pgfpathlineto{\pgfqpoint{3.646905in}{2.816324in}}%
\pgfpathlineto{\pgfqpoint{3.723062in}{2.793790in}}%
\pgfpathlineto{\pgfqpoint{3.799219in}{2.768896in}}%
\pgfpathlineto{\pgfqpoint{3.875377in}{2.742953in}}%
\pgfpathlineto{\pgfqpoint{3.951534in}{2.715744in}}%
\pgfpathlineto{\pgfqpoint{4.027691in}{2.687340in}}%
\pgfpathlineto{\pgfqpoint{4.103848in}{2.658335in}}%
\pgfpathlineto{\pgfqpoint{4.180005in}{2.628260in}}%
\pgfpathlineto{\pgfqpoint{4.256163in}{2.597456in}}%
\pgfpathlineto{\pgfqpoint{4.332320in}{2.565554in}}%
\pgfpathlineto{\pgfqpoint{4.408477in}{2.533184in}}%
\pgfpathlineto{\pgfqpoint{4.484634in}{2.503190in}}%
\pgfpathlineto{\pgfqpoint{4.560791in}{2.475639in}}%
\pgfpathlineto{\pgfqpoint{4.636949in}{2.447787in}}%
\pgfpathlineto{\pgfqpoint{4.713106in}{2.418678in}}%
\pgfpathlineto{\pgfqpoint{4.789263in}{2.389384in}}%
\pgfpathlineto{\pgfqpoint{4.865420in}{2.359805in}}%
\pgfpathlineto{\pgfqpoint{4.941577in}{2.328741in}}%
\pgfpathlineto{\pgfqpoint{5.017735in}{2.295747in}}%
\pgfpathlineto{\pgfqpoint{5.093892in}{2.262463in}}%
\pgfpathlineto{\pgfqpoint{5.170049in}{2.229979in}}%
\pgfpathlineto{\pgfqpoint{5.246206in}{2.199744in}}%
\pgfpathlineto{\pgfqpoint{5.322363in}{2.172533in}}%
\pgfpathlineto{\pgfqpoint{5.398521in}{2.147095in}}%
\pgfpathlineto{\pgfqpoint{5.474678in}{2.121954in}}%
\pgfpathlineto{\pgfqpoint{5.550835in}{2.096402in}}%
\pgfpathlineto{\pgfqpoint{5.626992in}{2.070352in}}%
\pgfpathlineto{\pgfqpoint{5.703149in}{2.043085in}}%
\pgfpathlineto{\pgfqpoint{5.779307in}{2.014747in}}%
\pgfpathlineto{\pgfqpoint{5.855464in}{1.985191in}}%
\pgfpathlineto{\pgfqpoint{5.931621in}{1.954853in}}%
\pgfpathlineto{\pgfqpoint{6.007778in}{1.923304in}}%
\pgfpathlineto{\pgfqpoint{6.083935in}{1.890958in}}%
\pgfpathlineto{\pgfqpoint{6.160093in}{1.860170in}}%
\pgfusepath{stroke}%
\end{pgfscope}%
\begin{pgfscope}%
\pgfsetrectcap%
\pgfsetmiterjoin%
\pgfsetlinewidth{0.803000pt}%
\definecolor{currentstroke}{rgb}{0.000000,0.000000,0.000000}%
\pgfsetstrokecolor{currentstroke}%
\pgfsetdash{}{0pt}%
\pgfpathmoveto{\pgfqpoint{1.362189in}{1.077744in}}%
\pgfpathlineto{\pgfqpoint{1.362189in}{6.400000in}}%
\pgfusepath{stroke}%
\end{pgfscope}%
\begin{pgfscope}%
\pgfsetrectcap%
\pgfsetmiterjoin%
\pgfsetlinewidth{0.803000pt}%
\definecolor{currentstroke}{rgb}{0.000000,0.000000,0.000000}%
\pgfsetstrokecolor{currentstroke}%
\pgfsetdash{}{0pt}%
\pgfpathmoveto{\pgfqpoint{7.378608in}{1.077744in}}%
\pgfpathlineto{\pgfqpoint{7.378608in}{6.400000in}}%
\pgfusepath{stroke}%
\end{pgfscope}%
\begin{pgfscope}%
\pgfsetrectcap%
\pgfsetmiterjoin%
\pgfsetlinewidth{0.803000pt}%
\definecolor{currentstroke}{rgb}{0.000000,0.000000,0.000000}%
\pgfsetstrokecolor{currentstroke}%
\pgfsetdash{}{0pt}%
\pgfpathmoveto{\pgfqpoint{1.362189in}{1.077744in}}%
\pgfpathlineto{\pgfqpoint{7.378608in}{1.077744in}}%
\pgfusepath{stroke}%
\end{pgfscope}%
\begin{pgfscope}%
\pgfsetrectcap%
\pgfsetmiterjoin%
\pgfsetlinewidth{0.803000pt}%
\definecolor{currentstroke}{rgb}{0.000000,0.000000,0.000000}%
\pgfsetstrokecolor{currentstroke}%
\pgfsetdash{}{0pt}%
\pgfpathmoveto{\pgfqpoint{1.362189in}{6.400000in}}%
\pgfpathlineto{\pgfqpoint{7.378608in}{6.400000in}}%
\pgfusepath{stroke}%
\end{pgfscope}%
\end{pgfpicture}%
\makeatother%
\endgroup%

%% file: Figs/snr/snr_sensor2_eps.pgf
\begingroup%
\makeatletter%
\begin{pgfpicture}%
\pgfpathrectangle{\pgfpointorigin}{\pgfqpoint{7.600000in}{6.400000in}}%
\pgfusepath{use as bounding box, clip}%
\begin{pgfscope}%
\pgfsetbuttcap%
\pgfsetmiterjoin%
\definecolor{currentfill}{rgb}{1.000000,1.000000,1.000000}%
\pgfsetfillcolor{currentfill}%
\pgfsetlinewidth{0.000000pt}%
\definecolor{currentstroke}{rgb}{1.000000,1.000000,1.000000}%
\pgfsetstrokecolor{currentstroke}%
\pgfsetdash{}{0pt}%
\pgfpathmoveto{\pgfqpoint{0.000000in}{0.000000in}}%
\pgfpathlineto{\pgfqpoint{7.600000in}{0.000000in}}%
\pgfpathlineto{\pgfqpoint{7.600000in}{6.400000in}}%
\pgfpathlineto{\pgfqpoint{0.000000in}{6.400000in}}%
\pgfpathlineto{\pgfqpoint{0.000000in}{0.000000in}}%
\pgfpathclose%
\pgfusepath{fill}%
\end{pgfscope}%
\begin{pgfscope}%
\pgfsetbuttcap%
\pgfsetmiterjoin%
\definecolor{currentfill}{rgb}{1.000000,1.000000,1.000000}%
\pgfsetfillcolor{currentfill}%
\pgfsetlinewidth{0.000000pt}%
\definecolor{currentstroke}{rgb}{0.000000,0.000000,0.000000}%
\pgfsetstrokecolor{currentstroke}%
\pgfsetstrokeopacity{0.000000}%
\pgfsetdash{}{0pt}%
\pgfpathmoveto{\pgfqpoint{1.214743in}{1.077744in}}%
\pgfpathlineto{\pgfqpoint{6.680908in}{1.077744in}}%
\pgfpathlineto{\pgfqpoint{6.680908in}{6.249898in}}%
\pgfpathlineto{\pgfqpoint{1.214743in}{6.249898in}}%
\pgfpathlineto{\pgfqpoint{1.214743in}{1.077744in}}%
\pgfpathclose%
\pgfusepath{fill}%
\end{pgfscope}%
\begin{pgfscope}%
\pgfpathrectangle{\pgfqpoint{1.214743in}{1.077744in}}{\pgfqpoint{5.466165in}{5.172154in}}%
\pgfusepath{clip}%
\pgfsys@transformshift{1.214743in}{1.077744in}%
\pgftext[left,bottom]{\includegraphics[interpolate=true,width=5.470000in,height=5.180000in]{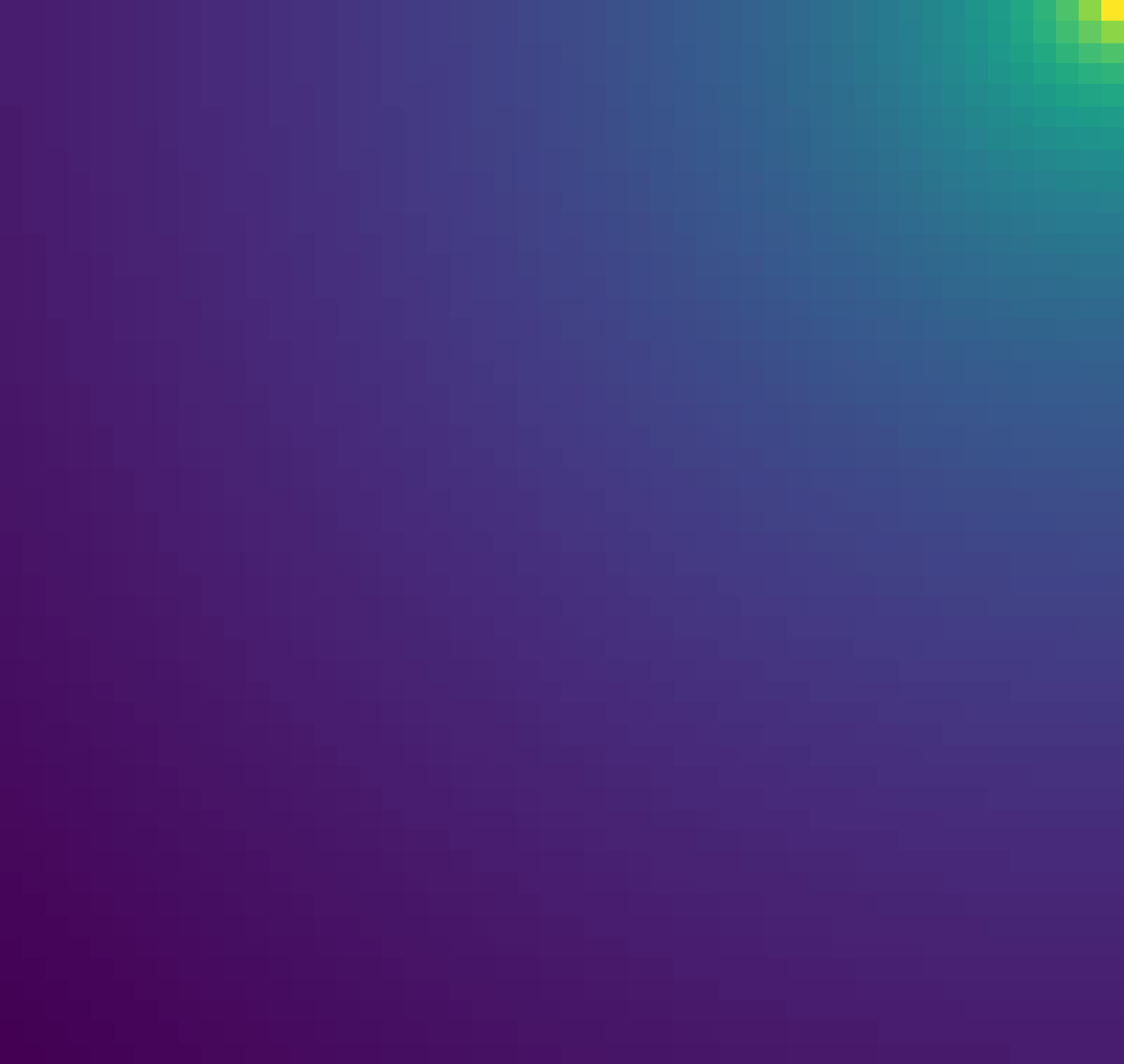}}%
\end{pgfscope}%
\begin{pgfscope}%
\pgfpathrectangle{\pgfqpoint{1.214743in}{1.077744in}}{\pgfqpoint{5.466165in}{5.172154in}}%
\pgfusepath{clip}%
\pgfsetbuttcap%
\pgfsetroundjoin%
\definecolor{currentfill}{rgb}{0.839216,0.152941,0.156863}%
\pgfsetfillcolor{currentfill}%
\pgfsetlinewidth{3.011250pt}%
\definecolor{currentstroke}{rgb}{0.839216,0.152941,0.156863}%
\pgfsetstrokecolor{currentstroke}%
\pgfsetdash{}{0pt}%
\pgfsys@defobject{currentmarker}{\pgfqpoint{-0.098209in}{-0.098209in}}{\pgfqpoint{0.098209in}{0.098209in}}{%
\pgfpathmoveto{\pgfqpoint{-0.098209in}{-0.098209in}}%
\pgfpathlineto{\pgfqpoint{0.098209in}{0.098209in}}%
\pgfpathmoveto{\pgfqpoint{-0.098209in}{0.098209in}}%
\pgfpathlineto{\pgfqpoint{0.098209in}{-0.098209in}}%
\pgfusepath{stroke,fill}%
}%
\begin{pgfscope}%
\pgfsys@transformshift{3.947826in}{2.112175in}%
\pgfsys@useobject{currentmarker}{}%
\end{pgfscope}%
\end{pgfscope}%
\begin{pgfscope}%
\pgfpathrectangle{\pgfqpoint{1.214743in}{1.077744in}}{\pgfqpoint{5.466165in}{5.172154in}}%
\pgfusepath{clip}%
\pgfsetbuttcap%
\pgfsetroundjoin%
\definecolor{currentfill}{rgb}{0.121569,0.466667,0.705882}%
\pgfsetfillcolor{currentfill}%
\pgfsetlinewidth{3.011250pt}%
\definecolor{currentstroke}{rgb}{0.121569,0.466667,0.705882}%
\pgfsetstrokecolor{currentstroke}%
\pgfsetdash{}{0pt}%
\pgfsys@defobject{currentmarker}{\pgfqpoint{-0.098209in}{-0.098209in}}{\pgfqpoint{0.098209in}{0.098209in}}{%
\pgfpathmoveto{\pgfqpoint{-0.098209in}{-0.098209in}}%
\pgfpathlineto{\pgfqpoint{0.098209in}{0.098209in}}%
\pgfpathmoveto{\pgfqpoint{-0.098209in}{0.098209in}}%
\pgfpathlineto{\pgfqpoint{0.098209in}{-0.098209in}}%
\pgfusepath{stroke,fill}%
}%
\begin{pgfscope}%
\pgfsys@transformshift{2.307976in}{3.663821in}%
\pgfsys@useobject{currentmarker}{}%
\end{pgfscope}%
\end{pgfscope}%
\begin{pgfscope}%
\pgfpathrectangle{\pgfqpoint{1.214743in}{1.077744in}}{\pgfqpoint{5.466165in}{5.172154in}}%
\pgfusepath{clip}%
\pgfsetbuttcap%
\pgfsetroundjoin%
\definecolor{currentfill}{rgb}{0.172549,0.627451,0.172549}%
\pgfsetfillcolor{currentfill}%
\pgfsetlinewidth{3.011250pt}%
\definecolor{currentstroke}{rgb}{0.172549,0.627451,0.172549}%
\pgfsetstrokecolor{currentstroke}%
\pgfsetdash{}{0pt}%
\pgfsys@defobject{currentmarker}{\pgfqpoint{-0.098209in}{-0.098209in}}{\pgfqpoint{0.098209in}{0.098209in}}{%
\pgfpathmoveto{\pgfqpoint{-0.098209in}{-0.098209in}}%
\pgfpathlineto{\pgfqpoint{0.098209in}{0.098209in}}%
\pgfpathmoveto{\pgfqpoint{-0.098209in}{0.098209in}}%
\pgfpathlineto{\pgfqpoint{0.098209in}{-0.098209in}}%
\pgfusepath{stroke,fill}%
}%
\begin{pgfscope}%
\pgfsys@transformshift{3.255820in}{3.136409in}%
\pgfsys@useobject{currentmarker}{}%
\end{pgfscope}%
\end{pgfscope}%
\begin{pgfscope}%
\pgfpathrectangle{\pgfqpoint{1.214743in}{1.077744in}}{\pgfqpoint{5.466165in}{5.172154in}}%
\pgfusepath{clip}%
\pgfsetbuttcap%
\pgfsetroundjoin%
\definecolor{currentfill}{rgb}{1.000000,0.498039,0.054902}%
\pgfsetfillcolor{currentfill}%
\pgfsetlinewidth{3.011250pt}%
\definecolor{currentstroke}{rgb}{1.000000,0.498039,0.054902}%
\pgfsetstrokecolor{currentstroke}%
\pgfsetdash{}{0pt}%
\pgfsys@defobject{currentmarker}{\pgfqpoint{-0.098209in}{-0.098209in}}{\pgfqpoint{0.098209in}{0.098209in}}{%
\pgfpathmoveto{\pgfqpoint{-0.098209in}{-0.098209in}}%
\pgfpathlineto{\pgfqpoint{0.098209in}{0.098209in}}%
\pgfpathmoveto{\pgfqpoint{-0.098209in}{0.098209in}}%
\pgfpathlineto{\pgfqpoint{0.098209in}{-0.098209in}}%
\pgfusepath{stroke,fill}%
}%
\begin{pgfscope}%
\pgfsys@transformshift{3.187765in}{3.773235in}%
\pgfsys@useobject{currentmarker}{}%
\end{pgfscope}%
\end{pgfscope}%
\begin{pgfscope}%
\pgfpathrectangle{\pgfqpoint{1.214743in}{1.077744in}}{\pgfqpoint{5.466165in}{5.172154in}}%
\pgfusepath{clip}%
\pgfsetbuttcap%
\pgfsetroundjoin%
\definecolor{currentfill}{rgb}{0.580392,0.403922,0.741176}%
\pgfsetfillcolor{currentfill}%
\pgfsetlinewidth{3.011250pt}%
\definecolor{currentstroke}{rgb}{0.580392,0.403922,0.741176}%
\pgfsetstrokecolor{currentstroke}%
\pgfsetdash{}{0pt}%
\pgfsys@defobject{currentmarker}{\pgfqpoint{-0.098209in}{-0.098209in}}{\pgfqpoint{0.098209in}{0.098209in}}{%
\pgfpathmoveto{\pgfqpoint{-0.098209in}{-0.098209in}}%
\pgfpathlineto{\pgfqpoint{0.098209in}{0.098209in}}%
\pgfpathmoveto{\pgfqpoint{-0.098209in}{0.098209in}}%
\pgfpathlineto{\pgfqpoint{0.098209in}{-0.098209in}}%
\pgfusepath{stroke,fill}%
}%
\begin{pgfscope}%
\pgfsys@transformshift{4.808004in}{3.490654in}%
\pgfsys@useobject{currentmarker}{}%
\end{pgfscope}%
\end{pgfscope}%
\begin{pgfscope}%
\pgfsetbuttcap%
\pgfsetroundjoin%
\definecolor{currentfill}{rgb}{0.000000,0.000000,0.000000}%
\pgfsetfillcolor{currentfill}%
\pgfsetlinewidth{0.803000pt}%
\definecolor{currentstroke}{rgb}{0.000000,0.000000,0.000000}%
\pgfsetstrokecolor{currentstroke}%
\pgfsetdash{}{0pt}%
\pgfsys@defobject{currentmarker}{\pgfqpoint{0.000000in}{-0.048611in}}{\pgfqpoint{0.000000in}{0.000000in}}{%
\pgfpathmoveto{\pgfqpoint{0.000000in}{0.000000in}}%
\pgfpathlineto{\pgfqpoint{0.000000in}{-0.048611in}}%
\pgfusepath{stroke,fill}%
}%
\begin{pgfscope}%
\pgfsys@transformshift{1.214743in}{1.077744in}%
\pgfsys@useobject{currentmarker}{}%
\end{pgfscope}%
\end{pgfscope}%
\begin{pgfscope}%
\definecolor{textcolor}{rgb}{0.000000,0.000000,0.000000}%
\pgfsetstrokecolor{textcolor}%
\pgfsetfillcolor{textcolor}%
\pgftext[x=1.214743in,y=0.980522in,,top]{\color{textcolor}\rmfamily\fontsize{32.000000}{38.400000}\selectfont 0}%
\end{pgfscope}%
\begin{pgfscope}%
\pgfsetbuttcap%
\pgfsetroundjoin%
\definecolor{currentfill}{rgb}{0.000000,0.000000,0.000000}%
\pgfsetfillcolor{currentfill}%
\pgfsetlinewidth{0.803000pt}%
\definecolor{currentstroke}{rgb}{0.000000,0.000000,0.000000}%
\pgfsetstrokecolor{currentstroke}%
\pgfsetdash{}{0pt}%
\pgfsys@defobject{currentmarker}{\pgfqpoint{0.000000in}{-0.048611in}}{\pgfqpoint{0.000000in}{0.000000in}}{%
\pgfpathmoveto{\pgfqpoint{0.000000in}{0.000000in}}%
\pgfpathlineto{\pgfqpoint{0.000000in}{-0.048611in}}%
\pgfusepath{stroke,fill}%
}%
\begin{pgfscope}%
\pgfsys@transformshift{2.307976in}{1.077744in}%
\pgfsys@useobject{currentmarker}{}%
\end{pgfscope}%
\end{pgfscope}%
\begin{pgfscope}%
\definecolor{textcolor}{rgb}{0.000000,0.000000,0.000000}%
\pgfsetstrokecolor{textcolor}%
\pgfsetfillcolor{textcolor}%
\pgftext[x=2.307976in,y=0.980522in,,top]{\color{textcolor}\rmfamily\fontsize{32.000000}{38.400000}\selectfont 10}%
\end{pgfscope}%
\begin{pgfscope}%
\pgfsetbuttcap%
\pgfsetroundjoin%
\definecolor{currentfill}{rgb}{0.000000,0.000000,0.000000}%
\pgfsetfillcolor{currentfill}%
\pgfsetlinewidth{0.803000pt}%
\definecolor{currentstroke}{rgb}{0.000000,0.000000,0.000000}%
\pgfsetstrokecolor{currentstroke}%
\pgfsetdash{}{0pt}%
\pgfsys@defobject{currentmarker}{\pgfqpoint{0.000000in}{-0.048611in}}{\pgfqpoint{0.000000in}{0.000000in}}{%
\pgfpathmoveto{\pgfqpoint{0.000000in}{0.000000in}}%
\pgfpathlineto{\pgfqpoint{0.000000in}{-0.048611in}}%
\pgfusepath{stroke,fill}%
}%
\begin{pgfscope}%
\pgfsys@transformshift{3.401209in}{1.077744in}%
\pgfsys@useobject{currentmarker}{}%
\end{pgfscope}%
\end{pgfscope}%
\begin{pgfscope}%
\definecolor{textcolor}{rgb}{0.000000,0.000000,0.000000}%
\pgfsetstrokecolor{textcolor}%
\pgfsetfillcolor{textcolor}%
\pgftext[x=3.401209in,y=0.980522in,,top]{\color{textcolor}\rmfamily\fontsize{32.000000}{38.400000}\selectfont 20}%
\end{pgfscope}%
\begin{pgfscope}%
\pgfsetbuttcap%
\pgfsetroundjoin%
\definecolor{currentfill}{rgb}{0.000000,0.000000,0.000000}%
\pgfsetfillcolor{currentfill}%
\pgfsetlinewidth{0.803000pt}%
\definecolor{currentstroke}{rgb}{0.000000,0.000000,0.000000}%
\pgfsetstrokecolor{currentstroke}%
\pgfsetdash{}{0pt}%
\pgfsys@defobject{currentmarker}{\pgfqpoint{0.000000in}{-0.048611in}}{\pgfqpoint{0.000000in}{0.000000in}}{%
\pgfpathmoveto{\pgfqpoint{0.000000in}{0.000000in}}%
\pgfpathlineto{\pgfqpoint{0.000000in}{-0.048611in}}%
\pgfusepath{stroke,fill}%
}%
\begin{pgfscope}%
\pgfsys@transformshift{4.494442in}{1.077744in}%
\pgfsys@useobject{currentmarker}{}%
\end{pgfscope}%
\end{pgfscope}%
\begin{pgfscope}%
\definecolor{textcolor}{rgb}{0.000000,0.000000,0.000000}%
\pgfsetstrokecolor{textcolor}%
\pgfsetfillcolor{textcolor}%
\pgftext[x=4.494442in,y=0.980522in,,top]{\color{textcolor}\rmfamily\fontsize{32.000000}{38.400000}\selectfont 30}%
\end{pgfscope}%
\begin{pgfscope}%
\pgfsetbuttcap%
\pgfsetroundjoin%
\definecolor{currentfill}{rgb}{0.000000,0.000000,0.000000}%
\pgfsetfillcolor{currentfill}%
\pgfsetlinewidth{0.803000pt}%
\definecolor{currentstroke}{rgb}{0.000000,0.000000,0.000000}%
\pgfsetstrokecolor{currentstroke}%
\pgfsetdash{}{0pt}%
\pgfsys@defobject{currentmarker}{\pgfqpoint{0.000000in}{-0.048611in}}{\pgfqpoint{0.000000in}{0.000000in}}{%
\pgfpathmoveto{\pgfqpoint{0.000000in}{0.000000in}}%
\pgfpathlineto{\pgfqpoint{0.000000in}{-0.048611in}}%
\pgfusepath{stroke,fill}%
}%
\begin{pgfscope}%
\pgfsys@transformshift{5.587675in}{1.077744in}%
\pgfsys@useobject{currentmarker}{}%
\end{pgfscope}%
\end{pgfscope}%
\begin{pgfscope}%
\definecolor{textcolor}{rgb}{0.000000,0.000000,0.000000}%
\pgfsetstrokecolor{textcolor}%
\pgfsetfillcolor{textcolor}%
\pgftext[x=5.587675in,y=0.980522in,,top]{\color{textcolor}\rmfamily\fontsize{32.000000}{38.400000}\selectfont 40}%
\end{pgfscope}%
\begin{pgfscope}%
\pgfsetbuttcap%
\pgfsetroundjoin%
\definecolor{currentfill}{rgb}{0.000000,0.000000,0.000000}%
\pgfsetfillcolor{currentfill}%
\pgfsetlinewidth{0.803000pt}%
\definecolor{currentstroke}{rgb}{0.000000,0.000000,0.000000}%
\pgfsetstrokecolor{currentstroke}%
\pgfsetdash{}{0pt}%
\pgfsys@defobject{currentmarker}{\pgfqpoint{0.000000in}{-0.048611in}}{\pgfqpoint{0.000000in}{0.000000in}}{%
\pgfpathmoveto{\pgfqpoint{0.000000in}{0.000000in}}%
\pgfpathlineto{\pgfqpoint{0.000000in}{-0.048611in}}%
\pgfusepath{stroke,fill}%
}%
\begin{pgfscope}%
\pgfsys@transformshift{6.680908in}{1.077744in}%
\pgfsys@useobject{currentmarker}{}%
\end{pgfscope}%
\end{pgfscope}%
\begin{pgfscope}%
\definecolor{textcolor}{rgb}{0.000000,0.000000,0.000000}%
\pgfsetstrokecolor{textcolor}%
\pgfsetfillcolor{textcolor}%
\pgftext[x=6.680908in,y=0.980522in,,top]{\color{textcolor}\rmfamily\fontsize{32.000000}{38.400000}\selectfont 50}%
\end{pgfscope}%
\begin{pgfscope}%
\definecolor{textcolor}{rgb}{0.000000,0.000000,0.000000}%
\pgfsetstrokecolor{textcolor}%
\pgfsetfillcolor{textcolor}%
\pgftext[x=3.947826in,y=0.534450in,,top]{\color{textcolor}\rmfamily\fontsize{36.000000}{43.200000}\selectfont \(\displaystyle x\) [m]}%
\end{pgfscope}%
\begin{pgfscope}%
\pgfsetbuttcap%
\pgfsetroundjoin%
\definecolor{currentfill}{rgb}{0.000000,0.000000,0.000000}%
\pgfsetfillcolor{currentfill}%
\pgfsetlinewidth{0.803000pt}%
\definecolor{currentstroke}{rgb}{0.000000,0.000000,0.000000}%
\pgfsetstrokecolor{currentstroke}%
\pgfsetdash{}{0pt}%
\pgfsys@defobject{currentmarker}{\pgfqpoint{-0.048611in}{0.000000in}}{\pgfqpoint{-0.000000in}{0.000000in}}{%
\pgfpathmoveto{\pgfqpoint{-0.000000in}{0.000000in}}%
\pgfpathlineto{\pgfqpoint{-0.048611in}{0.000000in}}%
\pgfusepath{stroke,fill}%
}%
\begin{pgfscope}%
\pgfsys@transformshift{1.214743in}{1.077744in}%
\pgfsys@useobject{currentmarker}{}%
\end{pgfscope}%
\end{pgfscope}%
\begin{pgfscope}%
\definecolor{textcolor}{rgb}{0.000000,0.000000,0.000000}%
\pgfsetstrokecolor{textcolor}%
\pgfsetfillcolor{textcolor}%
\pgftext[x=0.868765in, y=0.957759in, left, base]{\color{textcolor}\rmfamily\fontsize{32.000000}{38.400000}\selectfont 0}%
\end{pgfscope}%
\begin{pgfscope}%
\pgfsetbuttcap%
\pgfsetroundjoin%
\definecolor{currentfill}{rgb}{0.000000,0.000000,0.000000}%
\pgfsetfillcolor{currentfill}%
\pgfsetlinewidth{0.803000pt}%
\definecolor{currentstroke}{rgb}{0.000000,0.000000,0.000000}%
\pgfsetstrokecolor{currentstroke}%
\pgfsetdash{}{0pt}%
\pgfsys@defobject{currentmarker}{\pgfqpoint{-0.048611in}{0.000000in}}{\pgfqpoint{-0.000000in}{0.000000in}}{%
\pgfpathmoveto{\pgfqpoint{-0.000000in}{0.000000in}}%
\pgfpathlineto{\pgfqpoint{-0.048611in}{0.000000in}}%
\pgfusepath{stroke,fill}%
}%
\begin{pgfscope}%
\pgfsys@transformshift{1.214743in}{2.112175in}%
\pgfsys@useobject{currentmarker}{}%
\end{pgfscope}%
\end{pgfscope}%
\begin{pgfscope}%
\definecolor{textcolor}{rgb}{0.000000,0.000000,0.000000}%
\pgfsetstrokecolor{textcolor}%
\pgfsetfillcolor{textcolor}%
\pgftext[x=0.710287in, y=1.992190in, left, base]{\color{textcolor}\rmfamily\fontsize{32.000000}{38.400000}\selectfont 10}%
\end{pgfscope}%
\begin{pgfscope}%
\pgfsetbuttcap%
\pgfsetroundjoin%
\definecolor{currentfill}{rgb}{0.000000,0.000000,0.000000}%
\pgfsetfillcolor{currentfill}%
\pgfsetlinewidth{0.803000pt}%
\definecolor{currentstroke}{rgb}{0.000000,0.000000,0.000000}%
\pgfsetstrokecolor{currentstroke}%
\pgfsetdash{}{0pt}%
\pgfsys@defobject{currentmarker}{\pgfqpoint{-0.048611in}{0.000000in}}{\pgfqpoint{-0.000000in}{0.000000in}}{%
\pgfpathmoveto{\pgfqpoint{-0.000000in}{0.000000in}}%
\pgfpathlineto{\pgfqpoint{-0.048611in}{0.000000in}}%
\pgfusepath{stroke,fill}%
}%
\begin{pgfscope}%
\pgfsys@transformshift{1.214743in}{3.146606in}%
\pgfsys@useobject{currentmarker}{}%
\end{pgfscope}%
\end{pgfscope}%
\begin{pgfscope}%
\definecolor{textcolor}{rgb}{0.000000,0.000000,0.000000}%
\pgfsetstrokecolor{textcolor}%
\pgfsetfillcolor{textcolor}%
\pgftext[x=0.710287in, y=3.026621in, left, base]{\color{textcolor}\rmfamily\fontsize{32.000000}{38.400000}\selectfont 20}%
\end{pgfscope}%
\begin{pgfscope}%
\pgfsetbuttcap%
\pgfsetroundjoin%
\definecolor{currentfill}{rgb}{0.000000,0.000000,0.000000}%
\pgfsetfillcolor{currentfill}%
\pgfsetlinewidth{0.803000pt}%
\definecolor{currentstroke}{rgb}{0.000000,0.000000,0.000000}%
\pgfsetstrokecolor{currentstroke}%
\pgfsetdash{}{0pt}%
\pgfsys@defobject{currentmarker}{\pgfqpoint{-0.048611in}{0.000000in}}{\pgfqpoint{-0.000000in}{0.000000in}}{%
\pgfpathmoveto{\pgfqpoint{-0.000000in}{0.000000in}}%
\pgfpathlineto{\pgfqpoint{-0.048611in}{0.000000in}}%
\pgfusepath{stroke,fill}%
}%
\begin{pgfscope}%
\pgfsys@transformshift{1.214743in}{4.181037in}%
\pgfsys@useobject{currentmarker}{}%
\end{pgfscope}%
\end{pgfscope}%
\begin{pgfscope}%
\definecolor{textcolor}{rgb}{0.000000,0.000000,0.000000}%
\pgfsetstrokecolor{textcolor}%
\pgfsetfillcolor{textcolor}%
\pgftext[x=0.710287in, y=4.061052in, left, base]{\color{textcolor}\rmfamily\fontsize{32.000000}{38.400000}\selectfont 30}%
\end{pgfscope}%
\begin{pgfscope}%
\pgfsetbuttcap%
\pgfsetroundjoin%
\definecolor{currentfill}{rgb}{0.000000,0.000000,0.000000}%
\pgfsetfillcolor{currentfill}%
\pgfsetlinewidth{0.803000pt}%
\definecolor{currentstroke}{rgb}{0.000000,0.000000,0.000000}%
\pgfsetstrokecolor{currentstroke}%
\pgfsetdash{}{0pt}%
\pgfsys@defobject{currentmarker}{\pgfqpoint{-0.048611in}{0.000000in}}{\pgfqpoint{-0.000000in}{0.000000in}}{%
\pgfpathmoveto{\pgfqpoint{-0.000000in}{0.000000in}}%
\pgfpathlineto{\pgfqpoint{-0.048611in}{0.000000in}}%
\pgfusepath{stroke,fill}%
}%
\begin{pgfscope}%
\pgfsys@transformshift{1.214743in}{5.215467in}%
\pgfsys@useobject{currentmarker}{}%
\end{pgfscope}%
\end{pgfscope}%
\begin{pgfscope}%
\definecolor{textcolor}{rgb}{0.000000,0.000000,0.000000}%
\pgfsetstrokecolor{textcolor}%
\pgfsetfillcolor{textcolor}%
\pgftext[x=0.710287in, y=5.095483in, left, base]{\color{textcolor}\rmfamily\fontsize{32.000000}{38.400000}\selectfont 40}%
\end{pgfscope}%
\begin{pgfscope}%
\pgfsetbuttcap%
\pgfsetroundjoin%
\definecolor{currentfill}{rgb}{0.000000,0.000000,0.000000}%
\pgfsetfillcolor{currentfill}%
\pgfsetlinewidth{0.803000pt}%
\definecolor{currentstroke}{rgb}{0.000000,0.000000,0.000000}%
\pgfsetstrokecolor{currentstroke}%
\pgfsetdash{}{0pt}%
\pgfsys@defobject{currentmarker}{\pgfqpoint{-0.048611in}{0.000000in}}{\pgfqpoint{-0.000000in}{0.000000in}}{%
\pgfpathmoveto{\pgfqpoint{-0.000000in}{0.000000in}}%
\pgfpathlineto{\pgfqpoint{-0.048611in}{0.000000in}}%
\pgfusepath{stroke,fill}%
}%
\begin{pgfscope}%
\pgfsys@transformshift{1.214743in}{6.249898in}%
\pgfsys@useobject{currentmarker}{}%
\end{pgfscope}%
\end{pgfscope}%
\begin{pgfscope}%
\definecolor{textcolor}{rgb}{0.000000,0.000000,0.000000}%
\pgfsetstrokecolor{textcolor}%
\pgfsetfillcolor{textcolor}%
\pgftext[x=0.710287in, y=6.129914in, left, base]{\color{textcolor}\rmfamily\fontsize{32.000000}{38.400000}\selectfont 50}%
\end{pgfscope}%
\begin{pgfscope}%
\definecolor{textcolor}{rgb}{0.000000,0.000000,0.000000}%
\pgfsetstrokecolor{textcolor}%
\pgfsetfillcolor{textcolor}%
\pgftext[x=0.571398in,y=3.663821in,,bottom,rotate=90.000000]{\color{textcolor}\rmfamily\fontsize{36.000000}{43.200000}\selectfont \(\displaystyle y\) [m]}%
\end{pgfscope}%
\begin{pgfscope}%
\pgfpathrectangle{\pgfqpoint{1.214743in}{1.077744in}}{\pgfqpoint{5.466165in}{5.172154in}}%
\pgfusepath{clip}%
\pgfsetrectcap%
\pgfsetroundjoin%
\pgfsetlinewidth{3.011250pt}%
\definecolor{currentstroke}{rgb}{0.839216,0.152941,0.156863}%
\pgfsetstrokecolor{currentstroke}%
\pgfsetdash{}{0pt}%
\pgfpathmoveto{\pgfqpoint{3.947826in}{2.112175in}}%
\pgfpathlineto{\pgfqpoint{3.947425in}{2.163581in}}%
\pgfpathlineto{\pgfqpoint{3.945633in}{2.214272in}}%
\pgfpathlineto{\pgfqpoint{3.942969in}{2.265047in}}%
\pgfpathlineto{\pgfqpoint{3.939711in}{2.315778in}}%
\pgfpathlineto{\pgfqpoint{3.936186in}{2.366344in}}%
\pgfpathlineto{\pgfqpoint{3.933116in}{2.417843in}}%
\pgfpathlineto{\pgfqpoint{3.930098in}{2.469488in}}%
\pgfpathlineto{\pgfqpoint{3.927985in}{2.520836in}}%
\pgfpathlineto{\pgfqpoint{3.926557in}{2.573204in}}%
\pgfpathlineto{\pgfqpoint{3.924033in}{2.625803in}}%
\pgfpathlineto{\pgfqpoint{3.920613in}{2.678364in}}%
\pgfpathlineto{\pgfqpoint{3.916325in}{2.732177in}}%
\pgfpathlineto{\pgfqpoint{3.912511in}{2.786558in}}%
\pgfpathlineto{\pgfqpoint{3.909468in}{2.841300in}}%
\pgfpathlineto{\pgfqpoint{3.905093in}{2.897104in}}%
\pgfpathlineto{\pgfqpoint{3.898853in}{2.953769in}}%
\pgfpathlineto{\pgfqpoint{3.891283in}{3.010253in}}%
\pgfpathlineto{\pgfqpoint{3.883596in}{3.066853in}}%
\pgfpathlineto{\pgfqpoint{3.875528in}{3.123828in}}%
\pgfpathlineto{\pgfqpoint{3.866860in}{3.180814in}}%
\pgfpathlineto{\pgfqpoint{3.857316in}{3.237721in}}%
\pgfpathlineto{\pgfqpoint{3.846265in}{3.295013in}}%
\pgfpathlineto{\pgfqpoint{3.835019in}{3.352074in}}%
\pgfpathlineto{\pgfqpoint{3.824548in}{3.409508in}}%
\pgfpathlineto{\pgfqpoint{3.814112in}{3.468079in}}%
\pgfpathlineto{\pgfqpoint{3.802810in}{3.527216in}}%
\pgfpathlineto{\pgfqpoint{3.791145in}{3.587638in}}%
\pgfpathlineto{\pgfqpoint{3.779994in}{3.648903in}}%
\pgfpathlineto{\pgfqpoint{3.769211in}{3.710290in}}%
\pgfpathlineto{\pgfqpoint{3.758609in}{3.771677in}}%
\pgfpathlineto{\pgfqpoint{3.748115in}{3.832258in}}%
\pgfpathlineto{\pgfqpoint{3.736406in}{3.892289in}}%
\pgfpathlineto{\pgfqpoint{3.723413in}{3.952940in}}%
\pgfpathlineto{\pgfqpoint{3.709640in}{4.014463in}}%
\pgfpathlineto{\pgfqpoint{3.694764in}{4.076832in}}%
\pgfpathlineto{\pgfqpoint{3.679167in}{4.139978in}}%
\pgfpathlineto{\pgfqpoint{3.663933in}{4.204220in}}%
\pgfpathlineto{\pgfqpoint{3.650079in}{4.269386in}}%
\pgfpathlineto{\pgfqpoint{3.636943in}{4.335173in}}%
\pgfpathlineto{\pgfqpoint{3.624148in}{4.401177in}}%
\pgfpathlineto{\pgfqpoint{3.612226in}{4.466668in}}%
\pgfpathlineto{\pgfqpoint{3.601253in}{4.532420in}}%
\pgfpathlineto{\pgfqpoint{3.590806in}{4.598812in}}%
\pgfpathlineto{\pgfqpoint{3.581576in}{4.664851in}}%
\pgfpathlineto{\pgfqpoint{3.573033in}{4.731653in}}%
\pgfpathlineto{\pgfqpoint{3.564212in}{4.799147in}}%
\pgfpathlineto{\pgfqpoint{3.555348in}{4.866753in}}%
\pgfpathlineto{\pgfqpoint{3.546753in}{4.935101in}}%
\pgfpathlineto{\pgfqpoint{3.539281in}{5.003494in}}%
\pgfpathlineto{\pgfqpoint{3.532740in}{5.072257in}}%
\pgfpathlineto{\pgfqpoint{3.524935in}{5.141705in}}%
\pgfpathlineto{\pgfqpoint{3.515152in}{5.211349in}}%
\pgfpathlineto{\pgfqpoint{3.504777in}{5.281581in}}%
\pgfpathlineto{\pgfqpoint{3.493813in}{5.352294in}}%
\pgfpathlineto{\pgfqpoint{3.482387in}{5.422358in}}%
\pgfpathlineto{\pgfqpoint{3.471177in}{5.491881in}}%
\pgfpathlineto{\pgfqpoint{3.460601in}{5.561774in}}%
\pgfpathlineto{\pgfqpoint{3.451139in}{5.631890in}}%
\pgfpathlineto{\pgfqpoint{3.442060in}{5.702242in}}%
\pgfusepath{stroke}%
\end{pgfscope}%
\begin{pgfscope}%
\pgfpathrectangle{\pgfqpoint{1.214743in}{1.077744in}}{\pgfqpoint{5.466165in}{5.172154in}}%
\pgfusepath{clip}%
\pgfsetrectcap%
\pgfsetroundjoin%
\pgfsetlinewidth{3.011250pt}%
\definecolor{currentstroke}{rgb}{0.121569,0.466667,0.705882}%
\pgfsetstrokecolor{currentstroke}%
\pgfsetdash{}{0pt}%
\pgfpathmoveto{\pgfqpoint{2.307976in}{3.663821in}}%
\pgfpathlineto{\pgfqpoint{2.362800in}{3.664357in}}%
\pgfpathlineto{\pgfqpoint{2.417634in}{3.666348in}}%
\pgfpathlineto{\pgfqpoint{2.472684in}{3.669718in}}%
\pgfpathlineto{\pgfqpoint{2.527413in}{3.674479in}}%
\pgfpathlineto{\pgfqpoint{2.581168in}{3.680507in}}%
\pgfpathlineto{\pgfqpoint{2.634173in}{3.685928in}}%
\pgfpathlineto{\pgfqpoint{2.686457in}{3.690186in}}%
\pgfpathlineto{\pgfqpoint{2.738282in}{3.693915in}}%
\pgfpathlineto{\pgfqpoint{2.789454in}{3.696835in}}%
\pgfpathlineto{\pgfqpoint{2.840304in}{3.699397in}}%
\pgfpathlineto{\pgfqpoint{2.890607in}{3.701772in}}%
\pgfpathlineto{\pgfqpoint{2.939481in}{3.704059in}}%
\pgfpathlineto{\pgfqpoint{2.988511in}{3.706316in}}%
\pgfpathlineto{\pgfqpoint{3.038283in}{3.708307in}}%
\pgfpathlineto{\pgfqpoint{3.088111in}{3.710207in}}%
\pgfpathlineto{\pgfqpoint{3.138291in}{3.712176in}}%
\pgfpathlineto{\pgfqpoint{3.188184in}{3.714656in}}%
\pgfpathlineto{\pgfqpoint{3.238211in}{3.717708in}}%
\pgfpathlineto{\pgfqpoint{3.287596in}{3.720584in}}%
\pgfpathlineto{\pgfqpoint{3.335496in}{3.722410in}}%
\pgfpathlineto{\pgfqpoint{3.383268in}{3.724204in}}%
\pgfpathlineto{\pgfqpoint{3.430985in}{3.727139in}}%
\pgfpathlineto{\pgfqpoint{3.479012in}{3.729843in}}%
\pgfpathlineto{\pgfqpoint{3.526645in}{3.731306in}}%
\pgfpathlineto{\pgfqpoint{3.574209in}{3.732537in}}%
\pgfpathlineto{\pgfqpoint{3.622987in}{3.733802in}}%
\pgfpathlineto{\pgfqpoint{3.672935in}{3.735259in}}%
\pgfpathlineto{\pgfqpoint{3.722077in}{3.736894in}}%
\pgfpathlineto{\pgfqpoint{3.769770in}{3.738912in}}%
\pgfpathlineto{\pgfqpoint{3.817479in}{3.741058in}}%
\pgfpathlineto{\pgfqpoint{3.865339in}{3.742853in}}%
\pgfpathlineto{\pgfqpoint{3.913757in}{3.744823in}}%
\pgfpathlineto{\pgfqpoint{3.962070in}{3.747099in}}%
\pgfpathlineto{\pgfqpoint{4.009973in}{3.750561in}}%
\pgfpathlineto{\pgfqpoint{4.057103in}{3.754208in}}%
\pgfpathlineto{\pgfqpoint{4.103085in}{3.756477in}}%
\pgfpathlineto{\pgfqpoint{4.149232in}{3.757146in}}%
\pgfpathlineto{\pgfqpoint{4.195417in}{3.757303in}}%
\pgfpathlineto{\pgfqpoint{4.240721in}{3.757413in}}%
\pgfpathlineto{\pgfqpoint{4.284317in}{3.757581in}}%
\pgfpathlineto{\pgfqpoint{4.326196in}{3.758249in}}%
\pgfpathlineto{\pgfqpoint{4.367830in}{3.759008in}}%
\pgfpathlineto{\pgfqpoint{4.409525in}{3.760528in}}%
\pgfpathlineto{\pgfqpoint{4.451335in}{3.762924in}}%
\pgfpathlineto{\pgfqpoint{4.493525in}{3.766483in}}%
\pgfpathlineto{\pgfqpoint{4.535580in}{3.770322in}}%
\pgfpathlineto{\pgfqpoint{4.577984in}{3.773225in}}%
\pgfpathlineto{\pgfqpoint{4.621199in}{3.775951in}}%
\pgfpathlineto{\pgfqpoint{4.664142in}{3.779010in}}%
\pgfpathlineto{\pgfqpoint{4.707049in}{3.782174in}}%
\pgfpathlineto{\pgfqpoint{4.750572in}{3.785595in}}%
\pgfpathlineto{\pgfqpoint{4.795076in}{3.789695in}}%
\pgfpathlineto{\pgfqpoint{4.839888in}{3.794548in}}%
\pgfpathlineto{\pgfqpoint{4.883729in}{3.799851in}}%
\pgfpathlineto{\pgfqpoint{4.926849in}{3.805491in}}%
\pgfpathlineto{\pgfqpoint{4.969646in}{3.811428in}}%
\pgfpathlineto{\pgfqpoint{5.011744in}{3.817563in}}%
\pgfpathlineto{\pgfqpoint{5.054087in}{3.824047in}}%
\pgfpathlineto{\pgfqpoint{5.096849in}{3.830780in}}%
\pgfusepath{stroke}%
\end{pgfscope}%
\begin{pgfscope}%
\pgfpathrectangle{\pgfqpoint{1.214743in}{1.077744in}}{\pgfqpoint{5.466165in}{5.172154in}}%
\pgfusepath{clip}%
\pgfsetrectcap%
\pgfsetroundjoin%
\pgfsetlinewidth{3.011250pt}%
\definecolor{currentstroke}{rgb}{0.172549,0.627451,0.172549}%
\pgfsetstrokecolor{currentstroke}%
\pgfsetdash{}{0pt}%
\pgfpathmoveto{\pgfqpoint{3.255820in}{3.136409in}}%
\pgfpathlineto{\pgfqpoint{3.264186in}{3.175307in}}%
\pgfpathlineto{\pgfqpoint{3.273362in}{3.213731in}}%
\pgfpathlineto{\pgfqpoint{3.280889in}{3.251965in}}%
\pgfpathlineto{\pgfqpoint{3.286157in}{3.290266in}}%
\pgfpathlineto{\pgfqpoint{3.290601in}{3.329034in}}%
\pgfpathlineto{\pgfqpoint{3.294786in}{3.367501in}}%
\pgfpathlineto{\pgfqpoint{3.298687in}{3.405610in}}%
\pgfpathlineto{\pgfqpoint{3.301912in}{3.443510in}}%
\pgfpathlineto{\pgfqpoint{3.304481in}{3.479990in}}%
\pgfpathlineto{\pgfqpoint{3.307053in}{3.515856in}}%
\pgfpathlineto{\pgfqpoint{3.309647in}{3.551373in}}%
\pgfpathlineto{\pgfqpoint{3.312712in}{3.585446in}}%
\pgfpathlineto{\pgfqpoint{3.315562in}{3.618744in}}%
\pgfpathlineto{\pgfqpoint{3.317567in}{3.651579in}}%
\pgfpathlineto{\pgfqpoint{3.319349in}{3.684499in}}%
\pgfpathlineto{\pgfqpoint{3.319525in}{3.717409in}}%
\pgfpathlineto{\pgfqpoint{3.316969in}{3.750863in}}%
\pgfpathlineto{\pgfqpoint{3.313080in}{3.785194in}}%
\pgfpathlineto{\pgfqpoint{3.309277in}{3.820771in}}%
\pgfpathlineto{\pgfqpoint{3.306060in}{3.858189in}}%
\pgfpathlineto{\pgfqpoint{3.303957in}{3.895981in}}%
\pgfpathlineto{\pgfqpoint{3.302330in}{3.932791in}}%
\pgfpathlineto{\pgfqpoint{3.300658in}{3.968896in}}%
\pgfpathlineto{\pgfqpoint{3.300268in}{4.003927in}}%
\pgfpathlineto{\pgfqpoint{3.301000in}{4.038556in}}%
\pgfpathlineto{\pgfqpoint{3.302229in}{4.073628in}}%
\pgfpathlineto{\pgfqpoint{3.304139in}{4.107536in}}%
\pgfpathlineto{\pgfqpoint{3.306509in}{4.141154in}}%
\pgfpathlineto{\pgfqpoint{3.309058in}{4.175423in}}%
\pgfpathlineto{\pgfqpoint{3.312965in}{4.209141in}}%
\pgfpathlineto{\pgfqpoint{3.318191in}{4.242287in}}%
\pgfpathlineto{\pgfqpoint{3.322687in}{4.275333in}}%
\pgfpathlineto{\pgfqpoint{3.326051in}{4.308135in}}%
\pgfpathlineto{\pgfqpoint{3.328607in}{4.340713in}}%
\pgfpathlineto{\pgfqpoint{3.330815in}{4.373166in}}%
\pgfpathlineto{\pgfqpoint{3.332821in}{4.405632in}}%
\pgfpathlineto{\pgfqpoint{3.334604in}{4.437617in}}%
\pgfpathlineto{\pgfqpoint{3.336514in}{4.469564in}}%
\pgfpathlineto{\pgfqpoint{3.338572in}{4.501959in}}%
\pgfusepath{stroke}%
\end{pgfscope}%
\begin{pgfscope}%
\pgfpathrectangle{\pgfqpoint{1.214743in}{1.077744in}}{\pgfqpoint{5.466165in}{5.172154in}}%
\pgfusepath{clip}%
\pgfsetrectcap%
\pgfsetroundjoin%
\pgfsetlinewidth{3.011250pt}%
\definecolor{currentstroke}{rgb}{1.000000,0.498039,0.054902}%
\pgfsetstrokecolor{currentstroke}%
\pgfsetdash{}{0pt}%
\pgfpathmoveto{\pgfqpoint{3.187765in}{3.773235in}}%
\pgfpathlineto{\pgfqpoint{3.165500in}{3.782042in}}%
\pgfpathlineto{\pgfqpoint{3.144616in}{3.791793in}}%
\pgfpathlineto{\pgfqpoint{3.124794in}{3.801246in}}%
\pgfpathlineto{\pgfqpoint{3.105192in}{3.809468in}}%
\pgfpathlineto{\pgfqpoint{3.085175in}{3.817725in}}%
\pgfpathlineto{\pgfqpoint{3.065627in}{3.826093in}}%
\pgfpathlineto{\pgfqpoint{3.047002in}{3.835036in}}%
\pgfpathlineto{\pgfqpoint{3.027942in}{3.844992in}}%
\pgfpathlineto{\pgfqpoint{3.007813in}{3.855080in}}%
\pgfpathlineto{\pgfqpoint{2.986882in}{3.864237in}}%
\pgfpathlineto{\pgfqpoint{2.965100in}{3.872460in}}%
\pgfpathlineto{\pgfqpoint{2.942461in}{3.879546in}}%
\pgfpathlineto{\pgfqpoint{2.920331in}{3.886036in}}%
\pgfpathlineto{\pgfqpoint{2.898941in}{3.892768in}}%
\pgfpathlineto{\pgfqpoint{2.876645in}{3.899404in}}%
\pgfpathlineto{\pgfqpoint{2.853405in}{3.905762in}}%
\pgfpathlineto{\pgfqpoint{2.829774in}{3.911798in}}%
\pgfpathlineto{\pgfqpoint{2.806417in}{3.917547in}}%
\pgfpathlineto{\pgfqpoint{2.783816in}{3.923127in}}%
\pgfpathlineto{\pgfqpoint{2.761241in}{3.928742in}}%
\pgfpathlineto{\pgfqpoint{2.737880in}{3.934905in}}%
\pgfpathlineto{\pgfqpoint{2.713987in}{3.941944in}}%
\pgfpathlineto{\pgfqpoint{2.690826in}{3.949259in}}%
\pgfpathlineto{\pgfqpoint{2.668697in}{3.956275in}}%
\pgfpathlineto{\pgfqpoint{2.647455in}{3.963447in}}%
\pgfpathlineto{\pgfqpoint{2.626090in}{3.970756in}}%
\pgfpathlineto{\pgfqpoint{2.604451in}{3.977192in}}%
\pgfpathlineto{\pgfqpoint{2.583261in}{3.983646in}}%
\pgfpathlineto{\pgfqpoint{2.563272in}{3.989071in}}%
\pgfpathlineto{\pgfqpoint{2.544368in}{3.992922in}}%
\pgfpathlineto{\pgfqpoint{2.525113in}{3.997258in}}%
\pgfpathlineto{\pgfqpoint{2.504759in}{4.001501in}}%
\pgfpathlineto{\pgfqpoint{2.484100in}{4.006463in}}%
\pgfpathlineto{\pgfqpoint{2.464443in}{4.013234in}}%
\pgfpathlineto{\pgfqpoint{2.445171in}{4.020388in}}%
\pgfpathlineto{\pgfqpoint{2.424087in}{4.026921in}}%
\pgfpathlineto{\pgfqpoint{2.401034in}{4.033149in}}%
\pgfpathlineto{\pgfqpoint{2.376787in}{4.039887in}}%
\pgfpathlineto{\pgfqpoint{2.351567in}{4.046927in}}%
\pgfusepath{stroke}%
\end{pgfscope}%
\begin{pgfscope}%
\pgfpathrectangle{\pgfqpoint{1.214743in}{1.077744in}}{\pgfqpoint{5.466165in}{5.172154in}}%
\pgfusepath{clip}%
\pgfsetrectcap%
\pgfsetroundjoin%
\pgfsetlinewidth{3.011250pt}%
\definecolor{currentstroke}{rgb}{0.580392,0.403922,0.741176}%
\pgfsetstrokecolor{currentstroke}%
\pgfsetdash{}{0pt}%
\pgfpathmoveto{\pgfqpoint{4.808004in}{3.490654in}}%
\pgfpathlineto{\pgfqpoint{4.838085in}{3.461596in}}%
\pgfpathlineto{\pgfqpoint{4.868072in}{3.432689in}}%
\pgfpathlineto{\pgfqpoint{4.898584in}{3.403845in}}%
\pgfpathlineto{\pgfqpoint{4.929806in}{3.375048in}}%
\pgfpathlineto{\pgfqpoint{4.960908in}{3.347280in}}%
\pgfpathlineto{\pgfqpoint{4.991860in}{3.320615in}}%
\pgfpathlineto{\pgfqpoint{5.023545in}{3.294787in}}%
\pgfpathlineto{\pgfqpoint{5.056269in}{3.269151in}}%
\pgfpathlineto{\pgfqpoint{5.089370in}{3.243419in}}%
\pgfpathlineto{\pgfqpoint{5.122998in}{3.217574in}}%
\pgfpathlineto{\pgfqpoint{5.157303in}{3.191292in}}%
\pgfpathlineto{\pgfqpoint{5.192118in}{3.164187in}}%
\pgfpathlineto{\pgfqpoint{5.227728in}{3.136250in}}%
\pgfpathlineto{\pgfqpoint{5.263710in}{3.107800in}}%
\pgfpathlineto{\pgfqpoint{5.300259in}{3.078898in}}%
\pgfpathlineto{\pgfqpoint{5.336543in}{3.050327in}}%
\pgfpathlineto{\pgfqpoint{5.371072in}{3.021485in}}%
\pgfpathlineto{\pgfqpoint{5.403963in}{2.992063in}}%
\pgfpathlineto{\pgfqpoint{5.435479in}{2.965054in}}%
\pgfpathlineto{\pgfqpoint{5.466503in}{2.940083in}}%
\pgfpathlineto{\pgfqpoint{5.497554in}{2.914773in}}%
\pgfpathlineto{\pgfqpoint{5.528573in}{2.889381in}}%
\pgfpathlineto{\pgfqpoint{5.560152in}{2.864304in}}%
\pgfpathlineto{\pgfqpoint{5.592718in}{2.839267in}}%
\pgfpathlineto{\pgfqpoint{5.625540in}{2.813558in}}%
\pgfpathlineto{\pgfqpoint{5.658092in}{2.786734in}}%
\pgfpathlineto{\pgfqpoint{5.689260in}{2.759166in}}%
\pgfpathlineto{\pgfqpoint{5.718076in}{2.731954in}}%
\pgfpathlineto{\pgfqpoint{5.745674in}{2.704309in}}%
\pgfpathlineto{\pgfqpoint{5.772993in}{2.676250in}}%
\pgfpathlineto{\pgfqpoint{5.799760in}{2.649659in}}%
\pgfpathlineto{\pgfqpoint{5.825879in}{2.625062in}}%
\pgfpathlineto{\pgfqpoint{5.851738in}{2.602814in}}%
\pgfpathlineto{\pgfqpoint{5.877594in}{2.582099in}}%
\pgfpathlineto{\pgfqpoint{5.903989in}{2.561760in}}%
\pgfpathlineto{\pgfqpoint{5.930650in}{2.541817in}}%
\pgfpathlineto{\pgfqpoint{5.957730in}{2.522674in}}%
\pgfpathlineto{\pgfqpoint{5.984968in}{2.504377in}}%
\pgfpathlineto{\pgfqpoint{6.011216in}{2.485798in}}%
\pgfusepath{stroke}%
\end{pgfscope}%
\begin{pgfscope}%
\pgfsetrectcap%
\pgfsetmiterjoin%
\pgfsetlinewidth{0.803000pt}%
\definecolor{currentstroke}{rgb}{0.000000,0.000000,0.000000}%
\pgfsetstrokecolor{currentstroke}%
\pgfsetdash{}{0pt}%
\pgfpathmoveto{\pgfqpoint{1.214743in}{1.077744in}}%
\pgfpathlineto{\pgfqpoint{1.214743in}{6.249898in}}%
\pgfusepath{stroke}%
\end{pgfscope}%
\begin{pgfscope}%
\pgfsetrectcap%
\pgfsetmiterjoin%
\pgfsetlinewidth{0.803000pt}%
\definecolor{currentstroke}{rgb}{0.000000,0.000000,0.000000}%
\pgfsetstrokecolor{currentstroke}%
\pgfsetdash{}{0pt}%
\pgfpathmoveto{\pgfqpoint{6.680908in}{1.077744in}}%
\pgfpathlineto{\pgfqpoint{6.680908in}{6.249898in}}%
\pgfusepath{stroke}%
\end{pgfscope}%
\begin{pgfscope}%
\pgfsetrectcap%
\pgfsetmiterjoin%
\pgfsetlinewidth{0.803000pt}%
\definecolor{currentstroke}{rgb}{0.000000,0.000000,0.000000}%
\pgfsetstrokecolor{currentstroke}%
\pgfsetdash{}{0pt}%
\pgfpathmoveto{\pgfqpoint{1.214743in}{1.077744in}}%
\pgfpathlineto{\pgfqpoint{6.680908in}{1.077744in}}%
\pgfusepath{stroke}%
\end{pgfscope}%
\begin{pgfscope}%
\pgfsetrectcap%
\pgfsetmiterjoin%
\pgfsetlinewidth{0.803000pt}%
\definecolor{currentstroke}{rgb}{0.000000,0.000000,0.000000}%
\pgfsetstrokecolor{currentstroke}%
\pgfsetdash{}{0pt}%
\pgfpathmoveto{\pgfqpoint{1.214743in}{6.249898in}}%
\pgfpathlineto{\pgfqpoint{6.680908in}{6.249898in}}%
\pgfusepath{stroke}%
\end{pgfscope}%
\begin{pgfscope}%
\pgfsetbuttcap%
\pgfsetmiterjoin%
\definecolor{currentfill}{rgb}{1.000000,1.000000,1.000000}%
\pgfsetfillcolor{currentfill}%
\pgfsetlinewidth{0.000000pt}%
\definecolor{currentstroke}{rgb}{0.000000,0.000000,0.000000}%
\pgfsetstrokecolor{currentstroke}%
\pgfsetstrokeopacity{0.000000}%
\pgfsetdash{}{0pt}%
\pgfpathmoveto{\pgfqpoint{6.920127in}{1.077744in}}%
\pgfpathlineto{\pgfqpoint{7.178735in}{1.077744in}}%
\pgfpathlineto{\pgfqpoint{7.178735in}{6.249898in}}%
\pgfpathlineto{\pgfqpoint{6.920127in}{6.249898in}}%
\pgfpathlineto{\pgfqpoint{6.920127in}{1.077744in}}%
\pgfpathclose%
\pgfusepath{fill}%
\end{pgfscope}%
\begin{pgfscope}%
\pgfpathrectangle{\pgfqpoint{6.920127in}{1.077744in}}{\pgfqpoint{0.258608in}{5.172154in}}%
\pgfusepath{clip}%
\pgfsetbuttcap%
\pgfsetmiterjoin%
\definecolor{currentfill}{rgb}{1.000000,1.000000,1.000000}%
\pgfsetfillcolor{currentfill}%
\pgfsetlinewidth{0.010037pt}%
\definecolor{currentstroke}{rgb}{1.000000,1.000000,1.000000}%
\pgfsetstrokecolor{currentstroke}%
\pgfsetdash{}{0pt}%
\pgfusepath{stroke,fill}%
\end{pgfscope}%
\begin{pgfscope}%
\pgfsys@transformshift{6.920000in}{1.080000in}%
\pgftext[left,bottom]{\includegraphics[interpolate=true,width=0.260000in,height=5.170000in]{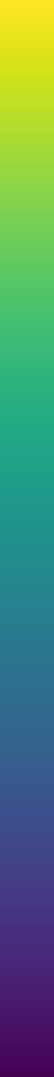}}%
\end{pgfscope}%
\begin{pgfscope}%
\pgfsetbuttcap%
\pgfsetroundjoin%
\definecolor{currentfill}{rgb}{0.000000,0.000000,0.000000}%
\pgfsetfillcolor{currentfill}%
\pgfsetlinewidth{0.803000pt}%
\definecolor{currentstroke}{rgb}{0.000000,0.000000,0.000000}%
\pgfsetstrokecolor{currentstroke}%
\pgfsetdash{}{0pt}%
\pgfsys@defobject{currentmarker}{\pgfqpoint{0.000000in}{0.000000in}}{\pgfqpoint{0.048611in}{0.000000in}}{%
\pgfpathmoveto{\pgfqpoint{0.000000in}{0.000000in}}%
\pgfpathlineto{\pgfqpoint{0.048611in}{0.000000in}}%
\pgfusepath{stroke,fill}%
}%
\begin{pgfscope}%
\pgfsys@transformshift{7.178735in}{2.243039in}%
\pgfsys@useobject{currentmarker}{}%
\end{pgfscope}%
\end{pgfscope}%
\begin{pgfscope}%
\definecolor{textcolor}{rgb}{0.000000,0.000000,0.000000}%
\pgfsetstrokecolor{textcolor}%
\pgfsetfillcolor{textcolor}%
\pgftext[x=7.275957in, y=2.123055in, left, base]{\color{textcolor}\rmfamily\fontsize{24.000000}{28.800000}\selectfont 0}%
\end{pgfscope}%
\begin{pgfscope}%
\pgfsetbuttcap%
\pgfsetroundjoin%
\definecolor{currentfill}{rgb}{0.000000,0.000000,0.000000}%
\pgfsetfillcolor{currentfill}%
\pgfsetlinewidth{0.803000pt}%
\definecolor{currentstroke}{rgb}{0.000000,0.000000,0.000000}%
\pgfsetstrokecolor{currentstroke}%
\pgfsetdash{}{0pt}%
\pgfsys@defobject{currentmarker}{\pgfqpoint{0.000000in}{0.000000in}}{\pgfqpoint{0.048611in}{0.000000in}}{%
\pgfpathmoveto{\pgfqpoint{0.000000in}{0.000000in}}%
\pgfpathlineto{\pgfqpoint{0.048611in}{0.000000in}}%
\pgfusepath{stroke,fill}%
}%
\begin{pgfscope}%
\pgfsys@transformshift{7.178735in}{3.538906in}%
\pgfsys@useobject{currentmarker}{}%
\end{pgfscope}%
\end{pgfscope}%
\begin{pgfscope}%
\definecolor{textcolor}{rgb}{0.000000,0.000000,0.000000}%
\pgfsetstrokecolor{textcolor}%
\pgfsetfillcolor{textcolor}%
\pgftext[x=7.275957in, y=3.418921in, left, base]{\color{textcolor}\rmfamily\fontsize{24.000000}{28.800000}\selectfont 20}%
\end{pgfscope}%
\begin{pgfscope}%
\pgfsetbuttcap%
\pgfsetroundjoin%
\definecolor{currentfill}{rgb}{0.000000,0.000000,0.000000}%
\pgfsetfillcolor{currentfill}%
\pgfsetlinewidth{0.803000pt}%
\definecolor{currentstroke}{rgb}{0.000000,0.000000,0.000000}%
\pgfsetstrokecolor{currentstroke}%
\pgfsetdash{}{0pt}%
\pgfsys@defobject{currentmarker}{\pgfqpoint{0.000000in}{0.000000in}}{\pgfqpoint{0.048611in}{0.000000in}}{%
\pgfpathmoveto{\pgfqpoint{0.000000in}{0.000000in}}%
\pgfpathlineto{\pgfqpoint{0.048611in}{0.000000in}}%
\pgfusepath{stroke,fill}%
}%
\begin{pgfscope}%
\pgfsys@transformshift{7.178735in}{4.834773in}%
\pgfsys@useobject{currentmarker}{}%
\end{pgfscope}%
\end{pgfscope}%
\begin{pgfscope}%
\definecolor{textcolor}{rgb}{0.000000,0.000000,0.000000}%
\pgfsetstrokecolor{textcolor}%
\pgfsetfillcolor{textcolor}%
\pgftext[x=7.275957in, y=4.714788in, left, base]{\color{textcolor}\rmfamily\fontsize{24.000000}{28.800000}\selectfont 40}%
\end{pgfscope}%
\begin{pgfscope}%
\pgfsetbuttcap%
\pgfsetroundjoin%
\definecolor{currentfill}{rgb}{0.000000,0.000000,0.000000}%
\pgfsetfillcolor{currentfill}%
\pgfsetlinewidth{0.803000pt}%
\definecolor{currentstroke}{rgb}{0.000000,0.000000,0.000000}%
\pgfsetstrokecolor{currentstroke}%
\pgfsetdash{}{0pt}%
\pgfsys@defobject{currentmarker}{\pgfqpoint{0.000000in}{0.000000in}}{\pgfqpoint{0.048611in}{0.000000in}}{%
\pgfpathmoveto{\pgfqpoint{0.000000in}{0.000000in}}%
\pgfpathlineto{\pgfqpoint{0.048611in}{0.000000in}}%
\pgfusepath{stroke,fill}%
}%
\begin{pgfscope}%
\pgfsys@transformshift{7.178735in}{6.130639in}%
\pgfsys@useobject{currentmarker}{}%
\end{pgfscope}%
\end{pgfscope}%
\begin{pgfscope}%
\definecolor{textcolor}{rgb}{0.000000,0.000000,0.000000}%
\pgfsetstrokecolor{textcolor}%
\pgfsetfillcolor{textcolor}%
\pgftext[x=7.275957in, y=6.010655in, left, base]{\color{textcolor}\rmfamily\fontsize{24.000000}{28.800000}\selectfont 60}%
\end{pgfscope}%
\begin{pgfscope}%
\pgfsetrectcap%
\pgfsetmiterjoin%
\pgfsetlinewidth{0.803000pt}%
\definecolor{currentstroke}{rgb}{0.000000,0.000000,0.000000}%
\pgfsetstrokecolor{currentstroke}%
\pgfsetdash{}{0pt}%
\pgfpathmoveto{\pgfqpoint{6.920127in}{1.077744in}}%
\pgfpathlineto{\pgfqpoint{7.049431in}{1.077744in}}%
\pgfpathlineto{\pgfqpoint{7.178735in}{1.077744in}}%
\pgfpathlineto{\pgfqpoint{7.178735in}{6.249898in}}%
\pgfpathlineto{\pgfqpoint{7.049431in}{6.249898in}}%
\pgfpathlineto{\pgfqpoint{6.920127in}{6.249898in}}%
\pgfpathlineto{\pgfqpoint{6.920127in}{1.077744in}}%
\pgfpathclose%
\pgfusepath{stroke}%
\end{pgfscope}%
\end{pgfpicture}%
\makeatother%
\endgroup%

%% file: Figs/snr/snr_vs_time_sensor2.pgf
\begingroup%
\makeatletter%
\begin{pgfpicture}%
\pgfpathrectangle{\pgfpointorigin}{\pgfqpoint{7.600000in}{6.400000in}}%
\pgfusepath{use as bounding box, clip}%
\begin{pgfscope}%
\pgfsetbuttcap%
\pgfsetmiterjoin%
\definecolor{currentfill}{rgb}{1.000000,1.000000,1.000000}%
\pgfsetfillcolor{currentfill}%
\pgfsetlinewidth{0.000000pt}%
\definecolor{currentstroke}{rgb}{1.000000,1.000000,1.000000}%
\pgfsetstrokecolor{currentstroke}%
\pgfsetdash{}{0pt}%
\pgfpathmoveto{\pgfqpoint{0.000000in}{0.000000in}}%
\pgfpathlineto{\pgfqpoint{7.600000in}{0.000000in}}%
\pgfpathlineto{\pgfqpoint{7.600000in}{6.400000in}}%
\pgfpathlineto{\pgfqpoint{0.000000in}{6.400000in}}%
\pgfpathlineto{\pgfqpoint{0.000000in}{0.000000in}}%
\pgfpathclose%
\pgfusepath{fill}%
\end{pgfscope}%
\begin{pgfscope}%
\pgfsetbuttcap%
\pgfsetmiterjoin%
\definecolor{currentfill}{rgb}{1.000000,1.000000,1.000000}%
\pgfsetfillcolor{currentfill}%
\pgfsetlinewidth{0.000000pt}%
\definecolor{currentstroke}{rgb}{0.000000,0.000000,0.000000}%
\pgfsetstrokecolor{currentstroke}%
\pgfsetstrokeopacity{0.000000}%
\pgfsetdash{}{0pt}%
\pgfpathmoveto{\pgfqpoint{1.362189in}{1.077744in}}%
\pgfpathlineto{\pgfqpoint{7.378608in}{1.077744in}}%
\pgfpathlineto{\pgfqpoint{7.378608in}{6.400000in}}%
\pgfpathlineto{\pgfqpoint{1.362189in}{6.400000in}}%
\pgfpathlineto{\pgfqpoint{1.362189in}{1.077744in}}%
\pgfpathclose%
\pgfusepath{fill}%
\end{pgfscope}%
\begin{pgfscope}%
\pgfpathrectangle{\pgfqpoint{1.362189in}{1.077744in}}{\pgfqpoint{6.016419in}{5.322256in}}%
\pgfusepath{clip}%
\pgfsetrectcap%
\pgfsetroundjoin%
\pgfsetlinewidth{0.803000pt}%
\definecolor{currentstroke}{rgb}{0.690196,0.690196,0.690196}%
\pgfsetstrokecolor{currentstroke}%
\pgfsetdash{}{0pt}%
\pgfpathmoveto{\pgfqpoint{2.809176in}{1.077744in}}%
\pgfpathlineto{\pgfqpoint{2.809176in}{6.400000in}}%
\pgfusepath{stroke}%
\end{pgfscope}%
\begin{pgfscope}%
\pgfsetbuttcap%
\pgfsetroundjoin%
\definecolor{currentfill}{rgb}{0.000000,0.000000,0.000000}%
\pgfsetfillcolor{currentfill}%
\pgfsetlinewidth{0.803000pt}%
\definecolor{currentstroke}{rgb}{0.000000,0.000000,0.000000}%
\pgfsetstrokecolor{currentstroke}%
\pgfsetdash{}{0pt}%
\pgfsys@defobject{currentmarker}{\pgfqpoint{0.000000in}{-0.048611in}}{\pgfqpoint{0.000000in}{0.000000in}}{%
\pgfpathmoveto{\pgfqpoint{0.000000in}{0.000000in}}%
\pgfpathlineto{\pgfqpoint{0.000000in}{-0.048611in}}%
\pgfusepath{stroke,fill}%
}%
\begin{pgfscope}%
\pgfsys@transformshift{2.809176in}{1.077744in}%
\pgfsys@useobject{currentmarker}{}%
\end{pgfscope}%
\end{pgfscope}%
\begin{pgfscope}%
\definecolor{textcolor}{rgb}{0.000000,0.000000,0.000000}%
\pgfsetstrokecolor{textcolor}%
\pgfsetfillcolor{textcolor}%
\pgftext[x=2.809176in,y=0.980522in,,top]{\color{textcolor}\rmfamily\fontsize{32.000000}{38.400000}\selectfont 20}%
\end{pgfscope}%
\begin{pgfscope}%
\pgfpathrectangle{\pgfqpoint{1.362189in}{1.077744in}}{\pgfqpoint{6.016419in}{5.322256in}}%
\pgfusepath{clip}%
\pgfsetrectcap%
\pgfsetroundjoin%
\pgfsetlinewidth{0.803000pt}%
\definecolor{currentstroke}{rgb}{0.690196,0.690196,0.690196}%
\pgfsetstrokecolor{currentstroke}%
\pgfsetdash{}{0pt}%
\pgfpathmoveto{\pgfqpoint{4.332320in}{1.077744in}}%
\pgfpathlineto{\pgfqpoint{4.332320in}{6.400000in}}%
\pgfusepath{stroke}%
\end{pgfscope}%
\begin{pgfscope}%
\pgfsetbuttcap%
\pgfsetroundjoin%
\definecolor{currentfill}{rgb}{0.000000,0.000000,0.000000}%
\pgfsetfillcolor{currentfill}%
\pgfsetlinewidth{0.803000pt}%
\definecolor{currentstroke}{rgb}{0.000000,0.000000,0.000000}%
\pgfsetstrokecolor{currentstroke}%
\pgfsetdash{}{0pt}%
\pgfsys@defobject{currentmarker}{\pgfqpoint{0.000000in}{-0.048611in}}{\pgfqpoint{0.000000in}{0.000000in}}{%
\pgfpathmoveto{\pgfqpoint{0.000000in}{0.000000in}}%
\pgfpathlineto{\pgfqpoint{0.000000in}{-0.048611in}}%
\pgfusepath{stroke,fill}%
}%
\begin{pgfscope}%
\pgfsys@transformshift{4.332320in}{1.077744in}%
\pgfsys@useobject{currentmarker}{}%
\end{pgfscope}%
\end{pgfscope}%
\begin{pgfscope}%
\definecolor{textcolor}{rgb}{0.000000,0.000000,0.000000}%
\pgfsetstrokecolor{textcolor}%
\pgfsetfillcolor{textcolor}%
\pgftext[x=4.332320in,y=0.980522in,,top]{\color{textcolor}\rmfamily\fontsize{32.000000}{38.400000}\selectfont 40}%
\end{pgfscope}%
\begin{pgfscope}%
\pgfpathrectangle{\pgfqpoint{1.362189in}{1.077744in}}{\pgfqpoint{6.016419in}{5.322256in}}%
\pgfusepath{clip}%
\pgfsetrectcap%
\pgfsetroundjoin%
\pgfsetlinewidth{0.803000pt}%
\definecolor{currentstroke}{rgb}{0.690196,0.690196,0.690196}%
\pgfsetstrokecolor{currentstroke}%
\pgfsetdash{}{0pt}%
\pgfpathmoveto{\pgfqpoint{5.855464in}{1.077744in}}%
\pgfpathlineto{\pgfqpoint{5.855464in}{6.400000in}}%
\pgfusepath{stroke}%
\end{pgfscope}%
\begin{pgfscope}%
\pgfsetbuttcap%
\pgfsetroundjoin%
\definecolor{currentfill}{rgb}{0.000000,0.000000,0.000000}%
\pgfsetfillcolor{currentfill}%
\pgfsetlinewidth{0.803000pt}%
\definecolor{currentstroke}{rgb}{0.000000,0.000000,0.000000}%
\pgfsetstrokecolor{currentstroke}%
\pgfsetdash{}{0pt}%
\pgfsys@defobject{currentmarker}{\pgfqpoint{0.000000in}{-0.048611in}}{\pgfqpoint{0.000000in}{0.000000in}}{%
\pgfpathmoveto{\pgfqpoint{0.000000in}{0.000000in}}%
\pgfpathlineto{\pgfqpoint{0.000000in}{-0.048611in}}%
\pgfusepath{stroke,fill}%
}%
\begin{pgfscope}%
\pgfsys@transformshift{5.855464in}{1.077744in}%
\pgfsys@useobject{currentmarker}{}%
\end{pgfscope}%
\end{pgfscope}%
\begin{pgfscope}%
\definecolor{textcolor}{rgb}{0.000000,0.000000,0.000000}%
\pgfsetstrokecolor{textcolor}%
\pgfsetfillcolor{textcolor}%
\pgftext[x=5.855464in,y=0.980522in,,top]{\color{textcolor}\rmfamily\fontsize{32.000000}{38.400000}\selectfont 60}%
\end{pgfscope}%
\begin{pgfscope}%
\pgfpathrectangle{\pgfqpoint{1.362189in}{1.077744in}}{\pgfqpoint{6.016419in}{5.322256in}}%
\pgfusepath{clip}%
\pgfsetrectcap%
\pgfsetroundjoin%
\pgfsetlinewidth{0.803000pt}%
\definecolor{currentstroke}{rgb}{0.690196,0.690196,0.690196}%
\pgfsetstrokecolor{currentstroke}%
\pgfsetdash{}{0pt}%
\pgfpathmoveto{\pgfqpoint{7.378608in}{1.077744in}}%
\pgfpathlineto{\pgfqpoint{7.378608in}{6.400000in}}%
\pgfusepath{stroke}%
\end{pgfscope}%
\begin{pgfscope}%
\pgfsetbuttcap%
\pgfsetroundjoin%
\definecolor{currentfill}{rgb}{0.000000,0.000000,0.000000}%
\pgfsetfillcolor{currentfill}%
\pgfsetlinewidth{0.803000pt}%
\definecolor{currentstroke}{rgb}{0.000000,0.000000,0.000000}%
\pgfsetstrokecolor{currentstroke}%
\pgfsetdash{}{0pt}%
\pgfsys@defobject{currentmarker}{\pgfqpoint{0.000000in}{-0.048611in}}{\pgfqpoint{0.000000in}{0.000000in}}{%
\pgfpathmoveto{\pgfqpoint{0.000000in}{0.000000in}}%
\pgfpathlineto{\pgfqpoint{0.000000in}{-0.048611in}}%
\pgfusepath{stroke,fill}%
}%
\begin{pgfscope}%
\pgfsys@transformshift{7.378608in}{1.077744in}%
\pgfsys@useobject{currentmarker}{}%
\end{pgfscope}%
\end{pgfscope}%
\begin{pgfscope}%
\definecolor{textcolor}{rgb}{0.000000,0.000000,0.000000}%
\pgfsetstrokecolor{textcolor}%
\pgfsetfillcolor{textcolor}%
\pgftext[x=7.378608in,y=0.980522in,,top]{\color{textcolor}\rmfamily\fontsize{32.000000}{38.400000}\selectfont 80}%
\end{pgfscope}%
\begin{pgfscope}%
\definecolor{textcolor}{rgb}{0.000000,0.000000,0.000000}%
\pgfsetstrokecolor{textcolor}%
\pgfsetfillcolor{textcolor}%
\pgftext[x=4.370398in,y=0.534450in,,top]{\color{textcolor}\rmfamily\fontsize{36.000000}{43.200000}\selectfont time step \(\displaystyle k\)}%
\end{pgfscope}%
\begin{pgfscope}%
\pgfpathrectangle{\pgfqpoint{1.362189in}{1.077744in}}{\pgfqpoint{6.016419in}{5.322256in}}%
\pgfusepath{clip}%
\pgfsetrectcap%
\pgfsetroundjoin%
\pgfsetlinewidth{0.803000pt}%
\definecolor{currentstroke}{rgb}{0.690196,0.690196,0.690196}%
\pgfsetstrokecolor{currentstroke}%
\pgfsetdash{}{0pt}%
\pgfpathmoveto{\pgfqpoint{1.362189in}{1.901636in}}%
\pgfpathlineto{\pgfqpoint{7.378608in}{1.901636in}}%
\pgfusepath{stroke}%
\end{pgfscope}%
\begin{pgfscope}%
\pgfsetbuttcap%
\pgfsetroundjoin%
\definecolor{currentfill}{rgb}{0.000000,0.000000,0.000000}%
\pgfsetfillcolor{currentfill}%
\pgfsetlinewidth{0.803000pt}%
\definecolor{currentstroke}{rgb}{0.000000,0.000000,0.000000}%
\pgfsetstrokecolor{currentstroke}%
\pgfsetdash{}{0pt}%
\pgfsys@defobject{currentmarker}{\pgfqpoint{-0.048611in}{0.000000in}}{\pgfqpoint{-0.000000in}{0.000000in}}{%
\pgfpathmoveto{\pgfqpoint{-0.000000in}{0.000000in}}%
\pgfpathlineto{\pgfqpoint{-0.048611in}{0.000000in}}%
\pgfusepath{stroke,fill}%
}%
\begin{pgfscope}%
\pgfsys@transformshift{1.362189in}{1.901636in}%
\pgfsys@useobject{currentmarker}{}%
\end{pgfscope}%
\end{pgfscope}%
\begin{pgfscope}%
\definecolor{textcolor}{rgb}{0.000000,0.000000,0.000000}%
\pgfsetstrokecolor{textcolor}%
\pgfsetfillcolor{textcolor}%
\pgftext[x=0.753414in, y=1.781651in, left, base]{\color{textcolor}\rmfamily\fontsize{32.000000}{38.400000}\selectfont -10}%
\end{pgfscope}%
\begin{pgfscope}%
\pgfpathrectangle{\pgfqpoint{1.362189in}{1.077744in}}{\pgfqpoint{6.016419in}{5.322256in}}%
\pgfusepath{clip}%
\pgfsetrectcap%
\pgfsetroundjoin%
\pgfsetlinewidth{0.803000pt}%
\definecolor{currentstroke}{rgb}{0.690196,0.690196,0.690196}%
\pgfsetstrokecolor{currentstroke}%
\pgfsetdash{}{0pt}%
\pgfpathmoveto{\pgfqpoint{1.362189in}{3.193072in}}%
\pgfpathlineto{\pgfqpoint{7.378608in}{3.193072in}}%
\pgfusepath{stroke}%
\end{pgfscope}%
\begin{pgfscope}%
\pgfsetbuttcap%
\pgfsetroundjoin%
\definecolor{currentfill}{rgb}{0.000000,0.000000,0.000000}%
\pgfsetfillcolor{currentfill}%
\pgfsetlinewidth{0.803000pt}%
\definecolor{currentstroke}{rgb}{0.000000,0.000000,0.000000}%
\pgfsetstrokecolor{currentstroke}%
\pgfsetdash{}{0pt}%
\pgfsys@defobject{currentmarker}{\pgfqpoint{-0.048611in}{0.000000in}}{\pgfqpoint{-0.000000in}{0.000000in}}{%
\pgfpathmoveto{\pgfqpoint{-0.000000in}{0.000000in}}%
\pgfpathlineto{\pgfqpoint{-0.048611in}{0.000000in}}%
\pgfusepath{stroke,fill}%
}%
\begin{pgfscope}%
\pgfsys@transformshift{1.362189in}{3.193072in}%
\pgfsys@useobject{currentmarker}{}%
\end{pgfscope}%
\end{pgfscope}%
\begin{pgfscope}%
\definecolor{textcolor}{rgb}{0.000000,0.000000,0.000000}%
\pgfsetstrokecolor{textcolor}%
\pgfsetfillcolor{textcolor}%
\pgftext[x=0.911892in, y=3.073087in, left, base]{\color{textcolor}\rmfamily\fontsize{32.000000}{38.400000}\selectfont -7}%
\end{pgfscope}%
\begin{pgfscope}%
\pgfpathrectangle{\pgfqpoint{1.362189in}{1.077744in}}{\pgfqpoint{6.016419in}{5.322256in}}%
\pgfusepath{clip}%
\pgfsetrectcap%
\pgfsetroundjoin%
\pgfsetlinewidth{0.803000pt}%
\definecolor{currentstroke}{rgb}{0.690196,0.690196,0.690196}%
\pgfsetstrokecolor{currentstroke}%
\pgfsetdash{}{0pt}%
\pgfpathmoveto{\pgfqpoint{1.362189in}{4.484507in}}%
\pgfpathlineto{\pgfqpoint{7.378608in}{4.484507in}}%
\pgfusepath{stroke}%
\end{pgfscope}%
\begin{pgfscope}%
\pgfsetbuttcap%
\pgfsetroundjoin%
\definecolor{currentfill}{rgb}{0.000000,0.000000,0.000000}%
\pgfsetfillcolor{currentfill}%
\pgfsetlinewidth{0.803000pt}%
\definecolor{currentstroke}{rgb}{0.000000,0.000000,0.000000}%
\pgfsetstrokecolor{currentstroke}%
\pgfsetdash{}{0pt}%
\pgfsys@defobject{currentmarker}{\pgfqpoint{-0.048611in}{0.000000in}}{\pgfqpoint{-0.000000in}{0.000000in}}{%
\pgfpathmoveto{\pgfqpoint{-0.000000in}{0.000000in}}%
\pgfpathlineto{\pgfqpoint{-0.048611in}{0.000000in}}%
\pgfusepath{stroke,fill}%
}%
\begin{pgfscope}%
\pgfsys@transformshift{1.362189in}{4.484507in}%
\pgfsys@useobject{currentmarker}{}%
\end{pgfscope}%
\end{pgfscope}%
\begin{pgfscope}%
\definecolor{textcolor}{rgb}{0.000000,0.000000,0.000000}%
\pgfsetstrokecolor{textcolor}%
\pgfsetfillcolor{textcolor}%
\pgftext[x=0.911892in, y=4.364523in, left, base]{\color{textcolor}\rmfamily\fontsize{32.000000}{38.400000}\selectfont -5}%
\end{pgfscope}%
\begin{pgfscope}%
\pgfpathrectangle{\pgfqpoint{1.362189in}{1.077744in}}{\pgfqpoint{6.016419in}{5.322256in}}%
\pgfusepath{clip}%
\pgfsetrectcap%
\pgfsetroundjoin%
\pgfsetlinewidth{0.803000pt}%
\definecolor{currentstroke}{rgb}{0.690196,0.690196,0.690196}%
\pgfsetstrokecolor{currentstroke}%
\pgfsetdash{}{0pt}%
\pgfpathmoveto{\pgfqpoint{1.362189in}{5.775943in}}%
\pgfpathlineto{\pgfqpoint{7.378608in}{5.775943in}}%
\pgfusepath{stroke}%
\end{pgfscope}%
\begin{pgfscope}%
\pgfsetbuttcap%
\pgfsetroundjoin%
\definecolor{currentfill}{rgb}{0.000000,0.000000,0.000000}%
\pgfsetfillcolor{currentfill}%
\pgfsetlinewidth{0.803000pt}%
\definecolor{currentstroke}{rgb}{0.000000,0.000000,0.000000}%
\pgfsetstrokecolor{currentstroke}%
\pgfsetdash{}{0pt}%
\pgfsys@defobject{currentmarker}{\pgfqpoint{-0.048611in}{0.000000in}}{\pgfqpoint{-0.000000in}{0.000000in}}{%
\pgfpathmoveto{\pgfqpoint{-0.000000in}{0.000000in}}%
\pgfpathlineto{\pgfqpoint{-0.048611in}{0.000000in}}%
\pgfusepath{stroke,fill}%
}%
\begin{pgfscope}%
\pgfsys@transformshift{1.362189in}{5.775943in}%
\pgfsys@useobject{currentmarker}{}%
\end{pgfscope}%
\end{pgfscope}%
\begin{pgfscope}%
\definecolor{textcolor}{rgb}{0.000000,0.000000,0.000000}%
\pgfsetstrokecolor{textcolor}%
\pgfsetfillcolor{textcolor}%
\pgftext[x=0.911892in, y=5.655959in, left, base]{\color{textcolor}\rmfamily\fontsize{32.000000}{38.400000}\selectfont -2}%
\end{pgfscope}%
\begin{pgfscope}%
\definecolor{textcolor}{rgb}{0.000000,0.000000,0.000000}%
\pgfsetstrokecolor{textcolor}%
\pgfsetfillcolor{textcolor}%
\pgftext[x=0.614525in,y=3.738872in,,bottom,rotate=90.000000]{\color{textcolor}\rmfamily\fontsize{36.000000}{43.200000}\selectfont SNR [dB]}%
\end{pgfscope}%
\begin{pgfscope}%
\pgfpathrectangle{\pgfqpoint{1.362189in}{1.077744in}}{\pgfqpoint{6.016419in}{5.322256in}}%
\pgfusepath{clip}%
\pgfsetrectcap%
\pgfsetroundjoin%
\pgfsetlinewidth{3.011250pt}%
\definecolor{currentstroke}{rgb}{0.839216,0.152941,0.156863}%
\pgfsetstrokecolor{currentstroke}%
\pgfsetdash{}{0pt}%
\pgfpathmoveto{\pgfqpoint{2.047604in}{1.319665in}}%
\pgfpathlineto{\pgfqpoint{2.123761in}{1.399677in}}%
\pgfpathlineto{\pgfqpoint{2.199918in}{1.477654in}}%
\pgfpathlineto{\pgfqpoint{2.276075in}{1.555271in}}%
\pgfpathlineto{\pgfqpoint{2.352232in}{1.632551in}}%
\pgfpathlineto{\pgfqpoint{2.428390in}{1.709590in}}%
\pgfpathlineto{\pgfqpoint{2.504547in}{1.788863in}}%
\pgfpathlineto{\pgfqpoint{2.580704in}{1.868720in}}%
\pgfpathlineto{\pgfqpoint{2.656861in}{1.949352in}}%
\pgfpathlineto{\pgfqpoint{2.733018in}{2.032693in}}%
\pgfpathlineto{\pgfqpoint{2.809176in}{2.115538in}}%
\pgfpathlineto{\pgfqpoint{2.885333in}{2.197591in}}%
\pgfpathlineto{\pgfqpoint{2.961490in}{2.280933in}}%
\pgfpathlineto{\pgfqpoint{3.037647in}{2.365962in}}%
\pgfpathlineto{\pgfqpoint{3.113804in}{2.452717in}}%
\pgfpathlineto{\pgfqpoint{3.189962in}{2.539821in}}%
\pgfpathlineto{\pgfqpoint{3.266119in}{2.626158in}}%
\pgfpathlineto{\pgfqpoint{3.342276in}{2.710512in}}%
\pgfpathlineto{\pgfqpoint{3.418433in}{2.794824in}}%
\pgfpathlineto{\pgfqpoint{3.494590in}{2.879128in}}%
\pgfpathlineto{\pgfqpoint{3.570748in}{2.962448in}}%
\pgfpathlineto{\pgfqpoint{3.646905in}{3.044150in}}%
\pgfpathlineto{\pgfqpoint{3.723062in}{3.123979in}}%
\pgfpathlineto{\pgfqpoint{3.799219in}{3.202643in}}%
\pgfpathlineto{\pgfqpoint{3.875377in}{3.282517in}}%
\pgfpathlineto{\pgfqpoint{3.951534in}{3.363756in}}%
\pgfpathlineto{\pgfqpoint{4.027691in}{3.443939in}}%
\pgfpathlineto{\pgfqpoint{4.103848in}{3.524886in}}%
\pgfpathlineto{\pgfqpoint{4.180005in}{3.607166in}}%
\pgfpathlineto{\pgfqpoint{4.256163in}{3.689324in}}%
\pgfpathlineto{\pgfqpoint{4.332320in}{3.770800in}}%
\pgfpathlineto{\pgfqpoint{4.408477in}{3.850117in}}%
\pgfpathlineto{\pgfqpoint{4.484634in}{3.925289in}}%
\pgfpathlineto{\pgfqpoint{4.560791in}{3.997836in}}%
\pgfpathlineto{\pgfqpoint{4.636949in}{4.068823in}}%
\pgfpathlineto{\pgfqpoint{4.713106in}{4.137373in}}%
\pgfpathlineto{\pgfqpoint{4.789263in}{4.203843in}}%
\pgfpathlineto{\pgfqpoint{4.865420in}{4.270596in}}%
\pgfpathlineto{\pgfqpoint{4.941577in}{4.339274in}}%
\pgfpathlineto{\pgfqpoint{5.017735in}{4.408067in}}%
\pgfpathlineto{\pgfqpoint{5.093892in}{4.475533in}}%
\pgfpathlineto{\pgfqpoint{5.170049in}{4.541695in}}%
\pgfpathlineto{\pgfqpoint{5.246206in}{4.607787in}}%
\pgfpathlineto{\pgfqpoint{5.322363in}{4.673338in}}%
\pgfpathlineto{\pgfqpoint{5.398521in}{4.738501in}}%
\pgfpathlineto{\pgfqpoint{5.474678in}{4.803502in}}%
\pgfpathlineto{\pgfqpoint{5.550835in}{4.865887in}}%
\pgfpathlineto{\pgfqpoint{5.626992in}{4.925285in}}%
\pgfpathlineto{\pgfqpoint{5.703149in}{4.982997in}}%
\pgfpathlineto{\pgfqpoint{5.779307in}{5.040145in}}%
\pgfpathlineto{\pgfqpoint{5.855464in}{5.096574in}}%
\pgfpathlineto{\pgfqpoint{5.931621in}{5.147013in}}%
\pgfpathlineto{\pgfqpoint{6.007778in}{5.188838in}}%
\pgfpathlineto{\pgfqpoint{6.083935in}{5.225610in}}%
\pgfpathlineto{\pgfqpoint{6.160093in}{5.256990in}}%
\pgfpathlineto{\pgfqpoint{6.236250in}{5.282237in}}%
\pgfpathlineto{\pgfqpoint{6.312407in}{5.303214in}}%
\pgfpathlineto{\pgfqpoint{6.388564in}{5.321655in}}%
\pgfpathlineto{\pgfqpoint{6.464722in}{5.338671in}}%
\pgfpathlineto{\pgfqpoint{6.540879in}{5.352274in}}%
\pgfusepath{stroke}%
\end{pgfscope}%
\begin{pgfscope}%
\pgfpathrectangle{\pgfqpoint{1.362189in}{1.077744in}}{\pgfqpoint{6.016419in}{5.322256in}}%
\pgfusepath{clip}%
\pgfsetrectcap%
\pgfsetroundjoin%
\pgfsetlinewidth{3.011250pt}%
\definecolor{currentstroke}{rgb}{0.121569,0.466667,0.705882}%
\pgfsetstrokecolor{currentstroke}%
\pgfsetdash{}{0pt}%
\pgfpathmoveto{\pgfqpoint{2.047604in}{1.319665in}}%
\pgfpathlineto{\pgfqpoint{2.123761in}{1.401322in}}%
\pgfpathlineto{\pgfqpoint{2.199918in}{1.484925in}}%
\pgfpathlineto{\pgfqpoint{2.276075in}{1.570784in}}%
\pgfpathlineto{\pgfqpoint{2.352232in}{1.658183in}}%
\pgfpathlineto{\pgfqpoint{2.428390in}{1.746064in}}%
\pgfpathlineto{\pgfqpoint{2.504547in}{1.832785in}}%
\pgfpathlineto{\pgfqpoint{2.580704in}{1.917727in}}%
\pgfpathlineto{\pgfqpoint{2.656861in}{2.001919in}}%
\pgfpathlineto{\pgfqpoint{2.733018in}{2.084701in}}%
\pgfpathlineto{\pgfqpoint{2.809176in}{2.167049in}}%
\pgfpathlineto{\pgfqpoint{2.885333in}{2.248777in}}%
\pgfpathlineto{\pgfqpoint{2.961490in}{2.328582in}}%
\pgfpathlineto{\pgfqpoint{3.037647in}{2.409015in}}%
\pgfpathlineto{\pgfqpoint{3.113804in}{2.490712in}}%
\pgfpathlineto{\pgfqpoint{3.189962in}{2.572795in}}%
\pgfpathlineto{\pgfqpoint{3.266119in}{2.655930in}}%
\pgfpathlineto{\pgfqpoint{3.342276in}{2.739673in}}%
\pgfpathlineto{\pgfqpoint{3.418433in}{2.824812in}}%
\pgfpathlineto{\pgfqpoint{3.494590in}{2.909103in}}%
\pgfpathlineto{\pgfqpoint{3.570748in}{2.989922in}}%
\pgfpathlineto{\pgfqpoint{3.646905in}{3.070822in}}%
\pgfpathlineto{\pgfqpoint{3.723062in}{3.153588in}}%
\pgfpathlineto{\pgfqpoint{3.799219in}{3.236905in}}%
\pgfpathlineto{\pgfqpoint{3.875377in}{3.318082in}}%
\pgfpathlineto{\pgfqpoint{3.951534in}{3.399075in}}%
\pgfpathlineto{\pgfqpoint{4.027691in}{3.482399in}}%
\pgfpathlineto{\pgfqpoint{4.103848in}{3.568234in}}%
\pgfpathlineto{\pgfqpoint{4.180005in}{3.653256in}}%
\pgfpathlineto{\pgfqpoint{4.256163in}{3.736701in}}%
\pgfpathlineto{\pgfqpoint{4.332320in}{3.820624in}}%
\pgfpathlineto{\pgfqpoint{4.408477in}{3.904441in}}%
\pgfpathlineto{\pgfqpoint{4.484634in}{3.989696in}}%
\pgfpathlineto{\pgfqpoint{4.560791in}{4.075492in}}%
\pgfpathlineto{\pgfqpoint{4.636949in}{4.162883in}}%
\pgfpathlineto{\pgfqpoint{4.713106in}{4.249531in}}%
\pgfpathlineto{\pgfqpoint{4.789263in}{4.331917in}}%
\pgfpathlineto{\pgfqpoint{4.865420in}{4.411705in}}%
\pgfpathlineto{\pgfqpoint{4.941577in}{4.490568in}}%
\pgfpathlineto{\pgfqpoint{5.017735in}{4.567771in}}%
\pgfpathlineto{\pgfqpoint{5.093892in}{4.642091in}}%
\pgfpathlineto{\pgfqpoint{5.170049in}{4.714389in}}%
\pgfpathlineto{\pgfqpoint{5.246206in}{4.786357in}}%
\pgfpathlineto{\pgfqpoint{5.322363in}{4.859887in}}%
\pgfpathlineto{\pgfqpoint{5.398521in}{4.935358in}}%
\pgfpathlineto{\pgfqpoint{5.474678in}{5.013892in}}%
\pgfpathlineto{\pgfqpoint{5.550835in}{5.092799in}}%
\pgfpathlineto{\pgfqpoint{5.626992in}{5.170223in}}%
\pgfpathlineto{\pgfqpoint{5.703149in}{5.248497in}}%
\pgfpathlineto{\pgfqpoint{5.779307in}{5.326922in}}%
\pgfpathlineto{\pgfqpoint{5.855464in}{5.405367in}}%
\pgfpathlineto{\pgfqpoint{5.931621in}{5.485248in}}%
\pgfpathlineto{\pgfqpoint{6.007778in}{5.568149in}}%
\pgfpathlineto{\pgfqpoint{6.083935in}{5.653173in}}%
\pgfpathlineto{\pgfqpoint{6.160093in}{5.737538in}}%
\pgfpathlineto{\pgfqpoint{6.236250in}{5.821403in}}%
\pgfpathlineto{\pgfqpoint{6.312407in}{5.905329in}}%
\pgfpathlineto{\pgfqpoint{6.388564in}{5.988467in}}%
\pgfpathlineto{\pgfqpoint{6.464722in}{6.072717in}}%
\pgfpathlineto{\pgfqpoint{6.540879in}{6.158079in}}%
\pgfusepath{stroke}%
\end{pgfscope}%
\begin{pgfscope}%
\pgfpathrectangle{\pgfqpoint{1.362189in}{1.077744in}}{\pgfqpoint{6.016419in}{5.322256in}}%
\pgfusepath{clip}%
\pgfsetrectcap%
\pgfsetroundjoin%
\pgfsetlinewidth{3.011250pt}%
\definecolor{currentstroke}{rgb}{0.172549,0.627451,0.172549}%
\pgfsetstrokecolor{currentstroke}%
\pgfsetdash{}{0pt}%
\pgfpathmoveto{\pgfqpoint{2.809176in}{2.057809in}}%
\pgfpathlineto{\pgfqpoint{2.885333in}{2.123141in}}%
\pgfpathlineto{\pgfqpoint{2.961490in}{2.189190in}}%
\pgfpathlineto{\pgfqpoint{3.037647in}{2.252912in}}%
\pgfpathlineto{\pgfqpoint{3.113804in}{2.313703in}}%
\pgfpathlineto{\pgfqpoint{3.189962in}{2.374075in}}%
\pgfpathlineto{\pgfqpoint{3.266119in}{2.433737in}}%
\pgfpathlineto{\pgfqpoint{3.342276in}{2.492551in}}%
\pgfpathlineto{\pgfqpoint{3.418433in}{2.550116in}}%
\pgfpathlineto{\pgfqpoint{3.494590in}{2.604732in}}%
\pgfpathlineto{\pgfqpoint{3.570748in}{2.658494in}}%
\pgfpathlineto{\pgfqpoint{3.646905in}{2.711793in}}%
\pgfpathlineto{\pgfqpoint{3.723062in}{2.763812in}}%
\pgfpathlineto{\pgfqpoint{3.799219in}{2.814412in}}%
\pgfpathlineto{\pgfqpoint{3.875377in}{2.863004in}}%
\pgfpathlineto{\pgfqpoint{3.951534in}{2.911302in}}%
\pgfpathlineto{\pgfqpoint{4.027691in}{2.956899in}}%
\pgfpathlineto{\pgfqpoint{4.103848in}{2.998605in}}%
\pgfpathlineto{\pgfqpoint{4.180005in}{3.039055in}}%
\pgfpathlineto{\pgfqpoint{4.256163in}{3.081040in}}%
\pgfpathlineto{\pgfqpoint{4.332320in}{3.126184in}}%
\pgfpathlineto{\pgfqpoint{4.408477in}{3.173403in}}%
\pgfpathlineto{\pgfqpoint{4.484634in}{3.219817in}}%
\pgfpathlineto{\pgfqpoint{4.560791in}{3.264915in}}%
\pgfpathlineto{\pgfqpoint{4.636949in}{3.310567in}}%
\pgfpathlineto{\pgfqpoint{4.713106in}{3.357445in}}%
\pgfpathlineto{\pgfqpoint{4.789263in}{3.405582in}}%
\pgfpathlineto{\pgfqpoint{4.865420in}{3.453229in}}%
\pgfpathlineto{\pgfqpoint{4.941577in}{3.501159in}}%
\pgfpathlineto{\pgfqpoint{5.017735in}{3.550075in}}%
\pgfpathlineto{\pgfqpoint{5.093892in}{3.600652in}}%
\pgfpathlineto{\pgfqpoint{5.170049in}{3.652865in}}%
\pgfpathlineto{\pgfqpoint{5.246206in}{3.703440in}}%
\pgfpathlineto{\pgfqpoint{5.322363in}{3.751360in}}%
\pgfpathlineto{\pgfqpoint{5.398521in}{3.797219in}}%
\pgfpathlineto{\pgfqpoint{5.474678in}{3.842000in}}%
\pgfpathlineto{\pgfqpoint{5.550835in}{3.886137in}}%
\pgfpathlineto{\pgfqpoint{5.626992in}{3.928961in}}%
\pgfpathlineto{\pgfqpoint{5.703149in}{3.971713in}}%
\pgfpathlineto{\pgfqpoint{5.779307in}{4.015013in}}%
\pgfusepath{stroke}%
\end{pgfscope}%
\begin{pgfscope}%
\pgfpathrectangle{\pgfqpoint{1.362189in}{1.077744in}}{\pgfqpoint{6.016419in}{5.322256in}}%
\pgfusepath{clip}%
\pgfsetrectcap%
\pgfsetroundjoin%
\pgfsetlinewidth{3.011250pt}%
\definecolor{currentstroke}{rgb}{1.000000,0.498039,0.054902}%
\pgfsetstrokecolor{currentstroke}%
\pgfsetdash{}{0pt}%
\pgfpathmoveto{\pgfqpoint{2.428390in}{2.815562in}}%
\pgfpathlineto{\pgfqpoint{2.504547in}{2.790339in}}%
\pgfpathlineto{\pgfqpoint{2.580704in}{2.768469in}}%
\pgfpathlineto{\pgfqpoint{2.656861in}{2.747827in}}%
\pgfpathlineto{\pgfqpoint{2.733018in}{2.725868in}}%
\pgfpathlineto{\pgfqpoint{2.809176in}{2.703162in}}%
\pgfpathlineto{\pgfqpoint{2.885333in}{2.681265in}}%
\pgfpathlineto{\pgfqpoint{2.961490in}{2.661498in}}%
\pgfpathlineto{\pgfqpoint{3.037647in}{2.642131in}}%
\pgfpathlineto{\pgfqpoint{3.113804in}{2.621022in}}%
\pgfpathlineto{\pgfqpoint{3.189962in}{2.597339in}}%
\pgfpathlineto{\pgfqpoint{3.266119in}{2.571024in}}%
\pgfpathlineto{\pgfqpoint{3.342276in}{2.541863in}}%
\pgfpathlineto{\pgfqpoint{3.418433in}{2.512794in}}%
\pgfpathlineto{\pgfqpoint{3.494590in}{2.485179in}}%
\pgfpathlineto{\pgfqpoint{3.570748in}{2.455903in}}%
\pgfpathlineto{\pgfqpoint{3.646905in}{2.424708in}}%
\pgfpathlineto{\pgfqpoint{3.723062in}{2.392471in}}%
\pgfpathlineto{\pgfqpoint{3.799219in}{2.360343in}}%
\pgfpathlineto{\pgfqpoint{3.875377in}{2.329256in}}%
\pgfpathlineto{\pgfqpoint{3.951534in}{2.298235in}}%
\pgfpathlineto{\pgfqpoint{4.027691in}{2.266500in}}%
\pgfpathlineto{\pgfqpoint{4.103848in}{2.234804in}}%
\pgfpathlineto{\pgfqpoint{4.180005in}{2.204568in}}%
\pgfpathlineto{\pgfqpoint{4.256163in}{2.175665in}}%
\pgfpathlineto{\pgfqpoint{4.332320in}{2.148343in}}%
\pgfpathlineto{\pgfqpoint{4.408477in}{2.120919in}}%
\pgfpathlineto{\pgfqpoint{4.484634in}{2.092120in}}%
\pgfpathlineto{\pgfqpoint{4.560791in}{2.064049in}}%
\pgfpathlineto{\pgfqpoint{4.636949in}{2.036887in}}%
\pgfpathlineto{\pgfqpoint{4.713106in}{2.009923in}}%
\pgfpathlineto{\pgfqpoint{4.789263in}{1.982882in}}%
\pgfpathlineto{\pgfqpoint{4.865420in}{1.953975in}}%
\pgfpathlineto{\pgfqpoint{4.941577in}{1.925288in}}%
\pgfpathlineto{\pgfqpoint{5.017735in}{1.899965in}}%
\pgfpathlineto{\pgfqpoint{5.093892in}{1.875604in}}%
\pgfpathlineto{\pgfqpoint{5.170049in}{1.847698in}}%
\pgfpathlineto{\pgfqpoint{5.246206in}{1.816318in}}%
\pgfpathlineto{\pgfqpoint{5.322363in}{1.783489in}}%
\pgfpathlineto{\pgfqpoint{5.398521in}{1.749372in}}%
\pgfusepath{stroke}%
\end{pgfscope}%
\begin{pgfscope}%
\pgfpathrectangle{\pgfqpoint{1.362189in}{1.077744in}}{\pgfqpoint{6.016419in}{5.322256in}}%
\pgfusepath{clip}%
\pgfsetrectcap%
\pgfsetroundjoin%
\pgfsetlinewidth{3.011250pt}%
\definecolor{currentstroke}{rgb}{0.580392,0.403922,0.741176}%
\pgfsetstrokecolor{currentstroke}%
\pgfsetdash{}{0pt}%
\pgfpathmoveto{\pgfqpoint{3.189962in}{4.885710in}}%
\pgfpathlineto{\pgfqpoint{3.266119in}{4.860279in}}%
\pgfpathlineto{\pgfqpoint{3.342276in}{4.833849in}}%
\pgfpathlineto{\pgfqpoint{3.418433in}{4.807066in}}%
\pgfpathlineto{\pgfqpoint{3.494590in}{4.780099in}}%
\pgfpathlineto{\pgfqpoint{3.570748in}{4.754184in}}%
\pgfpathlineto{\pgfqpoint{3.646905in}{4.729522in}}%
\pgfpathlineto{\pgfqpoint{3.723062in}{4.706590in}}%
\pgfpathlineto{\pgfqpoint{3.799219in}{4.684192in}}%
\pgfpathlineto{\pgfqpoint{3.875377in}{4.660815in}}%
\pgfpathlineto{\pgfqpoint{3.951534in}{4.636575in}}%
\pgfpathlineto{\pgfqpoint{4.027691in}{4.610823in}}%
\pgfpathlineto{\pgfqpoint{4.103848in}{4.582404in}}%
\pgfpathlineto{\pgfqpoint{4.180005in}{4.551549in}}%
\pgfpathlineto{\pgfqpoint{4.256163in}{4.518556in}}%
\pgfpathlineto{\pgfqpoint{4.332320in}{4.483742in}}%
\pgfpathlineto{\pgfqpoint{4.408477in}{4.448219in}}%
\pgfpathlineto{\pgfqpoint{4.484634in}{4.409275in}}%
\pgfpathlineto{\pgfqpoint{4.560791in}{4.366460in}}%
\pgfpathlineto{\pgfqpoint{4.636949in}{4.327409in}}%
\pgfpathlineto{\pgfqpoint{4.713106in}{4.292051in}}%
\pgfpathlineto{\pgfqpoint{4.789263in}{4.255095in}}%
\pgfpathlineto{\pgfqpoint{4.865420in}{4.217137in}}%
\pgfpathlineto{\pgfqpoint{4.941577in}{4.179585in}}%
\pgfpathlineto{\pgfqpoint{5.017735in}{4.142044in}}%
\pgfpathlineto{\pgfqpoint{5.093892in}{4.102260in}}%
\pgfpathlineto{\pgfqpoint{5.170049in}{4.058847in}}%
\pgfpathlineto{\pgfqpoint{5.246206in}{4.012076in}}%
\pgfpathlineto{\pgfqpoint{5.322363in}{3.964153in}}%
\pgfpathlineto{\pgfqpoint{5.398521in}{3.914022in}}%
\pgfpathlineto{\pgfqpoint{5.474678in}{3.862350in}}%
\pgfpathlineto{\pgfqpoint{5.550835in}{3.813499in}}%
\pgfpathlineto{\pgfqpoint{5.626992in}{3.768674in}}%
\pgfpathlineto{\pgfqpoint{5.703149in}{3.728909in}}%
\pgfpathlineto{\pgfqpoint{5.779307in}{3.692387in}}%
\pgfpathlineto{\pgfqpoint{5.855464in}{3.656628in}}%
\pgfpathlineto{\pgfqpoint{5.931621in}{3.621534in}}%
\pgfpathlineto{\pgfqpoint{6.007778in}{3.588099in}}%
\pgfpathlineto{\pgfqpoint{6.083935in}{3.556305in}}%
\pgfpathlineto{\pgfqpoint{6.160093in}{3.523100in}}%
\pgfusepath{stroke}%
\end{pgfscope}%
\begin{pgfscope}%
\pgfsetrectcap%
\pgfsetmiterjoin%
\pgfsetlinewidth{0.803000pt}%
\definecolor{currentstroke}{rgb}{0.000000,0.000000,0.000000}%
\pgfsetstrokecolor{currentstroke}%
\pgfsetdash{}{0pt}%
\pgfpathmoveto{\pgfqpoint{1.362189in}{1.077744in}}%
\pgfpathlineto{\pgfqpoint{1.362189in}{6.400000in}}%
\pgfusepath{stroke}%
\end{pgfscope}%
\begin{pgfscope}%
\pgfsetrectcap%
\pgfsetmiterjoin%
\pgfsetlinewidth{0.803000pt}%
\definecolor{currentstroke}{rgb}{0.000000,0.000000,0.000000}%
\pgfsetstrokecolor{currentstroke}%
\pgfsetdash{}{0pt}%
\pgfpathmoveto{\pgfqpoint{7.378608in}{1.077744in}}%
\pgfpathlineto{\pgfqpoint{7.378608in}{6.400000in}}%
\pgfusepath{stroke}%
\end{pgfscope}%
\begin{pgfscope}%
\pgfsetrectcap%
\pgfsetmiterjoin%
\pgfsetlinewidth{0.803000pt}%
\definecolor{currentstroke}{rgb}{0.000000,0.000000,0.000000}%
\pgfsetstrokecolor{currentstroke}%
\pgfsetdash{}{0pt}%
\pgfpathmoveto{\pgfqpoint{1.362189in}{1.077744in}}%
\pgfpathlineto{\pgfqpoint{7.378608in}{1.077744in}}%
\pgfusepath{stroke}%
\end{pgfscope}%
\begin{pgfscope}%
\pgfsetrectcap%
\pgfsetmiterjoin%
\pgfsetlinewidth{0.803000pt}%
\definecolor{currentstroke}{rgb}{0.000000,0.000000,0.000000}%
\pgfsetstrokecolor{currentstroke}%
\pgfsetdash{}{0pt}%
\pgfpathmoveto{\pgfqpoint{1.362189in}{6.400000in}}%
\pgfpathlineto{\pgfqpoint{7.378608in}{6.400000in}}%
\pgfusepath{stroke}%
\end{pgfscope}%
\end{pgfpicture}%
\makeatother%
\endgroup%

%% file: Figs/gospa_-6dB/gospa.pgf
\begingroup%
\makeatletter%
\begin{pgfpicture}%
\pgfpathrectangle{\pgfpointorigin}{\pgfqpoint{13.400000in}{6.600000in}}%
\pgfusepath{use as bounding box, clip}%
\begin{pgfscope}%
\pgfsetbuttcap%
\pgfsetmiterjoin%
\definecolor{currentfill}{rgb}{1.000000,1.000000,1.000000}%
\pgfsetfillcolor{currentfill}%
\pgfsetlinewidth{0.000000pt}%
\definecolor{currentstroke}{rgb}{1.000000,1.000000,1.000000}%
\pgfsetstrokecolor{currentstroke}%
\pgfsetdash{}{0pt}%
\pgfpathmoveto{\pgfqpoint{0.000000in}{0.000000in}}%
\pgfpathlineto{\pgfqpoint{13.400000in}{0.000000in}}%
\pgfpathlineto{\pgfqpoint{13.400000in}{6.600000in}}%
\pgfpathlineto{\pgfqpoint{0.000000in}{6.600000in}}%
\pgfpathlineto{\pgfqpoint{0.000000in}{0.000000in}}%
\pgfpathclose%
\pgfusepath{fill}%
\end{pgfscope}%
\begin{pgfscope}%
\pgfsetbuttcap%
\pgfsetmiterjoin%
\definecolor{currentfill}{rgb}{1.000000,1.000000,1.000000}%
\pgfsetfillcolor{currentfill}%
\pgfsetlinewidth{0.000000pt}%
\definecolor{currentstroke}{rgb}{0.000000,0.000000,0.000000}%
\pgfsetstrokecolor{currentstroke}%
\pgfsetstrokeopacity{0.000000}%
\pgfsetdash{}{0pt}%
\pgfpathmoveto{\pgfqpoint{0.866662in}{0.929220in}}%
\pgfpathlineto{\pgfqpoint{13.206282in}{0.929220in}}%
\pgfpathlineto{\pgfqpoint{13.206282in}{6.600000in}}%
\pgfpathlineto{\pgfqpoint{0.866662in}{6.600000in}}%
\pgfpathlineto{\pgfqpoint{0.866662in}{0.929220in}}%
\pgfpathclose%
\pgfusepath{fill}%
\end{pgfscope}%
\begin{pgfscope}%
\pgfpathrectangle{\pgfqpoint{0.866662in}{0.929220in}}{\pgfqpoint{12.339620in}{5.670780in}}%
\pgfusepath{clip}%
\pgfsetrectcap%
\pgfsetroundjoin%
\pgfsetlinewidth{0.803000pt}%
\definecolor{currentstroke}{rgb}{0.690196,0.690196,0.690196}%
\pgfsetstrokecolor{currentstroke}%
\pgfsetdash{}{0pt}%
\pgfpathmoveto{\pgfqpoint{2.272441in}{0.929220in}}%
\pgfpathlineto{\pgfqpoint{2.272441in}{6.600000in}}%
\pgfusepath{stroke}%
\end{pgfscope}%
\begin{pgfscope}%
\pgfsetbuttcap%
\pgfsetroundjoin%
\definecolor{currentfill}{rgb}{0.000000,0.000000,0.000000}%
\pgfsetfillcolor{currentfill}%
\pgfsetlinewidth{0.803000pt}%
\definecolor{currentstroke}{rgb}{0.000000,0.000000,0.000000}%
\pgfsetstrokecolor{currentstroke}%
\pgfsetdash{}{0pt}%
\pgfsys@defobject{currentmarker}{\pgfqpoint{0.000000in}{-0.048611in}}{\pgfqpoint{0.000000in}{0.000000in}}{%
\pgfpathmoveto{\pgfqpoint{0.000000in}{0.000000in}}%
\pgfpathlineto{\pgfqpoint{0.000000in}{-0.048611in}}%
\pgfusepath{stroke,fill}%
}%
\begin{pgfscope}%
\pgfsys@transformshift{2.272441in}{0.929220in}%
\pgfsys@useobject{currentmarker}{}%
\end{pgfscope}%
\end{pgfscope}%
\begin{pgfscope}%
\definecolor{textcolor}{rgb}{0.000000,0.000000,0.000000}%
\pgfsetstrokecolor{textcolor}%
\pgfsetfillcolor{textcolor}%
\pgftext[x=2.272441in,y=0.831998in,,top]{\color{textcolor}\rmfamily\fontsize{28.000000}{33.600000}\selectfont 10}%
\end{pgfscope}%
\begin{pgfscope}%
\pgfpathrectangle{\pgfqpoint{0.866662in}{0.929220in}}{\pgfqpoint{12.339620in}{5.670780in}}%
\pgfusepath{clip}%
\pgfsetrectcap%
\pgfsetroundjoin%
\pgfsetlinewidth{0.803000pt}%
\definecolor{currentstroke}{rgb}{0.690196,0.690196,0.690196}%
\pgfsetstrokecolor{currentstroke}%
\pgfsetdash{}{0pt}%
\pgfpathmoveto{\pgfqpoint{3.834418in}{0.929220in}}%
\pgfpathlineto{\pgfqpoint{3.834418in}{6.600000in}}%
\pgfusepath{stroke}%
\end{pgfscope}%
\begin{pgfscope}%
\pgfsetbuttcap%
\pgfsetroundjoin%
\definecolor{currentfill}{rgb}{0.000000,0.000000,0.000000}%
\pgfsetfillcolor{currentfill}%
\pgfsetlinewidth{0.803000pt}%
\definecolor{currentstroke}{rgb}{0.000000,0.000000,0.000000}%
\pgfsetstrokecolor{currentstroke}%
\pgfsetdash{}{0pt}%
\pgfsys@defobject{currentmarker}{\pgfqpoint{0.000000in}{-0.048611in}}{\pgfqpoint{0.000000in}{0.000000in}}{%
\pgfpathmoveto{\pgfqpoint{0.000000in}{0.000000in}}%
\pgfpathlineto{\pgfqpoint{0.000000in}{-0.048611in}}%
\pgfusepath{stroke,fill}%
}%
\begin{pgfscope}%
\pgfsys@transformshift{3.834418in}{0.929220in}%
\pgfsys@useobject{currentmarker}{}%
\end{pgfscope}%
\end{pgfscope}%
\begin{pgfscope}%
\definecolor{textcolor}{rgb}{0.000000,0.000000,0.000000}%
\pgfsetstrokecolor{textcolor}%
\pgfsetfillcolor{textcolor}%
\pgftext[x=3.834418in,y=0.831998in,,top]{\color{textcolor}\rmfamily\fontsize{28.000000}{33.600000}\selectfont 20}%
\end{pgfscope}%
\begin{pgfscope}%
\pgfpathrectangle{\pgfqpoint{0.866662in}{0.929220in}}{\pgfqpoint{12.339620in}{5.670780in}}%
\pgfusepath{clip}%
\pgfsetrectcap%
\pgfsetroundjoin%
\pgfsetlinewidth{0.803000pt}%
\definecolor{currentstroke}{rgb}{0.690196,0.690196,0.690196}%
\pgfsetstrokecolor{currentstroke}%
\pgfsetdash{}{0pt}%
\pgfpathmoveto{\pgfqpoint{5.396396in}{0.929220in}}%
\pgfpathlineto{\pgfqpoint{5.396396in}{6.600000in}}%
\pgfusepath{stroke}%
\end{pgfscope}%
\begin{pgfscope}%
\pgfsetbuttcap%
\pgfsetroundjoin%
\definecolor{currentfill}{rgb}{0.000000,0.000000,0.000000}%
\pgfsetfillcolor{currentfill}%
\pgfsetlinewidth{0.803000pt}%
\definecolor{currentstroke}{rgb}{0.000000,0.000000,0.000000}%
\pgfsetstrokecolor{currentstroke}%
\pgfsetdash{}{0pt}%
\pgfsys@defobject{currentmarker}{\pgfqpoint{0.000000in}{-0.048611in}}{\pgfqpoint{0.000000in}{0.000000in}}{%
\pgfpathmoveto{\pgfqpoint{0.000000in}{0.000000in}}%
\pgfpathlineto{\pgfqpoint{0.000000in}{-0.048611in}}%
\pgfusepath{stroke,fill}%
}%
\begin{pgfscope}%
\pgfsys@transformshift{5.396396in}{0.929220in}%
\pgfsys@useobject{currentmarker}{}%
\end{pgfscope}%
\end{pgfscope}%
\begin{pgfscope}%
\definecolor{textcolor}{rgb}{0.000000,0.000000,0.000000}%
\pgfsetstrokecolor{textcolor}%
\pgfsetfillcolor{textcolor}%
\pgftext[x=5.396396in,y=0.831998in,,top]{\color{textcolor}\rmfamily\fontsize{28.000000}{33.600000}\selectfont 30}%
\end{pgfscope}%
\begin{pgfscope}%
\pgfpathrectangle{\pgfqpoint{0.866662in}{0.929220in}}{\pgfqpoint{12.339620in}{5.670780in}}%
\pgfusepath{clip}%
\pgfsetrectcap%
\pgfsetroundjoin%
\pgfsetlinewidth{0.803000pt}%
\definecolor{currentstroke}{rgb}{0.690196,0.690196,0.690196}%
\pgfsetstrokecolor{currentstroke}%
\pgfsetdash{}{0pt}%
\pgfpathmoveto{\pgfqpoint{6.958373in}{0.929220in}}%
\pgfpathlineto{\pgfqpoint{6.958373in}{6.600000in}}%
\pgfusepath{stroke}%
\end{pgfscope}%
\begin{pgfscope}%
\pgfsetbuttcap%
\pgfsetroundjoin%
\definecolor{currentfill}{rgb}{0.000000,0.000000,0.000000}%
\pgfsetfillcolor{currentfill}%
\pgfsetlinewidth{0.803000pt}%
\definecolor{currentstroke}{rgb}{0.000000,0.000000,0.000000}%
\pgfsetstrokecolor{currentstroke}%
\pgfsetdash{}{0pt}%
\pgfsys@defobject{currentmarker}{\pgfqpoint{0.000000in}{-0.048611in}}{\pgfqpoint{0.000000in}{0.000000in}}{%
\pgfpathmoveto{\pgfqpoint{0.000000in}{0.000000in}}%
\pgfpathlineto{\pgfqpoint{0.000000in}{-0.048611in}}%
\pgfusepath{stroke,fill}%
}%
\begin{pgfscope}%
\pgfsys@transformshift{6.958373in}{0.929220in}%
\pgfsys@useobject{currentmarker}{}%
\end{pgfscope}%
\end{pgfscope}%
\begin{pgfscope}%
\definecolor{textcolor}{rgb}{0.000000,0.000000,0.000000}%
\pgfsetstrokecolor{textcolor}%
\pgfsetfillcolor{textcolor}%
\pgftext[x=6.958373in,y=0.831998in,,top]{\color{textcolor}\rmfamily\fontsize{28.000000}{33.600000}\selectfont 40}%
\end{pgfscope}%
\begin{pgfscope}%
\pgfpathrectangle{\pgfqpoint{0.866662in}{0.929220in}}{\pgfqpoint{12.339620in}{5.670780in}}%
\pgfusepath{clip}%
\pgfsetrectcap%
\pgfsetroundjoin%
\pgfsetlinewidth{0.803000pt}%
\definecolor{currentstroke}{rgb}{0.690196,0.690196,0.690196}%
\pgfsetstrokecolor{currentstroke}%
\pgfsetdash{}{0pt}%
\pgfpathmoveto{\pgfqpoint{8.520350in}{0.929220in}}%
\pgfpathlineto{\pgfqpoint{8.520350in}{6.600000in}}%
\pgfusepath{stroke}%
\end{pgfscope}%
\begin{pgfscope}%
\pgfsetbuttcap%
\pgfsetroundjoin%
\definecolor{currentfill}{rgb}{0.000000,0.000000,0.000000}%
\pgfsetfillcolor{currentfill}%
\pgfsetlinewidth{0.803000pt}%
\definecolor{currentstroke}{rgb}{0.000000,0.000000,0.000000}%
\pgfsetstrokecolor{currentstroke}%
\pgfsetdash{}{0pt}%
\pgfsys@defobject{currentmarker}{\pgfqpoint{0.000000in}{-0.048611in}}{\pgfqpoint{0.000000in}{0.000000in}}{%
\pgfpathmoveto{\pgfqpoint{0.000000in}{0.000000in}}%
\pgfpathlineto{\pgfqpoint{0.000000in}{-0.048611in}}%
\pgfusepath{stroke,fill}%
}%
\begin{pgfscope}%
\pgfsys@transformshift{8.520350in}{0.929220in}%
\pgfsys@useobject{currentmarker}{}%
\end{pgfscope}%
\end{pgfscope}%
\begin{pgfscope}%
\definecolor{textcolor}{rgb}{0.000000,0.000000,0.000000}%
\pgfsetstrokecolor{textcolor}%
\pgfsetfillcolor{textcolor}%
\pgftext[x=8.520350in,y=0.831998in,,top]{\color{textcolor}\rmfamily\fontsize{28.000000}{33.600000}\selectfont 50}%
\end{pgfscope}%
\begin{pgfscope}%
\pgfpathrectangle{\pgfqpoint{0.866662in}{0.929220in}}{\pgfqpoint{12.339620in}{5.670780in}}%
\pgfusepath{clip}%
\pgfsetrectcap%
\pgfsetroundjoin%
\pgfsetlinewidth{0.803000pt}%
\definecolor{currentstroke}{rgb}{0.690196,0.690196,0.690196}%
\pgfsetstrokecolor{currentstroke}%
\pgfsetdash{}{0pt}%
\pgfpathmoveto{\pgfqpoint{10.082327in}{0.929220in}}%
\pgfpathlineto{\pgfqpoint{10.082327in}{6.600000in}}%
\pgfusepath{stroke}%
\end{pgfscope}%
\begin{pgfscope}%
\pgfsetbuttcap%
\pgfsetroundjoin%
\definecolor{currentfill}{rgb}{0.000000,0.000000,0.000000}%
\pgfsetfillcolor{currentfill}%
\pgfsetlinewidth{0.803000pt}%
\definecolor{currentstroke}{rgb}{0.000000,0.000000,0.000000}%
\pgfsetstrokecolor{currentstroke}%
\pgfsetdash{}{0pt}%
\pgfsys@defobject{currentmarker}{\pgfqpoint{0.000000in}{-0.048611in}}{\pgfqpoint{0.000000in}{0.000000in}}{%
\pgfpathmoveto{\pgfqpoint{0.000000in}{0.000000in}}%
\pgfpathlineto{\pgfqpoint{0.000000in}{-0.048611in}}%
\pgfusepath{stroke,fill}%
}%
\begin{pgfscope}%
\pgfsys@transformshift{10.082327in}{0.929220in}%
\pgfsys@useobject{currentmarker}{}%
\end{pgfscope}%
\end{pgfscope}%
\begin{pgfscope}%
\definecolor{textcolor}{rgb}{0.000000,0.000000,0.000000}%
\pgfsetstrokecolor{textcolor}%
\pgfsetfillcolor{textcolor}%
\pgftext[x=10.082327in,y=0.831998in,,top]{\color{textcolor}\rmfamily\fontsize{28.000000}{33.600000}\selectfont 60}%
\end{pgfscope}%
\begin{pgfscope}%
\pgfpathrectangle{\pgfqpoint{0.866662in}{0.929220in}}{\pgfqpoint{12.339620in}{5.670780in}}%
\pgfusepath{clip}%
\pgfsetrectcap%
\pgfsetroundjoin%
\pgfsetlinewidth{0.803000pt}%
\definecolor{currentstroke}{rgb}{0.690196,0.690196,0.690196}%
\pgfsetstrokecolor{currentstroke}%
\pgfsetdash{}{0pt}%
\pgfpathmoveto{\pgfqpoint{11.644305in}{0.929220in}}%
\pgfpathlineto{\pgfqpoint{11.644305in}{6.600000in}}%
\pgfusepath{stroke}%
\end{pgfscope}%
\begin{pgfscope}%
\pgfsetbuttcap%
\pgfsetroundjoin%
\definecolor{currentfill}{rgb}{0.000000,0.000000,0.000000}%
\pgfsetfillcolor{currentfill}%
\pgfsetlinewidth{0.803000pt}%
\definecolor{currentstroke}{rgb}{0.000000,0.000000,0.000000}%
\pgfsetstrokecolor{currentstroke}%
\pgfsetdash{}{0pt}%
\pgfsys@defobject{currentmarker}{\pgfqpoint{0.000000in}{-0.048611in}}{\pgfqpoint{0.000000in}{0.000000in}}{%
\pgfpathmoveto{\pgfqpoint{0.000000in}{0.000000in}}%
\pgfpathlineto{\pgfqpoint{0.000000in}{-0.048611in}}%
\pgfusepath{stroke,fill}%
}%
\begin{pgfscope}%
\pgfsys@transformshift{11.644305in}{0.929220in}%
\pgfsys@useobject{currentmarker}{}%
\end{pgfscope}%
\end{pgfscope}%
\begin{pgfscope}%
\definecolor{textcolor}{rgb}{0.000000,0.000000,0.000000}%
\pgfsetstrokecolor{textcolor}%
\pgfsetfillcolor{textcolor}%
\pgftext[x=11.644305in,y=0.831998in,,top]{\color{textcolor}\rmfamily\fontsize{28.000000}{33.600000}\selectfont 70}%
\end{pgfscope}%
\begin{pgfscope}%
\pgfpathrectangle{\pgfqpoint{0.866662in}{0.929220in}}{\pgfqpoint{12.339620in}{5.670780in}}%
\pgfusepath{clip}%
\pgfsetrectcap%
\pgfsetroundjoin%
\pgfsetlinewidth{0.803000pt}%
\definecolor{currentstroke}{rgb}{0.690196,0.690196,0.690196}%
\pgfsetstrokecolor{currentstroke}%
\pgfsetdash{}{0pt}%
\pgfpathmoveto{\pgfqpoint{13.206282in}{0.929220in}}%
\pgfpathlineto{\pgfqpoint{13.206282in}{6.600000in}}%
\pgfusepath{stroke}%
\end{pgfscope}%
\begin{pgfscope}%
\pgfsetbuttcap%
\pgfsetroundjoin%
\definecolor{currentfill}{rgb}{0.000000,0.000000,0.000000}%
\pgfsetfillcolor{currentfill}%
\pgfsetlinewidth{0.803000pt}%
\definecolor{currentstroke}{rgb}{0.000000,0.000000,0.000000}%
\pgfsetstrokecolor{currentstroke}%
\pgfsetdash{}{0pt}%
\pgfsys@defobject{currentmarker}{\pgfqpoint{0.000000in}{-0.048611in}}{\pgfqpoint{0.000000in}{0.000000in}}{%
\pgfpathmoveto{\pgfqpoint{0.000000in}{0.000000in}}%
\pgfpathlineto{\pgfqpoint{0.000000in}{-0.048611in}}%
\pgfusepath{stroke,fill}%
}%
\begin{pgfscope}%
\pgfsys@transformshift{13.206282in}{0.929220in}%
\pgfsys@useobject{currentmarker}{}%
\end{pgfscope}%
\end{pgfscope}%
\begin{pgfscope}%
\definecolor{textcolor}{rgb}{0.000000,0.000000,0.000000}%
\pgfsetstrokecolor{textcolor}%
\pgfsetfillcolor{textcolor}%
\pgftext[x=13.206282in,y=0.831998in,,top]{\color{textcolor}\rmfamily\fontsize{28.000000}{33.600000}\selectfont 80}%
\end{pgfscope}%
\begin{pgfscope}%
\definecolor{textcolor}{rgb}{0.000000,0.000000,0.000000}%
\pgfsetstrokecolor{textcolor}%
\pgfsetfillcolor{textcolor}%
\pgftext[x=7.036472in,y=0.385926in,,top]{\color{textcolor}\rmfamily\fontsize{28.000000}{33.600000}\selectfont time step \(\displaystyle k\)}%
\end{pgfscope}%
\begin{pgfscope}%
\pgfpathrectangle{\pgfqpoint{0.866662in}{0.929220in}}{\pgfqpoint{12.339620in}{5.670780in}}%
\pgfusepath{clip}%
\pgfsetrectcap%
\pgfsetroundjoin%
\pgfsetlinewidth{0.803000pt}%
\definecolor{currentstroke}{rgb}{0.690196,0.690196,0.690196}%
\pgfsetstrokecolor{currentstroke}%
\pgfsetdash{}{0pt}%
\pgfpathmoveto{\pgfqpoint{0.866662in}{0.929220in}}%
\pgfpathlineto{\pgfqpoint{13.206282in}{0.929220in}}%
\pgfusepath{stroke}%
\end{pgfscope}%
\begin{pgfscope}%
\pgfsetbuttcap%
\pgfsetroundjoin%
\definecolor{currentfill}{rgb}{0.000000,0.000000,0.000000}%
\pgfsetfillcolor{currentfill}%
\pgfsetlinewidth{0.803000pt}%
\definecolor{currentstroke}{rgb}{0.000000,0.000000,0.000000}%
\pgfsetstrokecolor{currentstroke}%
\pgfsetdash{}{0pt}%
\pgfsys@defobject{currentmarker}{\pgfqpoint{-0.048611in}{0.000000in}}{\pgfqpoint{-0.000000in}{0.000000in}}{%
\pgfpathmoveto{\pgfqpoint{-0.000000in}{0.000000in}}%
\pgfpathlineto{\pgfqpoint{-0.048611in}{0.000000in}}%
\pgfusepath{stroke,fill}%
}%
\begin{pgfscope}%
\pgfsys@transformshift{0.866662in}{0.929220in}%
\pgfsys@useobject{currentmarker}{}%
\end{pgfscope}%
\end{pgfscope}%
\begin{pgfscope}%
\definecolor{textcolor}{rgb}{0.000000,0.000000,0.000000}%
\pgfsetstrokecolor{textcolor}%
\pgfsetfillcolor{textcolor}%
\pgftext[x=0.520683in, y=0.809236in, left, base]{\color{textcolor}\rmfamily\fontsize{28.000000}{33.600000}\selectfont 0}%
\end{pgfscope}%
\begin{pgfscope}%
\pgfpathrectangle{\pgfqpoint{0.866662in}{0.929220in}}{\pgfqpoint{12.339620in}{5.670780in}}%
\pgfusepath{clip}%
\pgfsetrectcap%
\pgfsetroundjoin%
\pgfsetlinewidth{0.803000pt}%
\definecolor{currentstroke}{rgb}{0.690196,0.690196,0.690196}%
\pgfsetstrokecolor{currentstroke}%
\pgfsetdash{}{0pt}%
\pgfpathmoveto{\pgfqpoint{0.866662in}{2.218034in}}%
\pgfpathlineto{\pgfqpoint{13.206282in}{2.218034in}}%
\pgfusepath{stroke}%
\end{pgfscope}%
\begin{pgfscope}%
\pgfsetbuttcap%
\pgfsetroundjoin%
\definecolor{currentfill}{rgb}{0.000000,0.000000,0.000000}%
\pgfsetfillcolor{currentfill}%
\pgfsetlinewidth{0.803000pt}%
\definecolor{currentstroke}{rgb}{0.000000,0.000000,0.000000}%
\pgfsetstrokecolor{currentstroke}%
\pgfsetdash{}{0pt}%
\pgfsys@defobject{currentmarker}{\pgfqpoint{-0.048611in}{0.000000in}}{\pgfqpoint{-0.000000in}{0.000000in}}{%
\pgfpathmoveto{\pgfqpoint{-0.000000in}{0.000000in}}%
\pgfpathlineto{\pgfqpoint{-0.048611in}{0.000000in}}%
\pgfusepath{stroke,fill}%
}%
\begin{pgfscope}%
\pgfsys@transformshift{0.866662in}{2.218034in}%
\pgfsys@useobject{currentmarker}{}%
\end{pgfscope}%
\end{pgfscope}%
\begin{pgfscope}%
\definecolor{textcolor}{rgb}{0.000000,0.000000,0.000000}%
\pgfsetstrokecolor{textcolor}%
\pgfsetfillcolor{textcolor}%
\pgftext[x=0.520683in, y=2.098049in, left, base]{\color{textcolor}\rmfamily\fontsize{28.000000}{33.600000}\selectfont 2}%
\end{pgfscope}%
\begin{pgfscope}%
\pgfpathrectangle{\pgfqpoint{0.866662in}{0.929220in}}{\pgfqpoint{12.339620in}{5.670780in}}%
\pgfusepath{clip}%
\pgfsetrectcap%
\pgfsetroundjoin%
\pgfsetlinewidth{0.803000pt}%
\definecolor{currentstroke}{rgb}{0.690196,0.690196,0.690196}%
\pgfsetstrokecolor{currentstroke}%
\pgfsetdash{}{0pt}%
\pgfpathmoveto{\pgfqpoint{0.866662in}{3.506847in}}%
\pgfpathlineto{\pgfqpoint{13.206282in}{3.506847in}}%
\pgfusepath{stroke}%
\end{pgfscope}%
\begin{pgfscope}%
\pgfsetbuttcap%
\pgfsetroundjoin%
\definecolor{currentfill}{rgb}{0.000000,0.000000,0.000000}%
\pgfsetfillcolor{currentfill}%
\pgfsetlinewidth{0.803000pt}%
\definecolor{currentstroke}{rgb}{0.000000,0.000000,0.000000}%
\pgfsetstrokecolor{currentstroke}%
\pgfsetdash{}{0pt}%
\pgfsys@defobject{currentmarker}{\pgfqpoint{-0.048611in}{0.000000in}}{\pgfqpoint{-0.000000in}{0.000000in}}{%
\pgfpathmoveto{\pgfqpoint{-0.000000in}{0.000000in}}%
\pgfpathlineto{\pgfqpoint{-0.048611in}{0.000000in}}%
\pgfusepath{stroke,fill}%
}%
\begin{pgfscope}%
\pgfsys@transformshift{0.866662in}{3.506847in}%
\pgfsys@useobject{currentmarker}{}%
\end{pgfscope}%
\end{pgfscope}%
\begin{pgfscope}%
\definecolor{textcolor}{rgb}{0.000000,0.000000,0.000000}%
\pgfsetstrokecolor{textcolor}%
\pgfsetfillcolor{textcolor}%
\pgftext[x=0.520683in, y=3.386863in, left, base]{\color{textcolor}\rmfamily\fontsize{28.000000}{33.600000}\selectfont 4}%
\end{pgfscope}%
\begin{pgfscope}%
\pgfpathrectangle{\pgfqpoint{0.866662in}{0.929220in}}{\pgfqpoint{12.339620in}{5.670780in}}%
\pgfusepath{clip}%
\pgfsetrectcap%
\pgfsetroundjoin%
\pgfsetlinewidth{0.803000pt}%
\definecolor{currentstroke}{rgb}{0.690196,0.690196,0.690196}%
\pgfsetstrokecolor{currentstroke}%
\pgfsetdash{}{0pt}%
\pgfpathmoveto{\pgfqpoint{0.866662in}{4.795661in}}%
\pgfpathlineto{\pgfqpoint{13.206282in}{4.795661in}}%
\pgfusepath{stroke}%
\end{pgfscope}%
\begin{pgfscope}%
\pgfsetbuttcap%
\pgfsetroundjoin%
\definecolor{currentfill}{rgb}{0.000000,0.000000,0.000000}%
\pgfsetfillcolor{currentfill}%
\pgfsetlinewidth{0.803000pt}%
\definecolor{currentstroke}{rgb}{0.000000,0.000000,0.000000}%
\pgfsetstrokecolor{currentstroke}%
\pgfsetdash{}{0pt}%
\pgfsys@defobject{currentmarker}{\pgfqpoint{-0.048611in}{0.000000in}}{\pgfqpoint{-0.000000in}{0.000000in}}{%
\pgfpathmoveto{\pgfqpoint{-0.000000in}{0.000000in}}%
\pgfpathlineto{\pgfqpoint{-0.048611in}{0.000000in}}%
\pgfusepath{stroke,fill}%
}%
\begin{pgfscope}%
\pgfsys@transformshift{0.866662in}{4.795661in}%
\pgfsys@useobject{currentmarker}{}%
\end{pgfscope}%
\end{pgfscope}%
\begin{pgfscope}%
\definecolor{textcolor}{rgb}{0.000000,0.000000,0.000000}%
\pgfsetstrokecolor{textcolor}%
\pgfsetfillcolor{textcolor}%
\pgftext[x=0.520683in, y=4.675676in, left, base]{\color{textcolor}\rmfamily\fontsize{28.000000}{33.600000}\selectfont 6}%
\end{pgfscope}%
\begin{pgfscope}%
\pgfpathrectangle{\pgfqpoint{0.866662in}{0.929220in}}{\pgfqpoint{12.339620in}{5.670780in}}%
\pgfusepath{clip}%
\pgfsetrectcap%
\pgfsetroundjoin%
\pgfsetlinewidth{0.803000pt}%
\definecolor{currentstroke}{rgb}{0.690196,0.690196,0.690196}%
\pgfsetstrokecolor{currentstroke}%
\pgfsetdash{}{0pt}%
\pgfpathmoveto{\pgfqpoint{0.866662in}{6.084475in}}%
\pgfpathlineto{\pgfqpoint{13.206282in}{6.084475in}}%
\pgfusepath{stroke}%
\end{pgfscope}%
\begin{pgfscope}%
\pgfsetbuttcap%
\pgfsetroundjoin%
\definecolor{currentfill}{rgb}{0.000000,0.000000,0.000000}%
\pgfsetfillcolor{currentfill}%
\pgfsetlinewidth{0.803000pt}%
\definecolor{currentstroke}{rgb}{0.000000,0.000000,0.000000}%
\pgfsetstrokecolor{currentstroke}%
\pgfsetdash{}{0pt}%
\pgfsys@defobject{currentmarker}{\pgfqpoint{-0.048611in}{0.000000in}}{\pgfqpoint{-0.000000in}{0.000000in}}{%
\pgfpathmoveto{\pgfqpoint{-0.000000in}{0.000000in}}%
\pgfpathlineto{\pgfqpoint{-0.048611in}{0.000000in}}%
\pgfusepath{stroke,fill}%
}%
\begin{pgfscope}%
\pgfsys@transformshift{0.866662in}{6.084475in}%
\pgfsys@useobject{currentmarker}{}%
\end{pgfscope}%
\end{pgfscope}%
\begin{pgfscope}%
\definecolor{textcolor}{rgb}{0.000000,0.000000,0.000000}%
\pgfsetstrokecolor{textcolor}%
\pgfsetfillcolor{textcolor}%
\pgftext[x=0.520683in, y=5.964490in, left, base]{\color{textcolor}\rmfamily\fontsize{28.000000}{33.600000}\selectfont 8}%
\end{pgfscope}%
\begin{pgfscope}%
\definecolor{textcolor}{rgb}{0.000000,0.000000,0.000000}%
\pgfsetstrokecolor{textcolor}%
\pgfsetfillcolor{textcolor}%
\pgftext[x=0.381795in,y=3.764610in,,bottom,rotate=90.000000]{\color{textcolor}\rmfamily\fontsize{28.000000}{33.600000}\selectfont GOSPA [m]}%
\end{pgfscope}%
\begin{pgfscope}%
\pgfpathrectangle{\pgfqpoint{0.866662in}{0.929220in}}{\pgfqpoint{12.339620in}{5.670780in}}%
\pgfusepath{clip}%
\pgfsetrectcap%
\pgfsetroundjoin%
\pgfsetlinewidth{1.505625pt}%
\definecolor{currentstroke}{rgb}{0.750000,0.750000,0.000000}%
\pgfsetstrokecolor{currentstroke}%
\pgfsetdash{}{0pt}%
\pgfpathmoveto{\pgfqpoint{0.866662in}{0.929220in}}%
\pgfpathlineto{\pgfqpoint{1.022859in}{0.929220in}}%
\pgfpathlineto{\pgfqpoint{1.179057in}{0.929220in}}%
\pgfpathlineto{\pgfqpoint{1.335255in}{0.929220in}}%
\pgfpathlineto{\pgfqpoint{1.491452in}{0.929220in}}%
\pgfpathlineto{\pgfqpoint{1.647650in}{0.929220in}}%
\pgfpathlineto{\pgfqpoint{1.803848in}{0.929220in}}%
\pgfpathlineto{\pgfqpoint{1.960046in}{0.929220in}}%
\pgfpathlineto{\pgfqpoint{2.116243in}{0.929220in}}%
\pgfpathlineto{\pgfqpoint{2.272441in}{1.729561in}}%
\pgfpathlineto{\pgfqpoint{2.428639in}{1.688445in}}%
\pgfpathlineto{\pgfqpoint{2.584837in}{1.843438in}}%
\pgfpathlineto{\pgfqpoint{2.741034in}{1.724819in}}%
\pgfpathlineto{\pgfqpoint{2.897232in}{1.830137in}}%
\pgfpathlineto{\pgfqpoint{3.053430in}{2.213328in}}%
\pgfpathlineto{\pgfqpoint{3.209627in}{2.364035in}}%
\pgfpathlineto{\pgfqpoint{3.365825in}{2.410691in}}%
\pgfpathlineto{\pgfqpoint{3.522023in}{2.412619in}}%
\pgfpathlineto{\pgfqpoint{3.678221in}{2.353238in}}%
\pgfpathlineto{\pgfqpoint{3.834418in}{2.874537in}}%
\pgfpathlineto{\pgfqpoint{3.990616in}{2.980577in}}%
\pgfpathlineto{\pgfqpoint{4.146814in}{2.995078in}}%
\pgfpathlineto{\pgfqpoint{4.303012in}{2.990421in}}%
\pgfpathlineto{\pgfqpoint{4.459209in}{3.010530in}}%
\pgfpathlineto{\pgfqpoint{4.615407in}{3.462190in}}%
\pgfpathlineto{\pgfqpoint{4.771605in}{3.525886in}}%
\pgfpathlineto{\pgfqpoint{4.927802in}{3.622234in}}%
\pgfpathlineto{\pgfqpoint{5.084000in}{3.687229in}}%
\pgfpathlineto{\pgfqpoint{5.240198in}{3.675862in}}%
\pgfpathlineto{\pgfqpoint{5.396396in}{3.696294in}}%
\pgfpathlineto{\pgfqpoint{5.552593in}{3.654033in}}%
\pgfpathlineto{\pgfqpoint{5.708791in}{3.525505in}}%
\pgfpathlineto{\pgfqpoint{5.864989in}{3.598618in}}%
\pgfpathlineto{\pgfqpoint{6.021187in}{3.591755in}}%
\pgfpathlineto{\pgfqpoint{6.177384in}{3.607842in}}%
\pgfpathlineto{\pgfqpoint{6.333582in}{3.659277in}}%
\pgfpathlineto{\pgfqpoint{6.489780in}{3.646314in}}%
\pgfpathlineto{\pgfqpoint{6.645977in}{3.601800in}}%
\pgfpathlineto{\pgfqpoint{6.802175in}{3.626002in}}%
\pgfpathlineto{\pgfqpoint{6.958373in}{3.661402in}}%
\pgfpathlineto{\pgfqpoint{7.114571in}{3.685732in}}%
\pgfpathlineto{\pgfqpoint{7.270768in}{3.640097in}}%
\pgfpathlineto{\pgfqpoint{7.426966in}{3.618101in}}%
\pgfpathlineto{\pgfqpoint{7.583164in}{3.653180in}}%
\pgfpathlineto{\pgfqpoint{7.739362in}{3.628276in}}%
\pgfpathlineto{\pgfqpoint{7.895559in}{3.626445in}}%
\pgfpathlineto{\pgfqpoint{8.051757in}{3.665222in}}%
\pgfpathlineto{\pgfqpoint{8.207955in}{3.648038in}}%
\pgfpathlineto{\pgfqpoint{8.364152in}{3.646915in}}%
\pgfpathlineto{\pgfqpoint{8.520350in}{3.629652in}}%
\pgfpathlineto{\pgfqpoint{8.676548in}{3.573124in}}%
\pgfpathlineto{\pgfqpoint{8.832746in}{3.622842in}}%
\pgfpathlineto{\pgfqpoint{8.988943in}{3.614722in}}%
\pgfpathlineto{\pgfqpoint{9.145141in}{3.608718in}}%
\pgfpathlineto{\pgfqpoint{9.301339in}{3.228265in}}%
\pgfpathlineto{\pgfqpoint{9.457537in}{3.130350in}}%
\pgfpathlineto{\pgfqpoint{9.613734in}{3.056600in}}%
\pgfpathlineto{\pgfqpoint{9.769932in}{3.079078in}}%
\pgfpathlineto{\pgfqpoint{9.926130in}{3.067411in}}%
\pgfpathlineto{\pgfqpoint{10.082327in}{2.685342in}}%
\pgfpathlineto{\pgfqpoint{10.238525in}{2.515126in}}%
\pgfpathlineto{\pgfqpoint{10.394723in}{2.504297in}}%
\pgfpathlineto{\pgfqpoint{10.550921in}{2.451660in}}%
\pgfpathlineto{\pgfqpoint{10.707118in}{2.447182in}}%
\pgfpathlineto{\pgfqpoint{10.863316in}{2.065449in}}%
\pgfpathlineto{\pgfqpoint{11.019514in}{1.949723in}}%
\pgfpathlineto{\pgfqpoint{11.175712in}{1.891744in}}%
\pgfpathlineto{\pgfqpoint{11.331909in}{1.842028in}}%
\pgfpathlineto{\pgfqpoint{11.488107in}{1.811496in}}%
\pgfpathlineto{\pgfqpoint{11.644305in}{1.041992in}}%
\pgfpathlineto{\pgfqpoint{11.800502in}{0.935664in}}%
\pgfpathlineto{\pgfqpoint{11.956700in}{0.929220in}}%
\pgfpathlineto{\pgfqpoint{12.112898in}{0.929220in}}%
\pgfpathlineto{\pgfqpoint{12.269096in}{0.929220in}}%
\pgfpathlineto{\pgfqpoint{12.425293in}{0.929220in}}%
\pgfpathlineto{\pgfqpoint{12.581491in}{0.929220in}}%
\pgfpathlineto{\pgfqpoint{12.737689in}{0.929220in}}%
\pgfpathlineto{\pgfqpoint{12.893887in}{0.929220in}}%
\pgfpathlineto{\pgfqpoint{13.050084in}{0.929220in}}%
\pgfpathlineto{\pgfqpoint{13.206282in}{0.929220in}}%
\pgfusepath{stroke}%
\end{pgfscope}%
\begin{pgfscope}%
\pgfpathrectangle{\pgfqpoint{0.866662in}{0.929220in}}{\pgfqpoint{12.339620in}{5.670780in}}%
\pgfusepath{clip}%
\pgfsetrectcap%
\pgfsetroundjoin%
\pgfsetlinewidth{1.505625pt}%
\definecolor{currentstroke}{rgb}{0.000000,0.000000,1.000000}%
\pgfsetstrokecolor{currentstroke}%
\pgfsetdash{}{0pt}%
\pgfpathmoveto{\pgfqpoint{0.866662in}{0.929220in}}%
\pgfpathlineto{\pgfqpoint{1.022859in}{0.938886in}}%
\pgfpathlineto{\pgfqpoint{1.179057in}{0.932442in}}%
\pgfpathlineto{\pgfqpoint{1.335255in}{0.932442in}}%
\pgfpathlineto{\pgfqpoint{1.491452in}{0.935664in}}%
\pgfpathlineto{\pgfqpoint{1.647650in}{0.945331in}}%
\pgfpathlineto{\pgfqpoint{1.803848in}{0.938886in}}%
\pgfpathlineto{\pgfqpoint{1.960046in}{0.938886in}}%
\pgfpathlineto{\pgfqpoint{2.116243in}{0.935664in}}%
\pgfpathlineto{\pgfqpoint{2.272441in}{1.573627in}}%
\pgfpathlineto{\pgfqpoint{2.428639in}{1.572692in}}%
\pgfpathlineto{\pgfqpoint{2.584837in}{1.625901in}}%
\pgfpathlineto{\pgfqpoint{2.741034in}{1.464846in}}%
\pgfpathlineto{\pgfqpoint{2.897232in}{1.396607in}}%
\pgfpathlineto{\pgfqpoint{3.053430in}{1.758906in}}%
\pgfpathlineto{\pgfqpoint{3.209627in}{1.840577in}}%
\pgfpathlineto{\pgfqpoint{3.365825in}{1.805128in}}%
\pgfpathlineto{\pgfqpoint{3.522023in}{1.804933in}}%
\pgfpathlineto{\pgfqpoint{3.678221in}{1.773526in}}%
\pgfpathlineto{\pgfqpoint{3.834418in}{2.147314in}}%
\pgfpathlineto{\pgfqpoint{3.990616in}{2.180138in}}%
\pgfpathlineto{\pgfqpoint{4.146814in}{2.185033in}}%
\pgfpathlineto{\pgfqpoint{4.303012in}{2.225611in}}%
\pgfpathlineto{\pgfqpoint{4.459209in}{2.239207in}}%
\pgfpathlineto{\pgfqpoint{4.615407in}{2.607804in}}%
\pgfpathlineto{\pgfqpoint{4.771605in}{2.636625in}}%
\pgfpathlineto{\pgfqpoint{4.927802in}{2.695777in}}%
\pgfpathlineto{\pgfqpoint{5.084000in}{2.755621in}}%
\pgfpathlineto{\pgfqpoint{5.240198in}{2.767836in}}%
\pgfpathlineto{\pgfqpoint{5.396396in}{2.794249in}}%
\pgfpathlineto{\pgfqpoint{5.552593in}{2.882844in}}%
\pgfpathlineto{\pgfqpoint{5.708791in}{2.909929in}}%
\pgfpathlineto{\pgfqpoint{5.864989in}{2.954429in}}%
\pgfpathlineto{\pgfqpoint{6.021187in}{2.939453in}}%
\pgfpathlineto{\pgfqpoint{6.177384in}{2.904573in}}%
\pgfpathlineto{\pgfqpoint{6.333582in}{2.874380in}}%
\pgfpathlineto{\pgfqpoint{6.489780in}{2.814873in}}%
\pgfpathlineto{\pgfqpoint{6.645977in}{2.755668in}}%
\pgfpathlineto{\pgfqpoint{6.802175in}{2.679869in}}%
\pgfpathlineto{\pgfqpoint{6.958373in}{2.670277in}}%
\pgfpathlineto{\pgfqpoint{7.114571in}{2.641233in}}%
\pgfpathlineto{\pgfqpoint{7.270768in}{2.663472in}}%
\pgfpathlineto{\pgfqpoint{7.426966in}{2.689024in}}%
\pgfpathlineto{\pgfqpoint{7.583164in}{2.735761in}}%
\pgfpathlineto{\pgfqpoint{7.739362in}{2.751330in}}%
\pgfpathlineto{\pgfqpoint{7.895559in}{2.760277in}}%
\pgfpathlineto{\pgfqpoint{8.051757in}{2.726060in}}%
\pgfpathlineto{\pgfqpoint{8.207955in}{2.703215in}}%
\pgfpathlineto{\pgfqpoint{8.364152in}{2.689888in}}%
\pgfpathlineto{\pgfqpoint{8.520350in}{2.676643in}}%
\pgfpathlineto{\pgfqpoint{8.676548in}{2.659584in}}%
\pgfpathlineto{\pgfqpoint{8.832746in}{2.668259in}}%
\pgfpathlineto{\pgfqpoint{8.988943in}{2.650835in}}%
\pgfpathlineto{\pgfqpoint{9.145141in}{2.598475in}}%
\pgfpathlineto{\pgfqpoint{9.301339in}{2.471188in}}%
\pgfpathlineto{\pgfqpoint{9.457537in}{2.431702in}}%
\pgfpathlineto{\pgfqpoint{9.613734in}{2.175926in}}%
\pgfpathlineto{\pgfqpoint{9.769932in}{2.163743in}}%
\pgfpathlineto{\pgfqpoint{9.926130in}{2.106565in}}%
\pgfpathlineto{\pgfqpoint{10.082327in}{2.122585in}}%
\pgfpathlineto{\pgfqpoint{10.238525in}{2.054056in}}%
\pgfpathlineto{\pgfqpoint{10.394723in}{1.730888in}}%
\pgfpathlineto{\pgfqpoint{10.550921in}{1.718186in}}%
\pgfpathlineto{\pgfqpoint{10.707118in}{1.707472in}}%
\pgfpathlineto{\pgfqpoint{10.863316in}{1.751348in}}%
\pgfpathlineto{\pgfqpoint{11.019514in}{1.712462in}}%
\pgfpathlineto{\pgfqpoint{11.175712in}{1.389327in}}%
\pgfpathlineto{\pgfqpoint{11.331909in}{1.361803in}}%
\pgfpathlineto{\pgfqpoint{11.488107in}{1.354459in}}%
\pgfpathlineto{\pgfqpoint{11.644305in}{1.573627in}}%
\pgfpathlineto{\pgfqpoint{11.800502in}{1.563961in}}%
\pgfpathlineto{\pgfqpoint{11.956700in}{0.987217in}}%
\pgfpathlineto{\pgfqpoint{12.112898in}{0.974329in}}%
\pgfpathlineto{\pgfqpoint{12.269096in}{0.961441in}}%
\pgfpathlineto{\pgfqpoint{12.425293in}{0.951775in}}%
\pgfpathlineto{\pgfqpoint{12.581491in}{0.945331in}}%
\pgfpathlineto{\pgfqpoint{12.737689in}{0.945331in}}%
\pgfpathlineto{\pgfqpoint{12.893887in}{0.938886in}}%
\pgfpathlineto{\pgfqpoint{13.050084in}{0.935664in}}%
\pgfpathlineto{\pgfqpoint{13.206282in}{0.932442in}}%
\pgfusepath{stroke}%
\end{pgfscope}%
\begin{pgfscope}%
\pgfpathrectangle{\pgfqpoint{0.866662in}{0.929220in}}{\pgfqpoint{12.339620in}{5.670780in}}%
\pgfusepath{clip}%
\pgfsetrectcap%
\pgfsetroundjoin%
\pgfsetlinewidth{1.505625pt}%
\definecolor{currentstroke}{rgb}{0.000000,0.500000,0.000000}%
\pgfsetstrokecolor{currentstroke}%
\pgfsetdash{}{0pt}%
\pgfpathmoveto{\pgfqpoint{0.866662in}{0.929220in}}%
\pgfpathlineto{\pgfqpoint{1.022859in}{0.932442in}}%
\pgfpathlineto{\pgfqpoint{1.179057in}{0.932442in}}%
\pgfpathlineto{\pgfqpoint{1.335255in}{0.932442in}}%
\pgfpathlineto{\pgfqpoint{1.491452in}{0.942108in}}%
\pgfpathlineto{\pgfqpoint{1.647650in}{0.948553in}}%
\pgfpathlineto{\pgfqpoint{1.803848in}{0.938886in}}%
\pgfpathlineto{\pgfqpoint{1.960046in}{0.938886in}}%
\pgfpathlineto{\pgfqpoint{2.116243in}{0.942108in}}%
\pgfpathlineto{\pgfqpoint{2.272441in}{1.574928in}}%
\pgfpathlineto{\pgfqpoint{2.428639in}{1.573105in}}%
\pgfpathlineto{\pgfqpoint{2.584837in}{1.628520in}}%
\pgfpathlineto{\pgfqpoint{2.741034in}{1.473137in}}%
\pgfpathlineto{\pgfqpoint{2.897232in}{1.515033in}}%
\pgfpathlineto{\pgfqpoint{3.053430in}{1.728458in}}%
\pgfpathlineto{\pgfqpoint{3.209627in}{1.744834in}}%
\pgfpathlineto{\pgfqpoint{3.365825in}{1.625217in}}%
\pgfpathlineto{\pgfqpoint{3.522023in}{1.636499in}}%
\pgfpathlineto{\pgfqpoint{3.678221in}{1.592124in}}%
\pgfpathlineto{\pgfqpoint{3.834418in}{1.930757in}}%
\pgfpathlineto{\pgfqpoint{3.990616in}{1.908021in}}%
\pgfpathlineto{\pgfqpoint{4.146814in}{1.802037in}}%
\pgfpathlineto{\pgfqpoint{4.303012in}{1.781039in}}%
\pgfpathlineto{\pgfqpoint{4.459209in}{1.749459in}}%
\pgfpathlineto{\pgfqpoint{4.615407in}{2.025811in}}%
\pgfpathlineto{\pgfqpoint{4.771605in}{2.023971in}}%
\pgfpathlineto{\pgfqpoint{4.927802in}{1.956293in}}%
\pgfpathlineto{\pgfqpoint{5.084000in}{1.960968in}}%
\pgfpathlineto{\pgfqpoint{5.240198in}{1.953508in}}%
\pgfpathlineto{\pgfqpoint{5.396396in}{1.922254in}}%
\pgfpathlineto{\pgfqpoint{5.552593in}{1.872980in}}%
\pgfpathlineto{\pgfqpoint{5.708791in}{1.859902in}}%
\pgfpathlineto{\pgfqpoint{5.864989in}{1.824944in}}%
\pgfpathlineto{\pgfqpoint{6.021187in}{1.810307in}}%
\pgfpathlineto{\pgfqpoint{6.177384in}{1.821712in}}%
\pgfpathlineto{\pgfqpoint{6.333582in}{1.832470in}}%
\pgfpathlineto{\pgfqpoint{6.489780in}{1.838166in}}%
\pgfpathlineto{\pgfqpoint{6.645977in}{1.850596in}}%
\pgfpathlineto{\pgfqpoint{6.802175in}{1.887009in}}%
\pgfpathlineto{\pgfqpoint{6.958373in}{1.890839in}}%
\pgfpathlineto{\pgfqpoint{7.114571in}{1.907230in}}%
\pgfpathlineto{\pgfqpoint{7.270768in}{1.926744in}}%
\pgfpathlineto{\pgfqpoint{7.426966in}{1.911261in}}%
\pgfpathlineto{\pgfqpoint{7.583164in}{1.914307in}}%
\pgfpathlineto{\pgfqpoint{7.739362in}{1.880209in}}%
\pgfpathlineto{\pgfqpoint{7.895559in}{1.829305in}}%
\pgfpathlineto{\pgfqpoint{8.051757in}{1.814947in}}%
\pgfpathlineto{\pgfqpoint{8.207955in}{1.805452in}}%
\pgfpathlineto{\pgfqpoint{8.364152in}{1.782592in}}%
\pgfpathlineto{\pgfqpoint{8.520350in}{1.774150in}}%
\pgfpathlineto{\pgfqpoint{8.676548in}{1.764929in}}%
\pgfpathlineto{\pgfqpoint{8.832746in}{1.757466in}}%
\pgfpathlineto{\pgfqpoint{8.988943in}{1.763964in}}%
\pgfpathlineto{\pgfqpoint{9.145141in}{1.754850in}}%
\pgfpathlineto{\pgfqpoint{9.301339in}{1.884591in}}%
\pgfpathlineto{\pgfqpoint{9.457537in}{1.858562in}}%
\pgfpathlineto{\pgfqpoint{9.613734in}{1.573327in}}%
\pgfpathlineto{\pgfqpoint{9.769932in}{1.566681in}}%
\pgfpathlineto{\pgfqpoint{9.926130in}{1.570442in}}%
\pgfpathlineto{\pgfqpoint{10.082327in}{1.711974in}}%
\pgfpathlineto{\pgfqpoint{10.238525in}{1.681839in}}%
\pgfpathlineto{\pgfqpoint{10.394723in}{1.373909in}}%
\pgfpathlineto{\pgfqpoint{10.550921in}{1.383879in}}%
\pgfpathlineto{\pgfqpoint{10.707118in}{1.380176in}}%
\pgfpathlineto{\pgfqpoint{10.863316in}{1.511346in}}%
\pgfpathlineto{\pgfqpoint{11.019514in}{1.517245in}}%
\pgfpathlineto{\pgfqpoint{11.175712in}{1.216650in}}%
\pgfpathlineto{\pgfqpoint{11.331909in}{1.204142in}}%
\pgfpathlineto{\pgfqpoint{11.488107in}{1.207695in}}%
\pgfpathlineto{\pgfqpoint{11.644305in}{1.573627in}}%
\pgfpathlineto{\pgfqpoint{11.800502in}{1.576849in}}%
\pgfpathlineto{\pgfqpoint{11.956700in}{0.971107in}}%
\pgfpathlineto{\pgfqpoint{12.112898in}{0.983995in}}%
\pgfpathlineto{\pgfqpoint{12.269096in}{0.971107in}}%
\pgfpathlineto{\pgfqpoint{12.425293in}{0.945331in}}%
\pgfpathlineto{\pgfqpoint{12.581491in}{0.932442in}}%
\pgfpathlineto{\pgfqpoint{12.737689in}{0.945331in}}%
\pgfpathlineto{\pgfqpoint{12.893887in}{0.945331in}}%
\pgfpathlineto{\pgfqpoint{13.050084in}{0.948553in}}%
\pgfpathlineto{\pgfqpoint{13.206282in}{0.945331in}}%
\pgfusepath{stroke}%
\end{pgfscope}%
\begin{pgfscope}%
\pgfpathrectangle{\pgfqpoint{0.866662in}{0.929220in}}{\pgfqpoint{12.339620in}{5.670780in}}%
\pgfusepath{clip}%
\pgfsetrectcap%
\pgfsetroundjoin%
\pgfsetlinewidth{1.505625pt}%
\definecolor{currentstroke}{rgb}{1.000000,0.000000,0.000000}%
\pgfsetstrokecolor{currentstroke}%
\pgfsetdash{}{0pt}%
\pgfpathmoveto{\pgfqpoint{0.866662in}{0.929220in}}%
\pgfpathlineto{\pgfqpoint{1.022859in}{0.929220in}}%
\pgfpathlineto{\pgfqpoint{1.179057in}{0.929220in}}%
\pgfpathlineto{\pgfqpoint{1.335255in}{0.929220in}}%
\pgfpathlineto{\pgfqpoint{1.491452in}{0.929220in}}%
\pgfpathlineto{\pgfqpoint{1.647650in}{0.929220in}}%
\pgfpathlineto{\pgfqpoint{1.803848in}{0.929220in}}%
\pgfpathlineto{\pgfqpoint{1.960046in}{0.929220in}}%
\pgfpathlineto{\pgfqpoint{2.116243in}{0.929220in}}%
\pgfpathlineto{\pgfqpoint{2.272441in}{1.039375in}}%
\pgfpathlineto{\pgfqpoint{2.428639in}{1.071341in}}%
\pgfpathlineto{\pgfqpoint{2.584837in}{1.066997in}}%
\pgfpathlineto{\pgfqpoint{2.741034in}{1.089166in}}%
\pgfpathlineto{\pgfqpoint{2.897232in}{1.092070in}}%
\pgfpathlineto{\pgfqpoint{3.053430in}{1.148632in}}%
\pgfpathlineto{\pgfqpoint{3.209627in}{1.218901in}}%
\pgfpathlineto{\pgfqpoint{3.365825in}{1.154972in}}%
\pgfpathlineto{\pgfqpoint{3.522023in}{1.140838in}}%
\pgfpathlineto{\pgfqpoint{3.678221in}{1.113620in}}%
\pgfpathlineto{\pgfqpoint{3.834418in}{1.331646in}}%
\pgfpathlineto{\pgfqpoint{3.990616in}{1.303314in}}%
\pgfpathlineto{\pgfqpoint{4.146814in}{1.257236in}}%
\pgfpathlineto{\pgfqpoint{4.303012in}{1.250356in}}%
\pgfpathlineto{\pgfqpoint{4.459209in}{1.252813in}}%
\pgfpathlineto{\pgfqpoint{4.615407in}{1.596534in}}%
\pgfpathlineto{\pgfqpoint{4.771605in}{1.544948in}}%
\pgfpathlineto{\pgfqpoint{4.927802in}{1.511055in}}%
\pgfpathlineto{\pgfqpoint{5.084000in}{1.558674in}}%
\pgfpathlineto{\pgfqpoint{5.240198in}{1.582428in}}%
\pgfpathlineto{\pgfqpoint{5.396396in}{1.612550in}}%
\pgfpathlineto{\pgfqpoint{5.552593in}{1.637956in}}%
\pgfpathlineto{\pgfqpoint{5.708791in}{1.648448in}}%
\pgfpathlineto{\pgfqpoint{5.864989in}{1.619242in}}%
\pgfpathlineto{\pgfqpoint{6.021187in}{1.609659in}}%
\pgfpathlineto{\pgfqpoint{6.177384in}{1.613963in}}%
\pgfpathlineto{\pgfqpoint{6.333582in}{1.625399in}}%
\pgfpathlineto{\pgfqpoint{6.489780in}{1.642195in}}%
\pgfpathlineto{\pgfqpoint{6.645977in}{1.659277in}}%
\pgfpathlineto{\pgfqpoint{6.802175in}{1.660992in}}%
\pgfpathlineto{\pgfqpoint{6.958373in}{1.652333in}}%
\pgfpathlineto{\pgfqpoint{7.114571in}{1.644102in}}%
\pgfpathlineto{\pgfqpoint{7.270768in}{1.630615in}}%
\pgfpathlineto{\pgfqpoint{7.426966in}{1.607048in}}%
\pgfpathlineto{\pgfqpoint{7.583164in}{1.593032in}}%
\pgfpathlineto{\pgfqpoint{7.739362in}{1.565286in}}%
\pgfpathlineto{\pgfqpoint{7.895559in}{1.533341in}}%
\pgfpathlineto{\pgfqpoint{8.051757in}{1.507685in}}%
\pgfpathlineto{\pgfqpoint{8.207955in}{1.476549in}}%
\pgfpathlineto{\pgfqpoint{8.364152in}{1.465511in}}%
\pgfpathlineto{\pgfqpoint{8.520350in}{1.441989in}}%
\pgfpathlineto{\pgfqpoint{8.676548in}{1.424317in}}%
\pgfpathlineto{\pgfqpoint{8.832746in}{1.405254in}}%
\pgfpathlineto{\pgfqpoint{8.988943in}{1.365460in}}%
\pgfpathlineto{\pgfqpoint{9.145141in}{1.326754in}}%
\pgfpathlineto{\pgfqpoint{9.301339in}{1.233072in}}%
\pgfpathlineto{\pgfqpoint{9.457537in}{1.216821in}}%
\pgfpathlineto{\pgfqpoint{9.613734in}{1.205964in}}%
\pgfpathlineto{\pgfqpoint{9.769932in}{1.196925in}}%
\pgfpathlineto{\pgfqpoint{9.926130in}{1.175632in}}%
\pgfpathlineto{\pgfqpoint{10.082327in}{1.111079in}}%
\pgfpathlineto{\pgfqpoint{10.238525in}{1.096032in}}%
\pgfpathlineto{\pgfqpoint{10.394723in}{1.088989in}}%
\pgfpathlineto{\pgfqpoint{10.550921in}{1.084563in}}%
\pgfpathlineto{\pgfqpoint{10.707118in}{1.077391in}}%
\pgfpathlineto{\pgfqpoint{10.863316in}{1.022402in}}%
\pgfpathlineto{\pgfqpoint{11.019514in}{1.014000in}}%
\pgfpathlineto{\pgfqpoint{11.175712in}{1.000672in}}%
\pgfpathlineto{\pgfqpoint{11.331909in}{1.000286in}}%
\pgfpathlineto{\pgfqpoint{11.488107in}{0.996644in}}%
\pgfpathlineto{\pgfqpoint{11.644305in}{0.929220in}}%
\pgfpathlineto{\pgfqpoint{11.800502in}{0.929220in}}%
\pgfpathlineto{\pgfqpoint{11.956700in}{0.929220in}}%
\pgfpathlineto{\pgfqpoint{12.112898in}{0.929220in}}%
\pgfpathlineto{\pgfqpoint{12.269096in}{0.929220in}}%
\pgfpathlineto{\pgfqpoint{12.425293in}{0.929220in}}%
\pgfpathlineto{\pgfqpoint{12.581491in}{0.929220in}}%
\pgfpathlineto{\pgfqpoint{12.737689in}{0.929220in}}%
\pgfpathlineto{\pgfqpoint{12.893887in}{0.929220in}}%
\pgfpathlineto{\pgfqpoint{13.050084in}{0.929220in}}%
\pgfpathlineto{\pgfqpoint{13.206282in}{0.929220in}}%
\pgfusepath{stroke}%
\end{pgfscope}%
\begin{pgfscope}%
\pgfsetrectcap%
\pgfsetmiterjoin%
\pgfsetlinewidth{0.803000pt}%
\definecolor{currentstroke}{rgb}{0.000000,0.000000,0.000000}%
\pgfsetstrokecolor{currentstroke}%
\pgfsetdash{}{0pt}%
\pgfpathmoveto{\pgfqpoint{0.866662in}{0.929220in}}%
\pgfpathlineto{\pgfqpoint{0.866662in}{6.600000in}}%
\pgfusepath{stroke}%
\end{pgfscope}%
\begin{pgfscope}%
\pgfsetrectcap%
\pgfsetmiterjoin%
\pgfsetlinewidth{0.803000pt}%
\definecolor{currentstroke}{rgb}{0.000000,0.000000,0.000000}%
\pgfsetstrokecolor{currentstroke}%
\pgfsetdash{}{0pt}%
\pgfpathmoveto{\pgfqpoint{13.206282in}{0.929220in}}%
\pgfpathlineto{\pgfqpoint{13.206282in}{6.600000in}}%
\pgfusepath{stroke}%
\end{pgfscope}%
\begin{pgfscope}%
\pgfsetrectcap%
\pgfsetmiterjoin%
\pgfsetlinewidth{0.803000pt}%
\definecolor{currentstroke}{rgb}{0.000000,0.000000,0.000000}%
\pgfsetstrokecolor{currentstroke}%
\pgfsetdash{}{0pt}%
\pgfpathmoveto{\pgfqpoint{0.866662in}{0.929220in}}%
\pgfpathlineto{\pgfqpoint{13.206282in}{0.929220in}}%
\pgfusepath{stroke}%
\end{pgfscope}%
\begin{pgfscope}%
\pgfsetrectcap%
\pgfsetmiterjoin%
\pgfsetlinewidth{0.803000pt}%
\definecolor{currentstroke}{rgb}{0.000000,0.000000,0.000000}%
\pgfsetstrokecolor{currentstroke}%
\pgfsetdash{}{0pt}%
\pgfpathmoveto{\pgfqpoint{0.866662in}{6.600000in}}%
\pgfpathlineto{\pgfqpoint{13.206282in}{6.600000in}}%
\pgfusepath{stroke}%
\end{pgfscope}%
\begin{pgfscope}%
\pgfsetbuttcap%
\pgfsetmiterjoin%
\definecolor{currentfill}{rgb}{1.000000,1.000000,1.000000}%
\pgfsetfillcolor{currentfill}%
\pgfsetfillopacity{0.800000}%
\pgfsetlinewidth{1.003750pt}%
\definecolor{currentstroke}{rgb}{0.800000,0.800000,0.800000}%
\pgfsetstrokecolor{currentstroke}%
\pgfsetstrokeopacity{0.800000}%
\pgfsetdash{}{0pt}%
\pgfpathmoveto{\pgfqpoint{8.160436in}{4.134482in}}%
\pgfpathlineto{\pgfqpoint{12.934060in}{4.134482in}}%
\pgfpathquadraticcurveto{\pgfqpoint{13.011838in}{4.134482in}}{\pgfqpoint{13.011838in}{4.212260in}}%
\pgfpathlineto{\pgfqpoint{13.011838in}{6.327778in}}%
\pgfpathquadraticcurveto{\pgfqpoint{13.011838in}{6.405556in}}{\pgfqpoint{12.934060in}{6.405556in}}%
\pgfpathlineto{\pgfqpoint{8.160436in}{6.405556in}}%
\pgfpathquadraticcurveto{\pgfqpoint{8.082658in}{6.405556in}}{\pgfqpoint{8.082658in}{6.327778in}}%
\pgfpathlineto{\pgfqpoint{8.082658in}{4.212260in}}%
\pgfpathquadraticcurveto{\pgfqpoint{8.082658in}{4.134482in}}{\pgfqpoint{8.160436in}{4.134482in}}%
\pgfpathlineto{\pgfqpoint{8.160436in}{4.134482in}}%
\pgfpathclose%
\pgfusepath{stroke,fill}%
\end{pgfscope}%
\begin{pgfscope}%
\pgfsetrectcap%
\pgfsetroundjoin%
\pgfsetlinewidth{1.505625pt}%
\definecolor{currentstroke}{rgb}{0.750000,0.750000,0.000000}%
\pgfsetstrokecolor{currentstroke}%
\pgfsetdash{}{0pt}%
\pgfpathmoveto{\pgfqpoint{8.238214in}{6.113889in}}%
\pgfpathlineto{\pgfqpoint{8.627103in}{6.113889in}}%
\pgfpathlineto{\pgfqpoint{9.015992in}{6.113889in}}%
\pgfusepath{stroke}%
\end{pgfscope}%
\begin{pgfscope}%
\definecolor{textcolor}{rgb}{0.000000,0.000000,0.000000}%
\pgfsetstrokecolor{textcolor}%
\pgfsetfillcolor{textcolor}%
\pgftext[x=9.327103in,y=5.977778in,left,base]{\color{textcolor}\rmfamily\fontsize{28.000000}{33.600000}\selectfont LMB-TkBD}%
\end{pgfscope}%
\begin{pgfscope}%
\pgfsetrectcap%
\pgfsetroundjoin%
\pgfsetlinewidth{1.505625pt}%
\definecolor{currentstroke}{rgb}{0.000000,0.000000,1.000000}%
\pgfsetstrokecolor{currentstroke}%
\pgfsetdash{}{0pt}%
\pgfpathmoveto{\pgfqpoint{8.238214in}{5.580009in}}%
\pgfpathlineto{\pgfqpoint{8.627103in}{5.580009in}}%
\pgfpathlineto{\pgfqpoint{9.015992in}{5.580009in}}%
\pgfusepath{stroke}%
\end{pgfscope}%
\begin{pgfscope}%
\definecolor{textcolor}{rgb}{0.000000,0.000000,0.000000}%
\pgfsetstrokecolor{textcolor}%
\pgfsetfillcolor{textcolor}%
\pgftext[x=9.327103in,y=5.443898in,left,base]{\color{textcolor}\rmfamily\fontsize{28.000000}{33.600000}\selectfont MP + BP-DTT}%
\end{pgfscope}%
\begin{pgfscope}%
\pgfsetrectcap%
\pgfsetroundjoin%
\pgfsetlinewidth{1.505625pt}%
\definecolor{currentstroke}{rgb}{0.000000,0.500000,0.000000}%
\pgfsetstrokecolor{currentstroke}%
\pgfsetdash{}{0pt}%
\pgfpathmoveto{\pgfqpoint{8.238214in}{5.046129in}}%
\pgfpathlineto{\pgfqpoint{8.627103in}{5.046129in}}%
\pgfpathlineto{\pgfqpoint{9.015992in}{5.046129in}}%
\pgfusepath{stroke}%
\end{pgfscope}%
\begin{pgfscope}%
\definecolor{textcolor}{rgb}{0.000000,0.000000,0.000000}%
\pgfsetstrokecolor{textcolor}%
\pgfsetfillcolor{textcolor}%
\pgftext[x=9.327103in,y=4.910017in,left,base]{\color{textcolor}\rmfamily\fontsize{28.000000}{33.600000}\selectfont SBL + BP-DTT}%
\end{pgfscope}%
\begin{pgfscope}%
\pgfsetrectcap%
\pgfsetroundjoin%
\pgfsetlinewidth{1.505625pt}%
\definecolor{currentstroke}{rgb}{1.000000,0.000000,0.000000}%
\pgfsetstrokecolor{currentstroke}%
\pgfsetdash{}{0pt}%
\pgfpathmoveto{\pgfqpoint{8.238214in}{4.512248in}}%
\pgfpathlineto{\pgfqpoint{8.627103in}{4.512248in}}%
\pgfpathlineto{\pgfqpoint{9.015992in}{4.512248in}}%
\pgfusepath{stroke}%
\end{pgfscope}%
\begin{pgfscope}%
\definecolor{textcolor}{rgb}{0.000000,0.000000,0.000000}%
\pgfsetstrokecolor{textcolor}%
\pgfsetfillcolor{textcolor}%
\pgftext[x=9.327103in,y=4.376137in,left,base]{\color{textcolor}\rmfamily\fontsize{28.000000}{33.600000}\selectfont BP-TkBD (proposed)~~~~~}%
\end{pgfscope}%
\end{pgfpicture}%
\makeatother%
\endgroup%

%% file: Figs/gospa_-10dB/gospa.pgf
\begingroup%
\makeatletter%
\begin{pgfpicture}%
\pgfpathrectangle{\pgfpointorigin}{\pgfqpoint{13.400000in}{6.600000in}}%
\pgfusepath{use as bounding box, clip}%
\begin{pgfscope}%
\pgfsetbuttcap%
\pgfsetmiterjoin%
\definecolor{currentfill}{rgb}{1.000000,1.000000,1.000000}%
\pgfsetfillcolor{currentfill}%
\pgfsetlinewidth{0.000000pt}%
\definecolor{currentstroke}{rgb}{1.000000,1.000000,1.000000}%
\pgfsetstrokecolor{currentstroke}%
\pgfsetdash{}{0pt}%
\pgfpathmoveto{\pgfqpoint{0.000000in}{0.000000in}}%
\pgfpathlineto{\pgfqpoint{13.400000in}{0.000000in}}%
\pgfpathlineto{\pgfqpoint{13.400000in}{6.600000in}}%
\pgfpathlineto{\pgfqpoint{0.000000in}{6.600000in}}%
\pgfpathlineto{\pgfqpoint{0.000000in}{0.000000in}}%
\pgfpathclose%
\pgfusepath{fill}%
\end{pgfscope}%
\begin{pgfscope}%
\pgfsetbuttcap%
\pgfsetmiterjoin%
\definecolor{currentfill}{rgb}{1.000000,1.000000,1.000000}%
\pgfsetfillcolor{currentfill}%
\pgfsetlinewidth{0.000000pt}%
\definecolor{currentstroke}{rgb}{0.000000,0.000000,0.000000}%
\pgfsetstrokecolor{currentstroke}%
\pgfsetstrokeopacity{0.000000}%
\pgfsetdash{}{0pt}%
\pgfpathmoveto{\pgfqpoint{0.866662in}{0.929220in}}%
\pgfpathlineto{\pgfqpoint{13.206282in}{0.929220in}}%
\pgfpathlineto{\pgfqpoint{13.206282in}{6.600000in}}%
\pgfpathlineto{\pgfqpoint{0.866662in}{6.600000in}}%
\pgfpathlineto{\pgfqpoint{0.866662in}{0.929220in}}%
\pgfpathclose%
\pgfusepath{fill}%
\end{pgfscope}%
\begin{pgfscope}%
\pgfpathrectangle{\pgfqpoint{0.866662in}{0.929220in}}{\pgfqpoint{12.339620in}{5.670780in}}%
\pgfusepath{clip}%
\pgfsetrectcap%
\pgfsetroundjoin%
\pgfsetlinewidth{0.803000pt}%
\definecolor{currentstroke}{rgb}{0.690196,0.690196,0.690196}%
\pgfsetstrokecolor{currentstroke}%
\pgfsetdash{}{0pt}%
\pgfpathmoveto{\pgfqpoint{2.272441in}{0.929220in}}%
\pgfpathlineto{\pgfqpoint{2.272441in}{6.600000in}}%
\pgfusepath{stroke}%
\end{pgfscope}%
\begin{pgfscope}%
\pgfsetbuttcap%
\pgfsetroundjoin%
\definecolor{currentfill}{rgb}{0.000000,0.000000,0.000000}%
\pgfsetfillcolor{currentfill}%
\pgfsetlinewidth{0.803000pt}%
\definecolor{currentstroke}{rgb}{0.000000,0.000000,0.000000}%
\pgfsetstrokecolor{currentstroke}%
\pgfsetdash{}{0pt}%
\pgfsys@defobject{currentmarker}{\pgfqpoint{0.000000in}{-0.048611in}}{\pgfqpoint{0.000000in}{0.000000in}}{%
\pgfpathmoveto{\pgfqpoint{0.000000in}{0.000000in}}%
\pgfpathlineto{\pgfqpoint{0.000000in}{-0.048611in}}%
\pgfusepath{stroke,fill}%
}%
\begin{pgfscope}%
\pgfsys@transformshift{2.272441in}{0.929220in}%
\pgfsys@useobject{currentmarker}{}%
\end{pgfscope}%
\end{pgfscope}%
\begin{pgfscope}%
\definecolor{textcolor}{rgb}{0.000000,0.000000,0.000000}%
\pgfsetstrokecolor{textcolor}%
\pgfsetfillcolor{textcolor}%
\pgftext[x=2.272441in,y=0.831998in,,top]{\color{textcolor}\rmfamily\fontsize{28.000000}{33.600000}\selectfont 10}%
\end{pgfscope}%
\begin{pgfscope}%
\pgfpathrectangle{\pgfqpoint{0.866662in}{0.929220in}}{\pgfqpoint{12.339620in}{5.670780in}}%
\pgfusepath{clip}%
\pgfsetrectcap%
\pgfsetroundjoin%
\pgfsetlinewidth{0.803000pt}%
\definecolor{currentstroke}{rgb}{0.690196,0.690196,0.690196}%
\pgfsetstrokecolor{currentstroke}%
\pgfsetdash{}{0pt}%
\pgfpathmoveto{\pgfqpoint{3.834418in}{0.929220in}}%
\pgfpathlineto{\pgfqpoint{3.834418in}{6.600000in}}%
\pgfusepath{stroke}%
\end{pgfscope}%
\begin{pgfscope}%
\pgfsetbuttcap%
\pgfsetroundjoin%
\definecolor{currentfill}{rgb}{0.000000,0.000000,0.000000}%
\pgfsetfillcolor{currentfill}%
\pgfsetlinewidth{0.803000pt}%
\definecolor{currentstroke}{rgb}{0.000000,0.000000,0.000000}%
\pgfsetstrokecolor{currentstroke}%
\pgfsetdash{}{0pt}%
\pgfsys@defobject{currentmarker}{\pgfqpoint{0.000000in}{-0.048611in}}{\pgfqpoint{0.000000in}{0.000000in}}{%
\pgfpathmoveto{\pgfqpoint{0.000000in}{0.000000in}}%
\pgfpathlineto{\pgfqpoint{0.000000in}{-0.048611in}}%
\pgfusepath{stroke,fill}%
}%
\begin{pgfscope}%
\pgfsys@transformshift{3.834418in}{0.929220in}%
\pgfsys@useobject{currentmarker}{}%
\end{pgfscope}%
\end{pgfscope}%
\begin{pgfscope}%
\definecolor{textcolor}{rgb}{0.000000,0.000000,0.000000}%
\pgfsetstrokecolor{textcolor}%
\pgfsetfillcolor{textcolor}%
\pgftext[x=3.834418in,y=0.831998in,,top]{\color{textcolor}\rmfamily\fontsize{28.000000}{33.600000}\selectfont 20}%
\end{pgfscope}%
\begin{pgfscope}%
\pgfpathrectangle{\pgfqpoint{0.866662in}{0.929220in}}{\pgfqpoint{12.339620in}{5.670780in}}%
\pgfusepath{clip}%
\pgfsetrectcap%
\pgfsetroundjoin%
\pgfsetlinewidth{0.803000pt}%
\definecolor{currentstroke}{rgb}{0.690196,0.690196,0.690196}%
\pgfsetstrokecolor{currentstroke}%
\pgfsetdash{}{0pt}%
\pgfpathmoveto{\pgfqpoint{5.396396in}{0.929220in}}%
\pgfpathlineto{\pgfqpoint{5.396396in}{6.600000in}}%
\pgfusepath{stroke}%
\end{pgfscope}%
\begin{pgfscope}%
\pgfsetbuttcap%
\pgfsetroundjoin%
\definecolor{currentfill}{rgb}{0.000000,0.000000,0.000000}%
\pgfsetfillcolor{currentfill}%
\pgfsetlinewidth{0.803000pt}%
\definecolor{currentstroke}{rgb}{0.000000,0.000000,0.000000}%
\pgfsetstrokecolor{currentstroke}%
\pgfsetdash{}{0pt}%
\pgfsys@defobject{currentmarker}{\pgfqpoint{0.000000in}{-0.048611in}}{\pgfqpoint{0.000000in}{0.000000in}}{%
\pgfpathmoveto{\pgfqpoint{0.000000in}{0.000000in}}%
\pgfpathlineto{\pgfqpoint{0.000000in}{-0.048611in}}%
\pgfusepath{stroke,fill}%
}%
\begin{pgfscope}%
\pgfsys@transformshift{5.396396in}{0.929220in}%
\pgfsys@useobject{currentmarker}{}%
\end{pgfscope}%
\end{pgfscope}%
\begin{pgfscope}%
\definecolor{textcolor}{rgb}{0.000000,0.000000,0.000000}%
\pgfsetstrokecolor{textcolor}%
\pgfsetfillcolor{textcolor}%
\pgftext[x=5.396396in,y=0.831998in,,top]{\color{textcolor}\rmfamily\fontsize{28.000000}{33.600000}\selectfont 30}%
\end{pgfscope}%
\begin{pgfscope}%
\pgfpathrectangle{\pgfqpoint{0.866662in}{0.929220in}}{\pgfqpoint{12.339620in}{5.670780in}}%
\pgfusepath{clip}%
\pgfsetrectcap%
\pgfsetroundjoin%
\pgfsetlinewidth{0.803000pt}%
\definecolor{currentstroke}{rgb}{0.690196,0.690196,0.690196}%
\pgfsetstrokecolor{currentstroke}%
\pgfsetdash{}{0pt}%
\pgfpathmoveto{\pgfqpoint{6.958373in}{0.929220in}}%
\pgfpathlineto{\pgfqpoint{6.958373in}{6.600000in}}%
\pgfusepath{stroke}%
\end{pgfscope}%
\begin{pgfscope}%
\pgfsetbuttcap%
\pgfsetroundjoin%
\definecolor{currentfill}{rgb}{0.000000,0.000000,0.000000}%
\pgfsetfillcolor{currentfill}%
\pgfsetlinewidth{0.803000pt}%
\definecolor{currentstroke}{rgb}{0.000000,0.000000,0.000000}%
\pgfsetstrokecolor{currentstroke}%
\pgfsetdash{}{0pt}%
\pgfsys@defobject{currentmarker}{\pgfqpoint{0.000000in}{-0.048611in}}{\pgfqpoint{0.000000in}{0.000000in}}{%
\pgfpathmoveto{\pgfqpoint{0.000000in}{0.000000in}}%
\pgfpathlineto{\pgfqpoint{0.000000in}{-0.048611in}}%
\pgfusepath{stroke,fill}%
}%
\begin{pgfscope}%
\pgfsys@transformshift{6.958373in}{0.929220in}%
\pgfsys@useobject{currentmarker}{}%
\end{pgfscope}%
\end{pgfscope}%
\begin{pgfscope}%
\definecolor{textcolor}{rgb}{0.000000,0.000000,0.000000}%
\pgfsetstrokecolor{textcolor}%
\pgfsetfillcolor{textcolor}%
\pgftext[x=6.958373in,y=0.831998in,,top]{\color{textcolor}\rmfamily\fontsize{28.000000}{33.600000}\selectfont 40}%
\end{pgfscope}%
\begin{pgfscope}%
\pgfpathrectangle{\pgfqpoint{0.866662in}{0.929220in}}{\pgfqpoint{12.339620in}{5.670780in}}%
\pgfusepath{clip}%
\pgfsetrectcap%
\pgfsetroundjoin%
\pgfsetlinewidth{0.803000pt}%
\definecolor{currentstroke}{rgb}{0.690196,0.690196,0.690196}%
\pgfsetstrokecolor{currentstroke}%
\pgfsetdash{}{0pt}%
\pgfpathmoveto{\pgfqpoint{8.520350in}{0.929220in}}%
\pgfpathlineto{\pgfqpoint{8.520350in}{6.600000in}}%
\pgfusepath{stroke}%
\end{pgfscope}%
\begin{pgfscope}%
\pgfsetbuttcap%
\pgfsetroundjoin%
\definecolor{currentfill}{rgb}{0.000000,0.000000,0.000000}%
\pgfsetfillcolor{currentfill}%
\pgfsetlinewidth{0.803000pt}%
\definecolor{currentstroke}{rgb}{0.000000,0.000000,0.000000}%
\pgfsetstrokecolor{currentstroke}%
\pgfsetdash{}{0pt}%
\pgfsys@defobject{currentmarker}{\pgfqpoint{0.000000in}{-0.048611in}}{\pgfqpoint{0.000000in}{0.000000in}}{%
\pgfpathmoveto{\pgfqpoint{0.000000in}{0.000000in}}%
\pgfpathlineto{\pgfqpoint{0.000000in}{-0.048611in}}%
\pgfusepath{stroke,fill}%
}%
\begin{pgfscope}%
\pgfsys@transformshift{8.520350in}{0.929220in}%
\pgfsys@useobject{currentmarker}{}%
\end{pgfscope}%
\end{pgfscope}%
\begin{pgfscope}%
\definecolor{textcolor}{rgb}{0.000000,0.000000,0.000000}%
\pgfsetstrokecolor{textcolor}%
\pgfsetfillcolor{textcolor}%
\pgftext[x=8.520350in,y=0.831998in,,top]{\color{textcolor}\rmfamily\fontsize{28.000000}{33.600000}\selectfont 50}%
\end{pgfscope}%
\begin{pgfscope}%
\pgfpathrectangle{\pgfqpoint{0.866662in}{0.929220in}}{\pgfqpoint{12.339620in}{5.670780in}}%
\pgfusepath{clip}%
\pgfsetrectcap%
\pgfsetroundjoin%
\pgfsetlinewidth{0.803000pt}%
\definecolor{currentstroke}{rgb}{0.690196,0.690196,0.690196}%
\pgfsetstrokecolor{currentstroke}%
\pgfsetdash{}{0pt}%
\pgfpathmoveto{\pgfqpoint{10.082327in}{0.929220in}}%
\pgfpathlineto{\pgfqpoint{10.082327in}{6.600000in}}%
\pgfusepath{stroke}%
\end{pgfscope}%
\begin{pgfscope}%
\pgfsetbuttcap%
\pgfsetroundjoin%
\definecolor{currentfill}{rgb}{0.000000,0.000000,0.000000}%
\pgfsetfillcolor{currentfill}%
\pgfsetlinewidth{0.803000pt}%
\definecolor{currentstroke}{rgb}{0.000000,0.000000,0.000000}%
\pgfsetstrokecolor{currentstroke}%
\pgfsetdash{}{0pt}%
\pgfsys@defobject{currentmarker}{\pgfqpoint{0.000000in}{-0.048611in}}{\pgfqpoint{0.000000in}{0.000000in}}{%
\pgfpathmoveto{\pgfqpoint{0.000000in}{0.000000in}}%
\pgfpathlineto{\pgfqpoint{0.000000in}{-0.048611in}}%
\pgfusepath{stroke,fill}%
}%
\begin{pgfscope}%
\pgfsys@transformshift{10.082327in}{0.929220in}%
\pgfsys@useobject{currentmarker}{}%
\end{pgfscope}%
\end{pgfscope}%
\begin{pgfscope}%
\definecolor{textcolor}{rgb}{0.000000,0.000000,0.000000}%
\pgfsetstrokecolor{textcolor}%
\pgfsetfillcolor{textcolor}%
\pgftext[x=10.082327in,y=0.831998in,,top]{\color{textcolor}\rmfamily\fontsize{28.000000}{33.600000}\selectfont 60}%
\end{pgfscope}%
\begin{pgfscope}%
\pgfpathrectangle{\pgfqpoint{0.866662in}{0.929220in}}{\pgfqpoint{12.339620in}{5.670780in}}%
\pgfusepath{clip}%
\pgfsetrectcap%
\pgfsetroundjoin%
\pgfsetlinewidth{0.803000pt}%
\definecolor{currentstroke}{rgb}{0.690196,0.690196,0.690196}%
\pgfsetstrokecolor{currentstroke}%
\pgfsetdash{}{0pt}%
\pgfpathmoveto{\pgfqpoint{11.644305in}{0.929220in}}%
\pgfpathlineto{\pgfqpoint{11.644305in}{6.600000in}}%
\pgfusepath{stroke}%
\end{pgfscope}%
\begin{pgfscope}%
\pgfsetbuttcap%
\pgfsetroundjoin%
\definecolor{currentfill}{rgb}{0.000000,0.000000,0.000000}%
\pgfsetfillcolor{currentfill}%
\pgfsetlinewidth{0.803000pt}%
\definecolor{currentstroke}{rgb}{0.000000,0.000000,0.000000}%
\pgfsetstrokecolor{currentstroke}%
\pgfsetdash{}{0pt}%
\pgfsys@defobject{currentmarker}{\pgfqpoint{0.000000in}{-0.048611in}}{\pgfqpoint{0.000000in}{0.000000in}}{%
\pgfpathmoveto{\pgfqpoint{0.000000in}{0.000000in}}%
\pgfpathlineto{\pgfqpoint{0.000000in}{-0.048611in}}%
\pgfusepath{stroke,fill}%
}%
\begin{pgfscope}%
\pgfsys@transformshift{11.644305in}{0.929220in}%
\pgfsys@useobject{currentmarker}{}%
\end{pgfscope}%
\end{pgfscope}%
\begin{pgfscope}%
\definecolor{textcolor}{rgb}{0.000000,0.000000,0.000000}%
\pgfsetstrokecolor{textcolor}%
\pgfsetfillcolor{textcolor}%
\pgftext[x=11.644305in,y=0.831998in,,top]{\color{textcolor}\rmfamily\fontsize{28.000000}{33.600000}\selectfont 70}%
\end{pgfscope}%
\begin{pgfscope}%
\pgfpathrectangle{\pgfqpoint{0.866662in}{0.929220in}}{\pgfqpoint{12.339620in}{5.670780in}}%
\pgfusepath{clip}%
\pgfsetrectcap%
\pgfsetroundjoin%
\pgfsetlinewidth{0.803000pt}%
\definecolor{currentstroke}{rgb}{0.690196,0.690196,0.690196}%
\pgfsetstrokecolor{currentstroke}%
\pgfsetdash{}{0pt}%
\pgfpathmoveto{\pgfqpoint{13.206282in}{0.929220in}}%
\pgfpathlineto{\pgfqpoint{13.206282in}{6.600000in}}%
\pgfusepath{stroke}%
\end{pgfscope}%
\begin{pgfscope}%
\pgfsetbuttcap%
\pgfsetroundjoin%
\definecolor{currentfill}{rgb}{0.000000,0.000000,0.000000}%
\pgfsetfillcolor{currentfill}%
\pgfsetlinewidth{0.803000pt}%
\definecolor{currentstroke}{rgb}{0.000000,0.000000,0.000000}%
\pgfsetstrokecolor{currentstroke}%
\pgfsetdash{}{0pt}%
\pgfsys@defobject{currentmarker}{\pgfqpoint{0.000000in}{-0.048611in}}{\pgfqpoint{0.000000in}{0.000000in}}{%
\pgfpathmoveto{\pgfqpoint{0.000000in}{0.000000in}}%
\pgfpathlineto{\pgfqpoint{0.000000in}{-0.048611in}}%
\pgfusepath{stroke,fill}%
}%
\begin{pgfscope}%
\pgfsys@transformshift{13.206282in}{0.929220in}%
\pgfsys@useobject{currentmarker}{}%
\end{pgfscope}%
\end{pgfscope}%
\begin{pgfscope}%
\definecolor{textcolor}{rgb}{0.000000,0.000000,0.000000}%
\pgfsetstrokecolor{textcolor}%
\pgfsetfillcolor{textcolor}%
\pgftext[x=13.206282in,y=0.831998in,,top]{\color{textcolor}\rmfamily\fontsize{28.000000}{33.600000}\selectfont 80}%
\end{pgfscope}%
\begin{pgfscope}%
\definecolor{textcolor}{rgb}{0.000000,0.000000,0.000000}%
\pgfsetstrokecolor{textcolor}%
\pgfsetfillcolor{textcolor}%
\pgftext[x=7.036472in,y=0.385926in,,top]{\color{textcolor}\rmfamily\fontsize{28.000000}{33.600000}\selectfont time step \(\displaystyle k\)}%
\end{pgfscope}%
\begin{pgfscope}%
\pgfpathrectangle{\pgfqpoint{0.866662in}{0.929220in}}{\pgfqpoint{12.339620in}{5.670780in}}%
\pgfusepath{clip}%
\pgfsetrectcap%
\pgfsetroundjoin%
\pgfsetlinewidth{0.803000pt}%
\definecolor{currentstroke}{rgb}{0.690196,0.690196,0.690196}%
\pgfsetstrokecolor{currentstroke}%
\pgfsetdash{}{0pt}%
\pgfpathmoveto{\pgfqpoint{0.866662in}{0.929220in}}%
\pgfpathlineto{\pgfqpoint{13.206282in}{0.929220in}}%
\pgfusepath{stroke}%
\end{pgfscope}%
\begin{pgfscope}%
\pgfsetbuttcap%
\pgfsetroundjoin%
\definecolor{currentfill}{rgb}{0.000000,0.000000,0.000000}%
\pgfsetfillcolor{currentfill}%
\pgfsetlinewidth{0.803000pt}%
\definecolor{currentstroke}{rgb}{0.000000,0.000000,0.000000}%
\pgfsetstrokecolor{currentstroke}%
\pgfsetdash{}{0pt}%
\pgfsys@defobject{currentmarker}{\pgfqpoint{-0.048611in}{0.000000in}}{\pgfqpoint{-0.000000in}{0.000000in}}{%
\pgfpathmoveto{\pgfqpoint{-0.000000in}{0.000000in}}%
\pgfpathlineto{\pgfqpoint{-0.048611in}{0.000000in}}%
\pgfusepath{stroke,fill}%
}%
\begin{pgfscope}%
\pgfsys@transformshift{0.866662in}{0.929220in}%
\pgfsys@useobject{currentmarker}{}%
\end{pgfscope}%
\end{pgfscope}%
\begin{pgfscope}%
\definecolor{textcolor}{rgb}{0.000000,0.000000,0.000000}%
\pgfsetstrokecolor{textcolor}%
\pgfsetfillcolor{textcolor}%
\pgftext[x=0.520683in, y=0.809236in, left, base]{\color{textcolor}\rmfamily\fontsize{28.000000}{33.600000}\selectfont 0}%
\end{pgfscope}%
\begin{pgfscope}%
\pgfpathrectangle{\pgfqpoint{0.866662in}{0.929220in}}{\pgfqpoint{12.339620in}{5.670780in}}%
\pgfusepath{clip}%
\pgfsetrectcap%
\pgfsetroundjoin%
\pgfsetlinewidth{0.803000pt}%
\definecolor{currentstroke}{rgb}{0.690196,0.690196,0.690196}%
\pgfsetstrokecolor{currentstroke}%
\pgfsetdash{}{0pt}%
\pgfpathmoveto{\pgfqpoint{0.866662in}{2.218034in}}%
\pgfpathlineto{\pgfqpoint{13.206282in}{2.218034in}}%
\pgfusepath{stroke}%
\end{pgfscope}%
\begin{pgfscope}%
\pgfsetbuttcap%
\pgfsetroundjoin%
\definecolor{currentfill}{rgb}{0.000000,0.000000,0.000000}%
\pgfsetfillcolor{currentfill}%
\pgfsetlinewidth{0.803000pt}%
\definecolor{currentstroke}{rgb}{0.000000,0.000000,0.000000}%
\pgfsetstrokecolor{currentstroke}%
\pgfsetdash{}{0pt}%
\pgfsys@defobject{currentmarker}{\pgfqpoint{-0.048611in}{0.000000in}}{\pgfqpoint{-0.000000in}{0.000000in}}{%
\pgfpathmoveto{\pgfqpoint{-0.000000in}{0.000000in}}%
\pgfpathlineto{\pgfqpoint{-0.048611in}{0.000000in}}%
\pgfusepath{stroke,fill}%
}%
\begin{pgfscope}%
\pgfsys@transformshift{0.866662in}{2.218034in}%
\pgfsys@useobject{currentmarker}{}%
\end{pgfscope}%
\end{pgfscope}%
\begin{pgfscope}%
\definecolor{textcolor}{rgb}{0.000000,0.000000,0.000000}%
\pgfsetstrokecolor{textcolor}%
\pgfsetfillcolor{textcolor}%
\pgftext[x=0.520683in, y=2.098049in, left, base]{\color{textcolor}\rmfamily\fontsize{28.000000}{33.600000}\selectfont 2}%
\end{pgfscope}%
\begin{pgfscope}%
\pgfpathrectangle{\pgfqpoint{0.866662in}{0.929220in}}{\pgfqpoint{12.339620in}{5.670780in}}%
\pgfusepath{clip}%
\pgfsetrectcap%
\pgfsetroundjoin%
\pgfsetlinewidth{0.803000pt}%
\definecolor{currentstroke}{rgb}{0.690196,0.690196,0.690196}%
\pgfsetstrokecolor{currentstroke}%
\pgfsetdash{}{0pt}%
\pgfpathmoveto{\pgfqpoint{0.866662in}{3.506847in}}%
\pgfpathlineto{\pgfqpoint{13.206282in}{3.506847in}}%
\pgfusepath{stroke}%
\end{pgfscope}%
\begin{pgfscope}%
\pgfsetbuttcap%
\pgfsetroundjoin%
\definecolor{currentfill}{rgb}{0.000000,0.000000,0.000000}%
\pgfsetfillcolor{currentfill}%
\pgfsetlinewidth{0.803000pt}%
\definecolor{currentstroke}{rgb}{0.000000,0.000000,0.000000}%
\pgfsetstrokecolor{currentstroke}%
\pgfsetdash{}{0pt}%
\pgfsys@defobject{currentmarker}{\pgfqpoint{-0.048611in}{0.000000in}}{\pgfqpoint{-0.000000in}{0.000000in}}{%
\pgfpathmoveto{\pgfqpoint{-0.000000in}{0.000000in}}%
\pgfpathlineto{\pgfqpoint{-0.048611in}{0.000000in}}%
\pgfusepath{stroke,fill}%
}%
\begin{pgfscope}%
\pgfsys@transformshift{0.866662in}{3.506847in}%
\pgfsys@useobject{currentmarker}{}%
\end{pgfscope}%
\end{pgfscope}%
\begin{pgfscope}%
\definecolor{textcolor}{rgb}{0.000000,0.000000,0.000000}%
\pgfsetstrokecolor{textcolor}%
\pgfsetfillcolor{textcolor}%
\pgftext[x=0.520683in, y=3.386863in, left, base]{\color{textcolor}\rmfamily\fontsize{28.000000}{33.600000}\selectfont 4}%
\end{pgfscope}%
\begin{pgfscope}%
\pgfpathrectangle{\pgfqpoint{0.866662in}{0.929220in}}{\pgfqpoint{12.339620in}{5.670780in}}%
\pgfusepath{clip}%
\pgfsetrectcap%
\pgfsetroundjoin%
\pgfsetlinewidth{0.803000pt}%
\definecolor{currentstroke}{rgb}{0.690196,0.690196,0.690196}%
\pgfsetstrokecolor{currentstroke}%
\pgfsetdash{}{0pt}%
\pgfpathmoveto{\pgfqpoint{0.866662in}{4.795661in}}%
\pgfpathlineto{\pgfqpoint{13.206282in}{4.795661in}}%
\pgfusepath{stroke}%
\end{pgfscope}%
\begin{pgfscope}%
\pgfsetbuttcap%
\pgfsetroundjoin%
\definecolor{currentfill}{rgb}{0.000000,0.000000,0.000000}%
\pgfsetfillcolor{currentfill}%
\pgfsetlinewidth{0.803000pt}%
\definecolor{currentstroke}{rgb}{0.000000,0.000000,0.000000}%
\pgfsetstrokecolor{currentstroke}%
\pgfsetdash{}{0pt}%
\pgfsys@defobject{currentmarker}{\pgfqpoint{-0.048611in}{0.000000in}}{\pgfqpoint{-0.000000in}{0.000000in}}{%
\pgfpathmoveto{\pgfqpoint{-0.000000in}{0.000000in}}%
\pgfpathlineto{\pgfqpoint{-0.048611in}{0.000000in}}%
\pgfusepath{stroke,fill}%
}%
\begin{pgfscope}%
\pgfsys@transformshift{0.866662in}{4.795661in}%
\pgfsys@useobject{currentmarker}{}%
\end{pgfscope}%
\end{pgfscope}%
\begin{pgfscope}%
\definecolor{textcolor}{rgb}{0.000000,0.000000,0.000000}%
\pgfsetstrokecolor{textcolor}%
\pgfsetfillcolor{textcolor}%
\pgftext[x=0.520683in, y=4.675676in, left, base]{\color{textcolor}\rmfamily\fontsize{28.000000}{33.600000}\selectfont 6}%
\end{pgfscope}%
\begin{pgfscope}%
\pgfpathrectangle{\pgfqpoint{0.866662in}{0.929220in}}{\pgfqpoint{12.339620in}{5.670780in}}%
\pgfusepath{clip}%
\pgfsetrectcap%
\pgfsetroundjoin%
\pgfsetlinewidth{0.803000pt}%
\definecolor{currentstroke}{rgb}{0.690196,0.690196,0.690196}%
\pgfsetstrokecolor{currentstroke}%
\pgfsetdash{}{0pt}%
\pgfpathmoveto{\pgfqpoint{0.866662in}{6.084475in}}%
\pgfpathlineto{\pgfqpoint{13.206282in}{6.084475in}}%
\pgfusepath{stroke}%
\end{pgfscope}%
\begin{pgfscope}%
\pgfsetbuttcap%
\pgfsetroundjoin%
\definecolor{currentfill}{rgb}{0.000000,0.000000,0.000000}%
\pgfsetfillcolor{currentfill}%
\pgfsetlinewidth{0.803000pt}%
\definecolor{currentstroke}{rgb}{0.000000,0.000000,0.000000}%
\pgfsetstrokecolor{currentstroke}%
\pgfsetdash{}{0pt}%
\pgfsys@defobject{currentmarker}{\pgfqpoint{-0.048611in}{0.000000in}}{\pgfqpoint{-0.000000in}{0.000000in}}{%
\pgfpathmoveto{\pgfqpoint{-0.000000in}{0.000000in}}%
\pgfpathlineto{\pgfqpoint{-0.048611in}{0.000000in}}%
\pgfusepath{stroke,fill}%
}%
\begin{pgfscope}%
\pgfsys@transformshift{0.866662in}{6.084475in}%
\pgfsys@useobject{currentmarker}{}%
\end{pgfscope}%
\end{pgfscope}%
\begin{pgfscope}%
\definecolor{textcolor}{rgb}{0.000000,0.000000,0.000000}%
\pgfsetstrokecolor{textcolor}%
\pgfsetfillcolor{textcolor}%
\pgftext[x=0.520683in, y=5.964490in, left, base]{\color{textcolor}\rmfamily\fontsize{28.000000}{33.600000}\selectfont 8}%
\end{pgfscope}%
\begin{pgfscope}%
\definecolor{textcolor}{rgb}{0.000000,0.000000,0.000000}%
\pgfsetstrokecolor{textcolor}%
\pgfsetfillcolor{textcolor}%
\pgftext[x=0.381795in,y=3.764610in,,bottom,rotate=90.000000]{\color{textcolor}\rmfamily\fontsize{28.000000}{33.600000}\selectfont GOSPA [m]}%
\end{pgfscope}%
\begin{pgfscope}%
\pgfpathrectangle{\pgfqpoint{0.866662in}{0.929220in}}{\pgfqpoint{12.339620in}{5.670780in}}%
\pgfusepath{clip}%
\pgfsetrectcap%
\pgfsetroundjoin%
\pgfsetlinewidth{1.505625pt}%
\definecolor{currentstroke}{rgb}{0.750000,0.750000,0.000000}%
\pgfsetstrokecolor{currentstroke}%
\pgfsetdash{}{0pt}%
\pgfpathmoveto{\pgfqpoint{0.866662in}{0.929220in}}%
\pgfpathlineto{\pgfqpoint{1.022859in}{0.929220in}}%
\pgfpathlineto{\pgfqpoint{1.179057in}{0.929220in}}%
\pgfpathlineto{\pgfqpoint{1.335255in}{0.929220in}}%
\pgfpathlineto{\pgfqpoint{1.491452in}{0.929220in}}%
\pgfpathlineto{\pgfqpoint{1.647650in}{0.929220in}}%
\pgfpathlineto{\pgfqpoint{1.803848in}{0.929220in}}%
\pgfpathlineto{\pgfqpoint{1.960046in}{0.929220in}}%
\pgfpathlineto{\pgfqpoint{2.116243in}{0.929220in}}%
\pgfpathlineto{\pgfqpoint{2.272441in}{1.731324in}}%
\pgfpathlineto{\pgfqpoint{2.428639in}{1.673870in}}%
\pgfpathlineto{\pgfqpoint{2.584837in}{1.818712in}}%
\pgfpathlineto{\pgfqpoint{2.741034in}{1.760678in}}%
\pgfpathlineto{\pgfqpoint{2.897232in}{1.808218in}}%
\pgfpathlineto{\pgfqpoint{3.053430in}{2.214763in}}%
\pgfpathlineto{\pgfqpoint{3.209627in}{2.342016in}}%
\pgfpathlineto{\pgfqpoint{3.365825in}{2.367222in}}%
\pgfpathlineto{\pgfqpoint{3.522023in}{2.399442in}}%
\pgfpathlineto{\pgfqpoint{3.678221in}{2.368030in}}%
\pgfpathlineto{\pgfqpoint{3.834418in}{2.878441in}}%
\pgfpathlineto{\pgfqpoint{3.990616in}{2.951943in}}%
\pgfpathlineto{\pgfqpoint{4.146814in}{3.011210in}}%
\pgfpathlineto{\pgfqpoint{4.303012in}{3.046913in}}%
\pgfpathlineto{\pgfqpoint{4.459209in}{3.035473in}}%
\pgfpathlineto{\pgfqpoint{4.615407in}{3.452053in}}%
\pgfpathlineto{\pgfqpoint{4.771605in}{3.535588in}}%
\pgfpathlineto{\pgfqpoint{4.927802in}{3.594460in}}%
\pgfpathlineto{\pgfqpoint{5.084000in}{3.646338in}}%
\pgfpathlineto{\pgfqpoint{5.240198in}{3.662605in}}%
\pgfpathlineto{\pgfqpoint{5.396396in}{3.679548in}}%
\pgfpathlineto{\pgfqpoint{5.552593in}{3.674169in}}%
\pgfpathlineto{\pgfqpoint{5.708791in}{3.645544in}}%
\pgfpathlineto{\pgfqpoint{5.864989in}{3.635753in}}%
\pgfpathlineto{\pgfqpoint{6.021187in}{3.649206in}}%
\pgfpathlineto{\pgfqpoint{6.177384in}{3.615570in}}%
\pgfpathlineto{\pgfqpoint{6.333582in}{3.606148in}}%
\pgfpathlineto{\pgfqpoint{6.489780in}{3.612324in}}%
\pgfpathlineto{\pgfqpoint{6.645977in}{3.647326in}}%
\pgfpathlineto{\pgfqpoint{6.802175in}{3.618409in}}%
\pgfpathlineto{\pgfqpoint{6.958373in}{3.679779in}}%
\pgfpathlineto{\pgfqpoint{7.114571in}{3.684592in}}%
\pgfpathlineto{\pgfqpoint{7.270768in}{3.697872in}}%
\pgfpathlineto{\pgfqpoint{7.426966in}{3.588174in}}%
\pgfpathlineto{\pgfqpoint{7.583164in}{3.616028in}}%
\pgfpathlineto{\pgfqpoint{7.739362in}{3.675223in}}%
\pgfpathlineto{\pgfqpoint{7.895559in}{3.570044in}}%
\pgfpathlineto{\pgfqpoint{8.051757in}{3.559694in}}%
\pgfpathlineto{\pgfqpoint{8.207955in}{3.562304in}}%
\pgfpathlineto{\pgfqpoint{8.364152in}{3.520324in}}%
\pgfpathlineto{\pgfqpoint{8.520350in}{3.548906in}}%
\pgfpathlineto{\pgfqpoint{8.676548in}{3.597038in}}%
\pgfpathlineto{\pgfqpoint{8.832746in}{3.593781in}}%
\pgfpathlineto{\pgfqpoint{8.988943in}{3.561397in}}%
\pgfpathlineto{\pgfqpoint{9.145141in}{3.611242in}}%
\pgfpathlineto{\pgfqpoint{9.301339in}{3.201775in}}%
\pgfpathlineto{\pgfqpoint{9.457537in}{3.139441in}}%
\pgfpathlineto{\pgfqpoint{9.613734in}{3.113290in}}%
\pgfpathlineto{\pgfqpoint{9.769932in}{3.106292in}}%
\pgfpathlineto{\pgfqpoint{9.926130in}{3.120087in}}%
\pgfpathlineto{\pgfqpoint{10.082327in}{2.727559in}}%
\pgfpathlineto{\pgfqpoint{10.238525in}{2.562127in}}%
\pgfpathlineto{\pgfqpoint{10.394723in}{2.489072in}}%
\pgfpathlineto{\pgfqpoint{10.550921in}{2.461727in}}%
\pgfpathlineto{\pgfqpoint{10.707118in}{2.463955in}}%
\pgfpathlineto{\pgfqpoint{10.863316in}{2.083556in}}%
\pgfpathlineto{\pgfqpoint{11.019514in}{1.933545in}}%
\pgfpathlineto{\pgfqpoint{11.175712in}{1.860796in}}%
\pgfpathlineto{\pgfqpoint{11.331909in}{1.801581in}}%
\pgfpathlineto{\pgfqpoint{11.488107in}{1.811773in}}%
\pgfpathlineto{\pgfqpoint{11.644305in}{1.012993in}}%
\pgfpathlineto{\pgfqpoint{11.800502in}{0.929220in}}%
\pgfpathlineto{\pgfqpoint{11.956700in}{0.929220in}}%
\pgfpathlineto{\pgfqpoint{12.112898in}{0.929220in}}%
\pgfpathlineto{\pgfqpoint{12.269096in}{0.929220in}}%
\pgfpathlineto{\pgfqpoint{12.425293in}{0.929220in}}%
\pgfpathlineto{\pgfqpoint{12.581491in}{0.929220in}}%
\pgfpathlineto{\pgfqpoint{12.737689in}{0.929220in}}%
\pgfpathlineto{\pgfqpoint{12.893887in}{0.929220in}}%
\pgfpathlineto{\pgfqpoint{13.050084in}{0.929220in}}%
\pgfpathlineto{\pgfqpoint{13.206282in}{0.929220in}}%
\pgfusepath{stroke}%
\end{pgfscope}%
\begin{pgfscope}%
\pgfpathrectangle{\pgfqpoint{0.866662in}{0.929220in}}{\pgfqpoint{12.339620in}{5.670780in}}%
\pgfusepath{clip}%
\pgfsetrectcap%
\pgfsetroundjoin%
\pgfsetlinewidth{1.505625pt}%
\definecolor{currentstroke}{rgb}{0.000000,0.000000,1.000000}%
\pgfsetstrokecolor{currentstroke}%
\pgfsetdash{}{0pt}%
\pgfpathmoveto{\pgfqpoint{0.866662in}{0.929220in}}%
\pgfpathlineto{\pgfqpoint{1.022859in}{0.938886in}}%
\pgfpathlineto{\pgfqpoint{1.179057in}{0.932442in}}%
\pgfpathlineto{\pgfqpoint{1.335255in}{0.932442in}}%
\pgfpathlineto{\pgfqpoint{1.491452in}{0.935664in}}%
\pgfpathlineto{\pgfqpoint{1.647650in}{0.945331in}}%
\pgfpathlineto{\pgfqpoint{1.803848in}{0.938886in}}%
\pgfpathlineto{\pgfqpoint{1.960046in}{0.938886in}}%
\pgfpathlineto{\pgfqpoint{2.116243in}{0.935664in}}%
\pgfpathlineto{\pgfqpoint{2.272441in}{1.573627in}}%
\pgfpathlineto{\pgfqpoint{2.428639in}{1.572544in}}%
\pgfpathlineto{\pgfqpoint{2.584837in}{1.622039in}}%
\pgfpathlineto{\pgfqpoint{2.741034in}{1.460737in}}%
\pgfpathlineto{\pgfqpoint{2.897232in}{1.429959in}}%
\pgfpathlineto{\pgfqpoint{3.053430in}{1.788156in}}%
\pgfpathlineto{\pgfqpoint{3.209627in}{1.883248in}}%
\pgfpathlineto{\pgfqpoint{3.365825in}{1.817836in}}%
\pgfpathlineto{\pgfqpoint{3.522023in}{1.809993in}}%
\pgfpathlineto{\pgfqpoint{3.678221in}{1.773290in}}%
\pgfpathlineto{\pgfqpoint{3.834418in}{2.131822in}}%
\pgfpathlineto{\pgfqpoint{3.990616in}{2.174282in}}%
\pgfpathlineto{\pgfqpoint{4.146814in}{2.167593in}}%
\pgfpathlineto{\pgfqpoint{4.303012in}{2.223157in}}%
\pgfpathlineto{\pgfqpoint{4.459209in}{2.226685in}}%
\pgfpathlineto{\pgfqpoint{4.615407in}{2.595125in}}%
\pgfpathlineto{\pgfqpoint{4.771605in}{2.623288in}}%
\pgfpathlineto{\pgfqpoint{4.927802in}{2.660071in}}%
\pgfpathlineto{\pgfqpoint{5.084000in}{2.701763in}}%
\pgfpathlineto{\pgfqpoint{5.240198in}{2.716674in}}%
\pgfpathlineto{\pgfqpoint{5.396396in}{2.773390in}}%
\pgfpathlineto{\pgfqpoint{5.552593in}{2.856800in}}%
\pgfpathlineto{\pgfqpoint{5.708791in}{2.913685in}}%
\pgfpathlineto{\pgfqpoint{5.864989in}{2.936771in}}%
\pgfpathlineto{\pgfqpoint{6.021187in}{2.923339in}}%
\pgfpathlineto{\pgfqpoint{6.177384in}{2.879122in}}%
\pgfpathlineto{\pgfqpoint{6.333582in}{2.918418in}}%
\pgfpathlineto{\pgfqpoint{6.489780in}{2.848826in}}%
\pgfpathlineto{\pgfqpoint{6.645977in}{2.802339in}}%
\pgfpathlineto{\pgfqpoint{6.802175in}{2.727579in}}%
\pgfpathlineto{\pgfqpoint{6.958373in}{2.686106in}}%
\pgfpathlineto{\pgfqpoint{7.114571in}{2.688781in}}%
\pgfpathlineto{\pgfqpoint{7.270768in}{2.693075in}}%
\pgfpathlineto{\pgfqpoint{7.426966in}{2.662471in}}%
\pgfpathlineto{\pgfqpoint{7.583164in}{2.740723in}}%
\pgfpathlineto{\pgfqpoint{7.739362in}{2.755667in}}%
\pgfpathlineto{\pgfqpoint{7.895559in}{2.756209in}}%
\pgfpathlineto{\pgfqpoint{8.051757in}{2.763495in}}%
\pgfpathlineto{\pgfqpoint{8.207955in}{2.728100in}}%
\pgfpathlineto{\pgfqpoint{8.364152in}{2.695708in}}%
\pgfpathlineto{\pgfqpoint{8.520350in}{2.672458in}}%
\pgfpathlineto{\pgfqpoint{8.676548in}{2.656319in}}%
\pgfpathlineto{\pgfqpoint{8.832746in}{2.635578in}}%
\pgfpathlineto{\pgfqpoint{8.988943in}{2.630592in}}%
\pgfpathlineto{\pgfqpoint{9.145141in}{2.613563in}}%
\pgfpathlineto{\pgfqpoint{9.301339in}{2.509619in}}%
\pgfpathlineto{\pgfqpoint{9.457537in}{2.439016in}}%
\pgfpathlineto{\pgfqpoint{9.613734in}{2.157856in}}%
\pgfpathlineto{\pgfqpoint{9.769932in}{2.123200in}}%
\pgfpathlineto{\pgfqpoint{9.926130in}{2.094431in}}%
\pgfpathlineto{\pgfqpoint{10.082327in}{2.079813in}}%
\pgfpathlineto{\pgfqpoint{10.238525in}{2.028708in}}%
\pgfpathlineto{\pgfqpoint{10.394723in}{1.719201in}}%
\pgfpathlineto{\pgfqpoint{10.550921in}{1.715710in}}%
\pgfpathlineto{\pgfqpoint{10.707118in}{1.687972in}}%
\pgfpathlineto{\pgfqpoint{10.863316in}{1.711984in}}%
\pgfpathlineto{\pgfqpoint{11.019514in}{1.697784in}}%
\pgfpathlineto{\pgfqpoint{11.175712in}{1.383022in}}%
\pgfpathlineto{\pgfqpoint{11.331909in}{1.360576in}}%
\pgfpathlineto{\pgfqpoint{11.488107in}{1.364793in}}%
\pgfpathlineto{\pgfqpoint{11.644305in}{1.570405in}}%
\pgfpathlineto{\pgfqpoint{11.800502in}{1.570405in}}%
\pgfpathlineto{\pgfqpoint{11.956700in}{0.987217in}}%
\pgfpathlineto{\pgfqpoint{12.112898in}{0.974329in}}%
\pgfpathlineto{\pgfqpoint{12.269096in}{0.961441in}}%
\pgfpathlineto{\pgfqpoint{12.425293in}{0.942108in}}%
\pgfpathlineto{\pgfqpoint{12.581491in}{0.938886in}}%
\pgfpathlineto{\pgfqpoint{12.737689in}{0.938886in}}%
\pgfpathlineto{\pgfqpoint{12.893887in}{0.932442in}}%
\pgfpathlineto{\pgfqpoint{13.050084in}{0.942108in}}%
\pgfpathlineto{\pgfqpoint{13.206282in}{0.929220in}}%
\pgfusepath{stroke}%
\end{pgfscope}%
\begin{pgfscope}%
\pgfpathrectangle{\pgfqpoint{0.866662in}{0.929220in}}{\pgfqpoint{12.339620in}{5.670780in}}%
\pgfusepath{clip}%
\pgfsetrectcap%
\pgfsetroundjoin%
\pgfsetlinewidth{1.505625pt}%
\definecolor{currentstroke}{rgb}{0.000000,0.500000,0.000000}%
\pgfsetstrokecolor{currentstroke}%
\pgfsetdash{}{0pt}%
\pgfpathmoveto{\pgfqpoint{0.866662in}{0.929220in}}%
\pgfpathlineto{\pgfqpoint{1.022859in}{0.932442in}}%
\pgfpathlineto{\pgfqpoint{1.179057in}{0.932442in}}%
\pgfpathlineto{\pgfqpoint{1.335255in}{0.932442in}}%
\pgfpathlineto{\pgfqpoint{1.491452in}{0.942108in}}%
\pgfpathlineto{\pgfqpoint{1.647650in}{0.948553in}}%
\pgfpathlineto{\pgfqpoint{1.803848in}{0.938886in}}%
\pgfpathlineto{\pgfqpoint{1.960046in}{0.938886in}}%
\pgfpathlineto{\pgfqpoint{2.116243in}{0.942108in}}%
\pgfpathlineto{\pgfqpoint{2.272441in}{1.573627in}}%
\pgfpathlineto{\pgfqpoint{2.428639in}{1.572778in}}%
\pgfpathlineto{\pgfqpoint{2.584837in}{1.641236in}}%
\pgfpathlineto{\pgfqpoint{2.741034in}{1.479685in}}%
\pgfpathlineto{\pgfqpoint{2.897232in}{1.533315in}}%
\pgfpathlineto{\pgfqpoint{3.053430in}{1.745724in}}%
\pgfpathlineto{\pgfqpoint{3.209627in}{1.773850in}}%
\pgfpathlineto{\pgfqpoint{3.365825in}{1.645300in}}%
\pgfpathlineto{\pgfqpoint{3.522023in}{1.656068in}}%
\pgfpathlineto{\pgfqpoint{3.678221in}{1.593128in}}%
\pgfpathlineto{\pgfqpoint{3.834418in}{1.929867in}}%
\pgfpathlineto{\pgfqpoint{3.990616in}{1.906392in}}%
\pgfpathlineto{\pgfqpoint{4.146814in}{1.805312in}}%
\pgfpathlineto{\pgfqpoint{4.303012in}{1.798228in}}%
\pgfpathlineto{\pgfqpoint{4.459209in}{1.782881in}}%
\pgfpathlineto{\pgfqpoint{4.615407in}{2.087768in}}%
\pgfpathlineto{\pgfqpoint{4.771605in}{2.083645in}}%
\pgfpathlineto{\pgfqpoint{4.927802in}{2.045621in}}%
\pgfpathlineto{\pgfqpoint{5.084000in}{2.029656in}}%
\pgfpathlineto{\pgfqpoint{5.240198in}{1.999173in}}%
\pgfpathlineto{\pgfqpoint{5.396396in}{1.959440in}}%
\pgfpathlineto{\pgfqpoint{5.552593in}{1.945435in}}%
\pgfpathlineto{\pgfqpoint{5.708791in}{1.938939in}}%
\pgfpathlineto{\pgfqpoint{5.864989in}{1.894243in}}%
\pgfpathlineto{\pgfqpoint{6.021187in}{1.892239in}}%
\pgfpathlineto{\pgfqpoint{6.177384in}{1.885899in}}%
\pgfpathlineto{\pgfqpoint{6.333582in}{1.895098in}}%
\pgfpathlineto{\pgfqpoint{6.489780in}{1.914325in}}%
\pgfpathlineto{\pgfqpoint{6.645977in}{1.958094in}}%
\pgfpathlineto{\pgfqpoint{6.802175in}{1.975016in}}%
\pgfpathlineto{\pgfqpoint{6.958373in}{1.992126in}}%
\pgfpathlineto{\pgfqpoint{7.114571in}{1.983406in}}%
\pgfpathlineto{\pgfqpoint{7.270768in}{1.996998in}}%
\pgfpathlineto{\pgfqpoint{7.426966in}{1.989759in}}%
\pgfpathlineto{\pgfqpoint{7.583164in}{1.977226in}}%
\pgfpathlineto{\pgfqpoint{7.739362in}{1.955942in}}%
\pgfpathlineto{\pgfqpoint{7.895559in}{1.897770in}}%
\pgfpathlineto{\pgfqpoint{8.051757in}{1.882040in}}%
\pgfpathlineto{\pgfqpoint{8.207955in}{1.843238in}}%
\pgfpathlineto{\pgfqpoint{8.364152in}{1.819459in}}%
\pgfpathlineto{\pgfqpoint{8.520350in}{1.825643in}}%
\pgfpathlineto{\pgfqpoint{8.676548in}{1.817809in}}%
\pgfpathlineto{\pgfqpoint{8.832746in}{1.810475in}}%
\pgfpathlineto{\pgfqpoint{8.988943in}{1.807799in}}%
\pgfpathlineto{\pgfqpoint{9.145141in}{1.776796in}}%
\pgfpathlineto{\pgfqpoint{9.301339in}{1.894281in}}%
\pgfpathlineto{\pgfqpoint{9.457537in}{1.874090in}}%
\pgfpathlineto{\pgfqpoint{9.613734in}{1.588889in}}%
\pgfpathlineto{\pgfqpoint{9.769932in}{1.581255in}}%
\pgfpathlineto{\pgfqpoint{9.926130in}{1.575628in}}%
\pgfpathlineto{\pgfqpoint{10.082327in}{1.734303in}}%
\pgfpathlineto{\pgfqpoint{10.238525in}{1.708256in}}%
\pgfpathlineto{\pgfqpoint{10.394723in}{1.394434in}}%
\pgfpathlineto{\pgfqpoint{10.550921in}{1.393482in}}%
\pgfpathlineto{\pgfqpoint{10.707118in}{1.380430in}}%
\pgfpathlineto{\pgfqpoint{10.863316in}{1.520533in}}%
\pgfpathlineto{\pgfqpoint{11.019514in}{1.529077in}}%
\pgfpathlineto{\pgfqpoint{11.175712in}{1.221136in}}%
\pgfpathlineto{\pgfqpoint{11.331909in}{1.219128in}}%
\pgfpathlineto{\pgfqpoint{11.488107in}{1.219702in}}%
\pgfpathlineto{\pgfqpoint{11.644305in}{1.573627in}}%
\pgfpathlineto{\pgfqpoint{11.800502in}{1.573627in}}%
\pgfpathlineto{\pgfqpoint{11.956700in}{0.967885in}}%
\pgfpathlineto{\pgfqpoint{12.112898in}{0.980773in}}%
\pgfpathlineto{\pgfqpoint{12.269096in}{0.967885in}}%
\pgfpathlineto{\pgfqpoint{12.425293in}{0.958219in}}%
\pgfpathlineto{\pgfqpoint{12.581491in}{0.935664in}}%
\pgfpathlineto{\pgfqpoint{12.737689in}{0.938886in}}%
\pgfpathlineto{\pgfqpoint{12.893887in}{0.938886in}}%
\pgfpathlineto{\pgfqpoint{13.050084in}{0.945331in}}%
\pgfpathlineto{\pgfqpoint{13.206282in}{0.948553in}}%
\pgfusepath{stroke}%
\end{pgfscope}%
\begin{pgfscope}%
\pgfpathrectangle{\pgfqpoint{0.866662in}{0.929220in}}{\pgfqpoint{12.339620in}{5.670780in}}%
\pgfusepath{clip}%
\pgfsetrectcap%
\pgfsetroundjoin%
\pgfsetlinewidth{1.505625pt}%
\definecolor{currentstroke}{rgb}{1.000000,0.000000,0.000000}%
\pgfsetstrokecolor{currentstroke}%
\pgfsetdash{}{0pt}%
\pgfpathmoveto{\pgfqpoint{0.866662in}{0.929220in}}%
\pgfpathlineto{\pgfqpoint{1.022859in}{0.929220in}}%
\pgfpathlineto{\pgfqpoint{1.179057in}{0.929220in}}%
\pgfpathlineto{\pgfqpoint{1.335255in}{0.929220in}}%
\pgfpathlineto{\pgfqpoint{1.491452in}{0.929220in}}%
\pgfpathlineto{\pgfqpoint{1.647650in}{0.929220in}}%
\pgfpathlineto{\pgfqpoint{1.803848in}{0.929220in}}%
\pgfpathlineto{\pgfqpoint{1.960046in}{0.929220in}}%
\pgfpathlineto{\pgfqpoint{2.116243in}{0.929220in}}%
\pgfpathlineto{\pgfqpoint{2.272441in}{1.100864in}}%
\pgfpathlineto{\pgfqpoint{2.428639in}{1.084169in}}%
\pgfpathlineto{\pgfqpoint{2.584837in}{1.082042in}}%
\pgfpathlineto{\pgfqpoint{2.741034in}{1.087575in}}%
\pgfpathlineto{\pgfqpoint{2.897232in}{1.100368in}}%
\pgfpathlineto{\pgfqpoint{3.053430in}{1.202132in}}%
\pgfpathlineto{\pgfqpoint{3.209627in}{1.225923in}}%
\pgfpathlineto{\pgfqpoint{3.365825in}{1.202265in}}%
\pgfpathlineto{\pgfqpoint{3.522023in}{1.172894in}}%
\pgfpathlineto{\pgfqpoint{3.678221in}{1.151181in}}%
\pgfpathlineto{\pgfqpoint{3.834418in}{1.399440in}}%
\pgfpathlineto{\pgfqpoint{3.990616in}{1.361840in}}%
\pgfpathlineto{\pgfqpoint{4.146814in}{1.349451in}}%
\pgfpathlineto{\pgfqpoint{4.303012in}{1.339689in}}%
\pgfpathlineto{\pgfqpoint{4.459209in}{1.346697in}}%
\pgfpathlineto{\pgfqpoint{4.615407in}{1.666503in}}%
\pgfpathlineto{\pgfqpoint{4.771605in}{1.618475in}}%
\pgfpathlineto{\pgfqpoint{4.927802in}{1.602847in}}%
\pgfpathlineto{\pgfqpoint{5.084000in}{1.618870in}}%
\pgfpathlineto{\pgfqpoint{5.240198in}{1.661385in}}%
\pgfpathlineto{\pgfqpoint{5.396396in}{1.681803in}}%
\pgfpathlineto{\pgfqpoint{5.552593in}{1.734690in}}%
\pgfpathlineto{\pgfqpoint{5.708791in}{1.763976in}}%
\pgfpathlineto{\pgfqpoint{5.864989in}{1.777475in}}%
\pgfpathlineto{\pgfqpoint{6.021187in}{1.763335in}}%
\pgfpathlineto{\pgfqpoint{6.177384in}{1.783123in}}%
\pgfpathlineto{\pgfqpoint{6.333582in}{1.772507in}}%
\pgfpathlineto{\pgfqpoint{6.489780in}{1.772490in}}%
\pgfpathlineto{\pgfqpoint{6.645977in}{1.815128in}}%
\pgfpathlineto{\pgfqpoint{6.802175in}{1.791391in}}%
\pgfpathlineto{\pgfqpoint{6.958373in}{1.802353in}}%
\pgfpathlineto{\pgfqpoint{7.114571in}{1.766955in}}%
\pgfpathlineto{\pgfqpoint{7.270768in}{1.788567in}}%
\pgfpathlineto{\pgfqpoint{7.426966in}{1.779444in}}%
\pgfpathlineto{\pgfqpoint{7.583164in}{1.759895in}}%
\pgfpathlineto{\pgfqpoint{7.739362in}{1.739410in}}%
\pgfpathlineto{\pgfqpoint{7.895559in}{1.718346in}}%
\pgfpathlineto{\pgfqpoint{8.051757in}{1.687252in}}%
\pgfpathlineto{\pgfqpoint{8.207955in}{1.677575in}}%
\pgfpathlineto{\pgfqpoint{8.364152in}{1.659337in}}%
\pgfpathlineto{\pgfqpoint{8.520350in}{1.639371in}}%
\pgfpathlineto{\pgfqpoint{8.676548in}{1.620528in}}%
\pgfpathlineto{\pgfqpoint{8.832746in}{1.588433in}}%
\pgfpathlineto{\pgfqpoint{8.988943in}{1.578903in}}%
\pgfpathlineto{\pgfqpoint{9.145141in}{1.527461in}}%
\pgfpathlineto{\pgfqpoint{9.301339in}{1.404275in}}%
\pgfpathlineto{\pgfqpoint{9.457537in}{1.359715in}}%
\pgfpathlineto{\pgfqpoint{9.613734in}{1.357313in}}%
\pgfpathlineto{\pgfqpoint{9.769932in}{1.335310in}}%
\pgfpathlineto{\pgfqpoint{9.926130in}{1.305046in}}%
\pgfpathlineto{\pgfqpoint{10.082327in}{1.179440in}}%
\pgfpathlineto{\pgfqpoint{10.238525in}{1.161201in}}%
\pgfpathlineto{\pgfqpoint{10.394723in}{1.156023in}}%
\pgfpathlineto{\pgfqpoint{10.550921in}{1.148225in}}%
\pgfpathlineto{\pgfqpoint{10.707118in}{1.136955in}}%
\pgfpathlineto{\pgfqpoint{10.863316in}{1.046453in}}%
\pgfpathlineto{\pgfqpoint{11.019514in}{1.036997in}}%
\pgfpathlineto{\pgfqpoint{11.175712in}{1.033711in}}%
\pgfpathlineto{\pgfqpoint{11.331909in}{1.035617in}}%
\pgfpathlineto{\pgfqpoint{11.488107in}{1.037781in}}%
\pgfpathlineto{\pgfqpoint{11.644305in}{0.929220in}}%
\pgfpathlineto{\pgfqpoint{11.800502in}{0.929220in}}%
\pgfpathlineto{\pgfqpoint{11.956700in}{0.929220in}}%
\pgfpathlineto{\pgfqpoint{12.112898in}{0.929220in}}%
\pgfpathlineto{\pgfqpoint{12.269096in}{0.929220in}}%
\pgfpathlineto{\pgfqpoint{12.425293in}{0.929220in}}%
\pgfpathlineto{\pgfqpoint{12.581491in}{0.929220in}}%
\pgfpathlineto{\pgfqpoint{12.737689in}{0.929220in}}%
\pgfpathlineto{\pgfqpoint{12.893887in}{0.929220in}}%
\pgfpathlineto{\pgfqpoint{13.050084in}{0.929220in}}%
\pgfpathlineto{\pgfqpoint{13.206282in}{0.929220in}}%
\pgfusepath{stroke}%
\end{pgfscope}%
\begin{pgfscope}%
\pgfsetrectcap%
\pgfsetmiterjoin%
\pgfsetlinewidth{0.803000pt}%
\definecolor{currentstroke}{rgb}{0.000000,0.000000,0.000000}%
\pgfsetstrokecolor{currentstroke}%
\pgfsetdash{}{0pt}%
\pgfpathmoveto{\pgfqpoint{0.866662in}{0.929220in}}%
\pgfpathlineto{\pgfqpoint{0.866662in}{6.600000in}}%
\pgfusepath{stroke}%
\end{pgfscope}%
\begin{pgfscope}%
\pgfsetrectcap%
\pgfsetmiterjoin%
\pgfsetlinewidth{0.803000pt}%
\definecolor{currentstroke}{rgb}{0.000000,0.000000,0.000000}%
\pgfsetstrokecolor{currentstroke}%
\pgfsetdash{}{0pt}%
\pgfpathmoveto{\pgfqpoint{13.206282in}{0.929220in}}%
\pgfpathlineto{\pgfqpoint{13.206282in}{6.600000in}}%
\pgfusepath{stroke}%
\end{pgfscope}%
\begin{pgfscope}%
\pgfsetrectcap%
\pgfsetmiterjoin%
\pgfsetlinewidth{0.803000pt}%
\definecolor{currentstroke}{rgb}{0.000000,0.000000,0.000000}%
\pgfsetstrokecolor{currentstroke}%
\pgfsetdash{}{0pt}%
\pgfpathmoveto{\pgfqpoint{0.866662in}{0.929220in}}%
\pgfpathlineto{\pgfqpoint{13.206282in}{0.929220in}}%
\pgfusepath{stroke}%
\end{pgfscope}%
\begin{pgfscope}%
\pgfsetrectcap%
\pgfsetmiterjoin%
\pgfsetlinewidth{0.803000pt}%
\definecolor{currentstroke}{rgb}{0.000000,0.000000,0.000000}%
\pgfsetstrokecolor{currentstroke}%
\pgfsetdash{}{0pt}%
\pgfpathmoveto{\pgfqpoint{0.866662in}{6.600000in}}%
\pgfpathlineto{\pgfqpoint{13.206282in}{6.600000in}}%
\pgfusepath{stroke}%
\end{pgfscope}%
\begin{pgfscope}%
\pgfsetbuttcap%
\pgfsetmiterjoin%
\definecolor{currentfill}{rgb}{1.000000,1.000000,1.000000}%
\pgfsetfillcolor{currentfill}%
\pgfsetfillopacity{0.800000}%
\pgfsetlinewidth{1.003750pt}%
\definecolor{currentstroke}{rgb}{0.800000,0.800000,0.800000}%
\pgfsetstrokecolor{currentstroke}%
\pgfsetstrokeopacity{0.800000}%
\pgfsetdash{}{0pt}%
\pgfpathmoveto{\pgfqpoint{8.160436in}{4.134482in}}%
\pgfpathlineto{\pgfqpoint{12.934060in}{4.134482in}}%
\pgfpathquadraticcurveto{\pgfqpoint{13.011838in}{4.134482in}}{\pgfqpoint{13.011838in}{4.212260in}}%
\pgfpathlineto{\pgfqpoint{13.011838in}{6.327778in}}%
\pgfpathquadraticcurveto{\pgfqpoint{13.011838in}{6.405556in}}{\pgfqpoint{12.934060in}{6.405556in}}%
\pgfpathlineto{\pgfqpoint{8.160436in}{6.405556in}}%
\pgfpathquadraticcurveto{\pgfqpoint{8.082658in}{6.405556in}}{\pgfqpoint{8.082658in}{6.327778in}}%
\pgfpathlineto{\pgfqpoint{8.082658in}{4.212260in}}%
\pgfpathquadraticcurveto{\pgfqpoint{8.082658in}{4.134482in}}{\pgfqpoint{8.160436in}{4.134482in}}%
\pgfpathlineto{\pgfqpoint{8.160436in}{4.134482in}}%
\pgfpathclose%
\pgfusepath{stroke,fill}%
\end{pgfscope}%
\begin{pgfscope}%
\pgfsetrectcap%
\pgfsetroundjoin%
\pgfsetlinewidth{1.505625pt}%
\definecolor{currentstroke}{rgb}{0.750000,0.750000,0.000000}%
\pgfsetstrokecolor{currentstroke}%
\pgfsetdash{}{0pt}%
\pgfpathmoveto{\pgfqpoint{8.238214in}{6.113889in}}%
\pgfpathlineto{\pgfqpoint{8.627103in}{6.113889in}}%
\pgfpathlineto{\pgfqpoint{9.015992in}{6.113889in}}%
\pgfusepath{stroke}%
\end{pgfscope}%
\begin{pgfscope}%
\definecolor{textcolor}{rgb}{0.000000,0.000000,0.000000}%
\pgfsetstrokecolor{textcolor}%
\pgfsetfillcolor{textcolor}%
\pgftext[x=9.327103in,y=5.977778in,left,base]{\color{textcolor}\rmfamily\fontsize{28.000000}{33.600000}\selectfont LMB-TkBD}%
\end{pgfscope}%
\begin{pgfscope}%
\pgfsetrectcap%
\pgfsetroundjoin%
\pgfsetlinewidth{1.505625pt}%
\definecolor{currentstroke}{rgb}{0.000000,0.000000,1.000000}%
\pgfsetstrokecolor{currentstroke}%
\pgfsetdash{}{0pt}%
\pgfpathmoveto{\pgfqpoint{8.238214in}{5.580009in}}%
\pgfpathlineto{\pgfqpoint{8.627103in}{5.580009in}}%
\pgfpathlineto{\pgfqpoint{9.015992in}{5.580009in}}%
\pgfusepath{stroke}%
\end{pgfscope}%
\begin{pgfscope}%
\definecolor{textcolor}{rgb}{0.000000,0.000000,0.000000}%
\pgfsetstrokecolor{textcolor}%
\pgfsetfillcolor{textcolor}%
\pgftext[x=9.327103in,y=5.443898in,left,base]{\color{textcolor}\rmfamily\fontsize{28.000000}{33.600000}\selectfont MP + BP-DTT}%
\end{pgfscope}%
\begin{pgfscope}%
\pgfsetrectcap%
\pgfsetroundjoin%
\pgfsetlinewidth{1.505625pt}%
\definecolor{currentstroke}{rgb}{0.000000,0.500000,0.000000}%
\pgfsetstrokecolor{currentstroke}%
\pgfsetdash{}{0pt}%
\pgfpathmoveto{\pgfqpoint{8.238214in}{5.046129in}}%
\pgfpathlineto{\pgfqpoint{8.627103in}{5.046129in}}%
\pgfpathlineto{\pgfqpoint{9.015992in}{5.046129in}}%
\pgfusepath{stroke}%
\end{pgfscope}%
\begin{pgfscope}%
\definecolor{textcolor}{rgb}{0.000000,0.000000,0.000000}%
\pgfsetstrokecolor{textcolor}%
\pgfsetfillcolor{textcolor}%
\pgftext[x=9.327103in,y=4.910017in,left,base]{\color{textcolor}\rmfamily\fontsize{28.000000}{33.600000}\selectfont SBL + BP-DTT}%
\end{pgfscope}%
\begin{pgfscope}%
\pgfsetrectcap%
\pgfsetroundjoin%
\pgfsetlinewidth{1.505625pt}%
\definecolor{currentstroke}{rgb}{1.000000,0.000000,0.000000}%
\pgfsetstrokecolor{currentstroke}%
\pgfsetdash{}{0pt}%
\pgfpathmoveto{\pgfqpoint{8.238214in}{4.512248in}}%
\pgfpathlineto{\pgfqpoint{8.627103in}{4.512248in}}%
\pgfpathlineto{\pgfqpoint{9.015992in}{4.512248in}}%
\pgfusepath{stroke}%
\end{pgfscope}%
\begin{pgfscope}%
\definecolor{textcolor}{rgb}{0.000000,0.000000,0.000000}%
\pgfsetstrokecolor{textcolor}%
\pgfsetfillcolor{textcolor}%
\pgftext[x=9.327103in,y=4.376137in,left,base]{\color{textcolor}\rmfamily\fontsize{28.000000}{33.600000}\selectfont BP-TkBD (proposed)~~~~~}%
\end{pgfscope}%
\end{pgfpicture}%
\makeatother%
\endgroup%

%% file: Figs/gospa_-10dB/gospa_total.pgf
\begingroup%
\makeatletter%
\begin{pgfpicture}%
\pgfpathrectangle{\pgfpointorigin}{\pgfqpoint{7.000000in}{6.000000in}}%
\pgfusepath{use as bounding box, clip}%
\begin{pgfscope}%
\pgfsetbuttcap%
\pgfsetmiterjoin%
\definecolor{currentfill}{rgb}{1.000000,1.000000,1.000000}%
\pgfsetfillcolor{currentfill}%
\pgfsetlinewidth{0.000000pt}%
\definecolor{currentstroke}{rgb}{1.000000,1.000000,1.000000}%
\pgfsetstrokecolor{currentstroke}%
\pgfsetdash{}{0pt}%
\pgfpathmoveto{\pgfqpoint{0.000000in}{0.000000in}}%
\pgfpathlineto{\pgfqpoint{7.000000in}{0.000000in}}%
\pgfpathlineto{\pgfqpoint{7.000000in}{6.000000in}}%
\pgfpathlineto{\pgfqpoint{0.000000in}{6.000000in}}%
\pgfpathlineto{\pgfqpoint{0.000000in}{0.000000in}}%
\pgfpathclose%
\pgfusepath{fill}%
\end{pgfscope}%
\begin{pgfscope}%
\pgfsetbuttcap%
\pgfsetmiterjoin%
\definecolor{currentfill}{rgb}{1.000000,1.000000,1.000000}%
\pgfsetfillcolor{currentfill}%
\pgfsetlinewidth{0.000000pt}%
\definecolor{currentstroke}{rgb}{0.000000,0.000000,0.000000}%
\pgfsetstrokecolor{currentstroke}%
\pgfsetstrokeopacity{0.000000}%
\pgfsetdash{}{0pt}%
\pgfpathmoveto{\pgfqpoint{0.866662in}{0.929220in}}%
\pgfpathlineto{\pgfqpoint{6.806282in}{0.929220in}}%
\pgfpathlineto{\pgfqpoint{6.806282in}{6.000000in}}%
\pgfpathlineto{\pgfqpoint{0.866662in}{6.000000in}}%
\pgfpathlineto{\pgfqpoint{0.866662in}{0.929220in}}%
\pgfpathclose%
\pgfusepath{fill}%
\end{pgfscope}%
\begin{pgfscope}%
\pgfpathrectangle{\pgfqpoint{0.866662in}{0.929220in}}{\pgfqpoint{5.939620in}{5.070780in}}%
\pgfusepath{clip}%
\pgfsetrectcap%
\pgfsetroundjoin%
\pgfsetlinewidth{0.803000pt}%
\definecolor{currentstroke}{rgb}{0.690196,0.690196,0.690196}%
\pgfsetstrokecolor{currentstroke}%
\pgfsetdash{}{0pt}%
\pgfpathmoveto{\pgfqpoint{2.295178in}{0.929220in}}%
\pgfpathlineto{\pgfqpoint{2.295178in}{6.000000in}}%
\pgfusepath{stroke}%
\end{pgfscope}%
\begin{pgfscope}%
\pgfsetbuttcap%
\pgfsetroundjoin%
\definecolor{currentfill}{rgb}{0.000000,0.000000,0.000000}%
\pgfsetfillcolor{currentfill}%
\pgfsetlinewidth{0.803000pt}%
\definecolor{currentstroke}{rgb}{0.000000,0.000000,0.000000}%
\pgfsetstrokecolor{currentstroke}%
\pgfsetdash{}{0pt}%
\pgfsys@defobject{currentmarker}{\pgfqpoint{0.000000in}{-0.048611in}}{\pgfqpoint{0.000000in}{0.000000in}}{%
\pgfpathmoveto{\pgfqpoint{0.000000in}{0.000000in}}%
\pgfpathlineto{\pgfqpoint{0.000000in}{-0.048611in}}%
\pgfusepath{stroke,fill}%
}%
\begin{pgfscope}%
\pgfsys@transformshift{2.295178in}{0.929220in}%
\pgfsys@useobject{currentmarker}{}%
\end{pgfscope}%
\end{pgfscope}%
\begin{pgfscope}%
\definecolor{textcolor}{rgb}{0.000000,0.000000,0.000000}%
\pgfsetstrokecolor{textcolor}%
\pgfsetfillcolor{textcolor}%
\pgftext[x=2.295178in,y=0.831998in,,top]{\color{textcolor}\rmfamily\fontsize{28.000000}{33.600000}\selectfont 20}%
\end{pgfscope}%
\begin{pgfscope}%
\pgfpathrectangle{\pgfqpoint{0.866662in}{0.929220in}}{\pgfqpoint{5.939620in}{5.070780in}}%
\pgfusepath{clip}%
\pgfsetrectcap%
\pgfsetroundjoin%
\pgfsetlinewidth{0.803000pt}%
\definecolor{currentstroke}{rgb}{0.690196,0.690196,0.690196}%
\pgfsetstrokecolor{currentstroke}%
\pgfsetdash{}{0pt}%
\pgfpathmoveto{\pgfqpoint{3.798879in}{0.929220in}}%
\pgfpathlineto{\pgfqpoint{3.798879in}{6.000000in}}%
\pgfusepath{stroke}%
\end{pgfscope}%
\begin{pgfscope}%
\pgfsetbuttcap%
\pgfsetroundjoin%
\definecolor{currentfill}{rgb}{0.000000,0.000000,0.000000}%
\pgfsetfillcolor{currentfill}%
\pgfsetlinewidth{0.803000pt}%
\definecolor{currentstroke}{rgb}{0.000000,0.000000,0.000000}%
\pgfsetstrokecolor{currentstroke}%
\pgfsetdash{}{0pt}%
\pgfsys@defobject{currentmarker}{\pgfqpoint{0.000000in}{-0.048611in}}{\pgfqpoint{0.000000in}{0.000000in}}{%
\pgfpathmoveto{\pgfqpoint{0.000000in}{0.000000in}}%
\pgfpathlineto{\pgfqpoint{0.000000in}{-0.048611in}}%
\pgfusepath{stroke,fill}%
}%
\begin{pgfscope}%
\pgfsys@transformshift{3.798879in}{0.929220in}%
\pgfsys@useobject{currentmarker}{}%
\end{pgfscope}%
\end{pgfscope}%
\begin{pgfscope}%
\definecolor{textcolor}{rgb}{0.000000,0.000000,0.000000}%
\pgfsetstrokecolor{textcolor}%
\pgfsetfillcolor{textcolor}%
\pgftext[x=3.798879in,y=0.831998in,,top]{\color{textcolor}\rmfamily\fontsize{28.000000}{33.600000}\selectfont 40}%
\end{pgfscope}%
\begin{pgfscope}%
\pgfpathrectangle{\pgfqpoint{0.866662in}{0.929220in}}{\pgfqpoint{5.939620in}{5.070780in}}%
\pgfusepath{clip}%
\pgfsetrectcap%
\pgfsetroundjoin%
\pgfsetlinewidth{0.803000pt}%
\definecolor{currentstroke}{rgb}{0.690196,0.690196,0.690196}%
\pgfsetstrokecolor{currentstroke}%
\pgfsetdash{}{0pt}%
\pgfpathmoveto{\pgfqpoint{5.302581in}{0.929220in}}%
\pgfpathlineto{\pgfqpoint{5.302581in}{6.000000in}}%
\pgfusepath{stroke}%
\end{pgfscope}%
\begin{pgfscope}%
\pgfsetbuttcap%
\pgfsetroundjoin%
\definecolor{currentfill}{rgb}{0.000000,0.000000,0.000000}%
\pgfsetfillcolor{currentfill}%
\pgfsetlinewidth{0.803000pt}%
\definecolor{currentstroke}{rgb}{0.000000,0.000000,0.000000}%
\pgfsetstrokecolor{currentstroke}%
\pgfsetdash{}{0pt}%
\pgfsys@defobject{currentmarker}{\pgfqpoint{0.000000in}{-0.048611in}}{\pgfqpoint{0.000000in}{0.000000in}}{%
\pgfpathmoveto{\pgfqpoint{0.000000in}{0.000000in}}%
\pgfpathlineto{\pgfqpoint{0.000000in}{-0.048611in}}%
\pgfusepath{stroke,fill}%
}%
\begin{pgfscope}%
\pgfsys@transformshift{5.302581in}{0.929220in}%
\pgfsys@useobject{currentmarker}{}%
\end{pgfscope}%
\end{pgfscope}%
\begin{pgfscope}%
\definecolor{textcolor}{rgb}{0.000000,0.000000,0.000000}%
\pgfsetstrokecolor{textcolor}%
\pgfsetfillcolor{textcolor}%
\pgftext[x=5.302581in,y=0.831998in,,top]{\color{textcolor}\rmfamily\fontsize{28.000000}{33.600000}\selectfont 60}%
\end{pgfscope}%
\begin{pgfscope}%
\pgfpathrectangle{\pgfqpoint{0.866662in}{0.929220in}}{\pgfqpoint{5.939620in}{5.070780in}}%
\pgfusepath{clip}%
\pgfsetrectcap%
\pgfsetroundjoin%
\pgfsetlinewidth{0.803000pt}%
\definecolor{currentstroke}{rgb}{0.690196,0.690196,0.690196}%
\pgfsetstrokecolor{currentstroke}%
\pgfsetdash{}{0pt}%
\pgfpathmoveto{\pgfqpoint{6.806282in}{0.929220in}}%
\pgfpathlineto{\pgfqpoint{6.806282in}{6.000000in}}%
\pgfusepath{stroke}%
\end{pgfscope}%
\begin{pgfscope}%
\pgfsetbuttcap%
\pgfsetroundjoin%
\definecolor{currentfill}{rgb}{0.000000,0.000000,0.000000}%
\pgfsetfillcolor{currentfill}%
\pgfsetlinewidth{0.803000pt}%
\definecolor{currentstroke}{rgb}{0.000000,0.000000,0.000000}%
\pgfsetstrokecolor{currentstroke}%
\pgfsetdash{}{0pt}%
\pgfsys@defobject{currentmarker}{\pgfqpoint{0.000000in}{-0.048611in}}{\pgfqpoint{0.000000in}{0.000000in}}{%
\pgfpathmoveto{\pgfqpoint{0.000000in}{0.000000in}}%
\pgfpathlineto{\pgfqpoint{0.000000in}{-0.048611in}}%
\pgfusepath{stroke,fill}%
}%
\begin{pgfscope}%
\pgfsys@transformshift{6.806282in}{0.929220in}%
\pgfsys@useobject{currentmarker}{}%
\end{pgfscope}%
\end{pgfscope}%
\begin{pgfscope}%
\definecolor{textcolor}{rgb}{0.000000,0.000000,0.000000}%
\pgfsetstrokecolor{textcolor}%
\pgfsetfillcolor{textcolor}%
\pgftext[x=6.806282in,y=0.831998in,,top]{\color{textcolor}\rmfamily\fontsize{28.000000}{33.600000}\selectfont 80}%
\end{pgfscope}%
\begin{pgfscope}%
\definecolor{textcolor}{rgb}{0.000000,0.000000,0.000000}%
\pgfsetstrokecolor{textcolor}%
\pgfsetfillcolor{textcolor}%
\pgftext[x=3.836472in,y=0.385926in,,top]{\color{textcolor}\rmfamily\fontsize{28.000000}{33.600000}\selectfont time step \(\displaystyle k\)}%
\end{pgfscope}%
\begin{pgfscope}%
\pgfpathrectangle{\pgfqpoint{0.866662in}{0.929220in}}{\pgfqpoint{5.939620in}{5.070780in}}%
\pgfusepath{clip}%
\pgfsetrectcap%
\pgfsetroundjoin%
\pgfsetlinewidth{0.803000pt}%
\definecolor{currentstroke}{rgb}{0.690196,0.690196,0.690196}%
\pgfsetstrokecolor{currentstroke}%
\pgfsetdash{}{0pt}%
\pgfpathmoveto{\pgfqpoint{0.866662in}{0.929220in}}%
\pgfpathlineto{\pgfqpoint{6.806282in}{0.929220in}}%
\pgfusepath{stroke}%
\end{pgfscope}%
\begin{pgfscope}%
\pgfsetbuttcap%
\pgfsetroundjoin%
\definecolor{currentfill}{rgb}{0.000000,0.000000,0.000000}%
\pgfsetfillcolor{currentfill}%
\pgfsetlinewidth{0.803000pt}%
\definecolor{currentstroke}{rgb}{0.000000,0.000000,0.000000}%
\pgfsetstrokecolor{currentstroke}%
\pgfsetdash{}{0pt}%
\pgfsys@defobject{currentmarker}{\pgfqpoint{-0.048611in}{0.000000in}}{\pgfqpoint{-0.000000in}{0.000000in}}{%
\pgfpathmoveto{\pgfqpoint{-0.000000in}{0.000000in}}%
\pgfpathlineto{\pgfqpoint{-0.048611in}{0.000000in}}%
\pgfusepath{stroke,fill}%
}%
\begin{pgfscope}%
\pgfsys@transformshift{0.866662in}{0.929220in}%
\pgfsys@useobject{currentmarker}{}%
\end{pgfscope}%
\end{pgfscope}%
\begin{pgfscope}%
\definecolor{textcolor}{rgb}{0.000000,0.000000,0.000000}%
\pgfsetstrokecolor{textcolor}%
\pgfsetfillcolor{textcolor}%
\pgftext[x=0.520683in, y=0.809236in, left, base]{\color{textcolor}\rmfamily\fontsize{28.000000}{33.600000}\selectfont 0}%
\end{pgfscope}%
\begin{pgfscope}%
\pgfpathrectangle{\pgfqpoint{0.866662in}{0.929220in}}{\pgfqpoint{5.939620in}{5.070780in}}%
\pgfusepath{clip}%
\pgfsetrectcap%
\pgfsetroundjoin%
\pgfsetlinewidth{0.803000pt}%
\definecolor{currentstroke}{rgb}{0.690196,0.690196,0.690196}%
\pgfsetstrokecolor{currentstroke}%
\pgfsetdash{}{0pt}%
\pgfpathmoveto{\pgfqpoint{0.866662in}{2.081670in}}%
\pgfpathlineto{\pgfqpoint{6.806282in}{2.081670in}}%
\pgfusepath{stroke}%
\end{pgfscope}%
\begin{pgfscope}%
\pgfsetbuttcap%
\pgfsetroundjoin%
\definecolor{currentfill}{rgb}{0.000000,0.000000,0.000000}%
\pgfsetfillcolor{currentfill}%
\pgfsetlinewidth{0.803000pt}%
\definecolor{currentstroke}{rgb}{0.000000,0.000000,0.000000}%
\pgfsetstrokecolor{currentstroke}%
\pgfsetdash{}{0pt}%
\pgfsys@defobject{currentmarker}{\pgfqpoint{-0.048611in}{0.000000in}}{\pgfqpoint{-0.000000in}{0.000000in}}{%
\pgfpathmoveto{\pgfqpoint{-0.000000in}{0.000000in}}%
\pgfpathlineto{\pgfqpoint{-0.048611in}{0.000000in}}%
\pgfusepath{stroke,fill}%
}%
\begin{pgfscope}%
\pgfsys@transformshift{0.866662in}{2.081670in}%
\pgfsys@useobject{currentmarker}{}%
\end{pgfscope}%
\end{pgfscope}%
\begin{pgfscope}%
\definecolor{textcolor}{rgb}{0.000000,0.000000,0.000000}%
\pgfsetstrokecolor{textcolor}%
\pgfsetfillcolor{textcolor}%
\pgftext[x=0.520683in, y=1.961686in, left, base]{\color{textcolor}\rmfamily\fontsize{28.000000}{33.600000}\selectfont 2}%
\end{pgfscope}%
\begin{pgfscope}%
\pgfpathrectangle{\pgfqpoint{0.866662in}{0.929220in}}{\pgfqpoint{5.939620in}{5.070780in}}%
\pgfusepath{clip}%
\pgfsetrectcap%
\pgfsetroundjoin%
\pgfsetlinewidth{0.803000pt}%
\definecolor{currentstroke}{rgb}{0.690196,0.690196,0.690196}%
\pgfsetstrokecolor{currentstroke}%
\pgfsetdash{}{0pt}%
\pgfpathmoveto{\pgfqpoint{0.866662in}{3.234120in}}%
\pgfpathlineto{\pgfqpoint{6.806282in}{3.234120in}}%
\pgfusepath{stroke}%
\end{pgfscope}%
\begin{pgfscope}%
\pgfsetbuttcap%
\pgfsetroundjoin%
\definecolor{currentfill}{rgb}{0.000000,0.000000,0.000000}%
\pgfsetfillcolor{currentfill}%
\pgfsetlinewidth{0.803000pt}%
\definecolor{currentstroke}{rgb}{0.000000,0.000000,0.000000}%
\pgfsetstrokecolor{currentstroke}%
\pgfsetdash{}{0pt}%
\pgfsys@defobject{currentmarker}{\pgfqpoint{-0.048611in}{0.000000in}}{\pgfqpoint{-0.000000in}{0.000000in}}{%
\pgfpathmoveto{\pgfqpoint{-0.000000in}{0.000000in}}%
\pgfpathlineto{\pgfqpoint{-0.048611in}{0.000000in}}%
\pgfusepath{stroke,fill}%
}%
\begin{pgfscope}%
\pgfsys@transformshift{0.866662in}{3.234120in}%
\pgfsys@useobject{currentmarker}{}%
\end{pgfscope}%
\end{pgfscope}%
\begin{pgfscope}%
\definecolor{textcolor}{rgb}{0.000000,0.000000,0.000000}%
\pgfsetstrokecolor{textcolor}%
\pgfsetfillcolor{textcolor}%
\pgftext[x=0.520683in, y=3.114136in, left, base]{\color{textcolor}\rmfamily\fontsize{28.000000}{33.600000}\selectfont 4}%
\end{pgfscope}%
\begin{pgfscope}%
\pgfpathrectangle{\pgfqpoint{0.866662in}{0.929220in}}{\pgfqpoint{5.939620in}{5.070780in}}%
\pgfusepath{clip}%
\pgfsetrectcap%
\pgfsetroundjoin%
\pgfsetlinewidth{0.803000pt}%
\definecolor{currentstroke}{rgb}{0.690196,0.690196,0.690196}%
\pgfsetstrokecolor{currentstroke}%
\pgfsetdash{}{0pt}%
\pgfpathmoveto{\pgfqpoint{0.866662in}{4.386570in}}%
\pgfpathlineto{\pgfqpoint{6.806282in}{4.386570in}}%
\pgfusepath{stroke}%
\end{pgfscope}%
\begin{pgfscope}%
\pgfsetbuttcap%
\pgfsetroundjoin%
\definecolor{currentfill}{rgb}{0.000000,0.000000,0.000000}%
\pgfsetfillcolor{currentfill}%
\pgfsetlinewidth{0.803000pt}%
\definecolor{currentstroke}{rgb}{0.000000,0.000000,0.000000}%
\pgfsetstrokecolor{currentstroke}%
\pgfsetdash{}{0pt}%
\pgfsys@defobject{currentmarker}{\pgfqpoint{-0.048611in}{0.000000in}}{\pgfqpoint{-0.000000in}{0.000000in}}{%
\pgfpathmoveto{\pgfqpoint{-0.000000in}{0.000000in}}%
\pgfpathlineto{\pgfqpoint{-0.048611in}{0.000000in}}%
\pgfusepath{stroke,fill}%
}%
\begin{pgfscope}%
\pgfsys@transformshift{0.866662in}{4.386570in}%
\pgfsys@useobject{currentmarker}{}%
\end{pgfscope}%
\end{pgfscope}%
\begin{pgfscope}%
\definecolor{textcolor}{rgb}{0.000000,0.000000,0.000000}%
\pgfsetstrokecolor{textcolor}%
\pgfsetfillcolor{textcolor}%
\pgftext[x=0.520683in, y=4.266585in, left, base]{\color{textcolor}\rmfamily\fontsize{28.000000}{33.600000}\selectfont 6}%
\end{pgfscope}%
\begin{pgfscope}%
\pgfpathrectangle{\pgfqpoint{0.866662in}{0.929220in}}{\pgfqpoint{5.939620in}{5.070780in}}%
\pgfusepath{clip}%
\pgfsetrectcap%
\pgfsetroundjoin%
\pgfsetlinewidth{0.803000pt}%
\definecolor{currentstroke}{rgb}{0.690196,0.690196,0.690196}%
\pgfsetstrokecolor{currentstroke}%
\pgfsetdash{}{0pt}%
\pgfpathmoveto{\pgfqpoint{0.866662in}{5.539020in}}%
\pgfpathlineto{\pgfqpoint{6.806282in}{5.539020in}}%
\pgfusepath{stroke}%
\end{pgfscope}%
\begin{pgfscope}%
\pgfsetbuttcap%
\pgfsetroundjoin%
\definecolor{currentfill}{rgb}{0.000000,0.000000,0.000000}%
\pgfsetfillcolor{currentfill}%
\pgfsetlinewidth{0.803000pt}%
\definecolor{currentstroke}{rgb}{0.000000,0.000000,0.000000}%
\pgfsetstrokecolor{currentstroke}%
\pgfsetdash{}{0pt}%
\pgfsys@defobject{currentmarker}{\pgfqpoint{-0.048611in}{0.000000in}}{\pgfqpoint{-0.000000in}{0.000000in}}{%
\pgfpathmoveto{\pgfqpoint{-0.000000in}{0.000000in}}%
\pgfpathlineto{\pgfqpoint{-0.048611in}{0.000000in}}%
\pgfusepath{stroke,fill}%
}%
\begin{pgfscope}%
\pgfsys@transformshift{0.866662in}{5.539020in}%
\pgfsys@useobject{currentmarker}{}%
\end{pgfscope}%
\end{pgfscope}%
\begin{pgfscope}%
\definecolor{textcolor}{rgb}{0.000000,0.000000,0.000000}%
\pgfsetstrokecolor{textcolor}%
\pgfsetfillcolor{textcolor}%
\pgftext[x=0.520683in, y=5.419035in, left, base]{\color{textcolor}\rmfamily\fontsize{28.000000}{33.600000}\selectfont 8}%
\end{pgfscope}%
\begin{pgfscope}%
\definecolor{textcolor}{rgb}{0.000000,0.000000,0.000000}%
\pgfsetstrokecolor{textcolor}%
\pgfsetfillcolor{textcolor}%
\pgftext[x=0.381795in,y=3.464610in,,bottom,rotate=90.000000]{\color{textcolor}\rmfamily\fontsize{28.000000}{33.600000}\selectfont GOSPA in m}%
\end{pgfscope}%
\begin{pgfscope}%
\pgfpathrectangle{\pgfqpoint{0.866662in}{0.929220in}}{\pgfqpoint{5.939620in}{5.070780in}}%
\pgfusepath{clip}%
\pgfsetrectcap%
\pgfsetroundjoin%
\pgfsetlinewidth{1.505625pt}%
\definecolor{currentstroke}{rgb}{0.750000,0.750000,0.000000}%
\pgfsetstrokecolor{currentstroke}%
\pgfsetdash{}{0pt}%
\pgfpathmoveto{\pgfqpoint{0.866662in}{0.929220in}}%
\pgfpathlineto{\pgfqpoint{0.941847in}{0.929220in}}%
\pgfpathlineto{\pgfqpoint{1.017032in}{0.929220in}}%
\pgfpathlineto{\pgfqpoint{1.092217in}{0.929220in}}%
\pgfpathlineto{\pgfqpoint{1.167402in}{0.929220in}}%
\pgfpathlineto{\pgfqpoint{1.242587in}{0.929220in}}%
\pgfpathlineto{\pgfqpoint{1.317772in}{0.929220in}}%
\pgfpathlineto{\pgfqpoint{1.392957in}{0.929220in}}%
\pgfpathlineto{\pgfqpoint{1.468142in}{0.929220in}}%
\pgfpathlineto{\pgfqpoint{1.543327in}{1.646457in}}%
\pgfpathlineto{\pgfqpoint{1.618512in}{1.595082in}}%
\pgfpathlineto{\pgfqpoint{1.693697in}{1.724599in}}%
\pgfpathlineto{\pgfqpoint{1.768882in}{1.672705in}}%
\pgfpathlineto{\pgfqpoint{1.844067in}{1.715215in}}%
\pgfpathlineto{\pgfqpoint{1.919252in}{2.078746in}}%
\pgfpathlineto{\pgfqpoint{1.994438in}{2.192535in}}%
\pgfpathlineto{\pgfqpoint{2.069623in}{2.215073in}}%
\pgfpathlineto{\pgfqpoint{2.144808in}{2.243885in}}%
\pgfpathlineto{\pgfqpoint{2.219993in}{2.215796in}}%
\pgfpathlineto{\pgfqpoint{2.295178in}{2.672203in}}%
\pgfpathlineto{\pgfqpoint{2.370363in}{2.737928in}}%
\pgfpathlineto{\pgfqpoint{2.445548in}{2.790924in}}%
\pgfpathlineto{\pgfqpoint{2.520733in}{2.822849in}}%
\pgfpathlineto{\pgfqpoint{2.595918in}{2.812620in}}%
\pgfpathlineto{\pgfqpoint{2.671103in}{3.185123in}}%
\pgfpathlineto{\pgfqpoint{2.746288in}{3.259820in}}%
\pgfpathlineto{\pgfqpoint{2.821473in}{3.312463in}}%
\pgfpathlineto{\pgfqpoint{2.896658in}{3.358852in}}%
\pgfpathlineto{\pgfqpoint{2.971843in}{3.373398in}}%
\pgfpathlineto{\pgfqpoint{3.047029in}{3.388548in}}%
\pgfpathlineto{\pgfqpoint{3.122214in}{3.383738in}}%
\pgfpathlineto{\pgfqpoint{3.197399in}{3.358142in}}%
\pgfpathlineto{\pgfqpoint{3.272584in}{3.349387in}}%
\pgfpathlineto{\pgfqpoint{3.347769in}{3.361416in}}%
\pgfpathlineto{\pgfqpoint{3.422954in}{3.331339in}}%
\pgfpathlineto{\pgfqpoint{3.498139in}{3.322914in}}%
\pgfpathlineto{\pgfqpoint{3.573324in}{3.328437in}}%
\pgfpathlineto{\pgfqpoint{3.648509in}{3.359735in}}%
\pgfpathlineto{\pgfqpoint{3.723694in}{3.333878in}}%
\pgfpathlineto{\pgfqpoint{3.798879in}{3.388754in}}%
\pgfpathlineto{\pgfqpoint{3.874064in}{3.393058in}}%
\pgfpathlineto{\pgfqpoint{3.949249in}{3.404933in}}%
\pgfpathlineto{\pgfqpoint{4.024434in}{3.306842in}}%
\pgfpathlineto{\pgfqpoint{4.099619in}{3.331749in}}%
\pgfpathlineto{\pgfqpoint{4.174805in}{3.384680in}}%
\pgfpathlineto{\pgfqpoint{4.249990in}{3.290630in}}%
\pgfpathlineto{\pgfqpoint{4.325175in}{3.281376in}}%
\pgfpathlineto{\pgfqpoint{4.400360in}{3.283709in}}%
\pgfpathlineto{\pgfqpoint{4.475545in}{3.246171in}}%
\pgfpathlineto{\pgfqpoint{4.550730in}{3.271729in}}%
\pgfpathlineto{\pgfqpoint{4.625915in}{3.314768in}}%
\pgfpathlineto{\pgfqpoint{4.701100in}{3.311855in}}%
\pgfpathlineto{\pgfqpoint{4.776285in}{3.282898in}}%
\pgfpathlineto{\pgfqpoint{4.851470in}{3.327469in}}%
\pgfpathlineto{\pgfqpoint{4.926655in}{2.961326in}}%
\pgfpathlineto{\pgfqpoint{5.001840in}{2.905587in}}%
\pgfpathlineto{\pgfqpoint{5.077025in}{2.882203in}}%
\pgfpathlineto{\pgfqpoint{5.152210in}{2.875946in}}%
\pgfpathlineto{\pgfqpoint{5.227396in}{2.888281in}}%
\pgfpathlineto{\pgfqpoint{5.302581in}{2.537285in}}%
\pgfpathlineto{\pgfqpoint{5.377766in}{2.389356in}}%
\pgfpathlineto{\pgfqpoint{5.452951in}{2.324031in}}%
\pgfpathlineto{\pgfqpoint{5.528136in}{2.299579in}}%
\pgfpathlineto{\pgfqpoint{5.603321in}{2.301571in}}%
\pgfpathlineto{\pgfqpoint{5.678506in}{1.961421in}}%
\pgfpathlineto{\pgfqpoint{5.753691in}{1.827282in}}%
\pgfpathlineto{\pgfqpoint{5.828876in}{1.762230in}}%
\pgfpathlineto{\pgfqpoint{5.904061in}{1.709280in}}%
\pgfpathlineto{\pgfqpoint{5.979246in}{1.718394in}}%
\pgfpathlineto{\pgfqpoint{6.054431in}{1.004130in}}%
\pgfpathlineto{\pgfqpoint{6.129616in}{0.929220in}}%
\pgfpathlineto{\pgfqpoint{6.204801in}{0.929220in}}%
\pgfpathlineto{\pgfqpoint{6.279987in}{0.929220in}}%
\pgfpathlineto{\pgfqpoint{6.355172in}{0.929220in}}%
\pgfpathlineto{\pgfqpoint{6.430357in}{0.929220in}}%
\pgfpathlineto{\pgfqpoint{6.505542in}{0.929220in}}%
\pgfpathlineto{\pgfqpoint{6.580727in}{0.929220in}}%
\pgfpathlineto{\pgfqpoint{6.655912in}{0.929220in}}%
\pgfpathlineto{\pgfqpoint{6.731097in}{0.929220in}}%
\pgfpathlineto{\pgfqpoint{6.806282in}{0.929220in}}%
\pgfusepath{stroke}%
\end{pgfscope}%
\begin{pgfscope}%
\pgfpathrectangle{\pgfqpoint{0.866662in}{0.929220in}}{\pgfqpoint{5.939620in}{5.070780in}}%
\pgfusepath{clip}%
\pgfsetrectcap%
\pgfsetroundjoin%
\pgfsetlinewidth{1.505625pt}%
\definecolor{currentstroke}{rgb}{0.000000,0.000000,1.000000}%
\pgfsetstrokecolor{currentstroke}%
\pgfsetdash{}{0pt}%
\pgfpathmoveto{\pgfqpoint{0.866662in}{0.929220in}}%
\pgfpathlineto{\pgfqpoint{0.941847in}{0.937864in}}%
\pgfpathlineto{\pgfqpoint{1.017032in}{0.932101in}}%
\pgfpathlineto{\pgfqpoint{1.092217in}{0.932101in}}%
\pgfpathlineto{\pgfqpoint{1.167402in}{0.934983in}}%
\pgfpathlineto{\pgfqpoint{1.242587in}{0.943626in}}%
\pgfpathlineto{\pgfqpoint{1.317772in}{0.937864in}}%
\pgfpathlineto{\pgfqpoint{1.392957in}{0.937864in}}%
\pgfpathlineto{\pgfqpoint{1.468142in}{0.934983in}}%
\pgfpathlineto{\pgfqpoint{1.543327in}{1.505445in}}%
\pgfpathlineto{\pgfqpoint{1.618512in}{1.504477in}}%
\pgfpathlineto{\pgfqpoint{1.693697in}{1.548735in}}%
\pgfpathlineto{\pgfqpoint{1.768882in}{1.404500in}}%
\pgfpathlineto{\pgfqpoint{1.844067in}{1.376978in}}%
\pgfpathlineto{\pgfqpoint{1.919252in}{1.697276in}}%
\pgfpathlineto{\pgfqpoint{1.994438in}{1.782307in}}%
\pgfpathlineto{\pgfqpoint{2.069623in}{1.723815in}}%
\pgfpathlineto{\pgfqpoint{2.144808in}{1.716802in}}%
\pgfpathlineto{\pgfqpoint{2.219993in}{1.683983in}}%
\pgfpathlineto{\pgfqpoint{2.295178in}{2.004580in}}%
\pgfpathlineto{\pgfqpoint{2.370363in}{2.042548in}}%
\pgfpathlineto{\pgfqpoint{2.445548in}{2.036566in}}%
\pgfpathlineto{\pgfqpoint{2.520733in}{2.086252in}}%
\pgfpathlineto{\pgfqpoint{2.595918in}{2.089406in}}%
\pgfpathlineto{\pgfqpoint{2.671103in}{2.418863in}}%
\pgfpathlineto{\pgfqpoint{2.746288in}{2.444046in}}%
\pgfpathlineto{\pgfqpoint{2.821473in}{2.476938in}}%
\pgfpathlineto{\pgfqpoint{2.896658in}{2.514218in}}%
\pgfpathlineto{\pgfqpoint{2.971843in}{2.527552in}}%
\pgfpathlineto{\pgfqpoint{3.047029in}{2.578266in}}%
\pgfpathlineto{\pgfqpoint{3.122214in}{2.652852in}}%
\pgfpathlineto{\pgfqpoint{3.197399in}{2.703717in}}%
\pgfpathlineto{\pgfqpoint{3.272584in}{2.724361in}}%
\pgfpathlineto{\pgfqpoint{3.347769in}{2.712350in}}%
\pgfpathlineto{\pgfqpoint{3.422954in}{2.672812in}}%
\pgfpathlineto{\pgfqpoint{3.498139in}{2.707950in}}%
\pgfpathlineto{\pgfqpoint{3.573324in}{2.645721in}}%
\pgfpathlineto{\pgfqpoint{3.648509in}{2.604152in}}%
\pgfpathlineto{\pgfqpoint{3.723694in}{2.537303in}}%
\pgfpathlineto{\pgfqpoint{3.798879in}{2.500218in}}%
\pgfpathlineto{\pgfqpoint{3.874064in}{2.502610in}}%
\pgfpathlineto{\pgfqpoint{3.949249in}{2.506449in}}%
\pgfpathlineto{\pgfqpoint{4.024434in}{2.479083in}}%
\pgfpathlineto{\pgfqpoint{4.099619in}{2.549056in}}%
\pgfpathlineto{\pgfqpoint{4.174805in}{2.562419in}}%
\pgfpathlineto{\pgfqpoint{4.249990in}{2.562903in}}%
\pgfpathlineto{\pgfqpoint{4.325175in}{2.569418in}}%
\pgfpathlineto{\pgfqpoint{4.400360in}{2.537769in}}%
\pgfpathlineto{\pgfqpoint{4.475545in}{2.508804in}}%
\pgfpathlineto{\pgfqpoint{4.550730in}{2.488013in}}%
\pgfpathlineto{\pgfqpoint{4.625915in}{2.473582in}}%
\pgfpathlineto{\pgfqpoint{4.701100in}{2.455036in}}%
\pgfpathlineto{\pgfqpoint{4.776285in}{2.450577in}}%
\pgfpathlineto{\pgfqpoint{4.851470in}{2.435350in}}%
\pgfpathlineto{\pgfqpoint{4.926655in}{2.342404in}}%
\pgfpathlineto{\pgfqpoint{5.001840in}{2.279272in}}%
\pgfpathlineto{\pgfqpoint{5.077025in}{2.027859in}}%
\pgfpathlineto{\pgfqpoint{5.152210in}{1.996870in}}%
\pgfpathlineto{\pgfqpoint{5.227396in}{1.971145in}}%
\pgfpathlineto{\pgfqpoint{5.302581in}{1.958074in}}%
\pgfpathlineto{\pgfqpoint{5.377766in}{1.912376in}}%
\pgfpathlineto{\pgfqpoint{5.452951in}{1.635617in}}%
\pgfpathlineto{\pgfqpoint{5.528136in}{1.632495in}}%
\pgfpathlineto{\pgfqpoint{5.603321in}{1.607692in}}%
\pgfpathlineto{\pgfqpoint{5.678506in}{1.629163in}}%
\pgfpathlineto{\pgfqpoint{5.753691in}{1.616466in}}%
\pgfpathlineto{\pgfqpoint{5.828876in}{1.335008in}}%
\pgfpathlineto{\pgfqpoint{5.904061in}{1.314936in}}%
\pgfpathlineto{\pgfqpoint{5.979246in}{1.318707in}}%
\pgfpathlineto{\pgfqpoint{6.054431in}{1.502564in}}%
\pgfpathlineto{\pgfqpoint{6.129616in}{1.502564in}}%
\pgfpathlineto{\pgfqpoint{6.204801in}{0.981081in}}%
\pgfpathlineto{\pgfqpoint{6.279987in}{0.969556in}}%
\pgfpathlineto{\pgfqpoint{6.355172in}{0.958032in}}%
\pgfpathlineto{\pgfqpoint{6.430357in}{0.940745in}}%
\pgfpathlineto{\pgfqpoint{6.505542in}{0.937864in}}%
\pgfpathlineto{\pgfqpoint{6.580727in}{0.937864in}}%
\pgfpathlineto{\pgfqpoint{6.655912in}{0.932101in}}%
\pgfpathlineto{\pgfqpoint{6.731097in}{0.940745in}}%
\pgfpathlineto{\pgfqpoint{6.806282in}{0.929220in}}%
\pgfusepath{stroke}%
\end{pgfscope}%
\begin{pgfscope}%
\pgfpathrectangle{\pgfqpoint{0.866662in}{0.929220in}}{\pgfqpoint{5.939620in}{5.070780in}}%
\pgfusepath{clip}%
\pgfsetrectcap%
\pgfsetroundjoin%
\pgfsetlinewidth{1.505625pt}%
\definecolor{currentstroke}{rgb}{0.000000,0.500000,0.000000}%
\pgfsetstrokecolor{currentstroke}%
\pgfsetdash{}{0pt}%
\pgfpathmoveto{\pgfqpoint{0.866662in}{0.929220in}}%
\pgfpathlineto{\pgfqpoint{0.941847in}{0.932101in}}%
\pgfpathlineto{\pgfqpoint{1.017032in}{0.932101in}}%
\pgfpathlineto{\pgfqpoint{1.092217in}{0.932101in}}%
\pgfpathlineto{\pgfqpoint{1.167402in}{0.940745in}}%
\pgfpathlineto{\pgfqpoint{1.242587in}{0.946507in}}%
\pgfpathlineto{\pgfqpoint{1.317772in}{0.937864in}}%
\pgfpathlineto{\pgfqpoint{1.392957in}{0.937864in}}%
\pgfpathlineto{\pgfqpoint{1.468142in}{0.940745in}}%
\pgfpathlineto{\pgfqpoint{1.543327in}{1.505445in}}%
\pgfpathlineto{\pgfqpoint{1.618512in}{1.504686in}}%
\pgfpathlineto{\pgfqpoint{1.693697in}{1.565901in}}%
\pgfpathlineto{\pgfqpoint{1.768882in}{1.421443in}}%
\pgfpathlineto{\pgfqpoint{1.844067in}{1.469399in}}%
\pgfpathlineto{\pgfqpoint{1.919252in}{1.659333in}}%
\pgfpathlineto{\pgfqpoint{1.994438in}{1.684484in}}%
\pgfpathlineto{\pgfqpoint{2.069623in}{1.569535in}}%
\pgfpathlineto{\pgfqpoint{2.144808in}{1.579164in}}%
\pgfpathlineto{\pgfqpoint{2.219993in}{1.522883in}}%
\pgfpathlineto{\pgfqpoint{2.295178in}{1.823993in}}%
\pgfpathlineto{\pgfqpoint{2.370363in}{1.803002in}}%
\pgfpathlineto{\pgfqpoint{2.445548in}{1.712617in}}%
\pgfpathlineto{\pgfqpoint{2.520733in}{1.706282in}}%
\pgfpathlineto{\pgfqpoint{2.595918in}{1.692559in}}%
\pgfpathlineto{\pgfqpoint{2.671103in}{1.965187in}}%
\pgfpathlineto{\pgfqpoint{2.746288in}{1.961501in}}%
\pgfpathlineto{\pgfqpoint{2.821473in}{1.927500in}}%
\pgfpathlineto{\pgfqpoint{2.896658in}{1.913224in}}%
\pgfpathlineto{\pgfqpoint{2.971843in}{1.885966in}}%
\pgfpathlineto{\pgfqpoint{3.047029in}{1.850437in}}%
\pgfpathlineto{\pgfqpoint{3.122214in}{1.837914in}}%
\pgfpathlineto{\pgfqpoint{3.197399in}{1.832105in}}%
\pgfpathlineto{\pgfqpoint{3.272584in}{1.792138in}}%
\pgfpathlineto{\pgfqpoint{3.347769in}{1.790346in}}%
\pgfpathlineto{\pgfqpoint{3.422954in}{1.784677in}}%
\pgfpathlineto{\pgfqpoint{3.498139in}{1.792903in}}%
\pgfpathlineto{\pgfqpoint{3.573324in}{1.810096in}}%
\pgfpathlineto{\pgfqpoint{3.648509in}{1.849234in}}%
\pgfpathlineto{\pgfqpoint{3.723694in}{1.864365in}}%
\pgfpathlineto{\pgfqpoint{3.798879in}{1.879665in}}%
\pgfpathlineto{\pgfqpoint{3.874064in}{1.871867in}}%
\pgfpathlineto{\pgfqpoint{3.949249in}{1.884021in}}%
\pgfpathlineto{\pgfqpoint{4.024434in}{1.877548in}}%
\pgfpathlineto{\pgfqpoint{4.099619in}{1.866341in}}%
\pgfpathlineto{\pgfqpoint{4.174805in}{1.847309in}}%
\pgfpathlineto{\pgfqpoint{4.249990in}{1.795292in}}%
\pgfpathlineto{\pgfqpoint{4.325175in}{1.781226in}}%
\pgfpathlineto{\pgfqpoint{4.400360in}{1.746530in}}%
\pgfpathlineto{\pgfqpoint{4.475545in}{1.725267in}}%
\pgfpathlineto{\pgfqpoint{4.550730in}{1.730796in}}%
\pgfpathlineto{\pgfqpoint{4.625915in}{1.723791in}}%
\pgfpathlineto{\pgfqpoint{4.701100in}{1.717233in}}%
\pgfpathlineto{\pgfqpoint{4.776285in}{1.714841in}}%
\pgfpathlineto{\pgfqpoint{4.851470in}{1.687118in}}%
\pgfpathlineto{\pgfqpoint{4.926655in}{1.792172in}}%
\pgfpathlineto{\pgfqpoint{5.001840in}{1.774118in}}%
\pgfpathlineto{\pgfqpoint{5.077025in}{1.519093in}}%
\pgfpathlineto{\pgfqpoint{5.152210in}{1.512266in}}%
\pgfpathlineto{\pgfqpoint{5.227396in}{1.507234in}}%
\pgfpathlineto{\pgfqpoint{5.302581in}{1.649121in}}%
\pgfpathlineto{\pgfqpoint{5.377766in}{1.625830in}}%
\pgfpathlineto{\pgfqpoint{5.452951in}{1.345212in}}%
\pgfpathlineto{\pgfqpoint{5.528136in}{1.344360in}}%
\pgfpathlineto{\pgfqpoint{5.603321in}{1.332689in}}%
\pgfpathlineto{\pgfqpoint{5.678506in}{1.457969in}}%
\pgfpathlineto{\pgfqpoint{5.753691in}{1.465609in}}%
\pgfpathlineto{\pgfqpoint{5.828876in}{1.190250in}}%
\pgfpathlineto{\pgfqpoint{5.904061in}{1.188454in}}%
\pgfpathlineto{\pgfqpoint{5.979246in}{1.188968in}}%
\pgfpathlineto{\pgfqpoint{6.054431in}{1.505445in}}%
\pgfpathlineto{\pgfqpoint{6.129616in}{1.505445in}}%
\pgfpathlineto{\pgfqpoint{6.204801in}{0.963794in}}%
\pgfpathlineto{\pgfqpoint{6.279987in}{0.975318in}}%
\pgfpathlineto{\pgfqpoint{6.355172in}{0.963794in}}%
\pgfpathlineto{\pgfqpoint{6.430357in}{0.955150in}}%
\pgfpathlineto{\pgfqpoint{6.505542in}{0.934983in}}%
\pgfpathlineto{\pgfqpoint{6.580727in}{0.937864in}}%
\pgfpathlineto{\pgfqpoint{6.655912in}{0.937864in}}%
\pgfpathlineto{\pgfqpoint{6.731097in}{0.943626in}}%
\pgfpathlineto{\pgfqpoint{6.806282in}{0.946507in}}%
\pgfusepath{stroke}%
\end{pgfscope}%
\begin{pgfscope}%
\pgfpathrectangle{\pgfqpoint{0.866662in}{0.929220in}}{\pgfqpoint{5.939620in}{5.070780in}}%
\pgfusepath{clip}%
\pgfsetrectcap%
\pgfsetroundjoin%
\pgfsetlinewidth{1.505625pt}%
\definecolor{currentstroke}{rgb}{1.000000,0.000000,0.000000}%
\pgfsetstrokecolor{currentstroke}%
\pgfsetdash{}{0pt}%
\pgfpathmoveto{\pgfqpoint{0.866662in}{0.929220in}}%
\pgfpathlineto{\pgfqpoint{0.941847in}{0.929220in}}%
\pgfpathlineto{\pgfqpoint{1.017032in}{0.929220in}}%
\pgfpathlineto{\pgfqpoint{1.092217in}{0.929220in}}%
\pgfpathlineto{\pgfqpoint{1.167402in}{0.929220in}}%
\pgfpathlineto{\pgfqpoint{1.242587in}{0.929220in}}%
\pgfpathlineto{\pgfqpoint{1.317772in}{0.929220in}}%
\pgfpathlineto{\pgfqpoint{1.392957in}{0.929220in}}%
\pgfpathlineto{\pgfqpoint{1.468142in}{0.929220in}}%
\pgfpathlineto{\pgfqpoint{1.543327in}{1.082703in}}%
\pgfpathlineto{\pgfqpoint{1.618512in}{1.067775in}}%
\pgfpathlineto{\pgfqpoint{1.693697in}{1.065873in}}%
\pgfpathlineto{\pgfqpoint{1.768882in}{1.070821in}}%
\pgfpathlineto{\pgfqpoint{1.844067in}{1.082260in}}%
\pgfpathlineto{\pgfqpoint{1.919252in}{1.173256in}}%
\pgfpathlineto{\pgfqpoint{1.994438in}{1.194530in}}%
\pgfpathlineto{\pgfqpoint{2.069623in}{1.173375in}}%
\pgfpathlineto{\pgfqpoint{2.144808in}{1.147112in}}%
\pgfpathlineto{\pgfqpoint{2.219993in}{1.127696in}}%
\pgfpathlineto{\pgfqpoint{2.295178in}{1.349688in}}%
\pgfpathlineto{\pgfqpoint{2.370363in}{1.316066in}}%
\pgfpathlineto{\pgfqpoint{2.445548in}{1.304988in}}%
\pgfpathlineto{\pgfqpoint{2.520733in}{1.296259in}}%
\pgfpathlineto{\pgfqpoint{2.595918in}{1.302526in}}%
\pgfpathlineto{\pgfqpoint{2.671103in}{1.588495in}}%
\pgfpathlineto{\pgfqpoint{2.746288in}{1.545548in}}%
\pgfpathlineto{\pgfqpoint{2.821473in}{1.531574in}}%
\pgfpathlineto{\pgfqpoint{2.896658in}{1.545901in}}%
\pgfpathlineto{\pgfqpoint{2.971843in}{1.583918in}}%
\pgfpathlineto{\pgfqpoint{3.047029in}{1.602176in}}%
\pgfpathlineto{\pgfqpoint{3.122214in}{1.649467in}}%
\pgfpathlineto{\pgfqpoint{3.197399in}{1.675654in}}%
\pgfpathlineto{\pgfqpoint{3.272584in}{1.687725in}}%
\pgfpathlineto{\pgfqpoint{3.347769in}{1.675081in}}%
\pgfpathlineto{\pgfqpoint{3.422954in}{1.692775in}}%
\pgfpathlineto{\pgfqpoint{3.498139in}{1.683283in}}%
\pgfpathlineto{\pgfqpoint{3.573324in}{1.683267in}}%
\pgfpathlineto{\pgfqpoint{3.648509in}{1.721394in}}%
\pgfpathlineto{\pgfqpoint{3.723694in}{1.700168in}}%
\pgfpathlineto{\pgfqpoint{3.798879in}{1.709971in}}%
\pgfpathlineto{\pgfqpoint{3.874064in}{1.678318in}}%
\pgfpathlineto{\pgfqpoint{3.949249in}{1.697643in}}%
\pgfpathlineto{\pgfqpoint{4.024434in}{1.689486in}}%
\pgfpathlineto{\pgfqpoint{4.099619in}{1.672005in}}%
\pgfpathlineto{\pgfqpoint{4.174805in}{1.653688in}}%
\pgfpathlineto{\pgfqpoint{4.249990in}{1.634852in}}%
\pgfpathlineto{\pgfqpoint{4.325175in}{1.607048in}}%
\pgfpathlineto{\pgfqpoint{4.400360in}{1.598395in}}%
\pgfpathlineto{\pgfqpoint{4.475545in}{1.582087in}}%
\pgfpathlineto{\pgfqpoint{4.550730in}{1.564233in}}%
\pgfpathlineto{\pgfqpoint{4.625915in}{1.547384in}}%
\pgfpathlineto{\pgfqpoint{4.701100in}{1.518685in}}%
\pgfpathlineto{\pgfqpoint{4.776285in}{1.510163in}}%
\pgfpathlineto{\pgfqpoint{4.851470in}{1.464164in}}%
\pgfpathlineto{\pgfqpoint{4.926655in}{1.354011in}}%
\pgfpathlineto{\pgfqpoint{5.001840in}{1.314166in}}%
\pgfpathlineto{\pgfqpoint{5.077025in}{1.312018in}}%
\pgfpathlineto{\pgfqpoint{5.152210in}{1.292344in}}%
\pgfpathlineto{\pgfqpoint{5.227396in}{1.265281in}}%
\pgfpathlineto{\pgfqpoint{5.302581in}{1.152965in}}%
\pgfpathlineto{\pgfqpoint{5.377766in}{1.136656in}}%
\pgfpathlineto{\pgfqpoint{5.452951in}{1.132026in}}%
\pgfpathlineto{\pgfqpoint{5.528136in}{1.125053in}}%
\pgfpathlineto{\pgfqpoint{5.603321in}{1.114976in}}%
\pgfpathlineto{\pgfqpoint{5.678506in}{1.034049in}}%
\pgfpathlineto{\pgfqpoint{5.753691in}{1.025593in}}%
\pgfpathlineto{\pgfqpoint{5.828876in}{1.022656in}}%
\pgfpathlineto{\pgfqpoint{5.904061in}{1.024360in}}%
\pgfpathlineto{\pgfqpoint{5.979246in}{1.026295in}}%
\pgfpathlineto{\pgfqpoint{6.054431in}{0.929220in}}%
\pgfpathlineto{\pgfqpoint{6.129616in}{0.929220in}}%
\pgfpathlineto{\pgfqpoint{6.204801in}{0.929220in}}%
\pgfpathlineto{\pgfqpoint{6.279987in}{0.929220in}}%
\pgfpathlineto{\pgfqpoint{6.355172in}{0.929220in}}%
\pgfpathlineto{\pgfqpoint{6.430357in}{0.929220in}}%
\pgfpathlineto{\pgfqpoint{6.505542in}{0.929220in}}%
\pgfpathlineto{\pgfqpoint{6.580727in}{0.929220in}}%
\pgfpathlineto{\pgfqpoint{6.655912in}{0.929220in}}%
\pgfpathlineto{\pgfqpoint{6.731097in}{0.929220in}}%
\pgfpathlineto{\pgfqpoint{6.806282in}{0.929220in}}%
\pgfusepath{stroke}%
\end{pgfscope}%
\begin{pgfscope}%
\pgfsetrectcap%
\pgfsetmiterjoin%
\pgfsetlinewidth{0.803000pt}%
\definecolor{currentstroke}{rgb}{0.000000,0.000000,0.000000}%
\pgfsetstrokecolor{currentstroke}%
\pgfsetdash{}{0pt}%
\pgfpathmoveto{\pgfqpoint{0.866662in}{0.929220in}}%
\pgfpathlineto{\pgfqpoint{0.866662in}{6.000000in}}%
\pgfusepath{stroke}%
\end{pgfscope}%
\begin{pgfscope}%
\pgfsetrectcap%
\pgfsetmiterjoin%
\pgfsetlinewidth{0.803000pt}%
\definecolor{currentstroke}{rgb}{0.000000,0.000000,0.000000}%
\pgfsetstrokecolor{currentstroke}%
\pgfsetdash{}{0pt}%
\pgfpathmoveto{\pgfqpoint{6.806282in}{0.929220in}}%
\pgfpathlineto{\pgfqpoint{6.806282in}{6.000000in}}%
\pgfusepath{stroke}%
\end{pgfscope}%
\begin{pgfscope}%
\pgfsetrectcap%
\pgfsetmiterjoin%
\pgfsetlinewidth{0.803000pt}%
\definecolor{currentstroke}{rgb}{0.000000,0.000000,0.000000}%
\pgfsetstrokecolor{currentstroke}%
\pgfsetdash{}{0pt}%
\pgfpathmoveto{\pgfqpoint{0.866662in}{0.929220in}}%
\pgfpathlineto{\pgfqpoint{6.806282in}{0.929220in}}%
\pgfusepath{stroke}%
\end{pgfscope}%
\begin{pgfscope}%
\pgfsetrectcap%
\pgfsetmiterjoin%
\pgfsetlinewidth{0.803000pt}%
\definecolor{currentstroke}{rgb}{0.000000,0.000000,0.000000}%
\pgfsetstrokecolor{currentstroke}%
\pgfsetdash{}{0pt}%
\pgfpathmoveto{\pgfqpoint{0.866662in}{6.000000in}}%
\pgfpathlineto{\pgfqpoint{6.806282in}{6.000000in}}%
\pgfusepath{stroke}%
\end{pgfscope}%
\begin{pgfscope}%
\pgfsetbuttcap%
\pgfsetmiterjoin%
\definecolor{currentfill}{rgb}{1.000000,1.000000,1.000000}%
\pgfsetfillcolor{currentfill}%
\pgfsetfillopacity{0.800000}%
\pgfsetlinewidth{1.003750pt}%
\definecolor{currentstroke}{rgb}{0.800000,0.800000,0.800000}%
\pgfsetstrokecolor{currentstroke}%
\pgfsetstrokeopacity{0.800000}%
\pgfsetdash{}{0pt}%
\pgfpathmoveto{\pgfqpoint{1.760436in}{3.534482in}}%
\pgfpathlineto{\pgfqpoint{6.534060in}{3.534482in}}%
\pgfpathquadraticcurveto{\pgfqpoint{6.611838in}{3.534482in}}{\pgfqpoint{6.611838in}{3.612260in}}%
\pgfpathlineto{\pgfqpoint{6.611838in}{5.727778in}}%
\pgfpathquadraticcurveto{\pgfqpoint{6.611838in}{5.805556in}}{\pgfqpoint{6.534060in}{5.805556in}}%
\pgfpathlineto{\pgfqpoint{1.760436in}{5.805556in}}%
\pgfpathquadraticcurveto{\pgfqpoint{1.682658in}{5.805556in}}{\pgfqpoint{1.682658in}{5.727778in}}%
\pgfpathlineto{\pgfqpoint{1.682658in}{3.612260in}}%
\pgfpathquadraticcurveto{\pgfqpoint{1.682658in}{3.534482in}}{\pgfqpoint{1.760436in}{3.534482in}}%
\pgfpathlineto{\pgfqpoint{1.760436in}{3.534482in}}%
\pgfpathclose%
\pgfusepath{stroke,fill}%
\end{pgfscope}%
\begin{pgfscope}%
\pgfsetrectcap%
\pgfsetroundjoin%
\pgfsetlinewidth{1.505625pt}%
\definecolor{currentstroke}{rgb}{0.750000,0.750000,0.000000}%
\pgfsetstrokecolor{currentstroke}%
\pgfsetdash{}{0pt}%
\pgfpathmoveto{\pgfqpoint{1.838214in}{5.513889in}}%
\pgfpathlineto{\pgfqpoint{2.227103in}{5.513889in}}%
\pgfpathlineto{\pgfqpoint{2.615992in}{5.513889in}}%
\pgfusepath{stroke}%
\end{pgfscope}%
\begin{pgfscope}%
\definecolor{textcolor}{rgb}{0.000000,0.000000,0.000000}%
\pgfsetstrokecolor{textcolor}%
\pgfsetfillcolor{textcolor}%
\pgftext[x=2.927103in,y=5.377778in,left,base]{\color{textcolor}\rmfamily\fontsize{28.000000}{33.600000}\selectfont LMB-TkBD}%
\end{pgfscope}%
\begin{pgfscope}%
\pgfsetrectcap%
\pgfsetroundjoin%
\pgfsetlinewidth{1.505625pt}%
\definecolor{currentstroke}{rgb}{0.000000,0.000000,1.000000}%
\pgfsetstrokecolor{currentstroke}%
\pgfsetdash{}{0pt}%
\pgfpathmoveto{\pgfqpoint{1.838214in}{4.980009in}}%
\pgfpathlineto{\pgfqpoint{2.227103in}{4.980009in}}%
\pgfpathlineto{\pgfqpoint{2.615992in}{4.980009in}}%
\pgfusepath{stroke}%
\end{pgfscope}%
\begin{pgfscope}%
\definecolor{textcolor}{rgb}{0.000000,0.000000,0.000000}%
\pgfsetstrokecolor{textcolor}%
\pgfsetfillcolor{textcolor}%
\pgftext[x=2.927103in,y=4.843898in,left,base]{\color{textcolor}\rmfamily\fontsize{28.000000}{33.600000}\selectfont MP + BP-DTT}%
\end{pgfscope}%
\begin{pgfscope}%
\pgfsetrectcap%
\pgfsetroundjoin%
\pgfsetlinewidth{1.505625pt}%
\definecolor{currentstroke}{rgb}{0.000000,0.500000,0.000000}%
\pgfsetstrokecolor{currentstroke}%
\pgfsetdash{}{0pt}%
\pgfpathmoveto{\pgfqpoint{1.838214in}{4.446129in}}%
\pgfpathlineto{\pgfqpoint{2.227103in}{4.446129in}}%
\pgfpathlineto{\pgfqpoint{2.615992in}{4.446129in}}%
\pgfusepath{stroke}%
\end{pgfscope}%
\begin{pgfscope}%
\definecolor{textcolor}{rgb}{0.000000,0.000000,0.000000}%
\pgfsetstrokecolor{textcolor}%
\pgfsetfillcolor{textcolor}%
\pgftext[x=2.927103in,y=4.310018in,left,base]{\color{textcolor}\rmfamily\fontsize{28.000000}{33.600000}\selectfont SBL + BP-DTT}%
\end{pgfscope}%
\begin{pgfscope}%
\pgfsetrectcap%
\pgfsetroundjoin%
\pgfsetlinewidth{1.505625pt}%
\definecolor{currentstroke}{rgb}{1.000000,0.000000,0.000000}%
\pgfsetstrokecolor{currentstroke}%
\pgfsetdash{}{0pt}%
\pgfpathmoveto{\pgfqpoint{1.838214in}{3.912248in}}%
\pgfpathlineto{\pgfqpoint{2.227103in}{3.912248in}}%
\pgfpathlineto{\pgfqpoint{2.615992in}{3.912248in}}%
\pgfusepath{stroke}%
\end{pgfscope}%
\begin{pgfscope}%
\definecolor{textcolor}{rgb}{0.000000,0.000000,0.000000}%
\pgfsetstrokecolor{textcolor}%
\pgfsetfillcolor{textcolor}%
\pgftext[x=2.927103in,y=3.776137in,left,base]{\color{textcolor}\rmfamily\fontsize{28.000000}{33.600000}\selectfont BP-TkBD (proposed)~~~~~}%
\end{pgfscope}%
\end{pgfpicture}%
\makeatother%
\endgroup%

%% file: Figs/gospa_-10dB/gospa_loc.pgf
\begingroup%
\makeatletter%
\begin{pgfpicture}%
\pgfpathrectangle{\pgfpointorigin}{\pgfqpoint{7.000000in}{6.000000in}}%
\pgfusepath{use as bounding box, clip}%
\begin{pgfscope}%
\pgfsetbuttcap%
\pgfsetmiterjoin%
\definecolor{currentfill}{rgb}{1.000000,1.000000,1.000000}%
\pgfsetfillcolor{currentfill}%
\pgfsetlinewidth{0.000000pt}%
\definecolor{currentstroke}{rgb}{1.000000,1.000000,1.000000}%
\pgfsetstrokecolor{currentstroke}%
\pgfsetdash{}{0pt}%
\pgfpathmoveto{\pgfqpoint{0.000000in}{0.000000in}}%
\pgfpathlineto{\pgfqpoint{7.000000in}{0.000000in}}%
\pgfpathlineto{\pgfqpoint{7.000000in}{6.000000in}}%
\pgfpathlineto{\pgfqpoint{0.000000in}{6.000000in}}%
\pgfpathlineto{\pgfqpoint{0.000000in}{0.000000in}}%
\pgfpathclose%
\pgfusepath{fill}%
\end{pgfscope}%
\begin{pgfscope}%
\pgfsetbuttcap%
\pgfsetmiterjoin%
\definecolor{currentfill}{rgb}{1.000000,1.000000,1.000000}%
\pgfsetfillcolor{currentfill}%
\pgfsetlinewidth{0.000000pt}%
\definecolor{currentstroke}{rgb}{0.000000,0.000000,0.000000}%
\pgfsetstrokecolor{currentstroke}%
\pgfsetstrokeopacity{0.000000}%
\pgfsetdash{}{0pt}%
\pgfpathmoveto{\pgfqpoint{1.157239in}{0.929220in}}%
\pgfpathlineto{\pgfqpoint{6.806282in}{0.929220in}}%
\pgfpathlineto{\pgfqpoint{6.806282in}{6.000000in}}%
\pgfpathlineto{\pgfqpoint{1.157239in}{6.000000in}}%
\pgfpathlineto{\pgfqpoint{1.157239in}{0.929220in}}%
\pgfpathclose%
\pgfusepath{fill}%
\end{pgfscope}%
\begin{pgfscope}%
\pgfpathrectangle{\pgfqpoint{1.157239in}{0.929220in}}{\pgfqpoint{5.649043in}{5.070780in}}%
\pgfusepath{clip}%
\pgfsetrectcap%
\pgfsetroundjoin%
\pgfsetlinewidth{0.803000pt}%
\definecolor{currentstroke}{rgb}{0.690196,0.690196,0.690196}%
\pgfsetstrokecolor{currentstroke}%
\pgfsetdash{}{0pt}%
\pgfpathmoveto{\pgfqpoint{2.515869in}{0.929220in}}%
\pgfpathlineto{\pgfqpoint{2.515869in}{6.000000in}}%
\pgfusepath{stroke}%
\end{pgfscope}%
\begin{pgfscope}%
\pgfsetbuttcap%
\pgfsetroundjoin%
\definecolor{currentfill}{rgb}{0.000000,0.000000,0.000000}%
\pgfsetfillcolor{currentfill}%
\pgfsetlinewidth{0.803000pt}%
\definecolor{currentstroke}{rgb}{0.000000,0.000000,0.000000}%
\pgfsetstrokecolor{currentstroke}%
\pgfsetdash{}{0pt}%
\pgfsys@defobject{currentmarker}{\pgfqpoint{0.000000in}{-0.048611in}}{\pgfqpoint{0.000000in}{0.000000in}}{%
\pgfpathmoveto{\pgfqpoint{0.000000in}{0.000000in}}%
\pgfpathlineto{\pgfqpoint{0.000000in}{-0.048611in}}%
\pgfusepath{stroke,fill}%
}%
\begin{pgfscope}%
\pgfsys@transformshift{2.515869in}{0.929220in}%
\pgfsys@useobject{currentmarker}{}%
\end{pgfscope}%
\end{pgfscope}%
\begin{pgfscope}%
\definecolor{textcolor}{rgb}{0.000000,0.000000,0.000000}%
\pgfsetstrokecolor{textcolor}%
\pgfsetfillcolor{textcolor}%
\pgftext[x=2.515869in,y=0.831998in,,top]{\color{textcolor}\rmfamily\fontsize{28.000000}{33.600000}\selectfont 20}%
\end{pgfscope}%
\begin{pgfscope}%
\pgfpathrectangle{\pgfqpoint{1.157239in}{0.929220in}}{\pgfqpoint{5.649043in}{5.070780in}}%
\pgfusepath{clip}%
\pgfsetrectcap%
\pgfsetroundjoin%
\pgfsetlinewidth{0.803000pt}%
\definecolor{currentstroke}{rgb}{0.690196,0.690196,0.690196}%
\pgfsetstrokecolor{currentstroke}%
\pgfsetdash{}{0pt}%
\pgfpathmoveto{\pgfqpoint{3.946007in}{0.929220in}}%
\pgfpathlineto{\pgfqpoint{3.946007in}{6.000000in}}%
\pgfusepath{stroke}%
\end{pgfscope}%
\begin{pgfscope}%
\pgfsetbuttcap%
\pgfsetroundjoin%
\definecolor{currentfill}{rgb}{0.000000,0.000000,0.000000}%
\pgfsetfillcolor{currentfill}%
\pgfsetlinewidth{0.803000pt}%
\definecolor{currentstroke}{rgb}{0.000000,0.000000,0.000000}%
\pgfsetstrokecolor{currentstroke}%
\pgfsetdash{}{0pt}%
\pgfsys@defobject{currentmarker}{\pgfqpoint{0.000000in}{-0.048611in}}{\pgfqpoint{0.000000in}{0.000000in}}{%
\pgfpathmoveto{\pgfqpoint{0.000000in}{0.000000in}}%
\pgfpathlineto{\pgfqpoint{0.000000in}{-0.048611in}}%
\pgfusepath{stroke,fill}%
}%
\begin{pgfscope}%
\pgfsys@transformshift{3.946007in}{0.929220in}%
\pgfsys@useobject{currentmarker}{}%
\end{pgfscope}%
\end{pgfscope}%
\begin{pgfscope}%
\definecolor{textcolor}{rgb}{0.000000,0.000000,0.000000}%
\pgfsetstrokecolor{textcolor}%
\pgfsetfillcolor{textcolor}%
\pgftext[x=3.946007in,y=0.831998in,,top]{\color{textcolor}\rmfamily\fontsize{28.000000}{33.600000}\selectfont 40}%
\end{pgfscope}%
\begin{pgfscope}%
\pgfpathrectangle{\pgfqpoint{1.157239in}{0.929220in}}{\pgfqpoint{5.649043in}{5.070780in}}%
\pgfusepath{clip}%
\pgfsetrectcap%
\pgfsetroundjoin%
\pgfsetlinewidth{0.803000pt}%
\definecolor{currentstroke}{rgb}{0.690196,0.690196,0.690196}%
\pgfsetstrokecolor{currentstroke}%
\pgfsetdash{}{0pt}%
\pgfpathmoveto{\pgfqpoint{5.376144in}{0.929220in}}%
\pgfpathlineto{\pgfqpoint{5.376144in}{6.000000in}}%
\pgfusepath{stroke}%
\end{pgfscope}%
\begin{pgfscope}%
\pgfsetbuttcap%
\pgfsetroundjoin%
\definecolor{currentfill}{rgb}{0.000000,0.000000,0.000000}%
\pgfsetfillcolor{currentfill}%
\pgfsetlinewidth{0.803000pt}%
\definecolor{currentstroke}{rgb}{0.000000,0.000000,0.000000}%
\pgfsetstrokecolor{currentstroke}%
\pgfsetdash{}{0pt}%
\pgfsys@defobject{currentmarker}{\pgfqpoint{0.000000in}{-0.048611in}}{\pgfqpoint{0.000000in}{0.000000in}}{%
\pgfpathmoveto{\pgfqpoint{0.000000in}{0.000000in}}%
\pgfpathlineto{\pgfqpoint{0.000000in}{-0.048611in}}%
\pgfusepath{stroke,fill}%
}%
\begin{pgfscope}%
\pgfsys@transformshift{5.376144in}{0.929220in}%
\pgfsys@useobject{currentmarker}{}%
\end{pgfscope}%
\end{pgfscope}%
\begin{pgfscope}%
\definecolor{textcolor}{rgb}{0.000000,0.000000,0.000000}%
\pgfsetstrokecolor{textcolor}%
\pgfsetfillcolor{textcolor}%
\pgftext[x=5.376144in,y=0.831998in,,top]{\color{textcolor}\rmfamily\fontsize{28.000000}{33.600000}\selectfont 60}%
\end{pgfscope}%
\begin{pgfscope}%
\pgfpathrectangle{\pgfqpoint{1.157239in}{0.929220in}}{\pgfqpoint{5.649043in}{5.070780in}}%
\pgfusepath{clip}%
\pgfsetrectcap%
\pgfsetroundjoin%
\pgfsetlinewidth{0.803000pt}%
\definecolor{currentstroke}{rgb}{0.690196,0.690196,0.690196}%
\pgfsetstrokecolor{currentstroke}%
\pgfsetdash{}{0pt}%
\pgfpathmoveto{\pgfqpoint{6.806282in}{0.929220in}}%
\pgfpathlineto{\pgfqpoint{6.806282in}{6.000000in}}%
\pgfusepath{stroke}%
\end{pgfscope}%
\begin{pgfscope}%
\pgfsetbuttcap%
\pgfsetroundjoin%
\definecolor{currentfill}{rgb}{0.000000,0.000000,0.000000}%
\pgfsetfillcolor{currentfill}%
\pgfsetlinewidth{0.803000pt}%
\definecolor{currentstroke}{rgb}{0.000000,0.000000,0.000000}%
\pgfsetstrokecolor{currentstroke}%
\pgfsetdash{}{0pt}%
\pgfsys@defobject{currentmarker}{\pgfqpoint{0.000000in}{-0.048611in}}{\pgfqpoint{0.000000in}{0.000000in}}{%
\pgfpathmoveto{\pgfqpoint{0.000000in}{0.000000in}}%
\pgfpathlineto{\pgfqpoint{0.000000in}{-0.048611in}}%
\pgfusepath{stroke,fill}%
}%
\begin{pgfscope}%
\pgfsys@transformshift{6.806282in}{0.929220in}%
\pgfsys@useobject{currentmarker}{}%
\end{pgfscope}%
\end{pgfscope}%
\begin{pgfscope}%
\definecolor{textcolor}{rgb}{0.000000,0.000000,0.000000}%
\pgfsetstrokecolor{textcolor}%
\pgfsetfillcolor{textcolor}%
\pgftext[x=6.806282in,y=0.831998in,,top]{\color{textcolor}\rmfamily\fontsize{28.000000}{33.600000}\selectfont 80}%
\end{pgfscope}%
\begin{pgfscope}%
\definecolor{textcolor}{rgb}{0.000000,0.000000,0.000000}%
\pgfsetstrokecolor{textcolor}%
\pgfsetfillcolor{textcolor}%
\pgftext[x=3.981760in,y=0.385926in,,top]{\color{textcolor}\rmfamily\fontsize{28.000000}{33.600000}\selectfont time step \(\displaystyle k\)}%
\end{pgfscope}%
\begin{pgfscope}%
\pgfpathrectangle{\pgfqpoint{1.157239in}{0.929220in}}{\pgfqpoint{5.649043in}{5.070780in}}%
\pgfusepath{clip}%
\pgfsetrectcap%
\pgfsetroundjoin%
\pgfsetlinewidth{0.803000pt}%
\definecolor{currentstroke}{rgb}{0.690196,0.690196,0.690196}%
\pgfsetstrokecolor{currentstroke}%
\pgfsetdash{}{0pt}%
\pgfpathmoveto{\pgfqpoint{1.157239in}{0.929220in}}%
\pgfpathlineto{\pgfqpoint{6.806282in}{0.929220in}}%
\pgfusepath{stroke}%
\end{pgfscope}%
\begin{pgfscope}%
\pgfsetbuttcap%
\pgfsetroundjoin%
\definecolor{currentfill}{rgb}{0.000000,0.000000,0.000000}%
\pgfsetfillcolor{currentfill}%
\pgfsetlinewidth{0.803000pt}%
\definecolor{currentstroke}{rgb}{0.000000,0.000000,0.000000}%
\pgfsetstrokecolor{currentstroke}%
\pgfsetdash{}{0pt}%
\pgfsys@defobject{currentmarker}{\pgfqpoint{-0.048611in}{0.000000in}}{\pgfqpoint{-0.000000in}{0.000000in}}{%
\pgfpathmoveto{\pgfqpoint{-0.000000in}{0.000000in}}%
\pgfpathlineto{\pgfqpoint{-0.048611in}{0.000000in}}%
\pgfusepath{stroke,fill}%
}%
\begin{pgfscope}%
\pgfsys@transformshift{1.157239in}{0.929220in}%
\pgfsys@useobject{currentmarker}{}%
\end{pgfscope}%
\end{pgfscope}%
\begin{pgfscope}%
\definecolor{textcolor}{rgb}{0.000000,0.000000,0.000000}%
\pgfsetstrokecolor{textcolor}%
\pgfsetfillcolor{textcolor}%
\pgftext[x=0.811260in, y=0.809236in, left, base]{\color{textcolor}\rmfamily\fontsize{28.000000}{33.600000}\selectfont 0}%
\end{pgfscope}%
\begin{pgfscope}%
\pgfpathrectangle{\pgfqpoint{1.157239in}{0.929220in}}{\pgfqpoint{5.649043in}{5.070780in}}%
\pgfusepath{clip}%
\pgfsetrectcap%
\pgfsetroundjoin%
\pgfsetlinewidth{0.803000pt}%
\definecolor{currentstroke}{rgb}{0.690196,0.690196,0.690196}%
\pgfsetstrokecolor{currentstroke}%
\pgfsetdash{}{0pt}%
\pgfpathmoveto{\pgfqpoint{1.157239in}{2.369251in}}%
\pgfpathlineto{\pgfqpoint{6.806282in}{2.369251in}}%
\pgfusepath{stroke}%
\end{pgfscope}%
\begin{pgfscope}%
\pgfsetbuttcap%
\pgfsetroundjoin%
\definecolor{currentfill}{rgb}{0.000000,0.000000,0.000000}%
\pgfsetfillcolor{currentfill}%
\pgfsetlinewidth{0.803000pt}%
\definecolor{currentstroke}{rgb}{0.000000,0.000000,0.000000}%
\pgfsetstrokecolor{currentstroke}%
\pgfsetdash{}{0pt}%
\pgfsys@defobject{currentmarker}{\pgfqpoint{-0.048611in}{0.000000in}}{\pgfqpoint{-0.000000in}{0.000000in}}{%
\pgfpathmoveto{\pgfqpoint{-0.000000in}{0.000000in}}%
\pgfpathlineto{\pgfqpoint{-0.048611in}{0.000000in}}%
\pgfusepath{stroke,fill}%
}%
\begin{pgfscope}%
\pgfsys@transformshift{1.157239in}{2.369251in}%
\pgfsys@useobject{currentmarker}{}%
\end{pgfscope}%
\end{pgfscope}%
\begin{pgfscope}%
\definecolor{textcolor}{rgb}{0.000000,0.000000,0.000000}%
\pgfsetstrokecolor{textcolor}%
\pgfsetfillcolor{textcolor}%
\pgftext[x=0.566517in, y=2.249266in, left, base]{\color{textcolor}\rmfamily\fontsize{28.000000}{33.600000}\selectfont 0.5}%
\end{pgfscope}%
\begin{pgfscope}%
\pgfpathrectangle{\pgfqpoint{1.157239in}{0.929220in}}{\pgfqpoint{5.649043in}{5.070780in}}%
\pgfusepath{clip}%
\pgfsetrectcap%
\pgfsetroundjoin%
\pgfsetlinewidth{0.803000pt}%
\definecolor{currentstroke}{rgb}{0.690196,0.690196,0.690196}%
\pgfsetstrokecolor{currentstroke}%
\pgfsetdash{}{0pt}%
\pgfpathmoveto{\pgfqpoint{1.157239in}{3.809281in}}%
\pgfpathlineto{\pgfqpoint{6.806282in}{3.809281in}}%
\pgfusepath{stroke}%
\end{pgfscope}%
\begin{pgfscope}%
\pgfsetbuttcap%
\pgfsetroundjoin%
\definecolor{currentfill}{rgb}{0.000000,0.000000,0.000000}%
\pgfsetfillcolor{currentfill}%
\pgfsetlinewidth{0.803000pt}%
\definecolor{currentstroke}{rgb}{0.000000,0.000000,0.000000}%
\pgfsetstrokecolor{currentstroke}%
\pgfsetdash{}{0pt}%
\pgfsys@defobject{currentmarker}{\pgfqpoint{-0.048611in}{0.000000in}}{\pgfqpoint{-0.000000in}{0.000000in}}{%
\pgfpathmoveto{\pgfqpoint{-0.000000in}{0.000000in}}%
\pgfpathlineto{\pgfqpoint{-0.048611in}{0.000000in}}%
\pgfusepath{stroke,fill}%
}%
\begin{pgfscope}%
\pgfsys@transformshift{1.157239in}{3.809281in}%
\pgfsys@useobject{currentmarker}{}%
\end{pgfscope}%
\end{pgfscope}%
\begin{pgfscope}%
\definecolor{textcolor}{rgb}{0.000000,0.000000,0.000000}%
\pgfsetstrokecolor{textcolor}%
\pgfsetfillcolor{textcolor}%
\pgftext[x=0.811260in, y=3.689297in, left, base]{\color{textcolor}\rmfamily\fontsize{28.000000}{33.600000}\selectfont 1}%
\end{pgfscope}%
\begin{pgfscope}%
\pgfpathrectangle{\pgfqpoint{1.157239in}{0.929220in}}{\pgfqpoint{5.649043in}{5.070780in}}%
\pgfusepath{clip}%
\pgfsetrectcap%
\pgfsetroundjoin%
\pgfsetlinewidth{0.803000pt}%
\definecolor{currentstroke}{rgb}{0.690196,0.690196,0.690196}%
\pgfsetstrokecolor{currentstroke}%
\pgfsetdash{}{0pt}%
\pgfpathmoveto{\pgfqpoint{1.157239in}{5.249312in}}%
\pgfpathlineto{\pgfqpoint{6.806282in}{5.249312in}}%
\pgfusepath{stroke}%
\end{pgfscope}%
\begin{pgfscope}%
\pgfsetbuttcap%
\pgfsetroundjoin%
\definecolor{currentfill}{rgb}{0.000000,0.000000,0.000000}%
\pgfsetfillcolor{currentfill}%
\pgfsetlinewidth{0.803000pt}%
\definecolor{currentstroke}{rgb}{0.000000,0.000000,0.000000}%
\pgfsetstrokecolor{currentstroke}%
\pgfsetdash{}{0pt}%
\pgfsys@defobject{currentmarker}{\pgfqpoint{-0.048611in}{0.000000in}}{\pgfqpoint{-0.000000in}{0.000000in}}{%
\pgfpathmoveto{\pgfqpoint{-0.000000in}{0.000000in}}%
\pgfpathlineto{\pgfqpoint{-0.048611in}{0.000000in}}%
\pgfusepath{stroke,fill}%
}%
\begin{pgfscope}%
\pgfsys@transformshift{1.157239in}{5.249312in}%
\pgfsys@useobject{currentmarker}{}%
\end{pgfscope}%
\end{pgfscope}%
\begin{pgfscope}%
\definecolor{textcolor}{rgb}{0.000000,0.000000,0.000000}%
\pgfsetstrokecolor{textcolor}%
\pgfsetfillcolor{textcolor}%
\pgftext[x=0.566517in, y=5.129327in, left, base]{\color{textcolor}\rmfamily\fontsize{28.000000}{33.600000}\selectfont 1.5}%
\end{pgfscope}%
\begin{pgfscope}%
\definecolor{textcolor}{rgb}{0.000000,0.000000,0.000000}%
\pgfsetstrokecolor{textcolor}%
\pgfsetfillcolor{textcolor}%
\pgftext[x=0.427628in,y=3.464610in,,bottom,rotate=90.000000]{\color{textcolor}\rmfamily\fontsize{28.000000}{33.600000}\selectfont Localization Error [m]}%
\end{pgfscope}%
\begin{pgfscope}%
\pgfpathrectangle{\pgfqpoint{1.157239in}{0.929220in}}{\pgfqpoint{5.649043in}{5.070780in}}%
\pgfusepath{clip}%
\pgfsetrectcap%
\pgfsetroundjoin%
\pgfsetlinewidth{1.505625pt}%
\definecolor{currentstroke}{rgb}{0.750000,0.750000,0.000000}%
\pgfsetstrokecolor{currentstroke}%
\pgfsetdash{}{0pt}%
\pgfpathmoveto{\pgfqpoint{1.157239in}{0.929220in}}%
\pgfpathlineto{\pgfqpoint{1.228745in}{0.929220in}}%
\pgfpathlineto{\pgfqpoint{1.300252in}{0.929220in}}%
\pgfpathlineto{\pgfqpoint{1.371759in}{0.929220in}}%
\pgfpathlineto{\pgfqpoint{1.443266in}{0.929220in}}%
\pgfpathlineto{\pgfqpoint{1.514773in}{0.929220in}}%
\pgfpathlineto{\pgfqpoint{1.586280in}{0.929220in}}%
\pgfpathlineto{\pgfqpoint{1.657787in}{0.929220in}}%
\pgfpathlineto{\pgfqpoint{1.729294in}{0.929220in}}%
\pgfpathlineto{\pgfqpoint{1.800800in}{3.621261in}}%
\pgfpathlineto{\pgfqpoint{1.872307in}{2.903671in}}%
\pgfpathlineto{\pgfqpoint{1.943814in}{3.104607in}}%
\pgfpathlineto{\pgfqpoint{2.015321in}{3.104437in}}%
\pgfpathlineto{\pgfqpoint{2.086828in}{3.000104in}}%
\pgfpathlineto{\pgfqpoint{2.158335in}{3.132250in}}%
\pgfpathlineto{\pgfqpoint{2.229842in}{3.038571in}}%
\pgfpathlineto{\pgfqpoint{2.301349in}{3.266424in}}%
\pgfpathlineto{\pgfqpoint{2.372855in}{3.439229in}}%
\pgfpathlineto{\pgfqpoint{2.444362in}{3.183633in}}%
\pgfpathlineto{\pgfqpoint{2.515869in}{3.535185in}}%
\pgfpathlineto{\pgfqpoint{2.587376in}{3.446080in}}%
\pgfpathlineto{\pgfqpoint{2.658883in}{3.538158in}}%
\pgfpathlineto{\pgfqpoint{2.730390in}{3.352120in}}%
\pgfpathlineto{\pgfqpoint{2.801897in}{3.372992in}}%
\pgfpathlineto{\pgfqpoint{2.873404in}{3.866793in}}%
\pgfpathlineto{\pgfqpoint{2.944910in}{3.664126in}}%
\pgfpathlineto{\pgfqpoint{3.016417in}{3.480835in}}%
\pgfpathlineto{\pgfqpoint{3.087924in}{4.029502in}}%
\pgfpathlineto{\pgfqpoint{3.159431in}{3.497393in}}%
\pgfpathlineto{\pgfqpoint{3.230938in}{3.702716in}}%
\pgfpathlineto{\pgfqpoint{3.302445in}{3.621074in}}%
\pgfpathlineto{\pgfqpoint{3.373952in}{3.147534in}}%
\pgfpathlineto{\pgfqpoint{3.445459in}{3.391782in}}%
\pgfpathlineto{\pgfqpoint{3.516966in}{3.624709in}}%
\pgfpathlineto{\pgfqpoint{3.588472in}{3.402377in}}%
\pgfpathlineto{\pgfqpoint{3.659979in}{3.777877in}}%
\pgfpathlineto{\pgfqpoint{3.731486in}{3.647078in}}%
\pgfpathlineto{\pgfqpoint{3.802993in}{3.774710in}}%
\pgfpathlineto{\pgfqpoint{3.874500in}{3.789475in}}%
\pgfpathlineto{\pgfqpoint{3.946007in}{3.689347in}}%
\pgfpathlineto{\pgfqpoint{4.017514in}{4.099667in}}%
\pgfpathlineto{\pgfqpoint{4.089021in}{3.770211in}}%
\pgfpathlineto{\pgfqpoint{4.160527in}{3.726344in}}%
\pgfpathlineto{\pgfqpoint{4.232034in}{3.634831in}}%
\pgfpathlineto{\pgfqpoint{4.303541in}{3.755387in}}%
\pgfpathlineto{\pgfqpoint{4.375048in}{3.342912in}}%
\pgfpathlineto{\pgfqpoint{4.446555in}{3.555861in}}%
\pgfpathlineto{\pgfqpoint{4.518062in}{3.639525in}}%
\pgfpathlineto{\pgfqpoint{4.589569in}{4.013515in}}%
\pgfpathlineto{\pgfqpoint{4.661076in}{3.666048in}}%
\pgfpathlineto{\pgfqpoint{4.732582in}{3.823564in}}%
\pgfpathlineto{\pgfqpoint{4.804089in}{3.636203in}}%
\pgfpathlineto{\pgfqpoint{4.875596in}{3.736275in}}%
\pgfpathlineto{\pgfqpoint{4.947103in}{3.800645in}}%
\pgfpathlineto{\pgfqpoint{5.018610in}{3.036226in}}%
\pgfpathlineto{\pgfqpoint{5.090117in}{3.247245in}}%
\pgfpathlineto{\pgfqpoint{5.161624in}{3.187969in}}%
\pgfpathlineto{\pgfqpoint{5.233131in}{3.257497in}}%
\pgfpathlineto{\pgfqpoint{5.304638in}{3.347950in}}%
\pgfpathlineto{\pgfqpoint{5.376144in}{2.976046in}}%
\pgfpathlineto{\pgfqpoint{5.447651in}{2.711887in}}%
\pgfpathlineto{\pgfqpoint{5.519158in}{3.076598in}}%
\pgfpathlineto{\pgfqpoint{5.590665in}{3.026385in}}%
\pgfpathlineto{\pgfqpoint{5.662172in}{3.007540in}}%
\pgfpathlineto{\pgfqpoint{5.733679in}{2.214634in}}%
\pgfpathlineto{\pgfqpoint{5.805186in}{2.638612in}}%
\pgfpathlineto{\pgfqpoint{5.876693in}{2.918287in}}%
\pgfpathlineto{\pgfqpoint{5.948199in}{2.668035in}}%
\pgfpathlineto{\pgfqpoint{6.019706in}{2.627185in}}%
\pgfpathlineto{\pgfqpoint{6.091213in}{0.929220in}}%
\pgfpathlineto{\pgfqpoint{6.162720in}{0.929220in}}%
\pgfpathlineto{\pgfqpoint{6.234227in}{0.929220in}}%
\pgfpathlineto{\pgfqpoint{6.305734in}{0.929220in}}%
\pgfpathlineto{\pgfqpoint{6.377241in}{0.929220in}}%
\pgfpathlineto{\pgfqpoint{6.448748in}{0.929220in}}%
\pgfpathlineto{\pgfqpoint{6.520254in}{0.929220in}}%
\pgfpathlineto{\pgfqpoint{6.591761in}{0.929220in}}%
\pgfpathlineto{\pgfqpoint{6.663268in}{0.929220in}}%
\pgfpathlineto{\pgfqpoint{6.734775in}{0.929220in}}%
\pgfpathlineto{\pgfqpoint{6.806282in}{0.929220in}}%
\pgfusepath{stroke}%
\end{pgfscope}%
\begin{pgfscope}%
\pgfpathrectangle{\pgfqpoint{1.157239in}{0.929220in}}{\pgfqpoint{5.649043in}{5.070780in}}%
\pgfusepath{clip}%
\pgfsetrectcap%
\pgfsetroundjoin%
\pgfsetlinewidth{1.505625pt}%
\definecolor{currentstroke}{rgb}{0.000000,0.000000,1.000000}%
\pgfsetstrokecolor{currentstroke}%
\pgfsetdash{}{0pt}%
\pgfpathmoveto{\pgfqpoint{1.157239in}{0.929220in}}%
\pgfpathlineto{\pgfqpoint{1.228745in}{0.929220in}}%
\pgfpathlineto{\pgfqpoint{1.300252in}{0.929220in}}%
\pgfpathlineto{\pgfqpoint{1.371759in}{0.929220in}}%
\pgfpathlineto{\pgfqpoint{1.443266in}{0.929220in}}%
\pgfpathlineto{\pgfqpoint{1.514773in}{0.929220in}}%
\pgfpathlineto{\pgfqpoint{1.586280in}{0.929220in}}%
\pgfpathlineto{\pgfqpoint{1.657787in}{0.929220in}}%
\pgfpathlineto{\pgfqpoint{1.729294in}{0.929220in}}%
\pgfpathlineto{\pgfqpoint{1.800800in}{0.929220in}}%
\pgfpathlineto{\pgfqpoint{1.872307in}{1.125985in}}%
\pgfpathlineto{\pgfqpoint{1.943814in}{3.968049in}}%
\pgfpathlineto{\pgfqpoint{2.015321in}{3.290341in}}%
\pgfpathlineto{\pgfqpoint{2.086828in}{3.109582in}}%
\pgfpathlineto{\pgfqpoint{2.158335in}{3.313649in}}%
\pgfpathlineto{\pgfqpoint{2.229842in}{3.969051in}}%
\pgfpathlineto{\pgfqpoint{2.301349in}{4.425517in}}%
\pgfpathlineto{\pgfqpoint{2.372855in}{4.289665in}}%
\pgfpathlineto{\pgfqpoint{2.444362in}{4.053625in}}%
\pgfpathlineto{\pgfqpoint{2.515869in}{4.043184in}}%
\pgfpathlineto{\pgfqpoint{2.587376in}{4.290556in}}%
\pgfpathlineto{\pgfqpoint{2.658883in}{4.951874in}}%
\pgfpathlineto{\pgfqpoint{2.730390in}{5.200209in}}%
\pgfpathlineto{\pgfqpoint{2.801897in}{5.028772in}}%
\pgfpathlineto{\pgfqpoint{2.873404in}{4.990611in}}%
\pgfpathlineto{\pgfqpoint{2.944910in}{5.130882in}}%
\pgfpathlineto{\pgfqpoint{3.016417in}{5.511282in}}%
\pgfpathlineto{\pgfqpoint{3.087924in}{5.395209in}}%
\pgfpathlineto{\pgfqpoint{3.159431in}{5.562655in}}%
\pgfpathlineto{\pgfqpoint{3.230938in}{5.758534in}}%
\pgfpathlineto{\pgfqpoint{3.302445in}{5.512110in}}%
\pgfpathlineto{\pgfqpoint{3.373952in}{5.521539in}}%
\pgfpathlineto{\pgfqpoint{3.445459in}{5.394315in}}%
\pgfpathlineto{\pgfqpoint{3.516966in}{5.161480in}}%
\pgfpathlineto{\pgfqpoint{3.588472in}{4.791056in}}%
\pgfpathlineto{\pgfqpoint{3.659979in}{5.168288in}}%
\pgfpathlineto{\pgfqpoint{3.731486in}{5.289264in}}%
\pgfpathlineto{\pgfqpoint{3.802993in}{5.052699in}}%
\pgfpathlineto{\pgfqpoint{3.874500in}{5.092983in}}%
\pgfpathlineto{\pgfqpoint{3.946007in}{5.051629in}}%
\pgfpathlineto{\pgfqpoint{4.017514in}{4.919583in}}%
\pgfpathlineto{\pgfqpoint{4.089021in}{4.578764in}}%
\pgfpathlineto{\pgfqpoint{4.160527in}{4.614788in}}%
\pgfpathlineto{\pgfqpoint{4.232034in}{4.820519in}}%
\pgfpathlineto{\pgfqpoint{4.303541in}{4.455300in}}%
\pgfpathlineto{\pgfqpoint{4.375048in}{4.400120in}}%
\pgfpathlineto{\pgfqpoint{4.446555in}{4.475883in}}%
\pgfpathlineto{\pgfqpoint{4.518062in}{4.519300in}}%
\pgfpathlineto{\pgfqpoint{4.589569in}{4.878537in}}%
\pgfpathlineto{\pgfqpoint{4.661076in}{4.933028in}}%
\pgfpathlineto{\pgfqpoint{4.732582in}{5.148906in}}%
\pgfpathlineto{\pgfqpoint{4.804089in}{5.113810in}}%
\pgfpathlineto{\pgfqpoint{4.875596in}{5.105923in}}%
\pgfpathlineto{\pgfqpoint{4.947103in}{5.029818in}}%
\pgfpathlineto{\pgfqpoint{5.018610in}{4.666060in}}%
\pgfpathlineto{\pgfqpoint{5.090117in}{4.566519in}}%
\pgfpathlineto{\pgfqpoint{5.161624in}{4.591549in}}%
\pgfpathlineto{\pgfqpoint{5.233131in}{4.508661in}}%
\pgfpathlineto{\pgfqpoint{5.304638in}{4.596088in}}%
\pgfpathlineto{\pgfqpoint{5.376144in}{3.522734in}}%
\pgfpathlineto{\pgfqpoint{5.447651in}{3.596737in}}%
\pgfpathlineto{\pgfqpoint{5.519158in}{3.480678in}}%
\pgfpathlineto{\pgfqpoint{5.590665in}{3.537077in}}%
\pgfpathlineto{\pgfqpoint{5.662172in}{3.485109in}}%
\pgfpathlineto{\pgfqpoint{5.733679in}{2.670804in}}%
\pgfpathlineto{\pgfqpoint{5.805186in}{2.837747in}}%
\pgfpathlineto{\pgfqpoint{5.876693in}{2.813405in}}%
\pgfpathlineto{\pgfqpoint{5.948199in}{2.756286in}}%
\pgfpathlineto{\pgfqpoint{6.019706in}{2.803932in}}%
\pgfpathlineto{\pgfqpoint{6.091213in}{0.929220in}}%
\pgfpathlineto{\pgfqpoint{6.162720in}{0.929220in}}%
\pgfpathlineto{\pgfqpoint{6.234227in}{0.929220in}}%
\pgfpathlineto{\pgfqpoint{6.305734in}{0.929220in}}%
\pgfpathlineto{\pgfqpoint{6.377241in}{0.929220in}}%
\pgfpathlineto{\pgfqpoint{6.448748in}{0.929220in}}%
\pgfpathlineto{\pgfqpoint{6.520254in}{0.929220in}}%
\pgfpathlineto{\pgfqpoint{6.591761in}{0.929220in}}%
\pgfpathlineto{\pgfqpoint{6.663268in}{0.929220in}}%
\pgfpathlineto{\pgfqpoint{6.734775in}{0.929220in}}%
\pgfpathlineto{\pgfqpoint{6.806282in}{0.929220in}}%
\pgfusepath{stroke}%
\end{pgfscope}%
\begin{pgfscope}%
\pgfpathrectangle{\pgfqpoint{1.157239in}{0.929220in}}{\pgfqpoint{5.649043in}{5.070780in}}%
\pgfusepath{clip}%
\pgfsetrectcap%
\pgfsetroundjoin%
\pgfsetlinewidth{1.505625pt}%
\definecolor{currentstroke}{rgb}{0.000000,0.500000,0.000000}%
\pgfsetstrokecolor{currentstroke}%
\pgfsetdash{}{0pt}%
\pgfpathmoveto{\pgfqpoint{1.157239in}{0.929220in}}%
\pgfpathlineto{\pgfqpoint{1.228745in}{0.929220in}}%
\pgfpathlineto{\pgfqpoint{1.300252in}{0.929220in}}%
\pgfpathlineto{\pgfqpoint{1.371759in}{0.929220in}}%
\pgfpathlineto{\pgfqpoint{1.443266in}{0.929220in}}%
\pgfpathlineto{\pgfqpoint{1.514773in}{0.929220in}}%
\pgfpathlineto{\pgfqpoint{1.586280in}{0.929220in}}%
\pgfpathlineto{\pgfqpoint{1.657787in}{0.929220in}}%
\pgfpathlineto{\pgfqpoint{1.729294in}{0.929220in}}%
\pgfpathlineto{\pgfqpoint{1.800800in}{0.929220in}}%
\pgfpathlineto{\pgfqpoint{1.872307in}{1.155831in}}%
\pgfpathlineto{\pgfqpoint{1.943814in}{4.025047in}}%
\pgfpathlineto{\pgfqpoint{2.015321in}{3.331822in}}%
\pgfpathlineto{\pgfqpoint{2.086828in}{3.629115in}}%
\pgfpathlineto{\pgfqpoint{2.158335in}{3.138405in}}%
\pgfpathlineto{\pgfqpoint{2.229842in}{3.552118in}}%
\pgfpathlineto{\pgfqpoint{2.301349in}{4.129612in}}%
\pgfpathlineto{\pgfqpoint{2.372855in}{4.134537in}}%
\pgfpathlineto{\pgfqpoint{2.444362in}{3.896437in}}%
\pgfpathlineto{\pgfqpoint{2.515869in}{3.903799in}}%
\pgfpathlineto{\pgfqpoint{2.587376in}{4.158892in}}%
\pgfpathlineto{\pgfqpoint{2.658883in}{4.614352in}}%
\pgfpathlineto{\pgfqpoint{2.730390in}{4.597091in}}%
\pgfpathlineto{\pgfqpoint{2.801897in}{4.499698in}}%
\pgfpathlineto{\pgfqpoint{2.873404in}{4.364706in}}%
\pgfpathlineto{\pgfqpoint{2.944910in}{4.576685in}}%
\pgfpathlineto{\pgfqpoint{3.016417in}{5.184360in}}%
\pgfpathlineto{\pgfqpoint{3.087924in}{5.213807in}}%
\pgfpathlineto{\pgfqpoint{3.159431in}{5.235972in}}%
\pgfpathlineto{\pgfqpoint{3.230938in}{5.130393in}}%
\pgfpathlineto{\pgfqpoint{3.302445in}{5.010202in}}%
\pgfpathlineto{\pgfqpoint{3.373952in}{4.952367in}}%
\pgfpathlineto{\pgfqpoint{3.445459in}{4.795807in}}%
\pgfpathlineto{\pgfqpoint{3.516966in}{4.902054in}}%
\pgfpathlineto{\pgfqpoint{3.588472in}{4.758515in}}%
\pgfpathlineto{\pgfqpoint{3.659979in}{4.929232in}}%
\pgfpathlineto{\pgfqpoint{3.731486in}{4.957563in}}%
\pgfpathlineto{\pgfqpoint{3.802993in}{5.052379in}}%
\pgfpathlineto{\pgfqpoint{3.874500in}{4.955206in}}%
\pgfpathlineto{\pgfqpoint{3.946007in}{4.902071in}}%
\pgfpathlineto{\pgfqpoint{4.017514in}{4.647093in}}%
\pgfpathlineto{\pgfqpoint{4.089021in}{4.794244in}}%
\pgfpathlineto{\pgfqpoint{4.160527in}{4.848291in}}%
\pgfpathlineto{\pgfqpoint{4.232034in}{4.806676in}}%
\pgfpathlineto{\pgfqpoint{4.303541in}{4.812353in}}%
\pgfpathlineto{\pgfqpoint{4.375048in}{4.624369in}}%
\pgfpathlineto{\pgfqpoint{4.446555in}{4.482062in}}%
\pgfpathlineto{\pgfqpoint{4.518062in}{4.323046in}}%
\pgfpathlineto{\pgfqpoint{4.589569in}{4.187969in}}%
\pgfpathlineto{\pgfqpoint{4.661076in}{4.230005in}}%
\pgfpathlineto{\pgfqpoint{4.732582in}{4.324597in}}%
\pgfpathlineto{\pgfqpoint{4.804089in}{4.320618in}}%
\pgfpathlineto{\pgfqpoint{4.875596in}{4.193458in}}%
\pgfpathlineto{\pgfqpoint{4.947103in}{4.213298in}}%
\pgfpathlineto{\pgfqpoint{5.018610in}{3.629550in}}%
\pgfpathlineto{\pgfqpoint{5.090117in}{3.625715in}}%
\pgfpathlineto{\pgfqpoint{5.161624in}{3.589488in}}%
\pgfpathlineto{\pgfqpoint{5.233131in}{3.540966in}}%
\pgfpathlineto{\pgfqpoint{5.304638in}{3.472615in}}%
\pgfpathlineto{\pgfqpoint{5.376144in}{2.885759in}}%
\pgfpathlineto{\pgfqpoint{5.447651in}{2.826949in}}%
\pgfpathlineto{\pgfqpoint{5.519158in}{2.893206in}}%
\pgfpathlineto{\pgfqpoint{5.590665in}{2.831349in}}%
\pgfpathlineto{\pgfqpoint{5.662172in}{2.801817in}}%
\pgfpathlineto{\pgfqpoint{5.733679in}{2.203959in}}%
\pgfpathlineto{\pgfqpoint{5.805186in}{2.242145in}}%
\pgfpathlineto{\pgfqpoint{5.876693in}{2.233887in}}%
\pgfpathlineto{\pgfqpoint{5.948199in}{2.210512in}}%
\pgfpathlineto{\pgfqpoint{6.019706in}{2.169877in}}%
\pgfpathlineto{\pgfqpoint{6.091213in}{0.929220in}}%
\pgfpathlineto{\pgfqpoint{6.162720in}{0.929220in}}%
\pgfpathlineto{\pgfqpoint{6.234227in}{0.929220in}}%
\pgfpathlineto{\pgfqpoint{6.305734in}{0.929220in}}%
\pgfpathlineto{\pgfqpoint{6.377241in}{0.929220in}}%
\pgfpathlineto{\pgfqpoint{6.448748in}{0.929220in}}%
\pgfpathlineto{\pgfqpoint{6.520254in}{0.929220in}}%
\pgfpathlineto{\pgfqpoint{6.591761in}{0.929220in}}%
\pgfpathlineto{\pgfqpoint{6.663268in}{0.929220in}}%
\pgfpathlineto{\pgfqpoint{6.734775in}{0.929220in}}%
\pgfpathlineto{\pgfqpoint{6.806282in}{0.929220in}}%
\pgfusepath{stroke}%
\end{pgfscope}%
\begin{pgfscope}%
\pgfpathrectangle{\pgfqpoint{1.157239in}{0.929220in}}{\pgfqpoint{5.649043in}{5.070780in}}%
\pgfusepath{clip}%
\pgfsetrectcap%
\pgfsetroundjoin%
\pgfsetlinewidth{1.505625pt}%
\definecolor{currentstroke}{rgb}{1.000000,0.000000,0.000000}%
\pgfsetstrokecolor{currentstroke}%
\pgfsetdash{}{0pt}%
\pgfpathmoveto{\pgfqpoint{1.157239in}{0.929220in}}%
\pgfpathlineto{\pgfqpoint{1.228745in}{0.929220in}}%
\pgfpathlineto{\pgfqpoint{1.300252in}{0.929220in}}%
\pgfpathlineto{\pgfqpoint{1.371759in}{0.929220in}}%
\pgfpathlineto{\pgfqpoint{1.443266in}{0.929220in}}%
\pgfpathlineto{\pgfqpoint{1.514773in}{0.929220in}}%
\pgfpathlineto{\pgfqpoint{1.586280in}{0.929220in}}%
\pgfpathlineto{\pgfqpoint{1.657787in}{0.929220in}}%
\pgfpathlineto{\pgfqpoint{1.729294in}{0.929220in}}%
\pgfpathlineto{\pgfqpoint{1.800800in}{1.696352in}}%
\pgfpathlineto{\pgfqpoint{1.872307in}{1.506534in}}%
\pgfpathlineto{\pgfqpoint{1.943814in}{1.583428in}}%
\pgfpathlineto{\pgfqpoint{2.015321in}{1.636960in}}%
\pgfpathlineto{\pgfqpoint{2.086828in}{1.694135in}}%
\pgfpathlineto{\pgfqpoint{2.158335in}{2.134549in}}%
\pgfpathlineto{\pgfqpoint{2.229842in}{2.096876in}}%
\pgfpathlineto{\pgfqpoint{2.301349in}{2.034341in}}%
\pgfpathlineto{\pgfqpoint{2.372855in}{1.989477in}}%
\pgfpathlineto{\pgfqpoint{2.444362in}{1.892434in}}%
\pgfpathlineto{\pgfqpoint{2.515869in}{2.555573in}}%
\pgfpathlineto{\pgfqpoint{2.587376in}{2.603529in}}%
\pgfpathlineto{\pgfqpoint{2.658883in}{2.591360in}}%
\pgfpathlineto{\pgfqpoint{2.730390in}{2.634134in}}%
\pgfpathlineto{\pgfqpoint{2.801897in}{2.636654in}}%
\pgfpathlineto{\pgfqpoint{2.873404in}{3.504360in}}%
\pgfpathlineto{\pgfqpoint{2.944910in}{3.635315in}}%
\pgfpathlineto{\pgfqpoint{3.016417in}{3.579867in}}%
\pgfpathlineto{\pgfqpoint{3.087924in}{3.579476in}}%
\pgfpathlineto{\pgfqpoint{3.159431in}{3.697490in}}%
\pgfpathlineto{\pgfqpoint{3.230938in}{3.846346in}}%
\pgfpathlineto{\pgfqpoint{3.302445in}{3.823510in}}%
\pgfpathlineto{\pgfqpoint{3.373952in}{3.839194in}}%
\pgfpathlineto{\pgfqpoint{3.445459in}{3.870723in}}%
\pgfpathlineto{\pgfqpoint{3.516966in}{3.965933in}}%
\pgfpathlineto{\pgfqpoint{3.588472in}{3.881566in}}%
\pgfpathlineto{\pgfqpoint{3.659979in}{3.934924in}}%
\pgfpathlineto{\pgfqpoint{3.731486in}{3.949248in}}%
\pgfpathlineto{\pgfqpoint{3.802993in}{4.154210in}}%
\pgfpathlineto{\pgfqpoint{3.874500in}{4.192124in}}%
\pgfpathlineto{\pgfqpoint{3.946007in}{4.269918in}}%
\pgfpathlineto{\pgfqpoint{4.017514in}{4.097316in}}%
\pgfpathlineto{\pgfqpoint{4.089021in}{4.294706in}}%
\pgfpathlineto{\pgfqpoint{4.160527in}{4.153133in}}%
\pgfpathlineto{\pgfqpoint{4.232034in}{4.008161in}}%
\pgfpathlineto{\pgfqpoint{4.303541in}{3.902206in}}%
\pgfpathlineto{\pgfqpoint{4.375048in}{3.880064in}}%
\pgfpathlineto{\pgfqpoint{4.446555in}{3.784296in}}%
\pgfpathlineto{\pgfqpoint{4.518062in}{3.683447in}}%
\pgfpathlineto{\pgfqpoint{4.589569in}{3.529933in}}%
\pgfpathlineto{\pgfqpoint{4.661076in}{3.383097in}}%
\pgfpathlineto{\pgfqpoint{4.732582in}{3.370885in}}%
\pgfpathlineto{\pgfqpoint{4.804089in}{3.227442in}}%
\pgfpathlineto{\pgfqpoint{4.875596in}{3.213649in}}%
\pgfpathlineto{\pgfqpoint{4.947103in}{3.026940in}}%
\pgfpathlineto{\pgfqpoint{5.018610in}{2.505180in}}%
\pgfpathlineto{\pgfqpoint{5.090117in}{2.363628in}}%
\pgfpathlineto{\pgfqpoint{5.161624in}{2.352894in}}%
\pgfpathlineto{\pgfqpoint{5.233131in}{2.340958in}}%
\pgfpathlineto{\pgfqpoint{5.304638in}{2.306498in}}%
\pgfpathlineto{\pgfqpoint{5.376144in}{1.860329in}}%
\pgfpathlineto{\pgfqpoint{5.447651in}{1.793212in}}%
\pgfpathlineto{\pgfqpoint{5.519158in}{1.741269in}}%
\pgfpathlineto{\pgfqpoint{5.590665in}{1.720817in}}%
\pgfpathlineto{\pgfqpoint{5.662172in}{1.699250in}}%
\pgfpathlineto{\pgfqpoint{5.733679in}{1.366768in}}%
\pgfpathlineto{\pgfqpoint{5.805186in}{1.353306in}}%
\pgfpathlineto{\pgfqpoint{5.876693in}{1.353024in}}%
\pgfpathlineto{\pgfqpoint{5.948199in}{1.361541in}}%
\pgfpathlineto{\pgfqpoint{6.019706in}{1.356811in}}%
\pgfpathlineto{\pgfqpoint{6.091213in}{0.929220in}}%
\pgfpathlineto{\pgfqpoint{6.162720in}{0.929220in}}%
\pgfpathlineto{\pgfqpoint{6.234227in}{0.929220in}}%
\pgfpathlineto{\pgfqpoint{6.305734in}{0.929220in}}%
\pgfpathlineto{\pgfqpoint{6.377241in}{0.929220in}}%
\pgfpathlineto{\pgfqpoint{6.448748in}{0.929220in}}%
\pgfpathlineto{\pgfqpoint{6.520254in}{0.929220in}}%
\pgfpathlineto{\pgfqpoint{6.591761in}{0.929220in}}%
\pgfpathlineto{\pgfqpoint{6.663268in}{0.929220in}}%
\pgfpathlineto{\pgfqpoint{6.734775in}{0.929220in}}%
\pgfpathlineto{\pgfqpoint{6.806282in}{0.929220in}}%
\pgfusepath{stroke}%
\end{pgfscope}%
\begin{pgfscope}%
\pgfsetrectcap%
\pgfsetmiterjoin%
\pgfsetlinewidth{0.803000pt}%
\definecolor{currentstroke}{rgb}{0.000000,0.000000,0.000000}%
\pgfsetstrokecolor{currentstroke}%
\pgfsetdash{}{0pt}%
\pgfpathmoveto{\pgfqpoint{1.157239in}{0.929220in}}%
\pgfpathlineto{\pgfqpoint{1.157239in}{6.000000in}}%
\pgfusepath{stroke}%
\end{pgfscope}%
\begin{pgfscope}%
\pgfsetrectcap%
\pgfsetmiterjoin%
\pgfsetlinewidth{0.803000pt}%
\definecolor{currentstroke}{rgb}{0.000000,0.000000,0.000000}%
\pgfsetstrokecolor{currentstroke}%
\pgfsetdash{}{0pt}%
\pgfpathmoveto{\pgfqpoint{6.806282in}{0.929220in}}%
\pgfpathlineto{\pgfqpoint{6.806282in}{6.000000in}}%
\pgfusepath{stroke}%
\end{pgfscope}%
\begin{pgfscope}%
\pgfsetrectcap%
\pgfsetmiterjoin%
\pgfsetlinewidth{0.803000pt}%
\definecolor{currentstroke}{rgb}{0.000000,0.000000,0.000000}%
\pgfsetstrokecolor{currentstroke}%
\pgfsetdash{}{0pt}%
\pgfpathmoveto{\pgfqpoint{1.157239in}{0.929220in}}%
\pgfpathlineto{\pgfqpoint{6.806282in}{0.929220in}}%
\pgfusepath{stroke}%
\end{pgfscope}%
\begin{pgfscope}%
\pgfsetrectcap%
\pgfsetmiterjoin%
\pgfsetlinewidth{0.803000pt}%
\definecolor{currentstroke}{rgb}{0.000000,0.000000,0.000000}%
\pgfsetstrokecolor{currentstroke}%
\pgfsetdash{}{0pt}%
\pgfpathmoveto{\pgfqpoint{1.157239in}{6.000000in}}%
\pgfpathlineto{\pgfqpoint{6.806282in}{6.000000in}}%
\pgfusepath{stroke}%
\end{pgfscope}%
\end{pgfpicture}%
\makeatother%
\endgroup%

%% file: Figs/gospa_-10dB/gospa_false.pgf
\begingroup%
\makeatletter%
\begin{pgfpicture}%
\pgfpathrectangle{\pgfpointorigin}{\pgfqpoint{7.000000in}{6.000000in}}%
\pgfusepath{use as bounding box, clip}%
\begin{pgfscope}%
\pgfsetbuttcap%
\pgfsetmiterjoin%
\definecolor{currentfill}{rgb}{1.000000,1.000000,1.000000}%
\pgfsetfillcolor{currentfill}%
\pgfsetlinewidth{0.000000pt}%
\definecolor{currentstroke}{rgb}{1.000000,1.000000,1.000000}%
\pgfsetstrokecolor{currentstroke}%
\pgfsetdash{}{0pt}%
\pgfpathmoveto{\pgfqpoint{0.000000in}{0.000000in}}%
\pgfpathlineto{\pgfqpoint{7.000000in}{0.000000in}}%
\pgfpathlineto{\pgfqpoint{7.000000in}{6.000000in}}%
\pgfpathlineto{\pgfqpoint{0.000000in}{6.000000in}}%
\pgfpathlineto{\pgfqpoint{0.000000in}{0.000000in}}%
\pgfpathclose%
\pgfusepath{fill}%
\end{pgfscope}%
\begin{pgfscope}%
\pgfsetbuttcap%
\pgfsetmiterjoin%
\definecolor{currentfill}{rgb}{1.000000,1.000000,1.000000}%
\pgfsetfillcolor{currentfill}%
\pgfsetlinewidth{0.000000pt}%
\definecolor{currentstroke}{rgb}{0.000000,0.000000,0.000000}%
\pgfsetstrokecolor{currentstroke}%
\pgfsetstrokeopacity{0.000000}%
\pgfsetdash{}{0pt}%
\pgfpathmoveto{\pgfqpoint{1.157239in}{0.929220in}}%
\pgfpathlineto{\pgfqpoint{6.806282in}{0.929220in}}%
\pgfpathlineto{\pgfqpoint{6.806282in}{6.000000in}}%
\pgfpathlineto{\pgfqpoint{1.157239in}{6.000000in}}%
\pgfpathlineto{\pgfqpoint{1.157239in}{0.929220in}}%
\pgfpathclose%
\pgfusepath{fill}%
\end{pgfscope}%
\begin{pgfscope}%
\pgfpathrectangle{\pgfqpoint{1.157239in}{0.929220in}}{\pgfqpoint{5.649043in}{5.070780in}}%
\pgfusepath{clip}%
\pgfsetrectcap%
\pgfsetroundjoin%
\pgfsetlinewidth{0.803000pt}%
\definecolor{currentstroke}{rgb}{0.690196,0.690196,0.690196}%
\pgfsetstrokecolor{currentstroke}%
\pgfsetdash{}{0pt}%
\pgfpathmoveto{\pgfqpoint{2.515869in}{0.929220in}}%
\pgfpathlineto{\pgfqpoint{2.515869in}{6.000000in}}%
\pgfusepath{stroke}%
\end{pgfscope}%
\begin{pgfscope}%
\pgfsetbuttcap%
\pgfsetroundjoin%
\definecolor{currentfill}{rgb}{0.000000,0.000000,0.000000}%
\pgfsetfillcolor{currentfill}%
\pgfsetlinewidth{0.803000pt}%
\definecolor{currentstroke}{rgb}{0.000000,0.000000,0.000000}%
\pgfsetstrokecolor{currentstroke}%
\pgfsetdash{}{0pt}%
\pgfsys@defobject{currentmarker}{\pgfqpoint{0.000000in}{-0.048611in}}{\pgfqpoint{0.000000in}{0.000000in}}{%
\pgfpathmoveto{\pgfqpoint{0.000000in}{0.000000in}}%
\pgfpathlineto{\pgfqpoint{0.000000in}{-0.048611in}}%
\pgfusepath{stroke,fill}%
}%
\begin{pgfscope}%
\pgfsys@transformshift{2.515869in}{0.929220in}%
\pgfsys@useobject{currentmarker}{}%
\end{pgfscope}%
\end{pgfscope}%
\begin{pgfscope}%
\definecolor{textcolor}{rgb}{0.000000,0.000000,0.000000}%
\pgfsetstrokecolor{textcolor}%
\pgfsetfillcolor{textcolor}%
\pgftext[x=2.515869in,y=0.831998in,,top]{\color{textcolor}\rmfamily\fontsize{28.000000}{33.600000}\selectfont 20}%
\end{pgfscope}%
\begin{pgfscope}%
\pgfpathrectangle{\pgfqpoint{1.157239in}{0.929220in}}{\pgfqpoint{5.649043in}{5.070780in}}%
\pgfusepath{clip}%
\pgfsetrectcap%
\pgfsetroundjoin%
\pgfsetlinewidth{0.803000pt}%
\definecolor{currentstroke}{rgb}{0.690196,0.690196,0.690196}%
\pgfsetstrokecolor{currentstroke}%
\pgfsetdash{}{0pt}%
\pgfpathmoveto{\pgfqpoint{3.946007in}{0.929220in}}%
\pgfpathlineto{\pgfqpoint{3.946007in}{6.000000in}}%
\pgfusepath{stroke}%
\end{pgfscope}%
\begin{pgfscope}%
\pgfsetbuttcap%
\pgfsetroundjoin%
\definecolor{currentfill}{rgb}{0.000000,0.000000,0.000000}%
\pgfsetfillcolor{currentfill}%
\pgfsetlinewidth{0.803000pt}%
\definecolor{currentstroke}{rgb}{0.000000,0.000000,0.000000}%
\pgfsetstrokecolor{currentstroke}%
\pgfsetdash{}{0pt}%
\pgfsys@defobject{currentmarker}{\pgfqpoint{0.000000in}{-0.048611in}}{\pgfqpoint{0.000000in}{0.000000in}}{%
\pgfpathmoveto{\pgfqpoint{0.000000in}{0.000000in}}%
\pgfpathlineto{\pgfqpoint{0.000000in}{-0.048611in}}%
\pgfusepath{stroke,fill}%
}%
\begin{pgfscope}%
\pgfsys@transformshift{3.946007in}{0.929220in}%
\pgfsys@useobject{currentmarker}{}%
\end{pgfscope}%
\end{pgfscope}%
\begin{pgfscope}%
\definecolor{textcolor}{rgb}{0.000000,0.000000,0.000000}%
\pgfsetstrokecolor{textcolor}%
\pgfsetfillcolor{textcolor}%
\pgftext[x=3.946007in,y=0.831998in,,top]{\color{textcolor}\rmfamily\fontsize{28.000000}{33.600000}\selectfont 40}%
\end{pgfscope}%
\begin{pgfscope}%
\pgfpathrectangle{\pgfqpoint{1.157239in}{0.929220in}}{\pgfqpoint{5.649043in}{5.070780in}}%
\pgfusepath{clip}%
\pgfsetrectcap%
\pgfsetroundjoin%
\pgfsetlinewidth{0.803000pt}%
\definecolor{currentstroke}{rgb}{0.690196,0.690196,0.690196}%
\pgfsetstrokecolor{currentstroke}%
\pgfsetdash{}{0pt}%
\pgfpathmoveto{\pgfqpoint{5.376144in}{0.929220in}}%
\pgfpathlineto{\pgfqpoint{5.376144in}{6.000000in}}%
\pgfusepath{stroke}%
\end{pgfscope}%
\begin{pgfscope}%
\pgfsetbuttcap%
\pgfsetroundjoin%
\definecolor{currentfill}{rgb}{0.000000,0.000000,0.000000}%
\pgfsetfillcolor{currentfill}%
\pgfsetlinewidth{0.803000pt}%
\definecolor{currentstroke}{rgb}{0.000000,0.000000,0.000000}%
\pgfsetstrokecolor{currentstroke}%
\pgfsetdash{}{0pt}%
\pgfsys@defobject{currentmarker}{\pgfqpoint{0.000000in}{-0.048611in}}{\pgfqpoint{0.000000in}{0.000000in}}{%
\pgfpathmoveto{\pgfqpoint{0.000000in}{0.000000in}}%
\pgfpathlineto{\pgfqpoint{0.000000in}{-0.048611in}}%
\pgfusepath{stroke,fill}%
}%
\begin{pgfscope}%
\pgfsys@transformshift{5.376144in}{0.929220in}%
\pgfsys@useobject{currentmarker}{}%
\end{pgfscope}%
\end{pgfscope}%
\begin{pgfscope}%
\definecolor{textcolor}{rgb}{0.000000,0.000000,0.000000}%
\pgfsetstrokecolor{textcolor}%
\pgfsetfillcolor{textcolor}%
\pgftext[x=5.376144in,y=0.831998in,,top]{\color{textcolor}\rmfamily\fontsize{28.000000}{33.600000}\selectfont 60}%
\end{pgfscope}%
\begin{pgfscope}%
\pgfpathrectangle{\pgfqpoint{1.157239in}{0.929220in}}{\pgfqpoint{5.649043in}{5.070780in}}%
\pgfusepath{clip}%
\pgfsetrectcap%
\pgfsetroundjoin%
\pgfsetlinewidth{0.803000pt}%
\definecolor{currentstroke}{rgb}{0.690196,0.690196,0.690196}%
\pgfsetstrokecolor{currentstroke}%
\pgfsetdash{}{0pt}%
\pgfpathmoveto{\pgfqpoint{6.806282in}{0.929220in}}%
\pgfpathlineto{\pgfqpoint{6.806282in}{6.000000in}}%
\pgfusepath{stroke}%
\end{pgfscope}%
\begin{pgfscope}%
\pgfsetbuttcap%
\pgfsetroundjoin%
\definecolor{currentfill}{rgb}{0.000000,0.000000,0.000000}%
\pgfsetfillcolor{currentfill}%
\pgfsetlinewidth{0.803000pt}%
\definecolor{currentstroke}{rgb}{0.000000,0.000000,0.000000}%
\pgfsetstrokecolor{currentstroke}%
\pgfsetdash{}{0pt}%
\pgfsys@defobject{currentmarker}{\pgfqpoint{0.000000in}{-0.048611in}}{\pgfqpoint{0.000000in}{0.000000in}}{%
\pgfpathmoveto{\pgfqpoint{0.000000in}{0.000000in}}%
\pgfpathlineto{\pgfqpoint{0.000000in}{-0.048611in}}%
\pgfusepath{stroke,fill}%
}%
\begin{pgfscope}%
\pgfsys@transformshift{6.806282in}{0.929220in}%
\pgfsys@useobject{currentmarker}{}%
\end{pgfscope}%
\end{pgfscope}%
\begin{pgfscope}%
\definecolor{textcolor}{rgb}{0.000000,0.000000,0.000000}%
\pgfsetstrokecolor{textcolor}%
\pgfsetfillcolor{textcolor}%
\pgftext[x=6.806282in,y=0.831998in,,top]{\color{textcolor}\rmfamily\fontsize{28.000000}{33.600000}\selectfont 80}%
\end{pgfscope}%
\begin{pgfscope}%
\definecolor{textcolor}{rgb}{0.000000,0.000000,0.000000}%
\pgfsetstrokecolor{textcolor}%
\pgfsetfillcolor{textcolor}%
\pgftext[x=3.981760in,y=0.385926in,,top]{\color{textcolor}\rmfamily\fontsize{28.000000}{33.600000}\selectfont time step \(\displaystyle k\)}%
\end{pgfscope}%
\begin{pgfscope}%
\pgfpathrectangle{\pgfqpoint{1.157239in}{0.929220in}}{\pgfqpoint{5.649043in}{5.070780in}}%
\pgfusepath{clip}%
\pgfsetrectcap%
\pgfsetroundjoin%
\pgfsetlinewidth{0.803000pt}%
\definecolor{currentstroke}{rgb}{0.690196,0.690196,0.690196}%
\pgfsetstrokecolor{currentstroke}%
\pgfsetdash{}{0pt}%
\pgfpathmoveto{\pgfqpoint{1.157239in}{0.929220in}}%
\pgfpathlineto{\pgfqpoint{6.806282in}{0.929220in}}%
\pgfusepath{stroke}%
\end{pgfscope}%
\begin{pgfscope}%
\pgfsetbuttcap%
\pgfsetroundjoin%
\definecolor{currentfill}{rgb}{0.000000,0.000000,0.000000}%
\pgfsetfillcolor{currentfill}%
\pgfsetlinewidth{0.803000pt}%
\definecolor{currentstroke}{rgb}{0.000000,0.000000,0.000000}%
\pgfsetstrokecolor{currentstroke}%
\pgfsetdash{}{0pt}%
\pgfsys@defobject{currentmarker}{\pgfqpoint{-0.048611in}{0.000000in}}{\pgfqpoint{-0.000000in}{0.000000in}}{%
\pgfpathmoveto{\pgfqpoint{-0.000000in}{0.000000in}}%
\pgfpathlineto{\pgfqpoint{-0.048611in}{0.000000in}}%
\pgfusepath{stroke,fill}%
}%
\begin{pgfscope}%
\pgfsys@transformshift{1.157239in}{0.929220in}%
\pgfsys@useobject{currentmarker}{}%
\end{pgfscope}%
\end{pgfscope}%
\begin{pgfscope}%
\definecolor{textcolor}{rgb}{0.000000,0.000000,0.000000}%
\pgfsetstrokecolor{textcolor}%
\pgfsetfillcolor{textcolor}%
\pgftext[x=0.811260in, y=0.809236in, left, base]{\color{textcolor}\rmfamily\fontsize{28.000000}{33.600000}\selectfont 0}%
\end{pgfscope}%
\begin{pgfscope}%
\pgfpathrectangle{\pgfqpoint{1.157239in}{0.929220in}}{\pgfqpoint{5.649043in}{5.070780in}}%
\pgfusepath{clip}%
\pgfsetrectcap%
\pgfsetroundjoin%
\pgfsetlinewidth{0.803000pt}%
\definecolor{currentstroke}{rgb}{0.690196,0.690196,0.690196}%
\pgfsetstrokecolor{currentstroke}%
\pgfsetdash{}{0pt}%
\pgfpathmoveto{\pgfqpoint{1.157239in}{2.358011in}}%
\pgfpathlineto{\pgfqpoint{6.806282in}{2.358011in}}%
\pgfusepath{stroke}%
\end{pgfscope}%
\begin{pgfscope}%
\pgfsetbuttcap%
\pgfsetroundjoin%
\definecolor{currentfill}{rgb}{0.000000,0.000000,0.000000}%
\pgfsetfillcolor{currentfill}%
\pgfsetlinewidth{0.803000pt}%
\definecolor{currentstroke}{rgb}{0.000000,0.000000,0.000000}%
\pgfsetstrokecolor{currentstroke}%
\pgfsetdash{}{0pt}%
\pgfsys@defobject{currentmarker}{\pgfqpoint{-0.048611in}{0.000000in}}{\pgfqpoint{-0.000000in}{0.000000in}}{%
\pgfpathmoveto{\pgfqpoint{-0.000000in}{0.000000in}}%
\pgfpathlineto{\pgfqpoint{-0.048611in}{0.000000in}}%
\pgfusepath{stroke,fill}%
}%
\begin{pgfscope}%
\pgfsys@transformshift{1.157239in}{2.358011in}%
\pgfsys@useobject{currentmarker}{}%
\end{pgfscope}%
\end{pgfscope}%
\begin{pgfscope}%
\definecolor{textcolor}{rgb}{0.000000,0.000000,0.000000}%
\pgfsetstrokecolor{textcolor}%
\pgfsetfillcolor{textcolor}%
\pgftext[x=0.566517in, y=2.238027in, left, base]{\color{textcolor}\rmfamily\fontsize{28.000000}{33.600000}\selectfont 0.5}%
\end{pgfscope}%
\begin{pgfscope}%
\pgfpathrectangle{\pgfqpoint{1.157239in}{0.929220in}}{\pgfqpoint{5.649043in}{5.070780in}}%
\pgfusepath{clip}%
\pgfsetrectcap%
\pgfsetroundjoin%
\pgfsetlinewidth{0.803000pt}%
\definecolor{currentstroke}{rgb}{0.690196,0.690196,0.690196}%
\pgfsetstrokecolor{currentstroke}%
\pgfsetdash{}{0pt}%
\pgfpathmoveto{\pgfqpoint{1.157239in}{3.786803in}}%
\pgfpathlineto{\pgfqpoint{6.806282in}{3.786803in}}%
\pgfusepath{stroke}%
\end{pgfscope}%
\begin{pgfscope}%
\pgfsetbuttcap%
\pgfsetroundjoin%
\definecolor{currentfill}{rgb}{0.000000,0.000000,0.000000}%
\pgfsetfillcolor{currentfill}%
\pgfsetlinewidth{0.803000pt}%
\definecolor{currentstroke}{rgb}{0.000000,0.000000,0.000000}%
\pgfsetstrokecolor{currentstroke}%
\pgfsetdash{}{0pt}%
\pgfsys@defobject{currentmarker}{\pgfqpoint{-0.048611in}{0.000000in}}{\pgfqpoint{-0.000000in}{0.000000in}}{%
\pgfpathmoveto{\pgfqpoint{-0.000000in}{0.000000in}}%
\pgfpathlineto{\pgfqpoint{-0.048611in}{0.000000in}}%
\pgfusepath{stroke,fill}%
}%
\begin{pgfscope}%
\pgfsys@transformshift{1.157239in}{3.786803in}%
\pgfsys@useobject{currentmarker}{}%
\end{pgfscope}%
\end{pgfscope}%
\begin{pgfscope}%
\definecolor{textcolor}{rgb}{0.000000,0.000000,0.000000}%
\pgfsetstrokecolor{textcolor}%
\pgfsetfillcolor{textcolor}%
\pgftext[x=0.811260in, y=3.666818in, left, base]{\color{textcolor}\rmfamily\fontsize{28.000000}{33.600000}\selectfont 1}%
\end{pgfscope}%
\begin{pgfscope}%
\pgfpathrectangle{\pgfqpoint{1.157239in}{0.929220in}}{\pgfqpoint{5.649043in}{5.070780in}}%
\pgfusepath{clip}%
\pgfsetrectcap%
\pgfsetroundjoin%
\pgfsetlinewidth{0.803000pt}%
\definecolor{currentstroke}{rgb}{0.690196,0.690196,0.690196}%
\pgfsetstrokecolor{currentstroke}%
\pgfsetdash{}{0pt}%
\pgfpathmoveto{\pgfqpoint{1.157239in}{5.215594in}}%
\pgfpathlineto{\pgfqpoint{6.806282in}{5.215594in}}%
\pgfusepath{stroke}%
\end{pgfscope}%
\begin{pgfscope}%
\pgfsetbuttcap%
\pgfsetroundjoin%
\definecolor{currentfill}{rgb}{0.000000,0.000000,0.000000}%
\pgfsetfillcolor{currentfill}%
\pgfsetlinewidth{0.803000pt}%
\definecolor{currentstroke}{rgb}{0.000000,0.000000,0.000000}%
\pgfsetstrokecolor{currentstroke}%
\pgfsetdash{}{0pt}%
\pgfsys@defobject{currentmarker}{\pgfqpoint{-0.048611in}{0.000000in}}{\pgfqpoint{-0.000000in}{0.000000in}}{%
\pgfpathmoveto{\pgfqpoint{-0.000000in}{0.000000in}}%
\pgfpathlineto{\pgfqpoint{-0.048611in}{0.000000in}}%
\pgfusepath{stroke,fill}%
}%
\begin{pgfscope}%
\pgfsys@transformshift{1.157239in}{5.215594in}%
\pgfsys@useobject{currentmarker}{}%
\end{pgfscope}%
\end{pgfscope}%
\begin{pgfscope}%
\definecolor{textcolor}{rgb}{0.000000,0.000000,0.000000}%
\pgfsetstrokecolor{textcolor}%
\pgfsetfillcolor{textcolor}%
\pgftext[x=0.566517in, y=5.095609in, left, base]{\color{textcolor}\rmfamily\fontsize{28.000000}{33.600000}\selectfont 1.5}%
\end{pgfscope}%
\begin{pgfscope}%
\definecolor{textcolor}{rgb}{0.000000,0.000000,0.000000}%
\pgfsetstrokecolor{textcolor}%
\pgfsetfillcolor{textcolor}%
\pgftext[x=0.427628in,y=3.464610in,,bottom,rotate=90.000000]{\color{textcolor}\rmfamily\fontsize{28.000000}{33.600000}\selectfont False Objects Error [m]}%
\end{pgfscope}%
\begin{pgfscope}%
\pgfpathrectangle{\pgfqpoint{1.157239in}{0.929220in}}{\pgfqpoint{5.649043in}{5.070780in}}%
\pgfusepath{clip}%
\pgfsetrectcap%
\pgfsetroundjoin%
\pgfsetlinewidth{1.505625pt}%
\definecolor{currentstroke}{rgb}{0.750000,0.750000,0.000000}%
\pgfsetstrokecolor{currentstroke}%
\pgfsetdash{}{0pt}%
\pgfpathmoveto{\pgfqpoint{1.157239in}{0.929220in}}%
\pgfpathlineto{\pgfqpoint{1.228745in}{0.929220in}}%
\pgfpathlineto{\pgfqpoint{1.300252in}{0.929220in}}%
\pgfpathlineto{\pgfqpoint{1.371759in}{0.929220in}}%
\pgfpathlineto{\pgfqpoint{1.443266in}{0.929220in}}%
\pgfpathlineto{\pgfqpoint{1.514773in}{0.929220in}}%
\pgfpathlineto{\pgfqpoint{1.586280in}{0.929220in}}%
\pgfpathlineto{\pgfqpoint{1.657787in}{0.929220in}}%
\pgfpathlineto{\pgfqpoint{1.729294in}{0.929220in}}%
\pgfpathlineto{\pgfqpoint{1.800800in}{1.343570in}}%
\pgfpathlineto{\pgfqpoint{1.872307in}{1.600752in}}%
\pgfpathlineto{\pgfqpoint{1.943814in}{1.857935in}}%
\pgfpathlineto{\pgfqpoint{2.015321in}{1.729343in}}%
\pgfpathlineto{\pgfqpoint{2.086828in}{1.886510in}}%
\pgfpathlineto{\pgfqpoint{2.158335in}{2.529466in}}%
\pgfpathlineto{\pgfqpoint{2.229842in}{2.972392in}}%
\pgfpathlineto{\pgfqpoint{2.301349in}{2.972392in}}%
\pgfpathlineto{\pgfqpoint{2.372855in}{3.058119in}}%
\pgfpathlineto{\pgfqpoint{2.444362in}{3.100983in}}%
\pgfpathlineto{\pgfqpoint{2.515869in}{3.772515in}}%
\pgfpathlineto{\pgfqpoint{2.587376in}{4.172576in}}%
\pgfpathlineto{\pgfqpoint{2.658883in}{4.229728in}}%
\pgfpathlineto{\pgfqpoint{2.730390in}{4.429759in}}%
\pgfpathlineto{\pgfqpoint{2.801897in}{4.358319in}}%
\pgfpathlineto{\pgfqpoint{2.873404in}{4.644077in}}%
\pgfpathlineto{\pgfqpoint{2.944910in}{5.029851in}}%
\pgfpathlineto{\pgfqpoint{3.016417in}{5.329897in}}%
\pgfpathlineto{\pgfqpoint{3.087924in}{5.329897in}}%
\pgfpathlineto{\pgfqpoint{3.159431in}{5.629943in}}%
\pgfpathlineto{\pgfqpoint{3.230938in}{5.629943in}}%
\pgfpathlineto{\pgfqpoint{3.302445in}{5.558504in}}%
\pgfpathlineto{\pgfqpoint{3.373952in}{5.701383in}}%
\pgfpathlineto{\pgfqpoint{3.445459in}{5.429912in}}%
\pgfpathlineto{\pgfqpoint{3.516966in}{5.415624in}}%
\pgfpathlineto{\pgfqpoint{3.588472in}{5.401337in}}%
\pgfpathlineto{\pgfqpoint{3.659979in}{5.272745in}}%
\pgfpathlineto{\pgfqpoint{3.731486in}{5.315609in}}%
\pgfpathlineto{\pgfqpoint{3.802993in}{5.587079in}}%
\pgfpathlineto{\pgfqpoint{3.874500in}{5.487064in}}%
\pgfpathlineto{\pgfqpoint{3.946007in}{5.758534in}}%
\pgfpathlineto{\pgfqpoint{4.017514in}{5.601367in}}%
\pgfpathlineto{\pgfqpoint{4.089021in}{5.715671in}}%
\pgfpathlineto{\pgfqpoint{4.160527in}{5.387049in}}%
\pgfpathlineto{\pgfqpoint{4.232034in}{5.429912in}}%
\pgfpathlineto{\pgfqpoint{4.303541in}{5.529928in}}%
\pgfpathlineto{\pgfqpoint{4.375048in}{5.315609in}}%
\pgfpathlineto{\pgfqpoint{4.446555in}{5.215594in}}%
\pgfpathlineto{\pgfqpoint{4.518062in}{5.129866in}}%
\pgfpathlineto{\pgfqpoint{4.589569in}{4.872684in}}%
\pgfpathlineto{\pgfqpoint{4.661076in}{5.201306in}}%
\pgfpathlineto{\pgfqpoint{4.732582in}{5.272745in}}%
\pgfpathlineto{\pgfqpoint{4.804089in}{5.329897in}}%
\pgfpathlineto{\pgfqpoint{4.875596in}{5.158442in}}%
\pgfpathlineto{\pgfqpoint{4.947103in}{5.215594in}}%
\pgfpathlineto{\pgfqpoint{5.018610in}{5.101290in}}%
\pgfpathlineto{\pgfqpoint{5.090117in}{4.801244in}}%
\pgfpathlineto{\pgfqpoint{5.161624in}{4.815532in}}%
\pgfpathlineto{\pgfqpoint{5.233131in}{4.815532in}}%
\pgfpathlineto{\pgfqpoint{5.304638in}{4.772668in}}%
\pgfpathlineto{\pgfqpoint{5.376144in}{4.401183in}}%
\pgfpathlineto{\pgfqpoint{5.447651in}{3.943970in}}%
\pgfpathlineto{\pgfqpoint{5.519158in}{3.543908in}}%
\pgfpathlineto{\pgfqpoint{5.590665in}{3.501044in}}%
\pgfpathlineto{\pgfqpoint{5.662172in}{3.472469in}}%
\pgfpathlineto{\pgfqpoint{5.733679in}{3.258150in}}%
\pgfpathlineto{\pgfqpoint{5.805186in}{2.658058in}}%
\pgfpathlineto{\pgfqpoint{5.876693in}{2.257996in}}%
\pgfpathlineto{\pgfqpoint{5.948199in}{2.129405in}}%
\pgfpathlineto{\pgfqpoint{6.019706in}{2.229420in}}%
\pgfpathlineto{\pgfqpoint{6.091213in}{1.300706in}}%
\pgfpathlineto{\pgfqpoint{6.162720in}{0.929220in}}%
\pgfpathlineto{\pgfqpoint{6.234227in}{0.929220in}}%
\pgfpathlineto{\pgfqpoint{6.305734in}{0.929220in}}%
\pgfpathlineto{\pgfqpoint{6.377241in}{0.929220in}}%
\pgfpathlineto{\pgfqpoint{6.448748in}{0.929220in}}%
\pgfpathlineto{\pgfqpoint{6.520254in}{0.929220in}}%
\pgfpathlineto{\pgfqpoint{6.591761in}{0.929220in}}%
\pgfpathlineto{\pgfqpoint{6.663268in}{0.929220in}}%
\pgfpathlineto{\pgfqpoint{6.734775in}{0.929220in}}%
\pgfpathlineto{\pgfqpoint{6.806282in}{0.929220in}}%
\pgfusepath{stroke}%
\end{pgfscope}%
\begin{pgfscope}%
\pgfpathrectangle{\pgfqpoint{1.157239in}{0.929220in}}{\pgfqpoint{5.649043in}{5.070780in}}%
\pgfusepath{clip}%
\pgfsetrectcap%
\pgfsetroundjoin%
\pgfsetlinewidth{1.505625pt}%
\definecolor{currentstroke}{rgb}{0.000000,0.000000,1.000000}%
\pgfsetstrokecolor{currentstroke}%
\pgfsetdash{}{0pt}%
\pgfpathmoveto{\pgfqpoint{1.157239in}{0.929220in}}%
\pgfpathlineto{\pgfqpoint{1.228745in}{0.972084in}}%
\pgfpathlineto{\pgfqpoint{1.300252in}{0.943508in}}%
\pgfpathlineto{\pgfqpoint{1.371759in}{0.943508in}}%
\pgfpathlineto{\pgfqpoint{1.443266in}{0.957796in}}%
\pgfpathlineto{\pgfqpoint{1.514773in}{1.000660in}}%
\pgfpathlineto{\pgfqpoint{1.586280in}{0.972084in}}%
\pgfpathlineto{\pgfqpoint{1.657787in}{0.972084in}}%
\pgfpathlineto{\pgfqpoint{1.729294in}{0.957796in}}%
\pgfpathlineto{\pgfqpoint{1.800800in}{0.929220in}}%
\pgfpathlineto{\pgfqpoint{1.872307in}{0.929220in}}%
\pgfpathlineto{\pgfqpoint{1.943814in}{0.929220in}}%
\pgfpathlineto{\pgfqpoint{2.015321in}{0.929220in}}%
\pgfpathlineto{\pgfqpoint{2.086828in}{0.972084in}}%
\pgfpathlineto{\pgfqpoint{2.158335in}{0.943508in}}%
\pgfpathlineto{\pgfqpoint{2.229842in}{0.972084in}}%
\pgfpathlineto{\pgfqpoint{2.301349in}{1.172115in}}%
\pgfpathlineto{\pgfqpoint{2.372855in}{1.229266in}}%
\pgfpathlineto{\pgfqpoint{2.444362in}{1.257842in}}%
\pgfpathlineto{\pgfqpoint{2.515869in}{1.343570in}}%
\pgfpathlineto{\pgfqpoint{2.587376in}{1.415009in}}%
\pgfpathlineto{\pgfqpoint{2.658883in}{1.529313in}}%
\pgfpathlineto{\pgfqpoint{2.730390in}{1.557888in}}%
\pgfpathlineto{\pgfqpoint{2.801897in}{1.629328in}}%
\pgfpathlineto{\pgfqpoint{2.873404in}{1.700768in}}%
\pgfpathlineto{\pgfqpoint{2.944910in}{1.815071in}}%
\pgfpathlineto{\pgfqpoint{3.016417in}{1.986526in}}%
\pgfpathlineto{\pgfqpoint{3.087924in}{2.186557in}}%
\pgfpathlineto{\pgfqpoint{3.159431in}{2.143693in}}%
\pgfpathlineto{\pgfqpoint{3.230938in}{2.257996in}}%
\pgfpathlineto{\pgfqpoint{3.302445in}{2.558042in}}%
\pgfpathlineto{\pgfqpoint{3.373952in}{2.729497in}}%
\pgfpathlineto{\pgfqpoint{3.445459in}{2.815225in}}%
\pgfpathlineto{\pgfqpoint{3.516966in}{2.915240in}}%
\pgfpathlineto{\pgfqpoint{3.588472in}{2.843800in}}%
\pgfpathlineto{\pgfqpoint{3.659979in}{2.758073in}}%
\pgfpathlineto{\pgfqpoint{3.731486in}{2.343724in}}%
\pgfpathlineto{\pgfqpoint{3.802993in}{2.329436in}}%
\pgfpathlineto{\pgfqpoint{3.874500in}{2.043677in}}%
\pgfpathlineto{\pgfqpoint{3.946007in}{1.957950in}}%
\pgfpathlineto{\pgfqpoint{4.017514in}{1.972238in}}%
\pgfpathlineto{\pgfqpoint{4.089021in}{2.129405in}}%
\pgfpathlineto{\pgfqpoint{4.160527in}{2.043677in}}%
\pgfpathlineto{\pgfqpoint{4.232034in}{2.186557in}}%
\pgfpathlineto{\pgfqpoint{4.303541in}{2.415163in}}%
\pgfpathlineto{\pgfqpoint{4.375048in}{2.443739in}}%
\pgfpathlineto{\pgfqpoint{4.446555in}{2.543754in}}%
\pgfpathlineto{\pgfqpoint{4.518062in}{2.529466in}}%
\pgfpathlineto{\pgfqpoint{4.589569in}{2.329436in}}%
\pgfpathlineto{\pgfqpoint{4.661076in}{2.243708in}}%
\pgfpathlineto{\pgfqpoint{4.732582in}{2.172269in}}%
\pgfpathlineto{\pgfqpoint{4.804089in}{2.172269in}}%
\pgfpathlineto{\pgfqpoint{4.875596in}{2.172269in}}%
\pgfpathlineto{\pgfqpoint{4.947103in}{2.157981in}}%
\pgfpathlineto{\pgfqpoint{5.018610in}{2.872376in}}%
\pgfpathlineto{\pgfqpoint{5.090117in}{2.800937in}}%
\pgfpathlineto{\pgfqpoint{5.161624in}{1.715055in}}%
\pgfpathlineto{\pgfqpoint{5.233131in}{1.672192in}}%
\pgfpathlineto{\pgfqpoint{5.304638in}{1.572176in}}%
\pgfpathlineto{\pgfqpoint{5.376144in}{2.715209in}}%
\pgfpathlineto{\pgfqpoint{5.447651in}{2.543754in}}%
\pgfpathlineto{\pgfqpoint{5.519158in}{1.357858in}}%
\pgfpathlineto{\pgfqpoint{5.590665in}{1.286418in}}%
\pgfpathlineto{\pgfqpoint{5.662172in}{1.243554in}}%
\pgfpathlineto{\pgfqpoint{5.733679in}{2.415163in}}%
\pgfpathlineto{\pgfqpoint{5.805186in}{2.300860in}}%
\pgfpathlineto{\pgfqpoint{5.876693in}{1.029236in}}%
\pgfpathlineto{\pgfqpoint{5.948199in}{1.000660in}}%
\pgfpathlineto{\pgfqpoint{6.019706in}{0.986372in}}%
\pgfpathlineto{\pgfqpoint{6.091213in}{3.772515in}}%
\pgfpathlineto{\pgfqpoint{6.162720in}{3.772515in}}%
\pgfpathlineto{\pgfqpoint{6.234227in}{1.186403in}}%
\pgfpathlineto{\pgfqpoint{6.305734in}{1.129251in}}%
\pgfpathlineto{\pgfqpoint{6.377241in}{1.072099in}}%
\pgfpathlineto{\pgfqpoint{6.448748in}{0.986372in}}%
\pgfpathlineto{\pgfqpoint{6.520254in}{0.972084in}}%
\pgfpathlineto{\pgfqpoint{6.591761in}{0.972084in}}%
\pgfpathlineto{\pgfqpoint{6.663268in}{0.943508in}}%
\pgfpathlineto{\pgfqpoint{6.734775in}{0.986372in}}%
\pgfpathlineto{\pgfqpoint{6.806282in}{0.929220in}}%
\pgfusepath{stroke}%
\end{pgfscope}%
\begin{pgfscope}%
\pgfpathrectangle{\pgfqpoint{1.157239in}{0.929220in}}{\pgfqpoint{5.649043in}{5.070780in}}%
\pgfusepath{clip}%
\pgfsetrectcap%
\pgfsetroundjoin%
\pgfsetlinewidth{1.505625pt}%
\definecolor{currentstroke}{rgb}{0.000000,0.500000,0.000000}%
\pgfsetstrokecolor{currentstroke}%
\pgfsetdash{}{0pt}%
\pgfpathmoveto{\pgfqpoint{1.157239in}{0.929220in}}%
\pgfpathlineto{\pgfqpoint{1.228745in}{0.943508in}}%
\pgfpathlineto{\pgfqpoint{1.300252in}{0.943508in}}%
\pgfpathlineto{\pgfqpoint{1.371759in}{0.943508in}}%
\pgfpathlineto{\pgfqpoint{1.443266in}{0.986372in}}%
\pgfpathlineto{\pgfqpoint{1.514773in}{1.014948in}}%
\pgfpathlineto{\pgfqpoint{1.586280in}{0.972084in}}%
\pgfpathlineto{\pgfqpoint{1.657787in}{0.972084in}}%
\pgfpathlineto{\pgfqpoint{1.729294in}{0.986372in}}%
\pgfpathlineto{\pgfqpoint{1.800800in}{0.929220in}}%
\pgfpathlineto{\pgfqpoint{1.872307in}{0.929220in}}%
\pgfpathlineto{\pgfqpoint{1.943814in}{0.929220in}}%
\pgfpathlineto{\pgfqpoint{2.015321in}{0.957796in}}%
\pgfpathlineto{\pgfqpoint{2.086828in}{0.929220in}}%
\pgfpathlineto{\pgfqpoint{2.158335in}{0.929220in}}%
\pgfpathlineto{\pgfqpoint{2.229842in}{0.929220in}}%
\pgfpathlineto{\pgfqpoint{2.301349in}{0.929220in}}%
\pgfpathlineto{\pgfqpoint{2.372855in}{0.957796in}}%
\pgfpathlineto{\pgfqpoint{2.444362in}{0.929220in}}%
\pgfpathlineto{\pgfqpoint{2.515869in}{0.957796in}}%
\pgfpathlineto{\pgfqpoint{2.587376in}{0.957796in}}%
\pgfpathlineto{\pgfqpoint{2.658883in}{0.957796in}}%
\pgfpathlineto{\pgfqpoint{2.730390in}{0.957796in}}%
\pgfpathlineto{\pgfqpoint{2.801897in}{0.957796in}}%
\pgfpathlineto{\pgfqpoint{2.873404in}{0.986372in}}%
\pgfpathlineto{\pgfqpoint{2.944910in}{1.029236in}}%
\pgfpathlineto{\pgfqpoint{3.016417in}{1.072099in}}%
\pgfpathlineto{\pgfqpoint{3.087924in}{1.029236in}}%
\pgfpathlineto{\pgfqpoint{3.159431in}{0.986372in}}%
\pgfpathlineto{\pgfqpoint{3.230938in}{0.986372in}}%
\pgfpathlineto{\pgfqpoint{3.302445in}{1.014948in}}%
\pgfpathlineto{\pgfqpoint{3.373952in}{1.014948in}}%
\pgfpathlineto{\pgfqpoint{3.445459in}{1.014948in}}%
\pgfpathlineto{\pgfqpoint{3.516966in}{0.972084in}}%
\pgfpathlineto{\pgfqpoint{3.588472in}{1.043524in}}%
\pgfpathlineto{\pgfqpoint{3.659979in}{1.000660in}}%
\pgfpathlineto{\pgfqpoint{3.731486in}{0.986372in}}%
\pgfpathlineto{\pgfqpoint{3.802993in}{0.986372in}}%
\pgfpathlineto{\pgfqpoint{3.874500in}{1.000660in}}%
\pgfpathlineto{\pgfqpoint{3.946007in}{1.029236in}}%
\pgfpathlineto{\pgfqpoint{4.017514in}{1.014948in}}%
\pgfpathlineto{\pgfqpoint{4.089021in}{1.043524in}}%
\pgfpathlineto{\pgfqpoint{4.160527in}{1.043524in}}%
\pgfpathlineto{\pgfqpoint{4.232034in}{1.029236in}}%
\pgfpathlineto{\pgfqpoint{4.303541in}{1.000660in}}%
\pgfpathlineto{\pgfqpoint{4.375048in}{0.972084in}}%
\pgfpathlineto{\pgfqpoint{4.446555in}{1.029236in}}%
\pgfpathlineto{\pgfqpoint{4.518062in}{1.000660in}}%
\pgfpathlineto{\pgfqpoint{4.589569in}{1.029236in}}%
\pgfpathlineto{\pgfqpoint{4.661076in}{1.029236in}}%
\pgfpathlineto{\pgfqpoint{4.732582in}{1.000660in}}%
\pgfpathlineto{\pgfqpoint{4.804089in}{0.972084in}}%
\pgfpathlineto{\pgfqpoint{4.875596in}{1.029236in}}%
\pgfpathlineto{\pgfqpoint{4.947103in}{0.957796in}}%
\pgfpathlineto{\pgfqpoint{5.018610in}{2.243708in}}%
\pgfpathlineto{\pgfqpoint{5.090117in}{2.157981in}}%
\pgfpathlineto{\pgfqpoint{5.161624in}{0.972084in}}%
\pgfpathlineto{\pgfqpoint{5.233131in}{0.972084in}}%
\pgfpathlineto{\pgfqpoint{5.304638in}{0.943508in}}%
\pgfpathlineto{\pgfqpoint{5.376144in}{2.329436in}}%
\pgfpathlineto{\pgfqpoint{5.447651in}{2.300860in}}%
\pgfpathlineto{\pgfqpoint{5.519158in}{0.929220in}}%
\pgfpathlineto{\pgfqpoint{5.590665in}{0.929220in}}%
\pgfpathlineto{\pgfqpoint{5.662172in}{0.929220in}}%
\pgfpathlineto{\pgfqpoint{5.733679in}{2.257996in}}%
\pgfpathlineto{\pgfqpoint{5.805186in}{2.257996in}}%
\pgfpathlineto{\pgfqpoint{5.876693in}{0.929220in}}%
\pgfpathlineto{\pgfqpoint{5.948199in}{0.943508in}}%
\pgfpathlineto{\pgfqpoint{6.019706in}{0.957796in}}%
\pgfpathlineto{\pgfqpoint{6.091213in}{3.786803in}}%
\pgfpathlineto{\pgfqpoint{6.162720in}{3.786803in}}%
\pgfpathlineto{\pgfqpoint{6.234227in}{1.100675in}}%
\pgfpathlineto{\pgfqpoint{6.305734in}{1.157827in}}%
\pgfpathlineto{\pgfqpoint{6.377241in}{1.100675in}}%
\pgfpathlineto{\pgfqpoint{6.448748in}{1.057812in}}%
\pgfpathlineto{\pgfqpoint{6.520254in}{0.957796in}}%
\pgfpathlineto{\pgfqpoint{6.591761in}{0.972084in}}%
\pgfpathlineto{\pgfqpoint{6.663268in}{0.972084in}}%
\pgfpathlineto{\pgfqpoint{6.734775in}{1.000660in}}%
\pgfpathlineto{\pgfqpoint{6.806282in}{1.014948in}}%
\pgfusepath{stroke}%
\end{pgfscope}%
\begin{pgfscope}%
\pgfpathrectangle{\pgfqpoint{1.157239in}{0.929220in}}{\pgfqpoint{5.649043in}{5.070780in}}%
\pgfusepath{clip}%
\pgfsetrectcap%
\pgfsetroundjoin%
\pgfsetlinewidth{1.505625pt}%
\definecolor{currentstroke}{rgb}{1.000000,0.000000,0.000000}%
\pgfsetstrokecolor{currentstroke}%
\pgfsetdash{}{0pt}%
\pgfpathmoveto{\pgfqpoint{1.157239in}{0.929220in}}%
\pgfpathlineto{\pgfqpoint{1.228745in}{0.929220in}}%
\pgfpathlineto{\pgfqpoint{1.300252in}{0.929220in}}%
\pgfpathlineto{\pgfqpoint{1.371759in}{0.929220in}}%
\pgfpathlineto{\pgfqpoint{1.443266in}{0.929220in}}%
\pgfpathlineto{\pgfqpoint{1.514773in}{0.929220in}}%
\pgfpathlineto{\pgfqpoint{1.586280in}{0.929220in}}%
\pgfpathlineto{\pgfqpoint{1.657787in}{0.929220in}}%
\pgfpathlineto{\pgfqpoint{1.729294in}{0.929220in}}%
\pgfpathlineto{\pgfqpoint{1.800800in}{0.929220in}}%
\pgfpathlineto{\pgfqpoint{1.872307in}{0.972084in}}%
\pgfpathlineto{\pgfqpoint{1.943814in}{0.943508in}}%
\pgfpathlineto{\pgfqpoint{2.015321in}{0.929220in}}%
\pgfpathlineto{\pgfqpoint{2.086828in}{0.929220in}}%
\pgfpathlineto{\pgfqpoint{2.158335in}{0.943508in}}%
\pgfpathlineto{\pgfqpoint{2.229842in}{1.072099in}}%
\pgfpathlineto{\pgfqpoint{2.301349in}{1.029236in}}%
\pgfpathlineto{\pgfqpoint{2.372855in}{0.957796in}}%
\pgfpathlineto{\pgfqpoint{2.444362in}{0.957796in}}%
\pgfpathlineto{\pgfqpoint{2.515869in}{1.172115in}}%
\pgfpathlineto{\pgfqpoint{2.587376in}{1.072099in}}%
\pgfpathlineto{\pgfqpoint{2.658883in}{1.072099in}}%
\pgfpathlineto{\pgfqpoint{2.730390in}{0.986372in}}%
\pgfpathlineto{\pgfqpoint{2.801897in}{1.014948in}}%
\pgfpathlineto{\pgfqpoint{2.873404in}{1.200691in}}%
\pgfpathlineto{\pgfqpoint{2.944910in}{1.100675in}}%
\pgfpathlineto{\pgfqpoint{3.016417in}{1.072099in}}%
\pgfpathlineto{\pgfqpoint{3.087924in}{1.129251in}}%
\pgfpathlineto{\pgfqpoint{3.159431in}{1.157827in}}%
\pgfpathlineto{\pgfqpoint{3.230938in}{1.143539in}}%
\pgfpathlineto{\pgfqpoint{3.302445in}{1.257842in}}%
\pgfpathlineto{\pgfqpoint{3.373952in}{1.357858in}}%
\pgfpathlineto{\pgfqpoint{3.445459in}{1.357858in}}%
\pgfpathlineto{\pgfqpoint{3.516966in}{1.272130in}}%
\pgfpathlineto{\pgfqpoint{3.588472in}{1.386433in}}%
\pgfpathlineto{\pgfqpoint{3.659979in}{1.314994in}}%
\pgfpathlineto{\pgfqpoint{3.731486in}{1.272130in}}%
\pgfpathlineto{\pgfqpoint{3.802993in}{1.300706in}}%
\pgfpathlineto{\pgfqpoint{3.874500in}{1.214979in}}%
\pgfpathlineto{\pgfqpoint{3.946007in}{1.186403in}}%
\pgfpathlineto{\pgfqpoint{4.017514in}{1.172115in}}%
\pgfpathlineto{\pgfqpoint{4.089021in}{1.129251in}}%
\pgfpathlineto{\pgfqpoint{4.160527in}{1.172115in}}%
\pgfpathlineto{\pgfqpoint{4.232034in}{1.200691in}}%
\pgfpathlineto{\pgfqpoint{4.303541in}{1.214979in}}%
\pgfpathlineto{\pgfqpoint{4.375048in}{1.172115in}}%
\pgfpathlineto{\pgfqpoint{4.446555in}{1.143539in}}%
\pgfpathlineto{\pgfqpoint{4.518062in}{1.200691in}}%
\pgfpathlineto{\pgfqpoint{4.589569in}{1.229266in}}%
\pgfpathlineto{\pgfqpoint{4.661076in}{1.243554in}}%
\pgfpathlineto{\pgfqpoint{4.732582in}{1.229266in}}%
\pgfpathlineto{\pgfqpoint{4.804089in}{1.229266in}}%
\pgfpathlineto{\pgfqpoint{4.875596in}{1.229266in}}%
\pgfpathlineto{\pgfqpoint{4.947103in}{1.186403in}}%
\pgfpathlineto{\pgfqpoint{5.018610in}{1.257842in}}%
\pgfpathlineto{\pgfqpoint{5.090117in}{1.214979in}}%
\pgfpathlineto{\pgfqpoint{5.161624in}{1.229266in}}%
\pgfpathlineto{\pgfqpoint{5.233131in}{1.186403in}}%
\pgfpathlineto{\pgfqpoint{5.304638in}{1.114963in}}%
\pgfpathlineto{\pgfqpoint{5.376144in}{1.014948in}}%
\pgfpathlineto{\pgfqpoint{5.447651in}{1.014948in}}%
\pgfpathlineto{\pgfqpoint{5.519158in}{1.043524in}}%
\pgfpathlineto{\pgfqpoint{5.590665in}{1.014948in}}%
\pgfpathlineto{\pgfqpoint{5.662172in}{1.000660in}}%
\pgfpathlineto{\pgfqpoint{5.733679in}{1.000660in}}%
\pgfpathlineto{\pgfqpoint{5.805186in}{0.972084in}}%
\pgfpathlineto{\pgfqpoint{5.876693in}{0.957796in}}%
\pgfpathlineto{\pgfqpoint{5.948199in}{0.957796in}}%
\pgfpathlineto{\pgfqpoint{6.019706in}{0.972084in}}%
\pgfpathlineto{\pgfqpoint{6.091213in}{0.929220in}}%
\pgfpathlineto{\pgfqpoint{6.162720in}{0.929220in}}%
\pgfpathlineto{\pgfqpoint{6.234227in}{0.929220in}}%
\pgfpathlineto{\pgfqpoint{6.305734in}{0.929220in}}%
\pgfpathlineto{\pgfqpoint{6.377241in}{0.929220in}}%
\pgfpathlineto{\pgfqpoint{6.448748in}{0.929220in}}%
\pgfpathlineto{\pgfqpoint{6.520254in}{0.929220in}}%
\pgfpathlineto{\pgfqpoint{6.591761in}{0.929220in}}%
\pgfpathlineto{\pgfqpoint{6.663268in}{0.929220in}}%
\pgfpathlineto{\pgfqpoint{6.734775in}{0.929220in}}%
\pgfpathlineto{\pgfqpoint{6.806282in}{0.929220in}}%
\pgfusepath{stroke}%
\end{pgfscope}%
\begin{pgfscope}%
\pgfsetrectcap%
\pgfsetmiterjoin%
\pgfsetlinewidth{0.803000pt}%
\definecolor{currentstroke}{rgb}{0.000000,0.000000,0.000000}%
\pgfsetstrokecolor{currentstroke}%
\pgfsetdash{}{0pt}%
\pgfpathmoveto{\pgfqpoint{1.157239in}{0.929220in}}%
\pgfpathlineto{\pgfqpoint{1.157239in}{6.000000in}}%
\pgfusepath{stroke}%
\end{pgfscope}%
\begin{pgfscope}%
\pgfsetrectcap%
\pgfsetmiterjoin%
\pgfsetlinewidth{0.803000pt}%
\definecolor{currentstroke}{rgb}{0.000000,0.000000,0.000000}%
\pgfsetstrokecolor{currentstroke}%
\pgfsetdash{}{0pt}%
\pgfpathmoveto{\pgfqpoint{6.806282in}{0.929220in}}%
\pgfpathlineto{\pgfqpoint{6.806282in}{6.000000in}}%
\pgfusepath{stroke}%
\end{pgfscope}%
\begin{pgfscope}%
\pgfsetrectcap%
\pgfsetmiterjoin%
\pgfsetlinewidth{0.803000pt}%
\definecolor{currentstroke}{rgb}{0.000000,0.000000,0.000000}%
\pgfsetstrokecolor{currentstroke}%
\pgfsetdash{}{0pt}%
\pgfpathmoveto{\pgfqpoint{1.157239in}{0.929220in}}%
\pgfpathlineto{\pgfqpoint{6.806282in}{0.929220in}}%
\pgfusepath{stroke}%
\end{pgfscope}%
\begin{pgfscope}%
\pgfsetrectcap%
\pgfsetmiterjoin%
\pgfsetlinewidth{0.803000pt}%
\definecolor{currentstroke}{rgb}{0.000000,0.000000,0.000000}%
\pgfsetstrokecolor{currentstroke}%
\pgfsetdash{}{0pt}%
\pgfpathmoveto{\pgfqpoint{1.157239in}{6.000000in}}%
\pgfpathlineto{\pgfqpoint{6.806282in}{6.000000in}}%
\pgfusepath{stroke}%
\end{pgfscope}%
\end{pgfpicture}%
\makeatother%
\endgroup%

%% file: Figs/gospa_-10dB/gospa_miss.pgf
\begingroup%
\makeatletter%
\begin{pgfpicture}%
\pgfpathrectangle{\pgfpointorigin}{\pgfqpoint{7.000000in}{6.000000in}}%
\pgfusepath{use as bounding box, clip}%
\begin{pgfscope}%
\pgfsetbuttcap%
\pgfsetmiterjoin%
\definecolor{currentfill}{rgb}{1.000000,1.000000,1.000000}%
\pgfsetfillcolor{currentfill}%
\pgfsetlinewidth{0.000000pt}%
\definecolor{currentstroke}{rgb}{1.000000,1.000000,1.000000}%
\pgfsetstrokecolor{currentstroke}%
\pgfsetdash{}{0pt}%
\pgfpathmoveto{\pgfqpoint{0.000000in}{0.000000in}}%
\pgfpathlineto{\pgfqpoint{7.000000in}{0.000000in}}%
\pgfpathlineto{\pgfqpoint{7.000000in}{6.000000in}}%
\pgfpathlineto{\pgfqpoint{0.000000in}{6.000000in}}%
\pgfpathlineto{\pgfqpoint{0.000000in}{0.000000in}}%
\pgfpathclose%
\pgfusepath{fill}%
\end{pgfscope}%
\begin{pgfscope}%
\pgfsetbuttcap%
\pgfsetmiterjoin%
\definecolor{currentfill}{rgb}{1.000000,1.000000,1.000000}%
\pgfsetfillcolor{currentfill}%
\pgfsetlinewidth{0.000000pt}%
\definecolor{currentstroke}{rgb}{0.000000,0.000000,0.000000}%
\pgfsetstrokecolor{currentstroke}%
\pgfsetstrokeopacity{0.000000}%
\pgfsetdash{}{0pt}%
\pgfpathmoveto{\pgfqpoint{1.157239in}{0.929220in}}%
\pgfpathlineto{\pgfqpoint{6.806282in}{0.929220in}}%
\pgfpathlineto{\pgfqpoint{6.806282in}{6.000000in}}%
\pgfpathlineto{\pgfqpoint{1.157239in}{6.000000in}}%
\pgfpathlineto{\pgfqpoint{1.157239in}{0.929220in}}%
\pgfpathclose%
\pgfusepath{fill}%
\end{pgfscope}%
\begin{pgfscope}%
\pgfpathrectangle{\pgfqpoint{1.157239in}{0.929220in}}{\pgfqpoint{5.649043in}{5.070780in}}%
\pgfusepath{clip}%
\pgfsetrectcap%
\pgfsetroundjoin%
\pgfsetlinewidth{0.803000pt}%
\definecolor{currentstroke}{rgb}{0.690196,0.690196,0.690196}%
\pgfsetstrokecolor{currentstroke}%
\pgfsetdash{}{0pt}%
\pgfpathmoveto{\pgfqpoint{2.515869in}{0.929220in}}%
\pgfpathlineto{\pgfqpoint{2.515869in}{6.000000in}}%
\pgfusepath{stroke}%
\end{pgfscope}%
\begin{pgfscope}%
\pgfsetbuttcap%
\pgfsetroundjoin%
\definecolor{currentfill}{rgb}{0.000000,0.000000,0.000000}%
\pgfsetfillcolor{currentfill}%
\pgfsetlinewidth{0.803000pt}%
\definecolor{currentstroke}{rgb}{0.000000,0.000000,0.000000}%
\pgfsetstrokecolor{currentstroke}%
\pgfsetdash{}{0pt}%
\pgfsys@defobject{currentmarker}{\pgfqpoint{0.000000in}{-0.048611in}}{\pgfqpoint{0.000000in}{0.000000in}}{%
\pgfpathmoveto{\pgfqpoint{0.000000in}{0.000000in}}%
\pgfpathlineto{\pgfqpoint{0.000000in}{-0.048611in}}%
\pgfusepath{stroke,fill}%
}%
\begin{pgfscope}%
\pgfsys@transformshift{2.515869in}{0.929220in}%
\pgfsys@useobject{currentmarker}{}%
\end{pgfscope}%
\end{pgfscope}%
\begin{pgfscope}%
\definecolor{textcolor}{rgb}{0.000000,0.000000,0.000000}%
\pgfsetstrokecolor{textcolor}%
\pgfsetfillcolor{textcolor}%
\pgftext[x=2.515869in,y=0.831998in,,top]{\color{textcolor}\rmfamily\fontsize{28.000000}{33.600000}\selectfont 20}%
\end{pgfscope}%
\begin{pgfscope}%
\pgfpathrectangle{\pgfqpoint{1.157239in}{0.929220in}}{\pgfqpoint{5.649043in}{5.070780in}}%
\pgfusepath{clip}%
\pgfsetrectcap%
\pgfsetroundjoin%
\pgfsetlinewidth{0.803000pt}%
\definecolor{currentstroke}{rgb}{0.690196,0.690196,0.690196}%
\pgfsetstrokecolor{currentstroke}%
\pgfsetdash{}{0pt}%
\pgfpathmoveto{\pgfqpoint{3.946007in}{0.929220in}}%
\pgfpathlineto{\pgfqpoint{3.946007in}{6.000000in}}%
\pgfusepath{stroke}%
\end{pgfscope}%
\begin{pgfscope}%
\pgfsetbuttcap%
\pgfsetroundjoin%
\definecolor{currentfill}{rgb}{0.000000,0.000000,0.000000}%
\pgfsetfillcolor{currentfill}%
\pgfsetlinewidth{0.803000pt}%
\definecolor{currentstroke}{rgb}{0.000000,0.000000,0.000000}%
\pgfsetstrokecolor{currentstroke}%
\pgfsetdash{}{0pt}%
\pgfsys@defobject{currentmarker}{\pgfqpoint{0.000000in}{-0.048611in}}{\pgfqpoint{0.000000in}{0.000000in}}{%
\pgfpathmoveto{\pgfqpoint{0.000000in}{0.000000in}}%
\pgfpathlineto{\pgfqpoint{0.000000in}{-0.048611in}}%
\pgfusepath{stroke,fill}%
}%
\begin{pgfscope}%
\pgfsys@transformshift{3.946007in}{0.929220in}%
\pgfsys@useobject{currentmarker}{}%
\end{pgfscope}%
\end{pgfscope}%
\begin{pgfscope}%
\definecolor{textcolor}{rgb}{0.000000,0.000000,0.000000}%
\pgfsetstrokecolor{textcolor}%
\pgfsetfillcolor{textcolor}%
\pgftext[x=3.946007in,y=0.831998in,,top]{\color{textcolor}\rmfamily\fontsize{28.000000}{33.600000}\selectfont 40}%
\end{pgfscope}%
\begin{pgfscope}%
\pgfpathrectangle{\pgfqpoint{1.157239in}{0.929220in}}{\pgfqpoint{5.649043in}{5.070780in}}%
\pgfusepath{clip}%
\pgfsetrectcap%
\pgfsetroundjoin%
\pgfsetlinewidth{0.803000pt}%
\definecolor{currentstroke}{rgb}{0.690196,0.690196,0.690196}%
\pgfsetstrokecolor{currentstroke}%
\pgfsetdash{}{0pt}%
\pgfpathmoveto{\pgfqpoint{5.376144in}{0.929220in}}%
\pgfpathlineto{\pgfqpoint{5.376144in}{6.000000in}}%
\pgfusepath{stroke}%
\end{pgfscope}%
\begin{pgfscope}%
\pgfsetbuttcap%
\pgfsetroundjoin%
\definecolor{currentfill}{rgb}{0.000000,0.000000,0.000000}%
\pgfsetfillcolor{currentfill}%
\pgfsetlinewidth{0.803000pt}%
\definecolor{currentstroke}{rgb}{0.000000,0.000000,0.000000}%
\pgfsetstrokecolor{currentstroke}%
\pgfsetdash{}{0pt}%
\pgfsys@defobject{currentmarker}{\pgfqpoint{0.000000in}{-0.048611in}}{\pgfqpoint{0.000000in}{0.000000in}}{%
\pgfpathmoveto{\pgfqpoint{0.000000in}{0.000000in}}%
\pgfpathlineto{\pgfqpoint{0.000000in}{-0.048611in}}%
\pgfusepath{stroke,fill}%
}%
\begin{pgfscope}%
\pgfsys@transformshift{5.376144in}{0.929220in}%
\pgfsys@useobject{currentmarker}{}%
\end{pgfscope}%
\end{pgfscope}%
\begin{pgfscope}%
\definecolor{textcolor}{rgb}{0.000000,0.000000,0.000000}%
\pgfsetstrokecolor{textcolor}%
\pgfsetfillcolor{textcolor}%
\pgftext[x=5.376144in,y=0.831998in,,top]{\color{textcolor}\rmfamily\fontsize{28.000000}{33.600000}\selectfont 60}%
\end{pgfscope}%
\begin{pgfscope}%
\pgfpathrectangle{\pgfqpoint{1.157239in}{0.929220in}}{\pgfqpoint{5.649043in}{5.070780in}}%
\pgfusepath{clip}%
\pgfsetrectcap%
\pgfsetroundjoin%
\pgfsetlinewidth{0.803000pt}%
\definecolor{currentstroke}{rgb}{0.690196,0.690196,0.690196}%
\pgfsetstrokecolor{currentstroke}%
\pgfsetdash{}{0pt}%
\pgfpathmoveto{\pgfqpoint{6.806282in}{0.929220in}}%
\pgfpathlineto{\pgfqpoint{6.806282in}{6.000000in}}%
\pgfusepath{stroke}%
\end{pgfscope}%
\begin{pgfscope}%
\pgfsetbuttcap%
\pgfsetroundjoin%
\definecolor{currentfill}{rgb}{0.000000,0.000000,0.000000}%
\pgfsetfillcolor{currentfill}%
\pgfsetlinewidth{0.803000pt}%
\definecolor{currentstroke}{rgb}{0.000000,0.000000,0.000000}%
\pgfsetstrokecolor{currentstroke}%
\pgfsetdash{}{0pt}%
\pgfsys@defobject{currentmarker}{\pgfqpoint{0.000000in}{-0.048611in}}{\pgfqpoint{0.000000in}{0.000000in}}{%
\pgfpathmoveto{\pgfqpoint{0.000000in}{0.000000in}}%
\pgfpathlineto{\pgfqpoint{0.000000in}{-0.048611in}}%
\pgfusepath{stroke,fill}%
}%
\begin{pgfscope}%
\pgfsys@transformshift{6.806282in}{0.929220in}%
\pgfsys@useobject{currentmarker}{}%
\end{pgfscope}%
\end{pgfscope}%
\begin{pgfscope}%
\definecolor{textcolor}{rgb}{0.000000,0.000000,0.000000}%
\pgfsetstrokecolor{textcolor}%
\pgfsetfillcolor{textcolor}%
\pgftext[x=6.806282in,y=0.831998in,,top]{\color{textcolor}\rmfamily\fontsize{28.000000}{33.600000}\selectfont 80}%
\end{pgfscope}%
\begin{pgfscope}%
\definecolor{textcolor}{rgb}{0.000000,0.000000,0.000000}%
\pgfsetstrokecolor{textcolor}%
\pgfsetfillcolor{textcolor}%
\pgftext[x=3.981760in,y=0.385926in,,top]{\color{textcolor}\rmfamily\fontsize{28.000000}{33.600000}\selectfont time step \(\displaystyle k\)}%
\end{pgfscope}%
\begin{pgfscope}%
\pgfpathrectangle{\pgfqpoint{1.157239in}{0.929220in}}{\pgfqpoint{5.649043in}{5.070780in}}%
\pgfusepath{clip}%
\pgfsetrectcap%
\pgfsetroundjoin%
\pgfsetlinewidth{0.803000pt}%
\definecolor{currentstroke}{rgb}{0.690196,0.690196,0.690196}%
\pgfsetstrokecolor{currentstroke}%
\pgfsetdash{}{0pt}%
\pgfpathmoveto{\pgfqpoint{1.157239in}{0.929220in}}%
\pgfpathlineto{\pgfqpoint{6.806282in}{0.929220in}}%
\pgfusepath{stroke}%
\end{pgfscope}%
\begin{pgfscope}%
\pgfsetbuttcap%
\pgfsetroundjoin%
\definecolor{currentfill}{rgb}{0.000000,0.000000,0.000000}%
\pgfsetfillcolor{currentfill}%
\pgfsetlinewidth{0.803000pt}%
\definecolor{currentstroke}{rgb}{0.000000,0.000000,0.000000}%
\pgfsetstrokecolor{currentstroke}%
\pgfsetdash{}{0pt}%
\pgfsys@defobject{currentmarker}{\pgfqpoint{-0.048611in}{0.000000in}}{\pgfqpoint{-0.000000in}{0.000000in}}{%
\pgfpathmoveto{\pgfqpoint{-0.000000in}{0.000000in}}%
\pgfpathlineto{\pgfqpoint{-0.048611in}{0.000000in}}%
\pgfusepath{stroke,fill}%
}%
\begin{pgfscope}%
\pgfsys@transformshift{1.157239in}{0.929220in}%
\pgfsys@useobject{currentmarker}{}%
\end{pgfscope}%
\end{pgfscope}%
\begin{pgfscope}%
\definecolor{textcolor}{rgb}{0.000000,0.000000,0.000000}%
\pgfsetstrokecolor{textcolor}%
\pgfsetfillcolor{textcolor}%
\pgftext[x=0.811260in, y=0.809236in, left, base]{\color{textcolor}\rmfamily\fontsize{28.000000}{33.600000}\selectfont 0}%
\end{pgfscope}%
\begin{pgfscope}%
\pgfpathrectangle{\pgfqpoint{1.157239in}{0.929220in}}{\pgfqpoint{5.649043in}{5.070780in}}%
\pgfusepath{clip}%
\pgfsetrectcap%
\pgfsetroundjoin%
\pgfsetlinewidth{0.803000pt}%
\definecolor{currentstroke}{rgb}{0.690196,0.690196,0.690196}%
\pgfsetstrokecolor{currentstroke}%
\pgfsetdash{}{0pt}%
\pgfpathmoveto{\pgfqpoint{1.157239in}{2.289590in}}%
\pgfpathlineto{\pgfqpoint{6.806282in}{2.289590in}}%
\pgfusepath{stroke}%
\end{pgfscope}%
\begin{pgfscope}%
\pgfsetbuttcap%
\pgfsetroundjoin%
\definecolor{currentfill}{rgb}{0.000000,0.000000,0.000000}%
\pgfsetfillcolor{currentfill}%
\pgfsetlinewidth{0.803000pt}%
\definecolor{currentstroke}{rgb}{0.000000,0.000000,0.000000}%
\pgfsetstrokecolor{currentstroke}%
\pgfsetdash{}{0pt}%
\pgfsys@defobject{currentmarker}{\pgfqpoint{-0.048611in}{0.000000in}}{\pgfqpoint{-0.000000in}{0.000000in}}{%
\pgfpathmoveto{\pgfqpoint{-0.000000in}{0.000000in}}%
\pgfpathlineto{\pgfqpoint{-0.048611in}{0.000000in}}%
\pgfusepath{stroke,fill}%
}%
\begin{pgfscope}%
\pgfsys@transformshift{1.157239in}{2.289590in}%
\pgfsys@useobject{currentmarker}{}%
\end{pgfscope}%
\end{pgfscope}%
\begin{pgfscope}%
\definecolor{textcolor}{rgb}{0.000000,0.000000,0.000000}%
\pgfsetstrokecolor{textcolor}%
\pgfsetfillcolor{textcolor}%
\pgftext[x=0.566517in, y=2.169606in, left, base]{\color{textcolor}\rmfamily\fontsize{28.000000}{33.600000}\selectfont 0.5}%
\end{pgfscope}%
\begin{pgfscope}%
\pgfpathrectangle{\pgfqpoint{1.157239in}{0.929220in}}{\pgfqpoint{5.649043in}{5.070780in}}%
\pgfusepath{clip}%
\pgfsetrectcap%
\pgfsetroundjoin%
\pgfsetlinewidth{0.803000pt}%
\definecolor{currentstroke}{rgb}{0.690196,0.690196,0.690196}%
\pgfsetstrokecolor{currentstroke}%
\pgfsetdash{}{0pt}%
\pgfpathmoveto{\pgfqpoint{1.157239in}{3.649961in}}%
\pgfpathlineto{\pgfqpoint{6.806282in}{3.649961in}}%
\pgfusepath{stroke}%
\end{pgfscope}%
\begin{pgfscope}%
\pgfsetbuttcap%
\pgfsetroundjoin%
\definecolor{currentfill}{rgb}{0.000000,0.000000,0.000000}%
\pgfsetfillcolor{currentfill}%
\pgfsetlinewidth{0.803000pt}%
\definecolor{currentstroke}{rgb}{0.000000,0.000000,0.000000}%
\pgfsetstrokecolor{currentstroke}%
\pgfsetdash{}{0pt}%
\pgfsys@defobject{currentmarker}{\pgfqpoint{-0.048611in}{0.000000in}}{\pgfqpoint{-0.000000in}{0.000000in}}{%
\pgfpathmoveto{\pgfqpoint{-0.000000in}{0.000000in}}%
\pgfpathlineto{\pgfqpoint{-0.048611in}{0.000000in}}%
\pgfusepath{stroke,fill}%
}%
\begin{pgfscope}%
\pgfsys@transformshift{1.157239in}{3.649961in}%
\pgfsys@useobject{currentmarker}{}%
\end{pgfscope}%
\end{pgfscope}%
\begin{pgfscope}%
\definecolor{textcolor}{rgb}{0.000000,0.000000,0.000000}%
\pgfsetstrokecolor{textcolor}%
\pgfsetfillcolor{textcolor}%
\pgftext[x=0.811260in, y=3.529976in, left, base]{\color{textcolor}\rmfamily\fontsize{28.000000}{33.600000}\selectfont 1}%
\end{pgfscope}%
\begin{pgfscope}%
\pgfpathrectangle{\pgfqpoint{1.157239in}{0.929220in}}{\pgfqpoint{5.649043in}{5.070780in}}%
\pgfusepath{clip}%
\pgfsetrectcap%
\pgfsetroundjoin%
\pgfsetlinewidth{0.803000pt}%
\definecolor{currentstroke}{rgb}{0.690196,0.690196,0.690196}%
\pgfsetstrokecolor{currentstroke}%
\pgfsetdash{}{0pt}%
\pgfpathmoveto{\pgfqpoint{1.157239in}{5.010331in}}%
\pgfpathlineto{\pgfqpoint{6.806282in}{5.010331in}}%
\pgfusepath{stroke}%
\end{pgfscope}%
\begin{pgfscope}%
\pgfsetbuttcap%
\pgfsetroundjoin%
\definecolor{currentfill}{rgb}{0.000000,0.000000,0.000000}%
\pgfsetfillcolor{currentfill}%
\pgfsetlinewidth{0.803000pt}%
\definecolor{currentstroke}{rgb}{0.000000,0.000000,0.000000}%
\pgfsetstrokecolor{currentstroke}%
\pgfsetdash{}{0pt}%
\pgfsys@defobject{currentmarker}{\pgfqpoint{-0.048611in}{0.000000in}}{\pgfqpoint{-0.000000in}{0.000000in}}{%
\pgfpathmoveto{\pgfqpoint{-0.000000in}{0.000000in}}%
\pgfpathlineto{\pgfqpoint{-0.048611in}{0.000000in}}%
\pgfusepath{stroke,fill}%
}%
\begin{pgfscope}%
\pgfsys@transformshift{1.157239in}{5.010331in}%
\pgfsys@useobject{currentmarker}{}%
\end{pgfscope}%
\end{pgfscope}%
\begin{pgfscope}%
\definecolor{textcolor}{rgb}{0.000000,0.000000,0.000000}%
\pgfsetstrokecolor{textcolor}%
\pgfsetfillcolor{textcolor}%
\pgftext[x=0.566517in, y=4.890346in, left, base]{\color{textcolor}\rmfamily\fontsize{28.000000}{33.600000}\selectfont 1.5}%
\end{pgfscope}%
\begin{pgfscope}%
\definecolor{textcolor}{rgb}{0.000000,0.000000,0.000000}%
\pgfsetstrokecolor{textcolor}%
\pgfsetfillcolor{textcolor}%
\pgftext[x=0.427628in,y=3.464610in,,bottom,rotate=90.000000]{\color{textcolor}\rmfamily\fontsize{28.000000}{33.600000}\selectfont Missed Objects Error [m]}%
\end{pgfscope}%
\begin{pgfscope}%
\pgfpathrectangle{\pgfqpoint{1.157239in}{0.929220in}}{\pgfqpoint{5.649043in}{5.070780in}}%
\pgfusepath{clip}%
\pgfsetrectcap%
\pgfsetroundjoin%
\pgfsetlinewidth{1.505625pt}%
\definecolor{currentstroke}{rgb}{0.750000,0.750000,0.000000}%
\pgfsetstrokecolor{currentstroke}%
\pgfsetdash{}{0pt}%
\pgfpathmoveto{\pgfqpoint{1.157239in}{0.929220in}}%
\pgfpathlineto{\pgfqpoint{1.228745in}{0.929220in}}%
\pgfpathlineto{\pgfqpoint{1.300252in}{0.929220in}}%
\pgfpathlineto{\pgfqpoint{1.371759in}{0.929220in}}%
\pgfpathlineto{\pgfqpoint{1.443266in}{0.929220in}}%
\pgfpathlineto{\pgfqpoint{1.514773in}{0.929220in}}%
\pgfpathlineto{\pgfqpoint{1.586280in}{0.929220in}}%
\pgfpathlineto{\pgfqpoint{1.657787in}{0.929220in}}%
\pgfpathlineto{\pgfqpoint{1.729294in}{0.929220in}}%
\pgfpathlineto{\pgfqpoint{1.800800in}{1.378142in}}%
\pgfpathlineto{\pgfqpoint{1.872307in}{1.568594in}}%
\pgfpathlineto{\pgfqpoint{1.943814in}{1.745442in}}%
\pgfpathlineto{\pgfqpoint{2.015321in}{1.623009in}}%
\pgfpathlineto{\pgfqpoint{2.086828in}{1.772650in}}%
\pgfpathlineto{\pgfqpoint{2.158335in}{2.752116in}}%
\pgfpathlineto{\pgfqpoint{2.229842in}{2.956172in}}%
\pgfpathlineto{\pgfqpoint{2.301349in}{2.847342in}}%
\pgfpathlineto{\pgfqpoint{2.372855in}{2.738513in}}%
\pgfpathlineto{\pgfqpoint{2.444362in}{2.806531in}}%
\pgfpathlineto{\pgfqpoint{2.515869in}{3.990053in}}%
\pgfpathlineto{\pgfqpoint{2.587376in}{4.003657in}}%
\pgfpathlineto{\pgfqpoint{2.658883in}{4.112486in}}%
\pgfpathlineto{\pgfqpoint{2.730390in}{4.248523in}}%
\pgfpathlineto{\pgfqpoint{2.801897in}{4.248523in}}%
\pgfpathlineto{\pgfqpoint{2.873404in}{5.268801in}}%
\pgfpathlineto{\pgfqpoint{2.944910in}{5.445649in}}%
\pgfpathlineto{\pgfqpoint{3.016417in}{5.581686in}}%
\pgfpathlineto{\pgfqpoint{3.087924in}{5.282405in}}%
\pgfpathlineto{\pgfqpoint{3.159431in}{5.568082in}}%
\pgfpathlineto{\pgfqpoint{3.230938in}{5.445649in}}%
\pgfpathlineto{\pgfqpoint{3.302445in}{5.568082in}}%
\pgfpathlineto{\pgfqpoint{3.373952in}{5.758534in}}%
\pgfpathlineto{\pgfqpoint{3.445459in}{5.744931in}}%
\pgfpathlineto{\pgfqpoint{3.516966in}{5.595290in}}%
\pgfpathlineto{\pgfqpoint{3.588472in}{5.676912in}}%
\pgfpathlineto{\pgfqpoint{3.659979in}{5.404838in}}%
\pgfpathlineto{\pgfqpoint{3.731486in}{5.513668in}}%
\pgfpathlineto{\pgfqpoint{3.802993in}{5.282405in}}%
\pgfpathlineto{\pgfqpoint{3.874500in}{5.241594in}}%
\pgfpathlineto{\pgfqpoint{3.946007in}{5.336820in}}%
\pgfpathlineto{\pgfqpoint{4.017514in}{5.119160in}}%
\pgfpathlineto{\pgfqpoint{4.089021in}{5.377631in}}%
\pgfpathlineto{\pgfqpoint{4.160527in}{5.268801in}}%
\pgfpathlineto{\pgfqpoint{4.232034in}{5.432045in}}%
\pgfpathlineto{\pgfqpoint{4.303541in}{5.472857in}}%
\pgfpathlineto{\pgfqpoint{4.375048in}{5.622497in}}%
\pgfpathlineto{\pgfqpoint{4.446555in}{5.472857in}}%
\pgfpathlineto{\pgfqpoint{4.518062in}{5.486460in}}%
\pgfpathlineto{\pgfqpoint{4.589569in}{5.200783in}}%
\pgfpathlineto{\pgfqpoint{4.661076in}{5.336820in}}%
\pgfpathlineto{\pgfqpoint{4.732582in}{5.323216in}}%
\pgfpathlineto{\pgfqpoint{4.804089in}{5.432045in}}%
\pgfpathlineto{\pgfqpoint{4.875596in}{5.364027in}}%
\pgfpathlineto{\pgfqpoint{4.947103in}{5.459253in}}%
\pgfpathlineto{\pgfqpoint{5.018610in}{4.561409in}}%
\pgfpathlineto{\pgfqpoint{5.090117in}{4.384560in}}%
\pgfpathlineto{\pgfqpoint{5.161624in}{4.316542in}}%
\pgfpathlineto{\pgfqpoint{5.233131in}{4.221316in}}%
\pgfpathlineto{\pgfqpoint{5.304638in}{4.234920in}}%
\pgfpathlineto{\pgfqpoint{5.376144in}{3.282661in}}%
\pgfpathlineto{\pgfqpoint{5.447651in}{3.269057in}}%
\pgfpathlineto{\pgfqpoint{5.519158in}{2.996983in}}%
\pgfpathlineto{\pgfqpoint{5.590665in}{2.969776in}}%
\pgfpathlineto{\pgfqpoint{5.662172in}{3.024190in}}%
\pgfpathlineto{\pgfqpoint{5.733679in}{2.371213in}}%
\pgfpathlineto{\pgfqpoint{5.805186in}{1.908687in}}%
\pgfpathlineto{\pgfqpoint{5.876693in}{1.718235in}}%
\pgfpathlineto{\pgfqpoint{5.948199in}{1.827065in}}%
\pgfpathlineto{\pgfqpoint{6.019706in}{1.813461in}}%
\pgfpathlineto{\pgfqpoint{6.091213in}{0.929220in}}%
\pgfpathlineto{\pgfqpoint{6.162720in}{0.929220in}}%
\pgfpathlineto{\pgfqpoint{6.234227in}{0.929220in}}%
\pgfpathlineto{\pgfqpoint{6.305734in}{0.929220in}}%
\pgfpathlineto{\pgfqpoint{6.377241in}{0.929220in}}%
\pgfpathlineto{\pgfqpoint{6.448748in}{0.929220in}}%
\pgfpathlineto{\pgfqpoint{6.520254in}{0.929220in}}%
\pgfpathlineto{\pgfqpoint{6.591761in}{0.929220in}}%
\pgfpathlineto{\pgfqpoint{6.663268in}{0.929220in}}%
\pgfpathlineto{\pgfqpoint{6.734775in}{0.929220in}}%
\pgfpathlineto{\pgfqpoint{6.806282in}{0.929220in}}%
\pgfusepath{stroke}%
\end{pgfscope}%
\begin{pgfscope}%
\pgfpathrectangle{\pgfqpoint{1.157239in}{0.929220in}}{\pgfqpoint{5.649043in}{5.070780in}}%
\pgfusepath{clip}%
\pgfsetrectcap%
\pgfsetroundjoin%
\pgfsetlinewidth{1.505625pt}%
\definecolor{currentstroke}{rgb}{0.000000,0.000000,1.000000}%
\pgfsetstrokecolor{currentstroke}%
\pgfsetdash{}{0pt}%
\pgfpathmoveto{\pgfqpoint{1.157239in}{0.929220in}}%
\pgfpathlineto{\pgfqpoint{1.228745in}{0.929220in}}%
\pgfpathlineto{\pgfqpoint{1.300252in}{0.929220in}}%
\pgfpathlineto{\pgfqpoint{1.371759in}{0.929220in}}%
\pgfpathlineto{\pgfqpoint{1.443266in}{0.929220in}}%
\pgfpathlineto{\pgfqpoint{1.514773in}{0.929220in}}%
\pgfpathlineto{\pgfqpoint{1.586280in}{0.929220in}}%
\pgfpathlineto{\pgfqpoint{1.657787in}{0.929220in}}%
\pgfpathlineto{\pgfqpoint{1.729294in}{0.929220in}}%
\pgfpathlineto{\pgfqpoint{1.800800in}{3.649961in}}%
\pgfpathlineto{\pgfqpoint{1.872307in}{3.459509in}}%
\pgfpathlineto{\pgfqpoint{1.943814in}{0.983635in}}%
\pgfpathlineto{\pgfqpoint{2.015321in}{0.942824in}}%
\pgfpathlineto{\pgfqpoint{2.086828in}{0.942824in}}%
\pgfpathlineto{\pgfqpoint{2.158335in}{2.289590in}}%
\pgfpathlineto{\pgfqpoint{2.229842in}{2.044724in}}%
\pgfpathlineto{\pgfqpoint{2.301349in}{1.146880in}}%
\pgfpathlineto{\pgfqpoint{2.372855in}{1.187691in}}%
\pgfpathlineto{\pgfqpoint{2.444362in}{1.228502in}}%
\pgfpathlineto{\pgfqpoint{2.515869in}{2.670494in}}%
\pgfpathlineto{\pgfqpoint{2.587376in}{2.548061in}}%
\pgfpathlineto{\pgfqpoint{2.658883in}{1.786254in}}%
\pgfpathlineto{\pgfqpoint{2.730390in}{1.759046in}}%
\pgfpathlineto{\pgfqpoint{2.801897in}{1.867876in}}%
\pgfpathlineto{\pgfqpoint{2.873404in}{3.391490in}}%
\pgfpathlineto{\pgfqpoint{2.944910in}{3.269057in}}%
\pgfpathlineto{\pgfqpoint{3.016417in}{2.901757in}}%
\pgfpathlineto{\pgfqpoint{3.087924in}{2.996983in}}%
\pgfpathlineto{\pgfqpoint{3.159431in}{2.942568in}}%
\pgfpathlineto{\pgfqpoint{3.230938in}{2.888153in}}%
\pgfpathlineto{\pgfqpoint{3.302445in}{3.187435in}}%
\pgfpathlineto{\pgfqpoint{3.373952in}{3.255453in}}%
\pgfpathlineto{\pgfqpoint{3.445459in}{3.391490in}}%
\pgfpathlineto{\pgfqpoint{3.516966in}{3.459509in}}%
\pgfpathlineto{\pgfqpoint{3.588472in}{3.690772in}}%
\pgfpathlineto{\pgfqpoint{3.659979in}{3.581942in}}%
\pgfpathlineto{\pgfqpoint{3.731486in}{3.568338in}}%
\pgfpathlineto{\pgfqpoint{3.802993in}{3.609149in}}%
\pgfpathlineto{\pgfqpoint{3.874500in}{3.527527in}}%
\pgfpathlineto{\pgfqpoint{3.946007in}{3.473112in}}%
\pgfpathlineto{\pgfqpoint{4.017514in}{3.595546in}}%
\pgfpathlineto{\pgfqpoint{4.089021in}{3.785998in}}%
\pgfpathlineto{\pgfqpoint{4.160527in}{3.704375in}}%
\pgfpathlineto{\pgfqpoint{4.232034in}{3.704375in}}%
\pgfpathlineto{\pgfqpoint{4.303541in}{3.894827in}}%
\pgfpathlineto{\pgfqpoint{4.375048in}{3.922035in}}%
\pgfpathlineto{\pgfqpoint{4.446555in}{3.785998in}}%
\pgfpathlineto{\pgfqpoint{4.518062in}{3.609149in}}%
\pgfpathlineto{\pgfqpoint{4.589569in}{3.323472in}}%
\pgfpathlineto{\pgfqpoint{4.661076in}{3.255453in}}%
\pgfpathlineto{\pgfqpoint{4.732582in}{3.051398in}}%
\pgfpathlineto{\pgfqpoint{4.804089in}{2.996983in}}%
\pgfpathlineto{\pgfqpoint{4.875596in}{2.983379in}}%
\pgfpathlineto{\pgfqpoint{4.947103in}{2.996983in}}%
\pgfpathlineto{\pgfqpoint{5.018610in}{2.221572in}}%
\pgfpathlineto{\pgfqpoint{5.090117in}{2.085535in}}%
\pgfpathlineto{\pgfqpoint{5.161624in}{1.908687in}}%
\pgfpathlineto{\pgfqpoint{5.233131in}{1.881479in}}%
\pgfpathlineto{\pgfqpoint{5.304638in}{1.772650in}}%
\pgfpathlineto{\pgfqpoint{5.376144in}{1.636613in}}%
\pgfpathlineto{\pgfqpoint{5.447651in}{1.514179in}}%
\pgfpathlineto{\pgfqpoint{5.519158in}{1.446161in}}%
\pgfpathlineto{\pgfqpoint{5.590665in}{1.446161in}}%
\pgfpathlineto{\pgfqpoint{5.662172in}{1.418954in}}%
\pgfpathlineto{\pgfqpoint{5.733679in}{1.174087in}}%
\pgfpathlineto{\pgfqpoint{5.805186in}{1.065257in}}%
\pgfpathlineto{\pgfqpoint{5.876693in}{0.970031in}}%
\pgfpathlineto{\pgfqpoint{5.948199in}{0.956428in}}%
\pgfpathlineto{\pgfqpoint{6.019706in}{0.942824in}}%
\pgfpathlineto{\pgfqpoint{6.091213in}{0.929220in}}%
\pgfpathlineto{\pgfqpoint{6.162720in}{0.929220in}}%
\pgfpathlineto{\pgfqpoint{6.234227in}{0.929220in}}%
\pgfpathlineto{\pgfqpoint{6.305734in}{0.929220in}}%
\pgfpathlineto{\pgfqpoint{6.377241in}{0.929220in}}%
\pgfpathlineto{\pgfqpoint{6.448748in}{0.929220in}}%
\pgfpathlineto{\pgfqpoint{6.520254in}{0.929220in}}%
\pgfpathlineto{\pgfqpoint{6.591761in}{0.929220in}}%
\pgfpathlineto{\pgfqpoint{6.663268in}{0.929220in}}%
\pgfpathlineto{\pgfqpoint{6.734775in}{0.929220in}}%
\pgfpathlineto{\pgfqpoint{6.806282in}{0.929220in}}%
\pgfusepath{stroke}%
\end{pgfscope}%
\begin{pgfscope}%
\pgfpathrectangle{\pgfqpoint{1.157239in}{0.929220in}}{\pgfqpoint{5.649043in}{5.070780in}}%
\pgfusepath{clip}%
\pgfsetrectcap%
\pgfsetroundjoin%
\pgfsetlinewidth{1.505625pt}%
\definecolor{currentstroke}{rgb}{0.000000,0.500000,0.000000}%
\pgfsetstrokecolor{currentstroke}%
\pgfsetdash{}{0pt}%
\pgfpathmoveto{\pgfqpoint{1.157239in}{0.929220in}}%
\pgfpathlineto{\pgfqpoint{1.228745in}{0.929220in}}%
\pgfpathlineto{\pgfqpoint{1.300252in}{0.929220in}}%
\pgfpathlineto{\pgfqpoint{1.371759in}{0.929220in}}%
\pgfpathlineto{\pgfqpoint{1.443266in}{0.929220in}}%
\pgfpathlineto{\pgfqpoint{1.514773in}{0.929220in}}%
\pgfpathlineto{\pgfqpoint{1.586280in}{0.929220in}}%
\pgfpathlineto{\pgfqpoint{1.657787in}{0.929220in}}%
\pgfpathlineto{\pgfqpoint{1.729294in}{0.929220in}}%
\pgfpathlineto{\pgfqpoint{1.800800in}{3.649961in}}%
\pgfpathlineto{\pgfqpoint{1.872307in}{3.432301in}}%
\pgfpathlineto{\pgfqpoint{1.943814in}{1.010843in}}%
\pgfpathlineto{\pgfqpoint{2.015321in}{0.956428in}}%
\pgfpathlineto{\pgfqpoint{2.086828in}{0.929220in}}%
\pgfpathlineto{\pgfqpoint{2.158335in}{2.289590in}}%
\pgfpathlineto{\pgfqpoint{2.229842in}{2.017516in}}%
\pgfpathlineto{\pgfqpoint{2.301349in}{0.929220in}}%
\pgfpathlineto{\pgfqpoint{2.372855in}{0.942824in}}%
\pgfpathlineto{\pgfqpoint{2.444362in}{0.929220in}}%
\pgfpathlineto{\pgfqpoint{2.515869in}{2.316798in}}%
\pgfpathlineto{\pgfqpoint{2.587376in}{1.976705in}}%
\pgfpathlineto{\pgfqpoint{2.658883in}{1.119672in}}%
\pgfpathlineto{\pgfqpoint{2.730390in}{1.106068in}}%
\pgfpathlineto{\pgfqpoint{2.801897in}{1.133276in}}%
\pgfpathlineto{\pgfqpoint{2.873404in}{2.520853in}}%
\pgfpathlineto{\pgfqpoint{2.944910in}{2.262383in}}%
\pgfpathlineto{\pgfqpoint{3.016417in}{1.486972in}}%
\pgfpathlineto{\pgfqpoint{3.087924in}{1.432557in}}%
\pgfpathlineto{\pgfqpoint{3.159431in}{1.323728in}}%
\pgfpathlineto{\pgfqpoint{3.230938in}{1.255709in}}%
\pgfpathlineto{\pgfqpoint{3.302445in}{1.282917in}}%
\pgfpathlineto{\pgfqpoint{3.373952in}{1.310124in}}%
\pgfpathlineto{\pgfqpoint{3.445459in}{1.269313in}}%
\pgfpathlineto{\pgfqpoint{3.516966in}{1.201294in}}%
\pgfpathlineto{\pgfqpoint{3.588472in}{1.242105in}}%
\pgfpathlineto{\pgfqpoint{3.659979in}{1.160483in}}%
\pgfpathlineto{\pgfqpoint{3.731486in}{1.228502in}}%
\pgfpathlineto{\pgfqpoint{3.802993in}{1.323728in}}%
\pgfpathlineto{\pgfqpoint{3.874500in}{1.473368in}}%
\pgfpathlineto{\pgfqpoint{3.946007in}{1.568594in}}%
\pgfpathlineto{\pgfqpoint{4.017514in}{1.786254in}}%
\pgfpathlineto{\pgfqpoint{4.089021in}{1.677424in}}%
\pgfpathlineto{\pgfqpoint{4.160527in}{1.595802in}}%
\pgfpathlineto{\pgfqpoint{4.232034in}{1.595802in}}%
\pgfpathlineto{\pgfqpoint{4.303541in}{1.527783in}}%
\pgfpathlineto{\pgfqpoint{4.375048in}{1.486972in}}%
\pgfpathlineto{\pgfqpoint{4.446555in}{1.500576in}}%
\pgfpathlineto{\pgfqpoint{4.518062in}{1.514179in}}%
\pgfpathlineto{\pgfqpoint{4.589569in}{1.514179in}}%
\pgfpathlineto{\pgfqpoint{4.661076in}{1.500576in}}%
\pgfpathlineto{\pgfqpoint{4.732582in}{1.405350in}}%
\pgfpathlineto{\pgfqpoint{4.804089in}{1.405350in}}%
\pgfpathlineto{\pgfqpoint{4.875596in}{1.459765in}}%
\pgfpathlineto{\pgfqpoint{4.947103in}{1.378142in}}%
\pgfpathlineto{\pgfqpoint{5.018610in}{1.201294in}}%
\pgfpathlineto{\pgfqpoint{5.090117in}{1.201294in}}%
\pgfpathlineto{\pgfqpoint{5.161624in}{1.160483in}}%
\pgfpathlineto{\pgfqpoint{5.233131in}{1.174087in}}%
\pgfpathlineto{\pgfqpoint{5.304638in}{1.242105in}}%
\pgfpathlineto{\pgfqpoint{5.376144in}{1.146880in}}%
\pgfpathlineto{\pgfqpoint{5.447651in}{1.119672in}}%
\pgfpathlineto{\pgfqpoint{5.519158in}{1.038050in}}%
\pgfpathlineto{\pgfqpoint{5.590665in}{1.092465in}}%
\pgfpathlineto{\pgfqpoint{5.662172in}{1.065257in}}%
\pgfpathlineto{\pgfqpoint{5.733679in}{0.956428in}}%
\pgfpathlineto{\pgfqpoint{5.805186in}{0.956428in}}%
\pgfpathlineto{\pgfqpoint{5.876693in}{0.929220in}}%
\pgfpathlineto{\pgfqpoint{5.948199in}{0.929220in}}%
\pgfpathlineto{\pgfqpoint{6.019706in}{0.956428in}}%
\pgfpathlineto{\pgfqpoint{6.091213in}{0.929220in}}%
\pgfpathlineto{\pgfqpoint{6.162720in}{0.929220in}}%
\pgfpathlineto{\pgfqpoint{6.234227in}{0.929220in}}%
\pgfpathlineto{\pgfqpoint{6.305734in}{0.929220in}}%
\pgfpathlineto{\pgfqpoint{6.377241in}{0.929220in}}%
\pgfpathlineto{\pgfqpoint{6.448748in}{0.929220in}}%
\pgfpathlineto{\pgfqpoint{6.520254in}{0.929220in}}%
\pgfpathlineto{\pgfqpoint{6.591761in}{0.929220in}}%
\pgfpathlineto{\pgfqpoint{6.663268in}{0.929220in}}%
\pgfpathlineto{\pgfqpoint{6.734775in}{0.929220in}}%
\pgfpathlineto{\pgfqpoint{6.806282in}{0.929220in}}%
\pgfusepath{stroke}%
\end{pgfscope}%
\begin{pgfscope}%
\pgfpathrectangle{\pgfqpoint{1.157239in}{0.929220in}}{\pgfqpoint{5.649043in}{5.070780in}}%
\pgfusepath{clip}%
\pgfsetrectcap%
\pgfsetroundjoin%
\pgfsetlinewidth{1.505625pt}%
\definecolor{currentstroke}{rgb}{1.000000,0.000000,0.000000}%
\pgfsetstrokecolor{currentstroke}%
\pgfsetdash{}{0pt}%
\pgfpathmoveto{\pgfqpoint{1.157239in}{0.929220in}}%
\pgfpathlineto{\pgfqpoint{1.228745in}{0.929220in}}%
\pgfpathlineto{\pgfqpoint{1.300252in}{0.929220in}}%
\pgfpathlineto{\pgfqpoint{1.371759in}{0.929220in}}%
\pgfpathlineto{\pgfqpoint{1.443266in}{0.929220in}}%
\pgfpathlineto{\pgfqpoint{1.514773in}{0.929220in}}%
\pgfpathlineto{\pgfqpoint{1.586280in}{0.929220in}}%
\pgfpathlineto{\pgfqpoint{1.657787in}{0.929220in}}%
\pgfpathlineto{\pgfqpoint{1.729294in}{0.929220in}}%
\pgfpathlineto{\pgfqpoint{1.800800in}{0.929220in}}%
\pgfpathlineto{\pgfqpoint{1.872307in}{0.997239in}}%
\pgfpathlineto{\pgfqpoint{1.943814in}{0.942824in}}%
\pgfpathlineto{\pgfqpoint{2.015321in}{0.929220in}}%
\pgfpathlineto{\pgfqpoint{2.086828in}{0.929220in}}%
\pgfpathlineto{\pgfqpoint{2.158335in}{0.929220in}}%
\pgfpathlineto{\pgfqpoint{2.229842in}{0.942824in}}%
\pgfpathlineto{\pgfqpoint{2.301349in}{0.942824in}}%
\pgfpathlineto{\pgfqpoint{2.372855in}{0.929220in}}%
\pgfpathlineto{\pgfqpoint{2.444362in}{0.929220in}}%
\pgfpathlineto{\pgfqpoint{2.515869in}{1.146880in}}%
\pgfpathlineto{\pgfqpoint{2.587376in}{1.038050in}}%
\pgfpathlineto{\pgfqpoint{2.658883in}{0.997239in}}%
\pgfpathlineto{\pgfqpoint{2.730390in}{0.997239in}}%
\pgfpathlineto{\pgfqpoint{2.801897in}{0.997239in}}%
\pgfpathlineto{\pgfqpoint{2.873404in}{1.350935in}}%
\pgfpathlineto{\pgfqpoint{2.944910in}{1.119672in}}%
\pgfpathlineto{\pgfqpoint{3.016417in}{1.133276in}}%
\pgfpathlineto{\pgfqpoint{3.087924in}{1.146880in}}%
\pgfpathlineto{\pgfqpoint{3.159431in}{1.187691in}}%
\pgfpathlineto{\pgfqpoint{3.230938in}{1.146880in}}%
\pgfpathlineto{\pgfqpoint{3.302445in}{1.282917in}}%
\pgfpathlineto{\pgfqpoint{3.373952in}{1.296520in}}%
\pgfpathlineto{\pgfqpoint{3.445459in}{1.323728in}}%
\pgfpathlineto{\pgfqpoint{3.516966in}{1.255709in}}%
\pgfpathlineto{\pgfqpoint{3.588472in}{1.310124in}}%
\pgfpathlineto{\pgfqpoint{3.659979in}{1.282917in}}%
\pgfpathlineto{\pgfqpoint{3.731486in}{1.310124in}}%
\pgfpathlineto{\pgfqpoint{3.802993in}{1.269313in}}%
\pgfpathlineto{\pgfqpoint{3.874500in}{1.214898in}}%
\pgfpathlineto{\pgfqpoint{3.946007in}{1.214898in}}%
\pgfpathlineto{\pgfqpoint{4.017514in}{1.242105in}}%
\pgfpathlineto{\pgfqpoint{4.089021in}{1.187691in}}%
\pgfpathlineto{\pgfqpoint{4.160527in}{1.242105in}}%
\pgfpathlineto{\pgfqpoint{4.232034in}{1.269313in}}%
\pgfpathlineto{\pgfqpoint{4.303541in}{1.269313in}}%
\pgfpathlineto{\pgfqpoint{4.375048in}{1.242105in}}%
\pgfpathlineto{\pgfqpoint{4.446555in}{1.228502in}}%
\pgfpathlineto{\pgfqpoint{4.518062in}{1.228502in}}%
\pgfpathlineto{\pgfqpoint{4.589569in}{1.269313in}}%
\pgfpathlineto{\pgfqpoint{4.661076in}{1.310124in}}%
\pgfpathlineto{\pgfqpoint{4.732582in}{1.255709in}}%
\pgfpathlineto{\pgfqpoint{4.804089in}{1.255709in}}%
\pgfpathlineto{\pgfqpoint{4.875596in}{1.228502in}}%
\pgfpathlineto{\pgfqpoint{4.947103in}{1.228502in}}%
\pgfpathlineto{\pgfqpoint{5.018610in}{1.133276in}}%
\pgfpathlineto{\pgfqpoint{5.090117in}{1.119672in}}%
\pgfpathlineto{\pgfqpoint{5.161624in}{1.106068in}}%
\pgfpathlineto{\pgfqpoint{5.233131in}{1.065257in}}%
\pgfpathlineto{\pgfqpoint{5.304638in}{1.038050in}}%
\pgfpathlineto{\pgfqpoint{5.376144in}{1.024446in}}%
\pgfpathlineto{\pgfqpoint{5.447651in}{1.010843in}}%
\pgfpathlineto{\pgfqpoint{5.519158in}{1.010843in}}%
\pgfpathlineto{\pgfqpoint{5.590665in}{1.024446in}}%
\pgfpathlineto{\pgfqpoint{5.662172in}{1.010843in}}%
\pgfpathlineto{\pgfqpoint{5.733679in}{0.942824in}}%
\pgfpathlineto{\pgfqpoint{5.805186in}{0.942824in}}%
\pgfpathlineto{\pgfqpoint{5.876693in}{0.942824in}}%
\pgfpathlineto{\pgfqpoint{5.948199in}{0.942824in}}%
\pgfpathlineto{\pgfqpoint{6.019706in}{0.942824in}}%
\pgfpathlineto{\pgfqpoint{6.091213in}{0.929220in}}%
\pgfpathlineto{\pgfqpoint{6.162720in}{0.929220in}}%
\pgfpathlineto{\pgfqpoint{6.234227in}{0.929220in}}%
\pgfpathlineto{\pgfqpoint{6.305734in}{0.929220in}}%
\pgfpathlineto{\pgfqpoint{6.377241in}{0.929220in}}%
\pgfpathlineto{\pgfqpoint{6.448748in}{0.929220in}}%
\pgfpathlineto{\pgfqpoint{6.520254in}{0.929220in}}%
\pgfpathlineto{\pgfqpoint{6.591761in}{0.929220in}}%
\pgfpathlineto{\pgfqpoint{6.663268in}{0.929220in}}%
\pgfpathlineto{\pgfqpoint{6.734775in}{0.929220in}}%
\pgfpathlineto{\pgfqpoint{6.806282in}{0.929220in}}%
\pgfusepath{stroke}%
\end{pgfscope}%
\begin{pgfscope}%
\pgfsetrectcap%
\pgfsetmiterjoin%
\pgfsetlinewidth{0.803000pt}%
\definecolor{currentstroke}{rgb}{0.000000,0.000000,0.000000}%
\pgfsetstrokecolor{currentstroke}%
\pgfsetdash{}{0pt}%
\pgfpathmoveto{\pgfqpoint{1.157239in}{0.929220in}}%
\pgfpathlineto{\pgfqpoint{1.157239in}{6.000000in}}%
\pgfusepath{stroke}%
\end{pgfscope}%
\begin{pgfscope}%
\pgfsetrectcap%
\pgfsetmiterjoin%
\pgfsetlinewidth{0.803000pt}%
\definecolor{currentstroke}{rgb}{0.000000,0.000000,0.000000}%
\pgfsetstrokecolor{currentstroke}%
\pgfsetdash{}{0pt}%
\pgfpathmoveto{\pgfqpoint{6.806282in}{0.929220in}}%
\pgfpathlineto{\pgfqpoint{6.806282in}{6.000000in}}%
\pgfusepath{stroke}%
\end{pgfscope}%
\begin{pgfscope}%
\pgfsetrectcap%
\pgfsetmiterjoin%
\pgfsetlinewidth{0.803000pt}%
\definecolor{currentstroke}{rgb}{0.000000,0.000000,0.000000}%
\pgfsetstrokecolor{currentstroke}%
\pgfsetdash{}{0pt}%
\pgfpathmoveto{\pgfqpoint{1.157239in}{0.929220in}}%
\pgfpathlineto{\pgfqpoint{6.806282in}{0.929220in}}%
\pgfusepath{stroke}%
\end{pgfscope}%
\begin{pgfscope}%
\pgfsetrectcap%
\pgfsetmiterjoin%
\pgfsetlinewidth{0.803000pt}%
\definecolor{currentstroke}{rgb}{0.000000,0.000000,0.000000}%
\pgfsetstrokecolor{currentstroke}%
\pgfsetdash{}{0pt}%
\pgfpathmoveto{\pgfqpoint{1.157239in}{6.000000in}}%
\pgfpathlineto{\pgfqpoint{6.806282in}{6.000000in}}%
\pgfusepath{stroke}%
\end{pgfscope}%
\end{pgfpicture}%
\makeatother%
\endgroup%

%% file: Figs/noise_var_sensor1.pgf
\begingroup%
\makeatletter%
\begin{pgfpicture}%
\pgfpathrectangle{\pgfpointorigin}{\pgfqpoint{6.600000in}{6.200000in}}%
\pgfusepath{use as bounding box, clip}%
\begin{pgfscope}%
\pgfsetbuttcap%
\pgfsetmiterjoin%
\definecolor{currentfill}{rgb}{1.000000,1.000000,1.000000}%
\pgfsetfillcolor{currentfill}%
\pgfsetlinewidth{0.000000pt}%
\definecolor{currentstroke}{rgb}{1.000000,1.000000,1.000000}%
\pgfsetstrokecolor{currentstroke}%
\pgfsetdash{}{0pt}%
\pgfpathmoveto{\pgfqpoint{0.000000in}{0.000000in}}%
\pgfpathlineto{\pgfqpoint{6.600000in}{0.000000in}}%
\pgfpathlineto{\pgfqpoint{6.600000in}{6.200000in}}%
\pgfpathlineto{\pgfqpoint{0.000000in}{6.200000in}}%
\pgfpathlineto{\pgfqpoint{0.000000in}{0.000000in}}%
\pgfpathclose%
\pgfusepath{fill}%
\end{pgfscope}%
\begin{pgfscope}%
\pgfsetbuttcap%
\pgfsetmiterjoin%
\definecolor{currentfill}{rgb}{1.000000,1.000000,1.000000}%
\pgfsetfillcolor{currentfill}%
\pgfsetlinewidth{0.000000pt}%
\definecolor{currentstroke}{rgb}{0.000000,0.000000,0.000000}%
\pgfsetstrokecolor{currentstroke}%
\pgfsetstrokeopacity{0.000000}%
\pgfsetdash{}{0pt}%
\pgfpathmoveto{\pgfqpoint{1.009028in}{1.080000in}}%
\pgfpathlineto{\pgfqpoint{6.338611in}{1.080000in}}%
\pgfpathlineto{\pgfqpoint{6.338611in}{5.904444in}}%
\pgfpathlineto{\pgfqpoint{1.009028in}{5.904444in}}%
\pgfpathlineto{\pgfqpoint{1.009028in}{1.080000in}}%
\pgfpathclose%
\pgfusepath{fill}%
\end{pgfscope}%
\begin{pgfscope}%
\pgfpathrectangle{\pgfqpoint{1.009028in}{1.080000in}}{\pgfqpoint{5.329583in}{4.824444in}}%
\pgfusepath{clip}%
\pgfsetrectcap%
\pgfsetroundjoin%
\pgfsetlinewidth{0.803000pt}%
\definecolor{currentstroke}{rgb}{0.690196,0.690196,0.690196}%
\pgfsetstrokecolor{currentstroke}%
\pgfsetdash{}{0pt}%
\pgfpathmoveto{\pgfqpoint{2.290826in}{1.080000in}}%
\pgfpathlineto{\pgfqpoint{2.290826in}{5.904444in}}%
\pgfusepath{stroke}%
\end{pgfscope}%
\begin{pgfscope}%
\pgfsetbuttcap%
\pgfsetroundjoin%
\definecolor{currentfill}{rgb}{0.000000,0.000000,0.000000}%
\pgfsetfillcolor{currentfill}%
\pgfsetlinewidth{0.803000pt}%
\definecolor{currentstroke}{rgb}{0.000000,0.000000,0.000000}%
\pgfsetstrokecolor{currentstroke}%
\pgfsetdash{}{0pt}%
\pgfsys@defobject{currentmarker}{\pgfqpoint{0.000000in}{-0.048611in}}{\pgfqpoint{0.000000in}{0.000000in}}{%
\pgfpathmoveto{\pgfqpoint{0.000000in}{0.000000in}}%
\pgfpathlineto{\pgfqpoint{0.000000in}{-0.048611in}}%
\pgfusepath{stroke,fill}%
}%
\begin{pgfscope}%
\pgfsys@transformshift{2.290826in}{1.080000in}%
\pgfsys@useobject{currentmarker}{}%
\end{pgfscope}%
\end{pgfscope}%
\begin{pgfscope}%
\definecolor{textcolor}{rgb}{0.000000,0.000000,0.000000}%
\pgfsetstrokecolor{textcolor}%
\pgfsetfillcolor{textcolor}%
\pgftext[x=2.290826in,y=0.982778in,,top]{\color{textcolor}\rmfamily\fontsize{28.000000}{33.600000}\selectfont 20}%
\end{pgfscope}%
\begin{pgfscope}%
\pgfpathrectangle{\pgfqpoint{1.009028in}{1.080000in}}{\pgfqpoint{5.329583in}{4.824444in}}%
\pgfusepath{clip}%
\pgfsetrectcap%
\pgfsetroundjoin%
\pgfsetlinewidth{0.803000pt}%
\definecolor{currentstroke}{rgb}{0.690196,0.690196,0.690196}%
\pgfsetstrokecolor{currentstroke}%
\pgfsetdash{}{0pt}%
\pgfpathmoveto{\pgfqpoint{3.640088in}{1.080000in}}%
\pgfpathlineto{\pgfqpoint{3.640088in}{5.904444in}}%
\pgfusepath{stroke}%
\end{pgfscope}%
\begin{pgfscope}%
\pgfsetbuttcap%
\pgfsetroundjoin%
\definecolor{currentfill}{rgb}{0.000000,0.000000,0.000000}%
\pgfsetfillcolor{currentfill}%
\pgfsetlinewidth{0.803000pt}%
\definecolor{currentstroke}{rgb}{0.000000,0.000000,0.000000}%
\pgfsetstrokecolor{currentstroke}%
\pgfsetdash{}{0pt}%
\pgfsys@defobject{currentmarker}{\pgfqpoint{0.000000in}{-0.048611in}}{\pgfqpoint{0.000000in}{0.000000in}}{%
\pgfpathmoveto{\pgfqpoint{0.000000in}{0.000000in}}%
\pgfpathlineto{\pgfqpoint{0.000000in}{-0.048611in}}%
\pgfusepath{stroke,fill}%
}%
\begin{pgfscope}%
\pgfsys@transformshift{3.640088in}{1.080000in}%
\pgfsys@useobject{currentmarker}{}%
\end{pgfscope}%
\end{pgfscope}%
\begin{pgfscope}%
\definecolor{textcolor}{rgb}{0.000000,0.000000,0.000000}%
\pgfsetstrokecolor{textcolor}%
\pgfsetfillcolor{textcolor}%
\pgftext[x=3.640088in,y=0.982778in,,top]{\color{textcolor}\rmfamily\fontsize{28.000000}{33.600000}\selectfont 40}%
\end{pgfscope}%
\begin{pgfscope}%
\pgfpathrectangle{\pgfqpoint{1.009028in}{1.080000in}}{\pgfqpoint{5.329583in}{4.824444in}}%
\pgfusepath{clip}%
\pgfsetrectcap%
\pgfsetroundjoin%
\pgfsetlinewidth{0.803000pt}%
\definecolor{currentstroke}{rgb}{0.690196,0.690196,0.690196}%
\pgfsetstrokecolor{currentstroke}%
\pgfsetdash{}{0pt}%
\pgfpathmoveto{\pgfqpoint{4.989350in}{1.080000in}}%
\pgfpathlineto{\pgfqpoint{4.989350in}{5.904444in}}%
\pgfusepath{stroke}%
\end{pgfscope}%
\begin{pgfscope}%
\pgfsetbuttcap%
\pgfsetroundjoin%
\definecolor{currentfill}{rgb}{0.000000,0.000000,0.000000}%
\pgfsetfillcolor{currentfill}%
\pgfsetlinewidth{0.803000pt}%
\definecolor{currentstroke}{rgb}{0.000000,0.000000,0.000000}%
\pgfsetstrokecolor{currentstroke}%
\pgfsetdash{}{0pt}%
\pgfsys@defobject{currentmarker}{\pgfqpoint{0.000000in}{-0.048611in}}{\pgfqpoint{0.000000in}{0.000000in}}{%
\pgfpathmoveto{\pgfqpoint{0.000000in}{0.000000in}}%
\pgfpathlineto{\pgfqpoint{0.000000in}{-0.048611in}}%
\pgfusepath{stroke,fill}%
}%
\begin{pgfscope}%
\pgfsys@transformshift{4.989350in}{1.080000in}%
\pgfsys@useobject{currentmarker}{}%
\end{pgfscope}%
\end{pgfscope}%
\begin{pgfscope}%
\definecolor{textcolor}{rgb}{0.000000,0.000000,0.000000}%
\pgfsetstrokecolor{textcolor}%
\pgfsetfillcolor{textcolor}%
\pgftext[x=4.989350in,y=0.982778in,,top]{\color{textcolor}\rmfamily\fontsize{28.000000}{33.600000}\selectfont 60}%
\end{pgfscope}%
\begin{pgfscope}%
\pgfpathrectangle{\pgfqpoint{1.009028in}{1.080000in}}{\pgfqpoint{5.329583in}{4.824444in}}%
\pgfusepath{clip}%
\pgfsetrectcap%
\pgfsetroundjoin%
\pgfsetlinewidth{0.803000pt}%
\definecolor{currentstroke}{rgb}{0.690196,0.690196,0.690196}%
\pgfsetstrokecolor{currentstroke}%
\pgfsetdash{}{0pt}%
\pgfpathmoveto{\pgfqpoint{6.338611in}{1.080000in}}%
\pgfpathlineto{\pgfqpoint{6.338611in}{5.904444in}}%
\pgfusepath{stroke}%
\end{pgfscope}%
\begin{pgfscope}%
\pgfsetbuttcap%
\pgfsetroundjoin%
\definecolor{currentfill}{rgb}{0.000000,0.000000,0.000000}%
\pgfsetfillcolor{currentfill}%
\pgfsetlinewidth{0.803000pt}%
\definecolor{currentstroke}{rgb}{0.000000,0.000000,0.000000}%
\pgfsetstrokecolor{currentstroke}%
\pgfsetdash{}{0pt}%
\pgfsys@defobject{currentmarker}{\pgfqpoint{0.000000in}{-0.048611in}}{\pgfqpoint{0.000000in}{0.000000in}}{%
\pgfpathmoveto{\pgfqpoint{0.000000in}{0.000000in}}%
\pgfpathlineto{\pgfqpoint{0.000000in}{-0.048611in}}%
\pgfusepath{stroke,fill}%
}%
\begin{pgfscope}%
\pgfsys@transformshift{6.338611in}{1.080000in}%
\pgfsys@useobject{currentmarker}{}%
\end{pgfscope}%
\end{pgfscope}%
\begin{pgfscope}%
\definecolor{textcolor}{rgb}{0.000000,0.000000,0.000000}%
\pgfsetstrokecolor{textcolor}%
\pgfsetfillcolor{textcolor}%
\pgftext[x=6.338611in,y=0.982778in,,top]{\color{textcolor}\rmfamily\fontsize{28.000000}{33.600000}\selectfont 80}%
\end{pgfscope}%
\begin{pgfscope}%
\definecolor{textcolor}{rgb}{0.000000,0.000000,0.000000}%
\pgfsetstrokecolor{textcolor}%
\pgfsetfillcolor{textcolor}%
\pgftext[x=3.673819in,y=0.536706in,,top]{\color{textcolor}\rmfamily\fontsize{30.000000}{36.000000}\selectfont time step \(\displaystyle k\)}%
\end{pgfscope}%
\begin{pgfscope}%
\pgfpathrectangle{\pgfqpoint{1.009028in}{1.080000in}}{\pgfqpoint{5.329583in}{4.824444in}}%
\pgfusepath{clip}%
\pgfsetrectcap%
\pgfsetroundjoin%
\pgfsetlinewidth{0.803000pt}%
\definecolor{currentstroke}{rgb}{0.690196,0.690196,0.690196}%
\pgfsetstrokecolor{currentstroke}%
\pgfsetdash{}{0pt}%
\pgfpathmoveto{\pgfqpoint{1.009028in}{1.080000in}}%
\pgfpathlineto{\pgfqpoint{6.338611in}{1.080000in}}%
\pgfusepath{stroke}%
\end{pgfscope}%
\begin{pgfscope}%
\pgfsetbuttcap%
\pgfsetroundjoin%
\definecolor{currentfill}{rgb}{0.000000,0.000000,0.000000}%
\pgfsetfillcolor{currentfill}%
\pgfsetlinewidth{0.803000pt}%
\definecolor{currentstroke}{rgb}{0.000000,0.000000,0.000000}%
\pgfsetstrokecolor{currentstroke}%
\pgfsetdash{}{0pt}%
\pgfsys@defobject{currentmarker}{\pgfqpoint{-0.048611in}{0.000000in}}{\pgfqpoint{-0.000000in}{0.000000in}}{%
\pgfpathmoveto{\pgfqpoint{-0.000000in}{0.000000in}}%
\pgfpathlineto{\pgfqpoint{-0.048611in}{0.000000in}}%
\pgfusepath{stroke,fill}%
}%
\begin{pgfscope}%
\pgfsys@transformshift{1.009028in}{1.080000in}%
\pgfsys@useobject{currentmarker}{}%
\end{pgfscope}%
\end{pgfscope}%
\begin{pgfscope}%
\definecolor{textcolor}{rgb}{0.000000,0.000000,0.000000}%
\pgfsetstrokecolor{textcolor}%
\pgfsetfillcolor{textcolor}%
\pgftext[x=0.663050in, y=0.960015in, left, base]{\color{textcolor}\rmfamily\fontsize{28.000000}{33.600000}\selectfont 0}%
\end{pgfscope}%
\begin{pgfscope}%
\pgfpathrectangle{\pgfqpoint{1.009028in}{1.080000in}}{\pgfqpoint{5.329583in}{4.824444in}}%
\pgfusepath{clip}%
\pgfsetrectcap%
\pgfsetroundjoin%
\pgfsetlinewidth{0.803000pt}%
\definecolor{currentstroke}{rgb}{0.690196,0.690196,0.690196}%
\pgfsetstrokecolor{currentstroke}%
\pgfsetdash{}{0pt}%
\pgfpathmoveto{\pgfqpoint{1.009028in}{1.884074in}}%
\pgfpathlineto{\pgfqpoint{6.338611in}{1.884074in}}%
\pgfusepath{stroke}%
\end{pgfscope}%
\begin{pgfscope}%
\pgfsetbuttcap%
\pgfsetroundjoin%
\definecolor{currentfill}{rgb}{0.000000,0.000000,0.000000}%
\pgfsetfillcolor{currentfill}%
\pgfsetlinewidth{0.803000pt}%
\definecolor{currentstroke}{rgb}{0.000000,0.000000,0.000000}%
\pgfsetstrokecolor{currentstroke}%
\pgfsetdash{}{0pt}%
\pgfsys@defobject{currentmarker}{\pgfqpoint{-0.048611in}{0.000000in}}{\pgfqpoint{-0.000000in}{0.000000in}}{%
\pgfpathmoveto{\pgfqpoint{-0.000000in}{0.000000in}}%
\pgfpathlineto{\pgfqpoint{-0.048611in}{0.000000in}}%
\pgfusepath{stroke,fill}%
}%
\begin{pgfscope}%
\pgfsys@transformshift{1.009028in}{1.884074in}%
\pgfsys@useobject{currentmarker}{}%
\end{pgfscope}%
\end{pgfscope}%
\begin{pgfscope}%
\definecolor{textcolor}{rgb}{0.000000,0.000000,0.000000}%
\pgfsetstrokecolor{textcolor}%
\pgfsetfillcolor{textcolor}%
\pgftext[x=0.663050in, y=1.764089in, left, base]{\color{textcolor}\rmfamily\fontsize{28.000000}{33.600000}\selectfont 1}%
\end{pgfscope}%
\begin{pgfscope}%
\pgfpathrectangle{\pgfqpoint{1.009028in}{1.080000in}}{\pgfqpoint{5.329583in}{4.824444in}}%
\pgfusepath{clip}%
\pgfsetrectcap%
\pgfsetroundjoin%
\pgfsetlinewidth{0.803000pt}%
\definecolor{currentstroke}{rgb}{0.690196,0.690196,0.690196}%
\pgfsetstrokecolor{currentstroke}%
\pgfsetdash{}{0pt}%
\pgfpathmoveto{\pgfqpoint{1.009028in}{2.688148in}}%
\pgfpathlineto{\pgfqpoint{6.338611in}{2.688148in}}%
\pgfusepath{stroke}%
\end{pgfscope}%
\begin{pgfscope}%
\pgfsetbuttcap%
\pgfsetroundjoin%
\definecolor{currentfill}{rgb}{0.000000,0.000000,0.000000}%
\pgfsetfillcolor{currentfill}%
\pgfsetlinewidth{0.803000pt}%
\definecolor{currentstroke}{rgb}{0.000000,0.000000,0.000000}%
\pgfsetstrokecolor{currentstroke}%
\pgfsetdash{}{0pt}%
\pgfsys@defobject{currentmarker}{\pgfqpoint{-0.048611in}{0.000000in}}{\pgfqpoint{-0.000000in}{0.000000in}}{%
\pgfpathmoveto{\pgfqpoint{-0.000000in}{0.000000in}}%
\pgfpathlineto{\pgfqpoint{-0.048611in}{0.000000in}}%
\pgfusepath{stroke,fill}%
}%
\begin{pgfscope}%
\pgfsys@transformshift{1.009028in}{2.688148in}%
\pgfsys@useobject{currentmarker}{}%
\end{pgfscope}%
\end{pgfscope}%
\begin{pgfscope}%
\definecolor{textcolor}{rgb}{0.000000,0.000000,0.000000}%
\pgfsetstrokecolor{textcolor}%
\pgfsetfillcolor{textcolor}%
\pgftext[x=0.663050in, y=2.568163in, left, base]{\color{textcolor}\rmfamily\fontsize{28.000000}{33.600000}\selectfont 2}%
\end{pgfscope}%
\begin{pgfscope}%
\pgfpathrectangle{\pgfqpoint{1.009028in}{1.080000in}}{\pgfqpoint{5.329583in}{4.824444in}}%
\pgfusepath{clip}%
\pgfsetrectcap%
\pgfsetroundjoin%
\pgfsetlinewidth{0.803000pt}%
\definecolor{currentstroke}{rgb}{0.690196,0.690196,0.690196}%
\pgfsetstrokecolor{currentstroke}%
\pgfsetdash{}{0pt}%
\pgfpathmoveto{\pgfqpoint{1.009028in}{3.492222in}}%
\pgfpathlineto{\pgfqpoint{6.338611in}{3.492222in}}%
\pgfusepath{stroke}%
\end{pgfscope}%
\begin{pgfscope}%
\pgfsetbuttcap%
\pgfsetroundjoin%
\definecolor{currentfill}{rgb}{0.000000,0.000000,0.000000}%
\pgfsetfillcolor{currentfill}%
\pgfsetlinewidth{0.803000pt}%
\definecolor{currentstroke}{rgb}{0.000000,0.000000,0.000000}%
\pgfsetstrokecolor{currentstroke}%
\pgfsetdash{}{0pt}%
\pgfsys@defobject{currentmarker}{\pgfqpoint{-0.048611in}{0.000000in}}{\pgfqpoint{-0.000000in}{0.000000in}}{%
\pgfpathmoveto{\pgfqpoint{-0.000000in}{0.000000in}}%
\pgfpathlineto{\pgfqpoint{-0.048611in}{0.000000in}}%
\pgfusepath{stroke,fill}%
}%
\begin{pgfscope}%
\pgfsys@transformshift{1.009028in}{3.492222in}%
\pgfsys@useobject{currentmarker}{}%
\end{pgfscope}%
\end{pgfscope}%
\begin{pgfscope}%
\definecolor{textcolor}{rgb}{0.000000,0.000000,0.000000}%
\pgfsetstrokecolor{textcolor}%
\pgfsetfillcolor{textcolor}%
\pgftext[x=0.663050in, y=3.372238in, left, base]{\color{textcolor}\rmfamily\fontsize{28.000000}{33.600000}\selectfont 3}%
\end{pgfscope}%
\begin{pgfscope}%
\pgfpathrectangle{\pgfqpoint{1.009028in}{1.080000in}}{\pgfqpoint{5.329583in}{4.824444in}}%
\pgfusepath{clip}%
\pgfsetrectcap%
\pgfsetroundjoin%
\pgfsetlinewidth{0.803000pt}%
\definecolor{currentstroke}{rgb}{0.690196,0.690196,0.690196}%
\pgfsetstrokecolor{currentstroke}%
\pgfsetdash{}{0pt}%
\pgfpathmoveto{\pgfqpoint{1.009028in}{4.296296in}}%
\pgfpathlineto{\pgfqpoint{6.338611in}{4.296296in}}%
\pgfusepath{stroke}%
\end{pgfscope}%
\begin{pgfscope}%
\pgfsetbuttcap%
\pgfsetroundjoin%
\definecolor{currentfill}{rgb}{0.000000,0.000000,0.000000}%
\pgfsetfillcolor{currentfill}%
\pgfsetlinewidth{0.803000pt}%
\definecolor{currentstroke}{rgb}{0.000000,0.000000,0.000000}%
\pgfsetstrokecolor{currentstroke}%
\pgfsetdash{}{0pt}%
\pgfsys@defobject{currentmarker}{\pgfqpoint{-0.048611in}{0.000000in}}{\pgfqpoint{-0.000000in}{0.000000in}}{%
\pgfpathmoveto{\pgfqpoint{-0.000000in}{0.000000in}}%
\pgfpathlineto{\pgfqpoint{-0.048611in}{0.000000in}}%
\pgfusepath{stroke,fill}%
}%
\begin{pgfscope}%
\pgfsys@transformshift{1.009028in}{4.296296in}%
\pgfsys@useobject{currentmarker}{}%
\end{pgfscope}%
\end{pgfscope}%
\begin{pgfscope}%
\definecolor{textcolor}{rgb}{0.000000,0.000000,0.000000}%
\pgfsetstrokecolor{textcolor}%
\pgfsetfillcolor{textcolor}%
\pgftext[x=0.663050in, y=4.176312in, left, base]{\color{textcolor}\rmfamily\fontsize{28.000000}{33.600000}\selectfont 4}%
\end{pgfscope}%
\begin{pgfscope}%
\pgfpathrectangle{\pgfqpoint{1.009028in}{1.080000in}}{\pgfqpoint{5.329583in}{4.824444in}}%
\pgfusepath{clip}%
\pgfsetrectcap%
\pgfsetroundjoin%
\pgfsetlinewidth{0.803000pt}%
\definecolor{currentstroke}{rgb}{0.690196,0.690196,0.690196}%
\pgfsetstrokecolor{currentstroke}%
\pgfsetdash{}{0pt}%
\pgfpathmoveto{\pgfqpoint{1.009028in}{5.100370in}}%
\pgfpathlineto{\pgfqpoint{6.338611in}{5.100370in}}%
\pgfusepath{stroke}%
\end{pgfscope}%
\begin{pgfscope}%
\pgfsetbuttcap%
\pgfsetroundjoin%
\definecolor{currentfill}{rgb}{0.000000,0.000000,0.000000}%
\pgfsetfillcolor{currentfill}%
\pgfsetlinewidth{0.803000pt}%
\definecolor{currentstroke}{rgb}{0.000000,0.000000,0.000000}%
\pgfsetstrokecolor{currentstroke}%
\pgfsetdash{}{0pt}%
\pgfsys@defobject{currentmarker}{\pgfqpoint{-0.048611in}{0.000000in}}{\pgfqpoint{-0.000000in}{0.000000in}}{%
\pgfpathmoveto{\pgfqpoint{-0.000000in}{0.000000in}}%
\pgfpathlineto{\pgfqpoint{-0.048611in}{0.000000in}}%
\pgfusepath{stroke,fill}%
}%
\begin{pgfscope}%
\pgfsys@transformshift{1.009028in}{5.100370in}%
\pgfsys@useobject{currentmarker}{}%
\end{pgfscope}%
\end{pgfscope}%
\begin{pgfscope}%
\definecolor{textcolor}{rgb}{0.000000,0.000000,0.000000}%
\pgfsetstrokecolor{textcolor}%
\pgfsetfillcolor{textcolor}%
\pgftext[x=0.663050in, y=4.980386in, left, base]{\color{textcolor}\rmfamily\fontsize{28.000000}{33.600000}\selectfont 5}%
\end{pgfscope}%
\begin{pgfscope}%
\pgfpathrectangle{\pgfqpoint{1.009028in}{1.080000in}}{\pgfqpoint{5.329583in}{4.824444in}}%
\pgfusepath{clip}%
\pgfsetrectcap%
\pgfsetroundjoin%
\pgfsetlinewidth{0.803000pt}%
\definecolor{currentstroke}{rgb}{0.690196,0.690196,0.690196}%
\pgfsetstrokecolor{currentstroke}%
\pgfsetdash{}{0pt}%
\pgfpathmoveto{\pgfqpoint{1.009028in}{5.904444in}}%
\pgfpathlineto{\pgfqpoint{6.338611in}{5.904444in}}%
\pgfusepath{stroke}%
\end{pgfscope}%
\begin{pgfscope}%
\pgfsetbuttcap%
\pgfsetroundjoin%
\definecolor{currentfill}{rgb}{0.000000,0.000000,0.000000}%
\pgfsetfillcolor{currentfill}%
\pgfsetlinewidth{0.803000pt}%
\definecolor{currentstroke}{rgb}{0.000000,0.000000,0.000000}%
\pgfsetstrokecolor{currentstroke}%
\pgfsetdash{}{0pt}%
\pgfsys@defobject{currentmarker}{\pgfqpoint{-0.048611in}{0.000000in}}{\pgfqpoint{-0.000000in}{0.000000in}}{%
\pgfpathmoveto{\pgfqpoint{-0.000000in}{0.000000in}}%
\pgfpathlineto{\pgfqpoint{-0.048611in}{0.000000in}}%
\pgfusepath{stroke,fill}%
}%
\begin{pgfscope}%
\pgfsys@transformshift{1.009028in}{5.904444in}%
\pgfsys@useobject{currentmarker}{}%
\end{pgfscope}%
\end{pgfscope}%
\begin{pgfscope}%
\definecolor{textcolor}{rgb}{0.000000,0.000000,0.000000}%
\pgfsetstrokecolor{textcolor}%
\pgfsetfillcolor{textcolor}%
\pgftext[x=0.663050in, y=5.784460in, left, base]{\color{textcolor}\rmfamily\fontsize{28.000000}{33.600000}\selectfont 6}%
\end{pgfscope}%
\begin{pgfscope}%
\definecolor{textcolor}{rgb}{0.000000,0.000000,0.000000}%
\pgfsetstrokecolor{textcolor}%
\pgfsetfillcolor{textcolor}%
\pgftext[x=0.524161in,y=3.492222in,,bottom,rotate=90.000000]{\color{textcolor}\rmfamily\fontsize{30.000000}{36.000000}\selectfont noise variance}%
\end{pgfscope}%
\begin{pgfscope}%
\definecolor{textcolor}{rgb}{0.000000,0.000000,0.000000}%
\pgfsetstrokecolor{textcolor}%
\pgfsetfillcolor{textcolor}%
\pgftext[x=1.009028in,y=5.946111in,left,base]{\color{textcolor}\rmfamily\fontsize{22.000000}{26.400000}\selectfont 1e\ensuremath{-}14}%
\end{pgfscope}%
\begin{pgfscope}%
\pgfpathrectangle{\pgfqpoint{1.009028in}{1.080000in}}{\pgfqpoint{5.329583in}{4.824444in}}%
\pgfusepath{clip}%
\pgfsetrectcap%
\pgfsetroundjoin%
\pgfsetlinewidth{1.505625pt}%
\definecolor{currentstroke}{rgb}{1.000000,0.000000,0.000000}%
\pgfsetstrokecolor{currentstroke}%
\pgfsetdash{}{0pt}%
\pgfpathmoveto{\pgfqpoint{1.316946in}{5.914444in}}%
\pgfpathlineto{\pgfqpoint{1.346343in}{5.330899in}}%
\pgfpathlineto{\pgfqpoint{1.413806in}{4.348906in}}%
\pgfpathlineto{\pgfqpoint{1.481269in}{3.688054in}}%
\pgfpathlineto{\pgfqpoint{1.548732in}{3.256025in}}%
\pgfpathlineto{\pgfqpoint{1.616195in}{2.992287in}}%
\pgfpathlineto{\pgfqpoint{1.683659in}{2.873292in}}%
\pgfpathlineto{\pgfqpoint{1.751122in}{2.779774in}}%
\pgfpathlineto{\pgfqpoint{1.818585in}{2.731735in}}%
\pgfpathlineto{\pgfqpoint{1.886048in}{2.713664in}}%
\pgfpathlineto{\pgfqpoint{1.953511in}{2.699723in}}%
\pgfpathlineto{\pgfqpoint{2.020974in}{2.706356in}}%
\pgfpathlineto{\pgfqpoint{2.088437in}{2.695383in}}%
\pgfpathlineto{\pgfqpoint{2.155900in}{2.696311in}}%
\pgfpathlineto{\pgfqpoint{2.223363in}{2.688632in}}%
\pgfpathlineto{\pgfqpoint{2.290826in}{2.700595in}}%
\pgfpathlineto{\pgfqpoint{2.358289in}{2.711242in}}%
\pgfpathlineto{\pgfqpoint{2.425752in}{2.694219in}}%
\pgfpathlineto{\pgfqpoint{2.493216in}{2.693459in}}%
\pgfpathlineto{\pgfqpoint{2.560679in}{2.703353in}}%
\pgfpathlineto{\pgfqpoint{2.628142in}{2.702886in}}%
\pgfpathlineto{\pgfqpoint{2.695605in}{2.705647in}}%
\pgfpathlineto{\pgfqpoint{2.763068in}{2.699426in}}%
\pgfpathlineto{\pgfqpoint{2.830531in}{2.706382in}}%
\pgfpathlineto{\pgfqpoint{2.897994in}{2.695322in}}%
\pgfpathlineto{\pgfqpoint{2.965457in}{2.696366in}}%
\pgfpathlineto{\pgfqpoint{3.032920in}{2.703049in}}%
\pgfpathlineto{\pgfqpoint{3.100383in}{2.691528in}}%
\pgfpathlineto{\pgfqpoint{3.167846in}{2.691249in}}%
\pgfpathlineto{\pgfqpoint{3.235309in}{2.697455in}}%
\pgfpathlineto{\pgfqpoint{3.302773in}{2.700900in}}%
\pgfpathlineto{\pgfqpoint{3.370236in}{2.695866in}}%
\pgfpathlineto{\pgfqpoint{3.437699in}{2.690829in}}%
\pgfpathlineto{\pgfqpoint{3.505162in}{2.691775in}}%
\pgfpathlineto{\pgfqpoint{3.572625in}{2.694433in}}%
\pgfpathlineto{\pgfqpoint{3.640088in}{2.688242in}}%
\pgfpathlineto{\pgfqpoint{3.707551in}{2.692342in}}%
\pgfpathlineto{\pgfqpoint{3.775014in}{2.693804in}}%
\pgfpathlineto{\pgfqpoint{3.842477in}{2.693718in}}%
\pgfpathlineto{\pgfqpoint{3.909940in}{2.694062in}}%
\pgfpathlineto{\pgfqpoint{3.977403in}{2.699034in}}%
\pgfpathlineto{\pgfqpoint{4.044866in}{2.697062in}}%
\pgfpathlineto{\pgfqpoint{4.112329in}{2.691849in}}%
\pgfpathlineto{\pgfqpoint{4.179793in}{2.694583in}}%
\pgfpathlineto{\pgfqpoint{4.247256in}{2.699720in}}%
\pgfpathlineto{\pgfqpoint{4.314719in}{2.690293in}}%
\pgfpathlineto{\pgfqpoint{4.382182in}{2.691487in}}%
\pgfpathlineto{\pgfqpoint{4.449645in}{2.690294in}}%
\pgfpathlineto{\pgfqpoint{4.517108in}{2.698330in}}%
\pgfpathlineto{\pgfqpoint{4.584571in}{2.686837in}}%
\pgfpathlineto{\pgfqpoint{4.652034in}{2.690364in}}%
\pgfpathlineto{\pgfqpoint{4.719497in}{2.694034in}}%
\pgfpathlineto{\pgfqpoint{4.786960in}{2.687515in}}%
\pgfpathlineto{\pgfqpoint{4.854423in}{2.688711in}}%
\pgfpathlineto{\pgfqpoint{4.921886in}{2.690933in}}%
\pgfpathlineto{\pgfqpoint{4.989350in}{2.686022in}}%
\pgfpathlineto{\pgfqpoint{5.056813in}{2.689159in}}%
\pgfpathlineto{\pgfqpoint{5.124276in}{2.685915in}}%
\pgfpathlineto{\pgfqpoint{5.191739in}{2.687325in}}%
\pgfpathlineto{\pgfqpoint{5.259202in}{2.688217in}}%
\pgfpathlineto{\pgfqpoint{5.326665in}{2.687360in}}%
\pgfpathlineto{\pgfqpoint{5.394128in}{2.683934in}}%
\pgfpathlineto{\pgfqpoint{5.461591in}{2.687728in}}%
\pgfpathlineto{\pgfqpoint{5.529054in}{2.688646in}}%
\pgfpathlineto{\pgfqpoint{5.596517in}{2.691620in}}%
\pgfpathlineto{\pgfqpoint{5.663980in}{2.692696in}}%
\pgfpathlineto{\pgfqpoint{5.731443in}{2.688482in}}%
\pgfpathlineto{\pgfqpoint{5.798906in}{2.687091in}}%
\pgfpathlineto{\pgfqpoint{5.866370in}{2.685700in}}%
\pgfpathlineto{\pgfqpoint{5.933833in}{2.691410in}}%
\pgfpathlineto{\pgfqpoint{6.001296in}{2.694321in}}%
\pgfpathlineto{\pgfqpoint{6.068759in}{2.691692in}}%
\pgfpathlineto{\pgfqpoint{6.136222in}{2.691288in}}%
\pgfpathlineto{\pgfqpoint{6.203685in}{2.692516in}}%
\pgfpathlineto{\pgfqpoint{6.271148in}{2.683352in}}%
\pgfpathlineto{\pgfqpoint{6.338611in}{2.690896in}}%
\pgfusepath{stroke}%
\end{pgfscope}%
\begin{pgfscope}%
\pgfpathrectangle{\pgfqpoint{1.009028in}{1.080000in}}{\pgfqpoint{5.329583in}{4.824444in}}%
\pgfusepath{clip}%
\pgfsetbuttcap%
\pgfsetroundjoin%
\pgfsetlinewidth{2.007500pt}%
\definecolor{currentstroke}{rgb}{0.000000,0.000000,0.000000}%
\pgfsetstrokecolor{currentstroke}%
\pgfsetdash{{7.400000pt}{3.200000pt}}{0.000000pt}%
\pgfpathmoveto{\pgfqpoint{1.009028in}{2.688148in}}%
\pgfpathlineto{\pgfqpoint{1.076491in}{2.688148in}}%
\pgfpathlineto{\pgfqpoint{1.143954in}{2.688148in}}%
\pgfpathlineto{\pgfqpoint{1.211417in}{2.688148in}}%
\pgfpathlineto{\pgfqpoint{1.278880in}{2.688148in}}%
\pgfpathlineto{\pgfqpoint{1.346343in}{2.688148in}}%
\pgfpathlineto{\pgfqpoint{1.413806in}{2.688148in}}%
\pgfpathlineto{\pgfqpoint{1.481269in}{2.688148in}}%
\pgfpathlineto{\pgfqpoint{1.548732in}{2.688148in}}%
\pgfpathlineto{\pgfqpoint{1.616195in}{2.688148in}}%
\pgfpathlineto{\pgfqpoint{1.683659in}{2.688148in}}%
\pgfpathlineto{\pgfqpoint{1.751122in}{2.688148in}}%
\pgfpathlineto{\pgfqpoint{1.818585in}{2.688148in}}%
\pgfpathlineto{\pgfqpoint{1.886048in}{2.688148in}}%
\pgfpathlineto{\pgfqpoint{1.953511in}{2.688148in}}%
\pgfpathlineto{\pgfqpoint{2.020974in}{2.688148in}}%
\pgfpathlineto{\pgfqpoint{2.088437in}{2.688148in}}%
\pgfpathlineto{\pgfqpoint{2.155900in}{2.688148in}}%
\pgfpathlineto{\pgfqpoint{2.223363in}{2.688148in}}%
\pgfpathlineto{\pgfqpoint{2.290826in}{2.688148in}}%
\pgfpathlineto{\pgfqpoint{2.358289in}{2.688148in}}%
\pgfpathlineto{\pgfqpoint{2.425752in}{2.688148in}}%
\pgfpathlineto{\pgfqpoint{2.493216in}{2.688148in}}%
\pgfpathlineto{\pgfqpoint{2.560679in}{2.688148in}}%
\pgfpathlineto{\pgfqpoint{2.628142in}{2.688148in}}%
\pgfpathlineto{\pgfqpoint{2.695605in}{2.688148in}}%
\pgfpathlineto{\pgfqpoint{2.763068in}{2.688148in}}%
\pgfpathlineto{\pgfqpoint{2.830531in}{2.688148in}}%
\pgfpathlineto{\pgfqpoint{2.897994in}{2.688148in}}%
\pgfpathlineto{\pgfqpoint{2.965457in}{2.688148in}}%
\pgfpathlineto{\pgfqpoint{3.032920in}{2.688148in}}%
\pgfpathlineto{\pgfqpoint{3.100383in}{2.688148in}}%
\pgfpathlineto{\pgfqpoint{3.167846in}{2.688148in}}%
\pgfpathlineto{\pgfqpoint{3.235309in}{2.688148in}}%
\pgfpathlineto{\pgfqpoint{3.302773in}{2.688148in}}%
\pgfpathlineto{\pgfqpoint{3.370236in}{2.688148in}}%
\pgfpathlineto{\pgfqpoint{3.437699in}{2.688148in}}%
\pgfpathlineto{\pgfqpoint{3.505162in}{2.688148in}}%
\pgfpathlineto{\pgfqpoint{3.572625in}{2.688148in}}%
\pgfpathlineto{\pgfqpoint{3.640088in}{2.688148in}}%
\pgfpathlineto{\pgfqpoint{3.707551in}{2.688148in}}%
\pgfpathlineto{\pgfqpoint{3.775014in}{2.688148in}}%
\pgfpathlineto{\pgfqpoint{3.842477in}{2.688148in}}%
\pgfpathlineto{\pgfqpoint{3.909940in}{2.688148in}}%
\pgfpathlineto{\pgfqpoint{3.977403in}{2.688148in}}%
\pgfpathlineto{\pgfqpoint{4.044866in}{2.688148in}}%
\pgfpathlineto{\pgfqpoint{4.112329in}{2.688148in}}%
\pgfpathlineto{\pgfqpoint{4.179793in}{2.688148in}}%
\pgfpathlineto{\pgfqpoint{4.247256in}{2.688148in}}%
\pgfpathlineto{\pgfqpoint{4.314719in}{2.688148in}}%
\pgfpathlineto{\pgfqpoint{4.382182in}{2.688148in}}%
\pgfpathlineto{\pgfqpoint{4.449645in}{2.688148in}}%
\pgfpathlineto{\pgfqpoint{4.517108in}{2.688148in}}%
\pgfpathlineto{\pgfqpoint{4.584571in}{2.688148in}}%
\pgfpathlineto{\pgfqpoint{4.652034in}{2.688148in}}%
\pgfpathlineto{\pgfqpoint{4.719497in}{2.688148in}}%
\pgfpathlineto{\pgfqpoint{4.786960in}{2.688148in}}%
\pgfpathlineto{\pgfqpoint{4.854423in}{2.688148in}}%
\pgfpathlineto{\pgfqpoint{4.921886in}{2.688148in}}%
\pgfpathlineto{\pgfqpoint{4.989350in}{2.688148in}}%
\pgfpathlineto{\pgfqpoint{5.056813in}{2.688148in}}%
\pgfpathlineto{\pgfqpoint{5.124276in}{2.688148in}}%
\pgfpathlineto{\pgfqpoint{5.191739in}{2.688148in}}%
\pgfpathlineto{\pgfqpoint{5.259202in}{2.688148in}}%
\pgfpathlineto{\pgfqpoint{5.326665in}{2.688148in}}%
\pgfpathlineto{\pgfqpoint{5.394128in}{2.688148in}}%
\pgfpathlineto{\pgfqpoint{5.461591in}{2.688148in}}%
\pgfpathlineto{\pgfqpoint{5.529054in}{2.688148in}}%
\pgfpathlineto{\pgfqpoint{5.596517in}{2.688148in}}%
\pgfpathlineto{\pgfqpoint{5.663980in}{2.688148in}}%
\pgfpathlineto{\pgfqpoint{5.731443in}{2.688148in}}%
\pgfpathlineto{\pgfqpoint{5.798906in}{2.688148in}}%
\pgfpathlineto{\pgfqpoint{5.866370in}{2.688148in}}%
\pgfpathlineto{\pgfqpoint{5.933833in}{2.688148in}}%
\pgfpathlineto{\pgfqpoint{6.001296in}{2.688148in}}%
\pgfpathlineto{\pgfqpoint{6.068759in}{2.688148in}}%
\pgfpathlineto{\pgfqpoint{6.136222in}{2.688148in}}%
\pgfpathlineto{\pgfqpoint{6.203685in}{2.688148in}}%
\pgfpathlineto{\pgfqpoint{6.271148in}{2.688148in}}%
\pgfpathlineto{\pgfqpoint{6.338611in}{2.688148in}}%
\pgfusepath{stroke}%
\end{pgfscope}%
\begin{pgfscope}%
\pgfsetrectcap%
\pgfsetmiterjoin%
\pgfsetlinewidth{0.803000pt}%
\definecolor{currentstroke}{rgb}{0.000000,0.000000,0.000000}%
\pgfsetstrokecolor{currentstroke}%
\pgfsetdash{}{0pt}%
\pgfpathmoveto{\pgfqpoint{1.009028in}{1.080000in}}%
\pgfpathlineto{\pgfqpoint{1.009028in}{5.904444in}}%
\pgfusepath{stroke}%
\end{pgfscope}%
\begin{pgfscope}%
\pgfsetrectcap%
\pgfsetmiterjoin%
\pgfsetlinewidth{0.803000pt}%
\definecolor{currentstroke}{rgb}{0.000000,0.000000,0.000000}%
\pgfsetstrokecolor{currentstroke}%
\pgfsetdash{}{0pt}%
\pgfpathmoveto{\pgfqpoint{6.338611in}{1.080000in}}%
\pgfpathlineto{\pgfqpoint{6.338611in}{5.904444in}}%
\pgfusepath{stroke}%
\end{pgfscope}%
\begin{pgfscope}%
\pgfsetrectcap%
\pgfsetmiterjoin%
\pgfsetlinewidth{0.803000pt}%
\definecolor{currentstroke}{rgb}{0.000000,0.000000,0.000000}%
\pgfsetstrokecolor{currentstroke}%
\pgfsetdash{}{0pt}%
\pgfpathmoveto{\pgfqpoint{1.009028in}{1.080000in}}%
\pgfpathlineto{\pgfqpoint{6.338611in}{1.080000in}}%
\pgfusepath{stroke}%
\end{pgfscope}%
\begin{pgfscope}%
\pgfsetrectcap%
\pgfsetmiterjoin%
\pgfsetlinewidth{0.803000pt}%
\definecolor{currentstroke}{rgb}{0.000000,0.000000,0.000000}%
\pgfsetstrokecolor{currentstroke}%
\pgfsetdash{}{0pt}%
\pgfpathmoveto{\pgfqpoint{1.009028in}{5.904444in}}%
\pgfpathlineto{\pgfqpoint{6.338611in}{5.904444in}}%
\pgfusepath{stroke}%
\end{pgfscope}%
\begin{pgfscope}%
\pgfsetbuttcap%
\pgfsetmiterjoin%
\definecolor{currentfill}{rgb}{1.000000,1.000000,1.000000}%
\pgfsetfillcolor{currentfill}%
\pgfsetfillopacity{0.800000}%
\pgfsetlinewidth{1.003750pt}%
\definecolor{currentstroke}{rgb}{0.800000,0.800000,0.800000}%
\pgfsetstrokecolor{currentstroke}%
\pgfsetstrokeopacity{0.800000}%
\pgfsetdash{}{0pt}%
\pgfpathmoveto{\pgfqpoint{3.198808in}{4.525573in}}%
\pgfpathlineto{\pgfqpoint{6.066389in}{4.525573in}}%
\pgfpathquadraticcurveto{\pgfqpoint{6.144167in}{4.525573in}}{\pgfqpoint{6.144167in}{4.603351in}}%
\pgfpathlineto{\pgfqpoint{6.144167in}{5.632222in}}%
\pgfpathquadraticcurveto{\pgfqpoint{6.144167in}{5.710000in}}{\pgfqpoint{6.066389in}{5.710000in}}%
\pgfpathlineto{\pgfqpoint{3.198808in}{5.710000in}}%
\pgfpathquadraticcurveto{\pgfqpoint{3.121031in}{5.710000in}}{\pgfqpoint{3.121031in}{5.632222in}}%
\pgfpathlineto{\pgfqpoint{3.121031in}{4.603351in}}%
\pgfpathquadraticcurveto{\pgfqpoint{3.121031in}{4.525573in}}{\pgfqpoint{3.198808in}{4.525573in}}%
\pgfpathlineto{\pgfqpoint{3.198808in}{4.525573in}}%
\pgfpathclose%
\pgfusepath{stroke,fill}%
\end{pgfscope}%
\begin{pgfscope}%
\pgfsetrectcap%
\pgfsetroundjoin%
\pgfsetlinewidth{1.505625pt}%
\definecolor{currentstroke}{rgb}{1.000000,0.000000,0.000000}%
\pgfsetstrokecolor{currentstroke}%
\pgfsetdash{}{0pt}%
\pgfpathmoveto{\pgfqpoint{3.276586in}{5.418333in}}%
\pgfpathlineto{\pgfqpoint{3.665475in}{5.418333in}}%
\pgfpathlineto{\pgfqpoint{4.054364in}{5.418333in}}%
\pgfusepath{stroke}%
\end{pgfscope}%
\begin{pgfscope}%
\definecolor{textcolor}{rgb}{0.000000,0.000000,0.000000}%
\pgfsetstrokecolor{textcolor}%
\pgfsetfillcolor{textcolor}%
\pgftext[x=4.365475in,y=5.282222in,left,base]{\color{textcolor}\rmfamily\fontsize{28.000000}{33.600000}\selectfont Estimated~~}%
\end{pgfscope}%
\begin{pgfscope}%
\pgfsetbuttcap%
\pgfsetroundjoin%
\pgfsetlinewidth{2.007500pt}%
\definecolor{currentstroke}{rgb}{0.000000,0.000000,0.000000}%
\pgfsetstrokecolor{currentstroke}%
\pgfsetdash{{7.400000pt}{3.200000pt}}{0.000000pt}%
\pgfpathmoveto{\pgfqpoint{3.276586in}{4.884453in}}%
\pgfpathlineto{\pgfqpoint{3.665475in}{4.884453in}}%
\pgfpathlineto{\pgfqpoint{4.054364in}{4.884453in}}%
\pgfusepath{stroke}%
\end{pgfscope}%
\begin{pgfscope}%
\definecolor{textcolor}{rgb}{0.000000,0.000000,0.000000}%
\pgfsetstrokecolor{textcolor}%
\pgfsetfillcolor{textcolor}%
\pgftext[x=4.365475in,y=4.748342in,left,base]{\color{textcolor}\rmfamily\fontsize{28.000000}{33.600000}\selectfont True}%
\end{pgfscope}%
\end{pgfpicture}%
\makeatother%
\endgroup%

%% file: Figs/noise_var_sensor2.pgf
\begingroup%
\makeatletter%
\begin{pgfpicture}%
\pgfpathrectangle{\pgfpointorigin}{\pgfqpoint{6.600000in}{6.200000in}}%
\pgfusepath{use as bounding box, clip}%
\begin{pgfscope}%
\pgfsetbuttcap%
\pgfsetmiterjoin%
\definecolor{currentfill}{rgb}{1.000000,1.000000,1.000000}%
\pgfsetfillcolor{currentfill}%
\pgfsetlinewidth{0.000000pt}%
\definecolor{currentstroke}{rgb}{1.000000,1.000000,1.000000}%
\pgfsetstrokecolor{currentstroke}%
\pgfsetdash{}{0pt}%
\pgfpathmoveto{\pgfqpoint{0.000000in}{0.000000in}}%
\pgfpathlineto{\pgfqpoint{6.600000in}{0.000000in}}%
\pgfpathlineto{\pgfqpoint{6.600000in}{6.200000in}}%
\pgfpathlineto{\pgfqpoint{0.000000in}{6.200000in}}%
\pgfpathlineto{\pgfqpoint{0.000000in}{0.000000in}}%
\pgfpathclose%
\pgfusepath{fill}%
\end{pgfscope}%
\begin{pgfscope}%
\pgfsetbuttcap%
\pgfsetmiterjoin%
\definecolor{currentfill}{rgb}{1.000000,1.000000,1.000000}%
\pgfsetfillcolor{currentfill}%
\pgfsetlinewidth{0.000000pt}%
\definecolor{currentstroke}{rgb}{0.000000,0.000000,0.000000}%
\pgfsetstrokecolor{currentstroke}%
\pgfsetstrokeopacity{0.000000}%
\pgfsetdash{}{0pt}%
\pgfpathmoveto{\pgfqpoint{1.009028in}{1.080000in}}%
\pgfpathlineto{\pgfqpoint{6.338611in}{1.080000in}}%
\pgfpathlineto{\pgfqpoint{6.338611in}{5.904444in}}%
\pgfpathlineto{\pgfqpoint{1.009028in}{5.904444in}}%
\pgfpathlineto{\pgfqpoint{1.009028in}{1.080000in}}%
\pgfpathclose%
\pgfusepath{fill}%
\end{pgfscope}%
\begin{pgfscope}%
\pgfpathrectangle{\pgfqpoint{1.009028in}{1.080000in}}{\pgfqpoint{5.329583in}{4.824444in}}%
\pgfusepath{clip}%
\pgfsetrectcap%
\pgfsetroundjoin%
\pgfsetlinewidth{0.803000pt}%
\definecolor{currentstroke}{rgb}{0.690196,0.690196,0.690196}%
\pgfsetstrokecolor{currentstroke}%
\pgfsetdash{}{0pt}%
\pgfpathmoveto{\pgfqpoint{2.290826in}{1.080000in}}%
\pgfpathlineto{\pgfqpoint{2.290826in}{5.904444in}}%
\pgfusepath{stroke}%
\end{pgfscope}%
\begin{pgfscope}%
\pgfsetbuttcap%
\pgfsetroundjoin%
\definecolor{currentfill}{rgb}{0.000000,0.000000,0.000000}%
\pgfsetfillcolor{currentfill}%
\pgfsetlinewidth{0.803000pt}%
\definecolor{currentstroke}{rgb}{0.000000,0.000000,0.000000}%
\pgfsetstrokecolor{currentstroke}%
\pgfsetdash{}{0pt}%
\pgfsys@defobject{currentmarker}{\pgfqpoint{0.000000in}{-0.048611in}}{\pgfqpoint{0.000000in}{0.000000in}}{%
\pgfpathmoveto{\pgfqpoint{0.000000in}{0.000000in}}%
\pgfpathlineto{\pgfqpoint{0.000000in}{-0.048611in}}%
\pgfusepath{stroke,fill}%
}%
\begin{pgfscope}%
\pgfsys@transformshift{2.290826in}{1.080000in}%
\pgfsys@useobject{currentmarker}{}%
\end{pgfscope}%
\end{pgfscope}%
\begin{pgfscope}%
\definecolor{textcolor}{rgb}{0.000000,0.000000,0.000000}%
\pgfsetstrokecolor{textcolor}%
\pgfsetfillcolor{textcolor}%
\pgftext[x=2.290826in,y=0.982778in,,top]{\color{textcolor}\rmfamily\fontsize{28.000000}{33.600000}\selectfont 20}%
\end{pgfscope}%
\begin{pgfscope}%
\pgfpathrectangle{\pgfqpoint{1.009028in}{1.080000in}}{\pgfqpoint{5.329583in}{4.824444in}}%
\pgfusepath{clip}%
\pgfsetrectcap%
\pgfsetroundjoin%
\pgfsetlinewidth{0.803000pt}%
\definecolor{currentstroke}{rgb}{0.690196,0.690196,0.690196}%
\pgfsetstrokecolor{currentstroke}%
\pgfsetdash{}{0pt}%
\pgfpathmoveto{\pgfqpoint{3.640088in}{1.080000in}}%
\pgfpathlineto{\pgfqpoint{3.640088in}{5.904444in}}%
\pgfusepath{stroke}%
\end{pgfscope}%
\begin{pgfscope}%
\pgfsetbuttcap%
\pgfsetroundjoin%
\definecolor{currentfill}{rgb}{0.000000,0.000000,0.000000}%
\pgfsetfillcolor{currentfill}%
\pgfsetlinewidth{0.803000pt}%
\definecolor{currentstroke}{rgb}{0.000000,0.000000,0.000000}%
\pgfsetstrokecolor{currentstroke}%
\pgfsetdash{}{0pt}%
\pgfsys@defobject{currentmarker}{\pgfqpoint{0.000000in}{-0.048611in}}{\pgfqpoint{0.000000in}{0.000000in}}{%
\pgfpathmoveto{\pgfqpoint{0.000000in}{0.000000in}}%
\pgfpathlineto{\pgfqpoint{0.000000in}{-0.048611in}}%
\pgfusepath{stroke,fill}%
}%
\begin{pgfscope}%
\pgfsys@transformshift{3.640088in}{1.080000in}%
\pgfsys@useobject{currentmarker}{}%
\end{pgfscope}%
\end{pgfscope}%
\begin{pgfscope}%
\definecolor{textcolor}{rgb}{0.000000,0.000000,0.000000}%
\pgfsetstrokecolor{textcolor}%
\pgfsetfillcolor{textcolor}%
\pgftext[x=3.640088in,y=0.982778in,,top]{\color{textcolor}\rmfamily\fontsize{28.000000}{33.600000}\selectfont 40}%
\end{pgfscope}%
\begin{pgfscope}%
\pgfpathrectangle{\pgfqpoint{1.009028in}{1.080000in}}{\pgfqpoint{5.329583in}{4.824444in}}%
\pgfusepath{clip}%
\pgfsetrectcap%
\pgfsetroundjoin%
\pgfsetlinewidth{0.803000pt}%
\definecolor{currentstroke}{rgb}{0.690196,0.690196,0.690196}%
\pgfsetstrokecolor{currentstroke}%
\pgfsetdash{}{0pt}%
\pgfpathmoveto{\pgfqpoint{4.989350in}{1.080000in}}%
\pgfpathlineto{\pgfqpoint{4.989350in}{5.904444in}}%
\pgfusepath{stroke}%
\end{pgfscope}%
\begin{pgfscope}%
\pgfsetbuttcap%
\pgfsetroundjoin%
\definecolor{currentfill}{rgb}{0.000000,0.000000,0.000000}%
\pgfsetfillcolor{currentfill}%
\pgfsetlinewidth{0.803000pt}%
\definecolor{currentstroke}{rgb}{0.000000,0.000000,0.000000}%
\pgfsetstrokecolor{currentstroke}%
\pgfsetdash{}{0pt}%
\pgfsys@defobject{currentmarker}{\pgfqpoint{0.000000in}{-0.048611in}}{\pgfqpoint{0.000000in}{0.000000in}}{%
\pgfpathmoveto{\pgfqpoint{0.000000in}{0.000000in}}%
\pgfpathlineto{\pgfqpoint{0.000000in}{-0.048611in}}%
\pgfusepath{stroke,fill}%
}%
\begin{pgfscope}%
\pgfsys@transformshift{4.989350in}{1.080000in}%
\pgfsys@useobject{currentmarker}{}%
\end{pgfscope}%
\end{pgfscope}%
\begin{pgfscope}%
\definecolor{textcolor}{rgb}{0.000000,0.000000,0.000000}%
\pgfsetstrokecolor{textcolor}%
\pgfsetfillcolor{textcolor}%
\pgftext[x=4.989350in,y=0.982778in,,top]{\color{textcolor}\rmfamily\fontsize{28.000000}{33.600000}\selectfont 60}%
\end{pgfscope}%
\begin{pgfscope}%
\pgfpathrectangle{\pgfqpoint{1.009028in}{1.080000in}}{\pgfqpoint{5.329583in}{4.824444in}}%
\pgfusepath{clip}%
\pgfsetrectcap%
\pgfsetroundjoin%
\pgfsetlinewidth{0.803000pt}%
\definecolor{currentstroke}{rgb}{0.690196,0.690196,0.690196}%
\pgfsetstrokecolor{currentstroke}%
\pgfsetdash{}{0pt}%
\pgfpathmoveto{\pgfqpoint{6.338611in}{1.080000in}}%
\pgfpathlineto{\pgfqpoint{6.338611in}{5.904444in}}%
\pgfusepath{stroke}%
\end{pgfscope}%
\begin{pgfscope}%
\pgfsetbuttcap%
\pgfsetroundjoin%
\definecolor{currentfill}{rgb}{0.000000,0.000000,0.000000}%
\pgfsetfillcolor{currentfill}%
\pgfsetlinewidth{0.803000pt}%
\definecolor{currentstroke}{rgb}{0.000000,0.000000,0.000000}%
\pgfsetstrokecolor{currentstroke}%
\pgfsetdash{}{0pt}%
\pgfsys@defobject{currentmarker}{\pgfqpoint{0.000000in}{-0.048611in}}{\pgfqpoint{0.000000in}{0.000000in}}{%
\pgfpathmoveto{\pgfqpoint{0.000000in}{0.000000in}}%
\pgfpathlineto{\pgfqpoint{0.000000in}{-0.048611in}}%
\pgfusepath{stroke,fill}%
}%
\begin{pgfscope}%
\pgfsys@transformshift{6.338611in}{1.080000in}%
\pgfsys@useobject{currentmarker}{}%
\end{pgfscope}%
\end{pgfscope}%
\begin{pgfscope}%
\definecolor{textcolor}{rgb}{0.000000,0.000000,0.000000}%
\pgfsetstrokecolor{textcolor}%
\pgfsetfillcolor{textcolor}%
\pgftext[x=6.338611in,y=0.982778in,,top]{\color{textcolor}\rmfamily\fontsize{28.000000}{33.600000}\selectfont 80}%
\end{pgfscope}%
\begin{pgfscope}%
\definecolor{textcolor}{rgb}{0.000000,0.000000,0.000000}%
\pgfsetstrokecolor{textcolor}%
\pgfsetfillcolor{textcolor}%
\pgftext[x=3.673819in,y=0.536706in,,top]{\color{textcolor}\rmfamily\fontsize{30.000000}{36.000000}\selectfont time step \(\displaystyle k\)}%
\end{pgfscope}%
\begin{pgfscope}%
\pgfpathrectangle{\pgfqpoint{1.009028in}{1.080000in}}{\pgfqpoint{5.329583in}{4.824444in}}%
\pgfusepath{clip}%
\pgfsetrectcap%
\pgfsetroundjoin%
\pgfsetlinewidth{0.803000pt}%
\definecolor{currentstroke}{rgb}{0.690196,0.690196,0.690196}%
\pgfsetstrokecolor{currentstroke}%
\pgfsetdash{}{0pt}%
\pgfpathmoveto{\pgfqpoint{1.009028in}{1.080000in}}%
\pgfpathlineto{\pgfqpoint{6.338611in}{1.080000in}}%
\pgfusepath{stroke}%
\end{pgfscope}%
\begin{pgfscope}%
\pgfsetbuttcap%
\pgfsetroundjoin%
\definecolor{currentfill}{rgb}{0.000000,0.000000,0.000000}%
\pgfsetfillcolor{currentfill}%
\pgfsetlinewidth{0.803000pt}%
\definecolor{currentstroke}{rgb}{0.000000,0.000000,0.000000}%
\pgfsetstrokecolor{currentstroke}%
\pgfsetdash{}{0pt}%
\pgfsys@defobject{currentmarker}{\pgfqpoint{-0.048611in}{0.000000in}}{\pgfqpoint{-0.000000in}{0.000000in}}{%
\pgfpathmoveto{\pgfqpoint{-0.000000in}{0.000000in}}%
\pgfpathlineto{\pgfqpoint{-0.048611in}{0.000000in}}%
\pgfusepath{stroke,fill}%
}%
\begin{pgfscope}%
\pgfsys@transformshift{1.009028in}{1.080000in}%
\pgfsys@useobject{currentmarker}{}%
\end{pgfscope}%
\end{pgfscope}%
\begin{pgfscope}%
\definecolor{textcolor}{rgb}{0.000000,0.000000,0.000000}%
\pgfsetstrokecolor{textcolor}%
\pgfsetfillcolor{textcolor}%
\pgftext[x=0.663050in, y=0.960015in, left, base]{\color{textcolor}\rmfamily\fontsize{28.000000}{33.600000}\selectfont 0}%
\end{pgfscope}%
\begin{pgfscope}%
\pgfpathrectangle{\pgfqpoint{1.009028in}{1.080000in}}{\pgfqpoint{5.329583in}{4.824444in}}%
\pgfusepath{clip}%
\pgfsetrectcap%
\pgfsetroundjoin%
\pgfsetlinewidth{0.803000pt}%
\definecolor{currentstroke}{rgb}{0.690196,0.690196,0.690196}%
\pgfsetstrokecolor{currentstroke}%
\pgfsetdash{}{0pt}%
\pgfpathmoveto{\pgfqpoint{1.009028in}{1.884074in}}%
\pgfpathlineto{\pgfqpoint{6.338611in}{1.884074in}}%
\pgfusepath{stroke}%
\end{pgfscope}%
\begin{pgfscope}%
\pgfsetbuttcap%
\pgfsetroundjoin%
\definecolor{currentfill}{rgb}{0.000000,0.000000,0.000000}%
\pgfsetfillcolor{currentfill}%
\pgfsetlinewidth{0.803000pt}%
\definecolor{currentstroke}{rgb}{0.000000,0.000000,0.000000}%
\pgfsetstrokecolor{currentstroke}%
\pgfsetdash{}{0pt}%
\pgfsys@defobject{currentmarker}{\pgfqpoint{-0.048611in}{0.000000in}}{\pgfqpoint{-0.000000in}{0.000000in}}{%
\pgfpathmoveto{\pgfqpoint{-0.000000in}{0.000000in}}%
\pgfpathlineto{\pgfqpoint{-0.048611in}{0.000000in}}%
\pgfusepath{stroke,fill}%
}%
\begin{pgfscope}%
\pgfsys@transformshift{1.009028in}{1.884074in}%
\pgfsys@useobject{currentmarker}{}%
\end{pgfscope}%
\end{pgfscope}%
\begin{pgfscope}%
\definecolor{textcolor}{rgb}{0.000000,0.000000,0.000000}%
\pgfsetstrokecolor{textcolor}%
\pgfsetfillcolor{textcolor}%
\pgftext[x=0.663050in, y=1.764089in, left, base]{\color{textcolor}\rmfamily\fontsize{28.000000}{33.600000}\selectfont 1}%
\end{pgfscope}%
\begin{pgfscope}%
\pgfpathrectangle{\pgfqpoint{1.009028in}{1.080000in}}{\pgfqpoint{5.329583in}{4.824444in}}%
\pgfusepath{clip}%
\pgfsetrectcap%
\pgfsetroundjoin%
\pgfsetlinewidth{0.803000pt}%
\definecolor{currentstroke}{rgb}{0.690196,0.690196,0.690196}%
\pgfsetstrokecolor{currentstroke}%
\pgfsetdash{}{0pt}%
\pgfpathmoveto{\pgfqpoint{1.009028in}{2.688148in}}%
\pgfpathlineto{\pgfqpoint{6.338611in}{2.688148in}}%
\pgfusepath{stroke}%
\end{pgfscope}%
\begin{pgfscope}%
\pgfsetbuttcap%
\pgfsetroundjoin%
\definecolor{currentfill}{rgb}{0.000000,0.000000,0.000000}%
\pgfsetfillcolor{currentfill}%
\pgfsetlinewidth{0.803000pt}%
\definecolor{currentstroke}{rgb}{0.000000,0.000000,0.000000}%
\pgfsetstrokecolor{currentstroke}%
\pgfsetdash{}{0pt}%
\pgfsys@defobject{currentmarker}{\pgfqpoint{-0.048611in}{0.000000in}}{\pgfqpoint{-0.000000in}{0.000000in}}{%
\pgfpathmoveto{\pgfqpoint{-0.000000in}{0.000000in}}%
\pgfpathlineto{\pgfqpoint{-0.048611in}{0.000000in}}%
\pgfusepath{stroke,fill}%
}%
\begin{pgfscope}%
\pgfsys@transformshift{1.009028in}{2.688148in}%
\pgfsys@useobject{currentmarker}{}%
\end{pgfscope}%
\end{pgfscope}%
\begin{pgfscope}%
\definecolor{textcolor}{rgb}{0.000000,0.000000,0.000000}%
\pgfsetstrokecolor{textcolor}%
\pgfsetfillcolor{textcolor}%
\pgftext[x=0.663050in, y=2.568163in, left, base]{\color{textcolor}\rmfamily\fontsize{28.000000}{33.600000}\selectfont 2}%
\end{pgfscope}%
\begin{pgfscope}%
\pgfpathrectangle{\pgfqpoint{1.009028in}{1.080000in}}{\pgfqpoint{5.329583in}{4.824444in}}%
\pgfusepath{clip}%
\pgfsetrectcap%
\pgfsetroundjoin%
\pgfsetlinewidth{0.803000pt}%
\definecolor{currentstroke}{rgb}{0.690196,0.690196,0.690196}%
\pgfsetstrokecolor{currentstroke}%
\pgfsetdash{}{0pt}%
\pgfpathmoveto{\pgfqpoint{1.009028in}{3.492222in}}%
\pgfpathlineto{\pgfqpoint{6.338611in}{3.492222in}}%
\pgfusepath{stroke}%
\end{pgfscope}%
\begin{pgfscope}%
\pgfsetbuttcap%
\pgfsetroundjoin%
\definecolor{currentfill}{rgb}{0.000000,0.000000,0.000000}%
\pgfsetfillcolor{currentfill}%
\pgfsetlinewidth{0.803000pt}%
\definecolor{currentstroke}{rgb}{0.000000,0.000000,0.000000}%
\pgfsetstrokecolor{currentstroke}%
\pgfsetdash{}{0pt}%
\pgfsys@defobject{currentmarker}{\pgfqpoint{-0.048611in}{0.000000in}}{\pgfqpoint{-0.000000in}{0.000000in}}{%
\pgfpathmoveto{\pgfqpoint{-0.000000in}{0.000000in}}%
\pgfpathlineto{\pgfqpoint{-0.048611in}{0.000000in}}%
\pgfusepath{stroke,fill}%
}%
\begin{pgfscope}%
\pgfsys@transformshift{1.009028in}{3.492222in}%
\pgfsys@useobject{currentmarker}{}%
\end{pgfscope}%
\end{pgfscope}%
\begin{pgfscope}%
\definecolor{textcolor}{rgb}{0.000000,0.000000,0.000000}%
\pgfsetstrokecolor{textcolor}%
\pgfsetfillcolor{textcolor}%
\pgftext[x=0.663050in, y=3.372238in, left, base]{\color{textcolor}\rmfamily\fontsize{28.000000}{33.600000}\selectfont 3}%
\end{pgfscope}%
\begin{pgfscope}%
\pgfpathrectangle{\pgfqpoint{1.009028in}{1.080000in}}{\pgfqpoint{5.329583in}{4.824444in}}%
\pgfusepath{clip}%
\pgfsetrectcap%
\pgfsetroundjoin%
\pgfsetlinewidth{0.803000pt}%
\definecolor{currentstroke}{rgb}{0.690196,0.690196,0.690196}%
\pgfsetstrokecolor{currentstroke}%
\pgfsetdash{}{0pt}%
\pgfpathmoveto{\pgfqpoint{1.009028in}{4.296296in}}%
\pgfpathlineto{\pgfqpoint{6.338611in}{4.296296in}}%
\pgfusepath{stroke}%
\end{pgfscope}%
\begin{pgfscope}%
\pgfsetbuttcap%
\pgfsetroundjoin%
\definecolor{currentfill}{rgb}{0.000000,0.000000,0.000000}%
\pgfsetfillcolor{currentfill}%
\pgfsetlinewidth{0.803000pt}%
\definecolor{currentstroke}{rgb}{0.000000,0.000000,0.000000}%
\pgfsetstrokecolor{currentstroke}%
\pgfsetdash{}{0pt}%
\pgfsys@defobject{currentmarker}{\pgfqpoint{-0.048611in}{0.000000in}}{\pgfqpoint{-0.000000in}{0.000000in}}{%
\pgfpathmoveto{\pgfqpoint{-0.000000in}{0.000000in}}%
\pgfpathlineto{\pgfqpoint{-0.048611in}{0.000000in}}%
\pgfusepath{stroke,fill}%
}%
\begin{pgfscope}%
\pgfsys@transformshift{1.009028in}{4.296296in}%
\pgfsys@useobject{currentmarker}{}%
\end{pgfscope}%
\end{pgfscope}%
\begin{pgfscope}%
\definecolor{textcolor}{rgb}{0.000000,0.000000,0.000000}%
\pgfsetstrokecolor{textcolor}%
\pgfsetfillcolor{textcolor}%
\pgftext[x=0.663050in, y=4.176312in, left, base]{\color{textcolor}\rmfamily\fontsize{28.000000}{33.600000}\selectfont 4}%
\end{pgfscope}%
\begin{pgfscope}%
\pgfpathrectangle{\pgfqpoint{1.009028in}{1.080000in}}{\pgfqpoint{5.329583in}{4.824444in}}%
\pgfusepath{clip}%
\pgfsetrectcap%
\pgfsetroundjoin%
\pgfsetlinewidth{0.803000pt}%
\definecolor{currentstroke}{rgb}{0.690196,0.690196,0.690196}%
\pgfsetstrokecolor{currentstroke}%
\pgfsetdash{}{0pt}%
\pgfpathmoveto{\pgfqpoint{1.009028in}{5.100370in}}%
\pgfpathlineto{\pgfqpoint{6.338611in}{5.100370in}}%
\pgfusepath{stroke}%
\end{pgfscope}%
\begin{pgfscope}%
\pgfsetbuttcap%
\pgfsetroundjoin%
\definecolor{currentfill}{rgb}{0.000000,0.000000,0.000000}%
\pgfsetfillcolor{currentfill}%
\pgfsetlinewidth{0.803000pt}%
\definecolor{currentstroke}{rgb}{0.000000,0.000000,0.000000}%
\pgfsetstrokecolor{currentstroke}%
\pgfsetdash{}{0pt}%
\pgfsys@defobject{currentmarker}{\pgfqpoint{-0.048611in}{0.000000in}}{\pgfqpoint{-0.000000in}{0.000000in}}{%
\pgfpathmoveto{\pgfqpoint{-0.000000in}{0.000000in}}%
\pgfpathlineto{\pgfqpoint{-0.048611in}{0.000000in}}%
\pgfusepath{stroke,fill}%
}%
\begin{pgfscope}%
\pgfsys@transformshift{1.009028in}{5.100370in}%
\pgfsys@useobject{currentmarker}{}%
\end{pgfscope}%
\end{pgfscope}%
\begin{pgfscope}%
\definecolor{textcolor}{rgb}{0.000000,0.000000,0.000000}%
\pgfsetstrokecolor{textcolor}%
\pgfsetfillcolor{textcolor}%
\pgftext[x=0.663050in, y=4.980386in, left, base]{\color{textcolor}\rmfamily\fontsize{28.000000}{33.600000}\selectfont 5}%
\end{pgfscope}%
\begin{pgfscope}%
\pgfpathrectangle{\pgfqpoint{1.009028in}{1.080000in}}{\pgfqpoint{5.329583in}{4.824444in}}%
\pgfusepath{clip}%
\pgfsetrectcap%
\pgfsetroundjoin%
\pgfsetlinewidth{0.803000pt}%
\definecolor{currentstroke}{rgb}{0.690196,0.690196,0.690196}%
\pgfsetstrokecolor{currentstroke}%
\pgfsetdash{}{0pt}%
\pgfpathmoveto{\pgfqpoint{1.009028in}{5.904444in}}%
\pgfpathlineto{\pgfqpoint{6.338611in}{5.904444in}}%
\pgfusepath{stroke}%
\end{pgfscope}%
\begin{pgfscope}%
\pgfsetbuttcap%
\pgfsetroundjoin%
\definecolor{currentfill}{rgb}{0.000000,0.000000,0.000000}%
\pgfsetfillcolor{currentfill}%
\pgfsetlinewidth{0.803000pt}%
\definecolor{currentstroke}{rgb}{0.000000,0.000000,0.000000}%
\pgfsetstrokecolor{currentstroke}%
\pgfsetdash{}{0pt}%
\pgfsys@defobject{currentmarker}{\pgfqpoint{-0.048611in}{0.000000in}}{\pgfqpoint{-0.000000in}{0.000000in}}{%
\pgfpathmoveto{\pgfqpoint{-0.000000in}{0.000000in}}%
\pgfpathlineto{\pgfqpoint{-0.048611in}{0.000000in}}%
\pgfusepath{stroke,fill}%
}%
\begin{pgfscope}%
\pgfsys@transformshift{1.009028in}{5.904444in}%
\pgfsys@useobject{currentmarker}{}%
\end{pgfscope}%
\end{pgfscope}%
\begin{pgfscope}%
\definecolor{textcolor}{rgb}{0.000000,0.000000,0.000000}%
\pgfsetstrokecolor{textcolor}%
\pgfsetfillcolor{textcolor}%
\pgftext[x=0.663050in, y=5.784460in, left, base]{\color{textcolor}\rmfamily\fontsize{28.000000}{33.600000}\selectfont 6}%
\end{pgfscope}%
\begin{pgfscope}%
\definecolor{textcolor}{rgb}{0.000000,0.000000,0.000000}%
\pgfsetstrokecolor{textcolor}%
\pgfsetfillcolor{textcolor}%
\pgftext[x=0.524161in,y=3.492222in,,bottom,rotate=90.000000]{\color{textcolor}\rmfamily\fontsize{30.000000}{36.000000}\selectfont noise variance}%
\end{pgfscope}%
\begin{pgfscope}%
\definecolor{textcolor}{rgb}{0.000000,0.000000,0.000000}%
\pgfsetstrokecolor{textcolor}%
\pgfsetfillcolor{textcolor}%
\pgftext[x=1.009028in,y=5.946111in,left,base]{\color{textcolor}\rmfamily\fontsize{22.000000}{26.400000}\selectfont 1e\ensuremath{-}14}%
\end{pgfscope}%
\begin{pgfscope}%
\pgfpathrectangle{\pgfqpoint{1.009028in}{1.080000in}}{\pgfqpoint{5.329583in}{4.824444in}}%
\pgfusepath{clip}%
\pgfsetrectcap%
\pgfsetroundjoin%
\pgfsetlinewidth{1.505625pt}%
\definecolor{currentstroke}{rgb}{1.000000,0.000000,0.000000}%
\pgfsetstrokecolor{currentstroke}%
\pgfsetdash{}{0pt}%
\pgfpathmoveto{\pgfqpoint{1.368005in}{5.914444in}}%
\pgfpathlineto{\pgfqpoint{1.413806in}{5.061986in}}%
\pgfpathlineto{\pgfqpoint{1.481269in}{4.167438in}}%
\pgfpathlineto{\pgfqpoint{1.548732in}{3.557032in}}%
\pgfpathlineto{\pgfqpoint{1.616195in}{3.179545in}}%
\pgfpathlineto{\pgfqpoint{1.683659in}{2.951162in}}%
\pgfpathlineto{\pgfqpoint{1.751122in}{2.814904in}}%
\pgfpathlineto{\pgfqpoint{1.818585in}{2.733075in}}%
\pgfpathlineto{\pgfqpoint{1.886048in}{2.699209in}}%
\pgfpathlineto{\pgfqpoint{1.953511in}{2.686407in}}%
\pgfpathlineto{\pgfqpoint{2.020974in}{2.697436in}}%
\pgfpathlineto{\pgfqpoint{2.088437in}{2.684344in}}%
\pgfpathlineto{\pgfqpoint{2.155900in}{2.692488in}}%
\pgfpathlineto{\pgfqpoint{2.223363in}{2.687981in}}%
\pgfpathlineto{\pgfqpoint{2.290826in}{2.691031in}}%
\pgfpathlineto{\pgfqpoint{2.358289in}{2.703042in}}%
\pgfpathlineto{\pgfqpoint{2.425752in}{2.692432in}}%
\pgfpathlineto{\pgfqpoint{2.493216in}{2.690603in}}%
\pgfpathlineto{\pgfqpoint{2.560679in}{2.693895in}}%
\pgfpathlineto{\pgfqpoint{2.628142in}{2.694384in}}%
\pgfpathlineto{\pgfqpoint{2.695605in}{2.700540in}}%
\pgfpathlineto{\pgfqpoint{2.763068in}{2.690631in}}%
\pgfpathlineto{\pgfqpoint{2.830531in}{2.689987in}}%
\pgfpathlineto{\pgfqpoint{2.897994in}{2.688460in}}%
\pgfpathlineto{\pgfqpoint{2.965457in}{2.691244in}}%
\pgfpathlineto{\pgfqpoint{3.032920in}{2.695148in}}%
\pgfpathlineto{\pgfqpoint{3.100383in}{2.693014in}}%
\pgfpathlineto{\pgfqpoint{3.167846in}{2.691827in}}%
\pgfpathlineto{\pgfqpoint{3.235309in}{2.690226in}}%
\pgfpathlineto{\pgfqpoint{3.302773in}{2.694393in}}%
\pgfpathlineto{\pgfqpoint{3.370236in}{2.692553in}}%
\pgfpathlineto{\pgfqpoint{3.437699in}{2.689091in}}%
\pgfpathlineto{\pgfqpoint{3.505162in}{2.697339in}}%
\pgfpathlineto{\pgfqpoint{3.572625in}{2.693749in}}%
\pgfpathlineto{\pgfqpoint{3.640088in}{2.692121in}}%
\pgfpathlineto{\pgfqpoint{3.707551in}{2.693312in}}%
\pgfpathlineto{\pgfqpoint{3.775014in}{2.694166in}}%
\pgfpathlineto{\pgfqpoint{3.842477in}{2.694513in}}%
\pgfpathlineto{\pgfqpoint{3.909940in}{2.693032in}}%
\pgfpathlineto{\pgfqpoint{3.977403in}{2.696383in}}%
\pgfpathlineto{\pgfqpoint{4.044866in}{2.695976in}}%
\pgfpathlineto{\pgfqpoint{4.112329in}{2.692528in}}%
\pgfpathlineto{\pgfqpoint{4.179793in}{2.695625in}}%
\pgfpathlineto{\pgfqpoint{4.247256in}{2.699233in}}%
\pgfpathlineto{\pgfqpoint{4.314719in}{2.692113in}}%
\pgfpathlineto{\pgfqpoint{4.382182in}{2.690380in}}%
\pgfpathlineto{\pgfqpoint{4.449645in}{2.692018in}}%
\pgfpathlineto{\pgfqpoint{4.517108in}{2.693294in}}%
\pgfpathlineto{\pgfqpoint{4.584571in}{2.692480in}}%
\pgfpathlineto{\pgfqpoint{4.652034in}{2.694363in}}%
\pgfpathlineto{\pgfqpoint{4.719497in}{2.692598in}}%
\pgfpathlineto{\pgfqpoint{4.786960in}{2.691613in}}%
\pgfpathlineto{\pgfqpoint{4.854423in}{2.684453in}}%
\pgfpathlineto{\pgfqpoint{4.921886in}{2.685490in}}%
\pgfpathlineto{\pgfqpoint{4.989350in}{2.691704in}}%
\pgfpathlineto{\pgfqpoint{5.056813in}{2.690446in}}%
\pgfpathlineto{\pgfqpoint{5.124276in}{2.684286in}}%
\pgfpathlineto{\pgfqpoint{5.191739in}{2.693342in}}%
\pgfpathlineto{\pgfqpoint{5.259202in}{2.690797in}}%
\pgfpathlineto{\pgfqpoint{5.326665in}{2.689936in}}%
\pgfpathlineto{\pgfqpoint{5.394128in}{2.689245in}}%
\pgfpathlineto{\pgfqpoint{5.461591in}{2.690860in}}%
\pgfpathlineto{\pgfqpoint{5.529054in}{2.686002in}}%
\pgfpathlineto{\pgfqpoint{5.596517in}{2.692289in}}%
\pgfpathlineto{\pgfqpoint{5.663980in}{2.692946in}}%
\pgfpathlineto{\pgfqpoint{5.731443in}{2.686332in}}%
\pgfpathlineto{\pgfqpoint{5.798906in}{2.683981in}}%
\pgfpathlineto{\pgfqpoint{5.866370in}{2.684876in}}%
\pgfpathlineto{\pgfqpoint{5.933833in}{2.685952in}}%
\pgfpathlineto{\pgfqpoint{6.001296in}{2.685885in}}%
\pgfpathlineto{\pgfqpoint{6.068759in}{2.685588in}}%
\pgfpathlineto{\pgfqpoint{6.136222in}{2.688190in}}%
\pgfpathlineto{\pgfqpoint{6.203685in}{2.682335in}}%
\pgfpathlineto{\pgfqpoint{6.271148in}{2.689970in}}%
\pgfpathlineto{\pgfqpoint{6.338611in}{2.691607in}}%
\pgfusepath{stroke}%
\end{pgfscope}%
\begin{pgfscope}%
\pgfpathrectangle{\pgfqpoint{1.009028in}{1.080000in}}{\pgfqpoint{5.329583in}{4.824444in}}%
\pgfusepath{clip}%
\pgfsetbuttcap%
\pgfsetroundjoin%
\pgfsetlinewidth{2.007500pt}%
\definecolor{currentstroke}{rgb}{0.000000,0.000000,0.000000}%
\pgfsetstrokecolor{currentstroke}%
\pgfsetdash{{7.400000pt}{3.200000pt}}{0.000000pt}%
\pgfpathmoveto{\pgfqpoint{1.009028in}{2.688148in}}%
\pgfpathlineto{\pgfqpoint{1.076491in}{2.688148in}}%
\pgfpathlineto{\pgfqpoint{1.143954in}{2.688148in}}%
\pgfpathlineto{\pgfqpoint{1.211417in}{2.688148in}}%
\pgfpathlineto{\pgfqpoint{1.278880in}{2.688148in}}%
\pgfpathlineto{\pgfqpoint{1.346343in}{2.688148in}}%
\pgfpathlineto{\pgfqpoint{1.413806in}{2.688148in}}%
\pgfpathlineto{\pgfqpoint{1.481269in}{2.688148in}}%
\pgfpathlineto{\pgfqpoint{1.548732in}{2.688148in}}%
\pgfpathlineto{\pgfqpoint{1.616195in}{2.688148in}}%
\pgfpathlineto{\pgfqpoint{1.683659in}{2.688148in}}%
\pgfpathlineto{\pgfqpoint{1.751122in}{2.688148in}}%
\pgfpathlineto{\pgfqpoint{1.818585in}{2.688148in}}%
\pgfpathlineto{\pgfqpoint{1.886048in}{2.688148in}}%
\pgfpathlineto{\pgfqpoint{1.953511in}{2.688148in}}%
\pgfpathlineto{\pgfqpoint{2.020974in}{2.688148in}}%
\pgfpathlineto{\pgfqpoint{2.088437in}{2.688148in}}%
\pgfpathlineto{\pgfqpoint{2.155900in}{2.688148in}}%
\pgfpathlineto{\pgfqpoint{2.223363in}{2.688148in}}%
\pgfpathlineto{\pgfqpoint{2.290826in}{2.688148in}}%
\pgfpathlineto{\pgfqpoint{2.358289in}{2.688148in}}%
\pgfpathlineto{\pgfqpoint{2.425752in}{2.688148in}}%
\pgfpathlineto{\pgfqpoint{2.493216in}{2.688148in}}%
\pgfpathlineto{\pgfqpoint{2.560679in}{2.688148in}}%
\pgfpathlineto{\pgfqpoint{2.628142in}{2.688148in}}%
\pgfpathlineto{\pgfqpoint{2.695605in}{2.688148in}}%
\pgfpathlineto{\pgfqpoint{2.763068in}{2.688148in}}%
\pgfpathlineto{\pgfqpoint{2.830531in}{2.688148in}}%
\pgfpathlineto{\pgfqpoint{2.897994in}{2.688148in}}%
\pgfpathlineto{\pgfqpoint{2.965457in}{2.688148in}}%
\pgfpathlineto{\pgfqpoint{3.032920in}{2.688148in}}%
\pgfpathlineto{\pgfqpoint{3.100383in}{2.688148in}}%
\pgfpathlineto{\pgfqpoint{3.167846in}{2.688148in}}%
\pgfpathlineto{\pgfqpoint{3.235309in}{2.688148in}}%
\pgfpathlineto{\pgfqpoint{3.302773in}{2.688148in}}%
\pgfpathlineto{\pgfqpoint{3.370236in}{2.688148in}}%
\pgfpathlineto{\pgfqpoint{3.437699in}{2.688148in}}%
\pgfpathlineto{\pgfqpoint{3.505162in}{2.688148in}}%
\pgfpathlineto{\pgfqpoint{3.572625in}{2.688148in}}%
\pgfpathlineto{\pgfqpoint{3.640088in}{2.688148in}}%
\pgfpathlineto{\pgfqpoint{3.707551in}{2.688148in}}%
\pgfpathlineto{\pgfqpoint{3.775014in}{2.688148in}}%
\pgfpathlineto{\pgfqpoint{3.842477in}{2.688148in}}%
\pgfpathlineto{\pgfqpoint{3.909940in}{2.688148in}}%
\pgfpathlineto{\pgfqpoint{3.977403in}{2.688148in}}%
\pgfpathlineto{\pgfqpoint{4.044866in}{2.688148in}}%
\pgfpathlineto{\pgfqpoint{4.112329in}{2.688148in}}%
\pgfpathlineto{\pgfqpoint{4.179793in}{2.688148in}}%
\pgfpathlineto{\pgfqpoint{4.247256in}{2.688148in}}%
\pgfpathlineto{\pgfqpoint{4.314719in}{2.688148in}}%
\pgfpathlineto{\pgfqpoint{4.382182in}{2.688148in}}%
\pgfpathlineto{\pgfqpoint{4.449645in}{2.688148in}}%
\pgfpathlineto{\pgfqpoint{4.517108in}{2.688148in}}%
\pgfpathlineto{\pgfqpoint{4.584571in}{2.688148in}}%
\pgfpathlineto{\pgfqpoint{4.652034in}{2.688148in}}%
\pgfpathlineto{\pgfqpoint{4.719497in}{2.688148in}}%
\pgfpathlineto{\pgfqpoint{4.786960in}{2.688148in}}%
\pgfpathlineto{\pgfqpoint{4.854423in}{2.688148in}}%
\pgfpathlineto{\pgfqpoint{4.921886in}{2.688148in}}%
\pgfpathlineto{\pgfqpoint{4.989350in}{2.688148in}}%
\pgfpathlineto{\pgfqpoint{5.056813in}{2.688148in}}%
\pgfpathlineto{\pgfqpoint{5.124276in}{2.688148in}}%
\pgfpathlineto{\pgfqpoint{5.191739in}{2.688148in}}%
\pgfpathlineto{\pgfqpoint{5.259202in}{2.688148in}}%
\pgfpathlineto{\pgfqpoint{5.326665in}{2.688148in}}%
\pgfpathlineto{\pgfqpoint{5.394128in}{2.688148in}}%
\pgfpathlineto{\pgfqpoint{5.461591in}{2.688148in}}%
\pgfpathlineto{\pgfqpoint{5.529054in}{2.688148in}}%
\pgfpathlineto{\pgfqpoint{5.596517in}{2.688148in}}%
\pgfpathlineto{\pgfqpoint{5.663980in}{2.688148in}}%
\pgfpathlineto{\pgfqpoint{5.731443in}{2.688148in}}%
\pgfpathlineto{\pgfqpoint{5.798906in}{2.688148in}}%
\pgfpathlineto{\pgfqpoint{5.866370in}{2.688148in}}%
\pgfpathlineto{\pgfqpoint{5.933833in}{2.688148in}}%
\pgfpathlineto{\pgfqpoint{6.001296in}{2.688148in}}%
\pgfpathlineto{\pgfqpoint{6.068759in}{2.688148in}}%
\pgfpathlineto{\pgfqpoint{6.136222in}{2.688148in}}%
\pgfpathlineto{\pgfqpoint{6.203685in}{2.688148in}}%
\pgfpathlineto{\pgfqpoint{6.271148in}{2.688148in}}%
\pgfpathlineto{\pgfqpoint{6.338611in}{2.688148in}}%
\pgfusepath{stroke}%
\end{pgfscope}%
\begin{pgfscope}%
\pgfsetrectcap%
\pgfsetmiterjoin%
\pgfsetlinewidth{0.803000pt}%
\definecolor{currentstroke}{rgb}{0.000000,0.000000,0.000000}%
\pgfsetstrokecolor{currentstroke}%
\pgfsetdash{}{0pt}%
\pgfpathmoveto{\pgfqpoint{1.009028in}{1.080000in}}%
\pgfpathlineto{\pgfqpoint{1.009028in}{5.904444in}}%
\pgfusepath{stroke}%
\end{pgfscope}%
\begin{pgfscope}%
\pgfsetrectcap%
\pgfsetmiterjoin%
\pgfsetlinewidth{0.803000pt}%
\definecolor{currentstroke}{rgb}{0.000000,0.000000,0.000000}%
\pgfsetstrokecolor{currentstroke}%
\pgfsetdash{}{0pt}%
\pgfpathmoveto{\pgfqpoint{6.338611in}{1.080000in}}%
\pgfpathlineto{\pgfqpoint{6.338611in}{5.904444in}}%
\pgfusepath{stroke}%
\end{pgfscope}%
\begin{pgfscope}%
\pgfsetrectcap%
\pgfsetmiterjoin%
\pgfsetlinewidth{0.803000pt}%
\definecolor{currentstroke}{rgb}{0.000000,0.000000,0.000000}%
\pgfsetstrokecolor{currentstroke}%
\pgfsetdash{}{0pt}%
\pgfpathmoveto{\pgfqpoint{1.009028in}{1.080000in}}%
\pgfpathlineto{\pgfqpoint{6.338611in}{1.080000in}}%
\pgfusepath{stroke}%
\end{pgfscope}%
\begin{pgfscope}%
\pgfsetrectcap%
\pgfsetmiterjoin%
\pgfsetlinewidth{0.803000pt}%
\definecolor{currentstroke}{rgb}{0.000000,0.000000,0.000000}%
\pgfsetstrokecolor{currentstroke}%
\pgfsetdash{}{0pt}%
\pgfpathmoveto{\pgfqpoint{1.009028in}{5.904444in}}%
\pgfpathlineto{\pgfqpoint{6.338611in}{5.904444in}}%
\pgfusepath{stroke}%
\end{pgfscope}%
\begin{pgfscope}%
\pgfsetbuttcap%
\pgfsetmiterjoin%
\definecolor{currentfill}{rgb}{1.000000,1.000000,1.000000}%
\pgfsetfillcolor{currentfill}%
\pgfsetfillopacity{0.800000}%
\pgfsetlinewidth{1.003750pt}%
\definecolor{currentstroke}{rgb}{0.800000,0.800000,0.800000}%
\pgfsetstrokecolor{currentstroke}%
\pgfsetstrokeopacity{0.800000}%
\pgfsetdash{}{0pt}%
\pgfpathmoveto{\pgfqpoint{3.198808in}{4.525573in}}%
\pgfpathlineto{\pgfqpoint{6.066389in}{4.525573in}}%
\pgfpathquadraticcurveto{\pgfqpoint{6.144167in}{4.525573in}}{\pgfqpoint{6.144167in}{4.603351in}}%
\pgfpathlineto{\pgfqpoint{6.144167in}{5.632222in}}%
\pgfpathquadraticcurveto{\pgfqpoint{6.144167in}{5.710000in}}{\pgfqpoint{6.066389in}{5.710000in}}%
\pgfpathlineto{\pgfqpoint{3.198808in}{5.710000in}}%
\pgfpathquadraticcurveto{\pgfqpoint{3.121031in}{5.710000in}}{\pgfqpoint{3.121031in}{5.632222in}}%
\pgfpathlineto{\pgfqpoint{3.121031in}{4.603351in}}%
\pgfpathquadraticcurveto{\pgfqpoint{3.121031in}{4.525573in}}{\pgfqpoint{3.198808in}{4.525573in}}%
\pgfpathlineto{\pgfqpoint{3.198808in}{4.525573in}}%
\pgfpathclose%
\pgfusepath{stroke,fill}%
\end{pgfscope}%
\begin{pgfscope}%
\pgfsetrectcap%
\pgfsetroundjoin%
\pgfsetlinewidth{1.505625pt}%
\definecolor{currentstroke}{rgb}{1.000000,0.000000,0.000000}%
\pgfsetstrokecolor{currentstroke}%
\pgfsetdash{}{0pt}%
\pgfpathmoveto{\pgfqpoint{3.276586in}{5.418333in}}%
\pgfpathlineto{\pgfqpoint{3.665475in}{5.418333in}}%
\pgfpathlineto{\pgfqpoint{4.054364in}{5.418333in}}%
\pgfusepath{stroke}%
\end{pgfscope}%
\begin{pgfscope}%
\definecolor{textcolor}{rgb}{0.000000,0.000000,0.000000}%
\pgfsetstrokecolor{textcolor}%
\pgfsetfillcolor{textcolor}%
\pgftext[x=4.365475in,y=5.282222in,left,base]{\color{textcolor}\rmfamily\fontsize{28.000000}{33.600000}\selectfont Estimated~~}%
\end{pgfscope}%
\begin{pgfscope}%
\pgfsetbuttcap%
\pgfsetroundjoin%
\pgfsetlinewidth{2.007500pt}%
\definecolor{currentstroke}{rgb}{0.000000,0.000000,0.000000}%
\pgfsetstrokecolor{currentstroke}%
\pgfsetdash{{7.400000pt}{3.200000pt}}{0.000000pt}%
\pgfpathmoveto{\pgfqpoint{3.276586in}{4.884453in}}%
\pgfpathlineto{\pgfqpoint{3.665475in}{4.884453in}}%
\pgfpathlineto{\pgfqpoint{4.054364in}{4.884453in}}%
\pgfusepath{stroke}%
\end{pgfscope}%
\begin{pgfscope}%
\definecolor{textcolor}{rgb}{0.000000,0.000000,0.000000}%
\pgfsetstrokecolor{textcolor}%
\pgfsetfillcolor{textcolor}%
\pgftext[x=4.365475in,y=4.748342in,left,base]{\color{textcolor}\rmfamily\fontsize{28.000000}{33.600000}\selectfont True}%
\end{pgfscope}%
\end{pgfpicture}%
\makeatother%
\endgroup%

%% file: Figs/rawsignal/results_k246_eps.pgf
\begingroup%
\makeatletter%
\begin{pgfpicture}%
\pgfpathrectangle{\pgfpointorigin}{\pgfqpoint{13.400000in}{6.600000in}}%
\pgfusepath{use as bounding box, clip}%
\begin{pgfscope}%
\pgfsetbuttcap%
\pgfsetmiterjoin%
\definecolor{currentfill}{rgb}{1.000000,1.000000,1.000000}%
\pgfsetfillcolor{currentfill}%
\pgfsetlinewidth{0.000000pt}%
\definecolor{currentstroke}{rgb}{1.000000,1.000000,1.000000}%
\pgfsetstrokecolor{currentstroke}%
\pgfsetdash{}{0pt}%
\pgfpathmoveto{\pgfqpoint{0.000000in}{0.000000in}}%
\pgfpathlineto{\pgfqpoint{13.400000in}{0.000000in}}%
\pgfpathlineto{\pgfqpoint{13.400000in}{6.600000in}}%
\pgfpathlineto{\pgfqpoint{0.000000in}{6.600000in}}%
\pgfpathlineto{\pgfqpoint{0.000000in}{0.000000in}}%
\pgfpathclose%
\pgfusepath{fill}%
\end{pgfscope}%
\begin{pgfscope}%
\pgfsetbuttcap%
\pgfsetmiterjoin%
\definecolor{currentfill}{rgb}{1.000000,1.000000,1.000000}%
\pgfsetfillcolor{currentfill}%
\pgfsetlinewidth{0.000000pt}%
\definecolor{currentstroke}{rgb}{0.000000,0.000000,0.000000}%
\pgfsetstrokecolor{currentstroke}%
\pgfsetstrokeopacity{0.000000}%
\pgfsetdash{}{0pt}%
\pgfpathmoveto{\pgfqpoint{1.381353in}{1.061180in}}%
\pgfpathlineto{\pgfqpoint{12.073793in}{1.061180in}}%
\pgfpathlineto{\pgfqpoint{12.073793in}{6.439608in}}%
\pgfpathlineto{\pgfqpoint{1.381353in}{6.439608in}}%
\pgfpathlineto{\pgfqpoint{1.381353in}{1.061180in}}%
\pgfpathclose%
\pgfusepath{fill}%
\end{pgfscope}%
\begin{pgfscope}%
\pgfpathrectangle{\pgfqpoint{1.381353in}{1.061180in}}{\pgfqpoint{10.692439in}{5.378428in}}%
\pgfusepath{clip}%
\pgfsys@transformshift{1.381353in}{1.061180in}%
\pgftext[left,bottom]{\includegraphics[interpolate=true,width=10.700000in,height=5.380000in]{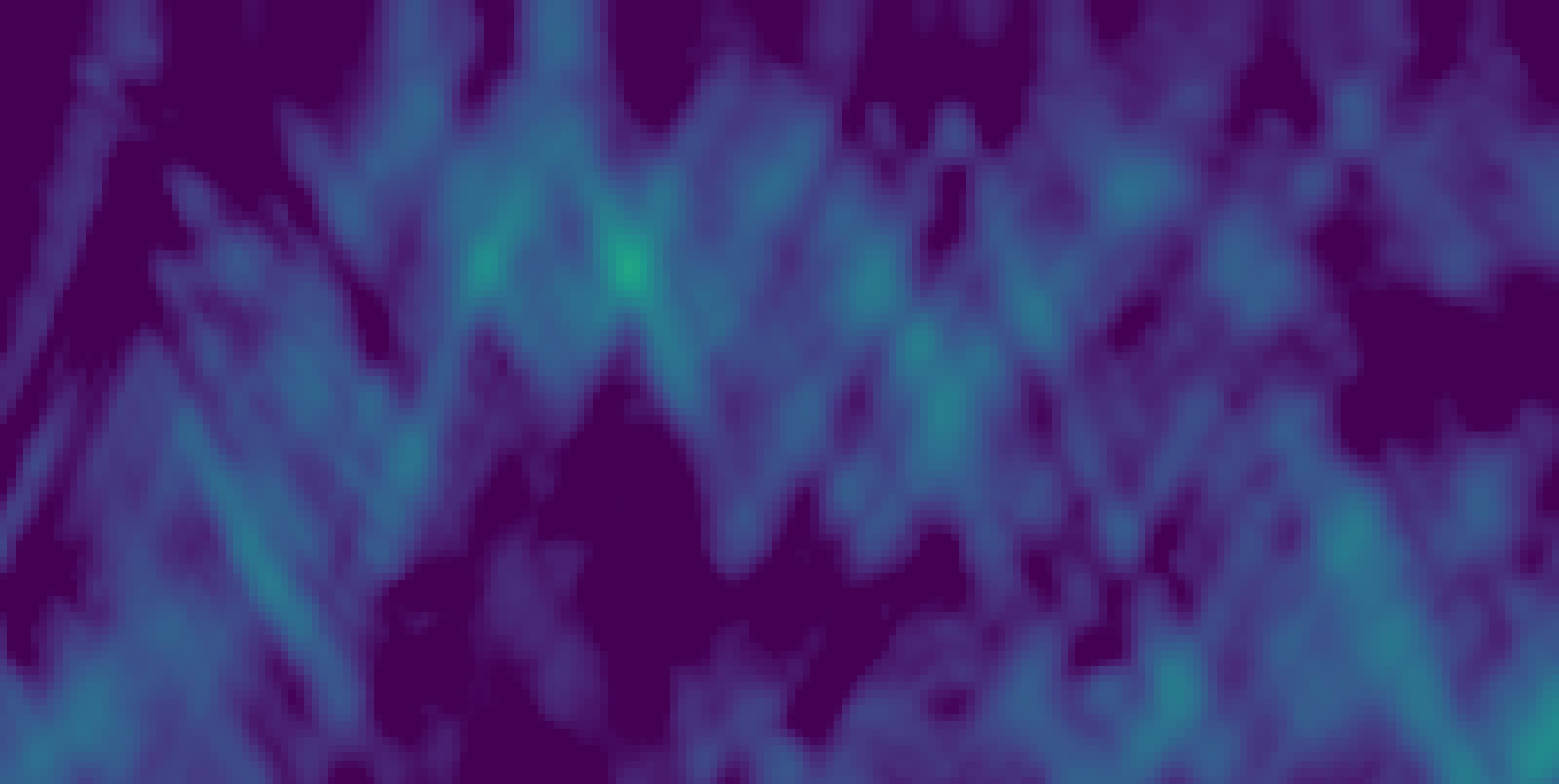}}%
\end{pgfscope}%
\begin{pgfscope}%
\pgfpathrectangle{\pgfqpoint{1.381353in}{1.061180in}}{\pgfqpoint{10.692439in}{5.378428in}}%
\pgfusepath{clip}%
\pgfsetbuttcap%
\pgfsetroundjoin%
\pgfsetlinewidth{3.011250pt}%
\definecolor{currentstroke}{rgb}{1.000000,0.000000,0.000000}%
\pgfsetstrokecolor{currentstroke}%
\pgfsetdash{}{0pt}%
\pgfpathmoveto{\pgfqpoint{5.741207in}{4.447165in}}%
\pgfpathcurveto{\pgfqpoint{5.767252in}{4.447165in}}{\pgfqpoint{5.792234in}{4.457513in}}{\pgfqpoint{5.810651in}{4.475930in}}%
\pgfpathcurveto{\pgfqpoint{5.829068in}{4.494347in}}{\pgfqpoint{5.839416in}{4.519329in}}{\pgfqpoint{5.839416in}{4.545375in}}%
\pgfpathcurveto{\pgfqpoint{5.839416in}{4.571420in}}{\pgfqpoint{5.829068in}{4.596402in}}{\pgfqpoint{5.810651in}{4.614819in}}%
\pgfpathcurveto{\pgfqpoint{5.792234in}{4.633236in}}{\pgfqpoint{5.767252in}{4.643584in}}{\pgfqpoint{5.741207in}{4.643584in}}%
\pgfpathcurveto{\pgfqpoint{5.715161in}{4.643584in}}{\pgfqpoint{5.690179in}{4.633236in}}{\pgfqpoint{5.671762in}{4.614819in}}%
\pgfpathcurveto{\pgfqpoint{5.653345in}{4.596402in}}{\pgfqpoint{5.642997in}{4.571420in}}{\pgfqpoint{5.642997in}{4.545375in}}%
\pgfpathcurveto{\pgfqpoint{5.642997in}{4.519329in}}{\pgfqpoint{5.653345in}{4.494347in}}{\pgfqpoint{5.671762in}{4.475930in}}%
\pgfpathcurveto{\pgfqpoint{5.690179in}{4.457513in}}{\pgfqpoint{5.715161in}{4.447165in}}{\pgfqpoint{5.741207in}{4.447165in}}%
\pgfpathlineto{\pgfqpoint{5.741207in}{4.447165in}}%
\pgfpathclose%
\pgfusepath{stroke}%
\end{pgfscope}%
\begin{pgfscope}%
\pgfpathrectangle{\pgfqpoint{1.381353in}{1.061180in}}{\pgfqpoint{10.692439in}{5.378428in}}%
\pgfusepath{clip}%
\pgfsetbuttcap%
\pgfsetroundjoin%
\pgfsetlinewidth{3.011250pt}%
\definecolor{currentstroke}{rgb}{1.000000,0.000000,0.000000}%
\pgfsetstrokecolor{currentstroke}%
\pgfsetdash{}{0pt}%
\pgfpathmoveto{\pgfqpoint{4.748114in}{4.480323in}}%
\pgfpathcurveto{\pgfqpoint{4.774160in}{4.480323in}}{\pgfqpoint{4.799142in}{4.490671in}}{\pgfqpoint{4.817559in}{4.509088in}}%
\pgfpathcurveto{\pgfqpoint{4.835975in}{4.527505in}}{\pgfqpoint{4.846323in}{4.552487in}}{\pgfqpoint{4.846323in}{4.578532in}}%
\pgfpathcurveto{\pgfqpoint{4.846323in}{4.604578in}}{\pgfqpoint{4.835975in}{4.629560in}}{\pgfqpoint{4.817559in}{4.647977in}}%
\pgfpathcurveto{\pgfqpoint{4.799142in}{4.666394in}}{\pgfqpoint{4.774160in}{4.676742in}}{\pgfqpoint{4.748114in}{4.676742in}}%
\pgfpathcurveto{\pgfqpoint{4.722069in}{4.676742in}}{\pgfqpoint{4.697087in}{4.666394in}}{\pgfqpoint{4.678670in}{4.647977in}}%
\pgfpathcurveto{\pgfqpoint{4.660253in}{4.629560in}}{\pgfqpoint{4.649905in}{4.604578in}}{\pgfqpoint{4.649905in}{4.578532in}}%
\pgfpathcurveto{\pgfqpoint{4.649905in}{4.552487in}}{\pgfqpoint{4.660253in}{4.527505in}}{\pgfqpoint{4.678670in}{4.509088in}}%
\pgfpathcurveto{\pgfqpoint{4.697087in}{4.490671in}}{\pgfqpoint{4.722069in}{4.480323in}}{\pgfqpoint{4.748114in}{4.480323in}}%
\pgfpathlineto{\pgfqpoint{4.748114in}{4.480323in}}%
\pgfpathclose%
\pgfusepath{stroke}%
\end{pgfscope}%
\begin{pgfscope}%
\pgfsetbuttcap%
\pgfsetroundjoin%
\definecolor{currentfill}{rgb}{0.000000,0.000000,0.000000}%
\pgfsetfillcolor{currentfill}%
\pgfsetlinewidth{0.803000pt}%
\definecolor{currentstroke}{rgb}{0.000000,0.000000,0.000000}%
\pgfsetstrokecolor{currentstroke}%
\pgfsetdash{}{0pt}%
\pgfsys@defobject{currentmarker}{\pgfqpoint{0.000000in}{-0.048611in}}{\pgfqpoint{0.000000in}{0.000000in}}{%
\pgfpathmoveto{\pgfqpoint{0.000000in}{0.000000in}}%
\pgfpathlineto{\pgfqpoint{0.000000in}{-0.048611in}}%
\pgfusepath{stroke,fill}%
}%
\begin{pgfscope}%
\pgfsys@transformshift{1.408084in}{1.061180in}%
\pgfsys@useobject{currentmarker}{}%
\end{pgfscope}%
\end{pgfscope}%
\begin{pgfscope}%
\definecolor{textcolor}{rgb}{0.000000,0.000000,0.000000}%
\pgfsetstrokecolor{textcolor}%
\pgfsetfillcolor{textcolor}%
\pgftext[x=1.408084in,y=0.963958in,,top]{\color{textcolor}\rmfamily\fontsize{30.000000}{36.000000}\selectfont 0}%
\end{pgfscope}%
\begin{pgfscope}%
\pgfsetbuttcap%
\pgfsetroundjoin%
\definecolor{currentfill}{rgb}{0.000000,0.000000,0.000000}%
\pgfsetfillcolor{currentfill}%
\pgfsetlinewidth{0.803000pt}%
\definecolor{currentstroke}{rgb}{0.000000,0.000000,0.000000}%
\pgfsetstrokecolor{currentstroke}%
\pgfsetdash{}{0pt}%
\pgfsys@defobject{currentmarker}{\pgfqpoint{0.000000in}{-0.048611in}}{\pgfqpoint{0.000000in}{0.000000in}}{%
\pgfpathmoveto{\pgfqpoint{0.000000in}{0.000000in}}%
\pgfpathlineto{\pgfqpoint{0.000000in}{-0.048611in}}%
\pgfusepath{stroke,fill}%
}%
\begin{pgfscope}%
\pgfsys@transformshift{3.546572in}{1.061180in}%
\pgfsys@useobject{currentmarker}{}%
\end{pgfscope}%
\end{pgfscope}%
\begin{pgfscope}%
\definecolor{textcolor}{rgb}{0.000000,0.000000,0.000000}%
\pgfsetstrokecolor{textcolor}%
\pgfsetfillcolor{textcolor}%
\pgftext[x=3.546572in,y=0.963958in,,top]{\color{textcolor}\rmfamily\fontsize{30.000000}{36.000000}\selectfont 1000}%
\end{pgfscope}%
\begin{pgfscope}%
\pgfsetbuttcap%
\pgfsetroundjoin%
\definecolor{currentfill}{rgb}{0.000000,0.000000,0.000000}%
\pgfsetfillcolor{currentfill}%
\pgfsetlinewidth{0.803000pt}%
\definecolor{currentstroke}{rgb}{0.000000,0.000000,0.000000}%
\pgfsetstrokecolor{currentstroke}%
\pgfsetdash{}{0pt}%
\pgfsys@defobject{currentmarker}{\pgfqpoint{0.000000in}{-0.048611in}}{\pgfqpoint{0.000000in}{0.000000in}}{%
\pgfpathmoveto{\pgfqpoint{0.000000in}{0.000000in}}%
\pgfpathlineto{\pgfqpoint{0.000000in}{-0.048611in}}%
\pgfusepath{stroke,fill}%
}%
\begin{pgfscope}%
\pgfsys@transformshift{5.685060in}{1.061180in}%
\pgfsys@useobject{currentmarker}{}%
\end{pgfscope}%
\end{pgfscope}%
\begin{pgfscope}%
\definecolor{textcolor}{rgb}{0.000000,0.000000,0.000000}%
\pgfsetstrokecolor{textcolor}%
\pgfsetfillcolor{textcolor}%
\pgftext[x=5.685060in,y=0.963958in,,top]{\color{textcolor}\rmfamily\fontsize{30.000000}{36.000000}\selectfont 2000}%
\end{pgfscope}%
\begin{pgfscope}%
\pgfsetbuttcap%
\pgfsetroundjoin%
\definecolor{currentfill}{rgb}{0.000000,0.000000,0.000000}%
\pgfsetfillcolor{currentfill}%
\pgfsetlinewidth{0.803000pt}%
\definecolor{currentstroke}{rgb}{0.000000,0.000000,0.000000}%
\pgfsetstrokecolor{currentstroke}%
\pgfsetdash{}{0pt}%
\pgfsys@defobject{currentmarker}{\pgfqpoint{0.000000in}{-0.048611in}}{\pgfqpoint{0.000000in}{0.000000in}}{%
\pgfpathmoveto{\pgfqpoint{0.000000in}{0.000000in}}%
\pgfpathlineto{\pgfqpoint{0.000000in}{-0.048611in}}%
\pgfusepath{stroke,fill}%
}%
\begin{pgfscope}%
\pgfsys@transformshift{7.823548in}{1.061180in}%
\pgfsys@useobject{currentmarker}{}%
\end{pgfscope}%
\end{pgfscope}%
\begin{pgfscope}%
\definecolor{textcolor}{rgb}{0.000000,0.000000,0.000000}%
\pgfsetstrokecolor{textcolor}%
\pgfsetfillcolor{textcolor}%
\pgftext[x=7.823548in,y=0.963958in,,top]{\color{textcolor}\rmfamily\fontsize{30.000000}{36.000000}\selectfont 3000}%
\end{pgfscope}%
\begin{pgfscope}%
\pgfsetbuttcap%
\pgfsetroundjoin%
\definecolor{currentfill}{rgb}{0.000000,0.000000,0.000000}%
\pgfsetfillcolor{currentfill}%
\pgfsetlinewidth{0.803000pt}%
\definecolor{currentstroke}{rgb}{0.000000,0.000000,0.000000}%
\pgfsetstrokecolor{currentstroke}%
\pgfsetdash{}{0pt}%
\pgfsys@defobject{currentmarker}{\pgfqpoint{0.000000in}{-0.048611in}}{\pgfqpoint{0.000000in}{0.000000in}}{%
\pgfpathmoveto{\pgfqpoint{0.000000in}{0.000000in}}%
\pgfpathlineto{\pgfqpoint{0.000000in}{-0.048611in}}%
\pgfusepath{stroke,fill}%
}%
\begin{pgfscope}%
\pgfsys@transformshift{9.962036in}{1.061180in}%
\pgfsys@useobject{currentmarker}{}%
\end{pgfscope}%
\end{pgfscope}%
\begin{pgfscope}%
\definecolor{textcolor}{rgb}{0.000000,0.000000,0.000000}%
\pgfsetstrokecolor{textcolor}%
\pgfsetfillcolor{textcolor}%
\pgftext[x=9.962036in,y=0.963958in,,top]{\color{textcolor}\rmfamily\fontsize{30.000000}{36.000000}\selectfont 4000}%
\end{pgfscope}%
\begin{pgfscope}%
\definecolor{textcolor}{rgb}{0.000000,0.000000,0.000000}%
\pgfsetstrokecolor{textcolor}%
\pgfsetfillcolor{textcolor}%
\pgftext[x=6.727573in,y=0.517886in,,top]{\color{textcolor}\rmfamily\fontsize{30.000000}{36.000000}\selectfont range [m]}%
\end{pgfscope}%
\begin{pgfscope}%
\pgfsetbuttcap%
\pgfsetroundjoin%
\definecolor{currentfill}{rgb}{0.000000,0.000000,0.000000}%
\pgfsetfillcolor{currentfill}%
\pgfsetlinewidth{0.803000pt}%
\definecolor{currentstroke}{rgb}{0.000000,0.000000,0.000000}%
\pgfsetstrokecolor{currentstroke}%
\pgfsetdash{}{0pt}%
\pgfsys@defobject{currentmarker}{\pgfqpoint{-0.048611in}{0.000000in}}{\pgfqpoint{-0.000000in}{0.000000in}}{%
\pgfpathmoveto{\pgfqpoint{-0.000000in}{0.000000in}}%
\pgfpathlineto{\pgfqpoint{-0.048611in}{0.000000in}}%
\pgfusepath{stroke,fill}%
}%
\begin{pgfscope}%
\pgfsys@transformshift{1.381353in}{6.412716in}%
\pgfsys@useobject{currentmarker}{}%
\end{pgfscope}%
\end{pgfscope}%
\begin{pgfscope}%
\definecolor{textcolor}{rgb}{0.000000,0.000000,0.000000}%
\pgfsetstrokecolor{textcolor}%
\pgfsetfillcolor{textcolor}%
\pgftext[x=1.007597in, y=6.292731in, left, base]{\color{textcolor}\rmfamily\fontsize{30.000000}{36.000000}\selectfont 0}%
\end{pgfscope}%
\begin{pgfscope}%
\pgfsetbuttcap%
\pgfsetroundjoin%
\definecolor{currentfill}{rgb}{0.000000,0.000000,0.000000}%
\pgfsetfillcolor{currentfill}%
\pgfsetlinewidth{0.803000pt}%
\definecolor{currentstroke}{rgb}{0.000000,0.000000,0.000000}%
\pgfsetstrokecolor{currentstroke}%
\pgfsetdash{}{0pt}%
\pgfsys@defobject{currentmarker}{\pgfqpoint{-0.048611in}{0.000000in}}{\pgfqpoint{-0.000000in}{0.000000in}}{%
\pgfpathmoveto{\pgfqpoint{-0.000000in}{0.000000in}}%
\pgfpathlineto{\pgfqpoint{-0.048611in}{0.000000in}}%
\pgfusepath{stroke,fill}%
}%
\begin{pgfscope}%
\pgfsys@transformshift{1.381353in}{5.068109in}%
\pgfsys@useobject{currentmarker}{}%
\end{pgfscope}%
\end{pgfscope}%
\begin{pgfscope}%
\definecolor{textcolor}{rgb}{0.000000,0.000000,0.000000}%
\pgfsetstrokecolor{textcolor}%
\pgfsetfillcolor{textcolor}%
\pgftext[x=0.849120in, y=4.948124in, left, base]{\color{textcolor}\rmfamily\fontsize{30.000000}{36.000000}\selectfont 50}%
\end{pgfscope}%
\begin{pgfscope}%
\pgfsetbuttcap%
\pgfsetroundjoin%
\definecolor{currentfill}{rgb}{0.000000,0.000000,0.000000}%
\pgfsetfillcolor{currentfill}%
\pgfsetlinewidth{0.803000pt}%
\definecolor{currentstroke}{rgb}{0.000000,0.000000,0.000000}%
\pgfsetstrokecolor{currentstroke}%
\pgfsetdash{}{0pt}%
\pgfsys@defobject{currentmarker}{\pgfqpoint{-0.048611in}{0.000000in}}{\pgfqpoint{-0.000000in}{0.000000in}}{%
\pgfpathmoveto{\pgfqpoint{-0.000000in}{0.000000in}}%
\pgfpathlineto{\pgfqpoint{-0.048611in}{0.000000in}}%
\pgfusepath{stroke,fill}%
}%
\begin{pgfscope}%
\pgfsys@transformshift{1.381353in}{3.723502in}%
\pgfsys@useobject{currentmarker}{}%
\end{pgfscope}%
\end{pgfscope}%
\begin{pgfscope}%
\definecolor{textcolor}{rgb}{0.000000,0.000000,0.000000}%
\pgfsetstrokecolor{textcolor}%
\pgfsetfillcolor{textcolor}%
\pgftext[x=0.690641in, y=3.603517in, left, base]{\color{textcolor}\rmfamily\fontsize{30.000000}{36.000000}\selectfont 100}%
\end{pgfscope}%
\begin{pgfscope}%
\pgfsetbuttcap%
\pgfsetroundjoin%
\definecolor{currentfill}{rgb}{0.000000,0.000000,0.000000}%
\pgfsetfillcolor{currentfill}%
\pgfsetlinewidth{0.803000pt}%
\definecolor{currentstroke}{rgb}{0.000000,0.000000,0.000000}%
\pgfsetstrokecolor{currentstroke}%
\pgfsetdash{}{0pt}%
\pgfsys@defobject{currentmarker}{\pgfqpoint{-0.048611in}{0.000000in}}{\pgfqpoint{-0.000000in}{0.000000in}}{%
\pgfpathmoveto{\pgfqpoint{-0.000000in}{0.000000in}}%
\pgfpathlineto{\pgfqpoint{-0.048611in}{0.000000in}}%
\pgfusepath{stroke,fill}%
}%
\begin{pgfscope}%
\pgfsys@transformshift{1.381353in}{2.378895in}%
\pgfsys@useobject{currentmarker}{}%
\end{pgfscope}%
\end{pgfscope}%
\begin{pgfscope}%
\definecolor{textcolor}{rgb}{0.000000,0.000000,0.000000}%
\pgfsetstrokecolor{textcolor}%
\pgfsetfillcolor{textcolor}%
\pgftext[x=0.690641in, y=2.258910in, left, base]{\color{textcolor}\rmfamily\fontsize{30.000000}{36.000000}\selectfont 150}%
\end{pgfscope}%
\begin{pgfscope}%
\definecolor{textcolor}{rgb}{0.000000,0.000000,0.000000}%
\pgfsetstrokecolor{textcolor}%
\pgfsetfillcolor{textcolor}%
\pgftext[x=0.551753in,y=3.750394in,,bottom,rotate=90.000000]{\color{textcolor}\rmfamily\fontsize{30.000000}{36.000000}\selectfont depth [m]}%
\end{pgfscope}%
\begin{pgfscope}%
\pgfsetrectcap%
\pgfsetmiterjoin%
\pgfsetlinewidth{0.803000pt}%
\definecolor{currentstroke}{rgb}{0.000000,0.000000,0.000000}%
\pgfsetstrokecolor{currentstroke}%
\pgfsetdash{}{0pt}%
\pgfpathmoveto{\pgfqpoint{1.381353in}{1.061180in}}%
\pgfpathlineto{\pgfqpoint{1.381353in}{6.439608in}}%
\pgfusepath{stroke}%
\end{pgfscope}%
\begin{pgfscope}%
\pgfsetrectcap%
\pgfsetmiterjoin%
\pgfsetlinewidth{0.803000pt}%
\definecolor{currentstroke}{rgb}{0.000000,0.000000,0.000000}%
\pgfsetstrokecolor{currentstroke}%
\pgfsetdash{}{0pt}%
\pgfpathmoveto{\pgfqpoint{12.073793in}{1.061180in}}%
\pgfpathlineto{\pgfqpoint{12.073793in}{6.439608in}}%
\pgfusepath{stroke}%
\end{pgfscope}%
\begin{pgfscope}%
\pgfsetrectcap%
\pgfsetmiterjoin%
\pgfsetlinewidth{0.803000pt}%
\definecolor{currentstroke}{rgb}{0.000000,0.000000,0.000000}%
\pgfsetstrokecolor{currentstroke}%
\pgfsetdash{}{0pt}%
\pgfpathmoveto{\pgfqpoint{1.381353in}{1.061180in}}%
\pgfpathlineto{\pgfqpoint{12.073793in}{1.061180in}}%
\pgfusepath{stroke}%
\end{pgfscope}%
\begin{pgfscope}%
\pgfsetrectcap%
\pgfsetmiterjoin%
\pgfsetlinewidth{0.803000pt}%
\definecolor{currentstroke}{rgb}{0.000000,0.000000,0.000000}%
\pgfsetstrokecolor{currentstroke}%
\pgfsetdash{}{0pt}%
\pgfpathmoveto{\pgfqpoint{1.381353in}{6.439608in}}%
\pgfpathlineto{\pgfqpoint{12.073793in}{6.439608in}}%
\pgfusepath{stroke}%
\end{pgfscope}%
\begin{pgfscope}%
\pgfsetbuttcap%
\pgfsetmiterjoin%
\definecolor{currentfill}{rgb}{1.000000,1.000000,1.000000}%
\pgfsetfillcolor{currentfill}%
\pgfsetfillopacity{0.800000}%
\pgfsetlinewidth{1.003750pt}%
\definecolor{currentstroke}{rgb}{0.800000,0.800000,0.800000}%
\pgfsetstrokecolor{currentstroke}%
\pgfsetstrokeopacity{0.800000}%
\pgfsetdash{}{0pt}%
\pgfpathmoveto{\pgfqpoint{7.864735in}{1.269513in}}%
\pgfpathlineto{\pgfqpoint{11.782126in}{1.269513in}}%
\pgfpathquadraticcurveto{\pgfqpoint{11.865459in}{1.269513in}}{\pgfqpoint{11.865459in}{1.352847in}}%
\pgfpathlineto{\pgfqpoint{11.865459in}{1.878393in}}%
\pgfpathquadraticcurveto{\pgfqpoint{11.865459in}{1.961727in}}{\pgfqpoint{11.782126in}{1.961727in}}%
\pgfpathlineto{\pgfqpoint{7.864735in}{1.961727in}}%
\pgfpathquadraticcurveto{\pgfqpoint{7.781402in}{1.961727in}}{\pgfqpoint{7.781402in}{1.878393in}}%
\pgfpathlineto{\pgfqpoint{7.781402in}{1.352847in}}%
\pgfpathquadraticcurveto{\pgfqpoint{7.781402in}{1.269513in}}{\pgfqpoint{7.864735in}{1.269513in}}%
\pgfpathlineto{\pgfqpoint{7.864735in}{1.269513in}}%
\pgfpathclose%
\pgfusepath{stroke,fill}%
\end{pgfscope}%
\begin{pgfscope}%
\pgfsetbuttcap%
\pgfsetroundjoin%
\pgfsetlinewidth{3.011250pt}%
\definecolor{currentstroke}{rgb}{1.000000,0.000000,0.000000}%
\pgfsetstrokecolor{currentstroke}%
\pgfsetdash{}{0pt}%
\pgfpathmoveto{\pgfqpoint{8.364735in}{1.514559in}}%
\pgfpathcurveto{\pgfqpoint{8.390780in}{1.514559in}}{\pgfqpoint{8.415763in}{1.524907in}}{\pgfqpoint{8.434180in}{1.543324in}}%
\pgfpathcurveto{\pgfqpoint{8.452596in}{1.561741in}}{\pgfqpoint{8.462944in}{1.586723in}}{\pgfqpoint{8.462944in}{1.612768in}}%
\pgfpathcurveto{\pgfqpoint{8.462944in}{1.638814in}}{\pgfqpoint{8.452596in}{1.663796in}}{\pgfqpoint{8.434180in}{1.682213in}}%
\pgfpathcurveto{\pgfqpoint{8.415763in}{1.700630in}}{\pgfqpoint{8.390780in}{1.710978in}}{\pgfqpoint{8.364735in}{1.710978in}}%
\pgfpathcurveto{\pgfqpoint{8.338690in}{1.710978in}}{\pgfqpoint{8.313708in}{1.700630in}}{\pgfqpoint{8.295291in}{1.682213in}}%
\pgfpathcurveto{\pgfqpoint{8.276874in}{1.663796in}}{\pgfqpoint{8.266526in}{1.638814in}}{\pgfqpoint{8.266526in}{1.612768in}}%
\pgfpathcurveto{\pgfqpoint{8.266526in}{1.586723in}}{\pgfqpoint{8.276874in}{1.561741in}}{\pgfqpoint{8.295291in}{1.543324in}}%
\pgfpathcurveto{\pgfqpoint{8.313708in}{1.524907in}}{\pgfqpoint{8.338690in}{1.514559in}}{\pgfqpoint{8.364735in}{1.514559in}}%
\pgfpathlineto{\pgfqpoint{8.364735in}{1.514559in}}%
\pgfpathclose%
\pgfusepath{stroke}%
\end{pgfscope}%
\begin{pgfscope}%
\definecolor{textcolor}{rgb}{0.000000,0.000000,0.000000}%
\pgfsetstrokecolor{textcolor}%
\pgfsetfillcolor{textcolor}%
\pgftext[x=9.114735in,y=1.503393in,left,base]{\color{textcolor}\rmfamily\fontsize{30.000000}{36.000000}\selectfont Object Position~~~~}%
\end{pgfscope}%
\begin{pgfscope}%
\pgfsetbuttcap%
\pgfsetmiterjoin%
\definecolor{currentfill}{rgb}{1.000000,1.000000,1.000000}%
\pgfsetfillcolor{currentfill}%
\pgfsetlinewidth{0.000000pt}%
\definecolor{currentstroke}{rgb}{0.000000,0.000000,0.000000}%
\pgfsetstrokecolor{currentstroke}%
\pgfsetstrokeopacity{0.000000}%
\pgfsetdash{}{0pt}%
\pgfpathmoveto{\pgfqpoint{12.541733in}{1.061180in}}%
\pgfpathlineto{\pgfqpoint{12.810655in}{1.061180in}}%
\pgfpathlineto{\pgfqpoint{12.810655in}{6.439608in}}%
\pgfpathlineto{\pgfqpoint{12.541733in}{6.439608in}}%
\pgfpathlineto{\pgfqpoint{12.541733in}{1.061180in}}%
\pgfpathclose%
\pgfusepath{fill}%
\end{pgfscope}%
\begin{pgfscope}%
\pgfpathrectangle{\pgfqpoint{12.541733in}{1.061180in}}{\pgfqpoint{0.268921in}{5.378428in}}%
\pgfusepath{clip}%
\pgfsetbuttcap%
\pgfsetmiterjoin%
\definecolor{currentfill}{rgb}{1.000000,1.000000,1.000000}%
\pgfsetfillcolor{currentfill}%
\pgfsetlinewidth{0.010037pt}%
\definecolor{currentstroke}{rgb}{1.000000,1.000000,1.000000}%
\pgfsetstrokecolor{currentstroke}%
\pgfsetdash{}{0pt}%
\pgfusepath{stroke,fill}%
\end{pgfscope}%
\begin{pgfscope}%
\pgfsys@transformshift{12.540000in}{1.060000in}%
\pgftext[left,bottom]{\includegraphics[interpolate=true,width=0.270000in,height=5.380000in]{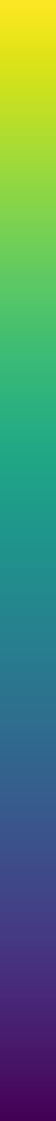}}%
\end{pgfscope}%
\begin{pgfscope}%
\pgfsetbuttcap%
\pgfsetroundjoin%
\definecolor{currentfill}{rgb}{0.000000,0.000000,0.000000}%
\pgfsetfillcolor{currentfill}%
\pgfsetlinewidth{0.803000pt}%
\definecolor{currentstroke}{rgb}{0.000000,0.000000,0.000000}%
\pgfsetstrokecolor{currentstroke}%
\pgfsetdash{}{0pt}%
\pgfsys@defobject{currentmarker}{\pgfqpoint{0.000000in}{0.000000in}}{\pgfqpoint{0.048611in}{0.000000in}}{%
\pgfpathmoveto{\pgfqpoint{0.000000in}{0.000000in}}%
\pgfpathlineto{\pgfqpoint{0.048611in}{0.000000in}}%
\pgfusepath{stroke,fill}%
}%
\begin{pgfscope}%
\pgfsys@transformshift{12.810655in}{1.419742in}%
\pgfsys@useobject{currentmarker}{}%
\end{pgfscope}%
\end{pgfscope}%
\begin{pgfscope}%
\definecolor{textcolor}{rgb}{0.000000,0.000000,0.000000}%
\pgfsetstrokecolor{textcolor}%
\pgfsetfillcolor{textcolor}%
\pgftext[x=12.907877in, y=1.299757in, left, base]{\color{textcolor}\rmfamily\fontsize{24.000000}{28.800000}\selectfont -14}%
\end{pgfscope}%
\begin{pgfscope}%
\pgfsetbuttcap%
\pgfsetroundjoin%
\definecolor{currentfill}{rgb}{0.000000,0.000000,0.000000}%
\pgfsetfillcolor{currentfill}%
\pgfsetlinewidth{0.803000pt}%
\definecolor{currentstroke}{rgb}{0.000000,0.000000,0.000000}%
\pgfsetstrokecolor{currentstroke}%
\pgfsetdash{}{0pt}%
\pgfsys@defobject{currentmarker}{\pgfqpoint{0.000000in}{0.000000in}}{\pgfqpoint{0.048611in}{0.000000in}}{%
\pgfpathmoveto{\pgfqpoint{0.000000in}{0.000000in}}%
\pgfpathlineto{\pgfqpoint{0.048611in}{0.000000in}}%
\pgfusepath{stroke,fill}%
}%
\begin{pgfscope}%
\pgfsys@transformshift{12.810655in}{2.136865in}%
\pgfsys@useobject{currentmarker}{}%
\end{pgfscope}%
\end{pgfscope}%
\begin{pgfscope}%
\definecolor{textcolor}{rgb}{0.000000,0.000000,0.000000}%
\pgfsetstrokecolor{textcolor}%
\pgfsetfillcolor{textcolor}%
\pgftext[x=12.907877in, y=2.016881in, left, base]{\color{textcolor}\rmfamily\fontsize{24.000000}{28.800000}\selectfont -12}%
\end{pgfscope}%
\begin{pgfscope}%
\pgfsetbuttcap%
\pgfsetroundjoin%
\definecolor{currentfill}{rgb}{0.000000,0.000000,0.000000}%
\pgfsetfillcolor{currentfill}%
\pgfsetlinewidth{0.803000pt}%
\definecolor{currentstroke}{rgb}{0.000000,0.000000,0.000000}%
\pgfsetstrokecolor{currentstroke}%
\pgfsetdash{}{0pt}%
\pgfsys@defobject{currentmarker}{\pgfqpoint{0.000000in}{0.000000in}}{\pgfqpoint{0.048611in}{0.000000in}}{%
\pgfpathmoveto{\pgfqpoint{0.000000in}{0.000000in}}%
\pgfpathlineto{\pgfqpoint{0.048611in}{0.000000in}}%
\pgfusepath{stroke,fill}%
}%
\begin{pgfscope}%
\pgfsys@transformshift{12.810655in}{2.853989in}%
\pgfsys@useobject{currentmarker}{}%
\end{pgfscope}%
\end{pgfscope}%
\begin{pgfscope}%
\definecolor{textcolor}{rgb}{0.000000,0.000000,0.000000}%
\pgfsetstrokecolor{textcolor}%
\pgfsetfillcolor{textcolor}%
\pgftext[x=12.907877in, y=2.734005in, left, base]{\color{textcolor}\rmfamily\fontsize{24.000000}{28.800000}\selectfont -10}%
\end{pgfscope}%
\begin{pgfscope}%
\pgfsetbuttcap%
\pgfsetroundjoin%
\definecolor{currentfill}{rgb}{0.000000,0.000000,0.000000}%
\pgfsetfillcolor{currentfill}%
\pgfsetlinewidth{0.803000pt}%
\definecolor{currentstroke}{rgb}{0.000000,0.000000,0.000000}%
\pgfsetstrokecolor{currentstroke}%
\pgfsetdash{}{0pt}%
\pgfsys@defobject{currentmarker}{\pgfqpoint{0.000000in}{0.000000in}}{\pgfqpoint{0.048611in}{0.000000in}}{%
\pgfpathmoveto{\pgfqpoint{0.000000in}{0.000000in}}%
\pgfpathlineto{\pgfqpoint{0.048611in}{0.000000in}}%
\pgfusepath{stroke,fill}%
}%
\begin{pgfscope}%
\pgfsys@transformshift{12.810655in}{3.571113in}%
\pgfsys@useobject{currentmarker}{}%
\end{pgfscope}%
\end{pgfscope}%
\begin{pgfscope}%
\definecolor{textcolor}{rgb}{0.000000,0.000000,0.000000}%
\pgfsetstrokecolor{textcolor}%
\pgfsetfillcolor{textcolor}%
\pgftext[x=12.907877in, y=3.451128in, left, base]{\color{textcolor}\rmfamily\fontsize{24.000000}{28.800000}\selectfont -8}%
\end{pgfscope}%
\begin{pgfscope}%
\pgfsetbuttcap%
\pgfsetroundjoin%
\definecolor{currentfill}{rgb}{0.000000,0.000000,0.000000}%
\pgfsetfillcolor{currentfill}%
\pgfsetlinewidth{0.803000pt}%
\definecolor{currentstroke}{rgb}{0.000000,0.000000,0.000000}%
\pgfsetstrokecolor{currentstroke}%
\pgfsetdash{}{0pt}%
\pgfsys@defobject{currentmarker}{\pgfqpoint{0.000000in}{0.000000in}}{\pgfqpoint{0.048611in}{0.000000in}}{%
\pgfpathmoveto{\pgfqpoint{0.000000in}{0.000000in}}%
\pgfpathlineto{\pgfqpoint{0.048611in}{0.000000in}}%
\pgfusepath{stroke,fill}%
}%
\begin{pgfscope}%
\pgfsys@transformshift{12.810655in}{4.288237in}%
\pgfsys@useobject{currentmarker}{}%
\end{pgfscope}%
\end{pgfscope}%
\begin{pgfscope}%
\definecolor{textcolor}{rgb}{0.000000,0.000000,0.000000}%
\pgfsetstrokecolor{textcolor}%
\pgfsetfillcolor{textcolor}%
\pgftext[x=12.907877in, y=4.168252in, left, base]{\color{textcolor}\rmfamily\fontsize{24.000000}{28.800000}\selectfont -6}%
\end{pgfscope}%
\begin{pgfscope}%
\pgfsetbuttcap%
\pgfsetroundjoin%
\definecolor{currentfill}{rgb}{0.000000,0.000000,0.000000}%
\pgfsetfillcolor{currentfill}%
\pgfsetlinewidth{0.803000pt}%
\definecolor{currentstroke}{rgb}{0.000000,0.000000,0.000000}%
\pgfsetstrokecolor{currentstroke}%
\pgfsetdash{}{0pt}%
\pgfsys@defobject{currentmarker}{\pgfqpoint{0.000000in}{0.000000in}}{\pgfqpoint{0.048611in}{0.000000in}}{%
\pgfpathmoveto{\pgfqpoint{0.000000in}{0.000000in}}%
\pgfpathlineto{\pgfqpoint{0.048611in}{0.000000in}}%
\pgfusepath{stroke,fill}%
}%
\begin{pgfscope}%
\pgfsys@transformshift{12.810655in}{5.005360in}%
\pgfsys@useobject{currentmarker}{}%
\end{pgfscope}%
\end{pgfscope}%
\begin{pgfscope}%
\definecolor{textcolor}{rgb}{0.000000,0.000000,0.000000}%
\pgfsetstrokecolor{textcolor}%
\pgfsetfillcolor{textcolor}%
\pgftext[x=12.907877in, y=4.885376in, left, base]{\color{textcolor}\rmfamily\fontsize{24.000000}{28.800000}\selectfont -4}%
\end{pgfscope}%
\begin{pgfscope}%
\pgfsetbuttcap%
\pgfsetroundjoin%
\definecolor{currentfill}{rgb}{0.000000,0.000000,0.000000}%
\pgfsetfillcolor{currentfill}%
\pgfsetlinewidth{0.803000pt}%
\definecolor{currentstroke}{rgb}{0.000000,0.000000,0.000000}%
\pgfsetstrokecolor{currentstroke}%
\pgfsetdash{}{0pt}%
\pgfsys@defobject{currentmarker}{\pgfqpoint{0.000000in}{0.000000in}}{\pgfqpoint{0.048611in}{0.000000in}}{%
\pgfpathmoveto{\pgfqpoint{0.000000in}{0.000000in}}%
\pgfpathlineto{\pgfqpoint{0.048611in}{0.000000in}}%
\pgfusepath{stroke,fill}%
}%
\begin{pgfscope}%
\pgfsys@transformshift{12.810655in}{5.722484in}%
\pgfsys@useobject{currentmarker}{}%
\end{pgfscope}%
\end{pgfscope}%
\begin{pgfscope}%
\definecolor{textcolor}{rgb}{0.000000,0.000000,0.000000}%
\pgfsetstrokecolor{textcolor}%
\pgfsetfillcolor{textcolor}%
\pgftext[x=12.907877in, y=5.602500in, left, base]{\color{textcolor}\rmfamily\fontsize{24.000000}{28.800000}\selectfont -2}%
\end{pgfscope}%
\begin{pgfscope}%
\pgfsetbuttcap%
\pgfsetroundjoin%
\definecolor{currentfill}{rgb}{0.000000,0.000000,0.000000}%
\pgfsetfillcolor{currentfill}%
\pgfsetlinewidth{0.803000pt}%
\definecolor{currentstroke}{rgb}{0.000000,0.000000,0.000000}%
\pgfsetstrokecolor{currentstroke}%
\pgfsetdash{}{0pt}%
\pgfsys@defobject{currentmarker}{\pgfqpoint{0.000000in}{0.000000in}}{\pgfqpoint{0.048611in}{0.000000in}}{%
\pgfpathmoveto{\pgfqpoint{0.000000in}{0.000000in}}%
\pgfpathlineto{\pgfqpoint{0.048611in}{0.000000in}}%
\pgfusepath{stroke,fill}%
}%
\begin{pgfscope}%
\pgfsys@transformshift{12.810655in}{6.439608in}%
\pgfsys@useobject{currentmarker}{}%
\end{pgfscope}%
\end{pgfscope}%
\begin{pgfscope}%
\definecolor{textcolor}{rgb}{0.000000,0.000000,0.000000}%
\pgfsetstrokecolor{textcolor}%
\pgfsetfillcolor{textcolor}%
\pgftext[x=12.907877in, y=6.319623in, left, base]{\color{textcolor}\rmfamily\fontsize{24.000000}{28.800000}\selectfont 0}%
\end{pgfscope}%
\begin{pgfscope}%
\pgfsetrectcap%
\pgfsetmiterjoin%
\pgfsetlinewidth{0.803000pt}%
\definecolor{currentstroke}{rgb}{0.000000,0.000000,0.000000}%
\pgfsetstrokecolor{currentstroke}%
\pgfsetdash{}{0pt}%
\pgfpathmoveto{\pgfqpoint{12.541733in}{1.061180in}}%
\pgfpathlineto{\pgfqpoint{12.676194in}{1.061180in}}%
\pgfpathlineto{\pgfqpoint{12.810655in}{1.061180in}}%
\pgfpathlineto{\pgfqpoint{12.810655in}{6.439608in}}%
\pgfpathlineto{\pgfqpoint{12.676194in}{6.439608in}}%
\pgfpathlineto{\pgfqpoint{12.541733in}{6.439608in}}%
\pgfpathlineto{\pgfqpoint{12.541733in}{1.061180in}}%
\pgfpathclose%
\pgfusepath{stroke}%
\end{pgfscope}%
\end{pgfpicture}%
\makeatother%
\endgroup%

%% file: Figs/rawsignal/results_k001_eps.pgf
\begingroup%
\makeatletter%
\begin{pgfpicture}%
\pgfpathrectangle{\pgfpointorigin}{\pgfqpoint{13.400000in}{6.600000in}}%
\pgfusepath{use as bounding box, clip}%
\begin{pgfscope}%
\pgfsetbuttcap%
\pgfsetmiterjoin%
\definecolor{currentfill}{rgb}{1.000000,1.000000,1.000000}%
\pgfsetfillcolor{currentfill}%
\pgfsetlinewidth{0.000000pt}%
\definecolor{currentstroke}{rgb}{1.000000,1.000000,1.000000}%
\pgfsetstrokecolor{currentstroke}%
\pgfsetdash{}{0pt}%
\pgfpathmoveto{\pgfqpoint{0.000000in}{0.000000in}}%
\pgfpathlineto{\pgfqpoint{13.400000in}{0.000000in}}%
\pgfpathlineto{\pgfqpoint{13.400000in}{6.600000in}}%
\pgfpathlineto{\pgfqpoint{0.000000in}{6.600000in}}%
\pgfpathlineto{\pgfqpoint{0.000000in}{0.000000in}}%
\pgfpathclose%
\pgfusepath{fill}%
\end{pgfscope}%
\begin{pgfscope}%
\pgfsetbuttcap%
\pgfsetmiterjoin%
\definecolor{currentfill}{rgb}{1.000000,1.000000,1.000000}%
\pgfsetfillcolor{currentfill}%
\pgfsetlinewidth{0.000000pt}%
\definecolor{currentstroke}{rgb}{0.000000,0.000000,0.000000}%
\pgfsetstrokecolor{currentstroke}%
\pgfsetstrokeopacity{0.000000}%
\pgfsetdash{}{0pt}%
\pgfpathmoveto{\pgfqpoint{1.381353in}{1.061180in}}%
\pgfpathlineto{\pgfqpoint{12.073793in}{1.061180in}}%
\pgfpathlineto{\pgfqpoint{12.073793in}{6.443025in}}%
\pgfpathlineto{\pgfqpoint{1.381353in}{6.443025in}}%
\pgfpathlineto{\pgfqpoint{1.381353in}{1.061180in}}%
\pgfpathclose%
\pgfusepath{fill}%
\end{pgfscope}%
\begin{pgfscope}%
\pgfpathrectangle{\pgfqpoint{1.381353in}{1.061180in}}{\pgfqpoint{10.692439in}{5.381845in}}%
\pgfusepath{clip}%
\pgfsys@transformshift{1.381353in}{1.061180in}%
\pgftext[left,bottom]{\includegraphics[interpolate=true,width=10.700000in,height=5.390000in]{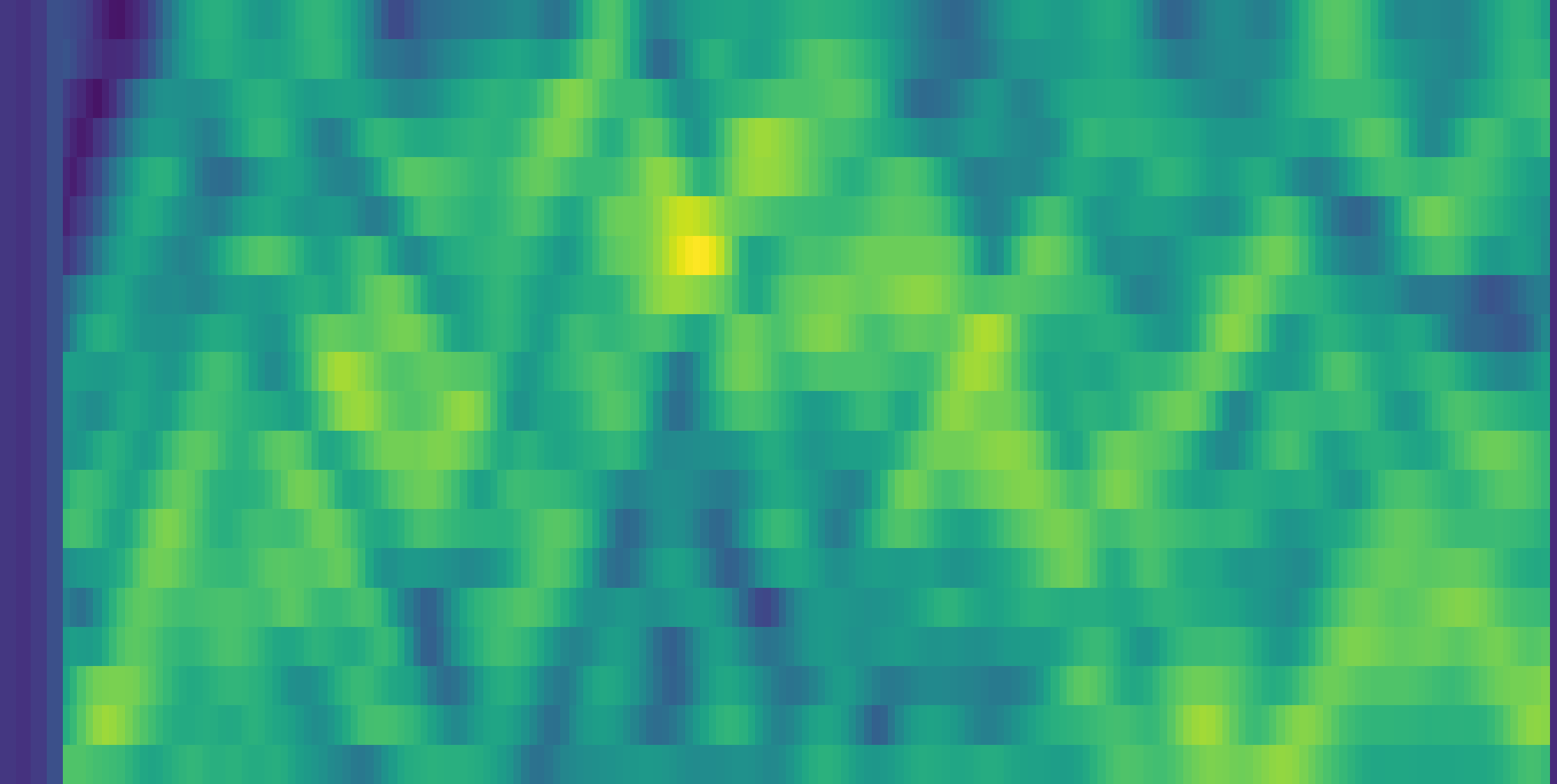}}%
\end{pgfscope}%
\begin{pgfscope}%
\pgfpathrectangle{\pgfqpoint{1.381353in}{1.061180in}}{\pgfqpoint{10.692439in}{5.381845in}}%
\pgfusepath{clip}%
\pgfsetbuttcap%
\pgfsetroundjoin%
\pgfsetlinewidth{3.011250pt}%
\definecolor{currentstroke}{rgb}{1.000000,0.000000,0.000000}%
\pgfsetstrokecolor{currentstroke}%
\pgfsetdash{}{0pt}%
\pgfpathmoveto{\pgfqpoint{6.440743in}{4.408376in}}%
\pgfpathcurveto{\pgfqpoint{6.466788in}{4.408376in}}{\pgfqpoint{6.491770in}{4.418724in}}{\pgfqpoint{6.510187in}{4.437141in}}%
\pgfpathcurveto{\pgfqpoint{6.528604in}{4.455558in}}{\pgfqpoint{6.538952in}{4.480540in}}{\pgfqpoint{6.538952in}{4.506585in}}%
\pgfpathcurveto{\pgfqpoint{6.538952in}{4.532631in}}{\pgfqpoint{6.528604in}{4.557613in}}{\pgfqpoint{6.510187in}{4.576030in}}%
\pgfpathcurveto{\pgfqpoint{6.491770in}{4.594447in}}{\pgfqpoint{6.466788in}{4.604794in}}{\pgfqpoint{6.440743in}{4.604794in}}%
\pgfpathcurveto{\pgfqpoint{6.414697in}{4.604794in}}{\pgfqpoint{6.389715in}{4.594447in}}{\pgfqpoint{6.371298in}{4.576030in}}%
\pgfpathcurveto{\pgfqpoint{6.352882in}{4.557613in}}{\pgfqpoint{6.342534in}{4.532631in}}{\pgfqpoint{6.342534in}{4.506585in}}%
\pgfpathcurveto{\pgfqpoint{6.342534in}{4.480540in}}{\pgfqpoint{6.352882in}{4.455558in}}{\pgfqpoint{6.371298in}{4.437141in}}%
\pgfpathcurveto{\pgfqpoint{6.389715in}{4.418724in}}{\pgfqpoint{6.414697in}{4.408376in}}{\pgfqpoint{6.440743in}{4.408376in}}%
\pgfpathlineto{\pgfqpoint{6.440743in}{4.408376in}}%
\pgfpathclose%
\pgfusepath{stroke}%
\end{pgfscope}%
\begin{pgfscope}%
\pgfsetbuttcap%
\pgfsetroundjoin%
\definecolor{currentfill}{rgb}{0.000000,0.000000,0.000000}%
\pgfsetfillcolor{currentfill}%
\pgfsetlinewidth{0.803000pt}%
\definecolor{currentstroke}{rgb}{0.000000,0.000000,0.000000}%
\pgfsetstrokecolor{currentstroke}%
\pgfsetdash{}{0pt}%
\pgfsys@defobject{currentmarker}{\pgfqpoint{0.000000in}{-0.048611in}}{\pgfqpoint{0.000000in}{0.000000in}}{%
\pgfpathmoveto{\pgfqpoint{0.000000in}{0.000000in}}%
\pgfpathlineto{\pgfqpoint{0.000000in}{-0.048611in}}%
\pgfusepath{stroke,fill}%
}%
\begin{pgfscope}%
\pgfsys@transformshift{1.408084in}{1.061180in}%
\pgfsys@useobject{currentmarker}{}%
\end{pgfscope}%
\end{pgfscope}%
\begin{pgfscope}%
\definecolor{textcolor}{rgb}{0.000000,0.000000,0.000000}%
\pgfsetstrokecolor{textcolor}%
\pgfsetfillcolor{textcolor}%
\pgftext[x=1.408084in,y=0.963958in,,top]{\color{textcolor}\rmfamily\fontsize{30.000000}{36.000000}\selectfont 0}%
\end{pgfscope}%
\begin{pgfscope}%
\pgfsetbuttcap%
\pgfsetroundjoin%
\definecolor{currentfill}{rgb}{0.000000,0.000000,0.000000}%
\pgfsetfillcolor{currentfill}%
\pgfsetlinewidth{0.803000pt}%
\definecolor{currentstroke}{rgb}{0.000000,0.000000,0.000000}%
\pgfsetstrokecolor{currentstroke}%
\pgfsetdash{}{0pt}%
\pgfsys@defobject{currentmarker}{\pgfqpoint{0.000000in}{-0.048611in}}{\pgfqpoint{0.000000in}{0.000000in}}{%
\pgfpathmoveto{\pgfqpoint{0.000000in}{0.000000in}}%
\pgfpathlineto{\pgfqpoint{0.000000in}{-0.048611in}}%
\pgfusepath{stroke,fill}%
}%
\begin{pgfscope}%
\pgfsys@transformshift{3.546572in}{1.061180in}%
\pgfsys@useobject{currentmarker}{}%
\end{pgfscope}%
\end{pgfscope}%
\begin{pgfscope}%
\definecolor{textcolor}{rgb}{0.000000,0.000000,0.000000}%
\pgfsetstrokecolor{textcolor}%
\pgfsetfillcolor{textcolor}%
\pgftext[x=3.546572in,y=0.963958in,,top]{\color{textcolor}\rmfamily\fontsize{30.000000}{36.000000}\selectfont 1000}%
\end{pgfscope}%
\begin{pgfscope}%
\pgfsetbuttcap%
\pgfsetroundjoin%
\definecolor{currentfill}{rgb}{0.000000,0.000000,0.000000}%
\pgfsetfillcolor{currentfill}%
\pgfsetlinewidth{0.803000pt}%
\definecolor{currentstroke}{rgb}{0.000000,0.000000,0.000000}%
\pgfsetstrokecolor{currentstroke}%
\pgfsetdash{}{0pt}%
\pgfsys@defobject{currentmarker}{\pgfqpoint{0.000000in}{-0.048611in}}{\pgfqpoint{0.000000in}{0.000000in}}{%
\pgfpathmoveto{\pgfqpoint{0.000000in}{0.000000in}}%
\pgfpathlineto{\pgfqpoint{0.000000in}{-0.048611in}}%
\pgfusepath{stroke,fill}%
}%
\begin{pgfscope}%
\pgfsys@transformshift{5.685060in}{1.061180in}%
\pgfsys@useobject{currentmarker}{}%
\end{pgfscope}%
\end{pgfscope}%
\begin{pgfscope}%
\definecolor{textcolor}{rgb}{0.000000,0.000000,0.000000}%
\pgfsetstrokecolor{textcolor}%
\pgfsetfillcolor{textcolor}%
\pgftext[x=5.685060in,y=0.963958in,,top]{\color{textcolor}\rmfamily\fontsize{30.000000}{36.000000}\selectfont 2000}%
\end{pgfscope}%
\begin{pgfscope}%
\pgfsetbuttcap%
\pgfsetroundjoin%
\definecolor{currentfill}{rgb}{0.000000,0.000000,0.000000}%
\pgfsetfillcolor{currentfill}%
\pgfsetlinewidth{0.803000pt}%
\definecolor{currentstroke}{rgb}{0.000000,0.000000,0.000000}%
\pgfsetstrokecolor{currentstroke}%
\pgfsetdash{}{0pt}%
\pgfsys@defobject{currentmarker}{\pgfqpoint{0.000000in}{-0.048611in}}{\pgfqpoint{0.000000in}{0.000000in}}{%
\pgfpathmoveto{\pgfqpoint{0.000000in}{0.000000in}}%
\pgfpathlineto{\pgfqpoint{0.000000in}{-0.048611in}}%
\pgfusepath{stroke,fill}%
}%
\begin{pgfscope}%
\pgfsys@transformshift{7.823548in}{1.061180in}%
\pgfsys@useobject{currentmarker}{}%
\end{pgfscope}%
\end{pgfscope}%
\begin{pgfscope}%
\definecolor{textcolor}{rgb}{0.000000,0.000000,0.000000}%
\pgfsetstrokecolor{textcolor}%
\pgfsetfillcolor{textcolor}%
\pgftext[x=7.823548in,y=0.963958in,,top]{\color{textcolor}\rmfamily\fontsize{30.000000}{36.000000}\selectfont 3000}%
\end{pgfscope}%
\begin{pgfscope}%
\pgfsetbuttcap%
\pgfsetroundjoin%
\definecolor{currentfill}{rgb}{0.000000,0.000000,0.000000}%
\pgfsetfillcolor{currentfill}%
\pgfsetlinewidth{0.803000pt}%
\definecolor{currentstroke}{rgb}{0.000000,0.000000,0.000000}%
\pgfsetstrokecolor{currentstroke}%
\pgfsetdash{}{0pt}%
\pgfsys@defobject{currentmarker}{\pgfqpoint{0.000000in}{-0.048611in}}{\pgfqpoint{0.000000in}{0.000000in}}{%
\pgfpathmoveto{\pgfqpoint{0.000000in}{0.000000in}}%
\pgfpathlineto{\pgfqpoint{0.000000in}{-0.048611in}}%
\pgfusepath{stroke,fill}%
}%
\begin{pgfscope}%
\pgfsys@transformshift{9.962036in}{1.061180in}%
\pgfsys@useobject{currentmarker}{}%
\end{pgfscope}%
\end{pgfscope}%
\begin{pgfscope}%
\definecolor{textcolor}{rgb}{0.000000,0.000000,0.000000}%
\pgfsetstrokecolor{textcolor}%
\pgfsetfillcolor{textcolor}%
\pgftext[x=9.962036in,y=0.963958in,,top]{\color{textcolor}\rmfamily\fontsize{30.000000}{36.000000}\selectfont 4000}%
\end{pgfscope}%
\begin{pgfscope}%
\definecolor{textcolor}{rgb}{0.000000,0.000000,0.000000}%
\pgfsetstrokecolor{textcolor}%
\pgfsetfillcolor{textcolor}%
\pgftext[x=6.727573in,y=0.517886in,,top]{\color{textcolor}\rmfamily\fontsize{30.000000}{36.000000}\selectfont range [m]}%
\end{pgfscope}%
\begin{pgfscope}%
\pgfsetbuttcap%
\pgfsetroundjoin%
\definecolor{currentfill}{rgb}{0.000000,0.000000,0.000000}%
\pgfsetfillcolor{currentfill}%
\pgfsetlinewidth{0.803000pt}%
\definecolor{currentstroke}{rgb}{0.000000,0.000000,0.000000}%
\pgfsetstrokecolor{currentstroke}%
\pgfsetdash{}{0pt}%
\pgfsys@defobject{currentmarker}{\pgfqpoint{-0.048611in}{0.000000in}}{\pgfqpoint{-0.000000in}{0.000000in}}{%
\pgfpathmoveto{\pgfqpoint{-0.000000in}{0.000000in}}%
\pgfpathlineto{\pgfqpoint{-0.048611in}{0.000000in}}%
\pgfusepath{stroke,fill}%
}%
\begin{pgfscope}%
\pgfsys@transformshift{1.381353in}{6.308479in}%
\pgfsys@useobject{currentmarker}{}%
\end{pgfscope}%
\end{pgfscope}%
\begin{pgfscope}%
\definecolor{textcolor}{rgb}{0.000000,0.000000,0.000000}%
\pgfsetstrokecolor{textcolor}%
\pgfsetfillcolor{textcolor}%
\pgftext[x=1.007597in, y=6.188494in, left, base]{\color{textcolor}\rmfamily\fontsize{30.000000}{36.000000}\selectfont 0}%
\end{pgfscope}%
\begin{pgfscope}%
\pgfsetbuttcap%
\pgfsetroundjoin%
\definecolor{currentfill}{rgb}{0.000000,0.000000,0.000000}%
\pgfsetfillcolor{currentfill}%
\pgfsetlinewidth{0.803000pt}%
\definecolor{currentstroke}{rgb}{0.000000,0.000000,0.000000}%
\pgfsetstrokecolor{currentstroke}%
\pgfsetdash{}{0pt}%
\pgfsys@defobject{currentmarker}{\pgfqpoint{-0.048611in}{0.000000in}}{\pgfqpoint{-0.000000in}{0.000000in}}{%
\pgfpathmoveto{\pgfqpoint{-0.000000in}{0.000000in}}%
\pgfpathlineto{\pgfqpoint{-0.048611in}{0.000000in}}%
\pgfusepath{stroke,fill}%
}%
\begin{pgfscope}%
\pgfsys@transformshift{1.381353in}{4.963018in}%
\pgfsys@useobject{currentmarker}{}%
\end{pgfscope}%
\end{pgfscope}%
\begin{pgfscope}%
\definecolor{textcolor}{rgb}{0.000000,0.000000,0.000000}%
\pgfsetstrokecolor{textcolor}%
\pgfsetfillcolor{textcolor}%
\pgftext[x=0.849120in, y=4.843033in, left, base]{\color{textcolor}\rmfamily\fontsize{30.000000}{36.000000}\selectfont 50}%
\end{pgfscope}%
\begin{pgfscope}%
\pgfsetbuttcap%
\pgfsetroundjoin%
\definecolor{currentfill}{rgb}{0.000000,0.000000,0.000000}%
\pgfsetfillcolor{currentfill}%
\pgfsetlinewidth{0.803000pt}%
\definecolor{currentstroke}{rgb}{0.000000,0.000000,0.000000}%
\pgfsetstrokecolor{currentstroke}%
\pgfsetdash{}{0pt}%
\pgfsys@defobject{currentmarker}{\pgfqpoint{-0.048611in}{0.000000in}}{\pgfqpoint{-0.000000in}{0.000000in}}{%
\pgfpathmoveto{\pgfqpoint{-0.000000in}{0.000000in}}%
\pgfpathlineto{\pgfqpoint{-0.048611in}{0.000000in}}%
\pgfusepath{stroke,fill}%
}%
\begin{pgfscope}%
\pgfsys@transformshift{1.381353in}{3.617556in}%
\pgfsys@useobject{currentmarker}{}%
\end{pgfscope}%
\end{pgfscope}%
\begin{pgfscope}%
\definecolor{textcolor}{rgb}{0.000000,0.000000,0.000000}%
\pgfsetstrokecolor{textcolor}%
\pgfsetfillcolor{textcolor}%
\pgftext[x=0.690641in, y=3.497572in, left, base]{\color{textcolor}\rmfamily\fontsize{30.000000}{36.000000}\selectfont 100}%
\end{pgfscope}%
\begin{pgfscope}%
\pgfsetbuttcap%
\pgfsetroundjoin%
\definecolor{currentfill}{rgb}{0.000000,0.000000,0.000000}%
\pgfsetfillcolor{currentfill}%
\pgfsetlinewidth{0.803000pt}%
\definecolor{currentstroke}{rgb}{0.000000,0.000000,0.000000}%
\pgfsetstrokecolor{currentstroke}%
\pgfsetdash{}{0pt}%
\pgfsys@defobject{currentmarker}{\pgfqpoint{-0.048611in}{0.000000in}}{\pgfqpoint{-0.000000in}{0.000000in}}{%
\pgfpathmoveto{\pgfqpoint{-0.000000in}{0.000000in}}%
\pgfpathlineto{\pgfqpoint{-0.048611in}{0.000000in}}%
\pgfusepath{stroke,fill}%
}%
\begin{pgfscope}%
\pgfsys@transformshift{1.381353in}{2.272095in}%
\pgfsys@useobject{currentmarker}{}%
\end{pgfscope}%
\end{pgfscope}%
\begin{pgfscope}%
\definecolor{textcolor}{rgb}{0.000000,0.000000,0.000000}%
\pgfsetstrokecolor{textcolor}%
\pgfsetfillcolor{textcolor}%
\pgftext[x=0.690641in, y=2.152110in, left, base]{\color{textcolor}\rmfamily\fontsize{30.000000}{36.000000}\selectfont 150}%
\end{pgfscope}%
\begin{pgfscope}%
\definecolor{textcolor}{rgb}{0.000000,0.000000,0.000000}%
\pgfsetstrokecolor{textcolor}%
\pgfsetfillcolor{textcolor}%
\pgftext[x=0.551753in,y=3.752102in,,bottom,rotate=90.000000]{\color{textcolor}\rmfamily\fontsize{30.000000}{36.000000}\selectfont depth [m]}%
\end{pgfscope}%
\begin{pgfscope}%
\pgfsetrectcap%
\pgfsetmiterjoin%
\pgfsetlinewidth{0.803000pt}%
\definecolor{currentstroke}{rgb}{0.000000,0.000000,0.000000}%
\pgfsetstrokecolor{currentstroke}%
\pgfsetdash{}{0pt}%
\pgfpathmoveto{\pgfqpoint{1.381353in}{1.061180in}}%
\pgfpathlineto{\pgfqpoint{1.381353in}{6.443025in}}%
\pgfusepath{stroke}%
\end{pgfscope}%
\begin{pgfscope}%
\pgfsetrectcap%
\pgfsetmiterjoin%
\pgfsetlinewidth{0.803000pt}%
\definecolor{currentstroke}{rgb}{0.000000,0.000000,0.000000}%
\pgfsetstrokecolor{currentstroke}%
\pgfsetdash{}{0pt}%
\pgfpathmoveto{\pgfqpoint{12.073793in}{1.061180in}}%
\pgfpathlineto{\pgfqpoint{12.073793in}{6.443025in}}%
\pgfusepath{stroke}%
\end{pgfscope}%
\begin{pgfscope}%
\pgfsetrectcap%
\pgfsetmiterjoin%
\pgfsetlinewidth{0.803000pt}%
\definecolor{currentstroke}{rgb}{0.000000,0.000000,0.000000}%
\pgfsetstrokecolor{currentstroke}%
\pgfsetdash{}{0pt}%
\pgfpathmoveto{\pgfqpoint{1.381353in}{1.061180in}}%
\pgfpathlineto{\pgfqpoint{12.073793in}{1.061180in}}%
\pgfusepath{stroke}%
\end{pgfscope}%
\begin{pgfscope}%
\pgfsetrectcap%
\pgfsetmiterjoin%
\pgfsetlinewidth{0.803000pt}%
\definecolor{currentstroke}{rgb}{0.000000,0.000000,0.000000}%
\pgfsetstrokecolor{currentstroke}%
\pgfsetdash{}{0pt}%
\pgfpathmoveto{\pgfqpoint{1.381353in}{6.443025in}}%
\pgfpathlineto{\pgfqpoint{12.073793in}{6.443025in}}%
\pgfusepath{stroke}%
\end{pgfscope}%
\begin{pgfscope}%
\pgfsetbuttcap%
\pgfsetmiterjoin%
\definecolor{currentfill}{rgb}{1.000000,1.000000,1.000000}%
\pgfsetfillcolor{currentfill}%
\pgfsetfillopacity{0.800000}%
\pgfsetlinewidth{1.003750pt}%
\definecolor{currentstroke}{rgb}{0.800000,0.800000,0.800000}%
\pgfsetstrokecolor{currentstroke}%
\pgfsetstrokeopacity{0.800000}%
\pgfsetdash{}{0pt}%
\pgfpathmoveto{\pgfqpoint{7.864735in}{1.269513in}}%
\pgfpathlineto{\pgfqpoint{11.782126in}{1.269513in}}%
\pgfpathquadraticcurveto{\pgfqpoint{11.865459in}{1.269513in}}{\pgfqpoint{11.865459in}{1.352847in}}%
\pgfpathlineto{\pgfqpoint{11.865459in}{1.878393in}}%
\pgfpathquadraticcurveto{\pgfqpoint{11.865459in}{1.961727in}}{\pgfqpoint{11.782126in}{1.961727in}}%
\pgfpathlineto{\pgfqpoint{7.864735in}{1.961727in}}%
\pgfpathquadraticcurveto{\pgfqpoint{7.781402in}{1.961727in}}{\pgfqpoint{7.781402in}{1.878393in}}%
\pgfpathlineto{\pgfqpoint{7.781402in}{1.352847in}}%
\pgfpathquadraticcurveto{\pgfqpoint{7.781402in}{1.269513in}}{\pgfqpoint{7.864735in}{1.269513in}}%
\pgfpathlineto{\pgfqpoint{7.864735in}{1.269513in}}%
\pgfpathclose%
\pgfusepath{stroke,fill}%
\end{pgfscope}%
\begin{pgfscope}%
\pgfsetbuttcap%
\pgfsetroundjoin%
\pgfsetlinewidth{3.011250pt}%
\definecolor{currentstroke}{rgb}{1.000000,0.000000,0.000000}%
\pgfsetstrokecolor{currentstroke}%
\pgfsetdash{}{0pt}%
\pgfpathmoveto{\pgfqpoint{8.364735in}{1.514559in}}%
\pgfpathcurveto{\pgfqpoint{8.390780in}{1.514559in}}{\pgfqpoint{8.415763in}{1.524907in}}{\pgfqpoint{8.434180in}{1.543324in}}%
\pgfpathcurveto{\pgfqpoint{8.452596in}{1.561741in}}{\pgfqpoint{8.462944in}{1.586723in}}{\pgfqpoint{8.462944in}{1.612768in}}%
\pgfpathcurveto{\pgfqpoint{8.462944in}{1.638814in}}{\pgfqpoint{8.452596in}{1.663796in}}{\pgfqpoint{8.434180in}{1.682213in}}%
\pgfpathcurveto{\pgfqpoint{8.415763in}{1.700630in}}{\pgfqpoint{8.390780in}{1.710978in}}{\pgfqpoint{8.364735in}{1.710978in}}%
\pgfpathcurveto{\pgfqpoint{8.338690in}{1.710978in}}{\pgfqpoint{8.313708in}{1.700630in}}{\pgfqpoint{8.295291in}{1.682213in}}%
\pgfpathcurveto{\pgfqpoint{8.276874in}{1.663796in}}{\pgfqpoint{8.266526in}{1.638814in}}{\pgfqpoint{8.266526in}{1.612768in}}%
\pgfpathcurveto{\pgfqpoint{8.266526in}{1.586723in}}{\pgfqpoint{8.276874in}{1.561741in}}{\pgfqpoint{8.295291in}{1.543324in}}%
\pgfpathcurveto{\pgfqpoint{8.313708in}{1.524907in}}{\pgfqpoint{8.338690in}{1.514559in}}{\pgfqpoint{8.364735in}{1.514559in}}%
\pgfpathlineto{\pgfqpoint{8.364735in}{1.514559in}}%
\pgfpathclose%
\pgfusepath{stroke}%
\end{pgfscope}%
\begin{pgfscope}%
\definecolor{textcolor}{rgb}{0.000000,0.000000,0.000000}%
\pgfsetstrokecolor{textcolor}%
\pgfsetfillcolor{textcolor}%
\pgftext[x=9.114735in,y=1.503393in,left,base]{\color{textcolor}\rmfamily\fontsize{30.000000}{36.000000}\selectfont Object Position~~~~}%
\end{pgfscope}%
\begin{pgfscope}%
\pgfsetbuttcap%
\pgfsetmiterjoin%
\definecolor{currentfill}{rgb}{1.000000,1.000000,1.000000}%
\pgfsetfillcolor{currentfill}%
\pgfsetlinewidth{0.000000pt}%
\definecolor{currentstroke}{rgb}{0.000000,0.000000,0.000000}%
\pgfsetstrokecolor{currentstroke}%
\pgfsetstrokeopacity{0.000000}%
\pgfsetdash{}{0pt}%
\pgfpathmoveto{\pgfqpoint{12.541733in}{1.061180in}}%
\pgfpathlineto{\pgfqpoint{12.810825in}{1.061180in}}%
\pgfpathlineto{\pgfqpoint{12.810825in}{6.443025in}}%
\pgfpathlineto{\pgfqpoint{12.541733in}{6.443025in}}%
\pgfpathlineto{\pgfqpoint{12.541733in}{1.061180in}}%
\pgfpathclose%
\pgfusepath{fill}%
\end{pgfscope}%
\begin{pgfscope}%
\pgfpathrectangle{\pgfqpoint{12.541733in}{1.061180in}}{\pgfqpoint{0.269092in}{5.381845in}}%
\pgfusepath{clip}%
\pgfsetbuttcap%
\pgfsetmiterjoin%
\definecolor{currentfill}{rgb}{1.000000,1.000000,1.000000}%
\pgfsetfillcolor{currentfill}%
\pgfsetlinewidth{0.010037pt}%
\definecolor{currentstroke}{rgb}{1.000000,1.000000,1.000000}%
\pgfsetstrokecolor{currentstroke}%
\pgfsetdash{}{0pt}%
\pgfusepath{stroke,fill}%
\end{pgfscope}%
\begin{pgfscope}%
\pgfsys@transformshift{12.540000in}{1.060000in}%
\pgftext[left,bottom]{\includegraphics[interpolate=true,width=0.270000in,height=5.380000in]{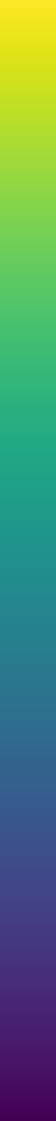}}%
\end{pgfscope}%
\begin{pgfscope}%
\pgfsetbuttcap%
\pgfsetroundjoin%
\definecolor{currentfill}{rgb}{0.000000,0.000000,0.000000}%
\pgfsetfillcolor{currentfill}%
\pgfsetlinewidth{0.803000pt}%
\definecolor{currentstroke}{rgb}{0.000000,0.000000,0.000000}%
\pgfsetstrokecolor{currentstroke}%
\pgfsetdash{}{0pt}%
\pgfsys@defobject{currentmarker}{\pgfqpoint{0.000000in}{0.000000in}}{\pgfqpoint{0.048611in}{0.000000in}}{%
\pgfpathmoveto{\pgfqpoint{0.000000in}{0.000000in}}%
\pgfpathlineto{\pgfqpoint{0.048611in}{0.000000in}}%
\pgfusepath{stroke,fill}%
}%
\begin{pgfscope}%
\pgfsys@transformshift{12.810825in}{1.419970in}%
\pgfsys@useobject{currentmarker}{}%
\end{pgfscope}%
\end{pgfscope}%
\begin{pgfscope}%
\definecolor{textcolor}{rgb}{0.000000,0.000000,0.000000}%
\pgfsetstrokecolor{textcolor}%
\pgfsetfillcolor{textcolor}%
\pgftext[x=12.908048in, y=1.299985in, left, base]{\color{textcolor}\rmfamily\fontsize{24.000000}{28.800000}\selectfont -14}%
\end{pgfscope}%
\begin{pgfscope}%
\pgfsetbuttcap%
\pgfsetroundjoin%
\definecolor{currentfill}{rgb}{0.000000,0.000000,0.000000}%
\pgfsetfillcolor{currentfill}%
\pgfsetlinewidth{0.803000pt}%
\definecolor{currentstroke}{rgb}{0.000000,0.000000,0.000000}%
\pgfsetstrokecolor{currentstroke}%
\pgfsetdash{}{0pt}%
\pgfsys@defobject{currentmarker}{\pgfqpoint{0.000000in}{0.000000in}}{\pgfqpoint{0.048611in}{0.000000in}}{%
\pgfpathmoveto{\pgfqpoint{0.000000in}{0.000000in}}%
\pgfpathlineto{\pgfqpoint{0.048611in}{0.000000in}}%
\pgfusepath{stroke,fill}%
}%
\begin{pgfscope}%
\pgfsys@transformshift{12.810825in}{2.137549in}%
\pgfsys@useobject{currentmarker}{}%
\end{pgfscope}%
\end{pgfscope}%
\begin{pgfscope}%
\definecolor{textcolor}{rgb}{0.000000,0.000000,0.000000}%
\pgfsetstrokecolor{textcolor}%
\pgfsetfillcolor{textcolor}%
\pgftext[x=12.908048in, y=2.017564in, left, base]{\color{textcolor}\rmfamily\fontsize{24.000000}{28.800000}\selectfont -12}%
\end{pgfscope}%
\begin{pgfscope}%
\pgfsetbuttcap%
\pgfsetroundjoin%
\definecolor{currentfill}{rgb}{0.000000,0.000000,0.000000}%
\pgfsetfillcolor{currentfill}%
\pgfsetlinewidth{0.803000pt}%
\definecolor{currentstroke}{rgb}{0.000000,0.000000,0.000000}%
\pgfsetstrokecolor{currentstroke}%
\pgfsetdash{}{0pt}%
\pgfsys@defobject{currentmarker}{\pgfqpoint{0.000000in}{0.000000in}}{\pgfqpoint{0.048611in}{0.000000in}}{%
\pgfpathmoveto{\pgfqpoint{0.000000in}{0.000000in}}%
\pgfpathlineto{\pgfqpoint{0.048611in}{0.000000in}}%
\pgfusepath{stroke,fill}%
}%
\begin{pgfscope}%
\pgfsys@transformshift{12.810825in}{2.855128in}%
\pgfsys@useobject{currentmarker}{}%
\end{pgfscope}%
\end{pgfscope}%
\begin{pgfscope}%
\definecolor{textcolor}{rgb}{0.000000,0.000000,0.000000}%
\pgfsetstrokecolor{textcolor}%
\pgfsetfillcolor{textcolor}%
\pgftext[x=12.908048in, y=2.735144in, left, base]{\color{textcolor}\rmfamily\fontsize{24.000000}{28.800000}\selectfont -10}%
\end{pgfscope}%
\begin{pgfscope}%
\pgfsetbuttcap%
\pgfsetroundjoin%
\definecolor{currentfill}{rgb}{0.000000,0.000000,0.000000}%
\pgfsetfillcolor{currentfill}%
\pgfsetlinewidth{0.803000pt}%
\definecolor{currentstroke}{rgb}{0.000000,0.000000,0.000000}%
\pgfsetstrokecolor{currentstroke}%
\pgfsetdash{}{0pt}%
\pgfsys@defobject{currentmarker}{\pgfqpoint{0.000000in}{0.000000in}}{\pgfqpoint{0.048611in}{0.000000in}}{%
\pgfpathmoveto{\pgfqpoint{0.000000in}{0.000000in}}%
\pgfpathlineto{\pgfqpoint{0.048611in}{0.000000in}}%
\pgfusepath{stroke,fill}%
}%
\begin{pgfscope}%
\pgfsys@transformshift{12.810825in}{3.572708in}%
\pgfsys@useobject{currentmarker}{}%
\end{pgfscope}%
\end{pgfscope}%
\begin{pgfscope}%
\definecolor{textcolor}{rgb}{0.000000,0.000000,0.000000}%
\pgfsetstrokecolor{textcolor}%
\pgfsetfillcolor{textcolor}%
\pgftext[x=12.908048in, y=3.452723in, left, base]{\color{textcolor}\rmfamily\fontsize{24.000000}{28.800000}\selectfont -8}%
\end{pgfscope}%
\begin{pgfscope}%
\pgfsetbuttcap%
\pgfsetroundjoin%
\definecolor{currentfill}{rgb}{0.000000,0.000000,0.000000}%
\pgfsetfillcolor{currentfill}%
\pgfsetlinewidth{0.803000pt}%
\definecolor{currentstroke}{rgb}{0.000000,0.000000,0.000000}%
\pgfsetstrokecolor{currentstroke}%
\pgfsetdash{}{0pt}%
\pgfsys@defobject{currentmarker}{\pgfqpoint{0.000000in}{0.000000in}}{\pgfqpoint{0.048611in}{0.000000in}}{%
\pgfpathmoveto{\pgfqpoint{0.000000in}{0.000000in}}%
\pgfpathlineto{\pgfqpoint{0.048611in}{0.000000in}}%
\pgfusepath{stroke,fill}%
}%
\begin{pgfscope}%
\pgfsys@transformshift{12.810825in}{4.290287in}%
\pgfsys@useobject{currentmarker}{}%
\end{pgfscope}%
\end{pgfscope}%
\begin{pgfscope}%
\definecolor{textcolor}{rgb}{0.000000,0.000000,0.000000}%
\pgfsetstrokecolor{textcolor}%
\pgfsetfillcolor{textcolor}%
\pgftext[x=12.908048in, y=4.170302in, left, base]{\color{textcolor}\rmfamily\fontsize{24.000000}{28.800000}\selectfont -6}%
\end{pgfscope}%
\begin{pgfscope}%
\pgfsetbuttcap%
\pgfsetroundjoin%
\definecolor{currentfill}{rgb}{0.000000,0.000000,0.000000}%
\pgfsetfillcolor{currentfill}%
\pgfsetlinewidth{0.803000pt}%
\definecolor{currentstroke}{rgb}{0.000000,0.000000,0.000000}%
\pgfsetstrokecolor{currentstroke}%
\pgfsetdash{}{0pt}%
\pgfsys@defobject{currentmarker}{\pgfqpoint{0.000000in}{0.000000in}}{\pgfqpoint{0.048611in}{0.000000in}}{%
\pgfpathmoveto{\pgfqpoint{0.000000in}{0.000000in}}%
\pgfpathlineto{\pgfqpoint{0.048611in}{0.000000in}}%
\pgfusepath{stroke,fill}%
}%
\begin{pgfscope}%
\pgfsys@transformshift{12.810825in}{5.007866in}%
\pgfsys@useobject{currentmarker}{}%
\end{pgfscope}%
\end{pgfscope}%
\begin{pgfscope}%
\definecolor{textcolor}{rgb}{0.000000,0.000000,0.000000}%
\pgfsetstrokecolor{textcolor}%
\pgfsetfillcolor{textcolor}%
\pgftext[x=12.908048in, y=4.887882in, left, base]{\color{textcolor}\rmfamily\fontsize{24.000000}{28.800000}\selectfont -4}%
\end{pgfscope}%
\begin{pgfscope}%
\pgfsetbuttcap%
\pgfsetroundjoin%
\definecolor{currentfill}{rgb}{0.000000,0.000000,0.000000}%
\pgfsetfillcolor{currentfill}%
\pgfsetlinewidth{0.803000pt}%
\definecolor{currentstroke}{rgb}{0.000000,0.000000,0.000000}%
\pgfsetstrokecolor{currentstroke}%
\pgfsetdash{}{0pt}%
\pgfsys@defobject{currentmarker}{\pgfqpoint{0.000000in}{0.000000in}}{\pgfqpoint{0.048611in}{0.000000in}}{%
\pgfpathmoveto{\pgfqpoint{0.000000in}{0.000000in}}%
\pgfpathlineto{\pgfqpoint{0.048611in}{0.000000in}}%
\pgfusepath{stroke,fill}%
}%
\begin{pgfscope}%
\pgfsys@transformshift{12.810825in}{5.725446in}%
\pgfsys@useobject{currentmarker}{}%
\end{pgfscope}%
\end{pgfscope}%
\begin{pgfscope}%
\definecolor{textcolor}{rgb}{0.000000,0.000000,0.000000}%
\pgfsetstrokecolor{textcolor}%
\pgfsetfillcolor{textcolor}%
\pgftext[x=12.908048in, y=5.605461in, left, base]{\color{textcolor}\rmfamily\fontsize{24.000000}{28.800000}\selectfont -2}%
\end{pgfscope}%
\begin{pgfscope}%
\pgfsetbuttcap%
\pgfsetroundjoin%
\definecolor{currentfill}{rgb}{0.000000,0.000000,0.000000}%
\pgfsetfillcolor{currentfill}%
\pgfsetlinewidth{0.803000pt}%
\definecolor{currentstroke}{rgb}{0.000000,0.000000,0.000000}%
\pgfsetstrokecolor{currentstroke}%
\pgfsetdash{}{0pt}%
\pgfsys@defobject{currentmarker}{\pgfqpoint{0.000000in}{0.000000in}}{\pgfqpoint{0.048611in}{0.000000in}}{%
\pgfpathmoveto{\pgfqpoint{0.000000in}{0.000000in}}%
\pgfpathlineto{\pgfqpoint{0.048611in}{0.000000in}}%
\pgfusepath{stroke,fill}%
}%
\begin{pgfscope}%
\pgfsys@transformshift{12.810825in}{6.443025in}%
\pgfsys@useobject{currentmarker}{}%
\end{pgfscope}%
\end{pgfscope}%
\begin{pgfscope}%
\definecolor{textcolor}{rgb}{0.000000,0.000000,0.000000}%
\pgfsetstrokecolor{textcolor}%
\pgfsetfillcolor{textcolor}%
\pgftext[x=12.908048in, y=6.323040in, left, base]{\color{textcolor}\rmfamily\fontsize{24.000000}{28.800000}\selectfont 0}%
\end{pgfscope}%
\begin{pgfscope}%
\pgfsetrectcap%
\pgfsetmiterjoin%
\pgfsetlinewidth{0.803000pt}%
\definecolor{currentstroke}{rgb}{0.000000,0.000000,0.000000}%
\pgfsetstrokecolor{currentstroke}%
\pgfsetdash{}{0pt}%
\pgfpathmoveto{\pgfqpoint{12.541733in}{1.061180in}}%
\pgfpathlineto{\pgfqpoint{12.676279in}{1.061180in}}%
\pgfpathlineto{\pgfqpoint{12.810825in}{1.061180in}}%
\pgfpathlineto{\pgfqpoint{12.810825in}{6.443025in}}%
\pgfpathlineto{\pgfqpoint{12.676279in}{6.443025in}}%
\pgfpathlineto{\pgfqpoint{12.541733in}{6.443025in}}%
\pgfpathlineto{\pgfqpoint{12.541733in}{1.061180in}}%
\pgfpathclose%
\pgfusepath{stroke}%
\end{pgfscope}%
\end{pgfpicture}%
\makeatother%
\endgroup%

%% file: Figs/gospa_sot.pgf
\begingroup%
\makeatletter%
\begin{pgfpicture}%
\pgfpathrectangle{\pgfpointorigin}{\pgfqpoint{13.501560in}{6.588415in}}%
\pgfusepath{use as bounding box, clip}%
\begin{pgfscope}%
\pgfsetbuttcap%
\pgfsetmiterjoin%
\definecolor{currentfill}{rgb}{1.000000,1.000000,1.000000}%
\pgfsetfillcolor{currentfill}%
\pgfsetlinewidth{0.000000pt}%
\definecolor{currentstroke}{rgb}{1.000000,1.000000,1.000000}%
\pgfsetstrokecolor{currentstroke}%
\pgfsetdash{}{0pt}%
\pgfpathmoveto{\pgfqpoint{0.000000in}{0.000000in}}%
\pgfpathlineto{\pgfqpoint{13.501560in}{0.000000in}}%
\pgfpathlineto{\pgfqpoint{13.501560in}{6.588415in}}%
\pgfpathlineto{\pgfqpoint{0.000000in}{6.588415in}}%
\pgfpathlineto{\pgfqpoint{0.000000in}{0.000000in}}%
\pgfpathclose%
\pgfusepath{fill}%
\end{pgfscope}%
\begin{pgfscope}%
\pgfsetbuttcap%
\pgfsetmiterjoin%
\definecolor{currentfill}{rgb}{1.000000,1.000000,1.000000}%
\pgfsetfillcolor{currentfill}%
\pgfsetlinewidth{0.000000pt}%
\definecolor{currentstroke}{rgb}{0.000000,0.000000,0.000000}%
\pgfsetstrokecolor{currentstroke}%
\pgfsetstrokeopacity{0.000000}%
\pgfsetdash{}{0pt}%
\pgfpathmoveto{\pgfqpoint{1.316223in}{0.950477in}}%
\pgfpathlineto{\pgfqpoint{13.243083in}{0.950477in}}%
\pgfpathlineto{\pgfqpoint{13.243083in}{6.488415in}}%
\pgfpathlineto{\pgfqpoint{1.316223in}{6.488415in}}%
\pgfpathlineto{\pgfqpoint{1.316223in}{0.950477in}}%
\pgfpathclose%
\pgfusepath{fill}%
\end{pgfscope}%
\begin{pgfscope}%
\pgfpathrectangle{\pgfqpoint{1.316223in}{0.950477in}}{\pgfqpoint{11.926860in}{5.537938in}}%
\pgfusepath{clip}%
\pgfsetrectcap%
\pgfsetroundjoin%
\pgfsetlinewidth{0.803000pt}%
\definecolor{currentstroke}{rgb}{0.690196,0.690196,0.690196}%
\pgfsetstrokecolor{currentstroke}%
\pgfsetdash{}{0pt}%
\pgfpathmoveto{\pgfqpoint{2.400483in}{0.950477in}}%
\pgfpathlineto{\pgfqpoint{2.400483in}{6.488415in}}%
\pgfusepath{stroke}%
\end{pgfscope}%
\begin{pgfscope}%
\pgfsetbuttcap%
\pgfsetroundjoin%
\definecolor{currentfill}{rgb}{0.000000,0.000000,0.000000}%
\pgfsetfillcolor{currentfill}%
\pgfsetlinewidth{0.803000pt}%
\definecolor{currentstroke}{rgb}{0.000000,0.000000,0.000000}%
\pgfsetstrokecolor{currentstroke}%
\pgfsetdash{}{0pt}%
\pgfsys@defobject{currentmarker}{\pgfqpoint{0.000000in}{-0.048611in}}{\pgfqpoint{0.000000in}{0.000000in}}{%
\pgfpathmoveto{\pgfqpoint{0.000000in}{0.000000in}}%
\pgfpathlineto{\pgfqpoint{0.000000in}{-0.048611in}}%
\pgfusepath{stroke,fill}%
}%
\begin{pgfscope}%
\pgfsys@transformshift{2.400483in}{0.950477in}%
\pgfsys@useobject{currentmarker}{}%
\end{pgfscope}%
\end{pgfscope}%
\begin{pgfscope}%
\definecolor{textcolor}{rgb}{0.000000,0.000000,0.000000}%
\pgfsetstrokecolor{textcolor}%
\pgfsetfillcolor{textcolor}%
\pgftext[x=2.400483in,y=0.853254in,,top]{\color{textcolor}\rmfamily\fontsize{30.000000}{36.000000}\selectfont 5}%
\end{pgfscope}%
\begin{pgfscope}%
\pgfpathrectangle{\pgfqpoint{1.316223in}{0.950477in}}{\pgfqpoint{11.926860in}{5.537938in}}%
\pgfusepath{clip}%
\pgfsetrectcap%
\pgfsetroundjoin%
\pgfsetlinewidth{0.803000pt}%
\definecolor{currentstroke}{rgb}{0.690196,0.690196,0.690196}%
\pgfsetstrokecolor{currentstroke}%
\pgfsetdash{}{0pt}%
\pgfpathmoveto{\pgfqpoint{3.755808in}{0.950477in}}%
\pgfpathlineto{\pgfqpoint{3.755808in}{6.488415in}}%
\pgfusepath{stroke}%
\end{pgfscope}%
\begin{pgfscope}%
\pgfsetbuttcap%
\pgfsetroundjoin%
\definecolor{currentfill}{rgb}{0.000000,0.000000,0.000000}%
\pgfsetfillcolor{currentfill}%
\pgfsetlinewidth{0.803000pt}%
\definecolor{currentstroke}{rgb}{0.000000,0.000000,0.000000}%
\pgfsetstrokecolor{currentstroke}%
\pgfsetdash{}{0pt}%
\pgfsys@defobject{currentmarker}{\pgfqpoint{0.000000in}{-0.048611in}}{\pgfqpoint{0.000000in}{0.000000in}}{%
\pgfpathmoveto{\pgfqpoint{0.000000in}{0.000000in}}%
\pgfpathlineto{\pgfqpoint{0.000000in}{-0.048611in}}%
\pgfusepath{stroke,fill}%
}%
\begin{pgfscope}%
\pgfsys@transformshift{3.755808in}{0.950477in}%
\pgfsys@useobject{currentmarker}{}%
\end{pgfscope}%
\end{pgfscope}%
\begin{pgfscope}%
\definecolor{textcolor}{rgb}{0.000000,0.000000,0.000000}%
\pgfsetstrokecolor{textcolor}%
\pgfsetfillcolor{textcolor}%
\pgftext[x=3.755808in,y=0.853254in,,top]{\color{textcolor}\rmfamily\fontsize{30.000000}{36.000000}\selectfont 10}%
\end{pgfscope}%
\begin{pgfscope}%
\pgfpathrectangle{\pgfqpoint{1.316223in}{0.950477in}}{\pgfqpoint{11.926860in}{5.537938in}}%
\pgfusepath{clip}%
\pgfsetrectcap%
\pgfsetroundjoin%
\pgfsetlinewidth{0.803000pt}%
\definecolor{currentstroke}{rgb}{0.690196,0.690196,0.690196}%
\pgfsetstrokecolor{currentstroke}%
\pgfsetdash{}{0pt}%
\pgfpathmoveto{\pgfqpoint{5.111133in}{0.950477in}}%
\pgfpathlineto{\pgfqpoint{5.111133in}{6.488415in}}%
\pgfusepath{stroke}%
\end{pgfscope}%
\begin{pgfscope}%
\pgfsetbuttcap%
\pgfsetroundjoin%
\definecolor{currentfill}{rgb}{0.000000,0.000000,0.000000}%
\pgfsetfillcolor{currentfill}%
\pgfsetlinewidth{0.803000pt}%
\definecolor{currentstroke}{rgb}{0.000000,0.000000,0.000000}%
\pgfsetstrokecolor{currentstroke}%
\pgfsetdash{}{0pt}%
\pgfsys@defobject{currentmarker}{\pgfqpoint{0.000000in}{-0.048611in}}{\pgfqpoint{0.000000in}{0.000000in}}{%
\pgfpathmoveto{\pgfqpoint{0.000000in}{0.000000in}}%
\pgfpathlineto{\pgfqpoint{0.000000in}{-0.048611in}}%
\pgfusepath{stroke,fill}%
}%
\begin{pgfscope}%
\pgfsys@transformshift{5.111133in}{0.950477in}%
\pgfsys@useobject{currentmarker}{}%
\end{pgfscope}%
\end{pgfscope}%
\begin{pgfscope}%
\definecolor{textcolor}{rgb}{0.000000,0.000000,0.000000}%
\pgfsetstrokecolor{textcolor}%
\pgfsetfillcolor{textcolor}%
\pgftext[x=5.111133in,y=0.853254in,,top]{\color{textcolor}\rmfamily\fontsize{30.000000}{36.000000}\selectfont 15}%
\end{pgfscope}%
\begin{pgfscope}%
\pgfpathrectangle{\pgfqpoint{1.316223in}{0.950477in}}{\pgfqpoint{11.926860in}{5.537938in}}%
\pgfusepath{clip}%
\pgfsetrectcap%
\pgfsetroundjoin%
\pgfsetlinewidth{0.803000pt}%
\definecolor{currentstroke}{rgb}{0.690196,0.690196,0.690196}%
\pgfsetstrokecolor{currentstroke}%
\pgfsetdash{}{0pt}%
\pgfpathmoveto{\pgfqpoint{6.466458in}{0.950477in}}%
\pgfpathlineto{\pgfqpoint{6.466458in}{6.488415in}}%
\pgfusepath{stroke}%
\end{pgfscope}%
\begin{pgfscope}%
\pgfsetbuttcap%
\pgfsetroundjoin%
\definecolor{currentfill}{rgb}{0.000000,0.000000,0.000000}%
\pgfsetfillcolor{currentfill}%
\pgfsetlinewidth{0.803000pt}%
\definecolor{currentstroke}{rgb}{0.000000,0.000000,0.000000}%
\pgfsetstrokecolor{currentstroke}%
\pgfsetdash{}{0pt}%
\pgfsys@defobject{currentmarker}{\pgfqpoint{0.000000in}{-0.048611in}}{\pgfqpoint{0.000000in}{0.000000in}}{%
\pgfpathmoveto{\pgfqpoint{0.000000in}{0.000000in}}%
\pgfpathlineto{\pgfqpoint{0.000000in}{-0.048611in}}%
\pgfusepath{stroke,fill}%
}%
\begin{pgfscope}%
\pgfsys@transformshift{6.466458in}{0.950477in}%
\pgfsys@useobject{currentmarker}{}%
\end{pgfscope}%
\end{pgfscope}%
\begin{pgfscope}%
\definecolor{textcolor}{rgb}{0.000000,0.000000,0.000000}%
\pgfsetstrokecolor{textcolor}%
\pgfsetfillcolor{textcolor}%
\pgftext[x=6.466458in,y=0.853254in,,top]{\color{textcolor}\rmfamily\fontsize{30.000000}{36.000000}\selectfont 20}%
\end{pgfscope}%
\begin{pgfscope}%
\pgfpathrectangle{\pgfqpoint{1.316223in}{0.950477in}}{\pgfqpoint{11.926860in}{5.537938in}}%
\pgfusepath{clip}%
\pgfsetrectcap%
\pgfsetroundjoin%
\pgfsetlinewidth{0.803000pt}%
\definecolor{currentstroke}{rgb}{0.690196,0.690196,0.690196}%
\pgfsetstrokecolor{currentstroke}%
\pgfsetdash{}{0pt}%
\pgfpathmoveto{\pgfqpoint{7.821783in}{0.950477in}}%
\pgfpathlineto{\pgfqpoint{7.821783in}{6.488415in}}%
\pgfusepath{stroke}%
\end{pgfscope}%
\begin{pgfscope}%
\pgfsetbuttcap%
\pgfsetroundjoin%
\definecolor{currentfill}{rgb}{0.000000,0.000000,0.000000}%
\pgfsetfillcolor{currentfill}%
\pgfsetlinewidth{0.803000pt}%
\definecolor{currentstroke}{rgb}{0.000000,0.000000,0.000000}%
\pgfsetstrokecolor{currentstroke}%
\pgfsetdash{}{0pt}%
\pgfsys@defobject{currentmarker}{\pgfqpoint{0.000000in}{-0.048611in}}{\pgfqpoint{0.000000in}{0.000000in}}{%
\pgfpathmoveto{\pgfqpoint{0.000000in}{0.000000in}}%
\pgfpathlineto{\pgfqpoint{0.000000in}{-0.048611in}}%
\pgfusepath{stroke,fill}%
}%
\begin{pgfscope}%
\pgfsys@transformshift{7.821783in}{0.950477in}%
\pgfsys@useobject{currentmarker}{}%
\end{pgfscope}%
\end{pgfscope}%
\begin{pgfscope}%
\definecolor{textcolor}{rgb}{0.000000,0.000000,0.000000}%
\pgfsetstrokecolor{textcolor}%
\pgfsetfillcolor{textcolor}%
\pgftext[x=7.821783in,y=0.853254in,,top]{\color{textcolor}\rmfamily\fontsize{30.000000}{36.000000}\selectfont 25}%
\end{pgfscope}%
\begin{pgfscope}%
\pgfpathrectangle{\pgfqpoint{1.316223in}{0.950477in}}{\pgfqpoint{11.926860in}{5.537938in}}%
\pgfusepath{clip}%
\pgfsetrectcap%
\pgfsetroundjoin%
\pgfsetlinewidth{0.803000pt}%
\definecolor{currentstroke}{rgb}{0.690196,0.690196,0.690196}%
\pgfsetstrokecolor{currentstroke}%
\pgfsetdash{}{0pt}%
\pgfpathmoveto{\pgfqpoint{9.177108in}{0.950477in}}%
\pgfpathlineto{\pgfqpoint{9.177108in}{6.488415in}}%
\pgfusepath{stroke}%
\end{pgfscope}%
\begin{pgfscope}%
\pgfsetbuttcap%
\pgfsetroundjoin%
\definecolor{currentfill}{rgb}{0.000000,0.000000,0.000000}%
\pgfsetfillcolor{currentfill}%
\pgfsetlinewidth{0.803000pt}%
\definecolor{currentstroke}{rgb}{0.000000,0.000000,0.000000}%
\pgfsetstrokecolor{currentstroke}%
\pgfsetdash{}{0pt}%
\pgfsys@defobject{currentmarker}{\pgfqpoint{0.000000in}{-0.048611in}}{\pgfqpoint{0.000000in}{0.000000in}}{%
\pgfpathmoveto{\pgfqpoint{0.000000in}{0.000000in}}%
\pgfpathlineto{\pgfqpoint{0.000000in}{-0.048611in}}%
\pgfusepath{stroke,fill}%
}%
\begin{pgfscope}%
\pgfsys@transformshift{9.177108in}{0.950477in}%
\pgfsys@useobject{currentmarker}{}%
\end{pgfscope}%
\end{pgfscope}%
\begin{pgfscope}%
\definecolor{textcolor}{rgb}{0.000000,0.000000,0.000000}%
\pgfsetstrokecolor{textcolor}%
\pgfsetfillcolor{textcolor}%
\pgftext[x=9.177108in,y=0.853254in,,top]{\color{textcolor}\rmfamily\fontsize{30.000000}{36.000000}\selectfont 30}%
\end{pgfscope}%
\begin{pgfscope}%
\pgfpathrectangle{\pgfqpoint{1.316223in}{0.950477in}}{\pgfqpoint{11.926860in}{5.537938in}}%
\pgfusepath{clip}%
\pgfsetrectcap%
\pgfsetroundjoin%
\pgfsetlinewidth{0.803000pt}%
\definecolor{currentstroke}{rgb}{0.690196,0.690196,0.690196}%
\pgfsetstrokecolor{currentstroke}%
\pgfsetdash{}{0pt}%
\pgfpathmoveto{\pgfqpoint{10.532433in}{0.950477in}}%
\pgfpathlineto{\pgfqpoint{10.532433in}{6.488415in}}%
\pgfusepath{stroke}%
\end{pgfscope}%
\begin{pgfscope}%
\pgfsetbuttcap%
\pgfsetroundjoin%
\definecolor{currentfill}{rgb}{0.000000,0.000000,0.000000}%
\pgfsetfillcolor{currentfill}%
\pgfsetlinewidth{0.803000pt}%
\definecolor{currentstroke}{rgb}{0.000000,0.000000,0.000000}%
\pgfsetstrokecolor{currentstroke}%
\pgfsetdash{}{0pt}%
\pgfsys@defobject{currentmarker}{\pgfqpoint{0.000000in}{-0.048611in}}{\pgfqpoint{0.000000in}{0.000000in}}{%
\pgfpathmoveto{\pgfqpoint{0.000000in}{0.000000in}}%
\pgfpathlineto{\pgfqpoint{0.000000in}{-0.048611in}}%
\pgfusepath{stroke,fill}%
}%
\begin{pgfscope}%
\pgfsys@transformshift{10.532433in}{0.950477in}%
\pgfsys@useobject{currentmarker}{}%
\end{pgfscope}%
\end{pgfscope}%
\begin{pgfscope}%
\definecolor{textcolor}{rgb}{0.000000,0.000000,0.000000}%
\pgfsetstrokecolor{textcolor}%
\pgfsetfillcolor{textcolor}%
\pgftext[x=10.532433in,y=0.853254in,,top]{\color{textcolor}\rmfamily\fontsize{30.000000}{36.000000}\selectfont 35}%
\end{pgfscope}%
\begin{pgfscope}%
\pgfpathrectangle{\pgfqpoint{1.316223in}{0.950477in}}{\pgfqpoint{11.926860in}{5.537938in}}%
\pgfusepath{clip}%
\pgfsetrectcap%
\pgfsetroundjoin%
\pgfsetlinewidth{0.803000pt}%
\definecolor{currentstroke}{rgb}{0.690196,0.690196,0.690196}%
\pgfsetstrokecolor{currentstroke}%
\pgfsetdash{}{0pt}%
\pgfpathmoveto{\pgfqpoint{11.887758in}{0.950477in}}%
\pgfpathlineto{\pgfqpoint{11.887758in}{6.488415in}}%
\pgfusepath{stroke}%
\end{pgfscope}%
\begin{pgfscope}%
\pgfsetbuttcap%
\pgfsetroundjoin%
\definecolor{currentfill}{rgb}{0.000000,0.000000,0.000000}%
\pgfsetfillcolor{currentfill}%
\pgfsetlinewidth{0.803000pt}%
\definecolor{currentstroke}{rgb}{0.000000,0.000000,0.000000}%
\pgfsetstrokecolor{currentstroke}%
\pgfsetdash{}{0pt}%
\pgfsys@defobject{currentmarker}{\pgfqpoint{0.000000in}{-0.048611in}}{\pgfqpoint{0.000000in}{0.000000in}}{%
\pgfpathmoveto{\pgfqpoint{0.000000in}{0.000000in}}%
\pgfpathlineto{\pgfqpoint{0.000000in}{-0.048611in}}%
\pgfusepath{stroke,fill}%
}%
\begin{pgfscope}%
\pgfsys@transformshift{11.887758in}{0.950477in}%
\pgfsys@useobject{currentmarker}{}%
\end{pgfscope}%
\end{pgfscope}%
\begin{pgfscope}%
\definecolor{textcolor}{rgb}{0.000000,0.000000,0.000000}%
\pgfsetstrokecolor{textcolor}%
\pgfsetfillcolor{textcolor}%
\pgftext[x=11.887758in,y=0.853254in,,top]{\color{textcolor}\rmfamily\fontsize{30.000000}{36.000000}\selectfont 40}%
\end{pgfscope}%
\begin{pgfscope}%
\pgfpathrectangle{\pgfqpoint{1.316223in}{0.950477in}}{\pgfqpoint{11.926860in}{5.537938in}}%
\pgfusepath{clip}%
\pgfsetrectcap%
\pgfsetroundjoin%
\pgfsetlinewidth{0.803000pt}%
\definecolor{currentstroke}{rgb}{0.690196,0.690196,0.690196}%
\pgfsetstrokecolor{currentstroke}%
\pgfsetdash{}{0pt}%
\pgfpathmoveto{\pgfqpoint{13.243083in}{0.950477in}}%
\pgfpathlineto{\pgfqpoint{13.243083in}{6.488415in}}%
\pgfusepath{stroke}%
\end{pgfscope}%
\begin{pgfscope}%
\pgfsetbuttcap%
\pgfsetroundjoin%
\definecolor{currentfill}{rgb}{0.000000,0.000000,0.000000}%
\pgfsetfillcolor{currentfill}%
\pgfsetlinewidth{0.803000pt}%
\definecolor{currentstroke}{rgb}{0.000000,0.000000,0.000000}%
\pgfsetstrokecolor{currentstroke}%
\pgfsetdash{}{0pt}%
\pgfsys@defobject{currentmarker}{\pgfqpoint{0.000000in}{-0.048611in}}{\pgfqpoint{0.000000in}{0.000000in}}{%
\pgfpathmoveto{\pgfqpoint{0.000000in}{0.000000in}}%
\pgfpathlineto{\pgfqpoint{0.000000in}{-0.048611in}}%
\pgfusepath{stroke,fill}%
}%
\begin{pgfscope}%
\pgfsys@transformshift{13.243083in}{0.950477in}%
\pgfsys@useobject{currentmarker}{}%
\end{pgfscope}%
\end{pgfscope}%
\begin{pgfscope}%
\definecolor{textcolor}{rgb}{0.000000,0.000000,0.000000}%
\pgfsetstrokecolor{textcolor}%
\pgfsetfillcolor{textcolor}%
\pgftext[x=13.243083in,y=0.853254in,,top]{\color{textcolor}\rmfamily\fontsize{30.000000}{36.000000}\selectfont 45}%
\end{pgfscope}%
\begin{pgfscope}%
\definecolor{textcolor}{rgb}{0.000000,0.000000,0.000000}%
\pgfsetstrokecolor{textcolor}%
\pgfsetfillcolor{textcolor}%
\pgftext[x=7.279653in,y=0.407183in,,top]{\color{textcolor}\rmfamily\fontsize{30.000000}{36.000000}\selectfont time step \(\displaystyle k\)}%
\end{pgfscope}%
\begin{pgfscope}%
\pgfpathrectangle{\pgfqpoint{1.316223in}{0.950477in}}{\pgfqpoint{11.926860in}{5.537938in}}%
\pgfusepath{clip}%
\pgfsetrectcap%
\pgfsetroundjoin%
\pgfsetlinewidth{0.803000pt}%
\definecolor{currentstroke}{rgb}{0.690196,0.690196,0.690196}%
\pgfsetstrokecolor{currentstroke}%
\pgfsetdash{}{0pt}%
\pgfpathmoveto{\pgfqpoint{1.316223in}{0.950477in}}%
\pgfpathlineto{\pgfqpoint{13.243083in}{0.950477in}}%
\pgfusepath{stroke}%
\end{pgfscope}%
\begin{pgfscope}%
\pgfsetbuttcap%
\pgfsetroundjoin%
\definecolor{currentfill}{rgb}{0.000000,0.000000,0.000000}%
\pgfsetfillcolor{currentfill}%
\pgfsetlinewidth{0.803000pt}%
\definecolor{currentstroke}{rgb}{0.000000,0.000000,0.000000}%
\pgfsetstrokecolor{currentstroke}%
\pgfsetdash{}{0pt}%
\pgfsys@defobject{currentmarker}{\pgfqpoint{-0.048611in}{0.000000in}}{\pgfqpoint{-0.000000in}{0.000000in}}{%
\pgfpathmoveto{\pgfqpoint{-0.000000in}{0.000000in}}%
\pgfpathlineto{\pgfqpoint{-0.048611in}{0.000000in}}%
\pgfusepath{stroke,fill}%
}%
\begin{pgfscope}%
\pgfsys@transformshift{1.316223in}{0.950477in}%
\pgfsys@useobject{currentmarker}{}%
\end{pgfscope}%
\end{pgfscope}%
\begin{pgfscope}%
\definecolor{textcolor}{rgb}{0.000000,0.000000,0.000000}%
\pgfsetstrokecolor{textcolor}%
\pgfsetfillcolor{textcolor}%
\pgftext[x=0.900800in, y=0.830492in, left, base]{\color{textcolor}\rmfamily\fontsize{30.000000}{36.000000}\selectfont 0}%
\end{pgfscope}%
\begin{pgfscope}%
\pgfpathrectangle{\pgfqpoint{1.316223in}{0.950477in}}{\pgfqpoint{11.926860in}{5.537938in}}%
\pgfusepath{clip}%
\pgfsetrectcap%
\pgfsetroundjoin%
\pgfsetlinewidth{0.803000pt}%
\definecolor{currentstroke}{rgb}{0.690196,0.690196,0.690196}%
\pgfsetstrokecolor{currentstroke}%
\pgfsetdash{}{0pt}%
\pgfpathmoveto{\pgfqpoint{1.316223in}{2.532745in}}%
\pgfpathlineto{\pgfqpoint{13.243083in}{2.532745in}}%
\pgfusepath{stroke}%
\end{pgfscope}%
\begin{pgfscope}%
\pgfsetbuttcap%
\pgfsetroundjoin%
\definecolor{currentfill}{rgb}{0.000000,0.000000,0.000000}%
\pgfsetfillcolor{currentfill}%
\pgfsetlinewidth{0.803000pt}%
\definecolor{currentstroke}{rgb}{0.000000,0.000000,0.000000}%
\pgfsetstrokecolor{currentstroke}%
\pgfsetdash{}{0pt}%
\pgfsys@defobject{currentmarker}{\pgfqpoint{-0.048611in}{0.000000in}}{\pgfqpoint{-0.000000in}{0.000000in}}{%
\pgfpathmoveto{\pgfqpoint{-0.000000in}{0.000000in}}%
\pgfpathlineto{\pgfqpoint{-0.048611in}{0.000000in}}%
\pgfusepath{stroke,fill}%
}%
\begin{pgfscope}%
\pgfsys@transformshift{1.316223in}{2.532745in}%
\pgfsys@useobject{currentmarker}{}%
\end{pgfscope}%
\end{pgfscope}%
\begin{pgfscope}%
\definecolor{textcolor}{rgb}{0.000000,0.000000,0.000000}%
\pgfsetstrokecolor{textcolor}%
\pgfsetfillcolor{textcolor}%
\pgftext[x=0.583844in, y=2.412760in, left, base]{\color{textcolor}\rmfamily\fontsize{30.000000}{36.000000}\selectfont 200}%
\end{pgfscope}%
\begin{pgfscope}%
\pgfpathrectangle{\pgfqpoint{1.316223in}{0.950477in}}{\pgfqpoint{11.926860in}{5.537938in}}%
\pgfusepath{clip}%
\pgfsetrectcap%
\pgfsetroundjoin%
\pgfsetlinewidth{0.803000pt}%
\definecolor{currentstroke}{rgb}{0.690196,0.690196,0.690196}%
\pgfsetstrokecolor{currentstroke}%
\pgfsetdash{}{0pt}%
\pgfpathmoveto{\pgfqpoint{1.316223in}{4.115013in}}%
\pgfpathlineto{\pgfqpoint{13.243083in}{4.115013in}}%
\pgfusepath{stroke}%
\end{pgfscope}%
\begin{pgfscope}%
\pgfsetbuttcap%
\pgfsetroundjoin%
\definecolor{currentfill}{rgb}{0.000000,0.000000,0.000000}%
\pgfsetfillcolor{currentfill}%
\pgfsetlinewidth{0.803000pt}%
\definecolor{currentstroke}{rgb}{0.000000,0.000000,0.000000}%
\pgfsetstrokecolor{currentstroke}%
\pgfsetdash{}{0pt}%
\pgfsys@defobject{currentmarker}{\pgfqpoint{-0.048611in}{0.000000in}}{\pgfqpoint{-0.000000in}{0.000000in}}{%
\pgfpathmoveto{\pgfqpoint{-0.000000in}{0.000000in}}%
\pgfpathlineto{\pgfqpoint{-0.048611in}{0.000000in}}%
\pgfusepath{stroke,fill}%
}%
\begin{pgfscope}%
\pgfsys@transformshift{1.316223in}{4.115013in}%
\pgfsys@useobject{currentmarker}{}%
\end{pgfscope}%
\end{pgfscope}%
\begin{pgfscope}%
\definecolor{textcolor}{rgb}{0.000000,0.000000,0.000000}%
\pgfsetstrokecolor{textcolor}%
\pgfsetfillcolor{textcolor}%
\pgftext[x=0.583844in, y=3.995028in, left, base]{\color{textcolor}\rmfamily\fontsize{30.000000}{36.000000}\selectfont 400}%
\end{pgfscope}%
\begin{pgfscope}%
\pgfpathrectangle{\pgfqpoint{1.316223in}{0.950477in}}{\pgfqpoint{11.926860in}{5.537938in}}%
\pgfusepath{clip}%
\pgfsetrectcap%
\pgfsetroundjoin%
\pgfsetlinewidth{0.803000pt}%
\definecolor{currentstroke}{rgb}{0.690196,0.690196,0.690196}%
\pgfsetstrokecolor{currentstroke}%
\pgfsetdash{}{0pt}%
\pgfpathmoveto{\pgfqpoint{1.316223in}{5.697281in}}%
\pgfpathlineto{\pgfqpoint{13.243083in}{5.697281in}}%
\pgfusepath{stroke}%
\end{pgfscope}%
\begin{pgfscope}%
\pgfsetbuttcap%
\pgfsetroundjoin%
\definecolor{currentfill}{rgb}{0.000000,0.000000,0.000000}%
\pgfsetfillcolor{currentfill}%
\pgfsetlinewidth{0.803000pt}%
\definecolor{currentstroke}{rgb}{0.000000,0.000000,0.000000}%
\pgfsetstrokecolor{currentstroke}%
\pgfsetdash{}{0pt}%
\pgfsys@defobject{currentmarker}{\pgfqpoint{-0.048611in}{0.000000in}}{\pgfqpoint{-0.000000in}{0.000000in}}{%
\pgfpathmoveto{\pgfqpoint{-0.000000in}{0.000000in}}%
\pgfpathlineto{\pgfqpoint{-0.048611in}{0.000000in}}%
\pgfusepath{stroke,fill}%
}%
\begin{pgfscope}%
\pgfsys@transformshift{1.316223in}{5.697281in}%
\pgfsys@useobject{currentmarker}{}%
\end{pgfscope}%
\end{pgfscope}%
\begin{pgfscope}%
\definecolor{textcolor}{rgb}{0.000000,0.000000,0.000000}%
\pgfsetstrokecolor{textcolor}%
\pgfsetfillcolor{textcolor}%
\pgftext[x=0.583844in, y=5.577296in, left, base]{\color{textcolor}\rmfamily\fontsize{30.000000}{36.000000}\selectfont 600}%
\end{pgfscope}%
\begin{pgfscope}%
\definecolor{textcolor}{rgb}{0.000000,0.000000,0.000000}%
\pgfsetstrokecolor{textcolor}%
\pgfsetfillcolor{textcolor}%
\pgftext[x=0.444955in,y=3.719446in,,bottom,rotate=90.000000]{\color{textcolor}\rmfamily\fontsize{30.000000}{36.000000}\selectfont GOSPA [m]}%
\end{pgfscope}%
\begin{pgfscope}%
\pgfpathrectangle{\pgfqpoint{1.316223in}{0.950477in}}{\pgfqpoint{11.926860in}{5.537938in}}%
\pgfusepath{clip}%
\pgfsetrectcap%
\pgfsetroundjoin%
\pgfsetlinewidth{1.505625pt}%
\definecolor{currentstroke}{rgb}{0.000000,0.000000,1.000000}%
\pgfsetstrokecolor{currentstroke}%
\pgfsetdash{}{0pt}%
\pgfpathmoveto{\pgfqpoint{1.316223in}{2.928312in}}%
\pgfpathlineto{\pgfqpoint{1.587288in}{1.495775in}}%
\pgfpathlineto{\pgfqpoint{1.858353in}{1.456209in}}%
\pgfpathlineto{\pgfqpoint{2.129418in}{1.474008in}}%
\pgfpathlineto{\pgfqpoint{2.400483in}{1.481157in}}%
\pgfpathlineto{\pgfqpoint{2.671548in}{1.501866in}}%
\pgfpathlineto{\pgfqpoint{2.942613in}{1.550848in}}%
\pgfpathlineto{\pgfqpoint{3.213678in}{1.564639in}}%
\pgfpathlineto{\pgfqpoint{3.484743in}{3.474960in}}%
\pgfpathlineto{\pgfqpoint{3.755808in}{3.550689in}}%
\pgfpathlineto{\pgfqpoint{4.026873in}{3.609520in}}%
\pgfpathlineto{\pgfqpoint{4.297938in}{3.559647in}}%
\pgfpathlineto{\pgfqpoint{4.569003in}{1.692567in}}%
\pgfpathlineto{\pgfqpoint{4.840068in}{1.791204in}}%
\pgfpathlineto{\pgfqpoint{5.111133in}{1.739844in}}%
\pgfpathlineto{\pgfqpoint{5.382198in}{3.783255in}}%
\pgfpathlineto{\pgfqpoint{5.653263in}{1.841743in}}%
\pgfpathlineto{\pgfqpoint{5.924328in}{3.716066in}}%
\pgfpathlineto{\pgfqpoint{6.195393in}{1.781203in}}%
\pgfpathlineto{\pgfqpoint{6.466458in}{3.541180in}}%
\pgfpathlineto{\pgfqpoint{6.737523in}{3.507409in}}%
\pgfpathlineto{\pgfqpoint{7.008588in}{3.260783in}}%
\pgfpathlineto{\pgfqpoint{7.279653in}{3.251535in}}%
\pgfpathlineto{\pgfqpoint{7.550718in}{3.237554in}}%
\pgfpathlineto{\pgfqpoint{7.821783in}{3.246040in}}%
\pgfpathlineto{\pgfqpoint{8.092848in}{1.646122in}}%
\pgfpathlineto{\pgfqpoint{8.363913in}{1.607288in}}%
\pgfpathlineto{\pgfqpoint{8.634978in}{1.560670in}}%
\pgfpathlineto{\pgfqpoint{8.906043in}{1.548280in}}%
\pgfpathlineto{\pgfqpoint{9.177108in}{3.514561in}}%
\pgfpathlineto{\pgfqpoint{9.448173in}{3.383120in}}%
\pgfpathlineto{\pgfqpoint{9.719238in}{3.303269in}}%
\pgfpathlineto{\pgfqpoint{9.990303in}{1.200489in}}%
\pgfpathlineto{\pgfqpoint{10.261368in}{1.129201in}}%
\pgfpathlineto{\pgfqpoint{10.532433in}{3.101353in}}%
\pgfpathlineto{\pgfqpoint{10.803498in}{1.109952in}}%
\pgfpathlineto{\pgfqpoint{11.074563in}{1.286549in}}%
\pgfpathlineto{\pgfqpoint{11.345628in}{1.286999in}}%
\pgfpathlineto{\pgfqpoint{11.616693in}{1.278597in}}%
\pgfpathlineto{\pgfqpoint{11.887758in}{1.201880in}}%
\pgfpathlineto{\pgfqpoint{12.158823in}{3.147093in}}%
\pgfpathlineto{\pgfqpoint{12.429888in}{1.173548in}}%
\pgfpathlineto{\pgfqpoint{12.700953in}{2.928312in}}%
\pgfpathlineto{\pgfqpoint{12.972018in}{2.845062in}}%
\pgfpathlineto{\pgfqpoint{13.243083in}{2.928312in}}%
\pgfusepath{stroke}%
\end{pgfscope}%
\begin{pgfscope}%
\pgfpathrectangle{\pgfqpoint{1.316223in}{0.950477in}}{\pgfqpoint{11.926860in}{5.537938in}}%
\pgfusepath{clip}%
\pgfsetrectcap%
\pgfsetroundjoin%
\pgfsetlinewidth{1.505625pt}%
\definecolor{currentstroke}{rgb}{0.000000,0.500000,0.000000}%
\pgfsetstrokecolor{currentstroke}%
\pgfsetdash{}{0pt}%
\pgfpathmoveto{\pgfqpoint{1.316223in}{2.928312in}}%
\pgfpathlineto{\pgfqpoint{1.587288in}{1.482709in}}%
\pgfpathlineto{\pgfqpoint{1.858353in}{1.548380in}}%
\pgfpathlineto{\pgfqpoint{2.129418in}{1.543854in}}%
\pgfpathlineto{\pgfqpoint{2.400483in}{1.517314in}}%
\pgfpathlineto{\pgfqpoint{2.671548in}{1.517495in}}%
\pgfpathlineto{\pgfqpoint{2.942613in}{1.558071in}}%
\pgfpathlineto{\pgfqpoint{3.213678in}{1.566794in}}%
\pgfpathlineto{\pgfqpoint{3.484743in}{1.589781in}}%
\pgfpathlineto{\pgfqpoint{3.755808in}{1.628046in}}%
\pgfpathlineto{\pgfqpoint{4.026873in}{1.666218in}}%
\pgfpathlineto{\pgfqpoint{4.297938in}{1.749532in}}%
\pgfpathlineto{\pgfqpoint{4.569003in}{1.691772in}}%
\pgfpathlineto{\pgfqpoint{4.840068in}{1.791367in}}%
\pgfpathlineto{\pgfqpoint{5.111133in}{3.519524in}}%
\pgfpathlineto{\pgfqpoint{5.382198in}{1.940747in}}%
\pgfpathlineto{\pgfqpoint{5.653263in}{1.905266in}}%
\pgfpathlineto{\pgfqpoint{5.924328in}{1.870381in}}%
\pgfpathlineto{\pgfqpoint{6.195393in}{1.843170in}}%
\pgfpathlineto{\pgfqpoint{6.466458in}{1.710386in}}%
\pgfpathlineto{\pgfqpoint{6.737523in}{1.726441in}}%
\pgfpathlineto{\pgfqpoint{7.008588in}{1.623690in}}%
\pgfpathlineto{\pgfqpoint{7.279653in}{1.652311in}}%
\pgfpathlineto{\pgfqpoint{7.550718in}{1.559214in}}%
\pgfpathlineto{\pgfqpoint{7.821783in}{1.569756in}}%
\pgfpathlineto{\pgfqpoint{8.092848in}{1.655497in}}%
\pgfpathlineto{\pgfqpoint{8.363913in}{1.600586in}}%
\pgfpathlineto{\pgfqpoint{8.634978in}{1.664703in}}%
\pgfpathlineto{\pgfqpoint{8.906043in}{1.609243in}}%
\pgfpathlineto{\pgfqpoint{9.177108in}{1.564116in}}%
\pgfpathlineto{\pgfqpoint{9.448173in}{1.527182in}}%
\pgfpathlineto{\pgfqpoint{9.719238in}{1.619541in}}%
\pgfpathlineto{\pgfqpoint{9.990303in}{1.481595in}}%
\pgfpathlineto{\pgfqpoint{10.261368in}{1.493577in}}%
\pgfpathlineto{\pgfqpoint{10.532433in}{3.395023in}}%
\pgfpathlineto{\pgfqpoint{10.803498in}{5.191348in}}%
\pgfpathlineto{\pgfqpoint{11.074563in}{3.291528in}}%
\pgfpathlineto{\pgfqpoint{11.345628in}{3.242179in}}%
\pgfpathlineto{\pgfqpoint{11.616693in}{1.127670in}}%
\pgfpathlineto{\pgfqpoint{11.887758in}{1.079981in}}%
\pgfpathlineto{\pgfqpoint{12.158823in}{4.887251in}}%
\pgfpathlineto{\pgfqpoint{12.429888in}{5.026693in}}%
\pgfpathlineto{\pgfqpoint{12.700953in}{2.979808in}}%
\pgfpathlineto{\pgfqpoint{12.972018in}{4.937835in}}%
\pgfpathlineto{\pgfqpoint{13.243083in}{3.002773in}}%
\pgfusepath{stroke}%
\end{pgfscope}%
\begin{pgfscope}%
\pgfpathrectangle{\pgfqpoint{1.316223in}{0.950477in}}{\pgfqpoint{11.926860in}{5.537938in}}%
\pgfusepath{clip}%
\pgfsetrectcap%
\pgfsetroundjoin%
\pgfsetlinewidth{1.505625pt}%
\definecolor{currentstroke}{rgb}{0.750000,0.750000,0.000000}%
\pgfsetstrokecolor{currentstroke}%
\pgfsetdash{}{0pt}%
\pgfpathmoveto{\pgfqpoint{1.316223in}{1.620845in}}%
\pgfpathlineto{\pgfqpoint{1.587288in}{1.644565in}}%
\pgfpathlineto{\pgfqpoint{1.858353in}{1.637222in}}%
\pgfpathlineto{\pgfqpoint{2.129418in}{1.627885in}}%
\pgfpathlineto{\pgfqpoint{2.400483in}{1.603866in}}%
\pgfpathlineto{\pgfqpoint{2.671548in}{1.599051in}}%
\pgfpathlineto{\pgfqpoint{2.942613in}{1.632207in}}%
\pgfpathlineto{\pgfqpoint{3.213678in}{1.637192in}}%
\pgfpathlineto{\pgfqpoint{3.484743in}{1.656261in}}%
\pgfpathlineto{\pgfqpoint{3.755808in}{1.692071in}}%
\pgfpathlineto{\pgfqpoint{4.026873in}{1.727497in}}%
\pgfpathlineto{\pgfqpoint{4.297938in}{1.694492in}}%
\pgfpathlineto{\pgfqpoint{4.569003in}{1.678688in}}%
\pgfpathlineto{\pgfqpoint{4.840068in}{1.692728in}}%
\pgfpathlineto{\pgfqpoint{5.111133in}{1.700891in}}%
\pgfpathlineto{\pgfqpoint{5.382198in}{1.689227in}}%
\pgfpathlineto{\pgfqpoint{5.653263in}{1.671301in}}%
\pgfpathlineto{\pgfqpoint{5.924328in}{1.648060in}}%
\pgfpathlineto{\pgfqpoint{6.195393in}{1.622977in}}%
\pgfpathlineto{\pgfqpoint{6.466458in}{1.595649in}}%
\pgfpathlineto{\pgfqpoint{6.737523in}{1.551790in}}%
\pgfpathlineto{\pgfqpoint{7.008588in}{1.507415in}}%
\pgfpathlineto{\pgfqpoint{7.279653in}{1.443889in}}%
\pgfpathlineto{\pgfqpoint{7.550718in}{1.391635in}}%
\pgfpathlineto{\pgfqpoint{7.821783in}{1.413188in}}%
\pgfpathlineto{\pgfqpoint{8.092848in}{1.375390in}}%
\pgfpathlineto{\pgfqpoint{8.363913in}{1.347231in}}%
\pgfpathlineto{\pgfqpoint{8.634978in}{1.303375in}}%
\pgfpathlineto{\pgfqpoint{8.906043in}{1.280715in}}%
\pgfpathlineto{\pgfqpoint{9.177108in}{1.247016in}}%
\pgfpathlineto{\pgfqpoint{9.448173in}{1.198779in}}%
\pgfpathlineto{\pgfqpoint{9.719238in}{1.150031in}}%
\pgfpathlineto{\pgfqpoint{9.990303in}{1.129129in}}%
\pgfpathlineto{\pgfqpoint{10.261368in}{1.090076in}}%
\pgfpathlineto{\pgfqpoint{10.532433in}{1.065043in}}%
\pgfpathlineto{\pgfqpoint{10.803498in}{1.002765in}}%
\pgfpathlineto{\pgfqpoint{11.074563in}{1.006201in}}%
\pgfpathlineto{\pgfqpoint{11.345628in}{1.213790in}}%
\pgfpathlineto{\pgfqpoint{11.616693in}{1.373190in}}%
\pgfpathlineto{\pgfqpoint{11.887758in}{1.490123in}}%
\pgfpathlineto{\pgfqpoint{12.158823in}{1.593394in}}%
\pgfpathlineto{\pgfqpoint{12.429888in}{1.714572in}}%
\pgfpathlineto{\pgfqpoint{12.700953in}{1.693622in}}%
\pgfpathlineto{\pgfqpoint{12.972018in}{2.928312in}}%
\pgfpathlineto{\pgfqpoint{13.243083in}{2.928312in}}%
\pgfusepath{stroke}%
\end{pgfscope}%
\begin{pgfscope}%
\pgfpathrectangle{\pgfqpoint{1.316223in}{0.950477in}}{\pgfqpoint{11.926860in}{5.537938in}}%
\pgfusepath{clip}%
\pgfsetbuttcap%
\pgfsetroundjoin%
\pgfsetlinewidth{1.505625pt}%
\definecolor{currentstroke}{rgb}{1.000000,0.000000,0.000000}%
\pgfsetstrokecolor{currentstroke}%
\pgfsetdash{{9.000000pt}{7.500000pt}{0.000000pt}{0.000000pt}}{0.000000pt}%
\pgfpathmoveto{\pgfqpoint{1.316223in}{1.620845in}}%
\pgfpathlineto{\pgfqpoint{1.587288in}{1.644534in}}%
\pgfpathlineto{\pgfqpoint{1.858353in}{1.633693in}}%
\pgfpathlineto{\pgfqpoint{2.129418in}{1.624612in}}%
\pgfpathlineto{\pgfqpoint{2.400483in}{1.603554in}}%
\pgfpathlineto{\pgfqpoint{2.671548in}{1.595036in}}%
\pgfpathlineto{\pgfqpoint{2.942613in}{1.627003in}}%
\pgfpathlineto{\pgfqpoint{3.213678in}{1.635630in}}%
\pgfpathlineto{\pgfqpoint{3.484743in}{1.655659in}}%
\pgfpathlineto{\pgfqpoint{3.755808in}{1.690880in}}%
\pgfpathlineto{\pgfqpoint{4.026873in}{1.726689in}}%
\pgfpathlineto{\pgfqpoint{4.297938in}{1.693360in}}%
\pgfpathlineto{\pgfqpoint{4.569003in}{1.676521in}}%
\pgfpathlineto{\pgfqpoint{4.840068in}{1.690609in}}%
\pgfpathlineto{\pgfqpoint{5.111133in}{1.699015in}}%
\pgfpathlineto{\pgfqpoint{5.382198in}{1.687146in}}%
\pgfpathlineto{\pgfqpoint{5.653263in}{1.668563in}}%
\pgfpathlineto{\pgfqpoint{5.924328in}{1.645806in}}%
\pgfpathlineto{\pgfqpoint{6.195393in}{1.621427in}}%
\pgfpathlineto{\pgfqpoint{6.466458in}{1.596312in}}%
\pgfpathlineto{\pgfqpoint{6.737523in}{1.553974in}}%
\pgfpathlineto{\pgfqpoint{7.008588in}{1.510469in}}%
\pgfpathlineto{\pgfqpoint{7.279653in}{1.448570in}}%
\pgfpathlineto{\pgfqpoint{7.550718in}{1.397241in}}%
\pgfpathlineto{\pgfqpoint{7.821783in}{1.420469in}}%
\pgfpathlineto{\pgfqpoint{8.092848in}{1.382872in}}%
\pgfpathlineto{\pgfqpoint{8.363913in}{1.354729in}}%
\pgfpathlineto{\pgfqpoint{8.634978in}{1.310144in}}%
\pgfpathlineto{\pgfqpoint{8.906043in}{1.287686in}}%
\pgfpathlineto{\pgfqpoint{9.177108in}{1.253864in}}%
\pgfpathlineto{\pgfqpoint{9.448173in}{1.205872in}}%
\pgfpathlineto{\pgfqpoint{9.719238in}{1.156837in}}%
\pgfpathlineto{\pgfqpoint{9.990303in}{1.135588in}}%
\pgfpathlineto{\pgfqpoint{10.261368in}{1.096435in}}%
\pgfpathlineto{\pgfqpoint{10.532433in}{1.071378in}}%
\pgfpathlineto{\pgfqpoint{10.803498in}{1.009282in}}%
\pgfpathlineto{\pgfqpoint{11.074563in}{1.011714in}}%
\pgfpathlineto{\pgfqpoint{11.345628in}{0.990096in}}%
\pgfpathlineto{\pgfqpoint{11.616693in}{0.955775in}}%
\pgfpathlineto{\pgfqpoint{11.887758in}{0.965114in}}%
\pgfpathlineto{\pgfqpoint{12.158823in}{1.210198in}}%
\pgfpathlineto{\pgfqpoint{12.429888in}{1.163293in}}%
\pgfpathlineto{\pgfqpoint{12.700953in}{1.344788in}}%
\pgfpathlineto{\pgfqpoint{12.972018in}{2.928312in}}%
\pgfpathlineto{\pgfqpoint{13.243083in}{2.928312in}}%
\pgfusepath{stroke}%
\end{pgfscope}%
\begin{pgfscope}%
\pgfsetrectcap%
\pgfsetmiterjoin%
\pgfsetlinewidth{0.803000pt}%
\definecolor{currentstroke}{rgb}{0.000000,0.000000,0.000000}%
\pgfsetstrokecolor{currentstroke}%
\pgfsetdash{}{0pt}%
\pgfpathmoveto{\pgfqpoint{1.316223in}{0.950477in}}%
\pgfpathlineto{\pgfqpoint{1.316223in}{6.488415in}}%
\pgfusepath{stroke}%
\end{pgfscope}%
\begin{pgfscope}%
\pgfsetrectcap%
\pgfsetmiterjoin%
\pgfsetlinewidth{0.803000pt}%
\definecolor{currentstroke}{rgb}{0.000000,0.000000,0.000000}%
\pgfsetstrokecolor{currentstroke}%
\pgfsetdash{}{0pt}%
\pgfpathmoveto{\pgfqpoint{13.243083in}{0.950477in}}%
\pgfpathlineto{\pgfqpoint{13.243083in}{6.488415in}}%
\pgfusepath{stroke}%
\end{pgfscope}%
\begin{pgfscope}%
\pgfsetrectcap%
\pgfsetmiterjoin%
\pgfsetlinewidth{0.803000pt}%
\definecolor{currentstroke}{rgb}{0.000000,0.000000,0.000000}%
\pgfsetstrokecolor{currentstroke}%
\pgfsetdash{}{0pt}%
\pgfpathmoveto{\pgfqpoint{1.316223in}{0.950477in}}%
\pgfpathlineto{\pgfqpoint{13.243083in}{0.950477in}}%
\pgfusepath{stroke}%
\end{pgfscope}%
\begin{pgfscope}%
\pgfsetrectcap%
\pgfsetmiterjoin%
\pgfsetlinewidth{0.803000pt}%
\definecolor{currentstroke}{rgb}{0.000000,0.000000,0.000000}%
\pgfsetstrokecolor{currentstroke}%
\pgfsetdash{}{0pt}%
\pgfpathmoveto{\pgfqpoint{1.316223in}{6.488415in}}%
\pgfpathlineto{\pgfqpoint{13.243083in}{6.488415in}}%
\pgfusepath{stroke}%
\end{pgfscope}%
\begin{pgfscope}%
\pgfsetbuttcap%
\pgfsetmiterjoin%
\definecolor{currentfill}{rgb}{1.000000,1.000000,1.000000}%
\pgfsetfillcolor{currentfill}%
\pgfsetfillopacity{0.800000}%
\pgfsetlinewidth{1.003750pt}%
\definecolor{currentstroke}{rgb}{0.800000,0.800000,0.800000}%
\pgfsetstrokecolor{currentstroke}%
\pgfsetstrokeopacity{0.800000}%
\pgfsetdash{}{0pt}%
\pgfpathmoveto{\pgfqpoint{1.607889in}{3.867341in}}%
\pgfpathlineto{\pgfqpoint{6.574721in}{3.867341in}}%
\pgfpathquadraticcurveto{\pgfqpoint{6.658054in}{3.867341in}}{\pgfqpoint{6.658054in}{3.950675in}}%
\pgfpathlineto{\pgfqpoint{6.658054in}{6.196748in}}%
\pgfpathquadraticcurveto{\pgfqpoint{6.658054in}{6.280082in}}{\pgfqpoint{6.574721in}{6.280082in}}%
\pgfpathlineto{\pgfqpoint{1.607889in}{6.280082in}}%
\pgfpathquadraticcurveto{\pgfqpoint{1.524556in}{6.280082in}}{\pgfqpoint{1.524556in}{6.196748in}}%
\pgfpathlineto{\pgfqpoint{1.524556in}{3.950675in}}%
\pgfpathquadraticcurveto{\pgfqpoint{1.524556in}{3.867341in}}{\pgfqpoint{1.607889in}{3.867341in}}%
\pgfpathlineto{\pgfqpoint{1.607889in}{3.867341in}}%
\pgfpathclose%
\pgfusepath{stroke,fill}%
\end{pgfscope}%
\begin{pgfscope}%
\pgfsetrectcap%
\pgfsetroundjoin%
\pgfsetlinewidth{1.505625pt}%
\definecolor{currentstroke}{rgb}{0.000000,0.000000,1.000000}%
\pgfsetstrokecolor{currentstroke}%
\pgfsetdash{}{0pt}%
\pgfpathmoveto{\pgfqpoint{1.691223in}{5.967582in}}%
\pgfpathlineto{\pgfqpoint{2.107889in}{5.967582in}}%
\pgfpathlineto{\pgfqpoint{2.524556in}{5.967582in}}%
\pgfusepath{stroke}%
\end{pgfscope}%
\begin{pgfscope}%
\definecolor{textcolor}{rgb}{0.000000,0.000000,0.000000}%
\pgfsetstrokecolor{textcolor}%
\pgfsetfillcolor{textcolor}%
\pgftext[x=2.857889in,y=5.821748in,left,base]{\color{textcolor}\rmfamily\fontsize{30.000000}{36.000000}\selectfont MP + BP-DTT}%
\end{pgfscope}%
\begin{pgfscope}%
\pgfsetrectcap%
\pgfsetroundjoin%
\pgfsetlinewidth{1.505625pt}%
\definecolor{currentstroke}{rgb}{0.000000,0.500000,0.000000}%
\pgfsetstrokecolor{currentstroke}%
\pgfsetdash{}{0pt}%
\pgfpathmoveto{\pgfqpoint{1.691223in}{5.400368in}}%
\pgfpathlineto{\pgfqpoint{2.107889in}{5.400368in}}%
\pgfpathlineto{\pgfqpoint{2.524556in}{5.400368in}}%
\pgfusepath{stroke}%
\end{pgfscope}%
\begin{pgfscope}%
\definecolor{textcolor}{rgb}{0.000000,0.000000,0.000000}%
\pgfsetstrokecolor{textcolor}%
\pgfsetfillcolor{textcolor}%
\pgftext[x=2.857889in,y=5.254535in,left,base]{\color{textcolor}\rmfamily\fontsize{30.000000}{36.000000}\selectfont SBL + BP-DTT}%
\end{pgfscope}%
\begin{pgfscope}%
\pgfsetrectcap%
\pgfsetroundjoin%
\pgfsetlinewidth{1.505625pt}%
\definecolor{currentstroke}{rgb}{0.750000,0.750000,0.000000}%
\pgfsetstrokecolor{currentstroke}%
\pgfsetdash{}{0pt}%
\pgfpathmoveto{\pgfqpoint{1.691223in}{4.833155in}}%
\pgfpathlineto{\pgfqpoint{2.107889in}{4.833155in}}%
\pgfpathlineto{\pgfqpoint{2.524556in}{4.833155in}}%
\pgfusepath{stroke}%
\end{pgfscope}%
\begin{pgfscope}%
\definecolor{textcolor}{rgb}{0.000000,0.000000,0.000000}%
\pgfsetstrokecolor{textcolor}%
\pgfsetfillcolor{textcolor}%
\pgftext[x=2.857889in,y=4.687321in,left,base]{\color{textcolor}\rmfamily\fontsize{30.000000}{36.000000}\selectfont SOT-TkBD}%
\end{pgfscope}%
\begin{pgfscope}%
\pgfsetbuttcap%
\pgfsetroundjoin%
\pgfsetlinewidth{1.505625pt}%
\definecolor{currentstroke}{rgb}{1.000000,0.000000,0.000000}%
\pgfsetstrokecolor{currentstroke}%
\pgfsetdash{{9.000000pt}{7.500000pt}{0.000000pt}{0.000000pt}}{0.000000pt}%
\pgfpathmoveto{\pgfqpoint{1.691223in}{4.265941in}}%
\pgfpathlineto{\pgfqpoint{2.107889in}{4.265941in}}%
\pgfpathlineto{\pgfqpoint{2.524556in}{4.265941in}}%
\pgfusepath{stroke}%
\end{pgfscope}%
\begin{pgfscope}%
\definecolor{textcolor}{rgb}{0.000000,0.000000,0.000000}%
\pgfsetstrokecolor{textcolor}%
\pgfsetfillcolor{textcolor}%
\pgftext[x=2.857889in,y=4.120108in,left,base]{\color{textcolor}\rmfamily\fontsize{30.000000}{36.000000}\selectfont BP-TkBD (proposed)~~~~~~}%
\end{pgfscope}%
\end{pgfpicture}%
\makeatother%
\endgroup%